\documentclass[
  preprint,
  amsmath,amssymb,
  aps,
  superscriptaddress,
  longbibliography,
  floatfix]{revtex4-2}

\usepackage[T1]{fontenc}
\usepackage{graphicx}
\usepackage{dcolumn}
\usepackage{bm}
\usepackage{hyperref}
\usepackage{xcolor}
\usepackage{xcolor}
\usepackage{upgreek}
\hypersetup{colorlinks=true, linkcolor=blue, citecolor=teal, urlcolor=blue}
\usepackage{enumitem}
\usepackage{caption}
\usepackage{siunitx}

\begin{document}

\title{The 2026 Skyrmionics Roadmap}

\author{Sabri~Koraltan}
\email{sabri.koraltan@tuwien.ac.at}
\affiliation{Institute of Applied Physics, TU Wien, Wiedner Hauptstraße 8-10, Vienna, 1040, Austria}

\author{Claas~Abert}
\affiliation{Physics of Functional Materials, Faculty of Physics, University of Vienna, Vienna, Austria}

\author{Manfred~Albrecht}
\affiliation{Institute of Physics, University of Augsburg, 86135 Augsburg, Germany}

\author{Maria~Azhar}
\affiliation{Faculty of Physics and Center for Nanointegration Duisburg-Essen (CENIDE), University of Duisburg-Essen, 47057 Duisburg, Germany}

\author{Christian~Back}
\affiliation{Department of Physics, School of Natural Sciences, Technical University of Munich, Munich, Germany}
\affiliation{Center for Quantum Engineering (ZQE), Technical University of Munich, Munich, Germany}

\author{Hélène~Béa}
\affiliation{Univ. Grenoble Alpes, CNRS, CEA, SPINTEC, 38054 Grenoble, France}
\affiliation{Institut Universitaire de France (IUF), 75000 Paris, France}

\author{Max~T.~Birch}
\affiliation{RIKEN Center for Emergent Matter Science (CEMS), Wako, Saitama 351-0198, Japan}

\author{Stefan~Bl\"ugel}
\affiliation{Peter Gr\"unberg Institute (PGI-1), Forschungszentrum J\"ulich and JARA, 52425 J\"ulich, Germany}
\affiliation{Institute for Theoretical Physics, RWTH Aachen University, 52056 Aachen, Germany}

\author{Olivier~Boulle}
\affiliation{Univ. Grenoble Alpes, CNRS, CEA, SPINTEC, 38054 Grenoble, France}

\author{Felix~Büttner}
\affiliation{Experimental Physics V, Center for Electronic Correlations and Magnetism, University of Augsburg, 86159 Augsburg, Germany}
\affiliation{Helmholtz-Zentrum Berlin für Materialien und Energie, 12489 Berlin, Germany}

\author{Ping~Che}
\affiliation{Laboratoire Albert Fert, CNRS, Thales, Université Paris-Saclay, 91767 Palaiseau, France}

\author{Vincent~Cros}
\affiliation{Laboratoire Albert Fert, CNRS, Thales, Université Paris-Saclay, 91767 Palaiseau, France}

\author{Emily~Darwin}
\affiliation{Empa, Swiss Federal Laboratories for Materials Science and Technology, Ueberlandstrasse 129, 8600 Dübendorf, Switzerland}

\author{Louise~Desplat}
\affiliation{Univ. Grenoble Alpes, CNRS, CEA, SPINTEC, 38054 Grenoble, France}

\author{Claire~Donnelly}
\affiliation{Max Planck Institute for Chemical Physics of Solids, Nöthnitzer Str. 40, 01187 Dresden, Germany}
\affiliation{International Institute for Sustainability with Knotted Chiral Meta Matter (WPI-SKCM2), Hiroshima University, Hiroshima, 739–8526 Japan}

\author{Haifeng~Du}
\affiliation{Anhui Province Key Laboratory of Low-Energy Quantum Materials and Devices, High Magnetic Field Laboratory, HFIPS, Chinese Academy of Sciences, Hefei 230031, China}

\author{Karin~Everschor-Sitte}
\affiliation{Faculty of Physics and Center for Nanointegration Duisburg-Essen (CENIDE), University of Duisburg-Essen, 47057 Duisburg, Germany}

\author{Amalio~Fernández-Pacheco}
\affiliation{Institute of Applied Physics, TU Wien, Wiedner Hauptstraße 8-10, Vienna, 1040, Austria}

\author{Simone~Finizio}
\affiliation{Paul Scherrer Institut, 5232 Villigen PSI, Switzerland}

\author{Giovanni~Finocchio}
\affiliation{Department of Mathematical and Computer Sciences, Physical Sciences and Earth Sciences, University of Messina, Messina 98166,Italy}

\author{Markus~Garst}
\affiliation{Institute of Theoretical Solid State Physics, Karlsruhe Institute of Technology, 76131 Karlsruhe, Germany}
\affiliation{Institute for Quantum Materials and Technology, Karlsruhe Institute of Technology, 76131 Karlsruhe, Germany}

\author{Raphael~Gruber}
\affiliation{Institute of Physics, Johannes Gutenberg University Mainz, 55099 Mainz, Germany}

\author{Dirk~Grundler}
\affiliation{Laboratory of Nanoscale Magnetic Materials and Magnonics, Institute of Materials (IMX),École Polytechnique F\'ed\'erale de Lausanne (EPFL), 1015 Lausanne, Switzerland}
\affiliation{Institute of Electrical and Micro Engineering (IEM), École Polytechnique F\'ed\'erale de Lausanne (EPFL), 1015 Lausanne, Switzerland}

\author{Satoru~Hayami}
\affiliation{Graduate School of Science, Hokkaido University, Sapporo, Japan}

\author{Thorsten~Hesjedal}
\affiliation{Diamond Light Source, Harwell Science and Innovation Campus, Didcot OX11~0DE, United Kingdom}
\affiliation{Clarendon Laboratory, Department of Physics, University of Oxford, Oxford, OX1~3PU, United Kingdom}

\author{Axel~Hoffmann}
\affiliation{Materials Research Laboratory and Department of Materials Science and Engineering, The Grainger College of Engineering, University of Illinois Urbana-Champaign, Urbana, Illinois 61801, USA}

\author{Ale\v{s}~Hrabec}
\affiliation{Laboratory for Mesoscopic Systems, Department of Materials, ETH Zurich, 8093 Zurich, Switzerland}
\affiliation{Paul Scherrer Institut, 5232 Villigen PSI, Switzerland}

\author{Hans~Josef~Hug}
\affiliation{Empa, Swiss Federal Laboratories for Materials Science and Technology, Ueberlandstrasse 129, 8600 Dübendorf, Switzerland}
\affiliation{Department of Physics, University of Basel, Klingelbergstrasse 82, 4056 Basel, Switzerland}

\author{Hariom~Jani}
\affiliation{Clarendon Laboratory, Department of Physics, University of Oxford, Oxford, OX1~3PU, United Kingdom}

\author{Jagannath~Jena}
\affiliation{Materials Science Division, Argonne National Laboratory, IL, Lemont, USA}
\affiliation{Max Planck Institute of Microstructure Physics, Halle (Saale), Germany}

\author{Wanjun~Jiang}
\affiliation{State Key Laboratory of Low-Dimensional Quantum Physics and Department of Physics, Tsinghua University, Beijing 100084, China}
\affiliation{Frontier Science Center for Quantum Information, Tsinghua University, Beijing 100084, China}

\author{Javier~Junquera}
\affiliation{Departamento de Ciencias de la Tierra y F\'{\i}sica de la Materia Condensada, Universidad de Cantabria, Avenida de los Castros s/n, 39005 Santander, Spain}

\author{Kosuke~Karube}
\affiliation{RIKEN Center for Emergent Matter Science (CEMS), Wako, Saitama 351-0198, Japan}

\author{Lisa-Marie~Kern}
\affiliation{Department of Materials Science and Engineering, Massachusetts Institute of Technology, Cambridge, Massachusetts 02139, USA}

\author{Joo-Von~Kim}
\affiliation{Centre de Nanosciences et de Nanotechnologies, CNRS, Universit{\'e} Paris-Saclay, 91120 Palaiseau, France}

\author{Mathias~Kl\"aui}
\affiliation{Institute of Physics, Johannes Gutenberg University Mainz, 55099 Mainz, Germany}
\affiliation{Center for Quantum Spintronics, Department of Physics, Norwegian University of Science and Technology,
7491 Trondheim, Norway}

\author{Hidekazu~Kurebayashi}
\affiliation{Department of Electronic and Electrical Engineering, University College London, London, WC1E 7JE, UK}
\affiliation{London Centre for Nanotechnology, University College London, 17-19 Gordon
Street, London, WC1H 0AH, United Kingdom}
\affiliation{WPI Advanced Institute for Materials Research, Tohoku University, 2-1-1,
Katahira, Sendai 980-8577, Japan}
\affiliation{Center for Science and Innovation in Spintronics, Tohoku University, 2-1-1,
Katahira, Sendai, 980-8577 Japan}

\author{Kai~Litzius}
\affiliation{Experimental Physics V, Center for Electronic Correlations and Magnetism, University of Augsburg, 86159 Augsburg, Germany}

\author{Yizhou~Liu}
\affiliation{Anhui Province Key Laboratory of Low-Energy Quantum Materials and Devices, High Magnetic Field Laboratory, HFIPS, Chinese Academy of Sciences, Hefei 230031, China}

\author{Martin~Lonsky}
\affiliation{Institute of Physics, Goethe University Frankfurt, 60438 Frankfurt, Germany}

\author{Christopher~H.~Marrows}
\affiliation{School of Physics \& Astronomy, University of Leeds, Leeds LS2 9JT, United Kingdom}

\author{Jan~Masell}
\affiliation{Institute of Theoretical Solid State Physics, Karlsruhe Institute of Technology, 76131 Karlsruhe, Germany}
\affiliation{RIKEN Center for Emergent Matter Science (CEMS), Wako, Saitama 351-0198, Japan}

\author{Stefan~Mathias}
\affiliation{$\rm 1^{st}$ Institute of Physics, University of Göttingen, 37077 Göttingen, Germany}
\affiliation{International Center for Advanced Studies of Energy Conversion (ICASEC), University of Göttingen, 37077 Göttingen, Germany}

\author{Yuriy~Mokrousov}
\affiliation{Peter Gr\"unberg Institute (PGI-1), Forschungszentrum J\"ulich and JARA, 52425 J\"ulich, Germany}
\affiliation{Institute of Physics, Johannes Gutenberg University Mainz, 55099 Mainz, Germany}

\author{Stuart~S.~P.~Parkin}
\affiliation{Max Planck Institute of Microstructure Physics, Halle (Saale), Germany}

\author{Bastian~Pfau}
\affiliation{Max Born Institute for Nonlinear Optics and Short Pulse Spectroscopy, 12489 Berlin, Germany}

\author{Paolo~G.~Radaelli}
\affiliation{Clarendon Laboratory, Department of Physics, University of Oxford, Oxford, OX1~3PU, United Kingdom}

\author{Florin~Radu}
\affiliation{Helmholtz-Zentrum Berlin für Materialien und Energie, 12489 Berlin, Germany}

\author{Ramamoorthy~Ramesh}
\affiliation{Department of Physics and Department of Materials Science \& Engineering, University of California, Berkeley, CA 94720}

\author{Nicolas~Reyren}
\affiliation{Laboratoire Albert Fert, CNRS, Thales, Université Paris-Saclay, 91767 Palaiseau, France}

\author{Stanislas~Rohart}
\affiliation{Université Paris-Saclay, CNRS, Laboratoire de Physique des Solides, 91405 Orsay, France}

\author{Shinichiro~Seki}
\affiliation{Department of Applied Physics, University of Tokyo, Bunkyo, Tokyo, Japan}
\affiliation{Research Center for Advanced Science and Technology, University of Tokyo, Tokyo, Japan}

\author{Ivan~I.~Smalyukh}
\affiliation{Department of Physics and Chemical Physics Program, University of Colorado, Boulder, CO, USA}
\affiliation{Department of Electrical, Computer, and Energy Engineering, Materials Science and Engineering Program and Soft Materials Research Center, University of Colorado, Boulder, CO, USA}
\affiliation{Renewable and Sustainable Energy Institute, National Renewable Energy Laboratory and University of Colorado, Boulder, CO, USA}
\affiliation{International Institute for Sustainability with Knotted Chiral Meta Matter (WPI-SKCM2), Hiroshima University, Higashihiroshima, Japan}

\author{Sopheak~Sorn}
\affiliation{Institute of Theoretical Solid State Physics, Karlsruhe Institute of Technology, 76131 Karlsruhe, Germany}
\affiliation{Institute for Quantum Materials and Technology, Karlsruhe Institute of Technology, 76131 Karlsruhe, Germany}

\author{Daniel~Steil}
\affiliation{$\rm 1^{st}$ Institute of Physics, University of Göttingen, 37077 Göttingen, Germany}

\author{Dieter~Suess}
\affiliation{Physics of Functional Materials, Faculty of Physics, University of Vienna, Vienna, Austria}

\author{Mykola~Tasinkevych}
\affiliation{International Institute for Sustainability with Knotted Chiral Meta Matter (WPI-SKCM2), Hiroshima University, Higashihiroshima, Japan}
\affiliation{SOFT Group, School of Science and Technology, Nottingham Trent University, Clifton Lane, Nottingham NG11 8NS, United Kingdom}

\author{Yoshinori~Tokura}
\affiliation{RIKEN Center for Emergent Matter Science (CEMS), Wako, Saitama 351-0198, Japan}
\affiliation{Department of Applied Physics, The University of Tokyo, Bunkyo-ku, Tokyo 113-8656, Japan}
\affiliation{Tokyo College, The University of Tokyo, Bunkyo-ku, Tokyo 113-8656, Japan}

\author{Riccardo~Tomasello}
\affiliation{Department of Electrical and Information Engineering,
Politecnico di Bari, 70125 Bari, Italy}

\author{Victor~Ukleev}
\affiliation{Helmholtz-Zentrum Berlin für Materialien und Energie, 12489 Berlin, Germany}

\author{Hyunsoo~Yang}
\affiliation{Department of Electrical and Computer Engineering, National University of Singapore, Singapore 117576, Singapore}

\author{Fehmi~Sami~Yasin}
\affiliation{Center for Nanophase Materials Sciences, Oak Ridge National Laboratory, Oak Ridge, Tennessee 37830, United States}

\author{Xiuzhen~Yu}
\affiliation{RIKEN Center for Emergent Matter Science (CEMS), Wako, Saitama 351-0198, Japan}

\author{Chenhui~Zhang}
\affiliation{Department of Electrical and Computer Engineering, National University of Singapore, Singapore 117576, Singapore}

\author{Shilei~Zhang}
\affiliation{School of Physical Science and Technology, ShanghaiTech University, Shanghai 201210, China}
\affiliation{ShanghaiTech Laboratory for Topological Physics, ShanghaiTech University, Shanghai 201210, China}
\affiliation{Center for Transformative Science, ShanghaiTech University,  Shanghai 201210, China}

\author{Le~Zhao}
\affiliation{Institute of Applied Physics, TU Wien, Wiedner Hauptstraße 8-10, Vienna, 1040, Austria}

\author{Sebastian~Wintz}
\email{sebastian.wintz@helmholtz-berlin.de}
\affiliation{Helmholtz-Zentrum Berlin für Materialien und Energie, 12489 Berlin, Germany}

\date{\today}

\begin{abstract}
Magnetic skyrmions and related topological spin textures have emerged as a central topic in condensed-matter physics, combining fundamental significance with potential for transformative applications in spintronics, magnonics, and beyond. Over the past decade, advances in material platforms, imaging techniques, theoretical modeling, and device concepts have established skyrmionics as a rapidly expanding field. At the same time, challenges remain in stabilizing, controlling, and integrating such textures into functional architectures, while novel phenomena such as antiskyrmions, higher-order skyrmions, hopfions, and antiferromagnetic textures arise. The 2026 Skyrmionics Roadmap represents a collective effort of many authors, providing a comprehensive perspective on the current state-of-the-art and the outlook for the coming years. In 33 focused sections, each co-authored by two researchers, we chart progress in theory and modeling, material systems, skyrmion dynamics, and skyrmion technologies. By offering a consolidated vision, this Roadmap aims to guide both fundamental research and application-driven efforts, accelerating the transition of skyrmionics from conceptual breakthroughs toward practical technologies.
\end{abstract}

\maketitle



\newpage

\section{Introduction}
\begingroup
    \let\section\subsection
    \let\subsection\subsubsection
    \let\subsubsection\paragraph
    \let\paragraph\subparagraph
Sabri Koraltan$^1$, and Sebastian Wintz $^2$
\vspace{0.5cm}

\noindent
\textit{$^1$ Institute of Applied Physics, TU Wien, Wiedner Hauptstraße 8-10, Vienna, 1040, Austria\\
$^2$ Helmholtz-Zentrum Berlin für Materialien und Energie, D-12489 Berlin, Germany}\\

\vspace{2cm}

In its original sense, the term \emph{Skyrmion} was not related to concepts in condensed matter physics. Instead, in the early 1960s, it referred to nuclei and baryons represented as collective excitations of pionic degrees of freedom in the so-called \emph{Skyrme} model, named after its founder, T.H.R. Skyrme~\cite{skyrme1962unified}. In this context, nuclei were modeled as topological solitons. Generally, a soliton describes a wave packet that propagates with finite velocity and maintains its shape during this process. A soliton becomes topological when a uniform and continuous force is not sufficient to deform it~\cite{manton2004topological}. Similarly, localized magnetic excitations can be referred to as magnetic solitons. Despite earlier concepts of magnetic bubbles investigated both theoretically and experimentally, especially in the context of magnetic bubble memory, it was Bogdanov and Yablonskii~\cite{Bogdanov1989} who in 1989 first predicted theoretically that \emph{vortex-like textures} are thermodynamically stable in single-crystal magnets with crystal symmetries $C_{nv}, D_n, D_{2d}$ and $S_4$, where $n = 3,4,6$. These local excitations were determined to behave as topological solitons. It is noteworthy that the term \textit{skyrmion} was mentioned in magnetic systems already by Pokrovsky in 1979~\cite{pokrovsky1979properties}. However, only around the mid-2000s, the condensed matter community began to refer to these magnetic spin textures as skyrmions, with the seminal work of R\"oßler, Bogdanov, and Pfleiderer explicitly introducing the term in the context of chiral magnets~\cite{roessler2006spontaneous}. Today, we define a magnetic skyrmion as a localized, topologically protected spin texture, which is characterized by an integer topological charge $Q$:

\begin{equation}
    Q = \dfrac{1}{4\pi}\iint \mathbf{m} \cdot 
    \left( \frac{\partial \mathbf{m}}{\partial x} \times 
    \frac{\partial \mathbf{m}}{\partial y} \right)\, dx\, dy,
\end{equation}

where $\mathbf{m}(x,y)$ is the normalized magnetization vector field with $|\mathbf{m}| = 1$, and $\frac{\partial \mathbf{m}}{\partial x}, \frac{\partial \mathbf{m}}{\partial y}$ denote the derivatives with respect to the in-plane coordinates. In a skyrmion, the orientation of spins maps the two-dimensional real space onto the unit sphere of spin directions. For a skyrmion with topological charge $Q = \pm 1$, the spin field covers the sphere exactly once.

Beyond magnetism, skyrmionic textures have also been discovered in nonmagnetic systems. For example, liquid crystals can host particle-like skyrmions stabilized by elastic interactions and confinement effects~\cite{smalyukh2010threedimensional}. In ferroic materials, polar skyrmions have been realized in ferroelectric superlattices, where the polarization field forms skyrmion-like whirls~\cite{Das2019}. These findings underscore that skyrmions represent a universal class of topological structures, which extend well beyond magnetism. Nevertheless, it is within magnetic materials that skyrmions were first experimentally demonstrated and where their study has since flourished.

The first experimental demonstration of a magnetic skyrmion was obtained by M\"uhlbauer et al.~\cite{muehlbauer2009skyrmion} in 2009, where a skyrmion lattice was observed in the chiral B20 crystal MnSi by means of neutron scattering. The first real-space observation of a skyrmion lattice was realized by Yu et al.~\cite{yu2010real}, who imaged a two-dimensional hexagonal skyrmion lattice using Lorentz transmission electron microscopy (LTEM) in $\rm Fe_{0.5}Co_{0.5}Si$. As later classified, these spin textures contain only Bloch walls and can, in principle, rotate either clockwise or anticlockwise. Today, we know them as \emph{Bloch skyrmions.} The term \emph{chiral skyrmion lattice} was introduced because the skyrmions reported by Yu et al.~\cite{yu2010real} in $\rm Fe_{0.5}Co_{0.5}Si$ show only an anticlockwise sense of rotation, as revealed by induction maps. Thus, the skyrmions possess the same chirality, and the lattice is homochiral. The energy term that is responsible for the stabilization of the skyrmions and the homochirality is called the Dzyaloshinskii-Moriya interaction (DMI).

Later, multiple systems were shown to host skyrmions. One of the early pioneering works by Heinze et al.~\cite{heinze2011spontaneous} demonstrated that a skyrmion lattice can be stabilized in monolayers of Fe grown on the Ir [111] surface at low temperatures and high magnetic fields. Unlike the Bloch skyrmions, their spin configuration formed \emph{Néel skyrmions}, where the magnetization follows a radial distribution. Another key finding was that when soft magnetic thin films are grown on materials with strong spin–orbit coupling, the heavy metal layer induces interfacial DMI. Jiang et. al~\cite{jiang2015blowing} harnessed this effect to generate skyrmions in heavy metal/ferromagnet bilayers. Later, this mechanism was further employed by Boulle et al.~\cite{boulle2016room} and Moreau-Louchaire et al.~\cite{moreau-luchaire2016additive} to realize magnetic multilayers consisting of Pt/Co/MgO and Ir/Co/Pt layers, respectively. Today, Néel skyrmions can be stabilized using this strategy in a wide variety of materials ranging from Ta/CoFeB/MgO~\cite{qin2018stabilization}, Pt/Co/Ta~\cite{zhang2018creation, wang2019construction}, to W/CoFeB/MgO~\cite{jaiswal2017investigation}, among others.

All magnetic systems mentioned above make use of broken inversion symmetry due to the presence of DMI, which, in combination with exchange and dipolar interactions and sometimes additional perpendicular magnetic anisotropy, leads to the stabilization of chiral skyrmions. Following the pioneering work of Montoya et al.~\cite{montoya2017tailoring}, it was shown that Bloch skyrmions with diameters ranging from a few tens to hundreds of nanometers can be stabilized purely by the competition between weak perpendicular anisotropy and dipolar interactions. However, the homochirality introduced by DMI is lost, and Bloch skyrmions with both senses of rotation were observed. This established that DMI is not the only stabilization mechanism for skyrmionic textures. In particular, recent theoretical and experimental studies show that magnetic frustration can also stabilize skyrmions~\cite{Okubo_PhysRevLett.108.017206, kurumaji2019skyrmion}.

The fact that skyrmions can be investigated across different material platforms using a wide variety of scientific instrumentation makes them a popular focus of research in nanomagnetism. Skyrmions can be imaged directly by magneto–optical Kerr microscopy in thin films \cite{jiang2015blowing}, scanning probe methods such as spin–polarized scanning tunneling microscopy \cite{heinze2011spontaneous} and magnetic force microscopy \cite{milde2013unwinding}, X-ray–based microscopies including (scanning) transmission X-ray microscopy [(S)TXM] \cite{woo2018current}, XMCD–PEEM \cite{boulle2016room}, coherent X-ray imaging (ptychography~\cite{li2019anatomy}, holography~\cite{turnbull2020tilted, grelier2023xray}
and X-ray magnetic tomography \cite{donnelly2017three, donnelly2016high}). Electron microscopy approaches provide complementary high-fidelity real-space views via Lorentz TEM \cite{yu2010real} and off-axis electron holography or DPC-STEM phase imaging \cite{mcvitie2018transmission, denneulin2021off}. In addition to these real-space probes, reciprocal-space scattering techniques offer bulk and element-specific sensitivity: small-angle neutron scattering reveals skyrmion-lattice order in chiral magnets \cite{muehlbauer2009skyrmion}, while resonant elastic soft X-ray scattering provides element-resolved diffraction signatures of skyrmion lattices \cite{langner2014coupled}. Taken together, this toolbox of real-space imaging and scattering methods enables a comprehensive \emph{static} characterization of skyrmions across a wide range of materials and geometries.

Beyond imaging, skyrmions can also be detected via their characteristic electrical signatures, which is essential for technological applications. The most prominent example is the \emph{topological Hall effect} (THE)~\cite{nagaosa2013topological}, arising from the real-space Berry curvature of the skyrmion spin texture, which is often observed as an additional contribution to the anomalous Hall effect (AHE) \cite{neubauer2009topological}. Thermoelectric counterparts such as the anomalous Nernst effect (ANE) have likewise been exploited to detect skyrmions \cite{shiomi2013topological}. In multilayer devices, skyrmions can be read out by tunneling magnetoresistance (TMR) effect in magnetic tunnel junctions (MTJs) \cite{koshibae2015memory, penthorn2019experimental, guang2023electrical}, where the skyrmion core modifies the local relative alignment of magnetic layers. These transport-based signatures provide scalable and integrable routes for skyrmion detection, a critical prerequisite for their use in information technologies.

In that respect, a further main motivation to study skyrmions and skyrmionic materials lies in their potential usage in spintronic devices. Three main applications have been in focus. First, following the initial suggestion of Parkin et al.~\cite{parkin2008magnetic} to use magnetic domain walls for a mobile 3D racetrack memory, the idea of skyrmion racetrack memory was proposed by Fert et al.~\cite{fert2013skyrmions}. The basic concept for storing data is that topologically robust skyrmions can encode the bit \texttt{1}, while their absence encodes the bit \texttt{0}. When the tracks are subjected to a charge current, this current gains a spin polarization and exerts a torque on the skyrmions, known as the spin-transfer torque (STT). The STT sets the skyrmions into motion, and the \textit{data} travel along the track in the sense of a shift register~\cite{zhao2024realization}. While this concept has not been demonstrated beyond prototypical realizations yet, it laid the foundation of one of the main attractions behind skyrmions. However, a major challenge is that a lateral current along the track leads to a transversal deflection of skyrmions~\cite{woo2018current}. Due to its analogy to the transverse voltage in the Hall effect, this phenomenon is known as the skyrmion Hall effect (SkHE)~\cite{litzius2017skyrmion, jiang2017direct}. Later, it was recognized that in the aforementioned systems, the vertically injected spin current is the main driving mechanism that sets the skyrmions into motion via spin–orbit torque (SOT). In practice, it is very challenging to distinguish between the two torques. A key differentiation is the direction of motion of te skyrmions. Nevertheless, both torques can, in principle, be exploited for steady skyrmion motion~\cite{tomasello2014strategy}. Several strategies have been proposed and experimentally realized to suppress or eliminate the SkHE, including the use of ferrimagnetic skyrmions~\cite{woo2018current}, antiferromagnetic skyrmions~\cite{legrand2020room}, and skyrmions stabilized in synthetic antiferromagnetic multilayer tracks~\cite{dohi2019formation, pham2024fast}.

\begin{figure}
\centering
\includegraphics[width=0.95\linewidth]{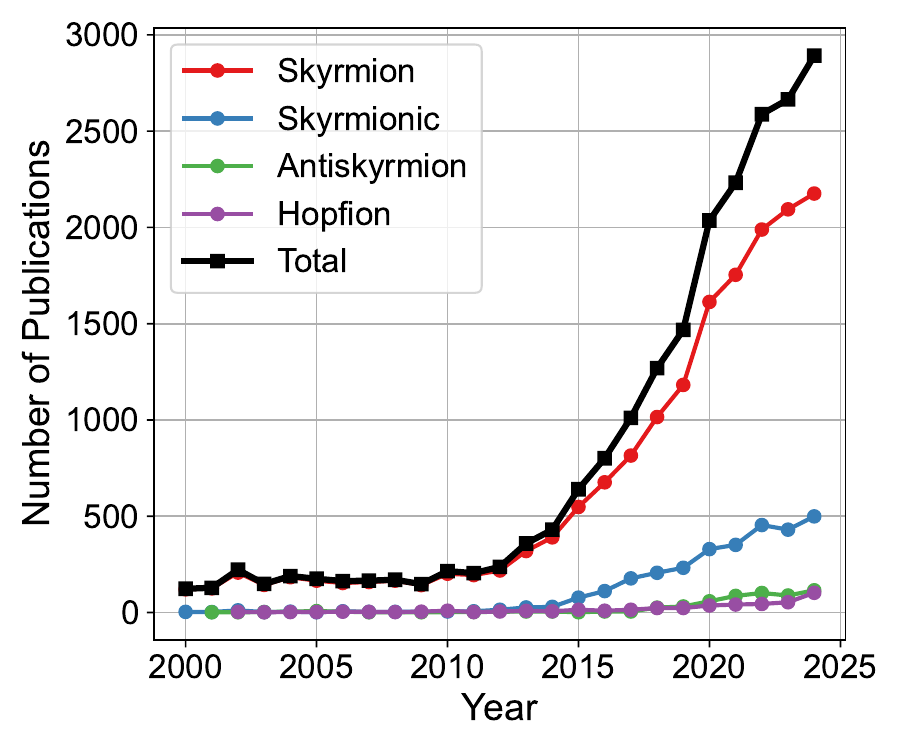}
\caption{Number of publications for each year between 2000 and 2024, where the keywords \textit{Skyrmion} (red), \textit{Skyrmionic} (blue), \textit{Antiskyrmion} (green), \textit{Hopfion} (purple) appear in the abstract, title, keywords or manuscript according to Scopus. The sum of all data is shown with the black curve as \textit{Total}.}
\label{fig:intro1}
\end{figure}

A second highly investigated application is the use of skyrmions in microwave devices~\cite{finocchio2015skyrmion}. Skyrmions can breathe and gyrate at frequencies in the MHz and GHz range in confined nanodisks. Their GHz dynamics open new possibilities for integration in filtering and microwave technologies~\cite{finocchio2015skyrmion}.

The third and more recent application is the use of skyrmions in unconventional computing devices~\cite{song2020skyrmionbased, pinna2020reservoir, lee2024taskadaptive, lee2023handwritten, dacamarasantaclaragomes2025neuromorphic}. Several concepts have been demonstrated theoretically and experimentally. The simplest one is to use a skyrmionic material as a reservoir fabric. When excited by microwave fields, a unique non-linear signal emerges which can be exploited for high-dimensional mapping to perform unconventional computing tasks such as digit recognition and sound classification.

Although early research on skyrmion technology (skyrmionics) centered mainly on ferromagnetic skyrmions, the field is rapidly flourishing. New concepts for stabilization of skyrmions, reduction of the SkHE, and interdisciplinary usage are being proposed across novel material systems. One key aspect is that the field of skyrmionics no longer deals exclusively with classical skyrmions. A diverse family of topological spin textures beyond fundamental skyrmions has gained significant attention in the past decade, especially since the discovery of antiskyrmions by Nayak et al.~\cite{nayak2017magnetic} in 2017 using noncentrosymmetric Heusler compounds with $D_{\rm 2d}$ symmetry. Antiskyrmions are the topological counterparts of skyrmions, carrying an opposite but integer topological charge. Later, Jena et al.~\cite{jena2020elliptical} showed that elliptical skyrmions can coexist with antiskyrmions in the same material, opening new concepts such as the use of multiple skyrmionic texture memory, where skyrmions encode bit \texttt{0} and antiskyrmions encode bit \texttt{1}~\cite{hoffmann2021skyrmion}. Antiskyrmions were initially thought to be unique to systems with anisotropic DMI, until Heigl et al.~\cite{heigl2021dipolar} demonstrated that ferrimagnetic multilayers can host a broad family of spin textures with different topological charges, stabilized purely by dipolar interactions. Together with theoretical and experimental progress on hopfions~\cite{battye1998knots, kent2021creation, rybakov2022magnetic, zheng2023hopfion}, this enrichment is giving the field new momentum, as it can be seen from the number of publications shown in Fig.~\ref{fig:intro1}. Although the number of publications mentioning the term skyrmion(s) continues to rise, the growth rate has slowed in recent years, suggesting that it is approaching a plateau. Further related keywords such as antiskyrmions, skyrmionic, or hopfion are increasing in a way that the total number of publications remains rather monotonically increasing.

While studies on skyrmions and skyrmionic materials are rapidly progressing, the flourishing of the field also brings new directions and further challenges. In this roadmap article, we bring together perspectives from the skyrmionics community to outline topical relevance, future directions, present challenges and envisioned applications. The roadmap is structured into four basic thematic sections, each highlighting a major frontier in skyrmionics: Theory and modeling, material systems, skyrmion dynamics, and skyrmion technology, as further discussed below. The main contents of each section are summarized in Fig.~\ref{fig:intro2}, which gives an overview of the roadmap.

\begin{figure*}
\centering
\includegraphics[width=0.75\linewidth]{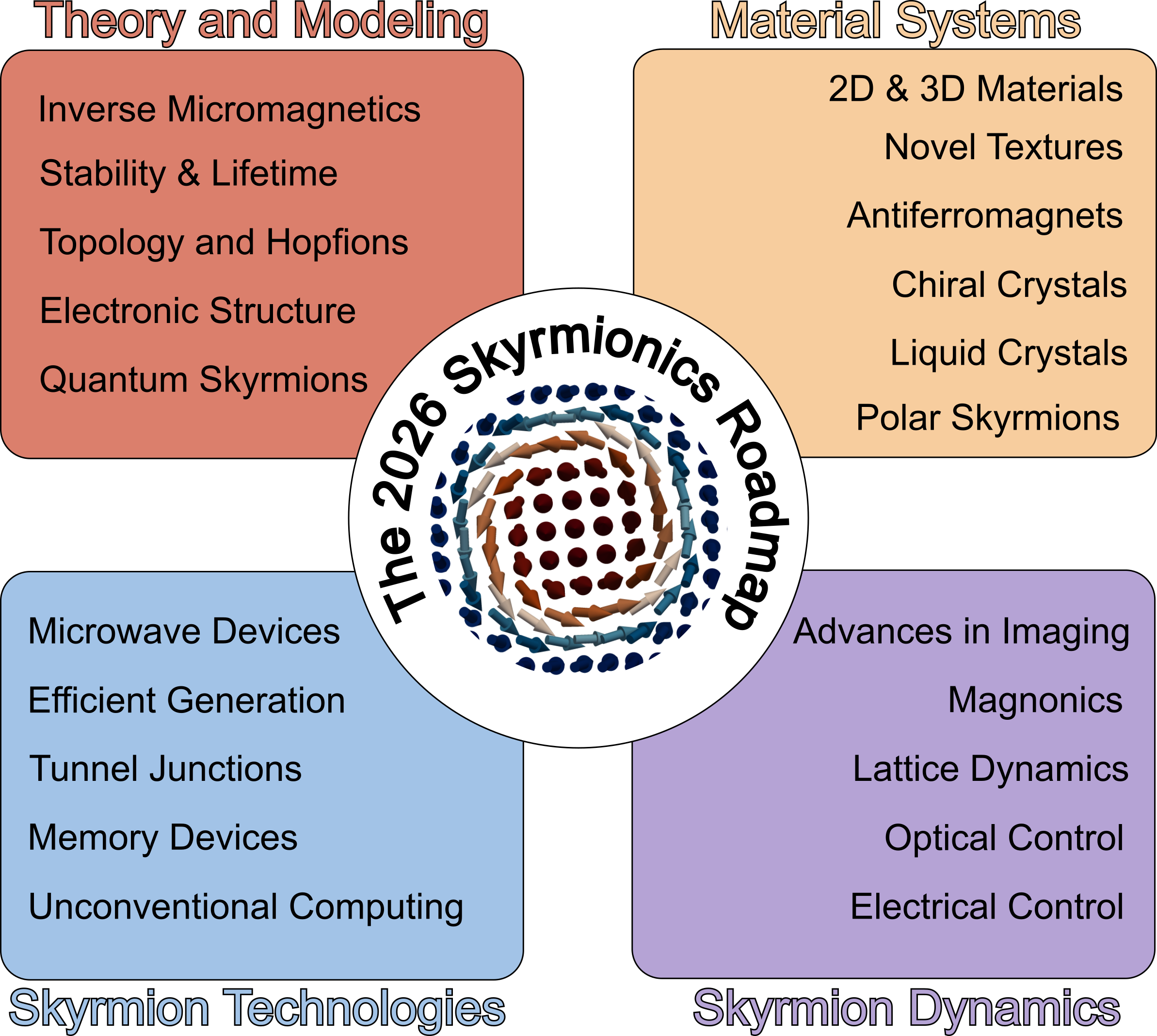}
\caption{The four main thematic sections addressed in this roadmap and the topical keywords discussed by the individual sections.}
\label{fig:intro2}
\end{figure*}

\paragraph{Theory and Modeling.}
This part outlines the latest advances in theoretical descriptions and computational tools for skyrmions and related textures. It covers analytical frameworks rooted in topology and micromagnetics, as well as emerging methods such as inverse modeling and machine-learning–based simulation. Particular emphasis is placed on the challenges of accurately capturing multiscale behavior, quantifying stability, and predicting novel textures beyond classical skyrmions. Together, these efforts provide the conceptual backbone for interpreting experiments and guiding materials design.

\paragraph{Material Systems.}
This section explores the platforms where skyrmions and related spin textures can be realized. It spans bulk chiral magnets, interfacial multilayers, centrosymmetric frustrated magnets, ferrimagnetic and antiferromagnetic systems, as well as emerging 3D and curvilinear geometries. Special attention is paid to the mechanisms of stabilization, the role of defects and interfaces, and opportunities for materials engineering. These advances broaden the range of conditions under which skyrmions can exist and point to pathways for discovering robust, application-ready systems.

\paragraph{Skyrmion Dynamics.}
Here, the focus shifts to how skyrmions behave under external stimuli such as electric currents, magnetic fields, strain, and ultrafast optical pulses. The section reviews progress in understanding current-driven motion, gyrotropic oscillations, collective interactions, and the suppression of the skyrmion Hall effect. Beyond quasi-static control, it highlights the ultrafast and nonlinear dynamics of skyrmions, and the open challenge of achieving reliable, low-power manipulation in realistic device architectures.

\paragraph{Skyrmionic Technologies.}
Finally, this section discusses the prospects of skyrmions as functional units in spintronic devices. It revisits memory concepts such as racetrack devices, considers skyrmion-based microwave oscillators and filters, and expands toward unconventional computing paradigms including reservoir and neuromorphic computing. Crucially, it addresses how skyrmions can be detected and interfaced electrically, ensuring compatibility with existing semiconductor technologies. The section closes with a perspective on the translational challenges---from reproducible nucleation and annihilation, to scalability and integration into CMOS environments---that must be overcome to bridge fundamental physics and technological impact.

\textit{Vienna,  Berlin, September 2025}

\endgroup

\newpage

\section{Inverse micromagnetics for material-parameter extraction and optimal design of skyrmionic devices}
\begingroup
    \let\section\subsection
    \let\subsection\subsubsection
    \let\subsubsection\paragraph
    \let\paragraph\subparagraph
Claas Abert$^1$, and Dieter Suess $^1$
\vspace{0.5cm}

\textit{$^1$ Physics of Functional Materials, Faculty of Physics, University of Vienna, Vienna, Austria}

\section*{Introduction}
The field of magnetic skyrmions has attracted considerable interest, not only because of their rich topological properties \cite{roessler2006spontaneous}, but also due to their accessibility to existing experimental and computational methods. Thin films that host skyrmions can be reliably fabricated using established techniques such as sputter deposition \cite{moreau-luchaire2016additive}. Once formed, skyrmions can be directly observed with widely available imaging tools like Lorentz Transmission Electron Microscopy (LTEM) \cite{yu2010real} or Magnetic Force Microscopy (MFM) \cite{bacani2019how}. In addition, their behavior can be realistically modeled using standard micromagnetic simulation software which is a cornerstone of computational skyrmionics, enabling the study of skyrmion stability, dynamics, and device functionality by means of solving the Landau-Lifshitz-Gilbert (LLG) equation\cite{abert2019micromagnetics}.
Several high-performance micromagnetic simulation codes have been developed to address the numerical challenges inherent to the Landau-Lifshitz-Gilbert (LLG) equation. Notable open-source frameworks include OOMMF~\cite{donahue1999oommf} and MuMax3~\cite{vansteenkiste2014design}, which offer efficient GPU and CPU-based solvers for micromagnetic problems. More recent developments have leveraged machine learning toolkits like PyTorch and JAX to enable efficient differentiation and GPU acceleration, as exemplified by the micromagnetic simulation libraries magnum.np\cite{bruckner2023magnum} and NeuralMag\cite{abert2025neuralmag}.

Simulating skyrmionic textures can be achieved with any standard micromagnetic simulation tool by implementing an antisymmetric exchange term like the DMI into the effective field. However, careful consideration is required when selecting an appropriate discretization size. In classical micromagnetic simulations without antisymmetric exchange, the exchange length $L_\text{ex} = \sqrt{\frac{A}{K}}$, where $A$ is the exchange stiffness and $K$ is the uniaxial anisotropy constant, typically offers a reliable estimate of the equilibrium domain-wall width, which can serve as a guideline for discretization. Nevertheless, systems with strong Dzyaloshinskii--Moriya interaction (DMI), characterized by the constant \( D \), demand a finer discretization.
 Specifically, the DMI introduces a characteristic length scale $L_D = 2A / |D|$~\cite{rohart2013skyrmion}, which may be significantly smaller than $L_\text{ex}$ and dictates the skyrmion or domain wall width. Failure to resolve $L_D$ can result in notable artifacts, such as inaccurate skyrmion energies and artificial pinning effects. Another pitfall in the simulation of the annihilation or generation of skyrmions due to applied fields or temperature is the change in the topological charge, which means that no smooth transition between the two states is possible. As a consequence, the critical fields or energy barriers become mesh-dependent \cite{suess2019spin}.  

Recently, the solution of inverse problems in the micromagnetic domain has garnered significant interest \cite{bruckner2017large, wang2021inverse, papp2021nanoscale, voronov2025inverse}. In this section, we will introduce the theoretical framework of inverse problems and discuss two highly promising application areas in skyrmionics: the physics-informed reconstruction of skyrmions from magnetic imaging data and the design and optimization of skyrmionic devices.

\section*{Numerical Methods and Micromagnetic Software\label{sec:numerics}}
Forward micromagnetic simulations solve the Landau–Lifshitz–Gilbert (LLG) equation to compute magnetization trajectories and equilibrium states for given material parameters, geometric configurations, and excitations. These simulations provide predictive insights into skyrmion stability, dynamics, and interaction with device structures.

In inverse modeling, the paradigm is reversed. Instead of predicting magnetization states from known inputs, the objective is to infer unknown system parameters from observed magnetic behavior. This includes material properties (e.g., anisotropy, exchange, or DMI constants), initial magnetization configurations, or even geometric features. Given some experimental or synthetic observation $d_\text{obs}$, the inverse problem can be formalized as identifying parameters $p$ such that the forward model $F$ reproduces the observed data:
\begin{equation}
    d_\text{obs} = F(p).
\end{equation}
This formulation appears naturally in measurement-driven magnetization reconstruction as well as in optimal device design.

Inverse problems are typically ill-posed: solutions may not exist, may not be unique, or may not depend continuously on the data. This challenge is further exacerbated by the high dimensionality of the parameter space in realistic micromagnetic systems. To address these difficulties, inverse problems are commonly reformulated as optimization tasks:
\begin{equation}
    \min_p \mathcal{J}(p) = \| F(p) - d_\text{obs} \|^2 + \mathcal{R}(p),
\end{equation}
where $\mathcal{R}(p)$ denotes a regularization term that incorporates prior knowledge or enforces physical constraints to improve well-posedness and solution stability.

The solution of such optimization problems requires the evaluation of gradients of the cost functional with respect to the model parameters. Efficient and accurate gradient computation is essential for the scalability of inverse modeling, particularly in high-dimensional or time-dependent settings. Building on the foundation of the adjoint method and backpropagation, recent developments have adopted modern machine learning frameworks to construct fully differentiable micromagnetic solvers. Codes such as magnum.np, NeuralMag, and SpinTorch embed micromagnetic models into autodiff-capable environments like PyTorch and JAX. This enables the use of backpropagation and automatic differentiation to compute gradients through the full simulation pipeline. These frameworks facilitate the application of gradient-based optimization to problems such as inverse design, parameter reconstruction, and magnetization imaging. For instance, SpinTorch~\cite{papp2021nanoscale} has been applied to the inverse design of spin-wave-based logic using differentiable dynamics, illustrating the power of this approach for functional device development. Similar strategies are increasingly being extended to skyrmionics and other topological magnetic systems.

The resulting workflow supports flexible formulations, including temporal regularization, spatial priors, or physical constraints, and can be run on GPU-accelerated hardware to scale to large geometries or high temporal resolution.

\section*{Inverse Micromagnetic Imaging and Material Characterization}
Spatially resolved magnetic imaging techniques such as magnetic force microscopy (MFM), nitrogen-vacancy (NV) magnetometry, and X-ray magnetic circular dichroism (XMCD) provide valuable information about magnetization distributions at the nanoscale. However, these methods only yield partial or projected information: MFM and NV magnetometry primarily measure components of the stray field, while XMCD-PEEM and STXM produce projections of the magnetization vector along fixed axes. As a result, the full three-dimensional magnetization texture—especially in complex chiral structures such as skyrmions—cannot be directly inferred from raw measurement data alone.

To overcome this limitation, physics-informed inverse approaches offer a promising direction. By incorporating prior knowledge about the micromagnetic behavior of the system into the reconstruction process, these methods make it possible to infer the underlying magnetization configuration in a way that is both consistent with the observed data and physically realistic.

\begin{figure}[h!]
    \centering
    \includegraphics[width=0.95\linewidth]{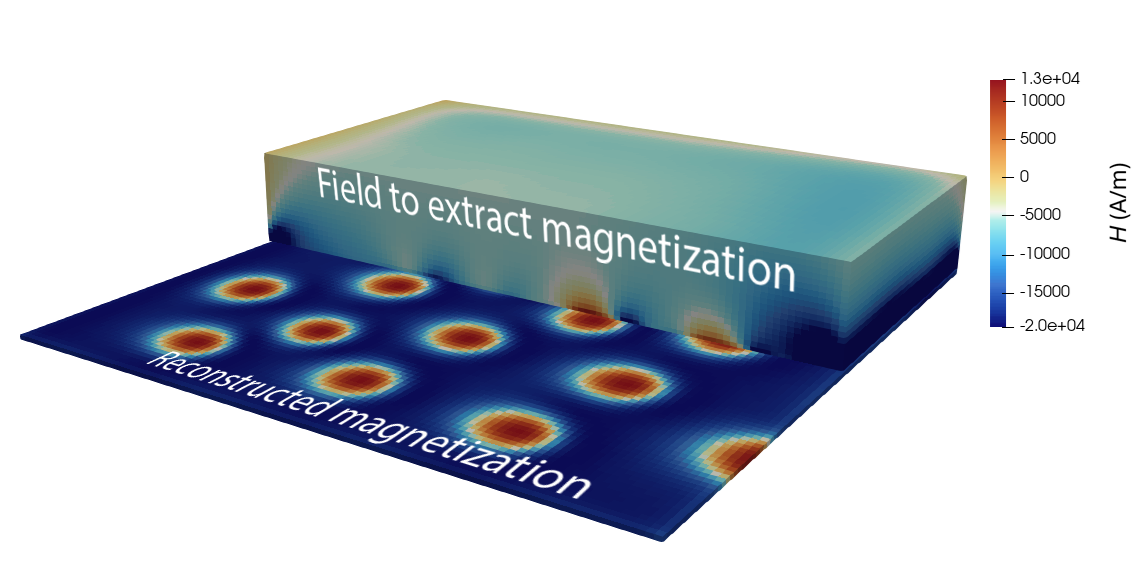}
    \caption{Reconstruction of a skyrmion structures from magnetic stray-field data. The top volume illustrates the measured vertical stray field component $H_z$, which is used as input to an inverse problem. The lower plane shows the reconstructed in-plane magnetization distribution obtained via a physics-informed optimization process. The inverse solution accurately recovers the characteristic circular structure of Néel-type skyrmions.}
    \label{fig:abert}
\end{figure}

A key enabling technique in this context is the adjoint method. In earlier work \cite{bruckner2017large}, the inverse problem of reconstructing the magnetization from stray-field measurements was solved by formulating a PDE-constrained optimization problem, where the magnetization configuration $\mathbf{m}$ is treated as the control variable. Building on this foundation, more general inverse formulations have emerged that incorporate additional micromagnetic constraints into the objective. For example, by solving for a magnetization conﬁguration $\mathbf{m}$ that generates
the measured signal and at the same time minimizes the Gibbs free energy of the
magnetic system $E(\mathbf{m})$ it is possible to restrict the solution space to
energetically stable states, such as static skyrmion textures.
A suitable objective functional in such a physics-informed reconstruction framework takes
the form
\begin{equation}
\mathcal{J}(\mathbf{m}) = \| A(\mathbf{m}) - d_\text{obs} \|^2 + \lambda E(\mathbf{m}),
\end{equation}
where $A(\mathbf{m})$ models the measurement process (e.g., a projection or stray-field operator), and $d_\text{obs}$ is the observed data. The parameter $\lambda$ controls the trade-off between data fidelity and physical plausibility.
Fig.~\ref{fig:abert} illustrates the inverse reconstruction of a Néel-type skyrmion lattice from stray-field data.

Importantly, this framework can be extended beyond magnetization reconstruction, allowing also for minimization with respect to $D$, $K$, $M_s$, and $A$. By treating material parameters such as the DMI constant $D$ as additional optimization variables, it becomes possible to estimate unknown material properties directly from imaging data. In this sense, inverse micromagnetic modeling provides a powerful tool not only for reconstructing detailed magnetization textures, but also for characterizing intrinsic magnetic properties of novel materials.

Compared to experimental tomographic methods, which require complex instrumentation and multiple measurement angles, the optimization-based strategy described here is more flexible and can be applied to single-view data. Moreover, the use of differentiable solvers facilitates efficient gradient computation, enabling high-resolution reconstructions even for large or strongly inhomogeneous systems.

\section*{Inverse Design of Skyrmion-Based Devices\label{sec:design}}
Skyrmions exhibit remarkable properties such as nanoscale size, topological protection, and low threshold for current-driven motion, making them promising candidates for a wide range of spintronic device concepts. Proposed applications include multi-turn sensors \cite{suess2024devices}, racetrack memory elements\cite{tomasello2014strategy}, logic gates, and probabilistic or neuromorphic computing platforms \cite{pinna2020reservoir,raab2022brownian},. Recent advances also point to the potential of skyrmions in reservoir computing, spintronic oscillators, and reconfigurable magnonic crystals—all of which rely on precise spatiotemporal control of the magnetic texture.

Designing such devices requires navigating a high-dimensional parameter space involving geometry, material composition, and excitation protocols. Forward micromagnetic simulations provide a valuable tool to assess candidate designs, but inverse design offers a more systematic and efficient path to engineer devices toward specific functional targets. In this context, time-dependent micromagnetic optimization—where the Landau–Lifshitz–Gilbert (LLG) equation serves as a dynamic constraint—has emerged as a powerful approach.
Recent work in the field of inverse magnonics demonstrates how such optimization frameworks can be employed to design functional magnetic structures. Papp et al. \cite{papp2021nanoscale} applied gradient-based methods to the inverse design of spin-wave-based logic using nonlinear spin interference. Wang et al. \cite{wang2021inverse} further extended these ideas to engineer field profiles and geometries that generate tailored spin-wave responses in nanoscale systems.  Voronov et al. \cite{voronov2025inverse} used level-set topology optimization to design high-efficiency spin-wave couplers and delay lines. 

The general principle underlying such approaches is to formulate a cost functional $\mathcal{J}$ that measures the deviation between a desired output and the simulated system response. A typical optimization problem can be written as:

\begin{equation}
    \min_{p(t)} \mathcal{J}(\mathbf{m},t) \quad \text{subject to} \quad \frac{\partial \mathbf{m}}{\partial t} = -\gamma \mathbf{m} \times \mathbf{H}_\text{eff}(\mathbf{m}, p(t)) + \alpha \mathbf{m} \times \frac{\partial \mathbf{m}}{\partial t},
\end{equation}

where $\mathbf{m}(\mathbf{x},t)$ is the magnetization vector, $\mathbf{H}_\text{eff}$ is the effective field, and $p(t)$ denotes time-dependent design variables such as applied fields, geometric masks, or current densities. The objective $\mathcal{J}$ may encode trajectory tracking, stability conditions, or final-state constraints, depending on the specific application.

By implementing such models in differentiable simulation frameworks—such as NeuralMag, magnum.np, or SpinTorch—it becomes possible to optimize over complex dynamic behavior using gradient-based methods. This enables the automatic design of devices that guide skyrmions along engineered tracks, nucleate specific topological states, or suppress unwanted pinning effects. Applications range from low-power memory to ultrafast signal processing and intelligent magnetic metasurfaces.

As the field matures, inverse modeling is expected to play a central role in the design of robust, adaptive, and energy-efficient skyrmion-based technologies.

\section*{Conclusion}
The convergence of differentiable micromagnetic simulation, automatic differentiation, and high-performance computing has laid the groundwork for a new paradigm in the computational design of skyrmionic systems. As machine learning frameworks mature and integrate more tightly with physics-based solvers, we foresee the emergence of scalable and flexible inverse modeling techniques that will enable not only high-fidelity reconstruction of magnetization textures from sparse experimental data, but also the automated design of next-generation skyrmion-based devices.

Looking ahead, future developments are expected to focus on extending inverse micromagnetics toward real-time device control, local extraction of material parmaters,  uncertainty quantification, and closed-loop experimental feedback. These capabilities could revolutionize how skyrmionic structures are characterized, tuned, and optimized across application domains ranging from memory and logic to spin-wave computing and beyond. In this vision, inverse modeling becomes not just a simulation tool, but a core design principle in the development of skyrmionic technologies.

\endgroup

\newpage

\section{Topology in 3D: From Skyrmions to Hopfions and Mixed Topology States - A Unified Framework for Characterizing Magnetic Textures}
\begingroup
    \let\section\subsection
    \let\subsection\subsubsection
    \let\subsubsection\paragraph
    \let\paragraph\subparagraph
Maria Azhar$^1$, and Karin Everschor-Sitte$^1$
\vspace{0.5cm}

\noindent
\textit{$^1$ Faculty of Physics and Center for Nanointegration Duisburg-Essen (CENIDE), University of Duisburg-Essen, 47057 Duisburg, Germany}

\section*{Introduction}
Magnetic textures with non-trivial topology, such as Skyrmions, are of central interest in condensed matter physics for their unique properties and relevance to energy-efficient spintronic applications.
Current capabilities in Skyrmion manipulation are already impressive, including multiple ways to create, control, move, transform, or erase topological textures on demand.
This level of control points toward scalable device architectures, but also underscores key limitations, most notably the constraint of being confined to two dimensions (2D).

Crucially, introducing a third dimension fundamentally enriches the landscape of magnetic textures~\cite{gubbiotti20252025}. In three dimensions (3D), Skyrmions can extend into string-like structures and form more complex configurations, such as twisted loops and knots. Among these, the Hopfion—a ring-like, 3D solitonic magnetic texture characterized by nested magnetic isosurfaces—stands out as a prime example of a topologically non-trivial 3D structure~\cite{Faddeev1997}. 

\begin{figure}[h!]
\begin{center}
\includegraphics[width=0.95\columnwidth]{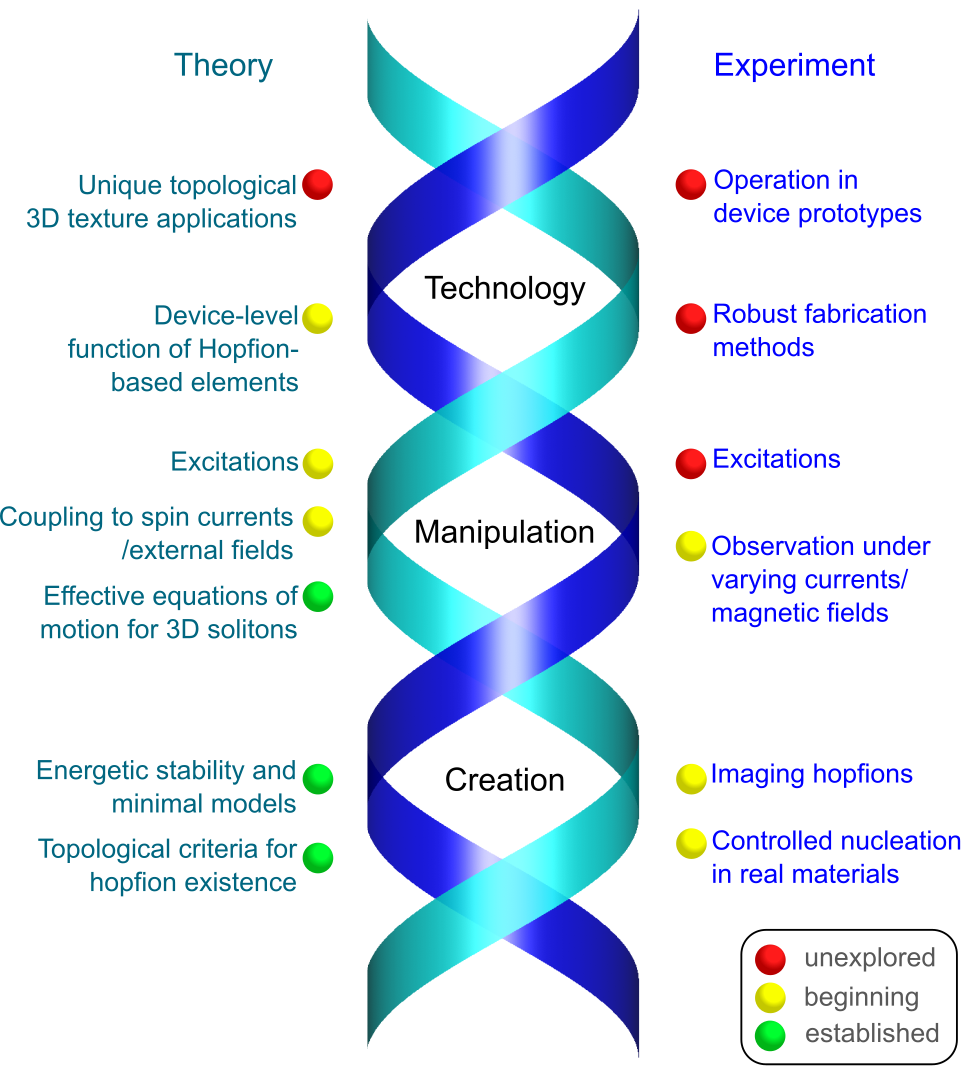}
\end{center}
\caption{Roadmap for 3D magnetic-texture-based technologies, illustrating the interplay between theoretical and experimental efforts across three key stages: Creation, Manipulation, and Technology. The double helix represents the intertwined progress of Theory (left) and Experiment (right). 
    Colored markers indicate stages of progress: green (established), yellow (beginning), and red (unexplored).}
	\label{fig:Karin1}
\end{figure}
The topology of such textures is usually classified by two topological indices--the Skyrmion number $N_\mathrm{sk}$ and the Hopf index $H$.
Both quantities can be computed from the emergent magnetic field 
$F^k=\frac{1}{8\pi}\epsilon^{ijk}\mathbf{m}\cdot(\partial_i\mathbf{m}\times\partial_j\mathbf{m})$, where $\epsilon^{ijk}$ is the Levi-Civita symbol, $i,j,k\in \{x,y,z\}$ and $\mathbf{m}$ describes the smoothly varying unit magnetization. In the absence of singularities like Bloch points, $\mathbf{F}$ is divergence-free.
The Skyrmion number is obtained by integrating this topological density over a (cross-sectional) area: 
$N^k_\mathrm{sk}=\int dx_idx_j F^k$.
Here, $k$ is the direction perpendicular to the surface over which the integration is performed.
The Hopf index can be computed via the volume integral known as the Whitehead integral~\cite{whitehead1947}: $H=\int \mathrm{d}V \, \mathbf{F}\cdot\mathbf{A}$ with $\mathbf{F}=\boldsymbol{\nabla}\times\mathbf{A}$. 
In systems with uniform magnetic backgrounds, these topological invariants are rigorously classified by homotopy theory, specifically,
$\pi_2(S^2) = \mathbb{Z}$ for Skyrmions and $\pi_3(S^2) = \mathbb{Z}$ for Hopfions-reflecting their integer quantization. This topological quantization promises additional stability and continues to drive intense research interest across theory and experiment.

However, compared to the rapidly advancing field of Skyrmion research, the study of 3D topological textures such as Hopfions remains in its early, theory-led stages. Figure 1 outlines a conceptual roadmap for advancing potential Hopfion-based technologies, highlighting the gap between theoretical predictions and experimental verification. 
Although theoretical predictions and advanced 3D simulations have proposed promising strategies for stabilizing Hopfions—such as engineered magnetic interactions and nanoscale confinement—their controlled creation and manipulation in experiments remain a major challenge.
Nonetheless, recent progress in microscopy and 3D imaging techniques, such as X-ray magnetic tomography—is encouraging, offering unprecedented access to complex 3D magnetic textures with spatial resolutions down to several tens of nanometers~\cite{Christensen_2024}.

Advances in both experiments and theory, particularly beyond the framework of homotopy groups of spheres, are reshaping our understanding of 3D magnetic textures.

\begin{figure*}[h!]
\includegraphics[width=\textwidth]{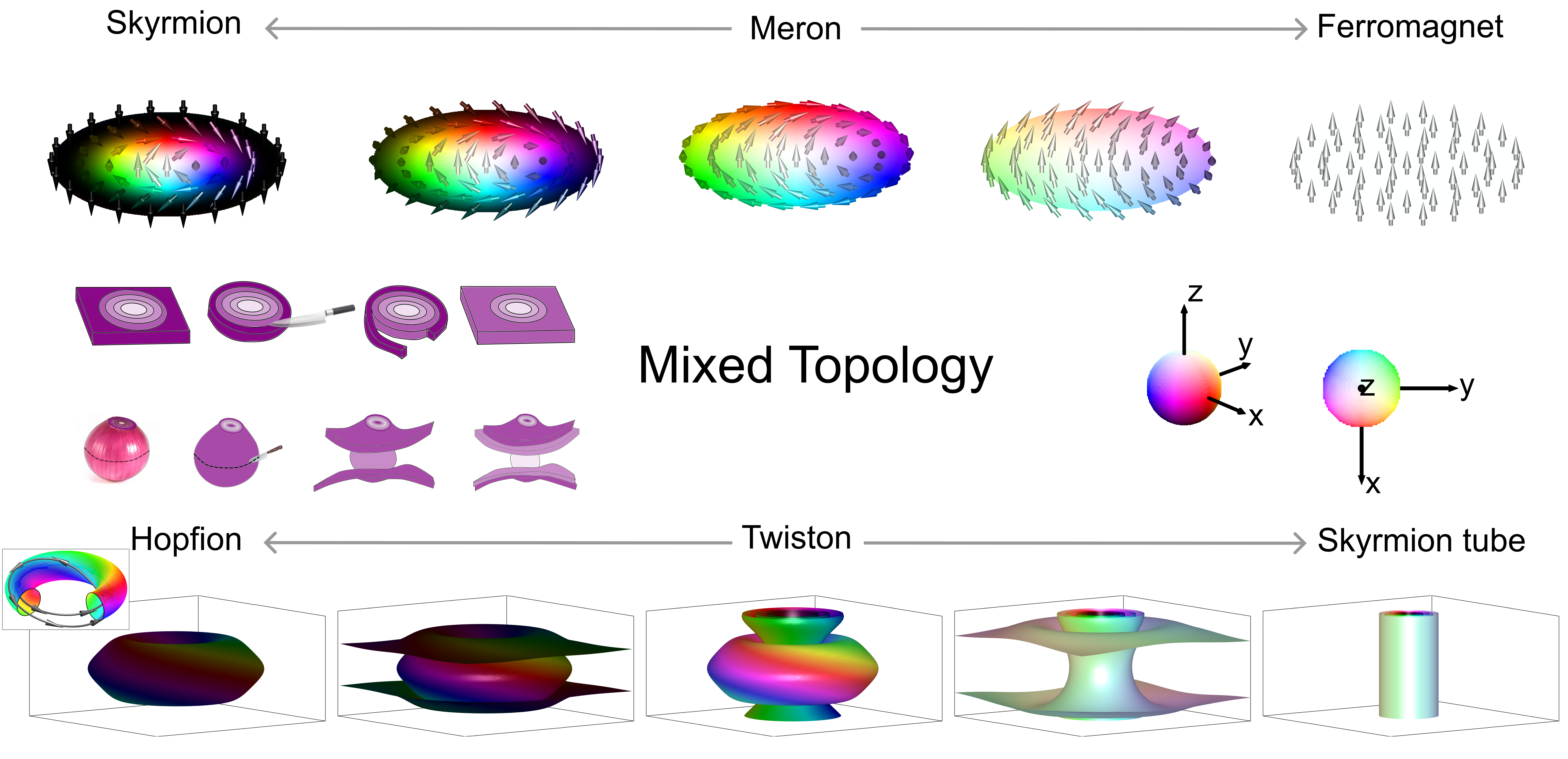}
    \caption{Topological transitions via mixed topology states. Colors represent the magnetization direction $\mathbf{m}$ as indicated in the legend at the right. 
Top row: Transition from a Skyrmion to a Ferromagnet via intermediate states such as a Meron.
Bottom row: Evolution from a Hopfion to a Skyrmion tube through a Twiston---a screw dislocation configuration~\cite{Azhar2022}. 
The inset on the far left depicts a Hopfion with a
more internal $m_z$ isosurface, illustrating the linking of two field lines of the emergent magnetic field $\mathbf{F}$.
These transitions can be conceptually understood as analogous to peeling an onion—removing successive ``layers" (isolines for Skyrmions and isosurfaces for Hopfions) while preserving the inner structure until reaching a topologically simpler configuration. 
}
	\label{fig:Karin2}
\end{figure*}

\section*{Relevance and Vision}

The promise of 3D magnetic textures lies in their potential to enable multidimensional information encoding, owing to their richer dynamical behavior compared to 2D structures. 
In particular, topological textures, such as Hopfions, promise to offer inherent robustness and increased stability due to their topological nature, while also enabling a rich variety of linked and knotted configurations.
These properties make them compelling candidates for a range of potential applications, including high-density 3D magnetic memory, where Hopfions act as stable topological bits; topological logic devices, which could exploit their constrained transformation rules for computation; magnetic sensors, taking advantage of their complex field responses for high-sensitivity detection; and neuromorphic computing, where their nonlinear dynamics may serve as a platform for brain-inspired information processing.
Advancing toward such applications will be significantly accelerated by a deeper theoretical understanding of the topology and dynamics of magnetic textures.

In realistic materials, non-uniform magnetic backgrounds—such as spiral, helical, vortex phases, or screw-dislocation structures—can emerge, forcing topological textures to adapt to the spatially varying background.
In such cases, the homotopy theory of spheres no longer applies directly, making topological characterization more challenging and enabling continuous transitions between topological sectors~\cite{azhar2024, wu2025photonic}.
Figure~\ref{fig:Karin2}a shows two such transitions: Skyrmion to ferromagnetic state and Hopfion to Skyrmion tube, each passing through intermediate mixed-topology states like a Meron and a Twiston.
These transitions can be conceptually likened to peeling an onion—removing successive isosurfaces ("layers") while preserving internal structure until a topologically simpler configuration is reached. 
Mixed-topology states can arise as higher-energy intermediates during topological transitions or bifurcations, involving gradual—and generally non-integer—changes in both the Skyrmion number $N_{\text{sk}}$ and the Hopf index $H$.
Transition pathways that conserve one topological index while varying the other might offer energetically favorable routes for switching, supporting more robust control and longer-lived metastable states for applications. Unlocking these possibilities requires a unified framework capable of describing topological transformations across both 2D and 3D textures.

A recently introduced framework for characterizing 3D topology via the Hopf index, equivalent to the Whitehead integral for configurations classified by $\pi_3(S^2) = \mathbb{Z}$, naturally captures mixed-topology states~\cite{azhar2024}. 
In this approach, the emergent magnetic field $\mathbf{F}$
is decomposed into $N_\Phi$ flux tubes labeled by $i=1,...N_\Phi$, and the Hopf index 
$H$ is expressed as the sum of their pairwise Gauss linking numbers $L_{ij}$, weighted by the flux $\Phi_i$ of each tube
\begin{equation}
    H=\sum_{i=1}^{N_{\Phi}} L_{ii}\Phi_i^2 + \sum_{\substack{i=1\\ j\neq i}}^{N_{\Phi}}L_{ij}\Phi_i\Phi_j.
    \label{eq:Hopf_decomposed}
\end{equation}
This inherently gauge-invariant formulation provides a direct and intuitive interpretation of the Hopf index and topological transitions, while remaining well-defined even in finite volumes or for non-compactifiable (i.e.~non-collinear) background magnetization. 
A key question, particularly for mixed-topology states, is whether these invariants are purely theoretical tools or whether they can be experimentally accessed and controlled.
A prominent example is the topological Hall Effect, in which conduction electrons traversing a non-coplanar spin texture experience the emergent field $\mathbf{F}$, leading to their transverse deflection and the generation of a measurable Hall voltage.
Similarly, the Hopf index can impose a lower bound on the energy of solitonic configurations~\cite{rybakov2022magnetic}.

\section*{Challenges}
Fundamental progress toward 3D magnetic-texture-based technologies requires overcoming both theoretical and experimental challenges.

A key theoretical challenge is to identify clear signatures of 3D topology—analogous to the Skyrmion or topological Hall effects in 2D—and to develop effective models, potentially incorporating higher-order energy terms consistent with the Derrick-Hobart scaling argument, that directly encode linking numbers or the Hopf index.
Numerically, simulating large 3D structures while avoiding boundary artifacts and accounting for real-material features like grains and defects remains difficult. Computing the Whitehead integral in finite volumes also requires care~\cite{Knapman2025}.
Further challenges to overcome arise in addressing more intricate textures, such as those featuring topological singularities (e.g., Bloch points, where the emergent magnetic field $\mathbf{F}$ is not divergence-free) or those governed by more sophisticated order parameters, such as $SO(3)$. The comprehensive classification of all possible 3D topological textures remains an open and actively pursued area of research.

Experimental challenges center on direct observation and control of 3D magnetic textures. Imaging techniques like LTEM and XMCD tomography still face limitations in spatial and temporal resolution, sample geometry, accessibility, and the accuracy of magnetization reconstruction from indirect measurements.
The direct measurement of topology and the manipulation of textures, especially while tracking their response to external stimuli, remain the main hurdles. Although recent breakthroughs have demonstrated the creation of magnetic textures with a non-zero Hopf index~\cite{kent2021creation, zheng2023hopfion} and imaged the dynamics of similar structures~\cite{Yu2023}, integration into functional devices is still an open frontier.

These challenges define the way forward, underscoring the importance of sustained synergistic progress of theoretical and experimental approaches.
The vision is clear: a future where 3D topological magnetic textures are not only observable but can be precisely engineered, enabling novel paradigms in unconventional computing, reconfigurable logic, and resilient memory technologies.

 \section*{Acknowledgements}
We acknowledge funding from the German Research Foundation (DFG) Project No.~403233384 (SPP2137 Skyrmionics), No.~278162697-SFB 1242 (project B10) and No.~505561633 (TOROID--co-funded by the French National Research Agency (ANR) under Contract No.~ANR-22-CE92-0032). M.~A.\ also acknowledges support from the UDE Postdoc Seed Funding.

\endgroup

\newpage

\section{Skyrmion stability and lifetimes under nonequilibrium conditions}
\begingroup
    \let\section\subsection
    \let\subsection\subsubsection
    \let\subsubsection\paragraph
    \let\paragraph\subparagraph
Louise Desplat$^1$, and Joo-Von Kim$^2$
\vspace{0.5cm}

\noindent
\textit{$^1$ Univ. Grenoble Alpes, CNRS, CEA, SPINTEC, 38054 Grenoble, France\\
$^2$ Centre de Nanosciences et de Nanotechnologies, CNRS, Universit{\'e} Paris-Saclay, 91120 Palaiseau, France}

\section*{Introduction}

The topological protection of magnetic skyrmions is often invoked as a key attribute for their use within magnetic storage. Skyrmions carry an integer topological charge which prevents them from being continuously unwound into the topologically trivial collinear magnetic background. However, this argument only applies within a continuum description of the magnetization within micromagnetics theory. When the atomic lattice is introduced, isolated skyrmions become metastable excitations of the collinear magnetic background. They are separated from it by finite energy barriers, which may be overcome at finite temperature under the effect of thermal fluctuations. The rate at which they decay can be described by the Arrhenius law,
\begin{equation}
    k=f_0  e^{-\beta \Delta E},
\end{equation}
in which $f_0$ is a prefactor, $\Delta E$ is the activation energy, and $\beta = 1/(k_B T)$ is the Boltzmann factor. The activation energy corresponds to the height of the barrier along the minimum energy path linking the initial and final states [Fig.~\ref{fig:Kim}(a)]. While this expression is often used empirically, in practice the rate prefactor contains a dynamical contribution, as well as an entropic contribution encompassing details of the fluctuations about the initial state and the barrier top. The physically relevant quantity is therefore the change in Helmholtz free energy, $\Delta F = \Delta E - T \Delta S$, in which $\Delta S$ is the activation entropy [Fig.~\ref{fig:Kim}(a)]. Neglecting it may lead to largely erroneous analyses.

In recent years, the stability of skyrmions in transition metal ultrathin films has been theoretically investigated within the atomistic framework in model systems~\cite{desplat2018thermal, desplat2019paths, desplat2020path}, as well as in ab initio-parametrized systems such as prototypical Pd/Fe/Ir(111)~\cite{bessarab2018lifetime, von2019skyrmion, goerzen2022atomistic}. This was achieved by combining a path finding scheme, such as the nudged elastic band or the string  method, with forms of reaction rate theory
. The atomistic framework, as opposed to a micromagnetic description, is required to properly resolve magnetic singularities around the transition state, where the angles between neighboring spins become large [Fig.~\ref{fig:Kim}(a)].

In addition to escape at sample boundaries, different collapse mechanisms of skyrmions were identified, namely isotropic collapse [Fig.~\ref{fig:Kim}(a)], including at defects~\cite{desplat2018thermal, desplat2019paths, bessarab2018lifetime}, and anisotropic, chimera collapse via the injection of the opposite topological charge in the presence of exchange frustration~\cite{desplat2019paths}, with recent experimental evidence of both mechanisms~\cite{muckel2021experimental}.

Within the framework of reaction rate theory, it was shown that the activation energy alone is not sufficient to estimate the lifetime of skyrmions. Skyrmions are stabilized not only by energy barriers, but also by the existence of low-energy, stable modes of deformation, which result in a large configurational entropy of the skyrmion state, compared to that of the transition state~\cite{desplat2018thermal, von2019skyrmion} [Fig.~\ref{fig:Kim}(a)]. As a consequence, the Arrhenius prefactor for skyrmion collapse assumes a wide range of values and varies by several orders of magnitude with material parameters and applied fields~\cite{desplat2018thermal, von2019skyrmion, desplat2020path}  [Fig.~\ref{fig:Kim}(b)]. This crucial role of entropy was also reported experimentally~\cite{wild2017entropy}, with an extreme case of entropy-enthalpy compensation. This finding implies that long-lived, small skyrmions may be achieved by tailoring a large activation entropy, in addition to a large activation barrier, which intrinsically decreases along with the skyrmion radius.

Another approach to rate calculation is to integrate the system’s dynamics at finite temperature in Langevin-type simulations, but this becomes computationally costly when the mean waiting time between transitions becomes large compared to the small timesteps required to integrate the magnetization dynamics. One can then resort to path sampling schemes, which are more efficient than the brute-force approach. One such scheme known as forward flux sampling was successfully applied to the problem of skyrmion collapse, and showed excellent agreement with reaction rate theory~\cite{desplat2020path} [Fig.~\ref{fig:Kim}(b)].

\begin{figure}[h!]
\includegraphics[width=.5\linewidth]{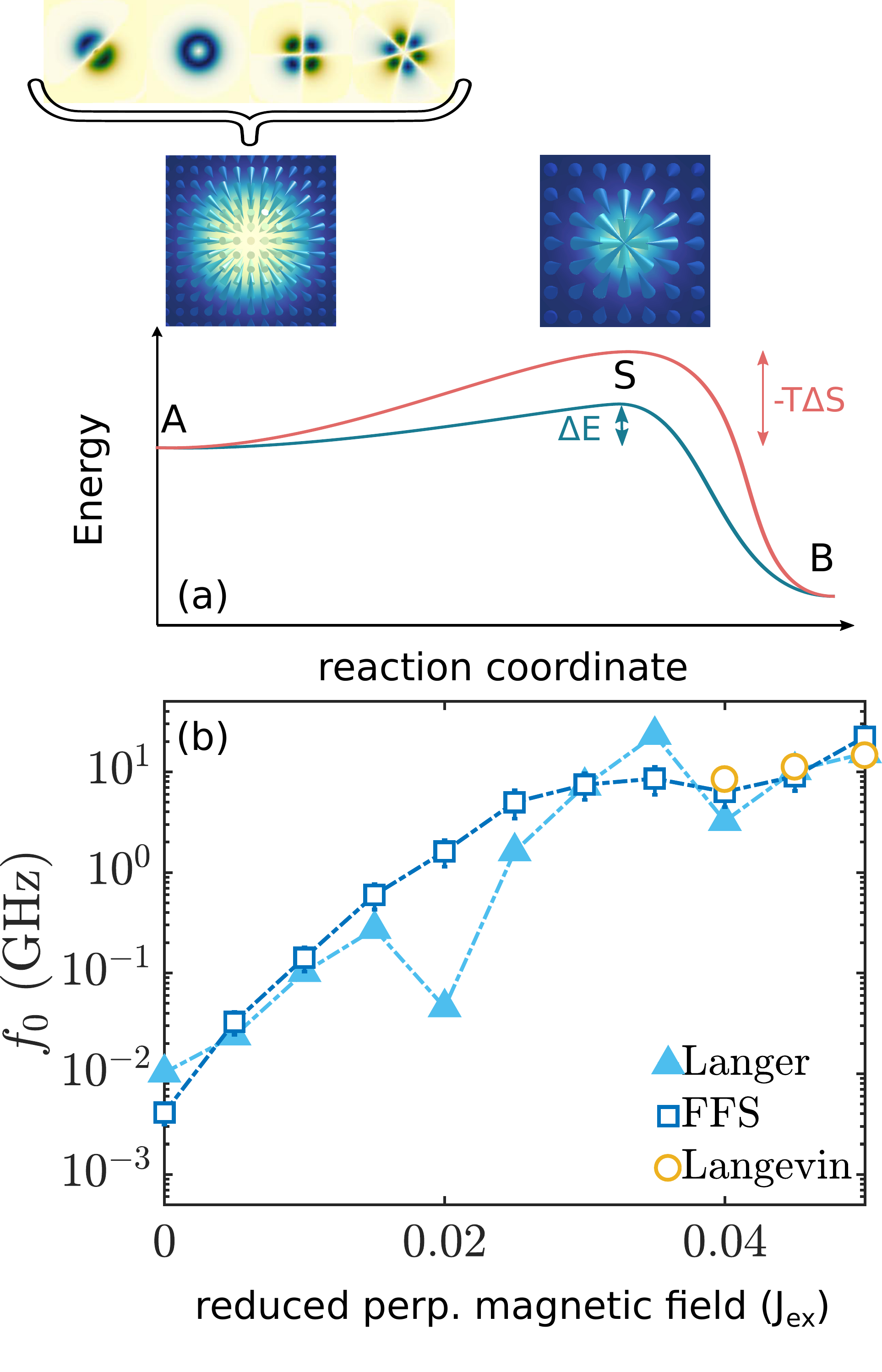}
\caption{(a) Sketch of the energy landscape for skyrmion isotropic collapse into the ferromagnetic state, with the activation energy, $\Delta E$, shown in blue, and the activation entropy, $-T\Delta S$, shown in red. The corresponding zoomed in spin configurations at the initial state, A, and at the transition state, S, are shown at the top, where the internal modes of the skyrmion are also shown in A  (adapted from~\cite{desplat2018thermal}). (b) Arrhenius prefactor for skyrmion isotropic collapse computed with reaction rate theory (Langer), forward flux sampling (FFS), and direct Langevin simulations, as a function of reduced perpendicular applied magnetic field given in units of the first-neighbor isotropic exchange constant, $J_{\mathrm{ex}}$ (adapted from~\cite{desplat2020path}).}
\label{fig:Kim}
\end{figure} 

 \section*{Relevance and Challenges}

Skyrmions have been envisioned as information carriers in novel spintronics devices, to, on the one hand, address the bottlenecks of the racetrack memory and logic gates~\cite{fert2013skyrmions}, and, on the other hand, open up new computing paradigms beyond von Neumann~\cite{prychynenko2018magnetic}. So far, the thermal stability of magnetic skyrmions has only been theoretically investigated at equilibrium, and never under a driving force. However, skyrmions are envisioned as dynamical objects for applications, in which they are created, manipulated, shuffled back and forth along tracks, and deleted by applied currents or fields.

Skyrmions can be nucleated and manipulated through electric currents~\cite{legrand2017room}. In this case, short current pulses are used to nucleate and displace the skyrmions. During the displacement, skyrmion nucleation and annihilation events are observed inside the sample~\cite{legrand2017room}. These are presumed to result from the combined effects of the current-induced Joule heating and spin torques.

The control of skyrmions via short pulses in electric fields within the regime of thermally-driven processes is an alternative, compelling approach to reduce power consumption. So far, the electric-field-mediated, thermally-activated writing and deleting of magnetic skyrmions was experimentally demonstrated at constant electric field in transition metal thin films~\cite{schott2017skyrmion}. It was empirically explained by a change in energy barriers induced by the electric field, based on a Néel-Brown macrospin model with a constant Arrhenius prefactor, which, as explained above, is missing key ingredients, such as the activation entropy, for a complete physical picture. At the microscopic level, an electric field applied to transition metal thin films can modify magnetic parameters--typically isotropic exchange, magnetic anisotropy, and the Dzyaloshinskii-Moriya interaction, through Rashba and magnetoelectric effects~\cite{goerzen2022atomistic}.

Meanwhile, only the thermally activated nucleation/annihilation of skyrmions under a constant electric field was theoretically investigated~\cite{goerzen2022atomistic}, whereas their nucleation probability and stability under electric currents was not. That is because reaction rate theory is set in the quasi-equilibrium approximation, where one assumes that the density of states remains Boltzmann-distributed everywhere but in a small region around the transition state
. Under current-induced spin torques, and/or time-dependent excitations, this assumption no longer holds. Skyrmions set in motion under a driving force experience an internal deformation of their profile, which can be seen as an additional dynamical magnetic field in the skyrmion center of mass reference frame~\cite{troncoso2014brownian}. Indeed, core deformations in antiskyrmions under spin-orbit torques underpin transitions to different dynamical regimes such as trochoidal motion~\cite{ritzmann2018trochoidal}, analogous to Walker breakdown for domain walls. Deformations should therefore have an impact on skyrmion stability, affecting not only the energy barrier but also the activation entropy, yet this effect has not been explored.

Additionally, reaction rate theory relies on sink boundary conditions past the barrier. In the nucleation problem, barrier recrossing events are \emph{a priori} non negligible, as the nucleation barrier is typically much larger than the annihilation barrier~\cite{desplat2018thermal, bessarab2018lifetime, von2019skyrmion} [Fig.~\ref{fig:Kim}(a)], so nucleation rates yielded by this method are typically overestimated. For dynamical sampling, frequent barrier recrossings are an issue as it becomes difficult to obtain sufficient sampling of the transition path trajectories within a reasonable computation time. The degeneracy of nucleation sites in finite-sized samples also makes the computation more tricky, as the choice of order parameter to track becomes less straightforward. 

Other unresolved questions concern the crossover regime around Goldstone modes. Goldstone modes are modes of zero-energy fluctuations that arise in the presence of continuous unbroken symmetries. In the case of skyrmions, the most common one is the translational invariance of the skyrmion position, which gives rise to a temperature dependence of the Arrhenius prefactor~\cite{bessarab2018lifetime,von2019skyrmion}. This applies to skyrmions that are large compared to the lattice constant, whereas smaller skyrmions experience a weak pinning to the lattice, thereby breaking the translation symmetry~\cite{desplat2018thermal,desplat2020path}. However, there does not exist a definite threshold between these regimes, leading to a gray area of intermediate skyrmion sizes where it is not clear whether the translation mode should be considered harmonic or Goldstone. Without a crossover theory, the computed rates may become discontinuous as a function of, e.g., perpendicular magnetic field, which is \textit{a priori} unphysical. 

\section*{Concluding Remarks}
The thermal stability of dynamical skyrmions under a non-equilibrium setting is as much a fundamental question as it is a technological issue. Lowering of the axial symmetry as a result of core deformations may impact paths towards collapse and internal mode frequencies, thereby impacting annihilation barriers and entropic contributions. Time-dependent forcing and dissipative torques also challenge basic assumptions of equilibrium noise driving the thermal activation processes. Addressing these questions will require extending theoretical and computational approaches to the nonequilibrium regime, which may also find useful applications in other emerging technologies like stochastic computing.

%

\endgroup

\newpage

\section{Interplay of Skyrmionic Properties and Electronic Structure: Mutual Influence and Emergent Phenomena}
\begingroup
    \let\section\subsection
    \let\subsection\subsubsection
    \let\subsubsection\paragraph
    \let\paragraph\subparagraph
Yuriy Mokrousov$^{1,2}$, and Stefan Bl\"ugel$^{1,3}$
\vspace{0.5cm}

\noindent
\textit{$^1$ Peter Gr\"unberg Institute (PGI-1), Forschungszentrum J\"ulich and JARA, 52425 J\"ulich, Germany\\
$^2$ Institute of Physics, Johannes Gutenberg University Mainz, 55099 Mainz, Germany\\
$^3$ Institute for Theoretical Physics, RWTH Aachen University, 52056 Aachen, Germany}

\section*{Introduction}
Magnetic skrymions and skyrmion lattices are fascinating objects for several reasons. Many of them arise directly from the topological  and solitonic nature of the magnetization texture. For example, topology contributes to the stability of otherwise metastable single skyrmions on  a ferro- or antiferromagnetic phase, to the skyrmion Hall  and the topological Hall effects, and the  small electric currents that can move them. Taken together, this promises to be of technological significance for information technology.
Even more intriguing than single  skyrmions are magnetic skyrmion lattices, as they represent a new type of ordered magnetic phase with collective properties that exhibit new features due to the topological nature of the skyrmions, such as non-reciprocal magnons or topologically protected magnon edge states. 

Interestingly, very little is known about the imprint of the topological nature in the magnetization space onto the electronic structure. This can be explained by the fact that most skyrmions investigated sofar are chiral skyrmions with the Dzyaloshinskii-Moriya interaction (DMI) as stabilizer. Since this interaction is small compared to the exchange interaction, we observe large skyrmions and skyrmion crystals with large lattice constants, typically larger than 30~nm, which can be easily generated by external magnetic field-control, whose imprint onto the  electronic structure on the other hand are small and difficult to observe.


\section*{Relevance and Vision}

The electronic properties and response characteristics of non-collinear spin systems are of  utmost importance for their implementation as functional and efficient bits of information. 
In this context, so-called multi-$q$ states, i.e.\ periodic arrangements of spins dominated by their lowest order Fourier modes, which can be used to describe the skyrmion lattices in chiral B20 materials such as MnSi, have emerged as a very rich class of magnetic textures~\cite{Shimizu2022,Hayami2021}. Their simple real-space form lends a shortcut to gaining a deeper insight into the electronic structure of noncollinear magnetic materials and the interplay of topological features in real- and reciprocal space. 
When the length scale of a skyrmionic multi-$q$ texture approaches the lattice constant of the host material, the strong non-collinearity driven variation in the effective potential leads to the formation of gapped topological states in the electronic system of the host material, manifesting for example as a quantized topological Hall effect~\cite{gobel2017antiferromagnetic, Nagaosa2015}, which is traditionally explained through the language of emergent magnetic fields. However, the emergence of topological electronic states which cannot be explained as a result of adiabatic dynamics or emergent fields calls the establishment of a unified paradigm that works across various length scales, electronic structure regimes and lattice types of the textures.

\section*{Challenges}

The challenges are two-fold. Recently, it was shown that all phenomena that root in emergent fields of multi-$q$ textures are encoded in their observables algebra $-$ an object known as a $C^*$-algebra containing all translations of the Hamiltonian and other observables which can be represented as bounded, self-adjoint operators on a physical Hilbert space~\cite{Lux2024}. For multi-$q$ systems, the observable algebra has a finite number of non-trivial equivalence classes of projections as described by operator $K$-theory. Correspondingly, the gaps in the energy spectrum and their corresponding Fermi projections and transport properties can be classified by a complete set of topological invariants, which opens a way to a complete topological classification of electronic states in multi-$q$ magnets without referring to ideas of texture smoothness or adiabatic limit. These topological invariants, including real, reciprocal and mixed Chern numbers, can be computed directly from the electronic structure of an incommensurate multi-$q$ texture~\cite{PascalSciPost}, and as a result, an exhaustive topological classification can be achieved that uniquely identifies the $K$-theory classes. The developed theory made it clear that the relationship between the momentum space Chern number and the texture’s real-space winding is much more intricate than traditionally assumed, see e.g. Fig.~\ref{fig:Mokrousov}. Moreover, the study of the evolution of the electronic topology across various phases  in real-space has shown  that the observability of the associated spectral flow is related to the subtleties of the observable algebra, and that the existence of emergent gapless modes is related to the topology of the point defects that arise across the transition. 

The novel approach based on non-commutative geometry and the properties of $C^*$-algebras opens new possibilities in attacking the long-standing properties of non-collinear textures subject to external fields. For example, the effect of the magnetic field, possibly spatially non-uniform, can be rigorously treated as an integral part of the non-commutative torus setup, which opens the door to addressing such challenging problems as linear and non-linear magneto-transport properties, topological nature of orbital susceptibility and orbital magnetism exhibited by spin textures~\cite{Lux2018}. On the other hand,
 the dynamical aspect of texture behavior is a quintessential element of texture applications. Traditionally, charge, spin and orbital current response of magnetic systems driven by time-dependent perturbations can be perceived from the viewpoint of geometrical properties of electronic states. Here, one can strive to tackle the non-trivial time-dependent responses by referring to the apparatus of 
non-commutative geometry, possibly establishing a way to incorporate the impact of  various types of excitations due to  e.g.\ spin fluctuations and lattice dynamics $-$ effects, which are important e.g.\ for applications of textures in quantum or neuromorphic computing. Finally, the main challenge of constructing a rigorous {\it ab initio} topological electronic structure theory of incommensurate spin textures can be finally addressed based on a solid topological footing.

A second challenge is to realize magnetic  materials with a multi-$q$ pitch of the order of a new nanometer, i.e.\ significantly smaller than that of the common B20 compounds, in order to experimentally explore the new physics predicted by the interaction of real and momentum space.  Multi-$q$ states typically arise from the mode coupling of symmetry-equivalent spiral magnetic ground  or excited states, so-called single-$q$ states, by external magnetic fields or higher-order spin interactions, e.g.\ the four-spin interaction, or a combination of both. Such single-$q$ states arise from  a competition of interactions, and comparable energy scales lead to shorter wavelengths, but if the effective energy scales are too large, a transition to a multi-$q$, or a skyrmion lattice phase, respectively, is no longer possible with lab magnetic fields.  The challenge is therefore to find materials with competing magnetic interactions whose energy scales lead to short spirals, but which can still be transitioned to a multi-$q$ state by appropriate magnetic fields.

\begin{figure*}[h!]
\centering
\includegraphics[angle=0, width=0.99\textwidth]{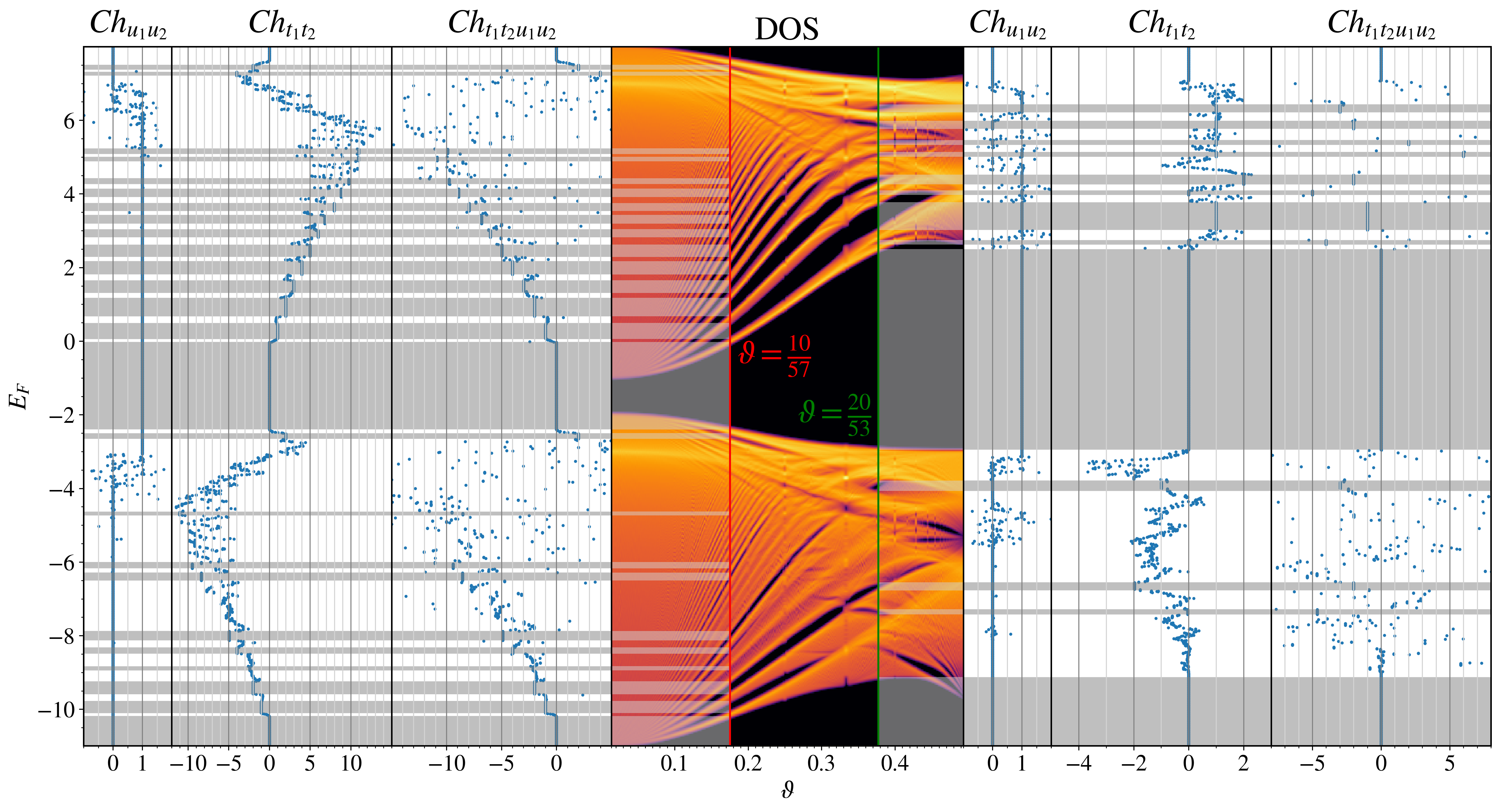}
\caption{
 Main Chern numbers and evolution of the electronic spectrum of a triple-$q$ state of $s$-electron model on a hexagonal lattice as a function of the pitch of the texture as characterized with parameter $\vartheta$. When starting from small values of $\vartheta$ (limit of large textures) the development of ``adiabatic" gaps in the spectrum is clearly visible. However, for pitch values falling into the ballpark of interatomic distance the emergence of so-called ``non-commutative" gaps in the spectrum $-$ gaps, which cannot be adiabatically connected to the limit of large textures $-$ can be observed. At $\vartheta=\frac{10}{57}$ and $\vartheta=\frac{20}{53}$ the values of the real, reciprocal and mixed  Chern numbers are shown as a function of band filling. Note a very different behavior in Chern numbers for these two values of the pitch.
The figure is adapted from Ref.~\cite{PascalSciPost}.}
\label{fig:Mokrousov}
\end{figure*}

Single-$q$ spirals can be categorized into Dzyaloshinskii  and Yoshimori spirals depending on the two alternative physical formation mechanisms.  In the past, most skyrmion lattices, e.g.\ those of B20 materials where formed by Dzyaloshinskii-spirals, spirals formed by the   competition of Heisenberg exchange interaction and DMI. Since the DMI is relativistic in nature and typically much smaller than the exchange interaction, we observe spirals of large wavelengths that can be easily controlled by a magnetic field. Yoshimori-spirals, on the other hand, arise from geometric or exchange frustration of competing Heisenberg antiferro- and ferromagnetic exchange interactions between  neighbors on complex lattices  or neighbors of  different distances. Typically, the pitches are much shorter, but cannot be  converted into a triple-$q$ state by  an external field. 

Recently a remarkable compromise of interaction strengths was discovered for the centrosymmetric layered Gd intermetallics. Experiments revealed the presence of
spin-spiral waves in materials such as Gd$_2$PdSi$_3$~\cite{kurumaji2019skyrmion,Paddison2022}, GdRu$_2$Si$_2$~\cite{khanh2020nanometric}, and others that transform into skyrmion lattices with a lattice constant of 2-3~nm under magnetic fields. The compromised interactions  arise from frustrated
Ruderman-Kittel-Kasuya-Yosida (RKKY) type exchange
interactions in which competing and distance-dependent
ferro- and antiferromagnetic couplings between magnetic
Gd ions arranged in two-dimensional (2D) sheets  play a central role~\cite{Bouaziz2022}. A thorough \textit{ab initio} analysis of the electronic structure of the skyrmion lattice phase  does not exist yet, which is basically a consequence of the system size and the associated computational effort of \textit{ab initio} methods, but a recently put-forward \textit{ab initio} modelling~\cite{chen2025topological} reveals for Gd$_2$PdSi$_3$ that the dominant contribution to the topological Hall effect (THE) of the skyrmion lattice arises from the crossing points between the folded bands along high-symmetry lines in the Brillouin zone. 

\section*{Concluding Remarks}
The imprint of multi-$q$ lattices on the electronic structure is an exciting subject promising new physics and properties. Centrosymmetric ternary intermetallic rare-earth compounds (RTX, RTX$_2$, RTX$_3$, RT$_2$X$_2$,  R$_2$TX$_3$), which tend to antiferromagnetism, and crystallize in structures consisting of layers containing a rare-earth element (R) separated from other layers by transition metal (T) and metalloid (X) elements, as well as magnetic two-dimensional (2D) van der Waals materials with their highly anisotropic and higher-order interactions represent huge classes of materials that can harbour magnetic  multi-$q$ states, skyrmion lattices or other non-trivial topological spin textures with small lateral features. Modern pre-exascale and exascale computing and \textit{ab initio} software infrastructures will enable the systematic analysis of the electronic structure and emergent properties imprinted by the topological magnetization textures at nanoscale sizes for real materials.

\section*{Acknowledgements}
S.B. acknowledges financial support from the MaX Center of Excellence funded by the EU through the H2020-INFRAEDI-2018 (project: GA 824143), by the European Research Council grant 856538 (project "3D MAGIC") and the Deutsche Forschungsgemeinschaft (DFG, German Research Foundation) through SFB 1238 (project C1).

\endgroup

\newpage

\section{Quantum Skyrmions}
\begingroup
    \let\section\subsection
    \let\subsection\subsubsection
    \let\subsubsection\paragraph
    \let\paragraph\subparagraph
\newcommand{\bea}{\begin{eqnarray}}
\newcommand{\eea}{\end{eqnarray}}
\newcommand{\nn}{\langle ij \rangle}
\newcommand{\nnn}{\langle\langle ij \rangle\rangle}
\newcommand{\bsigma}{\boldsymbol{\sigma}}
\newcommand{\bS}{\mathbf{S}}
\newcommand{\bR}{\mathbf{R}}
\newcommand{\bP}{\mathbf{P}}
\newcommand{\bb}{\mathbf{b}}
\newcommand{\br}{\mathbf{r}}
\newcommand{\hatz}{\hat{z}}
\newcommand{\ba}{\mathbf{a}}
\newcommand{\bs}{\mathbf{s}}
\newcommand{\bfm}{\mathbf{m}}
\newcommand{\bfM}{\mathbf{M}}
\newcommand{\bp}{\mathbf{p}}
\newcommand{\bn}{\mathbf{n}}
\newcommand{\bG}{\mathbf{G}}
\newcommand{\bOm}{\mathbf{\Omega}}

\newcommand{\td}{d}

\newcommand{\mcf}{\mathcal{F}}
\newcommand{\mb}{\mathfrak{b}}
\newcommand\bbone{\ensuremath{\mathbbm{1}}}
\newcommand{\vl}{v_{_L}}
\newcommand{\vc}{\mathbf}
\newcommand{\be}{\begin{equation}}
\newcommand{\ee}{\end{equation}}
\newcommand{\bk}{\mathbf{k}}
\newcommand{\bK}{{{\bf{K}}}}
\newcommand{\cE}{{{\cal E}}}
\newcommand{\bQ}{{{\bf{Q}}}}
\newcommand{\bg}{{{\bf{g}}}}
\newcommand{\hbr}{{\hat{\bf{r}}}}
\newcommand{\bq}{{\bf{q}}}
\newcommand{\hx}{{\hat{x}}}
\newcommand{\hy}{{\hat{y}}}
\newcommand{\ha}{{\hat{a}}}
\newcommand{\hb}{{\hat{b}}}
\newcommand{\hc}{{\hat{c}}}
\newcommand{\hd}{{\hat{\delta}}}
\newcommand{\beal}{\begin{align}}
\newcommand{\eeal}{\end{align}}
\newcommand{\ra}{\rangle}
\newcommand{\la}{\langle}
\renewcommand{\tt}{{\tilde{t}}}
\newcommand{\upa}{\uparrow}
\newcommand{\dna}{\downarrow}
\newcommand{\vS}{\vec{S}}
\newcommand{\dg}{{\dagger}}
\newcommand{\pdg}{{\phantom\dagger}}
\newcommand{\tphi}{{\tilde\phi}}
\newcommand{\cf}{{\cal F}}
\newcommand{\ca}{{\cal A}}
\renewcommand{\ni}{\noindent}
\newcommand{\ct}{{\cal T}}
\def\l{\ell}
\newcommand{\bu}{\mathbf{u}}

\newcommand{\I}{i}

\newcommand{\bx}{\mathbf{x}}
\newcommand{\by}{\mathbf{y}}
\newcommand{\vn}{\vec{n}}
\newcommand{\cl}{\mathcal{L}}
\newcommand{\cj}{\mathcal{J}}
\newcommand{\cz}{\mathcal{Z}}
\newcommand{\cd}{\mathcal{D}}
\newcommand{\co}{\mathcal{O}}



Sopheak Sorn$^{1,2}$, and Markus Garst$^{1,2}$
\vspace{0.5cm}

\noindent
\textit{$^1$ Institute of Theoretical Solid State Physics, Karlsruhe Institute of Technology, 76131 Karlsruhe, Germany\\
$^2$ Institute for Quantum Materials and Technology, Karlsruhe Institute of Technology, 76131 Karlsruhe, Germany}

\section*{Introduction}

Magnetic skyrmions are usually associated with topologically non-trivial solutions of the classical field equations of micromagnetism in two spatial dimensions, where the local magnetization is represented in terms of a classical unit vector field $\vec n(\vec r)$. In quantum magnetism, in contrast, the local spins are treated as quantum mechanical operators and magnetic ground or excited states are determined by eigenstates of the corresponding Hamiltonian operator. 
Is the concept of a magnetic skyrmion transferable to quantum magnetism, and what are the properties of such quantum skyrmions? What are quantum phases of skyrmions and how distinct is the phase diagram for classical and quantum many-skyrmion systems?  There exist two complementary approaches to address these questions, see Ref.~\cite{Petrovic2025} for a recent review and references therein.

In a first approach, the Hamiltonian operator for small systems of interacting spins is solved by exact methods and the skyrmionic content of quantum eigenstates is analyzed. Using exact diagonalization methods quantum skyrmions have been investigated consisting of $N_S$ spin-$\frac{1}{2}$ on a lattice with $N_S \lesssim 30$. Alternatively, the density matrix renormalization group (DMRG) method has been employed that is able to treat larger systems $N_S \sim 1000$ but usually gives access only to the ground state of the system.

The second approach starts from the classical limit of a skyrmion texture, that can be considered as a saddle point of the quantum action of the spin coherent state path integral. Quantum corrections to the classical solution are  obtained by considering fluctuations around this particular  saddle point. Quantum fluctuations with small amplitude just correspond to the magnon modes that can be systematically studied in an $1/s$ expansion, where $s$ is the spin density. Large amplitude fluctuations are usually parametrized in terms of collective coordinates whose choice depends on the problem at hand. Quantum nucleation and quantum collapse, i.e., the creation and annihilation of skyrmions by quantum fluctuations on a discrete lattice, have been studied by this method as well as quantum depinning, i.e., tunneling out of a pinning potential. 
Collective coordinates, like the helicity of a skyrmion stabilized by frustration, have been also proposed to serve as a macroscopic degree of freedom for the realization of qubits. 

In the following, we list a selection of open pertinent questions on the theory of quantum skyrmions and quantum phases of many-skyrmion systems. 

\section*{Open Questions and Visions}

\subsection*{
Quantum phase diagram of many-skyrmion systems}

An important collective coordinate is the 
dipole moment,
\begin{align} \label{eq:DipoleMoment}
\vec R(t) = \frac{1}{Q} \int d^2 r\,  \vec r\, \rho_{\rm top}(\vec r,t),
\end{align}
of the topological charge density $\rho_{\rm top} = \frac{1}{4\pi} \vec n (\partial_x \vec n  \times \partial_y \vec n)$ with the topological charge $Q = \int d^2 r \rho_{\rm top}(\vec r,t)$.
Being the total linear momentum of the magnetic sector, $P_i = 4\pi s Q \epsilon_{ji} R_j$ where $\epsilon_{xy} = - \epsilon_{yx} = 1$, it is a low-energy degree of freedom \cite{Papanicolaou1991}. The quantization of the topological dipole moment can be effectively obtained by replacing its classical Poisson bracket with the commutator \cite{Ochoa2019,Sorn2024},
\begin{align} \label{eq:Commutator}
[\hat R_j, \hat R_k ] = \frac{i \hbar}{4\pi s Q} \epsilon_{kj} .
\end{align}
This commutation relation serves as a basis for an effective quantum theory of many-skyrmion systems. 

The effective Hamiltonian of an interacting many-skyrmion system consisting of $N$ skyrmions with topological charge $Q$ is here given by 
\begin{align}
    H = \sum^N_{\alpha = 1}  V(\hat{\vec R}_{\alpha}) + \sum^{N_S}_{\alpha, \beta = 1; \alpha < \beta} V_{\rm int}(\hat{\vec R}_\alpha - \hat{\vec R}_{\beta}),
\end{align}
where $V$ is a single-skyrmion potential 
and $V_{\rm int}$ is an effective skyrmion-skyrmion interaction potential. The commutator of coordinates of the same skyrmion is given by Eq.~\eqref{eq:Commutator}, whereas coordinates of distinct skyrmions commute.
This theory strongly resembles the problem of electrons in the lowest Landau level where $\vec R_{\alpha}$ are identified with the guiding-center coordinates of electron orbits \cite{Arovas}. 
An important distinction from electrons, however, is that skyrmions are expected to behave as bosons. The quantum phases for many-skyrmion system that are predicted by this theory have only partially been investigated, and many open questions remain. 

Let us first consider the limit of a dilute skyrmion system where their interaction can be neglected in a first approximation $V_{\rm int} = 0$. 
For a random single-skyrmion potential $V$, it is known from the physics of the integer quantum Hall effect that the coordinates $\vec R_\alpha$ will be semiclassically encircling peaks and valleys of the disorder potential on closed orbits. Expanding the potential $V$ up to quadratic order around an extremum, the problem reduces to an effective one-dimensional harmonic oscillator with a frequency that depends on the shape of the potential thus leading to a wealth of low-energy magnetic resonances. Signatures of these resonances have neither been explored theoretically nor detected experimentally. 

For a periodic potential $V(\vec R)$ arising, e.g., from the underlying crystal lattice the Hamiltonian can be diagonalized for $V_{\rm int} = 0$. Its eigenvalue equation reduces to Harper's equation giving rise to a skyrmion band structure with band gaps determined by the strength of $V$. Identifying the band with lowest energy and projecting the skyrmion-skyrmion interaction $V_{\rm int}$ onto this band, a Bose-Einstein condensation of skyrmions occurs under certain circumstances resulting in a skyrmion liquid phase \cite{Balents2016,Haller2025}. 

In the opposite limit $V \ll V_{\rm int}$, the quantum phase diagram was not yet explored in the present context. Here the problem is similar to the fractional quantum Hall problem, where the Wigner crystal phase can be identified with the skyrmion crystal. The question then arises whether this crystal can be quantum-melted for certain parameters into a skyrmion liquid 
phase that is rather akin to a fractional quantum Hall (FQH) state. 

\subsection*{Skyrmion-fracton correspondence}

The conservation of the topological dipole moment \eqref{eq:DipoleMoment} is a consequence of a microscopic dipole conservation law \cite{Papanicolaou1991,Sorn2024}, which the topological density obeys in case of a translationally invariant magnetic theory,
\begin{align} \label{eq:DipoleConservationLaw}
\partial_t \rho_{\rm top} + \partial_i \partial_j J_{ij} = 0 .
\end{align}
Here, the current tensor $J_{ij}$ is related to the magnetic stress tensor. 

Such microscopic dipole conservation laws are characteristic for fracton matter \cite{Gromov2024} suggesting a close correspondence between quantum skyrmions in magnetic systems and fractons. A remarkable advancement would be to observe fractonic phenomena in skyrmion systems, given the scarcity of a realistic fracton-hosting platform. Two fractonic behaviors, intimately related with dipole conservation, are envisioned for skyrmions: (1) a characteristic slow-relaxation dynamics, e.g.~the subdiffusion of an out-of-equilibrium charge-density profile \cite{Gromov2024} and (2) Hilbert-space fragmentation which can lead to an unusual time evolution that strongly depends on the initial states \cite{Pollmann2020}. To guide an experimental exploration, quantum theories of relaxation dynamics for interacting many-skyrmion systems are needed first. One main challenge is to reliably solve for the full many-body spectrum for such dynamical studies, which commonly requires a large computational resource or a powerful analytical method.

\subsection*{Search for the analogue of GMP modes} 

On the quantum level, the topological density fulfills a closed operator algebra \cite{Sorn2024},
\begin{align} \label{eq:GMP}
[\hat \rho_{\rm top}(\vec r), \hat \rho_{\rm top}(\vec r\,') ] = \frac{i \hbar}{4\pi s} \epsilon_{kj} \partial_k \hat \rho_{\rm top}(\vec r) \partial_j \delta(\vec r - \vec r\,'),
\end{align}
which coincides with the Girvin-MacDonald-Platzman (GMP) algebra known from quantum Hall physics. 

In fractional quantum Hall states, the GMP algebra, together with the ground state and the interaction potential, governs the spectral properties of the low-lying charge-neutral density-wave excitations, known as the GMP modes. The latter have been identified in the long-wavelength limit as graviton modes  \cite{Haldane2012}. Given that the topological density also obeys the GMP algebra, a fascinating future direction is to explore how analogous GMP modes could arise in a FQH-like state of quantum skyrmions. Whether these modes can arise at all, how their lifetime depends on the inter-skyrmion interactions, and how feasible a skyrmion system can be tuned to observe GMP modes remain open questions. On a more general note, the relation between the GMP algebra \eqref{eq:GMP} valid for magnetic systems in the continuum limit and the stabilization of spin liquids on discrete lattices should be elucidated. 

\subsection*{Non-fermi liquids due to quantum skyrmions} 

Given the unusual properties of quantum skyrmions, their interaction with Fermi liquids deserves further investigations. In the adiabatic approximation, this interaction reduces to a coupling of electrons to an effective gauge field $\vec A$ whose curl just corresponds to the topological density $\epsilon_{ij}\partial_i A_j = \rho_{\rm top}$. The dipole conservation law \eqref{eq:DipoleConservationLaw} in turn implies an anomalous gauge field dynamics, that could potentially have a strong impact on the nature of itinerant electrons. A particular interesting question is whether the gauge coupling between electrons and a fracton density obeying a microscopic dipole conservation law \eqref{eq:DipoleConservationLaw} could lead to non-Fermi liquid behavior. This is especially relevant as the experimentally observed anomalous temperature dependence of the resistivity close to the magnetic quantum phase transition in skyrmion-hosting MnSi and FeGe \cite{Pfleiderer2001,Pedrazzini2007} still lack a theoretical understanding. 
    
\section*{Challenges}
The research on quantum skyrmions is still in a early stage with many open questions \cite{Petrovic2025}. Here, we focused mainly on open theoretical problems that might be addressed with a combination of effective field theoretical methods and numerical approaches. The experimental detection and identification of its signatures will be challenging and requires clean systems with skyrmion hosting phases close to zero temperature. 

\section*{Acknowledgment}
We would like to thank J\"org Schmalian for very helpful discussions. M.G. acknowledges support from Deutsche Forschungsgemeinschaft (DFG)
via Project-id 403030645 (SPP 2137 Skyrmionics) and Project-id 445312953.

\endgroup

\newpage

\section{Magnetic Force Microscopy for Skyrmion Imaging}
\begingroup
    \let\section\subsection
    \let\subsection\subsubsection
    \let\subsubsection\paragraph
    \let\paragraph\subparagraph
Emily Darwin$^1$, and Hans Josef Hug$^{1,2}$
\vspace{0.5cm}

\noindent
\textit{{$^1$ Empa, Swiss Federal Laboratories for Materials Science and Technology, Ueberlandstrasse 129, 8600 Dübendorf, Switzerland\\
$^2$ Department of Physics, University of Basel, Klingelbergstrasse 82, 4056 Basel, Switzerland}}

\section*{Introduction and Advantages of Magnetic Force Microscopy}
As a lab-based technique, Magnetic Force Microscopy (MFM) is widely available and offers rapid sample turn-around, making it ideal for the efficient screening of magnetic materials and multilayer systems with varying material combinations, thicknesses, and stacking sequences, as well as for the development and optimization of derived devices\cite{Hug2021chapter}. It supports multimodal imaging, simultaneously capturing topography, magnetic stray fields, and, local variations in surface potential via Kelvin probe force microscopy\cite{jaafar2011distinguishing}. This capability has been valuable since the first skyrmions were imaged in bulk-like systems (Figure~\ref{fig:Hug1}{\bf a}) \cite{milde2013unwinding}, and is now being used to investigate complex multilayer stacks engineered to host skyrmions\cite{soumyanarayanan2017tunable,mandru2020coexistence,grelier2022threedimensional}, where structural, magnetic, and electronic properties must be correlated with nanoscale precision (Figs.~\ref{fig:Hug1}{\bf b} to {\bf d}) \cite{soumyanarayanan2017tunable,mandru2020coexistence,grelier2022threedimensional}. MFM enables operando studies of skyrmion-based devices\cite{Liu2025, maccariello2018electrical}, where magnetic textures are imaged while electrical currents are applied, permitting insight into current-induced effects such as skyrmion motion, deformation, or annihilation (Figs.\ref{fig:Hug1}{\bf e} to {\bf i}) \cite{hrabec2017current,maccariello2018electrical,pham2024fast}. In advanced systems, in-plane and out-of-plane magnetic fields can be applied to study micromagnetic evolution, while variable-temperature operation allows exploration of thermal stability and phase transitions\cite{Raju2021}. MFM can probe buried magnetic layers, such as those found in magnetic tunnel junctions (MTJs)\cite{Liu2025} or device stacks, since it senses stray fields penetrating through capping layers.

\section*{Current Practice and Limitations}
Most MFM experiments are still performed under ambient conditions, where the mechanical quality factor $Q$ of the cantilever is low (around 200), significantly limiting sensitivity to small magnetic forces\cite{Feng2022}. Typically, the cantilever is driven at a frequency near its free resonance, and a two-pass (lift-mode) operation is employed: in the first pass, topography is recorded in intermittent contact (tapping) mode. In the second pass, the tip is lifted and retraces the same line to record phase shifts caused by long-range magnetic, electrostatic, and van der Waals force gradients\cite{meyer2021scanning}.

\begin{figure}[h!]
\centering
\includegraphics[width = 0.55\textwidth]{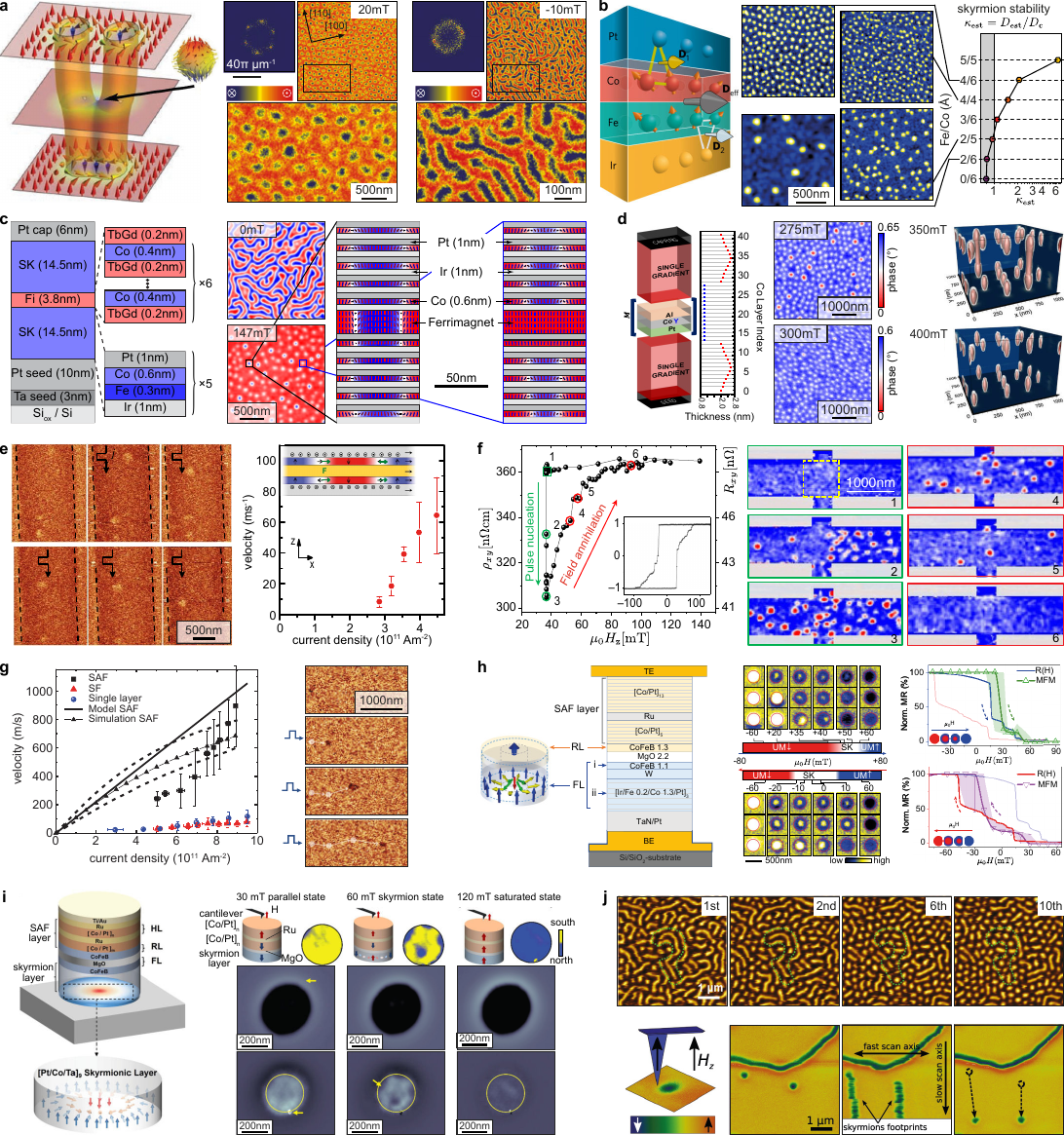}
\captionsetup{font=tiny,skip=-1pt,justification=raggedright,singlelinecheck=false}
\caption{\textbf{a} Left: Schematic illustration of the merging of two skyrmions. Right: MFM images recorded after field cooling to 10\,K show skyrmions at $+20$\,mT and stripe domains at $-10$\,mT. Modified from \cite{milde2013unwinding}.  
\textbf{b} Left: Schematic of a multilayer stack used to study additive interfacial DMI. Center: MFM images reveal different skyrmion densities depending on the Co/Fe thickness ratio. Right: Stability map based on the ratio $D_\mathrm{est}/D_\mathrm{c}$ shows transition from isolated skyrmions to lattices. Adapted from Fig.\,5 of \cite{soumyanarayanan2017tunable}.  
\textbf{c} Left: Structure of a ferromagnetic/ferrimagnetic multilayer designed to host multiple skyrmion states. Center: MFM images at 0\,mT and 147\,mT show stripe domains and two distinct skyrmion types. Right: Micromagnetic simulations of the layer-resolved spin textures. Based on \cite{mandru2020coexistence}.  
\textbf{d} Left: Stack for creating skyrmionic cocoons coexisting with tubular skyrmions. Center: MFM images show both types at 275\,mT and only cocoons at 400\,mT. Right: Simulated spin textures illustrate the distinct configurations. Modified from Fig.\,3 of \cite{grelier2022threedimensional}.  
\textbf{e} Left: MFM images showing skyrmion motion along a track induced by electrical current pulses. Right: Velocity as a function of pulse current density demonstrates motion up to 65\,m/s. Adapted from Fig.\,2 of \cite{hrabec2017current}.  
\textbf{f} Left: Hall resistivity drops (green arrow) and increases (red arrow) due to skyrmion nucleation and annihilation driven by current pulses and by increasing the field. Middle: MFM reveals skyrmions nucleated inside the Hall cross. Right: Skyrmion annihilation tracked via MFM at increasing fields. Adapted from Fig.\,3 of \cite{maccariello2018electrical}.  
\textbf{g} Left: Current-driven dynamics of synthetic antiferromagnetic skyrmions with velocities up to 900\,m/s plotted versus current density. Right: MFM images between pulses show nearly linear motion. Modified from Figs.\,4 and 3 of \cite{pham2024fast}.  
\textbf{h} Left: Schematic of an MTJ with a skyrmion-hosting free layer. Center: MFM images track skyrmion evolution with magnetic field. Right: Magnetoresistance curves agree with skyrmion states identified in MFM. Modified from Fig.\,2 of \cite{chen2024allelectrical}.  
\textbf{i} Left: Schematic of a skyrmion-based MTJ exhibiting a spin-torque-diode effect. Top: MFM at 30, 60, and 120\,mT shows south-saturated, skyrmion, and north-saturated states. Middle: Signal dominated by SAF stray fields. Bottom: Differential MFM isolates the skyrmion layer signal. Based on Figs.\,1 and 2 of \cite{Liu2025}.  
\textbf{j} Top: MFM images after successive scan passes show tip-induced transformation from stripe to skyrmion states. Bottom left: Schematic of the MFM tip over a skyrmion. Other panels: MFM images of skyrmion nucleation under different scan modes. Adapted from \cite{casiraghi2019individual}.}
\label{fig:Hug1}
\end{figure}

While robust and widely used, this lift-mode has several limitations. First, separation between topographic and magnetic signals is incomplete, complicating interpretation. This is vital for imaging different skyrmion phases and 3D skyrmionic structures, such as tubular skyrmions, as shown in Figs.~\ref{fig:Hug1}{\bf c} and \ref{fig:Hug1}{\bf d} \cite{mandru2020coexistence,grelier2022threedimensional}. Second, stable intermittent contact requires that mechanical energy stored in the cantilever exceed energy dissipated per cycle. This necessitates moderately stiff cantilevers (a few N/m) and large oscillation amplitudes ($ \sim 20\,\mathrm{nm}$), which increase tip-sample distance and reduce spatial resolution. Increased stiffness and low $Q$ also leads to enhanced thermal force noise\cite{Feng2022}. 
To compensate for low signal strength, MFM tips are often coated with thick magnetic layers to increase the magnetic moment. However, this degrades spatial resolution and increases the tip’s stray field, which can unintentionally alter the micromagnetic configuration of the sample. Although strong tip-sample interaction can be exploited for controlled manipulation (see Fig.~\ref{fig:Hug1}{\bf j}) \cite{casiraghi2019individual}, it compromises the integrity of magnetic imaging by perturbing the very textures being measured. For techniques aiming to faithfully map static magnetic states, such perturbations undermine the reliability and interpretability of the data.

\section*{Advanced Vacuum and Quantitative MFM}
Vacuum-based MFM, particularly at low temperatures, is currently pursued by only a few expert groups, as its operation remains technically demanding, time-consuming, and requires specialized expertise in scanning force microscopy instrumentation (Fig.~\ref{fig:Hug1}{\bf a}, and \ref{fig:Hug2}{\bf a}, to {\bf c}. Nonetheless, variable-temperature instruments have been successfully employed to detect nanoscale skyrmions in complex oxide materials under several Tesla of magnetic field (Fig.~\ref{fig:Hug2}{\bf a} and {\bf c}), where a topological Hall effect emerges\cite{meng2019observation}. In multilayer heterostructures, the combination of magnetic imaging and variable-temperature magnetotransport has demonstrated that the topological Hall resistivity scales with the density of isolated skyrmions over a wide range of temperature and magnetic field (Fig.~\ref{fig:Hug2}{\bf b}), confirming the role of the skyrmion Berry phase in electronic transport\cite{raju2019evolution}. Additionally, a power-law enhancement of the topological Hall signal has been observed near the phase boundary between isolated skyrmions and disordered skyrmion lattices, highlighting the critical influence of skyrmion stability and configuration\cite{Raju2021}. These findings underscore the value of low-temperature-compatible magnetic imaging techniques, such as MFM, for the quantitative investigation of skyrmion-related transport phenomena in technologically relevant materials.

While ultra-high vacuum and cryogenic MFM setups enable advanced magnetic imaging at variable temperatures and in the highest magnetic fields, their complexity limits widespread adoption. A more practical alternative is MFM under moderate vacuum conditions ($10^{-7}$ to $10^{-6}$ mbar), which offers a compelling balance between performance and accessibility. Moderate vacuum environments significantly reduce cantilever damping, increasing the quality factor $Q$ from around 200 in air to tens of thousands, and in optimized setups to over one million. When combined with low-stiffness cantilevers, this results in a favorable $k/f_0$ ratio and enhances sensitivity by more than two orders of magnitude compared to ambient MFM conditions\cite{Feng2022}.


This enhanced sensitivity allows the use of ultra-low moment tips, which minimizes tip-induced perturbations.
High sensitivity permits true non-contact, single-pass operation with advanced tip-sample distance control\cite{zhao2018magnetic}, crucial for drift-free measurements during experiments with temperature variation, and field application (Fig.~\ref{fig:Hug2}{\bf a}).

On flat samples, maintaining a constant average tip height remains the preferred approach, as it avoids the additional noise introduced by active distance control. 
Such stability also supports differential measurements with reversed tip magnetization, enabling unambiguous separation of magnetic and topographic signals and the removal of background contrast arising from variations in saturation magnetization, magnetic layer thickness, or anisotropy (Figs.~\ref{fig:Hug1}{\bf c} and {\bf d}) \cite{bacani2019how,meng2019observation}.
Most importantly, quantitative understanding of MFM contrast becomes possible under well-controlled conditions (Fig.~\ref{fig:Hug2}{\bf a}) \cite{meyer2021scanning,bacani2019how}. Imaging artifacts such as domain wall distortion, caused by the tip’s oscillation path, can be corrected, aiding the identification of spin textures such as skyrmions, antiskyrmions, bubbles, or higher-order topologies (Figs.~\ref{fig:Hug1}{\bf e}) \cite{koraltan2025signatures}. By calibrating the tip’s response to stray fields with known spatial frequency content, measured MFM contrast can be quantitatively matched to simulations of candidate magnetization patterns (Fig.~\ref{fig:Hug2}{\bf c}) \cite{bacani2019how,meng2019observation}. Such a quantitative approach may be particularly beneficial for interpreting recent MFM data on skyrmions in synthetic antiferromagnetic multilayers \cite{legrand2020room}, where the expected stray-field signal is inherently weak (Fig.~\ref{fig:Hug1}{\bf g} \cite{pham2024fast} and Fig.~\ref{fig:Hug2}{\bf f} \cite{Sim2025}).

However, local tip-sample distance control, combined with simultaneous Kelvin potential mapping\cite{jaafar2011distinguishing}, is essential for topographically complex or curved substrates. Measurements on 3D subjects, such 
MTJs (Fig.~\ref{fig:Hug1}{\bf h} and \ref{fig:Hug1}{\bf i})\cite{chen2024allelectrical, Liu2025} and 
curved surfaces (Fig.~\ref{fig:Hug2}{\bf g}) \cite{dugato2025curved} and , are becoming more relevant and in demand. However, note that the data shown in Fig.~\ref{fig:Hug2}{\bf g} has still been obtained by an MFM operating under ambient conditions where the tip's magnetic stray field can lead to significant modifications of the initial micromagnetic stucture.

\begin{figure}[h!]
\centering
\includegraphics[width=0.55\textwidth]{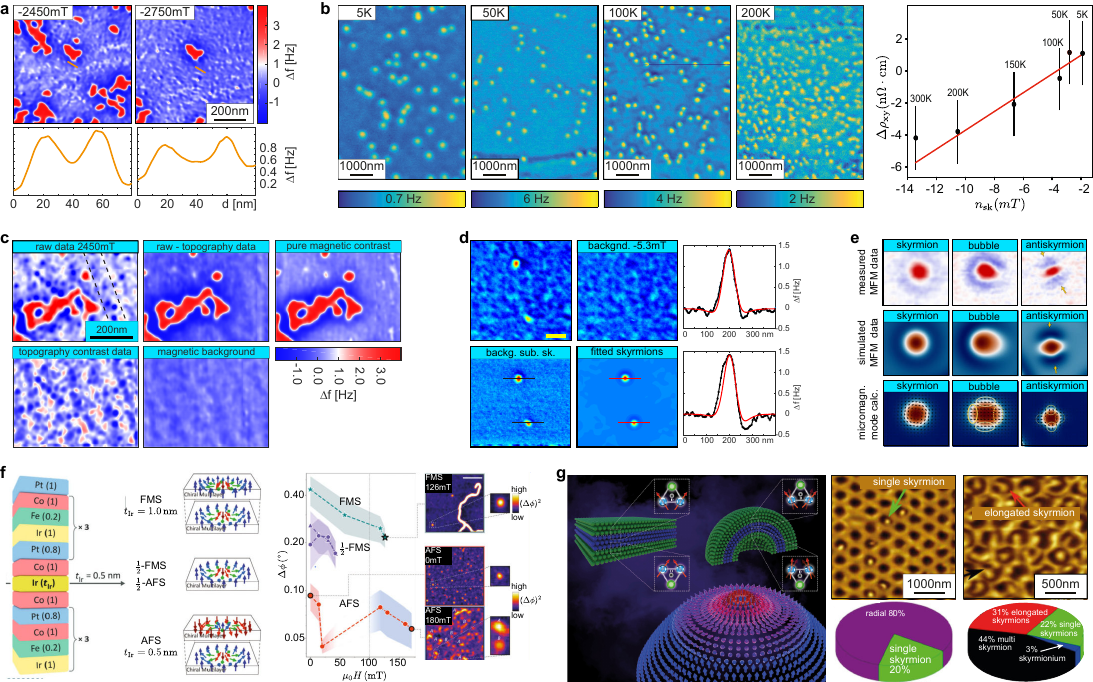}
\captionsetup{font=tiny,skip=-1pt,justification=raggedright,singlelinecheck=false}
\caption{\textbf{a} Magnetic force microscopy (MFM) images of a bilayer composed of 2\,unit cells (uc) SrIrO$_3$ on 10\,uc SrRuO$_3$, epitaxially grown on (001)-oriented SrTiO$_3$ substrates. Images were recorded at 10\,K under magnetic fields of $-2450$\,mT and $-2750$\,mT, following saturation at $+7000$\,mT. Cross-sections (bottom row) confirm the presence of nanoscale skyrmions \cite{meng2019observation}.  
\textbf{b} Temperature dependence of skyrmion density observed via MFM, and the corresponding transverse Hall resistivity $\Delta \rho_{x,y}$ after subtraction of ordinary and anomalous Hall contributions. The residual signal reflects the topological Hall effect. Modified from Fig.\,2 of \cite{raju2019evolution}.  
\textbf{c} The identification of skyrmions in epitaxial SrIrO$_3$/SrRuO$_3$ bilayers at 10\,K (top right panel) requires the removal of background contributions, as the frequency-shift signals from topography and long-range magnetic contrast are of similar or even larger magnitude than the skyrmion signal. Only after subtracting these background components can the skyrmions be reliably resolved in the MFM data \cite{meng2019observation}.
\textbf{d} Quantitative analysis of MFM signals from individual skyrmions. Bottom left: background-subtracted experimental MFM data. Bottom right: simulated MFM signal based on model magnetization patterns. Right: comparison of line profiles between measured and fitted frequency shift data \cite{bacani2019how}.  
\textbf{e} Identification of distinct spin textures using MFM. Top, middle, and bottom rows show measured MFM data, simulated MFM data, and corresponding magnetization configurations, respectively, for a skyrmion, a trivial bubble, and an antiskyrmion \cite{koraltan2025signatures}.  
\textbf{f} Skyrmions in synthetic antiferromagnets. Left: multilayer stack with three Ir/Fe/Co/Pt repetitions magnetically coupled either ferromagnetically or antiferromagnetically to three identical repetitions. Middle: Schematics of the spin textures and MFM images acquired for ferromagnetic (FMS) and antiferromagnetic (AFS) coupling. Modified from Fig.\,4 of \cite{Sim2025}.  
\textbf{g} Role of curvature-induced Dzyaloshinskii–Moriya interaction (DMI) in shaping spin textures. Left: schematic representation of curvature-modified DMI. Center: MFM images of Co/Pt multilayers with effective anisotropies $K_{\mathrm{eff}} \approx 75$\,kJ/m$^3$ and $\approx 10$\,kJ/m$^3$ show the formation of skyrmions and more complex textures. Right: pie charts quantify the distribution of different spin textures. Modified from Fig.\,4 and abstract figure of \cite{dugato2025curved}.}
\label{fig:Hug2}
\end{figure}

\section*{Future Developments}
Advanced vacuum-based MFM enables quantitative, high-resolution, and minimally invasive imaging of skyrmions in real devices. It offers unprecedented insight into material functionality under controlled external stimuli. Yet these capabilities remain limited to a few labs, as commercial instruments lack the necessary performance and flexibility. Broader availability would greatly facilitate the wider adoption of these techniques across the magnetism and spintronics communities.

Future developments include vacuum MFM measurements of samples with curved surfaces or substantial topographies and the integration of quantitative MFM with micromagnetic simulation packages, particularly if advanced matching of the MFM contrast of candidate micromagnetic structures and their dependence on the applied field with measured MFM data were implemented.

Reducing tip moment further could improve performance, for instance, using low-moment alloys or multipole tips.  Promising designs include antiferromagnetically coupled ferromagnetic layers, forming compact magnetic dipoles parallel to the surface. These could enhance resolution while minimizing perturbations.  Cantilevers with integrated coils to switch tip magnetization in situ would allow fast reversal for differential imaging, supporting the unambiguous separation of magnetic and topographic signals. 

In UHV, extreme sensitivity achievable with ultra-high quality factors may eventually enable detection of individual atomic spins via stray fields, using tips functionalized with single-spin molecules. Combined with multimodal operation, e.g. simultaneously using multiple flexural resonances, may allow atomic-resolution imaging while detecting individual magnetic moments. Additional applications include probing local piezoelectric phenomena, extending MFM’s reach to multiferroic and magnetoelectric materials.

Currently, MFM struggles with studying dynamic magnetic phenomena. Skyrmion motion induced by current pulses has been imaged through static snapshots of successive states, but real-time dynamics remain elusive. A potential path forward is to measure averaged responses to periodic excitation, such as local actuation of spin textures. Advanced schemes may use ultrafast tip magnetization switching triggered by laser pulses for stroboscopic imaging of magnetization dynamics at nanometer resolution.

\endgroup

\newpage

\section{Topological spin textures in three-dimensional nano-geometries}
\begingroup
    \let\section\subsection
    \let\subsection\subsubsection
    \let\subsubsection\paragraph
    \let\paragraph\subparagraph
Amalio Fernández-Pacheco$^1$, and Claire Donnelly$^{2,3}$
\vspace{0.5cm}

\textit{$^1$ Institute of Applied Physics, TU Wien, Wiedner Hauptstraße 8-10, Vienna, 1040, Austria\\
$^2$ Max Planck Institute for Chemical Physics of Solids, Nöthnitzer Str. 40, D-01187 Dresden, Germany\\
$^3$ International Institute for Sustainability with Knotted Chiral Meta Matter (WPI-SKCM2), Hiroshima University, Hiroshima, 739–8526 Japan}

\section*{Introduction}

Topological spin textures such as domain walls, skyrmions, and more recently, torons, hopfions and Bloch points are at the forefront of emerging spintronic technologies. These nontrivial magnetic configurations may offer transformative functionalities that go beyond the capabilities of conventional monodomain-based devices used in current magnetoresistive sensors, spin valves, and magnetic random-access memory (MRAM).

At the core of the functionality of topological spin textures lies their intrinsic noncollinear spin structure which couples efficiently to external stimuli like spin-polarized currents, enabling manipulation via spin-transfer and spin–orbit torques. This behavior underpins proposed applications such as racetrack memory, logic gates, and neuromorphic computing elements. In contrast to topologically trivial textures, their topology provides enhanced robustness against thermal and disorder-induced fluctuations, a critical feature for device stability at the nanoscale. Moreover, these textures may give rise to emergent electrodynamic phenomena—such as the topological Hall effect—originating from the Berry curvature associated with their real-space spin configuration. The nontrivial spin texture also directly interacts with magnons, modifying their dispersion relation and enabling nonreciprocal spin-wave transport, thus offering new pathways for transport phenomena and readout strategies in spintronic devices.

To date, the majority of realizations of topological spin textures have occurred in two-dimensional (2D) systems and bulk materials, where their formation is driven by bulk or interfacial Dzyaloshinskii–Moriya interaction (DMI), dipolar interactions, or competing exchange interactions (see other sections of this roadmap). Despite significant progress in understanding and controlling these planar and bulk systems, key challenges remain for their practical use, including deterministic nucleation and annihilation, precise manipulation, readout scalability, and overall energy efficiency. Furthermore, most experimental platforms rely on electrical and optical probes, whose resolution and compatibility with industrial processes require further optimization.

A new frontier is now opening in magnetism: the exploration of novel effects emerging from spin textures in three-dimensional (3D) nano-geometries\cite{fernandez-pacheco2017threedimensional}. This exciting direction is becoming possible thanks to key advances in 3D nanofabrication and deposition techniques~\cite{bhattacharya2025selfassembled, pip2020electroless}, including focused electron and ion beam-induced deposition, two-photon lithography, glancing-angle deposition, atomic-layer and electro-deposition of ferromagnetic thin films, and advanced 3D etching (see Figure~\ref{fig:amalio1}), which offer unprecedented control over complex magnetic architectures. Effects like curvature, torsion, and the topology of the underlying geometry directly couple to the spin structure, modifying effective anisotropies and DMI. These geometric interactions can stabilize new classes of textures, such as hopfions (three-dimensional analogs of skyrmions with a nonzero Hopf index), 3D skyrmion strings, torons, and Bloch point singularities. Moreover, 3D effects can be exploited to engineer unidirectional (nonreciprocal) spin-wave propagation~\cite{xu2025geometry} and diode-like effects in domain wall dynamics~\cite{farinha2025interplay}, introducing novel functionalities for magnonic and spintronic circuits.

\begin{figure*}[h!]
    \centering
    \includegraphics[width=0.95\linewidth]{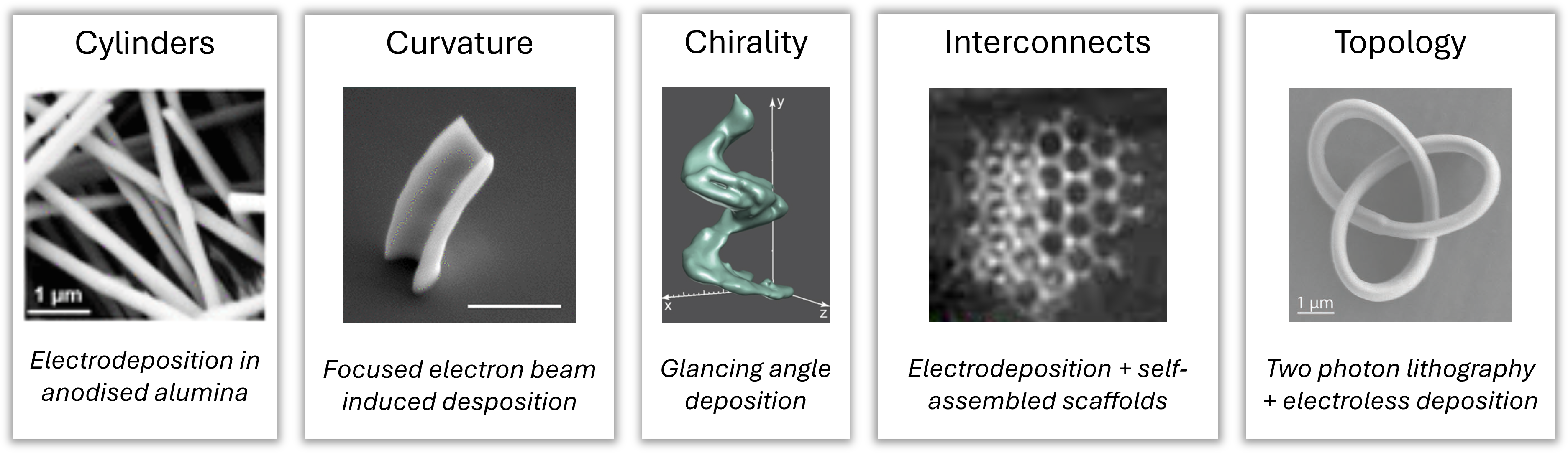}
    \caption{Nanofabrication of three-dimensional magnetic nanostructures. A wide range of geometries have been demonstrated using a variety of techniques. For instance (from left to right), cylindrical nanowires created with  electrodeposition in anodised alumina templates (reproduced from \cite{bhattacharya2025selfassembled}), to curved structures created with focused electron beam induced deposition (FEBID) (reproduced from \cite{skoric2020layerbylayer}), chiral geometries such as helices created with glancing angle deposition (reproduced from ~\cite{phatak2014visualization}), interconnected gyroid lattices created with electrodeposition and block-copolymer scaffolds (reproduced from ~\cite{llandro2020visualizing}), and topologically non-trivial trefoil knot structures, fabricated using a combination of two photon lithography and sputtering (Reproduced from ~\cite{pip2020electroless}).}
    \label{fig:amalio1}
\end{figure*}

\section*{Relevance And Vision}

Topological spin textures manifest in various forms, each characterized by distinct dimensionalities and topological indices. In all cases, geometrical nanoscale effects may play a crucial role, influencing their stability, dynamics, and interactions. Understanding these effects opens new possibilities for controlling magnetic behavior in curved and 3D systems. In Figure~\ref{fig:amalio2}, a series of proof-of-concept devices is included to highlight the influence of three dimensional geometries on topological spin textures of increasing dimensionality, from domain walls to vortices, skyrmions, Bloch points, and hopfions.

\textit{Domain walls in 3D nanowires.} In three-dimensional ferromagnetic nanowires, engineered geometrical gradients such as thickness variations along the wire length can induce domain wall automotion, enabling fast and robust all-magnetic interconnection between nanoelectronic planes, as demonstrated in spiral 3D-printed nanowires~\cite{skoric2022domain}. Moreover, introducing torsion in racetrack nanowires generates a local chiral field that stabilizes Bloch-type domain walls and results in handedness-dependent, non-reciprocal domain wall motion, effectively functioning as a diode-like shift register~\cite{farinha2025interplay}.

\textit{Vortex states in 3D.} Vortex spin states are also modified in 3D geometries. In curvilinear structures such as thick spherical caps, the magnetic configuration of vortices deviates from planar counterparts due to strong coupling between the geometric chirality of the vortex core string and the magnetic helicity of the surrounding flux-closure spin texture~\cite{volkov2023chirality}. Furthermore, 3D interconnected soft-magnetic structures like tetrapods and toroidal meshes use their shape and topology (specifically the number of loops and holes, described by the Euler characteristic) to control where magnetic vortices and antivortices form ~\cite{volkov2024threedimensional}, allowing precise control over complex magnetic patterns and enabling the design of magnetic soliton networks with potential applications in brain-inspired computing.

\textit{Skyrmions in curved systems.} The effect of curvature also impacts skyrmion textures, as curved nanostructures host effective chiral spin interactions that mimic the microscopic DMI, traditionally associated with noncentrosymmetric bulk materials or multilayers interfaced with heavy metals and oxides. Strongly curved caps have been both theoretically predicted~\cite{kravchuk2018multiplet} and experimentally shown to facilitate skyrmion nucleation~\cite{dugato2025curved}. Additionally, it is expected that curvature will fundamentally alter skyrmion dynamics by introducing an effective inertia or drift mass, deforming skyrmion shapes, breaking symmetry in their motion, and leading to non-reciprocal, direction-dependent dynamics~\cite{korniienko2020effect}.

\textit{Helical geometries and magnetic chirality.} Magnetic chirality can also be introduced by helical nano-geometries~\cite{fullertonfractional}. In double-helix nanowires, chiral spin textures such as helical domain walls~\cite{sanz2020artificial} and fractional skyrmion tubes can be stabilized~\cite{fullertonfractional}. Double helices may also give rise to nonconventional arrangements of dipolar-coupled domain wall pairs, resulting in vortices and antivortices distributed both inside the material and in the surrounding free space~\cite{donnelly2022complex}, with higher-order configurations emerging depending on the spatial arrangement and interaction of the helix strands~\cite{fullerton2025design}.

\textit{Bloch point singularities in 3D nanostructures.} Beyond well-known spin textures present in planar systems, three-dimensional geometries allow for the emergence of Bloch points—singular magnetization defects unique to 3D systems where the magnetization magnitude locally vanishes. These defects play a key role in topological transformations by enabling transitions impossible through continuous deformations. Bloch points can be nucleated in cylindrical nanowires, where the symmetry of the 3D geometry supports their controlled formation within stable domain walls~\cite{dacol2014observation}. By introducing curvature gradients into cylindrical nanowires that host such domain walls, Bloch point domain walls could be controlled by the definition of curvature-induced asymmetric energy landscapes~\cite{ruiz-gomez2025tailoring}. This curvature-induced asymmetry not only stabilizes the Bloch points but also directs their motion unidirectionally, resulting in non-reciprocal domain wall propagation and leading to the demonstration of curvature-defined magnetic shift registers~\cite{ruiz-gomez2025tailoring}.

\textit{Topologically non-trivial geometries.} As well as introducing curved and chiral geometries, changing the topology of the geometry itself has been shown to play a crucial role in the formation and stability of topological spin textures~\cite{castillo-sepulveda2021magnetic}. Micromagnetic simulations demonstrate that sufficiently large tori can stabilize two distinct hopfion configurations, characterized by abrupt energy transitions corresponding to changes in topological charge. Beyond static stabilization, hopfions are predicted to exhibit diverse internal dynamics, including rich resonant spin-wave modes and complex current-driven phenomena such as translation, rotation, and breathing motions under spin-transfer torque~\cite{wang2019current}. The inherent topological protection of hopfions gives rise to emergent electromagnetic effects, making them particularly attractive for developing robust, multifunctional spintronic components in three-dimensional device architectures~\cite{khodzhaev2022hopfion, bo2021spin}.

\begin{figure*}[h!]
    \centering
    \includegraphics[width=0.95\linewidth]{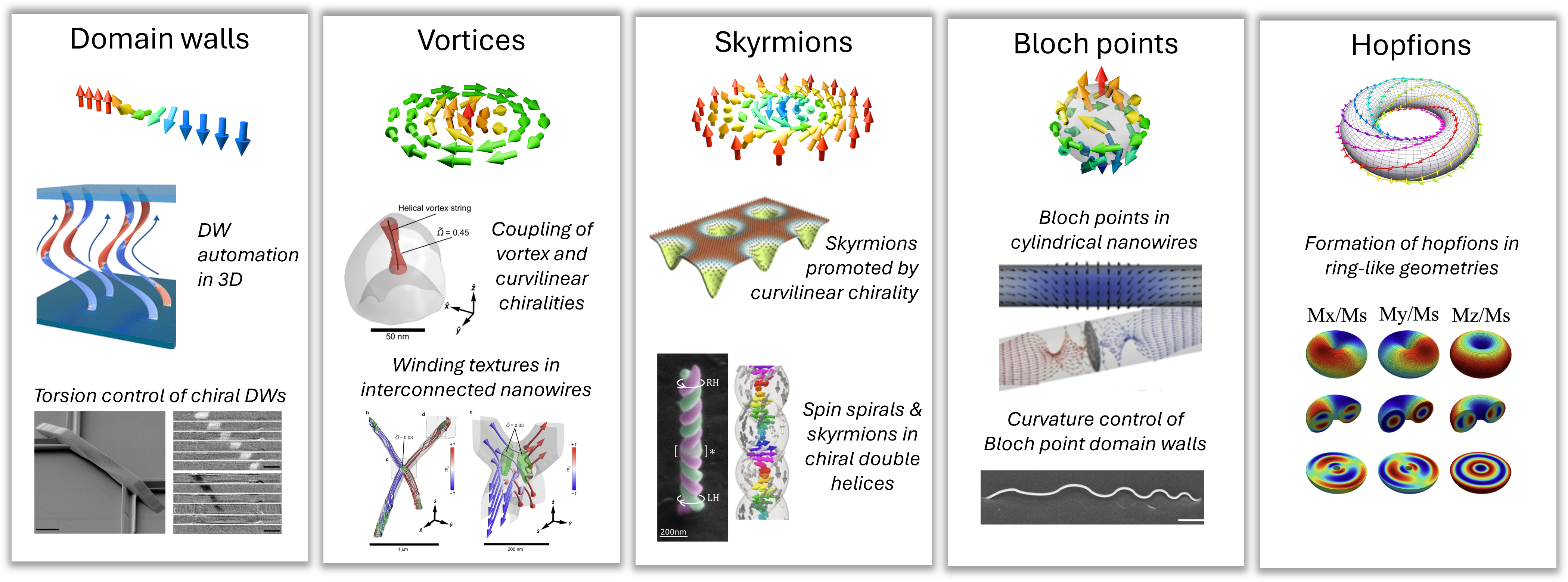}
    \caption{Topological textures in three-dimensional nano-geometries. From left to right: Domain walls: automotion in helical conduits (reproduced from \cite{skoric2022domain}) as well as the influence of torsion on chiral domain wall propagation (reproduced from \cite{farinha2025interplay}). Vortices: curved domes lead to a curvature influence on the chirality of the vortex core (reproduced  from \cite{volkov2023chirality}), as well as winding textures forming in interconnected nanowires (reproduced from \cite{volkov2024threedimensional}). Skyrmions: local curvature promotes the formation of skyrmions (reproduced from \cite{kravchuk2018multiplet}), while spin spirals and fractional skyrmions have been stabilised in double helix nanostructures (reproduced from \cite{fullertonfractional}). Bloch points: stabilisation in cylindrical nanowires (reproduced from \cite{dacol2014observation}) and curvature-induced control of their energy landscape (reproduced from \cite{ruiz-gomez2025tailoring}). Hopfions: formation of hopfions in patterned chiral nanomagnets (reproduced from~\cite{castillo-sepulveda2021magnetic}).}
    \label{fig:amalio2}
\end{figure*}

\section*{Outlook}

The integration of curvature, topology, and dimensionality in 3D nanomagnetic systems opens promising new avenues across spintronics and related fields. Recent breakthroughs have begun to demonstrate the feasibility of 3D topological control, including successful imprinting of skyrmionic textures in 3D nanostructures~\cite{fullertonfractional}, quantification of curvature-driven DMI and the mapping of topological indices in three dimensional space, and experimental verification of geometry-governed topological transformations. These initial achievements highlight the considerable potential that may be unlocked through further developments.

While many fundamental spintronic functionalities—such as unidirectional information transport, energy-efficient operations, topologically protected logic, and neuromorphic computing—can be realized in 2D systems, the transition to three dimensions could offer distinctive advantages. These include vastly enhanced interconnectivity capabilities, novel coupling mechanisms between geometric features and spin states, nonreciprocal transport effects, and geometrically-induced magnetic chirality that may prove difficult to achieve in planar configurations.

Emerging electromagnetic responses, such as predicted orbital Hall effects and magneto-optical phenomena driven by noncoplanar magnetization arrangements, suggest new physical mechanisms that could enable multifunctional device operation. Furthermore, the potential extension of these architectures to incorporate spin-wave diodes and topologically guided magnonic waveguides may pave the way for future wave-based logic and sensing technologies. Beyond conventional magnetic platforms, 3D geometries could enable previously inaccessible coupling mechanisms, including curvature-mediated anisotropy and chiral symmetry breaking effects that might be dynamically tuned through external stimuli such as strain and temperature. These capabilities appear especially promising for unconventional systems like 2D van der Waals magnets and antiferro- or altermagnetic materials, potentially offering novel pathways toward robust and reconfigurable spin manipulation.

\section*{Challenges And Requirements.} Progress in this area is contingent on advances in both experimental and theoretical fronts. On the characterization side, fourth-generation synchrotron sources and coherent imaging methods are pushing X-ray magnetic tomography resolution into sub-10 nm regimes, enabling non-destructive, three-dimensional mapping of complex spin textures. Techniques such as Lorentz ptychography and electron magnetic circular dichroism (EMCD) are approaching atomic-scale resolution, critical for resolving singularities like Bloch points, albeit often requiring large applied fields. Simultaneously, robust control over texture nucleation and annihilation in 3D remains an outstanding challenge. While micromagnetic simulations may predict deterministic control under ultrafast or strain-mediated conditions, experimental demonstrations remain limited. The development of energy-efficient electrical, optical, and strain-based nucleation strategies will be essential for functional devices.

Accurate three-dimensional micromagnetic and atomistic modelling must also evolve to fully capture the impact of curvature, torsion, and topological constraints in real geometries. Multi-scale approaches that seamlessly bridge atomic and micromagnetic simulations will be particularly critical for modelling Bloch point singularities, where the characteristic length scales of magnetic textures approach atomic dimensions and demand high-resolution treatment of the magnetization field. Simulations must also account for the influence of material heterogeneities, grain boundaries, and edge effects, which become increasingly pronounced at the nanoscale and are currently less controlled in 3D structures compared to their 2D counterparts.

To date, most investigations of 3D magnetic effects have been conducted in ferromagnetic metals and their alloys. However, the possibility of integrating exotic materials into 3D architectures promises to open exciting new avenues. Effective fabrication pathways are essential to incorporate advanced materials—such as low-damping and chiral systems, multilayer heterostructures, and tunable ferri- and antiferromagnets—into complex 3D nano-geometries, enabling both exploration and exploitation of their intrinsically stable topological textures and the novel functionalities that emerge from three-dimensional structuring.

Ultimately, future device architectures should be designed so that 3D spin textures serve both as active elements (directly performing information processing or logic functions) and as integral structural components, while the device's 3D geometry itself shapes, stabilizes, and controls the magnetic textures, effectively merging form and function within the device. This may include devices such as logic gates based on skyrmion or hopfion-mediated interactions, neuromorphic synapses exploiting stochastic domain wall dynamics, and reconfigurable magnonic channels shaped by engineered topology.

\section*{Acknowledgements}
AFP acknowledges support by the European Community under the Horizon 2020 Program, Contract No. 101001290 (3DNANOMAG). C.D. acknowledges funding from the Max Planck Society Lise Meitner Excellence Program and funding from the European Research Council (ERC) under the ERC Starting Grant No. 3DNANOQUANT 101116043. We acknowledge funding by the Austrian Science Fund (FWF) [10.55776/PIN1629824].

\endgroup

\newpage

\section{2D Skyrmion Lattice Phases}
\begingroup
    \let\section\subsection
    \let\subsection\subsubsection
    \let\subsubsection\paragraph
    \let\paragraph\subparagraph
Raphael Gruber$^1$, and Mathias Kl\"aui$^{1,2}$
\vspace{0.5cm}

\noindent
\textit{$^1$ Institute of Physics, Johannes Gutenberg University Mainz, 55099 Mainz, Germany\\
$^2$ Center for Quantum Spintronics, Department of Physics, Norwegian University of Science and Technology,
7491 Trondheim, Norway}

\section*{Introduction}

Phases and phase transitions in two dimensions (2D) are fundamentally distinct from those in other dimensions, a feature that has drawn significant scientific interest over decades~\cite{kosterlitz1973ordering}. While in 3D, the translationally ordered solid phase and the isotropic liquid phase are well-known, the situation is more complex in 2D, where an intermediate hexatic phase with only orientational order can arise~\cite{kosterlitz1973ordering,nelson1979dislocation}. The landmark theoretical framework developed by Kosterlitz, Thouless, Halperin, Nelson, and Young (KTHNY theory) predicts that topological defects play a pivotal role in mediating the phase transitions between those 2D phases~\cite{nelson1979dislocation}. In this picture, each site in a 2D lattice with a coordination number $N \neq 6$ constitutes a topological defect. While the solid phase features only tightly bound dislocation pairs (with a fivefold and a sevenfold coordinated site comprising a dislocation), the hexatic phase is characterized by the unbinding of these dislocations. Finally, in the isotropic liquid phase, dislocations further decompose into disclinations (isolated single defects), which then proliferate (Fig. ~\ref{fig:Klaui}a)~\cite{nelson1979dislocation}.

\begin{figure*}[h!]
\centering
\includegraphics[width=0.95\textwidth]{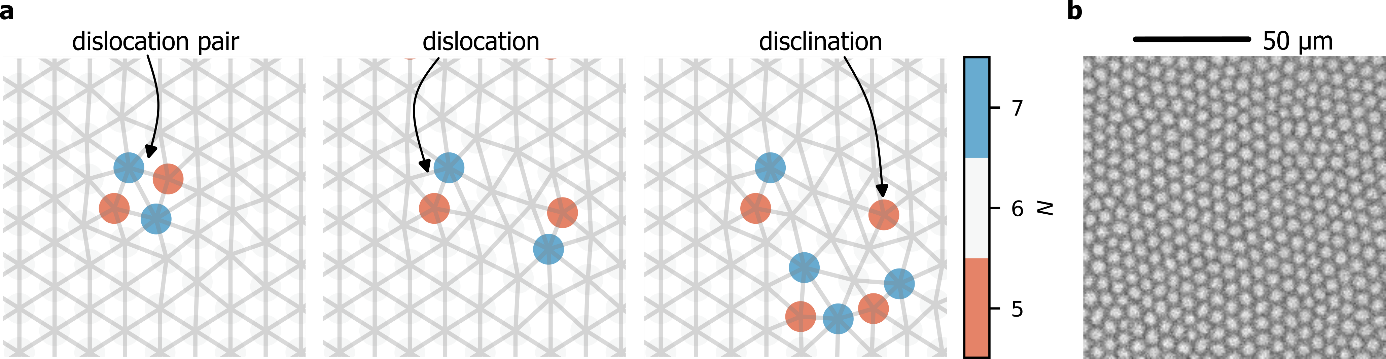}
\caption{(a) Schematic of topological lattice defects. Lattice neighbor connections are indicated by the grey lines. The colored dots indicated the number of lattice neighbors (coordination number) $N$. For slight distortions of an ideal lattice ($N=6$), dislocation pairs of two $N=5$ and $N=7$ defects form. With decreasing order across 2D phase transitions, they unbind into dislocations and disclinations. (b) Kerr microscopy snapshot of a skyrmion lattice in a CoFeB thin film. Black/white contrast corresponds to magnetization pointing up/down.}
\label{fig:Klaui}
\end{figure*}

This sequence of defect-mediated transitions has been directly observed in 2D skyrmion lattices (SkLs), realized in a variety of material platforms including magnetic monolayers (e.g., Fe/Ir(111)), van der Waals magnets (e.g., $\rm CrI_3$, $\rm Fe_2GeTe_2$), oxide heterostructures (e.g., $\rm Cu_2OSeO_3$), and amorphous or polycrystalline thin films (e.g., CoFeB, Fig. ~\ref{fig:Klaui}b)~\cite{huang2020melting, gruber2025imaging, zazvorka2020skyrmion}. Advances in materials synthesis techniques such as molecular beam epitaxy (MBE) and chemical vapor deposition (CVD) have enabled the engineering of key magnetic parameters—Dzyaloshinskii–Moriya interaction (DMI), anisotropy, and exchange stiffness—allowing precise stabilization of SkLs under ambient and tunable conditions.

As a result, 2D SkL phases have become highly relevant not only for fundamental studies in topological matter and phase transitions, but also for emerging spintronic applications. Their inherent topological stabilization, tunability, and compatibility with thin-film technologies position them as a versatile and promising class of physical systems.

\section*{Relevance}

At the heart of the relevance of 2D SkLs is their function as an accessible and tunable platform to investigate topological phase transitions and defect-mediated dynamics in two dimensions. Unlike colloidal or superconducting vortex lattices, SkLs allow exceptional control over parameters such as lattice spacing, interaction potential, and both skyrmion as well as defect mobility~\cite{huang2020melting, gruber2025imaging, gruber2023300}—making them uniquely suited to experimentally realize and manipulate the scenarios described by KTHNY theory~\cite{klaui2020freezing}.

In thin CoFeB films, for example, the skyrmion size and mobility can be modulated in situ via the applied magnetic field~\cite{gruber2023300}. In $\rm Cu_2OSeO_3$, the interaction potential among skyrmions is similarly field-dependent, enabling fine control over lattice rigidity~\cite{huang2020melting}. More generally, both temperature and external magnetic field govern SkL properties such as density and spacing, allowing the phase space of 2D transitions to be thoroughly explored~\cite{zazvorka2020skyrmion, gruber2022skyrmion} and enabling tuning of the dynamics~\cite{gruber2025imaging, gruber2023300}.

While nanoscale skyrmions are most relevant for device applications, micrometer-sized skyrmions offer distinct advantages for fundamental studies: they are optically detectable in real time and real space, permitting direct visualization of phase transitions and defect dynamics~\cite{gruber2025imaging, zazvorka2020skyrmion}. Kerr microscopy of SkLs in CoFeB thin films has confirmed the emergence of dislocations and disclinations consistent with KTHNY theory, and enabled the tracking of defect trajectories, revealing that dislocations—acting as higher-order quasiparticles—exhibit diffusion coefficients orders of magnitude larger than the skyrmions forming the underlying lattice~\cite{gruber2025imaging}.

Importantly, SkLs can extend KTHNY physics beyond equilibrium. Due to the chiral nature of skyrmions, they experience a transverse Magnus force, making them ideal for studying topological defect dynamics under non-conservative forces—a largely unexplored frontier~\cite{gruber2025imaging, reichhardt2022statics}. Furthermore, dynamic perturbations such as magnetic field oscillations or different parameter sweeps can drive SkLs out of equilibrium~\cite{gruber2025imaging, gruber2023300}, facilitating studies of, e.g., shock waves, lattice phonons, and dissipative phase transitions.

The system's versatility is further expanded by external control techniques. Electric gating and local irradiation can dynamically manipulate magnetic properties and even pin individual skyrmions to predetermined lattice sites, enabling artificial SkLs with engineered defect structures or novel phases~\cite{reichhardt2022statics, kern2022deterministic}.

Beyond fundamental insights, these properties translate into technological potential. For example, deliberately introducing and manipulating dislocations may provide new modes of low-power information transport~\cite{gruber2025imaging}. If a dislocation hops between two metastable configurations and this distortion can be electrically detected (e.g., via industrially established devices like magnetic tunnel junctions), it could serve as a logic state. Similarly, the nonlinear, history-dependent responses of SkLs to external fields position them as candidates for reservoir computing, where dynamic complexity is an asset~\cite{lee2023perspective}. Key observables include order parameters, elastic constants, noise spectra, and even thermally driven stochasticity, the latter enabling explorations of Brownian computing paradigms~\cite{lee2023perspective}.

\section*{Challenges}

Despite significant advances, several key challenges must be addressed to fully realize the promise of 2D SkLs.

First, spatial or time resolution are currently limiting factors. In many systems, skyrmions are only tens of nanometers or less in diameter, requiring high-resolution techniques such as Lorentz transmission electron microscopy (TEM), scanning tunneling microscopy (STM) or magnetic force microscopy (MFM)~\cite{huang2020melting}. These methods, however, offer limited time resolution, complicating the study of dynamics. By contrast, systems such as CoFeB with micrometer-scale skyrmions allow time-resolved optical imaging~\cite{gruber2025imaging} but suffer from pinning effects due to inhomogeneities and roughness in the film~\cite{gruber2022skyrmion}. Even with optimized growth conditions, such imperfections are unavoidable in real materials and hinder the formation of perfect SkLs~\cite{gruber2025imaging}.

Second, device integration remains a formidable challenge. For SkLs to be functional in practical architectures, they must be stable over large areas, reproducible, and compatible with CMOS technologies~\cite{lee2023perspective}. Precise control over skyrmion generation, annihilation, and manipulation within a lattice is essential, yet more complex than for isolated skyrmions due to collective behavior. Skyrmion-skyrmion interactions and the elastic properties of the lattice can restrict the motion of individual skyrmions~\cite{gruber2025imaging}, necessitating new schemes for local and reversible control—e.g., via tailored current pulses, localized heating, or strain fields—without destabilizing the lattice.

Third, readout and interfacing strategies must be refined. Detection techniques such as tunnel magnetoresistance (TMR) and manipulation by current-induced spin-transfer torque (STT) need to be combined with maintained skyrmion integrity, while providing robust signal contrast and low-power manipulation. Similarly, scalable, in situ, and ideally non-invasive methods for imaging SkLs—especially in buried interfaces and ambient conditions—remain a key requirement.

Finally, from a theoretical standpoint, accurate modeling of SkLs in realistic environments must incorporate disorder, thermal noise, complex interfacial couplings, and non-equilibrium dynamics~\cite{brems2025realizing}. Multiscale modeling frameworks that bridge atomistic simulations with micromagnetic and continuum approaches are essential to capture both local defect behavior and long-range lattice dynamics.

\section*{Conclusion}

In summary, 2D skyrmion lattice phases present an exciting frontier at the convergence of topological statistical physics, materials science, and spintronic device engineering. They not only offer an exceptional platform for studying fundamental aspects of 2D phase transitions and defect dynamics, but also hold promise for applications in reconfigurable, energy-efficient information processing. As we look toward 2026 and beyond, coordinated progress in materials development, experimental methodology, and multiscale modeling will be crucial to unlocking the full potential of SkLs in both scientific discovery and technological innovation.

\endgroup

\newpage

\section{Design and control of antiferromagnetic topological solitons}
\begingroup
    \let\section\subsection
    \let\subsection\subsubsection
    \let\subsubsection\paragraph
    \let\paragraph\subparagraph
Hariom Jani$^1$, and Paolo G. Radaelli$^1$
\vspace{0.5cm}

$^1$ Clarendon Laboratory, Department of Physics, University of Oxford, Oxford, OX1~3PU, United Kingdom

\section*{Status}

Over the past decade, antiferromagnetic materials have emerged as promising platforms for developing a new generation of spintronic and magnonic devices, offering unique advantages over ferromagnetic materials. Their atomically compensated spin structure enables three distinctive properties: immunity to stray fields, negligible internal demagnetising fields, and terahertz spin dynamics. Together, these properties could enable the development of robust, dense, energy-efficient, and ultrafast spin devices.

Driven by this potential, a strong research focus has been the exploration of topological solitons in antiferromagnets. These are nanoscale spin textures such as skyrmions, merons, and bimerons (Fig.~\ref{fig:Jani}), characterised by two topological invariants: winding number ($W$) and topological charge ($Q$). While these solitons have been theoretically studied since the 1980s~\cite{kosevich1990magnetic}, recent analytical and simulation work evidences their potential to overcome fundamental challenges faced today in ferromagnetic skyrmionics~\cite{barker2016static, shen2020currentinduced}.

Despite their immense potential, experimental realisation of antiferromagnetic solitons has been quite challenging due to the difficulties of manipulating and detecting compensated spins. Whilst light, field, and strain can influence antiferromagnetic states, they cannot be readily translated to controlling nanoscale textures. These challenges were initially addressed by recreating the topological physics of the celebrated Kibble-Zurek mechanism in an antiferromagnet. The key idea was to drive an antiferromagnet across a symmetry-breaking transition to create topological textures~\cite{chmiel2018observation}. This was realised in the common antiferromagnet $\rm \alpha-Fe_2O_3$ across its spin-reorientation transition, resulting in the reproducible creation of a family of topological solitons at room temperature~\cite{jani2021antiferromagnetic}. Subsequent advances have demonstrated soliton nucleation and control through electrical pulsing~\cite{amin2023antiferromagnetic}, local strain engineering~\cite{jani2024spatially, harrison2025}, field pulsing~\cite{harrison2024holographic, tan2024revealing}, and shape design~\cite{amin2024nanoscale}, across antiferromagnets like CuMnAs, $\rm \alpha-Fe_2O_3$, and MnTe, highlighting the tunability and potential of antiferromagnetic solitons.

\begin{figure}[h!]
    \centering
    \includegraphics[width=0.95\linewidth]{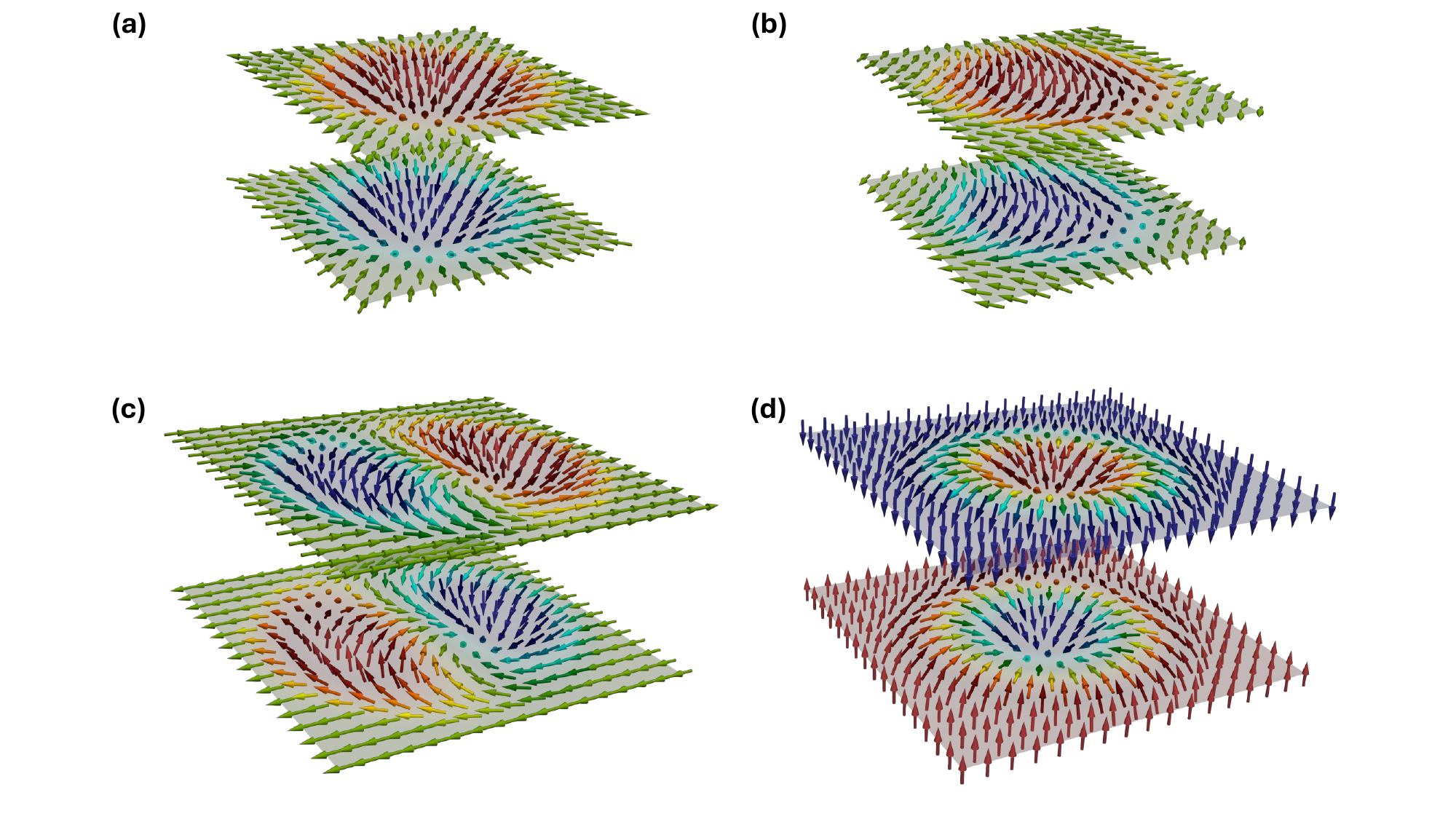}
    \caption{Compensated spin distribution of basic antiferromagnetic topological solitons and their topological invariants ($W$,$Q$): (a) meron (+1,+1/2), (b) antimeron (-1,-1/2), (c) bimeron (0,+1), and (d) skyrmion (+1,+1). (Anti)merons are equivalent to half-(anti)skyrmions and bimerons are equivalent to in-plane rotated skyrmions.}
    \label{fig:Jani}
\end{figure}

\section*{Future Prospects}

Building on the foundation of these advances, it is now crucial to make decisive progress toward dynamical control of solitons at the nanoscale and the development of novel soliton devices. Three promising device concepts are:

\textbf{Logic-in-memory racetracks:} Solitons, due to their compact dimensions, topological protection and particle-like behaviour, could serve as mobile, non-volatile information carriers in racetrack devices to perform logic operations. Ferromagnetic skyrmions have shown progress, but are hindered by field sensitivity, limited scalability, and slow gyrotropic motion. Conversely, antiferromagnetic solitons offer markedly superior stability, smaller sizes, negligible gyrotropic deflection, significantly faster dynamics, and efficient spin-torque control~\cite{barker2016static, shen2020currentinduced, gobel2017antiferromagnetic}. Some of these advantages have been demonstrated in synthetic antiferromagnets (SAFs), evidencing the underlying physics. However, weak interfacial exchange in SAFs is expected to result in soliton decoupling during ultrafast operation. In contrast, natural antiferromagnets, with strong exchange coupling, are ideally suited for soliton-based ultrafast logic-in-memory applications.

\textbf{High-frequency nano-oscillators: }Confined antiferromagnetic solitons, driven by spin torques, would exhibit naturally heterogeneous ultrafast dynamics, including precession, oscillation or gyration. Above a threshold drive, these solitons would exhibit auto-oscillations, which, in turn, could output high-frequency AC charge currents through spin-pumping and inverse spin-Hall effects~\cite{shen2020currentinduced, ovcharov2022spin}. This would produce (sub)terahertz signals, significantly faster than those of ferromagnetic oscillators. Moreover, unlike homogenous antiferromagnets, solitons would produce robust output signals with a lower threshold. Hence, antiferromagnetic solitons could serve as active elements in tuneable and compact sources of (sub)terahertz signals, for implementation in 6G telecommunications.

\textbf{Neuromorphic soliton reservoirs:} Physical reservoir computing is a recurrent neural network that utilises a dynamical platform with nonlinearity, fading/short-term memory, and a reproducible ground-state to perform neuromorphic tasks. To this end, networks of pinned antiferromagnetic solitons, exhibiting a rich spectrum of eigenmodes, nonlinear mutual interactions, and fading dynamics, could serve as a promising reservoir layer. While this idea is being explored with ferromagnetic solitons~\cite{lee2023perspective}, antiferromagnetic counterparts could offer field immunity, faster operation, and a wide frequency bandwidth due to the presence of low-/high-frequency quasi-ferromagnetic/quasi-antiferromagnetic modes. Ultimately, cascaded reservoirs with complementary memory and nonlinear properties could enable advanced neuromorphic hardware~\cite{lee2023perspective}.

\section*{Challenges Ahead And Advances Needed}

Although antiferromagnetic solitons enable unique opportunities, their compensated structure and nanoscale dimensions pose significant challenges for detection and control. The following areas have to be addressed urgently to advance this young field toward practical applications:

\textbf{Imaging solitons:} State-of-the-art approaches employing X-ray techniques, like linear dichroic photoemission microscopy~\cite{chmiel2018observation, amin2023antiferromagnetic, amin2024nanoscale}, provide valuable insights into quasi-static physics, but are ill-suited for in-situ studies due to limitations of high-voltage electron optics. Recent advances in designing topologically-rich membranes~\cite{jani2024spatially} extend antiferromagnetic imaging to transmission microscopy~\cite{jani2024spatially}, lens-less holography~\cite{harrison2024holographic}, ptychography and laminography, offering pathways to imaging dynamics with in-situ electrical/magnetic control. However, many materials cannot be designed as membranes, and membrane fabrication can introduce structural defects that may have deleterious effects on properties (e.g. mobility). Although some of these issues must be addressed via bespoke materials design, it is necessary to develop in parallel lab-scale techniques, like quantum magnetometry, that detect small stray fields~\cite{tan2024revealing} or noise fluctuations~\cite{finco2021imaging} from solitons in x-ray-opaque samples. Currently, these techniques can only access static or low-gigahertz physics. Moreover, since they probe stray fields, not the Néel vector directly, full vectorial reconstructions cannot be performed without additional information from other techniques. Thus, pump-probe X-ray methods at synchrotrons and free-electron lasers will remain essential for elucidating nanoscale soliton physics.

\textbf{Soliton stabilisation and nucleation:} Achieving targeted creation of antiferromagnetic solitons remains challenging due to the difficulties in implementing magnetic control and introducing interfacial Dzyaloshinskii-Moriya interactions (DMI), at least in the antiferromagnets studied thus far. The Kibble-Zurek mechanism reproducibly generates solitons, but does so randomly in a non-uniform magnetic background, limiting individual addressability of solitons~\cite{jani2021antiferromagnetic, jani2024spatially, tan2024revealing}. Conversely, current-driven generation via twisting of local Néel order within domain walls has enabled (anti)meron nucleation~\cite{amin2023antiferromagnetic}. However, the physics underpinning this spin-torque-driven phenomenon is not fully understood. Hence, advancing toward reproducible and targeted soliton creation will likely require innovations marrying strain engineering~\cite{harrison2025, amin2024nanoscale}, interface and geometric design~\cite{harrison2022route}, and spin-torque control~\cite{shen2020currentinduced}.

\textbf{Soliton motion and dynamics:} Thus far, a significant gap remains between theory and experimental realisation of ultrafast soliton motion. This is due in part to the challenges involved in in-situ soliton imaging, as discussed above. Encouragingly, spin-torque-induced soliton motion has been observed in CuMnAs~\cite{amin2023antiferromagnetic}, but the process may be intertwined with texture nucleation, and thus requires further exploration. Other topologically-rich systems host a network of solitons in complex, non-uniform backgrounds, likely inhibiting soliton motion~\cite{jani2021antiferromagnetic, tan2024revealing, amin2024nanoscale}. Furthermore, local/extended defects and magneto-elastic interactions could play a significant role in limiting antiferromagnetic soliton motion. Hence, hetero-structure and interfacial engineering will be crucial in enabling targeted ultrafast motion of solitons.

Apart from that, field-based approaches, employing waveguides, could be used to drive fast soliton dynamics. This is because many antiferromagnetic solitons host static and dynamic magnetisation in regions of non-collinearity, which respond to fields. Finally, topologically-rich antiferromagnet-ferromagnet hetero-structures~\cite{chmiel2018observation} offer a promising platform for electrical control, where spin injection into the ferromagnet could indirectly manipulate antiferromagnetic solitons via interfacial exchange. Such hetero-structures could operate in the ultrafast regime, much beyond what is possible in weakly-coupled SAFs, due to stronger exchange.

\textbf{Electrical detection:} To create practical devices, electrical read-out of solitons is crucial. Unlike ferromagnets, the topological Hall effect vanishes in antiferromagnets. However, sizable topological contributions are predicted in spin/orbital Hall effects~\cite{gobel2017antiferromagnetic}. Experimental detection of these novel signatures for nanoscale solitons will be challenging and is yet to be demonstrated. Alternatively, direct/inverse spin Hall-based detection has emerged as a reliable method for reading microscale antiferromagnetic states. Recent advances have extended this to detect antiferromagnetic spin waves using spin pumping~\cite{elkanj2023antiferromagnetic}, indicating its potential for detecting solitons. Another promising approach is to leverage interfacial exchange coupling to imprint antiferromagnetic solitons onto a ferromagnetic layer~\cite{chmiel2018observation}, which can be read easily with a magnetic tunnel junction.

\section*{Conclusions}

Topological solitons in antiferromagnets are expected to exhibit extraordinary properties, which could address some of the pressing issues in skyrmionics, whilst unlocking new opportunities in the ultrafast regime. Recent advances, leveraging concerted effort in materials design and antiferromagnetic imaging, have enabled the discovery and basic control of antiferromagnetic solitons. However, their nanoscale dimensions and compensation pose unique challenges in achieving targeted control of ultrafast soliton dynamics. Overcoming these challenges will pave the way for next-generation spintronic and magnonic devices, that could profoundly impact future computation and communication technologies.

\section{Acknowledgements}
H.J. acknowledges the support from the Royal Society URF Grant (URF/R1/241120). P.G.R. acknowledges the support from the Oxford-ShanghaiTech collaboration project.

\endgroup

\newpage

\section{Magnetic Spin Textures Beyond Simple Skyrmions}
\begingroup
    \let\section\subsection
    \let\subsection\subsubsection
    \let\subsubsection\paragraph
    \let\paragraph\subparagraph
Jagannath Jena$^{1,2}$, and Stuart S. P. Parkin$^2$
\vspace{0.5 cm}

\noindent
\textit{$^1$ Materials Science Division, Argonne National Laboratory, IL, Lemont, USA\\
$^2$ Max Planck Institute of Microstructure Physics, Halle (Saale), Germany}

\section*{Introduction}

In addition to the two well-known and extensively studied spin textures—Bloch and Néel skyrmions— antiskyrmions represent a distinct class of complex spin textures predicted and observed experimentally first in materials with $D_{2d}$-symmetry and later in materials with $S_4$ crystal symmetry. Unlike the azimuthally symmetric spin textures in conventional skyrmions, antiskyrmions feature crystal direction-dependent spin textures due to an anisotropic Dzyaloshinskii-Moriya interaction (DMI) that reflects the underlying crystal symmetry~\cite{bogdanov2002magnetic, meshcheriakova2014large}. This leads to structures with Bloch-type walls along the crystal directions [100] and [010], and Néel-type walls along the crystal directions [110] and [${1\bar{1}0}$] in $D_{2d}$-symmetric tetragonal Heusler compounds~\cite{meshcheriakova2014large, nayak2017magnetic, jena2020observation}. The antiskyrmion, just like conventional simple skyrmions, are topological spin textures with specific topological charge numbers ($N_{Sk}$), +1 for antiskyrmions but rather -1 for Bloch and Néel skyrmions~\cite{nagaosa2013topological}.

A particularly intriguing aspect of materials that display antiskyrmions is that these same materials have been shown to display other topological spin textures with distinct topological charge numbers~\cite{jena2020elliptical}. These can coexist at the same temperature but, at other temperatures, one or other of these spin textures dominates. Which type of skyrmion is observed depends on the interplay between various energy contributions.  For example, in the Heusler $D_{2d}$ compound, $\rm Mn_{1.4}Pt_{0.9}Pd_{0.1}Sn$, Bloch-type skyrmions emerge at low temperatures due to an increase in the strength of magnetic dipole–dipole interactions (DDI) as the saturation magnetization increases. The magnetostatic energy is lower for Bloch-like walls as compared to Néel-like walls, that have volume magnetic charges, so this results in spin textures that have an elliptical shape whose major axis is oriented along the [100] and [010] directions that, as mentioned earlier, are the directions along which the intrinsic DMI favors Bloch-like walls. The elongated Bloch skyrmions along [100] and [010] have opposite chiralities~\cite{jena2020elliptical}.  Thus, these elongated Bloch skyrmions are distinct from Bloch-like skyrmions that have been predominantly observed in the chiral $B_{20}$ cubic compounds such as MnSi and FeGe and which are typically circular in shape.

At room temperature and above, the DMI is dominant thereby stabilizing antiskyrmions. However, the combination of the DMI with the DDI can lead to square rather than circularly shaped objects (Fig.~\ref{fig:Parkin1})~\cite{jena2020elliptical}. The elliptical Bloch skyrmions can be transformed into antiskyrmions—and vice versa—via an intermediate state bullet-shaped, topologically trivial bubble state. This conversion can be induced either by thermal activation or by applying an in-plane magnetic field, which effectively modulates the magnetic spin texture stability~\cite{jena2024topological}. We note that it is very common in systems that display skyrmions that metastable objects can exist in temperature and field regimes where thermodynamically they would not be stable. For the $D_{2d}$ systems, this means that the temperature and magnetic field window can be substantially modified by the prior history.

Another significant finding is the exceptional stability of the spin textures in the $D_{2d}$ compounds. These can persist in layers that are up to several microns thick~\cite{saha2019intrinsic, ma2020tunable}, substantially exceeding the thickness limits observed in $B_{20}$ compounds where skyrmions are typically stable only within a few nanometers~\cite{yu2011near}. This enhanced stability is attributed to the absence of a component of the  Dzyaloshinskii–Moriya interaction (DMI) vector along the [001] direction, namely that of the tetragonal axis of the $D_{2d}$ compounds. This leads to  spin textures that adopt straight, tube-like configurations along [001] that, thereby, maintains coherence throughout the thickness of films whose out-of-plane axis is typically set to be along [001]. The in-plane anisotropic DMI also enables the stabilization of exotic spin configurations such as fractional antiskyrmions and fractional Bloch skyrmions (Fig.~\ref{fig:Parkin1}), which emerge near the sample edges~\cite{jena2022observation}. These textures, characterized by fractional topological charge numbers, preferentially form at relatively lower magnetic fields compared to their integer-valued counterparts. Their formation is driven by missing neighbors and resulting uncompensated DMI bonds at the edges, producing an edge twist that topologically protects the textures from annihilation. Intriguingly, this twist has been theoretically linked to the emergence of robust magnonic corner states, signaling a higher-order topological phase consistent with bulk-boundary correspondence principles~\cite{hirosawa2020magnonic}. We note that beyond the $D_{2d}$ systems, antiskyrmions and elliptical Bloch skyrmions have also been observed in compounds with $S_4$-symmetry, such as $\rm Fe_{1.9}Ni_{0.9}Pd_{0.2}P$, broadening the class of materials capable of stabilizing these exotic topological spin textures~\cite{karube2021room}.

\begin{figure*}[h!]
    \centering
    \includegraphics[width=0.75\textwidth]{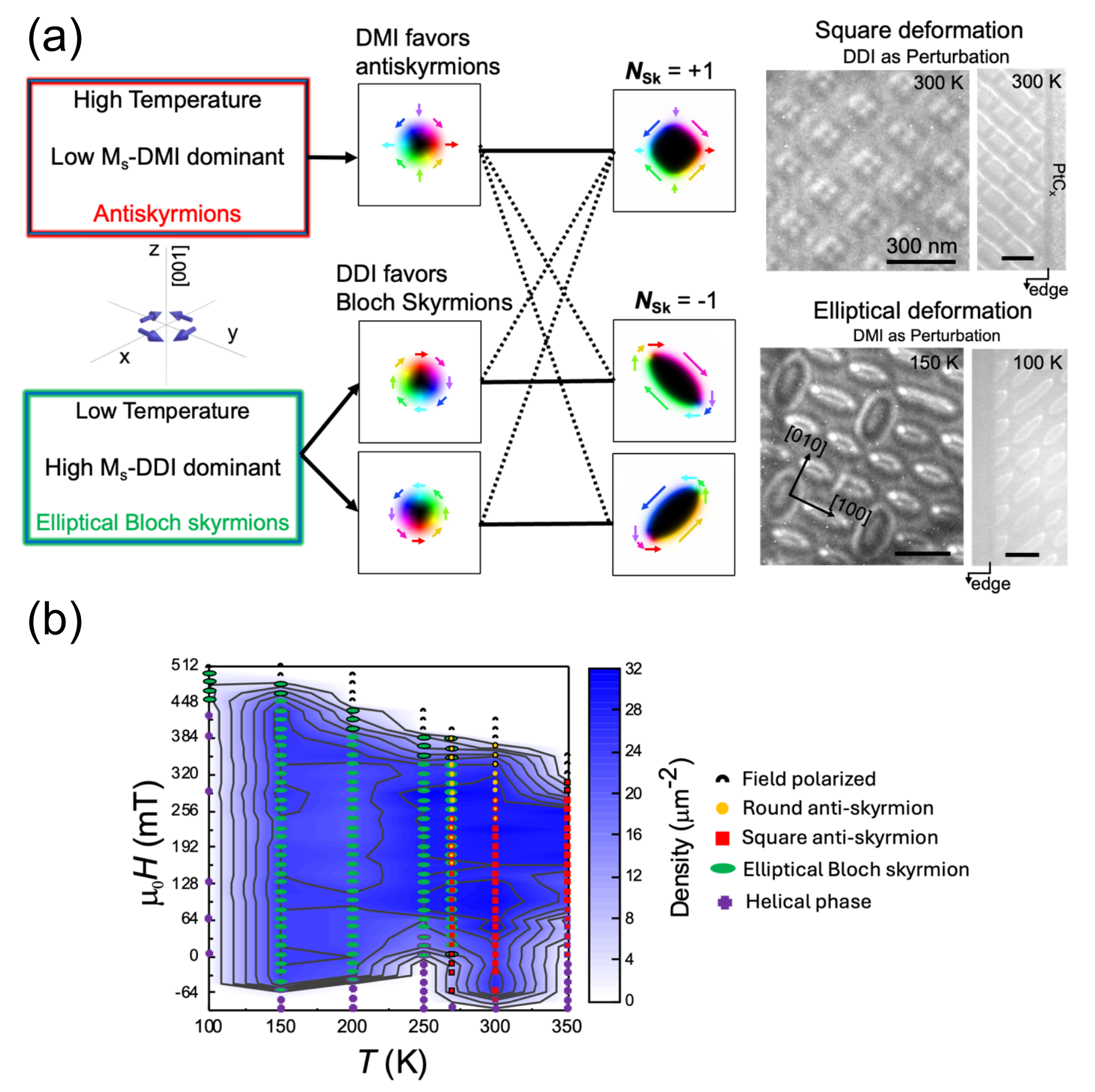}
    \captionsetup{font=tiny,skip=-1pt,justification=raggedright,singlelinecheck=false}
    \caption{Antiskyrmions and elliptical Bloch skyrmions in a $D_{2d}$-symmetric system. (a) Right upper panel: Experimental observation of square-shaped antiskyrmions and fractional antiskyrmions at the edge between $\rm PtC_x$ and $\rm Mn_{1.4}Pt_{0.9}Pd_{0.1}Sn$, recorded at 300 K under 224 mT and –48 mT, respectively, using Lorentz transmission electron microscopy imaging (LTEM). The scale bar in all LTEM images corresponds to 300 nm. Right lower panel: Elliptical Bloch skyrmion chiral twins and fractional Bloch skyrmions observed at the same edge under 128 mT at 150 K and –48 mT at 100 K, respectively. Left panel: Schematic illustration of the Dzyaloshinskii–Moriya interaction (DMI) in a system with $D_{2d}$-symmetry. The $z$-axis corresponds to the tetragonal axis [001], along which the DMI vector has no component. Micromagnetic simulations showing square-shaped antiskyrmions (topological charge $ N_{\rm sk} = +1$) and elliptical Bloch skyrmions ($N_{\rm sk} = +1$) with both clockwise and counter-clockwise chiralities. (b) The magnetic phase diagram corresponds to a 170 nm- thick $\rm Mn_{1.4}Pt_{0.9}Pd_{0.1}Sn$ lamella. Magnetic textures observed in the system are classified into five categories: field-polarized state, round and square antiskyrmions, elliptical Bloch skyrmions, and the helical phase, as indicated by the symbols in the figure. Mixed phases are denoted by symbols with appropriately colored edges. The background contour plot represents the density distribution of elliptical Bloch skyrmions and antiskyrmions. Reproduced from Refs.~\cite{jena2020elliptical,jena2022observation}.}
    \label{fig:Parkin1}
\end{figure*} 

In the B20 compounds, exotic spin textures beyond the simplest Bloch skyrmions that are typically found, have been observed.  These include skyrmion bags, which consist of multiple skyrmions enclosed within a larger topological boundary, and hopfions, that are three-dimensional topological solitons characterized by knotted spin configurations~\cite{tang2021magnetic, zheng2023hopfion}. More recently, a diversity of dipolar stabilized, higher-order spin textures—including multi-skyrmion clusters—have been reported in the same Co/Ni multilayer system~\cite{hassan2024dipolar}. Furthermore, thin-film multilayer systems such as Pt/Co/Al have shown evidence of cocoon-like spin structures~\cite{grelier2022threedimensional}, which coexist with more conventional tubular Néel skyrmions, indicating a rich interplay between material engineering, dimensionality, and magnetic interactions. These findings collectively underscore the growing diversity of spin textures and the need to understand their formation, interaction, and manipulation under external stimuli for future spintronic applications. These exotic textures present new opportunities for exploring three-dimensional spin dynamics and topology-driven functionalities. Of particular interest for future research is the study of their current-driven dynamics, which remains largely uncharted and could offer novel control mechanisms for information transport.

\section*{Relevance And Vision}

To date, only a limited number of materials have been experimentally confirmed to host antiskyrmions—specifically, two in the $D_{2d}$-symmetry group and one in the $S_4$ group. This highlights the need to discover and characterize new material systems capable of stabilizing antiskyrmions and other spin textures within a single host material. A broader understanding of their intrinsic properties, formation mechanisms, and dynamical behavior remains key to advancing the field. While isolated antiskyrmions have been successfully observed, controlled current-induced motion with significant movement without any skyrmion Hall effect has not yet been achieved. Understanding how coexisting spin textures respond to applied currents—particularly in systems where multiple textures such as skyrmions, antiskyrmions, and domain walls coexist—remains largely unexplored.

Although bulk, non-centrosymmetric multi-element compounds have long been investigated for hosting exotic spin textures, their integration into scalable, device-compatible platforms is hindered by limitations in the formation of the same materials by thin film deposition techniques such as sputtering or molecular beam epitaxy (MBE).  For example, the growth of $D_{2d}$-symmetric compounds in thin-film form is particularly challenging\cite{sharma2021nanoscale}, as these typically require three or more elemental constituents and high-temperature annealing and growth processes, often exceeding $\SI{500}{^\circ C}$. The fabrication of thin layers of these complex compounds would especially allow for the exploration of the current induced motion of the skyrmionic spin textures by spin-orbit torques (SOT).  In the thick layers, typically formed by focused ion beam milling of single crystals, only spin-transfer torque (STT) induced motion of the objects is possible and this appears to be very limited, perhaps due to the large spin-orbit coupling that gives rise to DMI, and which thereby depolarizes spin currents in the bulk of these compounds. The development of heterostructures that allow for SOT driven motion would not only be of great scientific interest but is also essential for large-scale production relevant to industrial applications, especially in the context of memory and logic devices.

In thin films with defects, surface roughness, or curvature, skyrmion motion under spin-polarized currents becomes non-uniform due to variations in the local energy landscape, causing fluctuations in velocity and dynamic changes in skyrmion spacing as well as pinning of these objects. This instability hampers reliable data encoding and transport in especially racetrack memory devices~\cite{parkin2008magnetic}. A potential solution is to utilize multiple topologically protected textures—such as skyrmions and antiskyrmions, as seen in $D_{2d}$-symmetric systems~\cite{jena2020evolution} — which are less susceptible to mutual annihilation due to differing topological charges and spin structures. Their coexistence may enhance robustness and control but also introduces complexity, as current-driven dynamics like nucleation, annihilation, and translation become more difficult to manage. Understanding how spin-polarized currents interact with each spin texture is therefore critical for enabling precise, multifunctional spintronic devices.

Overall, achieving efficient manipulation of exotic, complex spin textures via STT or SOT remains a key challenge. Questions such as how reliably data can be written, transported, and read using such spin textures, and how their velocity can be significantly increased and their movement without any skyrmion Hall effect achieved, are critical for practical deployment. Tailoring spin textures for application-specific demands, such as individual addressability and high-speed operation in racetrack memory devices, will be crucial for the next generation of spintronic technologies.

\begin{figure*}[h!]
    \centering
    \includegraphics[width=0.95\textwidth]{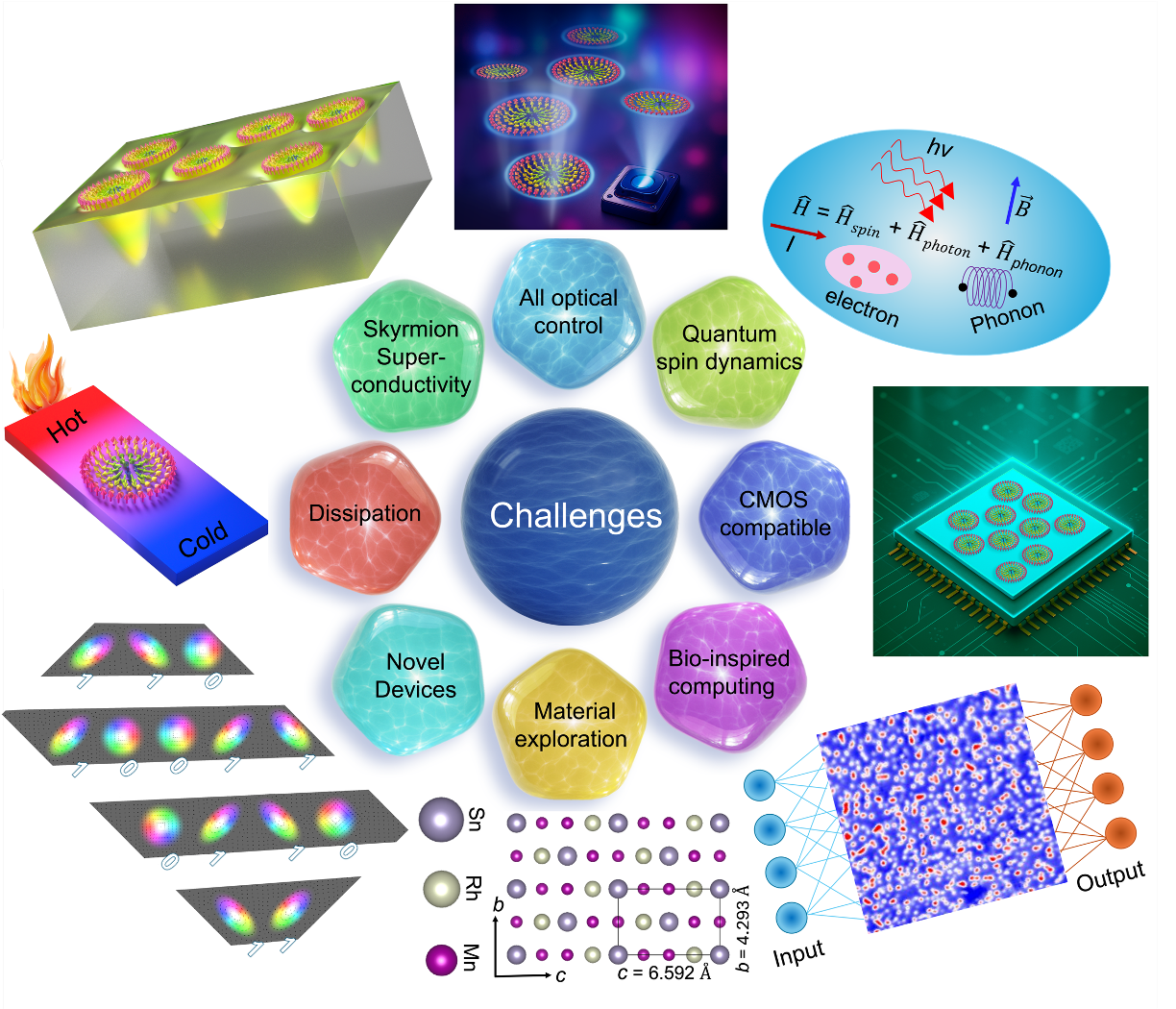}   
    \caption{Schematic illustration of potential challenges in future skyrmion-based devices and technological applications. Key challenges include finding materials hosting different skyrmion spintextures, understanding their stability, enabling control of spin textures via external stimuli such as heat, electric current, and light, and resolving discrepancies between theoretical models and experimental observations through detailed spin dynamics. Additional hurdles involve integrating skyrmions into unconventional computing paradigms, exploring their coupling with superconductivity for novel device architectures, and ensuring compatibility with existing CMOS technologies.}
    \label{fig:Parkin2}
\end{figure*}

\section*{Challenges}

Metastable, complex, spin configurations observed in several compounds and in some thin film heterostructures emerge from an interplay between long-range dipolar forces and competing exchange energies. When dipolar interactions are strong, adjacent spin textures interact more intensely, which limits their spatial density and compromises positional stability. To overcome these limitations in device applications, synthetic antiferromagnetic (SAF) thin film heterostructures are commonly employed~\cite{yang2015domainwall}. SAFs suppress stray fields, enabling higher packing densities, and uniquely support skyrmions with a vanishing skyrmion Hall angle.

Despite recent experimental progress in observing spin textures at the nanometer scale, achieving precise control over their creation, motion and detection in nanoscopic racetrack devices presents major technical hurdles~\cite{jeon2024multicore}. The field of spintronics is nevertheless steadily evolving, pushing into innovative domains such as unconventional computing, where spin textures are being explored for their potential in neuromorphic computing, reservoir computing, memristor, and even quantum information systems. Such investigations emphasize the unique benefits of spin-based devices, including their inherent non-volatility, robustness against perturbations due to topological protection, and potential for ultra-low power operation—all critical features for future computing technologies. These forward-looking perspectives also build upon the foundational understanding of spin configurations developed over the past decades (Fig.~\ref{fig:Parkin2}). However, realizing their full potential in real-world applications will depend on continued experimental breakthroughs and the translation of these conceptual frameworks into scalable technologies.

\section*{Acknowledgments}

We acknowledge funding from the European Research Council(ERC) under the European Union's Horizon 2020 research and innovation program (grant agreement SORBET No.670166). We also acknowledge funding from the DeutscheForschungsgemeinschaft (DFG, German Research Founda-tion) - Project No. 403505322, Priority Program SPP2137 and project No. 443406107 under the Priority Program SPP2244.

\endgroup

\newpage

\section{Polar vortices and skyrmions}
\begingroup
    \let\section\subsection
    \let\subsection\subsubsection
    \let\subsubsection\paragraph
    \let\paragraph\subparagraph
Javier Junquera$^1$, and Ramamoorthy Ramesh$^2$
\vspace{0.5cm}

\noindent
\textit{$^1$ Departamento de Ciencias de la Tierra y F\'{\i}sica de la Materia Condensada, Universidad de Cantabria, Avenida de los Castros s/n, 39005 Santander, Spain\\
$^2$ Department of Physics and Department of Materials Science \& Engineering, University of California, Berkeley, CA 94720 }

\section*{Introduction}
\label{sec:introduction}

In recent years, topologically non-trivial, polarization textures, such as polar skyrmions, vortices, dipolar waves, merons, or hopfions, have emerged as a dynamic field of research within ferroelectric materials. 
Originally theorized and observed in perovskite heterostructures such as PbTiO$_{3}$/SrTiO$_{3}$, these textures have revealed unexpected analogies to magnetic skyrmions, including particle-like behavior, emergent chirality and negative permittivity, their nonvolatile electric field switching, and dynamic responses to external fields.
In a previous review~\cite{Junquera-23} we presented a comprehensive review of the state-of-the-art at the time, theoretical foundations, atomically-precise synthesis techniques and the use of a wide range of characterization tools to probe the interplay between electrostatic, elastic, and gradient energies in stabilizing these exotic polar phases.

Since then, the field has rapidly diversified, with several groundbreaking studies expanding both (i) the material platforms, (ii) the physical mechanisms underpinning topological polarization textures, and (iii) different possibilities to control their dynamics. In the following we summarize a few milestone works along these three lines:

\emph{Beyond PbTiO$_{3}$-based nanostructures.}
Notably, the conceptual framework of ``twistronics'', previously explored in moiré graphene, has been successfully translated into ferroelectrics. In a seminal experimental study, S\'anchez-Santolino {\it et al.}~\cite{Santolino-24} demonstrated that a slight rotational misalignment in stacked BaTiO$_{3}$ membranes produces moiré-induced strain fields that stabilize a periodic array of polar vortex–antivortex pairs. This experimental verification, supported by first-principles calculations, identifies flexoelectric coupling as the key mechanism behind this new form of mechanically driven topology.

Parallel advances in van der Waals (vdW) materials have revealed entirely new avenues for polar topologies. In layered systems such as hexagonal boron nitride (hBN), interlayer charge redistribution gives rise to out-of-plane ferroelectricity, while relative twisting introduces in-plane polarization components. These coexisting components form meron–antimeron lattices, i. e. fractional topological structures predicted by Bennett {\it et al.}~\cite{Bennett-23} and experimentally observed via Piezoresponse Force Microscopy and STEM by Pan {\it et al.}~\cite{Pan-25}. These results underscore that two-dimensional vdW systems can host topological polarization textures even in the absence of conventional ferroelectric instabilities.

The existence of vortex/antivortex pairs have been recently extended to antiferroelectric materials like PbZrO$_{3}$~\cite{Liu-25}.

\emph{New sources for non-collinear polarization.} Recent theoretical developments have further expanded the foundational understanding of these systems. Bennett {\it et al.}~\cite{Bennett-23.2} formulated a gauge-invariant, real-space definition of local polarization suitable for large supercells, enabling accurate tracking of topological charge in moiré ferroelectrics. Meanwhile, a follow-up study~\cite{Bennett-24} revealed that asymmetric dynamical charges, particularly in vdW bilayers, can couple in-plane lattice sliding to polarization responses in orthogonal directions. These findings highlight fundamentally new electromechanical couplings unique to low-dimensional materials. These asymmetric dynamical charges may lead to asymmetric responses in functional properties, such as piezoelectricity or dielectric responses. 

\emph{Controlling the topology and the dynamics of the structures.} Importantly, the control and dynamics of these structures have also been explored. Using second-principles simulations, Aramberri and \'I\~niguez~\cite{Aramberri-24} demonstrated that electric skyrmion-like bubbles in PbTiO$_{3}$/SrTiO$_{3}$ heterostructures behave as mobile quasiparticles under external electric fields, exhibiting Brownian diffusion and long lifetimes, even at room temperature. This establishes their potential as stochastic computing elements. Complementing this, Gao {\it et al.}~\cite{Gao-24} theoretically showed that twisted light beams, carrying orbital angular momentum, can drive real-time topological switching between skyrmion (both of Bloch and N\'eel character) and non-skyrmion states in Pb(Zr$_{0.4}$Ti$_{0.6}$)O$_{3}$ ultrathin films. This ultrafast and contactless optical control represents a significant step toward functional topological devices.

\section*{Relevance and Vision} 
\label{sec:relevance}
The relevance of such polar textures is analogous to their magnetic counterparts.  Their emergent properties, including an electric field controllable chirality and negative permittivity-induced dramatic enhancements in the net capacitance upto frequencies of at least 100 GHz make them relevant to field tunable optical and dielectric elements. The broader vision would be to be able to use the emergent properties (chirality, negative permittivity, topology, etc) in future applications in computing, communications and sensing.

\section*{Challenges}
\label{sec:challenges}

Despite remarkable advances in the creation, control, and observation of topological polarization textures, several key challenges must be addressed before these structures can be systematically engineered and integrated into functional devices. The field has thrived on a fruitful interplay between theory and experiment, an essential synergy that must continue to evolve in tandem.

\emph{At the experimental level} open questions persist regarding the stabilization and characterization of topological signatures with direct functional impact, such as chirality. In particular, recent works~\cite{Zatterin-24} suggest that the Bloch component of the polarization at domains walls might be more sensitive to the boundary conditions than initially thought, especially to the strain imposed by the substrate. 4D-STEM imaging techniques has enabled the challenge of a direct imaging of nanoscale Bloch regions, buried in the ferroelectric layer of the PbTiO$_{3}$/SrTiO$_{3}$ heterostructures. But a broader adoption of complementary, more accessible methods is needed to facilitate reproducibility and cross-laboratory comparisons.

\emph{On the theoretical side}, progress in first-principles-based Hamiltonians and second-principles simulations has been substantial, but challenges remain. These include:
(i) the computational size of realistic systems, which often contain thousands to millions of atoms;
(ii) the need for high-accuracy methods to capture the delicate balance of electrostatic, elastic, flexoelectric, and interfacial forces, avoiding unphysical artifacts; and
(iii) simulations must be conducted under operational conditions, including finite electric fields and temperatures, to reflect real-world functionality.
A promising path forward lies in integrating machine learning (ML) and artificial intelligence (AI) into multiscale frameworks (second-principles and phase fields)~\cite{Ma-25}. Such approaches could accelerate materials discovery and enable large-scale, realistic simulations that are both accurate and efficient.

A further conceptual challenge is to uncover the interplay between real-space polarization textures and momentum-space electronic topology. While current studies emphasize atomic-scale structure, the impact of these textures on band structure, transport, and optoelectronic responses remains largely unexplored. 

Finally, more fundamental work to determine the topological features of these structures is another important aspect~\cite{Govinden-23}. In this sense, a step forward has been taken, borrowing concepts of fluid dynamics~\cite{Lukyanchuk-25}. However, the phenomenon of pattern formation outside of equilibrium or the eventual characterization of the phase transitions as Turing phases will require further studies. Much of the work so far has been on the PbTiO$_{3}$/SrTiO$_{3}$ system, with a few publications on the multiferroic BiFeO$_{3}$ system. It would be quite valuable to explore the formation of such polar textures across a broader spectrum of materials, including oxide/non-oxide as well as organic ferroelectrics.

\emph{At the functional level}, some discoveries might put the field on a different level. For example,the existence of a transverse dielectric response of the polar topologies or cross-coupling of dielectric responses would be of fundamental interest. On a similar note, there has been very little work on the field-induced dynamics of such polar textures as well as the effects on the dielectric and chiral responses, both in time-domain and frequency-domain. 

\section*{Conclusions}
\label{sec:conclusions}

Polar topological textures in ferroelectrics represent a rapidly maturing field with strong analogies to their magnetic counterparts, yet with distinctive advantages in electric-field tunability and integration into device architectures. Experimental discoveries in twisted oxide membranes and van der Waals systems have revealed new pathways for topological stabilization, while theoretical insights continue to uncover rich electromechanical coupling phenomena and suggest novel functionality. Despite significant progress, critical challenges remain, including the development of scalable and accessible imaging methods, the extension of accurate simulations including artificial intelligence, and the exploration of new materials classes beyond traditional perovskites. Furthermore, understanding how these real-space polarization textures influence momentum-space electronic properties remains an open frontier. Addressing these challenges through synergistic theory–experiment efforts and leveraging machine learning tools offers a promising route toward practical applications in neuromorphic computing, reconfigurable electronics, and high-frequency dielectrics. Continued exploration of topological phase transitions and dynamic responses under external fields will be essential in shaping the next phase of research in polar topologically non-trivial structures.

\section*{Acknowledgements}

J.J.~acknowledges financial support from Grant No.~PID2022-139776NB-C63 funded by MCIN/AEI/10.13039/501100011033 and by ERDF/EU. R. R. acknowledges support from the Army
Research Office under the ETHOS MURI via Cooperative Agreement No. W911NF-21-2-0162.

\endgroup

\newpage

\section{Dipolar-stabilized higher-order skyrmions and antiskyrmions}
\begingroup
    \let\section\subsection
    \let\subsection\subsubsection
    \let\subsubsection\paragraph
    \let\paragraph\subparagraph
Sabri Koraltan$^1$, and Manfred Albrecht $^2$
\vspace{0.5cm}

\noindent
\textit{$^1$ Institute of Applied Physics, TU Wien, Wiedner Hauptstraße 8-10, Vienna, 1040, Austria\\
$^2$ Experimental Physics IV, University of Augsburg, D-86135 Augsburg, Germany}

\section*{Introduction}
Magnetic materials offer an easily accessible platform to study topology. Topology in the magnetic sense means that the magnetization can be different than the saturated, uniform state of a typical ferromagnet, and can lead to exotic, non-uniform magnetic states. In this regard, magnetic skyrmions, localized topologically stable magnetic solitons, have attracted significant attention~\cite{back20202020, nagaosa2013topological}. Their topological stability, also commonly referred to as protection, suggests that a homogeneous and continuous force should not be able to annihilate these intricate spin objects without the possible creation of magnetic singularities known as Bloch points. Magnetic skyrmions are characterized by their topological charge $|Q| = 1$, 

\begin{equation}
    Q = \dfrac{1}{4\pi}\large{\iint}\mathbf{m}\cdot \left(\dfrac{\partial \mathbf{m}}{\partial \mathbf{x}}\times\dfrac{\partial \mathbf{m}}{\partial \mathbf{y}}\right)\mathrm{d}x\mathrm{d}y,
\end{equation}

where $\mathbf{m}$ denotes the magnetization of the magnetic specimen. The sign of $Q$ is determined by the relation between the skyrmion's core magnetization and the outer-region magnetization. In this section, we stick to the convention that skyrmions are defined by $Q = -1$.

In addition to regular skyrmions, their topological counterpart, antiskyrmions ($Q = +1$), were originally discovered in non-centrosymmetric chiral Heusler compounds~\cite{nayak2017magnetic}, and later in ferrimagnetic multilayers~\cite{heigl2021dipolar}. As these two types of spin objects are found in various systems, the idea of topology in magnetism can be further questioned: Can spin objects have topological charges higher than 1? Several systems have been reported in recent years, dealing with composite skyrmion bags or bundles~\cite{foster2019two, tang2021magnetic, zhang2024stable, kern2025controlled}; see Fig.~\ref{fig:Koraltan1}(a). A skyrmionium with a central skyrmion with $Q = -1$ and an outer skyrmion ($Q = +1$) has zero net charge and is the topologically trivial case. If the number of inner skyrmions increases, the net topological charge can take higher integer values. A skyrmion bag $Q = -5$ is illustrated in Fig.~\ref{fig:Koraltan1}(a). While skyrmion bags can be moved by electrical currents and also nucleated locally, the question remains whether single-walled spin objects can have higher topological charge. In the 1970s, Grundy and Rosencwaig discussed bubble objects in their works \textit{Hard Magnetic Bubbles}, which, unlike regular magnetic bubbles found in garnets, were exhibiting additional iterations of Bloch and Néel walls~\cite{rosencwaig1972new, grundy1977magnetic, slonczewski1973statics}. Unfortunately, only a few electron micrograph images existed at that time, which experimentally indicated the existence of these complex spin objects, even though the size of these bubbles exceeded several microns.

\begin{figure*}[h!]
    \centering
    \includegraphics[width=0.75\linewidth]{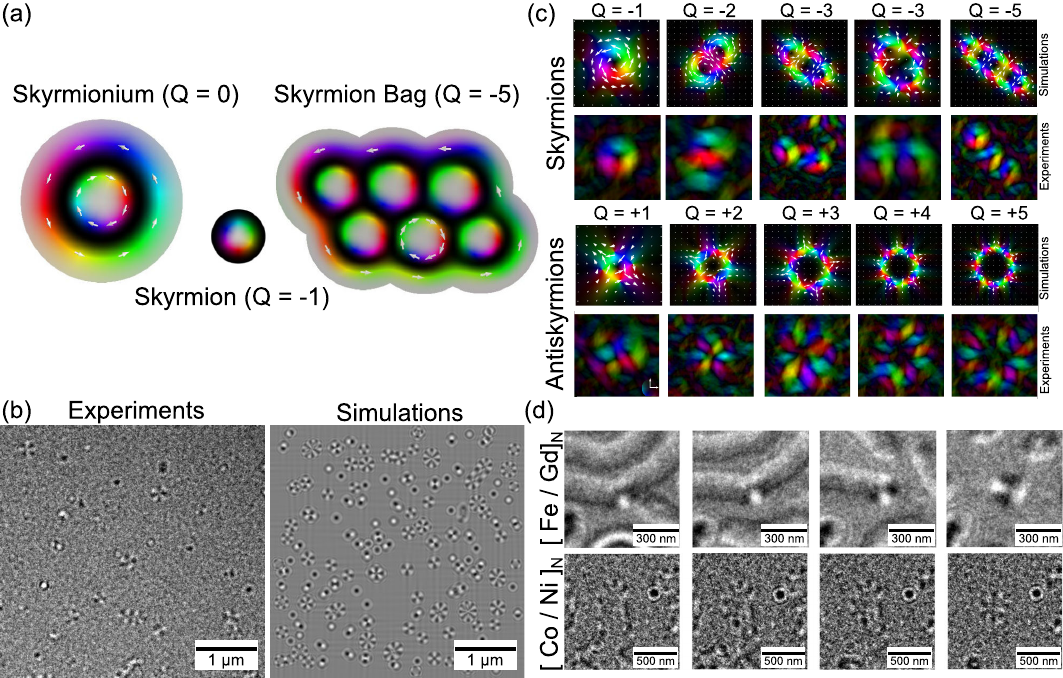}
    \caption{(a) Magnetization configuration of skyrmion bags with $Q = 0$ (skyrmionium) and $Q = -5$, where black (white) shows the oop component of the magnetization as up (down), and the colors represent the in-plane rotation of the magnetization, as indicated by the white arrows inside the objects. In addition, a skyrmion with $Q = -1$ is displayed. (b) Experimental and simulated LTEM images, which show the coexistence of a multitude of spin textures with arbitrary topological charges. (c) Experimental and simulated induction maps of higher-order skyrmions (upper panel) and antiskyrmions (lower panel) of isolated spin textures taken from experiments and large-size simulations. (d) The different nucleation mechanisms of antiskyrmions in Fe/Gd multilayers (upper panel) and higher-order antiskyrmions in Co/Ni multilayers (lower panel). Subfigure (a) was adapted from Ref.~\cite{tang2021magnetic}, (b,c) and the lower panel of (d) were adapted from Ref.~\cite{hassan2024dipolar}, while (d) upper panel was adapted from Ref.~\cite{heigl2021dipolar}.}
    \label{fig:Koraltan1}
\end{figure*}

Recently, we have shown that ferromagnetic multilayers consisting of [Co(0.2)/Ni(0.7)]$_N$ (all thicknesses in nanometers) can exhibit skyrmions and antiskyrmions with an arbitrary number of topological charges at room temperature~\cite{hassan2024dipolar}. Figure~\ref{fig:Koraltan1}(b) shows a comparison of the experimentally acquired and simulated Lorentz transmission electron microscopy (LTEM) images, where it is clear that different spin objects can coexist. Imaging higher-order spin textures using induction maps and comparing them to the simulated images (see Fig.~\ref{fig:Koraltan1}(c)) reveals key features such as alternating Bloch and Néel walls arranged in a circular shape. Calculation of the topological charge confirms that these spin objects have higher topological charges because of the additional rotations inside the skyrmion boundaries. They nucleate in a fundamentally different manner from the antiskyrmions found in ferrimagnetic multilayers (Fig.~\ref{fig:Koraltan1}(d), upper panel), where a chiral domain wall with a three-way junction shrinks down to an antiskyrmion after application of a perpendicular magnetic field (out of plane, oop). However, the key feature is the presence of the topological defect at the junction. This defect is known as a vertical Bloch line (vBL) and changes the chirality of the domain wall. The vBLs are present in very large numbers in Co/Ni multilayers. Thus, they facilitate the nucleation of higher-order skyrmions and antiskyrmions, which form from domain walls containing a large number of vBLs. The regions without vBLs are less stable against external perturbations from magnetic fields. Thus, spin objects up to a few hundreds of nanometers in size are stabilized when the oop magnetic field is strong enough.

\section*{Relevance and Vision}
One of the key features of higher-order skyrmions and antiskyrmions is that the lateral deflection of spin objects from the trajectory parallel to the current direction is diminished significantly with increasing topological charge, as predicted by micromagnetic simulations, see Fig.~\ref{fig:Koraltan2}(a,b). Furthermore, the angle of deflection depends on the sign of the topological charge. Hence, it is, in principle, possible to have a skyrmion and an antiskyrmion set into motion by an electrical current, where both are deflected towards opposing directions. A secondary effect observed mainly in micromagnetic simulations was that spin objects of lower topological charge can be nucleated from each other, as shown in Figs.~\ref{fig:Koraltan2}(c-f). Here, increasing the oop field reduced the topological charge from (c) $Q = -7$ to (d) $Q = -5$, to (e) $Q = -3$, and finally to (f) $Q = -1$. Since the sign of the topology does not change, these transformations are topologically allowed, yet irreversible, because the main mechanism relies on the annihilation of vBLs.

\begin{figure*}[h!]
    \centering
    \includegraphics[width=0.75\linewidth]{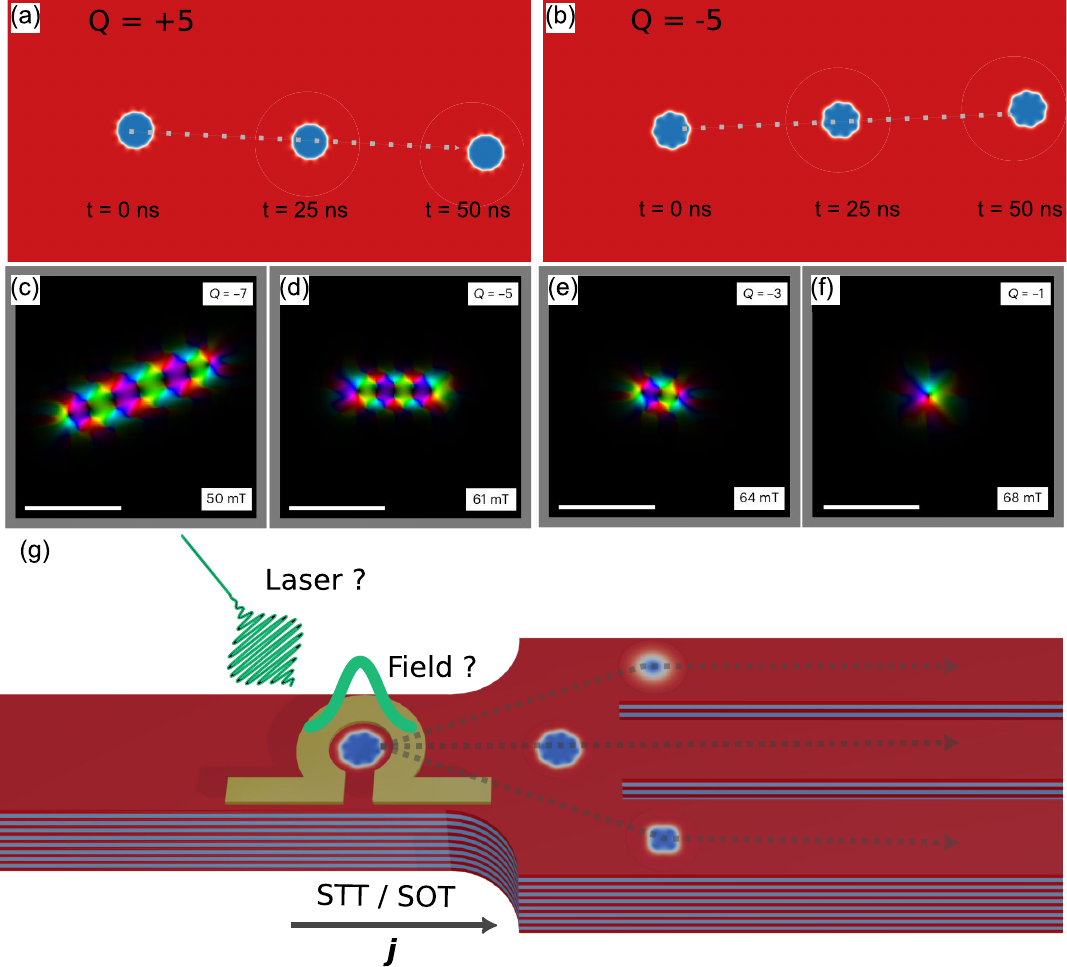}
    \caption{Current-induced motion of a $Q = +5$ antiskyrmion in (a) and of a $Q = -5$ skyrmion in (b), where the dotted lines shows opposing angles for the skyrmion Hall angle. In (c) to (f) the field-dependent annihilation of topological charge leads a $Q = -7$ skyrmion finally annihilating into a $Q = -1$ skyrmion. Subfigures (a) to (f) were adapted from Ref.~\cite{hassan2024dipolar}. (g) Schematic illustration of a possible application of higher-order skyrmions and antiskyrmions in a device where a higher-order spin texture gets transformed by a laser pulse, or localized magnetic field, and the new spin objects are sorted based on their topological charge due to different skyrmion Hall angles.}
    \label{fig:Koraltan2}
\end{figure*}

Adding the angle dependency of higher-order skyrmions into the big picture, we envision that a skyrmion sorting device might be utilized for potential computing and logic operations. Assume a spin object with $Q = -5$ propagated through a racetrack-like device by means of spin-transfer torque, or even spin-orbit torques~\cite{tomasello2014strategy}. A central decision region can be incorporated, where the magnetization could be read by means of magnetoresistive or anomalous Hall effects~\cite{guang2023electrical}. Then, a pulsed (oop) magnetic field from a microantenna~\cite{finizio2019deterministic, yang2022magnetic}, or a pulsed-laser shot~\cite{buttner2017fieldfree}, might enable topological switching. The current then further pushes the newly formed spin objects into different racetracks, which are then analyzed for further data processing. In this scenario, relating the input power or pulse duration to the resulting spin state might enable a reproducible signal which can be harnessed for either reservoir or stochastic computing~\cite{lee2023perspective}. For the latter, a randomized pulse sequence might lead to a certain type of (anti)skyrmion, which reaches a certain output.

\section*{Challenges}
However, there are several challenges to be addressed before any realistic device integration are considered. As can be seen in Fig.~\ref{fig:Koraltan1}(b), the higher-order spin textures appear rather randomly. A crucial challenge is to enable local control of the nucleation process of higher-order skyrmions and antiskyrmions. In this respect, the work of Kern et al. showed a highly promising route, where vertical Bloch lines might be favored in regions where the magnetic anisotropy has been altered by He$^+$-ion irradiation~\cite{kern2022deterministic}. Another approach is to couple them to magnetic vortice structures in magnetic microdisks, resulting in the nucleation of (anti)skyrmions~\cite{koraltan2023generation}.

Although the skyrmion Hall angle and its dependence on the topological charge were predicted by numerical investigations, it is important to show this experimentally. One of the key challenges here is that so far one had to rely on LTEM as the main imaging method to be able to differentiate between various spin objects. Recently, we have also demonstrated that magnetic force microscopy (MFM) can be used to discriminate between them~\cite{koraltan2025signatures}, which allows the integration of devices on Si/SiO$_x$ chips, which are much easier to handle than the ultrathin, electron-transparent membranes required for LTEM.

Equally important will be to investigate whether spin objects of arbitrary charge will have clearly distinguishable magnetoresistive signals. An option is to include an in-plane pinned layer into a giant magnetoresistance or tunneling magnetoresistance device, where the number of vBLs increases or decreases the none-skyrmion resistance found in the system.

Overall, we believe that the newly discovered higher-order skyrmions and antiskyrmions have great potential in the field of skyrmionics, enabling novel functionalities compared to conventional skyrmions.

\section*{Acknowledgements}
S.K. acknowledges funding support from the European Research Council (ERC) under the Horizon 2020 Program (Grant Agreement ID: 101001290, 3DNANOMAG). M.A.  acknowledge funding from the Deutsche Forschungsgemeinschaft (DFG, German Research Foundation) - Project No. 507821284.

\endgroup

\newpage

\section{Stabilization mechanisms of topological spin textures in
itinerant magnets}
\begingroup
    \let\section\subsection
    \let\subsection\subsubsection
    \let\subsubsection\paragraph
    \let\paragraph\subparagraph
\newcommand{\gguide}{{\it Preparing graphics for IOP Publishing journals}}
\newcommand{\redsout}[1]{\textcolor{red}{\sout{#1}}}
\newcommand{\red}[1]{\textcolor{red}{#1}}
\newcommand{\bluesout}[1]{\textcolor{blue}{\sout{#1}}}
\newcommand{\blue}[1]{\textcolor{blue}{#1}}

Satoru Hayami$^{1}$, and Shinichiro Seki$^{2,3}$
\vspace{0.5cm}

\noindent
\textit{$^{1}$ Graduate School of Science, Hokkaido University, Sapporo, Japan\\
$^{2}$ Department of Applied Physics, University of Tokyo, Bunkyo, Tokyo, Japan\\
$^{3}$ Research Center for Advanced Science and Technology, University of Tokyo, Tokyo, Japan\\}

\section*{Introduction} 
\label{sec:intro}
Magnetic skyrmions are topologically protected spin configurations with remarkable stability and nanoscale size, rendering them promising candidates for next-generation spintronic applications. 
They typically emerge in noncentrosymmetric systems where the Dzyaloshinskii-Moriya (DM) interaction, arising from broken inversion symmetry, plays a central role in stabilizing such topological spin textures.
However, the theoretical prediction of skyrmion crystals without relying on the DM interaction~\cite{Okubo_PhysRevLett.108.017206}, together with their experimental discovery in the centrosymmetric compound Gd$_2$PdSi$_3$~\cite{kurumaji2019skyrmion}, has established that the emergence of skyrmion crystals is a more general consequence of complex magnetic competition than previously assumed. 
Indeed, subsequent experiments have uncovered a plethora of topological spin textures in centrosymmetric itinerant magnets, including square-lattice skyrmion crystals in GdRu$_2$Si$_2$~\cite{khanh2020nanometric}, EuAl$_4$~\cite{takagi2022square}, and GdRu$_2$Ge$_2$~\cite{yoshimochi2024multi}, as well as hedgehog crystals in SrFeO$_3$~\cite{Ishiwata_PhysRevB.101.134406}. 
These findings underscore the key role of the interplay between spin and charge degrees of freedom in itinerant electron systems, which can give rise to long-range RKKY and complex multi-spin interactions mediated by Fermi surface instabilities~\cite{hayami2021topological}. 
Supporting this perspective, recent angle-resolved photoemission spectroscopy measurements have revealed pseudogap behavior associated with a nesting band in GdRu$_2$Si$_2$~\cite{dong2025pseudogap}. 

Since topological spin textures can be represented as superpositions of multiple magnetic ordering wave vectors, which are referred to as multiple-$Q$ states, a key to understanding the emergence of skyrmion crystals in centrosymmetric itinerant magnets lies in identifying the microscopic mechanisms that stabilize such multiple-$Q$ states~\cite{hayami2021topological}. 
To this end, an effective spin model with momentum-resolved interactions is introduced as~\cite{hayami2024stabilization}

\begin{eqnarray}
\label{eq: Ham}
\mathcal{H} = - \sum_{\bm{q}, \alpha_s, \beta_s} X^{\alpha_s \beta_s}_{\bm{q}} S^{\alpha_s}_{\bm{q}}  S^{\beta_s}_{-\bm{q}} + \sum_{\bm{q}_1,\bm{q}_2, \bm{q}_3 \bm{q}_4} Y^{\alpha_s \beta_s \alpha'_s \beta'_s}_{\bm{q}_1\bm{q}_2\bm{q}_3\bm{q}_4} S^{\alpha_s}_{\bm{q}_1}S^{\beta_s}_{\bm{q}_2}S^{\alpha'_s}_{\bm{q}_3}S^{\beta'_s}_{\bm{q}_4}  - \sum_{i} \bm{B} \cdot \bm{S}_i, 
\end{eqnarray}
where $S_{\bm{q}}^{\alpha_s}$ denotes the spin moment with cartesian spin coordinates $\alpha_s=(x_s, y_s, z_s)$ at wave vector $\bm{q}$, related to the real-space spin $S^{\alpha_s}_i$ via Fourier transformation.  
The coefficients $X^{\alpha_s \beta_s}_{\bm{q}}$ and $Y^{\alpha_s \beta_s \alpha'_s \beta'_s}_{\bm{q}_1\bm{q}_2\bm{q}_3\bm{q}_4}$ represent general two-spin and four-spin interactions ($\bm{q}_4= -\bm{q}_1-\bm{q}_2 -\bm{q}_3$), respectively, and their nonzero components are dictated by the crystal symmetry and the underlying wave vectors. 
When the ordering wave vectors are governed by the Fermi surface nesting, it is natural to restrict consideration to the interactions associated with symmetry-related nesting wave vectors among all possible interactions in the Brillouin zone. 
This simplification allows us to capture the essential ingredients for stabilizing skyrmion crystals while reducing computational cost. 
Consequently, several mechanisms have been identified through analyses of the effective spin model, such as biquadratic interactions, crystal-dependent anisotropic interactions, sublattice-dependent DM interactions, high-harmonic wave-vector interactions, in collaboration with the Zeeman coupling to an external magnetic field in the final term in Eq.~(\ref{eq: Ham})~\cite{hayami2024stabilization}.

\begin{figure}[h!]
\begin{center}
\includegraphics[width=10.0 cm]{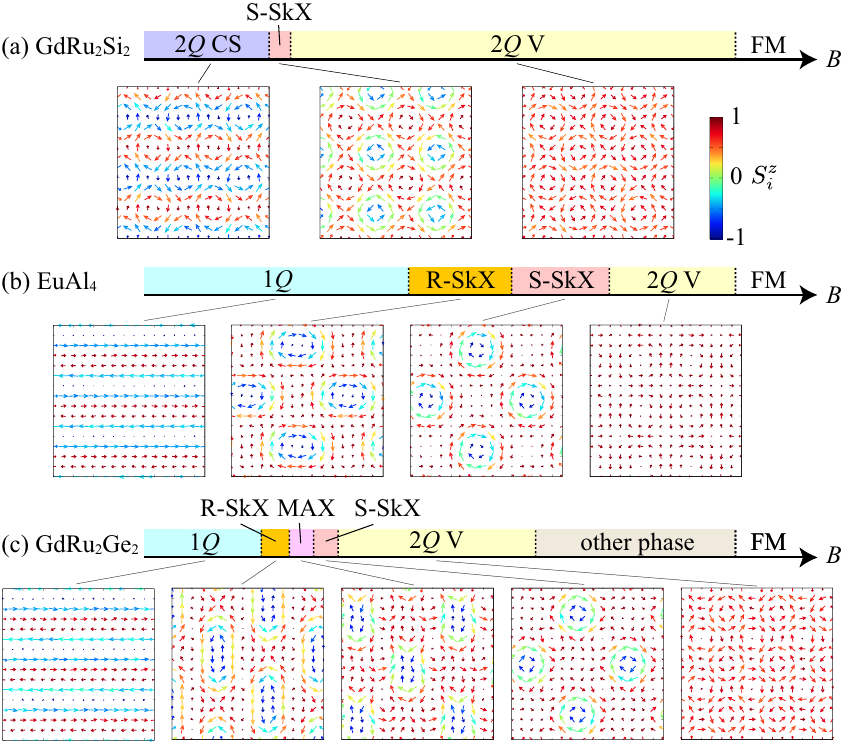}
\caption{
\label{fig:Satoru}
Experimental phase diagrams at low temperatures for (a) GdRu$_2$Si$_2$~\cite{khanh2020nanometric}, (b) EuAl$_4$~\cite{takagi2022square}, and (c) GdRu$_2$Ge$_2$ against the magnetic field $B$~\cite{yoshimochi2024multi}. 
1$Q$ and 2$Q$ denote the single-$Q$ and double-$Q$ states, respectively. 
CS, V, S-SkX, R-SkX, and MAX stand for the chiral stripe, vortex, square skyrmion crystal, rhombic skyrmion crystal, and meron-antimeron crystal, respectively. 
The bottom panel shows real-space spin configurations obtained from the theoretical spin model in Eq.~(\ref{eq: Ham}), where arrows indicate the spin directions and color represents the out-of-plane spin component. 
}
\end{center}
\end{figure}   
\unskip

The effective spin model in Eq.~(\ref{eq: Ham}) can also be applied to reproduce experimental phase diagrams of itinerant magnets, provided that the ordering wave vectors are identified, for example, through small-angle neutron or resonant elastic X-ray scattering measurements. 
One of the key achievements of this modeling approach is the elucidation of the microscopic origin of skyrmion crystals in centrosymmetric skyrmion-hosting materials with tetragonal symmetry, where neither the DM interaction nor geometrically frustrated exchange interactions are expected to play a significant role. 
Figure~\ref{fig:Satoru} presents the low-temperature phase diagrams of GdRu$_2$Si$_2$~\cite{khanh2020nanometric},  EuAl$_4$~\cite{takagi2022square}, and GdRu$_2$Ge$_2$~\cite{yoshimochi2024multi}, as determined experimentally. 
These phase diagrams feature a variety of complex magnetic phases, including distinct types of skyrmion crystals and double-$Q$ states. 
By tuning the parameters of symmetry-allowed interactions within the effective spin model, the experimental phase diagrams can be successfully reproduced; the real-space spin configurations, optimized using Monte Carlo simulations, are shown at the bottom of each panel. 
The analyses reveal that in GdRu$_2$Si$_2$, the magnetic phases are stabilized by a synergy among positive biquadratic, bond-dependent anisotropic, and easy-axis anisotropic interactions, while those in EuAl$_4$ and GdRu$_2$Ge$_2$ are stabilized by a competition of RKKY interactions between inequivalent wave vectors, reflecting momentum-space frustration~\cite{takagi2022square, yoshimochi2024multi}. 
Thus, the effective spin model serves not only as a powerful tool for interpreting topological spin textures observed in experiments, but also as a framework for predicting and designing new exotic spin textures theoretically.

\section*{Relevance and Vision} 
Understanding the microscopic stabilization mechanisms of topological spin textures in itinerant magnets is important for three main reasons. 
First, topological spin textures have been ubiquitously observed in itinerant magnets, regardless of differences in lattice structures,  crystal symmetries, or chemical compositions. 
Examples include hexagonal skyrmion crystals in Gd$_2$PdSi$_3$~\cite{kurumaji2019skyrmion}, square skyrmion crystals in GdRu$_2$Si$_2$~\cite{khanh2020nanometric}, EuAl$_4$~\cite{takagi2022square}, and GdRu$_2$Ge$_2$~\cite{yoshimochi2024multi}, as well as cubic hedgehog crystals in SrFeO$_3$~\cite{Ishiwata_PhysRevB.101.134406}. 
These experimental findings reflect the richness of magnetic interactions that result from the interplay between spin and charge degrees of freedom, intrinsic to itinerant electron systems~\cite{hayami2021topological}. 
A comprehensive understanding of these mechanisms offers guiding principles for engineering topological spin textures in itinerant magnets, thereby expanding the design scope for functional materials with topological properties. 
In this context, elucidating the relationship between the Fermi surface structure and magnetic interactions is of particular interest. 

Second, itinerant magnets provide a fertile platform for exploring a broader family of topological spin textures beyond conventional skyrmions and hedgehogs. 
Due to their diverse and tunable magnetic interactions depending on electronic structures, these systems are well suited for stabilizing exotic spin textures. 
One promising direction is the investigation of topological spin crystals composed of axions, hopfions, and skyrmioniums. 
Another intriguing avenue is to determine the stability conditions of isolated, nanoscale skyrmions, which hold promise for spintronic applications due to their small size and potential for high-speed control. 

Third, itinerant magnets offer a unique opportunity to realize topological spin textures with extremely short magnetic modulation periods, on the order of a few nanometers. 
This opens a pathway toward high-density magnetic information storage. 
In these systems, the characteristic length scale of spin textures is governed by the nesting vector of Fermi surfaces, which contrasts sharply with noncentrosymmetric magnets, where the skyrmion size is typically limited to tens or hundreds of nanometers by the small magnitude of the DM interaction. 
Accordingly, topological spin textures in itinerant electron systems can generate giant emergent electromagnetic fields, presenting another promising direction for future research.

\section*{Challenges} 

Several fundamental challenges must be addressed to accomplish the above objective.  
First, it is crucial to develop numerical methods capable of directly solving itinerant electron models without integrating out the itinerant electron degrees of freedom, as is commonly done in the effective spin model described in Eq.~(\ref{eq: Ham}). 
A major bottleneck in solving itinerant electron models, which incorporate both localized spins and itinerant electrons, lies in the high computational cost of standard diagonalization, which scales as $\mathcal{O}(N^3)$ with system size $N$. 
One promising approach to circumvent this challenge involves the use of an equivariant convolutional neural network architecture, which can efficiently predict thermodynamic spin configurations with significantly reduced computational effort~\cite{miyazaki2023machine}. 
Second, extracting the microscopic interactions that stabilize topological spin textures is equally important. 
In the current effective spin model [Eq.~(\ref{eq: Ham})], the model parameters used to reproduce experimental phase diagrams are typically adjusted manually. 
This motivates the development of machine learning techniques that can autonomously infer optimal parameter sets. 
Recently, Sharma {\it et al.} proposed a machine learning assisted protocol for deriving spin Hamiltonians from the Kondo lattice model~\cite{sharma2023machine}. 
Such advancements can provide valuable insights into the key magnetic interactions and anisotropies that underlie the formation of topological spin textures in real materials. 
Finally, advancing high-precision ab-initio band calculations is essential~\cite{Nomoto2023}. 
Since the stability of topological spin textures is expected to be strongly influenced by the electronic structures, accurate mapping of nesting features and electronic instabilities plays a crucial role in identifying the dominant magnetic interactions in itinerant magnets.  

\section*{Acknowledgements}
This research was supported by JSPS KAKENHI Grants Numbers JP21H01037, JP21H04990, JP22H00101, JP22H01183, JP22H04965, JP23H04869, JP23K03288, JP23K20827, JP24H02235, by JST CREST (JPMJCR23O4) and JST FOREST (JPMJFR2366), and by Asahi Glass
Foundation and Murata Science Foundation. 

\vspace{8mm}

\endgroup

\newpage

\section{Topological solitons in liquid crystals: Building blocks of novel active and driven matter}
\begingroup
    \let\section\subsection
    \let\subsection\subsubsection
    \let\subsubsection\paragraph
    \let\paragraph\subparagraph
Mykola Tasinkevych$^{1,2}$, and Ivan I. Smalyukh$^{1,3,4,5}$
\vspace{0.5cm}

\noindent
\textit{$^1$ International Institute for Sustainability with Knotted Chiral Meta Matter (WPI-SKCM2), Hiroshima University, Higashihiroshima, Japan\\
$^2$ SOFT Group, School of Science and Technology, Nottingham Trent University, Clifton Lane, Nottingham NG11 8NS, United Kingdom\\
$^3$ Department of Physics and Chemical Physics Program, University of Colorado, Boulder, CO, USA\\
$^4$ Department of Electrical, Computer, and Energy Engineering, Materials Science and Engineering Program and Soft Materials Research Center, University of Colorado, Boulder, CO, USA\\
$^5$ Renewable and Sustainable Energy Institute, National Renewable Energy Laboratory and University of Colorado, Boulder, CO, USA\\}

\section*{Introduction}

Liquid crystals (LCs) combine fluidity and various degrees of orientational and partial positional order in a single phase and as such exhibit very rich structural diversity and reconfigurable properties because of the facile response of the LC order parameter to external stimuli. Due to this behaviour, the LCs are used extensively in electro-optic, photonic and biomedical devices, and in reconfigurable on demand composite nanomaterials~\cite{mertelj2013ferromagnetism, mundoor2016triclinic}. On the other hand, being highly experimentally accessible LCs provide insights into other less accessible physical systems, such as elementary particles, the early Universe cosmology, high energy physics, and superfluids~\cite{smalyukh2020review}. For example, nematic LCs have been used to demonstrate the validity of the prediction of the Kibble-Zurek mechanism regarding the scaling behaviour of the defect density during continuous phase transitions~\cite{fowler2017kibble}. Nematic phase is the most common and simple realisation of a liquid crystalline order where the fluidity is combined with an orientational order of anisotropic molecules, characterised by the LC director field $\mathbf{n}(r)$. At no external bias, $\mathbf{n}$ tends to be uniform in space, but this preference may be altered by adding to the system a tiny fraction of a chiral component. This may render twist director perturbations energetically favourable resulting in the stabilisation of helical director configurations.

\begin{figure*}[h!]
    \centering
    \includegraphics[width=0.95\textwidth]{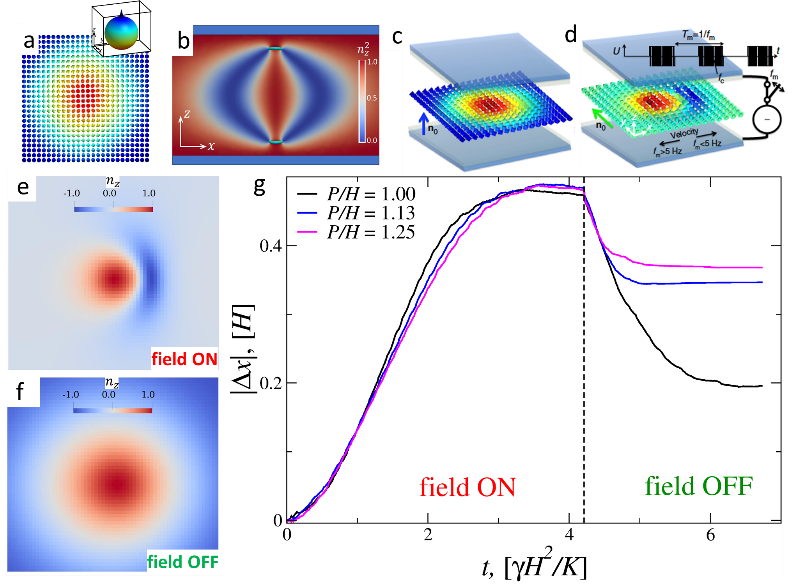}
    \caption{(a) Scheme of a 2D skyrmion vectorised director field colour-coded coded according to the legend at the right top; adapted with permission from~\cite{ackerman2017squirming}. (b) Cross sectional view of the 3D toron’s structure obtained between parallel plates imposing perpendicular boundary conditions. (c, d) By applying oscillating voltage across the cell, the toron moves in a certain direction with frequency-dependent velocity; adapted with permission from~\cite{ackerman2017squirming}.  (e, f) cross sectional, in the ($x,y$) plane, view of the toron at non-zero, e, and zero, (e), voltage across the cell. Colour corresponds to the $z-$component, $n_z$ of the director field. (g) shows the spatial displacement as a function of time of the central core region of the toron, coloured in red regions in (e) and (f), when the externa, electric field is changing from “on” to “off” states, as predicted by the minimization of the Frank-Oseen free energy; $H$ is the cell thickness, $K$ is the averaged elastic constant and $\gamma$ is the rotational viscosity.}
    \label{fig:Mykola1}
\end{figure*}

\section*{Relevance}

Endowing LCs with chirality opens exciting opportunities for fundamental exploration of manifestation of topology in a context of soft condensed matter physics linking it to other brunches of physics~\cite{smalyukh2020review}. Indeed, chiral LCs exhibit a particularly large variety of topologically nontrivial director distortions, or topological solitons, both without and with surface confinement and applied external fields ~\cite{wu2022hopfions}. These spatially localised features cannot be transformed continuously into a uniform/trivial state and reveal both field- and particle-like properties, thus named quasiparticles. For example, for weak perpendicular surface boundary conditions the so-called baby skyrmions were shown to be topologically nontrivial two dimensional (2D) translationally invariant structures, which are classified in terms of elements of the 2nd homotopy group of the order parameter space of nonpolar unit vectors (Fig.~\ref{fig:Mykola1}a). The director configuration of the baby skyrmion exhibits uniform twisting by  from its center in all radial directions~\cite{ackerman2017squirming}. These structures are low-dimensional analogue of Skyrme solitons proposed to describe sub-atomic particles with different baryon numbers in nuclear physics.  Under strong perpendicular boundary conditions, the skyrmion tubes terminate on point defects, which are also elements of the second homotopy group, and which allow embedding of the skyrmion tube in a uniform/trivial background director field in 3D. This configuration was called ``toron'' (Fig.~\ref{fig:Mykola1}b) and its mid-plane structure is identical to that of the 2D baby skyrmion (Fig.~\ref{fig:Mykola1}c). Torons were found experimentally in chiral LCs ~\cite{smalyukh2010threedimensional}, but later also observed in noncentrosymmetric magnets~\cite{donnelly2021experimental} and other systems, where the name coined within the LC field was also adopted and is now commonly used. In addition to skyrmions and torons more exotic solitons such as hopfions, heliknotons, and m\"obiusons, have been observed in chiral LCs experimentally~\cite{wu2022hopfions}.

From a practical standpoint, LC solitons are envisioned as elementary building blocks of reconfigurable on demand active soft materials (Fig.~\ref{fig:Mykola2}a) ~\cite{sohn2019schools}. By contrast to active colloids which are solid, LC solitons are soft as they lack physical interfaces and their motion is accompanied by the periodic expansion and contraction of topology-protected distorted LC regions, mimicking the behaviour of biological cells (Fig.~\ref{fig:Mykola1} e and f).

\begin{figure*}[h!]
    \centering
    \includegraphics[width=0.95\textwidth]{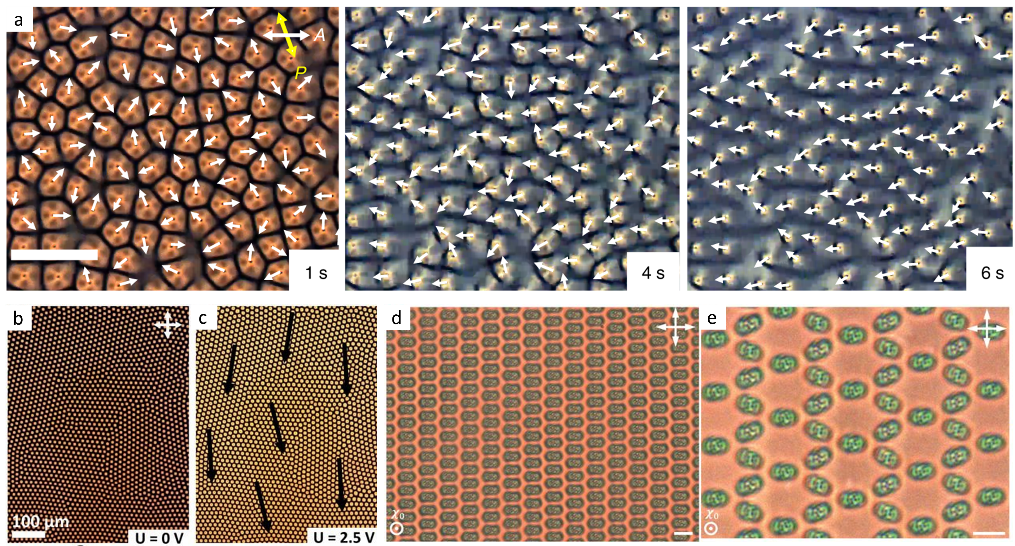}
    \caption{(a) Polarising micrograph showing a temporal evolution of skyrmion velocity vector orientations in a high number density region. The scale bar is ; adapted with permission from~\cite{sohn2019schools}. (b, c) Close-packed crystallites of torons are shown in polarising micrographs at zero voltage , (b), and at , (c). Grain boundaries separating different crystallites are clearly visible in (b). The black arrows in (c) denote the directions of the collective motion of each crystallite; adapted with permission from~\cite{sohn2020electrically}. (d) closed rhombic and (e) open heliknoton lattices obtained at applied voltage $U = \SI{1.9}{V}$ and $U = \SI{1.7}{V}$, respectively; the scale bars correspond to $\SI{10}{\mu m}$. Reproduced with permission from~\cite{tai2019threedimensional}.}
    \label{fig:Mykola2}
\end{figure*}

The motion of the solitons is powered by a time-dependent electric field applied to the LC in a direction normal to the confining surfacesa set up that resembles the one used in LC display technology (Fig.~\ref{fig:Mykola1}d)~\cite{ackerman2017squirming}. The basic physical mechanism of the soliton motion is related to the ``non-reciprocal'' rotational dynamics of the LC director field when the electric field is turned on and off (Fig.~\ref{fig:Mykola1} e-g), which was dubbed "squirming motion"~\cite{ackerman2017squirming}. Surprisingly, it is possible to control both the speed and the direction of the motion by varying the strength and the modulation frequency of the applied electric field ~\cite{ackerman2017squirming, sohn2019schools}. Additionally, soliton motion can be controlled by taking advantage of the unique optical properties of LCs~\cite{sohn2019lightcontrolled}. For example, the size and velocity of solitons, as well as their collective dynamics and self-assembly can be controlled by combining optical tweezer techniques and photo-patterning~\cite{sohn2019lightcontrolled}.

In multi-soliton systems, unexpected collective behaviour has been reported experimentally including light controlled soliton self-assembly and reconfigurable formation of large-scale soliton crystals \cite{sohn2019schools, sohn2019lightcontrolled, sohn2020electrically}. LC solitons exhibit effective elastic interactions that can be easily tuned in strength or switched from attractive to repulsive~\cite{sohn2019schools}. When the voltage modulation period is shorter than the LC response time, the soliton interactions are intrinsically nonequilibrium, resulting in remarkably rich emergent collective dynamics with reconfigurable out-of-equilibrium assemblies of solitons. At high packing fractions, hexagonal crystallites of tightly packed torons (Fig.~\ref{fig:Mykola1} b and c) can be brought into coherent motion along an arbitrary direction, which leads to an increased hexatic order parameter and is accompanied by the anisotropic deformation of the hexagonal soliton lattice~\cite{sohn2020electrically}. Another example is depicted in Fig.~\ref{fig:Mykola2} d and e, which show open and close lattices of ‘heliknotons' ~\cite{tai2019threedimensional} a novel type of 3D topological solitons with a dual nature, where the material director field  forms a non-singular Hopf soliton and an immaterial field of the helical axis  forms a disclination knot.

\section*{Challenges}

Experiments also show that the dynamic evolution of soliton textures that lack time-reversal symmetry can be used for transporting microparticles, as solitons can efficiently trap solid particles of comparable or smaller sizes due to the elastic properties of LCs. This provides a range of unique opportunities for controlled non-contact manipulation of colloids enabling the development of electro-optic responsive materials, because the experimental setups employed to drive solitons are similar to those used in LC display technologies~\cite{ackerman2017squirming}. Indeed, topological solitons are now becoming much more than just a playground for mathematical physics but also building blocks of meta matter where they are not only the fundamental particles but also active particles, with a host of potential technological uses. Despite the extensive body of experimental research, understanding the many-body nonequlibrium dynamics of LC solitons and soliton-colloid mixtures remains an outstanding challenge. Existing numerical investigations are limited to a small number of solitons and exploit relaxational dynamics to track the temporal evolution of the director field only, neglecting backflow effects and LC hydrodynamics. Addressing these open questions require novel multiscale approaches and coarse-graining strategies to access relevant length and time scales of the emergent collective dynamics of solitons and soliton-particle composites.

\endgroup

\newpage

\section{Competing interactions in skyrmion hosting materials}
\begingroup
    \let\section\subsection
    \let\subsection\subsubsection
    \let\subsubsection\paragraph
    \let\paragraph\subparagraph
Victor Ukleev$^1$, and Florin Radu$^1$
\vspace{0.5cm}

\noindent
\textit{$^1$ Helmholtz-Zentrum Berlin für Materialien und Energie, D-12489 Berlin, Germany}

\section*{Long-periodic Magnetic Textures in Non-centrosymmetric Materials}
Competition between symmetric exchange and antisymmetric Dzyaloshinskii-Moriya interaction (DMI) promotes topologically non-trivial magnetic states of high complexity~\cite{bogdanov1994thermodynamically}. DMI is a spin orbit interaction in nature and therefore has a small contribution to the low-energy Hamiltonian. It causes a local inversion symmetry breaking contributing to the stabilization of various spin textures, that are further modified by weak anisotropic interactions, dipole-dipole interactions, and external magnetic fields, thereby enriching the phase diagram of non-centrosymmetric materials.

Basic long-periodic magnetic textures that serve as building blocks for topological magnetic structures which occur in non-centrosymmetric magnets are illustrated in Fig. \ref{fig:Ukleev1}. The twisting topology of the spin textures is determined by the symmetry and by the entiometry of the underlying crystal lattice~\cite{cheong2022magnetic}. The ground state of archetypal skyrmion hosts -- cubic chiral magnets -- is a helical spiral with the propagation direction dictated by the balance between magneto-crystalline cubic anisotropy and anisotropic exchange interaction. A weak external magnetic field will re-orient the spiral and tilt the magnetic moments towards the field direction, independently of the orientation of the magnetic field and of the crystallographic axes, leading to a conical state. In the presence of an uniaxial or cubic anisotropy, an external magnetic field distorts the spiral, making it elliptic. This is referred to as chiral soliton lattice (CSL) state and it can easily be distinguished experimentally from spiral and conical states through the occurrence of higher-order magnetic diffraction peaks~\cite{togawa2012chiral}. For a finite external magnetic field and when the temperature is close to the ordering temperature ($T_\mathrm{C}$), a hexagonal skyrmion lattice (SkL) is formed, with its basal plane perpendicularly oriented with respect to the magnetic field direction~\cite{muehlbauer2009skyrmion}. For simplicity, a SkL can be viewed as a superposition of three spirals, with the skyrmion cores forming tubes along the direction of the external magnetic field. It is worth mentioning that in contrast to cubic systems, for magnets with polar $D_{2d}$, and $S_4$ symmetries, N\'eel-type skyrmions and anti-skyrmions are only stabilized when the field is oriented along the low-symmetry axis~\cite{bogdanov1994thermodynamically}.
\\
\begin{figure}[h!]
\includegraphics[width=.95\linewidth]{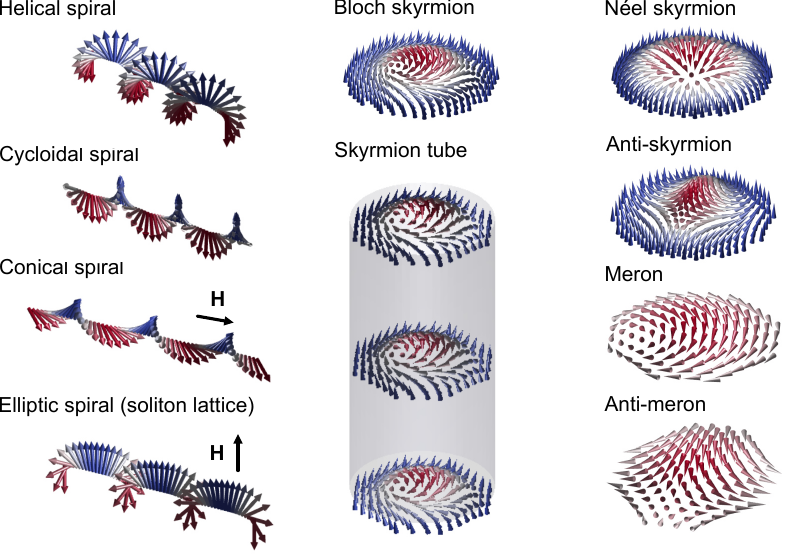}
\caption{Magnetic textures appearing in non-centrosymmetric materials.}
\label{fig:Ukleev1}
\end{figure}

\section*{Skyrmion Stability} 
Besides DMI, ferromagnetic exchange, and Zeeman interaction energy, two more key ingredients for skyrmion stability in bulk materials are identified, namely magnetic fluctuations \cite{muehlbauer2009skyrmion} and magnetically frustrated interactions \cite{leonov2015multiply}. In cubic chiral magnets, the skyrmion phase is typically confined in a few degrees’ proximity to the critical temperature \cite{muehlbauer2009skyrmion}. However, additional low-temperature (low-$T$) skyrmion phases have been observed in Co$_7$Zn$_7$Mn$_6$ \cite{karube2018disordered} and Cu$_2$OSeO$_3$ (Fig. \ref{fig:Ukleev2}) \cite{chacon2018observation,bannenberg2019multiple}. While the low-$T$ skyrmion phase in the frustrated $\beta$-Mn type structure that occurs in Co$_7$Zn$_7$Mn$_6$ was explained by the appearance of slow low-$T$ magnetic fluctuations \cite{ukleev2021frustration}, the stability of low-$T$ skyrmions as well as the additional tilted conical phase in Cu$_2$OSeO$_3$ (Fig. \ref{fig:Ukleev2}a) have been theoretically explained by the competition between the magnetocrystalline anisotropy and the anisotropic exchange interaction (AEI) \cite{bannenberg2019multiple}. The two interactions favor different directions of the spiral propagation vector, which under an  external magnetic field drives the system through magnetic frustration landscapes leading to stabilization of skyrmions far below $T_\mathrm{C}$.

\section*{Anisotropic Interactions.}
Recent advances in experimental methods allowed for observation and quantification of anisotropic exchange interaction through observing subtle helical/conical wavevector changes for continuous vectorial orientations of the external magnetic field with respect to crystal axes \cite{baral2023direct,ukleev2024competing}. Both small-angle neutron scattering (SANS) \cite{moody2021experimental} and resonant small-angle x-ray scattering (SAXS) \cite{baral2023direct} have thereby unveiled the importance of the AEI in Cu$_2$OSeO$_3$ at low $T$s (Fig. \ref{fig:Ukleev2}b), which supports the theoretical mechanism of skyrmion stability by the competition of magnetic anisotropies. 

Furthermore, chiral materials with intrinsic (magnetocrystalline) or extrinsic (e.g., strain-induced) uniaxial anisotropies were found to exhibit novel domain wall bimerons that appear at the boundaries between oppositely magnetized ferromagnetic domains \cite{nagase2021observation} and a double-$q$ square-coordinated meron-anti-meron lattices, as shown by means of Lorentz transmission electron microscopy (LTEM) (Fig. \ref{fig:Ukleev2}c) \cite{yu2018transformation}. In these systems, an easy-axis or an easy-plane anisotropy plays a critical role in stabilizing merons by favoring a directional locking of the magnetic moments along specific crystallographic or strain-induced axes. In $\beta-$Mn type chiral magnet Co$_8$Mn$_9$Zn$_3$~the balance between the DMI and easy-plane anisotropy is rather fragile, and therefore leads to a narrow stability window of the double-$q$ meron lattice phase, which transforms to a hexagonal SkL by a small change of the magnetic field \cite{yu2018transformation}. Particularly, the two phases coexist in bulk crystals as shown by means of SANS (Figs. \ref{fig:Ukleev2}d--f) \cite{white2025anisotropy}.

\section*{Anisotropic DMI and Dipolar Interactions}
Magnetic anti-skyrmions (aSk) are stabilized in non-centrosymmetric tetragonal systems with anisotropic DMI. Archetypal aSk hosts are Heusler alloys Mn$_{1.4}$PtSn \cite{nayak2017magnetic} and Fe$_{1.9}$Ni$_{0.9}$Pd$_{0.2}$P \cite{karube2021room} with $D_{2d}$ and $S_4$ symmetries, respectively. However, both of these material families only exhibit aSk in thin plate samples, indicating strong competition between anisotropic DMI, magnetic dipolar interactions, and uniaxial anisotropy. Due to the fragile balance between these interactions, a plethora of magnetic textures such as helices, elongated skyrmions, aSk, and CSL structures can be stabilized in samples with different thicknesses (control of the stray field), density of defects (control of the uniaxial anisotropy), temperature (control of the net magnetization), and eventually hydrostatic and shear pressure. These rich tunability options, as well as their ability to host topological textures at room temperature, allows for fine control of the topological response and their use in spintronic devices, correspondingly.

\begin{figure*}[h!]
\includegraphics[width=.95\linewidth]{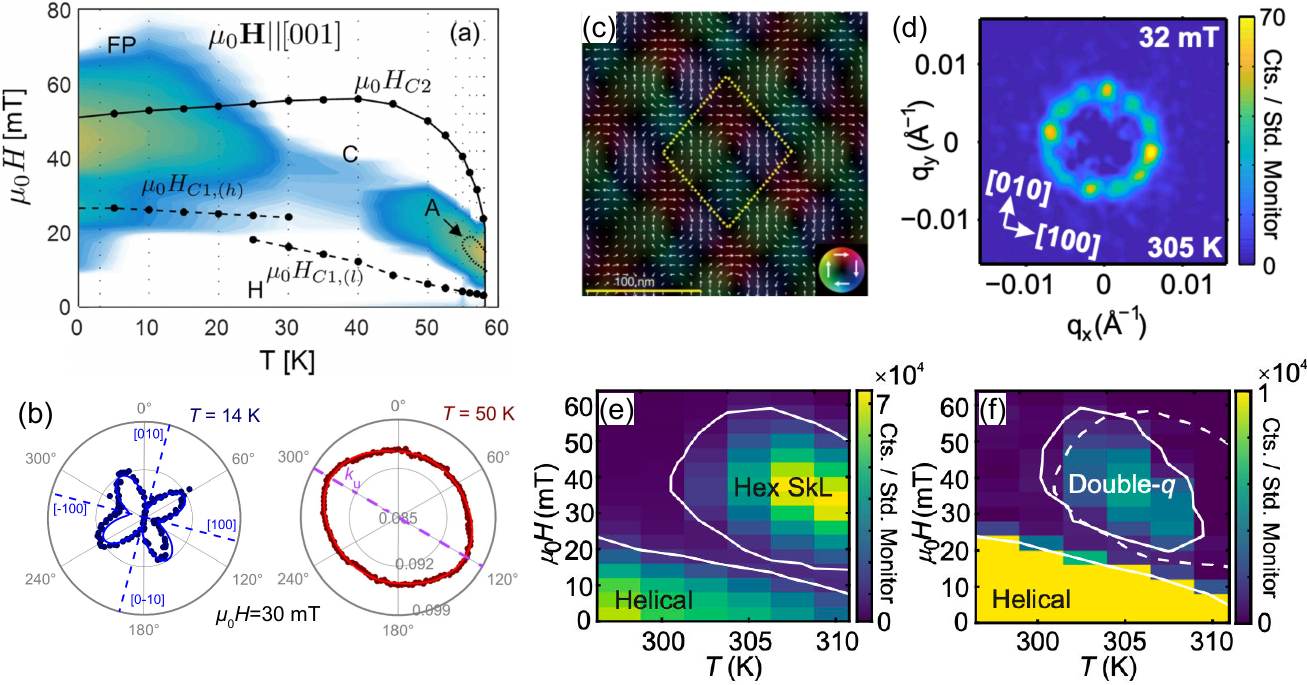}
\caption{(a) Magnetic phase diagram of Cu$_2$OSeO$_3$ constructed from SANS (reprinted by permission from npj Quant. Mat. \cite{bannenberg2019multiple}). (b,c) Azimuthal dependence of the conical wavevector in Cu$_2$OSeO$_3$ (001) plane at 14\,K and 50\,K, respectively \cite{baral2023direct}. The anisotropy of the conical wavevector manifests the impact of the AEI at low temperatures. (c) Double-$q$ meron-anti-meron lattice observed in a Co$_8$Zn$_9$Mn$_3$ thin plate by means of LTEM (reprinted by permission from Nature~\cite{yu2018transformation}). (d) Co-existing six-fold  and four-fold SANS patterns from a bulk Co$_8$Zn$_9$Mn$_3$, corresponding to a SkL and a double-$q$ texture, respectively. Magnetic phase diagram of bulk Co$_8$Zn$_9$Mn$_3$ measured by SANS highlighting stability regimes of a hexagonal SkL (e) and double-$q$ texture  (reprinted by permission from Adv. Mat. \cite{white2025anisotropy}).}
\label{fig:Ukleev2}
\end{figure*}

\section*{Future Directions}
The ultimate aim of future theoretical and experimental efforts is to deepen our understanding and control of these weak interactions and their interplay, enabling the design of tunable magnetic textures for spintronic applications. In particular, competing interactions may play a crucial role in stabilizing room-temperature topological spin structures with dimensions around 1-10 nm. Furthermore, the added degree of freedom provided by chirality control holds great promise for applications in quantum technologies \cite{psaroudaki2021skyrmion}. 

Finally, novel observation of theoretically predicted but not yet observed experimentally modulated chiral phases are to be addressed. They are a result of higher-order four-spin chiral exchange interactions in materials with $T_d$ group symmetries \cite{ado2021noncollinear,rybakov2021antichiral}. The anti-chiral spirals predicted in these materials provide a fertile soil for stabilization of field-induced modulated and topological spin textures.

\endgroup

\newpage

\section{Topological spin textures in S$_4$-symmetric antiskyrmion hosting magnets}
\begingroup
    \let\section\subsection
    \let\subsection\subsubsection
    \let\subsubsection\paragraph
    \let\paragraph\subparagraph
Fehmi Sami Yasin$^1$, and Jan Masell$^{2,3}$
\vspace{0.5cm}

\noindent
\textit{$^1$ Center for Nanophase Materials Sciences, Oak Ridge National Laboratory, Oak Ridge, Tennessee 37831, United States\\
$^2$ Institute of Theoretical Solid State Physics, Karlsruhe Institute of Technology, 76131 Karlsruhe, Germany\\
$^3$ RIKEN Center for Emergent Matter Science (CEMS), Wako 351-0198, Japan}

{\footnotesize Notice to the editor: This manuscript has been authored by UT-Battelle, LLC, under contract DE-AC05–00OR22725 with the US Department of Energy (DOE). The US government retains and the publisher, by accepting the article for publication, acknowledges that the US government retains a nonexclusive, paid-up, irrevocable, worldwide license to publish or reproduce the published form of this manuscript, or allow others to do so, for US government purposes. DOE will provide public access to these results of federally sponsored research in accordance with the DOE Public Access Plan (https://www.energy.gov/doe-public-access-plan).}

\section*{Introduction}
Spin-orbit coupling leads to anisotropic exchange interactions dictated by atomic lattice symmetries.
In an $S_4$-symmetric magnet, four-fold roto-inversion symmetry with respect to the tetragonal unit cell's c-axis enables uniaxial anisotropy, $E_{ani}=-K_1 (\hat{c}\cdot\vec{m})^2$.
The easy-axis type, $K_1>0$, stabilizes a labyrinth phase or magnetic bubbles.
Moreover, broken inversion symmetry leads to Dzyaloshinskii-Moriya interaction (DMI).
The Moriya vectors obey the $S_4$ symmetry which in the continuum limit results in $E_{DM}=D \,\vec{m}\cdot \left(\hat{d}_+\times\partial_+\vec{m}-\hat{d}_-\times\partial_-\vec{m}\right)$ with the two perpendicular vectors $\hat{d}_+ = \hat{d}_-\times\hat{c}$ in the a-b-plane and derivatives $\partial_{\pm}$ in the directions $\hat{d}_{\pm}$, respectively.
For $D>0$, right-handed helices and and left-handed helices stabilize along $\vec{d}_+$ and $\vec{d}_-$, respectively.
Unlike $D_{2d}$, $S_4$ symmetry does not fix the absolute direction of $\hat{d}_{\pm}$.~\cite{Bogdanov1989,nayak2017magnetic}
In the $S_4$-symmetric doped schreibersite $\rm (Fe_{0.63}Ni_{0.3}Pd_{0.07})_{3}P$, for example, the vectors $\hat{d}_{\pm}$ point along [110] and [$\overline{1}$10], pinning the stripe phase (one-dimensional soliton lattice) in these orthogonal directions depending on their handedness.~\cite{karube2021room}

\begin{figure*}[h!]
    \centering
    \includegraphics[width=0.99\textwidth]{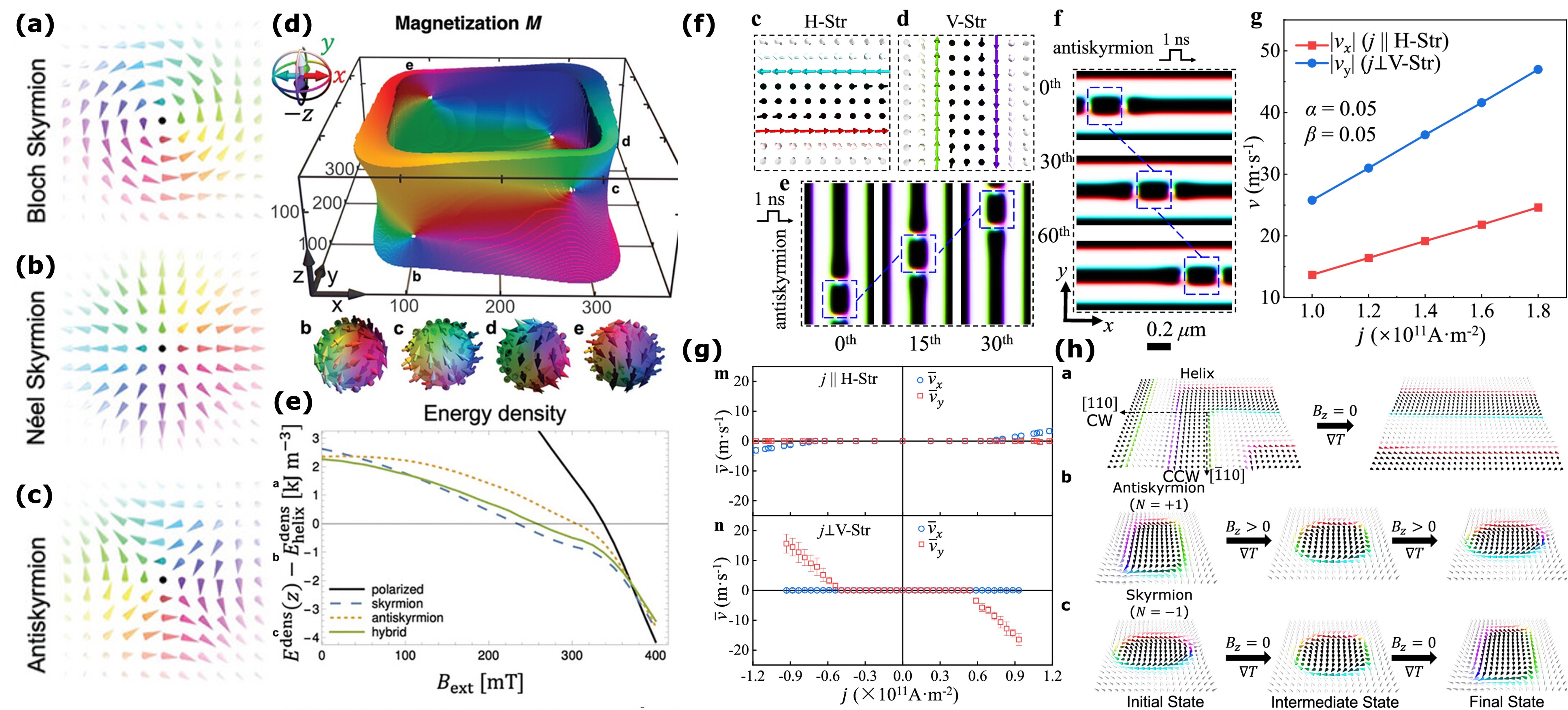}
    \caption{
        Magnetic textures in $S_4$-symmetric magnets and their collective dynamics.
        (a-c) Two-dimensional skyrmionic textures.
        (d) Magnetization $\left(|m_z| < 1/6\right)$ in a three-dimensional (3d) hybrid antiskyrmion with (anti-)Bloch points highlighted in sub-panels b-e.
        (e) Energy density of a hybrid antiskyrmion lattice relative to the stripe phase. 
        (f) Magnetization of the stripe phase (horizontal (h) and vertical (v) stripes, sub-panels c and d) and the simulated motion of antiskyrmions in the two stripe orientations (sub-panels e and f) via $1\,\textrm{ns}$ current pulses in the h-direction and their respective velocities (sub-panel g). 
        (g) Experimentally measured antiskyrmion velocity for current density parallel or perpendicular to the stripe's orientation.
        (h) Schematics of observed spin texture transitions induced by a temperature gradient. 
        (a-e) reproduced from Figs. 1 and 2 of Ref.~\cite{Yasin2024} with permission from Wiley‐VCH GmbH (https://creativecommons.org/licenses/by/4.0/). (f-g), (h) reproduced from Figs. 1 and 2 of Ref. \cite{guang2024confined} (https://creativecommons.org/licenses/by-nc-nd/4.0/) and Fig. 1 of Ref. \cite{Yasin2023} (https://creativecommons.org/licenses/by/4.0/) with permission from Springer Nature.
    }
    \label{fig:Yasin1}
\end{figure*}

The competition between the various interactions and  $S_4$-symmetric DMI results in exotic magnetic textures.
Without external field, the minimization of dipolar energy and uniaxial anisotropy leads to a one-dimensional soliton lattice (stripe phase) of \emph{domain walls} which have left- or right-handed Bloch helicity. 
Moreover, the $S_4$-symmetric DMI prefers a specific helicity depending on the orientation of the stripes, which leads to strong pinning, see Fig.~\ref{fig:Yasin1}(f) or (h). 
Under an external field, the stripes transform into bubbles, see Fig.~\ref{fig:Yasin1}(h). 
With an in-plane component, the bubbles are \emph{non-topological bubbles} with vanishing skyrmion winding number $W=\frac{1}{4 \pi} \int\!\!\int \mathbf{m} \cdot \left(\frac{\partial \mathbf{m}}{\partial x} \times \frac{\partial \mathbf{m}}{\partial y}\right)\ dx\ dy$ but finite in-plane magnetization and, thus, broken rotational symmetry.
The $S_4$-symmetric DMI in combination with predominantly Bloch-type winding of domain walls restricts these bubbles to four possible orientations, i.~e., the $\hat{d}_{\pm}$ directions.
If the field is perpendicular to the plane, symmetric solitons are favored. 
\emph{Elliptical skyrmions} with $W=-1$ emerge with Bloch-type (Fig.~\ref{fig:Yasin1}(a)) domain walls due to the strong dipolar interaction, with left- or right-handed chirality energetically degenerate. To maximize the DMI energy they distort elliptically, minimizing the side length where the helicity does not match the DMI  and pinning the skyrmion helicity to be distinct in orthogonal $\hat{d}_{\pm}$ directions..
Moreover, \emph{square-shaped antiskyrmions} ($W=+1$) with helicity fully compliant with the $S_4$-symmetric DMI~\cite{Bogdanov1989} form with shortened Néel-type parts (Fig.~\ref{fig:Yasin1}(b)) to minimize dipolar energy.~\cite{Camosi2018}
However, their antivortex character (Fig.~\ref{fig:Yasin1}(c)) leads to unavoidable large dipolar energy at the surfaces.
Such conditions result in the emergence of \emph{hybrid antiskyrmions}, which are antiskyrmions in the bulk which transition into skyrmions near the surface via pairs of \emph{(anti-) Bloch points} (BP), see Fig.~\ref{fig:Yasin1}(d-e).
Due to the attraction between BP and anti-BP, the hybrids spontaneously elongate in either of the $\hat{d}_{\pm}$ directions, leading to two flavors of rectangular shaped solitons.

The zoo of magnetic textures can be manipulated by various means, as has been shown recently, beyond the above-mentioned control via magnetic field.
An electric current can move the textures.
Besides motion of isolated textures or lattices, they can move in more complex settings, e.~g., an antiskyrmion embedded on a helical racetrack, Fig.~\ref{fig:Yasin1}(f-g).
This setting enhances the antiskyrmion's velocity and demonstates the relation between the direction of motion and the sign of the skyrmion charge $W$.~\cite{guang2024confined}
A temperature gradient and compressive strain have been used to demonstrate controlled orientation of the stripe phase.~\cite{Yasin2023, Mori2025}
The temperature gradient can also convert antiskyrmions into non-topological bubbles and subsequently into skyrmions or vice versa, depending on the external field, Fig.~\ref{fig:Yasin1}(h).

\section*{Relevance and Vision}

$S_4$-symmetric materials show promise across a range of spintronics applications. 
The large number of distinct spin textures - two skyrmions, one antiskyrmion, four distinct non-topological bubbles, and two distinct helical states - hints at the possibility of a higher numeral system-based memory and computing.
In 3d, two rectangular distortions of hybrid antiskyrmions exist due to strong magnetostatic surface effects which may also stabilize 3d topological states in other phases or hybrid textures that exhibit surface reconstruction on only one surface.
Moreover, the low symmetry of $S_4$ enables further symmetry breaking: Fourth-order on-site anisotropy couples the helicity and orientation of domain walls, competing with the DMI.~\cite{Hemmida2024}

What's more, hybrid textures offer the ability to engineer tunable electromagnetic responses and motion.
Textures with (anti-)Bloch points are mobile and their thickness-averaged topological winding number is tunable, e.~g., by the thickness and applied magnetic field.~\cite{Yasin2024} 
This may provide a method to steer these spin textures through current-driven devices or eliminate their Hall motion entirely. 
The topological Hall effect would be tunable and even reversible in sign. 
Such a tuning knob would have tremendous impact on spintronics memory and computing architectures.
One can envision heterostructures of materials with different symmetries and therefore different flavors of topological textures to engineer artificial hybrid solitons. 

Finally, unique transport signatures could emerge. 
For example, the monodomain stripe phase turns the crystal into a p-wave magnet, where the sign of the direction-dependent spin polarization depends on the helicity and thereby orientation of the stripe phase.

\section*{Challenges}

To date, only schreibersites and several chemically doped variations have been explored as topological magnetic texture-hosting $S_4$ symmetry candidates.~\cite{karube2022doping,Hemmida2024}
Other base materials are desirable, as they may show stronger competition between anisotropic interactions.
Moreover, $S_4$-symmetric insulators for magneto-electric control or magnonics remain elusive.

While 3d measurements have been performed,~\cite{Yasin2024} spatial resolution near the surfaces needs improvement.~\cite{Yu2022}
Further improvement of 3d imaging is required, both in the high-resolution static regime and in the dynamic regime, for the visualization of complex topology and dynamics.
Moreover, the crystal quality needs to be improved in terms of purity such that inclusions of dopants and other defects don't interfere with the magnetic textures.
Nanofabricated sample defects such as disordered "dead" layers which break the symmetry of the host crystal need to be reduced.

Although antiskyrmions embedded in a stripe phase are highly stable and can be manipulated, the pinned background restricts the breadth of possible dynamical phenomena.
The (anti-)skyrmion Hall motion and especially the effect of the Bloch point quadrupole present in hybrid antiskyrmions, for example, would be easier studied in a polarized spin background.
However, antiskyrmions are fragile under the magnetic field values required to polarize the background, although initial research concerning thermal gradient-driven antiskyrmions shows promise.~\cite{Peng2025}

Finally, the proper description of 3d magnetic textures in magnets with low symmetry remains a challenge. 
One issue is the extraction of interaction parameters from experiments on a system with highly competing interactions.
Simulating a 3d system with high precision is also a challenge due to the large number of sites required to discretize a multi-scale problem such as skyrmions which are large compared to their domain wall width or intrinsically multi-scale Bloch points.

\section*{Concluding remarks}
$S_4$-symmetric magnets offer many possibilities for competing magnetic interactions which can stabilize a multitude of 2d and 3d topological magnetic textures.
However, suitable materials are rare, imaging methods are complex and still in an early development stage, and theoretical understanding is hard to gain in a multi-scale system with many relevant parameters.
Artificial intelligence and machine learning methods should be coupled with real-space imaging to create automated magnetic phase mapping experimental workflows to improve the throughput from material genesis to spin texture identification and manipulation.
The creation of materials hosting (anti-)skyrmions with smaller sizes at higher temperatures should be explored.
In particular, creation, transportation, detection, and annihilation methods must be developed for all spin textures present in $S_4$-symmetric magnets.

\section*{Acknowledgments}
J.~M.~acknowledges funding by the Deutsche Forschungsgemeinschaft (DFG, German Research Foundation) under the Project No. 547968854. Research sponsored by the Laboratory Directed Research and Development Program of Oak Ridge National Laboratory, managed by UT-Battelle, LLC, for the US Department of Energy. Work conducted at the Center for Nanophase Materials Sciences (CNMS), which is a US Department of Energy Office of Science User Facility at Oak Ridge National Laboratory.
\endgroup

\newpage

\section{Skyrmions in 2D materials}
\begingroup
    \let\section\subsection
    \let\subsection\subsubsection
    \let\subsubsection\paragraph
    \let\paragraph\subparagraph
Chenhui Zhang$^1$, and Hyunsoo Yang$^1$
\vspace{0.5cm}

\noindent
\textit{$^1$ Department of Electrical and Computer Engineering, National University of Singapore, Singapore 117576, Singapore}

\section*{Status}
2D van der Waals (vdW) materials, represented by graphene and $\rm MoS_2$, have attracted extensive research interest in the past two decades. Among the huge family of 2D materials, a few members can maintain the long-range magnetic order against thermal fluctuations even down to atomic thickness scale, providing an extraordinary playground for spintronics. The interplay between 2D structures and magnetic interactions may give rise to topological spin textures, including magnetic skyrmions. Using 2D skyrmion-hosting materials, emerging spintronic logic and memory devices are expected to be developed for the post-Moore electronics.

To date, most of the reported 2D vdW magnets have a Curie temperature ($T_C$) far below room temperature. Thus, the observed skyrmion phase in 2D ferromagnets are usually located at cryogenic temperatures. For example, ferromagnetic insulator $\rm Cr_2Ge_2Te_6$ was the first reported 2D magnet that could host Bloch-type skyrmion bubbles~\cite{han2019topological}. As the $T_C$ of $\rm Cr_2Ge_2Te_6$ was only $\SI{65}{K}$, the skyrmion bubbles were observed around $\SI{20}{K}$~\cite{han2019topological}. $\rm Fe_3GeTe_2$ is a vdW ferromagnetic metal and exhibits strong perpendicular magnetic anisotropy, which is favorable for magnetic memory applications. It has a much higher $T_C$ of $\approx \SI{230}{K}$, while the existence of Fe deficiency can notably reduce the transition temperature~\cite{liu2023controllable}. Both Bloch-type skyrmion bubbles and Néel-type skyrmions were reported in $\rm Fe_3GeTe_2$ below 160 K~\cite{ding2020observation, park2021neeltype}. By increasing the Fe ratio, one can obtain another Fe-based 2D ferromagnet $\rm Fe_5GeTe_2$ which has a room-temperature $T_C$ of $\approx \SI{310}{K}$~\cite{may2019ferromagnetism}. (Anti)merons and skyrmion bubbles have been reported in $\rm Fe_5GeTe_2$, while their phase temperatures are still limited below 273 K~\cite{gao2020spontaneous, zhang2022magnetic}. Transition metal dichalcogenides ${\rm{Cr}_{1+\delta} \rm{Te_2}}$ is an interesting material whose $T_C$ can vary from 169 to \SI{333}{K} as the self-intercalate concentration $\delta$ increases from 0.37 to 0.60~\cite{zhang2023roomtemperature}. Remarkably, room-temperature skyrmion bubbles and topological Hall effect have been observed in this system \cite{zhang2023roomtemperature}. More recently, room-temperature Néel-type skyrmions were reported in a new 2D ferromagnet $\rm Fe_3GaTe_2$~\cite{li2024roomtemperature, zhang2024aboveroomtemperature}. The skyrmion lattice generated by a field-cooled process shows remarkable thermal stability and can persist up to 330 K, which is higher than all other known 2D magnets~\cite{zhang2024aboveroomtemperature}. Fig.~\ref{fig:Yang1} summarizes the recent advances in the exploration of emerging 2D skyrmion-hosting materials.

\begin{figure*}[h!]
    \centering
    \includegraphics[width=0.95\linewidth]{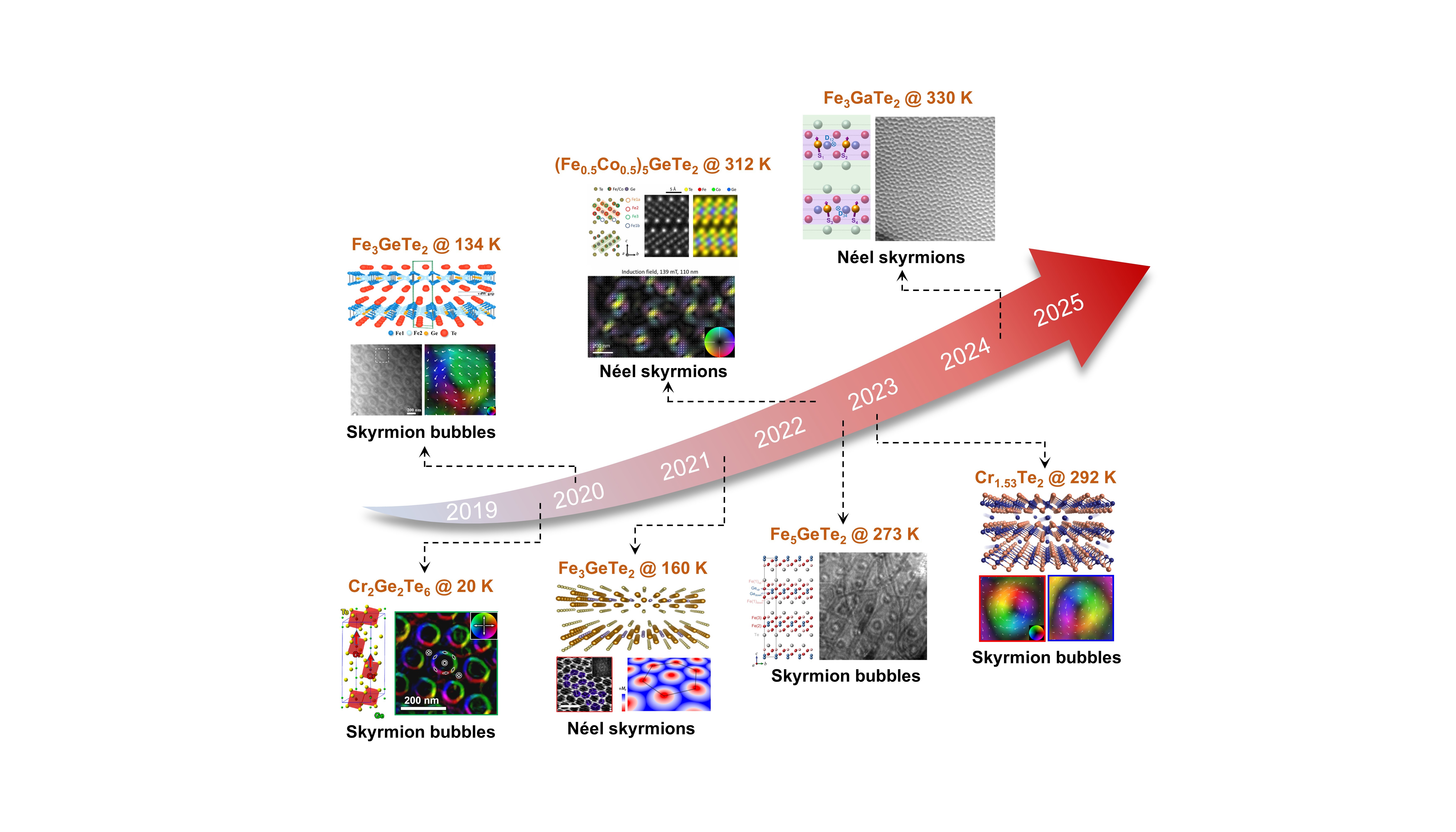}
    \caption{Recent advances in the exploration of emerging 2D skyrmion-hosting materials, including $\rm Cr_2Ge_2Te_6$ (Reprinted with permission from \cite{han2019topological}. Copyright © 2019 American Chemical Society), $\rm Fe_3GeTe_2$ (Reprinted with permission from \cite{ding2020observation}. Copyright © 2020 American Chemical Society; Reprinted with permission from \cite{park2021neeltype}, Copyright © 2021 by the American Physical Society), $\rm Fe_5GeTe_2$ (Reprinted with permission from \cite{zhang2022magnetic}. John Wiley \& Sons. © 2022 WILEY-VCH Verlag GmbH \& Co. KGaA, Weinheim), $\rm {(Fe_{0.5}Co_{0.5})_5GeTe_2}$, $\rm Cr_{1.53}Te_2$ (Reprinted with permission from \cite{zhang2023roomtemperature}. John Wiley \& Sons. © 2022 WILEY-VCH Verlag GmbH \& Co. KGaA, Weinheim), and $\rm Fe_3GaTe_2$.}
    \label{fig:Yang1}
\end{figure*}

It is known that the Dzyaloshinskii–Moriya interaction (DMI) plays an important role in skyrmion stabilization. The interplay among DMI, exchange interaction, magnetic anisotropy, and external magnetic field give rise to the helical spin texture of skyrmions. However, most of the discovered 2D magnets have centrosymmetric crystal structures, which prohibit the existence of effective DMI. Alternatively, the competition between the dipole–dipole interaction and uniaxial magnetic anisotropy can also stabilize Bloch-type skyrmion bubbles (i.e. dipole skyrmions), which are exactly observed in $\rm Cr_2Ge_2Te_6$ and $\rm Fe_5GeTe_2$~\cite{han2019topological, zhang2022magnetic}. Although these skyrmion bubbles have the same topological classification as the DMI-stabilized skyrmions in chiral B20 magnets and heavy metal/ferromagnet multilayers, the latter generally have a smaller skyrmion size, which is more favorable for high-density device integration. In addition, dipole skyrmion systems usually show two opposite skyrmion chiralities at the same time, while DMI can break the energy degeneracy and lead to homochiral skyrmions. The homochirality guarantees a consistent direction of skyrmion motion when driven by a spin-polarized current. More importantly, one may achieve an ideal Bloch-Néel hybridization via DMI modulation to obtain zero-skyrmion-Hall-angle skyrmions. These DMI-induced features can be very useful for skyrmion racetrack memories. Therefore, inducing DMI into 2D skyrmion systems is meaningful and merits particular attention.

VdW-gap-intercalation is a possible way to lower the crystal symmetry and induce an effective DMI. For example, $\rm CrTe_2$ is a centrosymmetric vdW magnet belonging to the space group $P\bar{3}m$. A small amount of Cr-intercalation, i.e. ${\rm{Cr}_{1+\delta} \rm{Te_2}}$ ($\delta = 0.3$), can reduce the crystal symmetry from $P\bar{3}m$ to $P3m1$, giving rise to Néel-type skyrmions but at low temperatures (100–200 K)~\cite{saha2022observation}. Although increasing the intercalate concentration can raise the skyrmion phase temperature, the crystal structure becomes centrosymmetric and DMI is dismissed \cite{zhang2023roomtemperature}. Chemical doping is another strategy to tune the structural and magnetic properties of 2D materials. $\rm {(Fe_{0.5}Co_{0.5})_5GeTe_2}$ exhibits an unexpectedly high $T_C$ of $\approx \SI{350}{K}$, and room-temperature Néel-type skyrmions can be achieved (up to 312 K) \cite{zhang2022roomtemperature}. The transition from Bloch-type skyrmion bubbles in $\rm Fe_5GeTe_2$ to Néel-type skyrmions in $\rm {(Fe_{0.5}Co_{0.5})_5GeTe_2}$ is attributed to the crystal inversion symmetry breaking induced by Co doping. In addition to the above-mentioned methods, recent investigations reveal that global DMI can also be induced by lattice distortions in 2D crystals. Room-temperature ferromagnet $\rm Fe_3GaTe_2$ was supposed to be centrosymmetric, while unexpected vertical displacements were observed at Fe sites in Fe-deficient $\rm Fe_{3-x}GaTe_2$ via atomic-resolution scanning transmission electron microscopy and single-crystal X-ray diffraction~\cite{li2024roomtemperature, zhang2024aboveroomtemperature}. Such displacements can change the crystal structure from centrosymmetric $P6_3/{mmc}$ to non-centrosymmetric $P3m1$ space group. First-principles calculations further demonstrate that an effective interfacial-type DMI can be induced through the Fe–Te–Fe path under the scheme of Fert–Lévy mechanis\cite{zhang2024aboveroomtemperature}. It is also expected that the DMI strength in $\rm Fe_{3-x}GaTe_2$ can be controlled by the deficiency level x, similar to the situation in $\rm Fe_{3-x}GaTe_2$ as they share the same crystal structure~\cite{liu2023controllable}. Therefore, Bloch-Néel hybrid skyrmions will emerge as a result of the interplay between DMI and dipole–dipole interaction~\cite{lv2024distinct}. In ideal circumstances, a nearly vanishing skyrmion Hall angle might be achieved by tuning the DMI strength carefully, which needs to be demonstrated in future studies.

\begin{figure}[h!]
    \centering
    \includegraphics[width=0.95\linewidth]{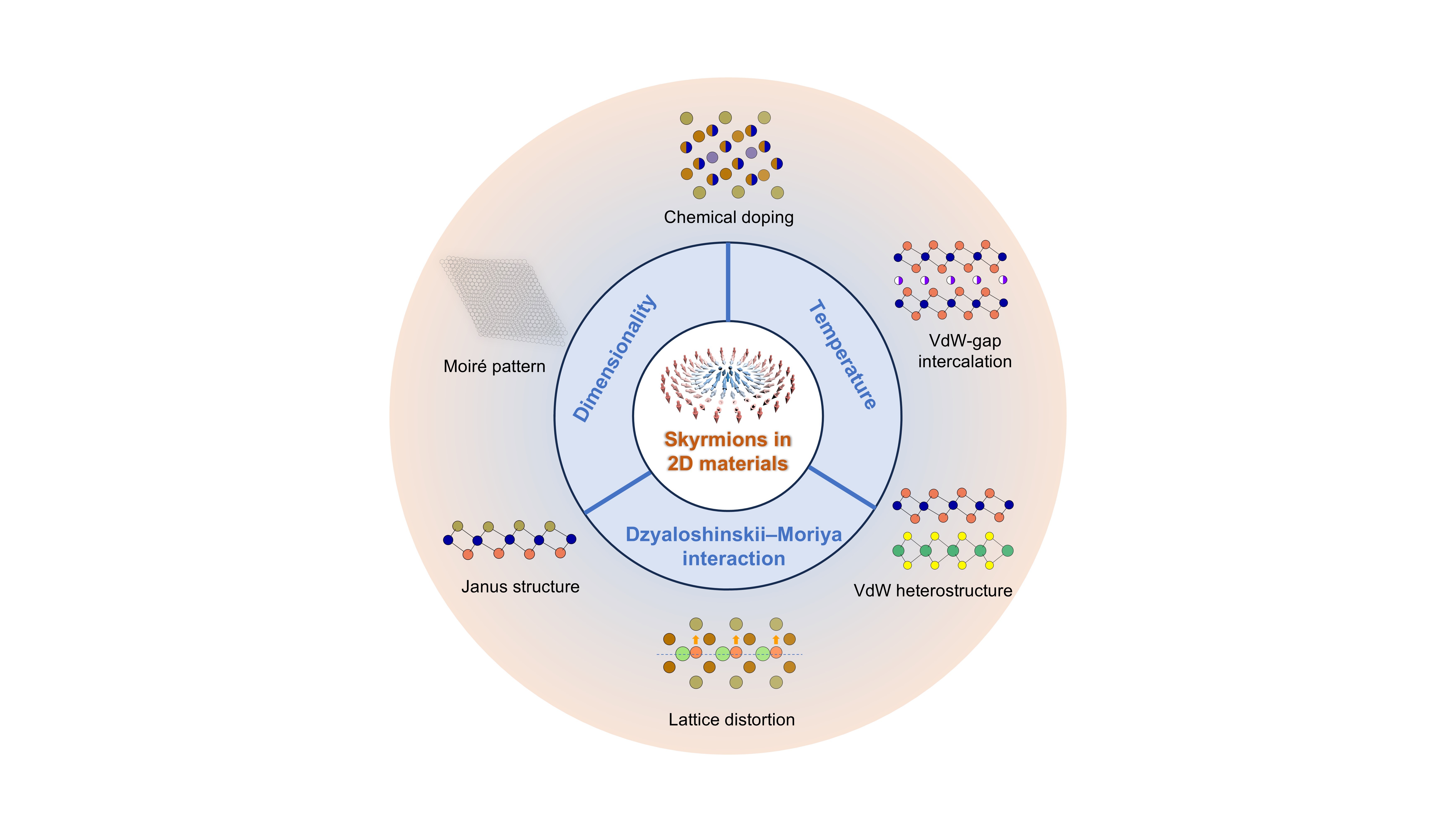}
    \caption{Challenges and opportunities in 2D skyrmion research. Reducing the dimensionality to approach real 2D, further increasing the skyrmion phase temperature, and inducing robust DMI are three main challenges. Future studies may leverage 2D materials to construct novel architectures, such as vdW 
    heterostructures, moiré patterns, and Janus structures, to explore exotic physical phenomena and device functionalities in 2D skyrmionics.}
    \label{fig:Yang2}
\end{figure}

\section*{Challenges}
However, despite extensive efforts and inspiring progress that have been made in the past few years, the 2D skyrmionics is still in its infancy. Many questions and challenges, both in terms of material and physical aspects, remain to be resolved, as summarized in Fig.~\ref{fig:Yang2}. First, the dimensionality of the reported vdW skyrmion-hosting materials is hard to define as truly 2D. In $\rm Fe_{3-x}GaTe_2$, the critical thickness for skyrmion formation is approximately 25 nm, below which only strip domains are observed~\cite{li2024roomtemperature}. The reason could be that the out-of-plane magnetic anisotropy in thin $\rm Fe_{3-x}GaTe_2$ is so strong that it cannot coordinate with other energy terms to stabilize skyrmions. Nevertheless, theoretical calculations predict that the lateral modulation of interlayer magnetic coupling by the locally different atomic registries in 2D moiré patterns can stabilize skyrmions~\cite{tong2018skyrmions}. This new mechanism for generating truly 2D skyrmions remains to be experimentally demonstrated. Second, further increasing the skyrmion phase temperature is necessary for practical applications. 2D materials with $T_C$ only slightly above room temperature may not be viable for practical devices, as the current-driven skyrmion motion will inevitably cause Joule heating, which can raise the device temperature by tens of kelvins. New 2D skyrmion-hosting materials with higher working temperatures are needed to address this challenge. Besides, chemical doping \cite{zhang2022roomtemperature} and utilizing vdW heterostructures to construct interfacial exchange coupling \cite{wang2020roomtemperature} are alternative ways to enhance $T_C$. Third, more innovative strategies should be implemented to induce robust DMI in 2D magnets. In 2D magnets such as $\rm Fe_{3-x}GaTe_2$ and $\rm Cr_{1.3}Te2$, the skyrmion size is highly sensitive to the sample thickness and external magnetic field, suggesting that skyrmions are primarily dipolar stabilized in spite of a finite DMI \cite{zhang2024aboveroomtemperature, saha2022observation}. An effective method to solve this problem is to build up an vdW interface with other 2D materials, and thus an extra interfacial DMI can be induced \cite{wu2020neeltype}. Moreover, theoretical works reveal that in magnetic Janus monolayers, such as Mn$XY$ ($X, Y =$ S, Se, or Te, $X \neq Y$), the distinct top and bottom surface atoms can break the inversion symmetry and induce a strong DMI \cite{liang2020very}. In this way, truly 2D skyrmions may also be realized.

\section*{Conclusion}
In summary, 2D magnetic materials provide an extraordinary platform for exploring nontrivial spin textures. To date, various 2D skyrmion-hosting materials have been demonstrated, some of which can even stabilize skyrmions at room temperature. Despite the progress achieved, sustained efforts are still needed to overcome the existing challenges. Specifically, 2D materials possess unique advantages for constructing novel architectures, such as vdW heterostructures, moiré patterns, and Janus structures, offering valuable opportunities to explore exotic physical and device properties beyond the reach of conventional magnetic systems.

\endgroup

\newpage

\section{Beyond Flatland: Unveiling the Three-Dimensional Skyrmion}
\begingroup
    \let\section\subsection
    \let\subsection\subsubsection
    \let\subsubsection\paragraph
    \let\paragraph\subparagraph
\newcommand*\mycommand[1]{\texttt{\emph{#1}}}
\newcommand{\THcomment}[1]{\noindent\fbox{\parbox{0.7798\textwidth}{\color{blue}#1}}}
	
Shilei Zhang$^{1,2,3}$, and Thorsten Hesjedal$^{4,5}$ 
\vspace{0.5cm}

\noindent
\textit{$^1$ School of Physical Science and Technology, ShanghaiTech University, Shanghai 201210, China\\
$^2$ ShanghaiTech Laboratory for Topological Physics, ShanghaiTech University, Shanghai 201210, China\\
$^3$ Center for Transformative Science, ShanghaiTech University,  Shanghai 201210, China\\
$^4$ Diamond Light Source, Harwell Science and Innovation Campus, Didcot OX11~0DE, United Kingdom\\
$^5$ Clarendon Laboratory, Department of Physics, University of Oxford, Oxford, OX1~3PU, United Kingdom\\}

\section*{Introduction}

The field of skyrmionics has been largely shaped by the study of two-dimensional (2D) skyrmion lattices, topologically non-trivial spin textures typically visualised as planar magnetic vortices \cite{yu2010real}. 
Their simple 2D geometry made them ideal testbeds for developing methods to generate, image, and manipulate skyrmions, thereby preparing the ground for potential device applications.
However, in magnetic materials where translational symmetry extends through the third dimension, skyrmions naturally form extended string-like configurations. 
Until recently, these skyrmion strings were often assumed to be homogeneous along their length, effectively treated as 2D textures extended along the third dimension. 
A growing number of theoretical predictions \cite{rybakov2013three} and experimental findings \cite{zhang2018direct, zhang2018reciprocal, zheng2018experimental, birch2020real, ran2021creation, ran2022axially, birch2022toggle, jin2023evolution, xie2023observation, yu2024skyrmion} revealed a far richer picture: in bulk or confined geometries, skyrmions exhibit complex three-dimensional (3D) modulations, including twisted, terminated, and even braided textures that go beyond purely planar descriptions.

\section*{Relevance and Vision}

A foundational breakthrough in this context was the direct mapping of depth-dependent spin structures within the skyrmion lattice phase of Cu$_2$OSeO$_3$ \cite{zhang2018reciprocal, zhang2018direct}. 
Using circular dichroism resonant elastic x-ray scattering (CD-REXS) tomography, we reconstructed the full 3D magnetisation profile and discovered a continuous transformation from pure N\'{e}el-type skyrmions at the surface to Bloch-type skyrmions in the bulk \cite{zhang2018reciprocal}. 
This smooth evolution spans over 100--200\,nm into the crystal, revealing that the skyrmion string's topology is not invariant along its length. 
CD-REXS enables this depth-sensitive imaging by tuning the photon energy near the absorption edge, which modifies the x-ray penetration depth and selectively probes different regions within the material. 
The underlying principle of the technique, along with key early results, is summarised in Figure~\ref{fig:hesjedal1}.
These observations highlight how surface effects exert long-range influence within the bulk, and show that skyrmions must be viewed as genuinely 3D topological solitons whose internal structure responds sensitively to geometry, confinement, and boundary conditions.

\begin{figure*}[h!]
	\centering
	\includegraphics[width = 0.75\linewidth]{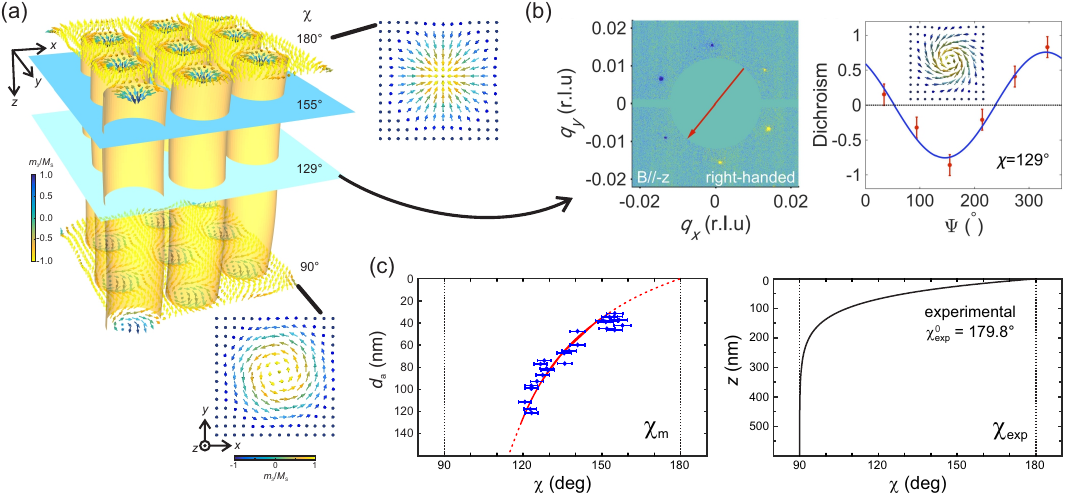}
    \captionsetup{font=tiny,skip=-1pt,justification=raggedright,singlelinecheck=false}
	\caption{
		\textbf{Surface-induced twist from N\'{e}el to Bloch: 3D reconstruction of skyrmion strings}. 
		(a) Schematic illustration of skyrmion strings in Cu$_2$OSeO$_3$, showing the evolution of the helicity angle $\chi$ from N\'{e}el-like ($180^\circ$) at the surface to Bloch-like ($90^\circ$) deeper in the bulk. 
		The gradual rotation of the spin texture is shown on selected cross-sections through the string, highlighting the influence of surface boundary conditions.
		(b) The CD-REXS pattern displays an extinction effect, i.e., the dichroism contrast vanishes at characteristic azimuthal angles (for this specific photon energy at $\chi = 129^\circ$), which serve as a direct fingerprint of the average helicity within the probed volume.
		(c)	As the photon energy is tuned across the Cu $L_3$ absorption edge (926-934\,eV), the x-ray attenuation length, and hence the effective probing depth, varies from $\sim$30\,nm (surface-sensitive) to over 120\,nm (bulk-sensitive).
		By systematically measuring the extinction angle as a function of energy, the depth profile of the measured helicity angle $\chi_m(d_a)$ is obtained (left panel), from which the continuous $\chi(z)$ twist of the skyrmion string from surface to bulk can be reconstructed (right panel).
		Panels adapted from Ref.~\cite{zhang2018direct} with permission from the American Physical Society 
		%
		%
		and Ref.~\cite{zhang2018reciprocal} (\textcopyright\ 2018 National Academy of Sciences).
		%
	}
	\label{fig:hesjedal1}
\end{figure*}

A second major development is the discovery of a robust surface-bound skyrmion lattice under in-plane magnetic fields \cite{zhang2020robust}, illustrated in Figure \ref{fig:Hesjedal2}(a,b). 
In contrast to the conventional Bloch-type skyrmions stabilised in the bulk, this surface state emerges as a distinct, thermodynamically stable phase, persisting across a broad temperature range well beyond the narrow stability pocket of the bulk skyrmion lattice. 
The state is characterised by a hexagonally ordered array of perpendicularly oriented skyrmions confined to the near-surface region of chiral magnets such as Cu$_2$OSeO$_3$. 
Using depth-resolving REXS, the surface origin of this lattice was unambiguously identified through its diffraction pattern: magnetic Bragg rods appear under surface-sensitive conditions, replacing the discrete peaks observed deeper in the bulk. 
The orientation of the skyrmion lattice remains pinned to the surface even under tilted magnetic fields, highlighting the dominant role of surface anisotropy and broken inversion symmetry in stabilising this uniquely 3D, yet geometrically confined, magnetic phase.
This surface-pinned character makes such states particularly attractive for integration into spintronic devices, where functionality is often confined to the top layers or interfaces of wafer-based architectures.

A particularly striking example of topological 3D complexity is the skyrmion screw, a magnetic configuration in which the spin structure helically winds along the skyrmion string, resembling a nanoscale corkscrew. 
First predicted in theoretical work exploring energetically favoured metastable states \cite{rybakov2015new}, this structure was recently observed experimentally using REXS \cite{zhang2018reciprocal, zhang2018direct} in both single- and double-confined chiral magnets \cite{xie2023observation}. 
The resulting diffraction patterns, featuring characteristic `X'-shaped motifs reminiscent of Photo 51, the iconic x-ray image that revealed the double-helix structure of DNA, offered unambiguous reciprocal-space signatures of a topological spin screw with a finite helix angle and uniform chirality \cite{xie2023observation, zhang2018direct}. 
Figure~\ref{fig:Hesjedal2}(c-e) illustrates both the experimental setup and the key results.
This analogy between molecular and magnetic helices marks a turning point in skyrmionics, where topology is no longer constrained to flatland but extends into a fully 3D parameter space.

\begin{figure*}[h!]
	\centering
	\includegraphics[width = 0.75\linewidth]{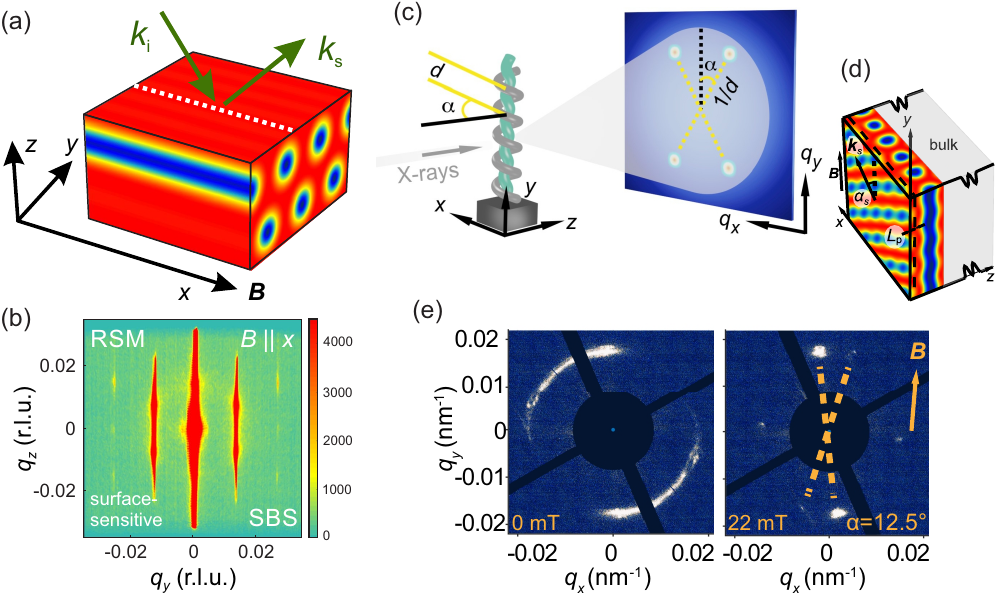}
	\caption{
		\textbf{Surface and screw states of 3D skyrmions}.
		(a) Surface-confined perpendicular skyrmion lattice stabilised under an in-plane magnetic field and schematic of the experimental configuration.
		(b) Experimental REXS pattern recorded under surface-sensitive conditions, revealing elongated rods in the $q_y$-$q_z$ plane.
		The rod-like scattering signature confirms the 2D character and limited penetration depth of the surface skyrmion lattice.
		(c) Transmission REXS geometry for probing the skyrmion screw state in laterally confined Cu$_2$OSeO$_3$.
		The resulting `X'-shaped diffraction pattern directly reflects a helical modulation of the skyrmion string along its axis, with characteristic helix angle $\alpha$ and screw pitch $d$.
		(d) Micromagnetic simulation of the confined screw state showing modulated skyrmion strings embedded beneath a spiral sideface skin. 
		(e) Experimental REXS patterns showing the emergence of the screw state. Left: At zero field, the pattern reflects an unmodulated helical or disordered background. 
		Right: Upon applying a moderate in-plane field (22\,mT), an `X'-shaped diffraction pattern emerges with a finite helix angle of $\alpha = 12.5^\circ$, unambiguously indicating the formation of a screw-type modulation along the skyrmion string. 
		The orientation and separation of the branches allow precise extraction of the helix angle and screw spacing, analogous to the double-helix pattern in x-ray diffraction of DNA.
		Panels adapted from Ref.~\cite{zhang2020robust} with permission from the American Chemical Society (\textcopyright\ 2020. Licensed under CC BY 4.0) 
		%
		%
		and Ref.~\cite{xie2023observation} with permission from the American Physical Society.
		%
	}
	\label{fig:Hesjedal2}
\end{figure*}

These results position 3D skyrmions not as a footnote to the well-studied 2D case, but as a central object of study with distinct physical properties. 
The presence of Bloch points, chirality reversals, and screw-like modulations within a single skyrmion string introduces new degrees of freedom that could be exploited for information encoding.
For instance, the handedness and pitch of the screw may serve as binary or multilevel logic states, while the response of the skyrmion to magnetic field gradients or current pulses could allow for dynamic control over these internal modes \cite{birch2022toggle}. 
Furthermore, the ability to couple or decouple multiple skyrmion configurations, e.g., via proximity effects in heterostructures or dipolar fields from adjacent layers \cite{ran2021creation, ran2022axially}, enables the design of entirely new magnetic computing architectures \cite{rybakov2022magnetic}.

\section*{Challenges}

Looking ahead, the long-term vision is to develop a fully 3D spintronic platform in which information is encoded not only in the presence or position of a skyrmion, but in its internal topological structure.
Devices based on skyrmion screws, sideface states, or surface-bound lattices could offer enhanced thermal stability and increased information density. 
Novel modes of control could become possible beyond what is achievable with conventional skyrmion racetracks. 
One can imagine a memory device in which each bit is encoded by the skyrmion twist angle, chirality, or termination state, or a neuromorphic circuit in which the coupling between adjacent skyrmion strings is modulated via 3D geometry, enabling complex logic operations in a compact footprint.

However, achieving this vision poses several fundamental challenges. 
First, the direct imaging of 3D spin textures at nanometre resolution remains technically demanding. 
Techniques such as Lorentz transmission electron microscopy or magnetic force microscopy offer exceptional 2D contrast but are largely blind to the full 3D variations. 
REXS, with its tunable depth sensitivity and polarisation selectivity, has emerged as a powerful tool, yet requires sophisticated analysis and synchrotron beamline access \cite{zhang2018reciprocal, zhang2018direct}. 
The development of next-generation tomographic and holographic methods, using either x-rays or electrons, will be essential for further progress \cite{donnelly2020time, birch2020real}.
Second, theoretical modelling of these complex structures requires new frameworks that account for spatial modulation, metastability, and the interplay of multiple interaction terms beyond minimal micromagnetic models. 
Unlike their 2D counterparts, skyrmion screws and sideface states cannot be easily described using minimal micromagnetic models. 
They often occupy metastable regions of the phase diagram, whose boundaries are sensitive to subtle changes in anisotropy, confinement, or dipolar interactions. 
Accurate predictive modelling will require new frameworks that go beyond static homotopy classifications to account for spatially modulated and dynamically reconfigurable topologies.
Third, materials discovery remains a bottleneck: 
The stabilisation of 3D skyrmion structures requires finely tuned balance between DMI, exchange stiffness, anisotropy, and dipolar coupling, and while materials such as Cu$_2$OSeO$_3$ and FeGe provide excellent testbeds, they often operate at low temperatures and in narrow field windows.
Extending 3D skyrmion phenomena to room-temperature, low-field materials is a crucial step toward applications.
The identification of new chiral or interfacial materials, guided by symmetry, DMI strength, and anisotropy landscape, will be critical to realising functional 3D skyrmionic devices. 
Finally, readout and control strategies must evolve:
If we are to harness the internal structure of skyrmion strings for computation, we must develop means of probing and manipulating these features locally. 
This might include MOKE-based readout of chirality, electrical detection of Bloch points via topological Hall signals, or controlled deformation using strain or current. 
Integration of these elements into functional architectures will determine the pace of future applications.

In conclusion, the discovery and characterisation of 3D skyrmion textures, including screws, sideface states, and depth-evolving lattices, opens a new frontier in topological magnetism. 
These structures challenge long-held assumptions about dimensionality and topology, and offer intriguing opportunities for future information technologies. 
As experimental and theoretical tools continue to mature, the skyrmion screw may well emerge as a foundational building block for next-generation 3D spintronic devices due to its intrinsic stability, geometric tunability, and potential for multi-level information encoding.



\section*{Acknowledgements}
S.Z.\ acknowledges financial support from the National Key R\&D Program of China under contract number 2020YFA0309400, the Science and Technology Commission of the Shanghai Municipality (21JC1405100), and the National Natural Science Foundation of China
(Grant No.\ 12074257). 
T.H.\ acknowledges financial support from the the Oxford-ShanghaiTech collaboration project, EPSRC (EP/N032128/1), Diamond Light Source, and the Science and Technology Facilities Council (UK).
We thank the following synchrotron radiation facilities for provision of beamtime: Diamond Light Source, SOLEIL, and ALBA.

\endgroup

\newpage

\section{Emergent Skyrmion Electrodynamics}
\begingroup
    \let\section\subsection
    \let\subsection\subsubsection
    \let\subsubsection\paragraph
    \let\paragraph\subparagraph
Max T. Birch$^1$, and Yoshinori Tokura$^{1,2,3}$
\vspace{0.5cm}

\noindent
\textit{$^1$ RIKEN Center for Emergent Matter Science (CEMS), Wako, Saitama 351-0198, Japan\\
$^2$ Department of Applied Physics, The University of Tokyo, Bunkyo-ku, Tokyo 113-8656, Japan\\
$^3$ Tokyo College, The University of Tokyo, Bunkyo-ku, Tokyo 113-8656, Japan}

\section*{Background}

The coupling of conduction electrons and magnetic spin textures leads to emergent quantum phenomena which may be elegantly described in terms of emergent electromagnetism\cite{nagaosa2012emergent}. When a conduction electron traverses a magnetic texture and its spin adiabatically follows the local magnetisation direction $\mathbf{n}$, its wavefunction may acquire an additional quantum phase known as the Berry phase. The curl of the related Berry connection can be considered as an emergent magnetic field~\cite{volovik2013spinmotive}, $\mathbf{b}$, expressed generically as

\begin{equation}
    b_i(\mathbf{r}) = \frac{h}{8\pi e}\epsilon_{ijk}\mathbf{n}\cdot\left(\partial_j\mathbf{n}\times \partial_k\mathbf{n} \right).
\end{equation}

This quantity is therefore finite only when the underlying magnetic texture has a net scalar spin chirality. In addition, should the Berry phase exhibit some time-dependence, for example due to the excited dynamics of the magnetic texture, then we can consider that the Berry connection is time varying, and this will give rise to an emergent electric field, $\mathbf{e}$. Once again for a generic spin texture, 

\begin{equation}
    e_i(\mathbf{r}) = \frac{h}{2\pi e}\mathbf{n}\cdot\left(\partial_i\mathbf{n}\times \partial_t\mathbf{n} \right),
\end{equation}
which can be considered as emergent electromagnetic induction, also known as a spin motive force~\cite{yang2009universal}.

\begin{figure}[h!]
    \centering
    \includegraphics[width=0.95\linewidth]{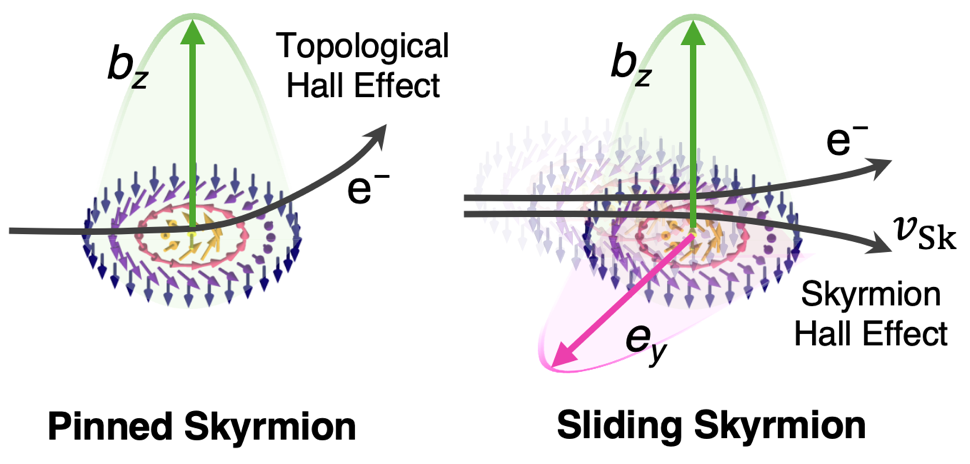}
    \caption{Schematic illustration of the emergent electrodynamics associated ith the coupled motion of conduction electrons and magnetic skyrmions. A skyrmion, formed under an applied external magnetic field Bext, possesses an emergent magnetic field b, which causes deflection of the conduction electrons, which we measure via the topological Hall effect (THE). Once the skyrmion achieves current-induced sliding motion, it exhibits the skyrmion Hall effect (SkHE) in response. The motion of b induces a transverse emergent electric field e, which opposes the THE voltage.}
    \label{fig:Birch1}
\end{figure}

The magnetic skyrmion lattice (SkL) provides an ideal case: the emergent magnetic flux is quantised due to the topology of the skyrmion, and b can therefore expressed in terms of the skyrmion number N: $\mathbf{b}= 2\pi N$. This $\mathbf{b}$ gives rise to a Lorentz force which acts upon moving conduction electrons, which may be measured as the topological Hall effect (THE)~\cite{neubauer2009topological}, visualised in Fig.~\ref{fig:Birch1}. The smaller the magnetic skyrmions, the tighter the topological winding, leading to an increase in the THE conductivity. Due to its incommensurability, above a certain current density $J$ threshold, the SkL will exhibit sliding motion as it is depinned by the spin transfer torque. The resulting coupled motion of SkL and conduction electrons gives rise to two further effects, also visualised in Fig~\ref{fig:Birch1}. Firstly, the SkL velocity $v_{Sk}$ itself will acquire a transverse component due to the counteraction of the THE, known as the skyrmion Hall effect. Secondly, the time varying Berry phase gives rise to an $\mathbf{e}$, which, in the case of a SkL, is simply the product of $\mathbf{b}$ and $v_{\rm Sk}$: $\mathbf{e} = -\mathbf{v}_{\rm Sk} \times \mathbf{b}$. This emergent electric field opposes the THE effect, and the measured THE voltage is therefore suppressed as the skyrmions gain velocity.

\section*{State of the Art}

Not long after the identification of the SkL, as well as the corresponding THE~\cite{neubauer2009topological}, skyrmion electrodynamics were observed in the chiral magnet MnSi, realised by the careful measurement of the current-dependent THE~\cite{schulz2012emergent}. However, in the decade since, there have been few follow up experiments due to the difficulty of identifying the typically small magnitude of the THE while simultaneously exciting the SkL motion. However, with the recent discovery of nanoscale skyrmions in the rare earth magnet systems, several systems have been found to exhibit a THE larger than that of MnSi, offering a fresh opportunity to study skyrmion electrodynamics.

This has recently been achieved using $\rm Gd_2PdSi_3$~\cite{birch2024dynamic}, which hosts skyrmions with size 2-3 nm, and which is notable for its extraordinarily large THE~\cite{kurumaji2019skyrmion} – two orders of magnitude larger than in MnSi. The high current densities required to move the SkL were achieved by fabricating a microscale device from the bulk single crystal using focused ion beam milling. The topological Hall effect measured as a function of a DC current density within the SkL phase is shown in Fig.~\ref{fig:Birch2}a, revealing a highly nonlinear response. Remarkably, this data can be used to directly calculate $v_{\rm Sk}$, which is plotted in Fig.~\ref{fig:Birch2}b. Three dynamical regimes may be distinguished, which are common for depinning phenomena: i) the pinned regime, where $v_{\rm Sk} = 0$, and the THE is at a maximum; ii) the creep regime, where $v_{\rm Sk} > 0$, the SkL motion is assisted by thermal fluctuations, and the THE decreases linearly; iii) the flow regime, where $v_{\rm Sk}$ is linearly proportional to $J$, and the THE appears to vanish. This total cancellation of the THE is notable, and implies the achievement of emergent Galilean relativity, where $v_{Sk}$ approaches the semiclassically defined electron drift velocity $v_e$~\cite{birch2024dynamic}.

\begin{figure}[h!]
    \centering
    \includegraphics[width=0.45\linewidth]{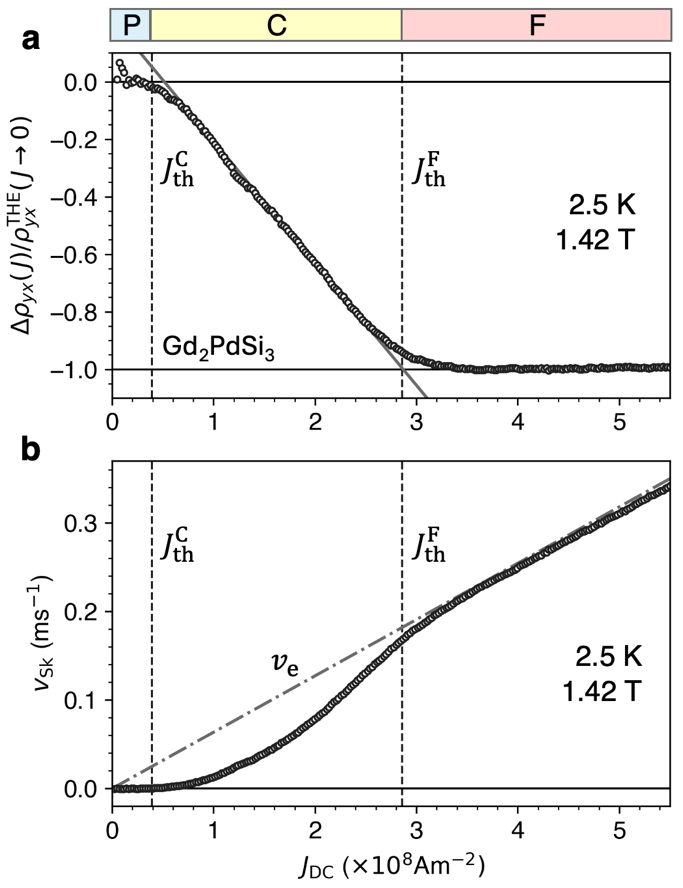}
    \caption{(a) The change in the Hall resistivity $\Delta\rho_{yx} = \rho_{yx}(J) - \rho_{yx}(J\rightarrow 0)$, normalized by the zero current
limit value of the Hall resistivity, $\rho_{yx}(J\rightarrow 0)$, measured as a function of the current density, $J$ , at 2.5 K and 1.42 T with a focused ion beam-fabricated device of $\rm {Gd_2PdSi_3}$. The pinned (P), creep (C) and flow (F) regimes are labelled above. The solid black line is a linear fit to the data in the C regime. (b) The $v_{\rm Sk}$ plotted as a function of $J$, calculated from the data in (a). In both panels, the threshold current densities associated with the dynamic transitions to the creep and flow regimes are labelled $J_{\rm th}^{\rm c}$ and $J_{\rm th}^{\rm F}$, respectively. Note that $v_e$ denoted in panel (b) is the electron drift velocity defined by $J/ne$ in the classical Drude model, and hence ill-defined in a real material with multiple spin-polarized bands. Therefore, the Galilean relativity implied by the cancellation of the THE shown in (a) should be carefully considered. On the other hand, the $v_{\rm Sk}$ is a well-defined quantity shows semi-quantitative accuracy. Data adapted from~\cite{birch2024dynamic}.}
    \label{fig:Birch2}
\end{figure}

As well as the transverse $\mathbf{e}$ generated by its current-induced velocity, it has also been shown that the dynamic deformation of a SkL, which also results in a time-dependent Berry phase, may act as another source of $\mathbf{e}$~\cite{littlehales2025emergent}. In addition, this deformation gives rise to inertial dynamics of the SkL within the creep regime, leading to a delay in its motion relative to the applied excitation. These results establish the SkL state as an ideal playground for the study of emergent electromagnetic induction, originally envisaged and observed in the current-driven dynamics of a spin spiral~\cite{yokouchi2020emergent}. While the THE is a manifestation of $\mathbf{b}$ in charge transport, the emergent electromagnetic signature of the SkL has also recently been identified in thermal transport, in the form of the topological Nernst effect~\cite{hirschberger2020topological}, as well as in the optical response, as the topological magnetooptical effect~\cite{kato2023topological}.

\section*{Outlook}

From a fundamental perspective, the SkL state will continue to be an ideal phase of matter for the study of real-space quantum Berry phase effects, which remain largely unexplored. On the other hand, the overwhelming majority of proposed skyrmion applications rely on the interplay of spin and charge currents with magnetic skyrmions and their dynamics for their read and write functionality or computation schemes. Therefore it is clear that the concepts of emergent electromagnetism are key from both foundational and practical perspectives, and their importance cannot be overstated.

In addition to the aforementioned topological charge, thermal and optical responses, theoretical works have envisaged that the $\mathbf{b}$ of the SkL state may give rise to the topological orbital Hall effect~\cite{gobel2025topological}, while the compensated $\mathbf{b}$ of an antiferromagnetic SkL would exhibit a topological spin Hall effect~\cite{gobel2017antiferromagnetic} – predictions which are awaiting experimental confirmation. Meanwhile, studies of $\mathbf{e}$ remain rather limited, and additional examples and detailed studies are required. In particular, the concept of emergent Galilean relativity touches upon a theme widely-explored by theory: the crossover between real-space and moment space-dominated Berry curvature~\cite{matsui2021skyrmionsize}, and its relation to the electron mean-free path and the skyrmion size. A systematic experimental study of such a crossover within a simple system would greatly improve our quantitative descriptions of these effects, and thus our fundamental understanding of quantum Berry phase phenomena.

\endgroup

\newpage

\section{Topological bulk skyrmion dynamics}
\begingroup
    \let\section\subsection
    \let\subsection\subsubsection
    \let\subsubsection\paragraph
    \let\paragraph\subparagraph
Ping Che$^1$, and Dirk Grundler$^{2,3}$
\vspace{0.5cm}

\noindent
\textit{$^1$ Laboratoire Albert Fert, CNRS, Thales, Université Paris-Saclay, 91767 Palaiseau, France
$^2$ Laboratory of Nanoscale Magnetic Materials and Magnonics, Institute of Materials (IMX), École Polytechnique F\'ed\'erale de Lausanne (EPFL), 1015 Lausanne, Switzerland\\
$^3$ Institute of Electrical and Micro Engineering (IEM), École Polytechnique F\'ed\'erale de Lausanne (EPFL), 1015 Lausanne, Switzerland}

\section*{Introduction}

Skyrmion lattices formed in a chiral bulk magnet have been shown to induce a characteristic magnon band structure thanks to their periodically arranged spin texture~\cite{garst2017collective}. When magnons propagate through topologically nontrivial skyrmions, the electron spins align with the local magnetic moments, leading to a Berry phase in the magnon band structure, as illustrated in Fig.~\ref{fig:Ping1}(a)~\cite{weber2022topological}. The Chern number $C_n$ serves as the topological invariant that quantifies the winding of the Berry curvature over the Brillouin zone~\cite{diaz2019topological}: $C_n = \frac{1}{2\pi}\int_{\mathrm{BZ}} {\boldsymbol{\Omega}_n (\mathbf{k}) d(\mathbf{k})}$, where $\boldsymbol{\Omega}_n (\mathrm{k})$ is the Berry curvature of the $n$-th magnon band. Figure~\ref{fig:Ping1}(b) displays the topological magnon band structure within the first Brillouin zone for the skyrmion lattice existing in MnSi below its critical temperature of about 30~K. The Chern number of each band is labeled on the right. Magnons in these bands feature distinct real-space wave functions realizing counterclockwise (CCW), sextupole, breathing (BR), clockwise (CW), and octupole modes, as shown in Fig.~\ref{fig:Ping1}(c). 

Among all minibands, only the ones belonging to CCW, BR, and CW modes are dipole-active. In chiral magnetic metals, semiconductors and insulators their resonance behaviors have been observed near the $\Gamma$-point via coupling to dynamic magnetic fields generated by radio-frequency currents in microwave antennas~\cite{schwarze2015universal}. Magneto-optical coupling has been used to reveal both the three basic dipole-active modes and multipole modes at higher magnon wavevectors $k$ inside the first Brillouin zone~\cite{ogawa2015ultrafast,che2025short}. The multipole magnons possess nonzero density of states in the high-$k$ regime and become directly accessible via optical detection using Brillouin light scattering~\cite{che2025short}, without relying on mode hybridization~\cite{takagi2021hybridized}. In a multiferroic chiral magnet such as insulating Cu$_2$OSeO$_3$, electrically active modes have been detected through magnetoelectric (ME) coupling~\cite{okamura2013microwave}. Outside the first Brillouin zone, inelastic neutron scattering was used to detect the high-energy (meV-scale) topological magnon bands of skyrmion lattices in metallic MnSi ~\cite{weber2022topological}.

\begin{figure}[h!]
	\centering
	\includegraphics[width=0.55\columnwidth]{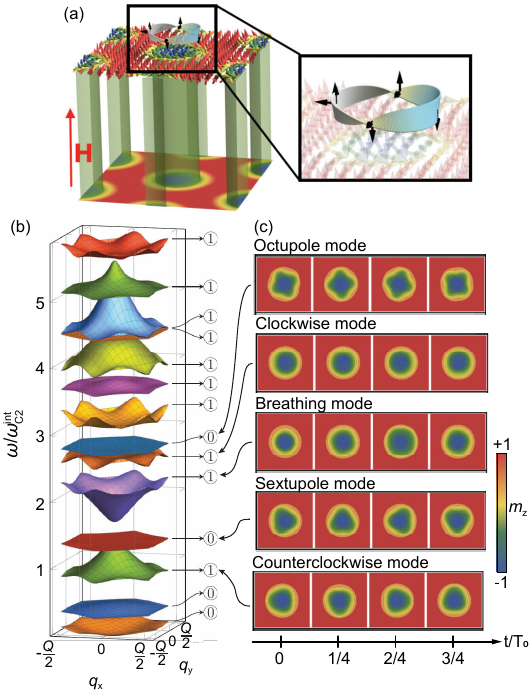} 
	\caption{\textbf{Topological magnon band structure of a skyrmion lattice in chiral magnets.} (a) Sketch of a skyrmion lattice in a magnetic field H in which magnons accumulate a Berry phase when following a trajectory depicted in the inset. Black arrows indicate the local coordinate system. From Ref.~\cite{weber2022topological}. Reprinted with permission from AAAS. (b) Topological magnon band structure of the skyrmion lattice in MnSi in the first Brillouin zone. The circled numbers on the right represent Chern numbers. Reprinted from Ref.~\cite{garst2017collective}. CC BY 3.0. (c) Evolution of the spin wave function of the  counterclockwise, sextupole, breathing, clockwise and octupole mode (bottom to top) calculated as a function of time $t$ at the $\Gamma$-point of the Brillouin zone. $T_0$ is the period of the spin-precessional motion and typically in the sub-ns regime. The colors indicate the out-of-plane $m_z$ component (legend). Reprinted figure with permission from Ref.~\cite{takagi2021hybridized}, Copyright (2014) by the American Physical Society.}
	\label{fig:Ping1}
\end{figure}

\section*{Relevance and Vision}

\subsection*{Topological chiral edge magnon modes}

Topological chiral edge magnons are edge-localized excitations that arise from the nontrivial topology of bulk magnon modes~\cite{diaz2019topological}. Analogous to the quantum Hall effect, the nonzero Chern numbers of bulk magnonic bands suggest the existence of quantized magnonic states at the edges (edge states). The bulk-boundary correspondence guarantees the presence of edge magnons in the band gap of the bulk states shown in Fig.~\ref{fig:Ping1}(b). They exhibit a unidirectional propagation property and robustness against disorders without backscattering. Topological chiral edge magnons provide an ideal platform for exploring ultrashort-waved magnons and quantum magnonic phenomena, as they offer resilience against decoherence, low-dissipation propagation, and strong confinement at the nanoscale. The different microscopic mechanisms and materials ~\cite{schwarze2015universal} available for stabilization of magnetic skyrmions (Bloch- and Néel-type skyrmions, antiskyrmions) as well as hexagonal and square lattice arrangements of skyrmions with circular or square shapes provide a rich landscape for studying the interaction between magnons and topology.

A distinguishing feature of topological magnons and the edge states is their bosonic nature, which sets them apart from fermionic topological electron modes. Unlike fermions, magnons can occupy the same quantum state and do not obey the particle number conservation law. This bosonic character enables nonlinear effects such as three-magnon scattering. While the topological protection of edge modes typically suppresses such decoherence processes due to the unidirectional propagation property, a theoretical predication suggests that these nonlinear interactions can potentially open the gap at the Dirac cone of magnon bands and lead to edge mode formation \textcolor{red}{\cite{habel2024breakdown}}. Thus, understanding the role of nonlinear magnon scattering for edge magnon generation, stabilization and resilience, as well as the control of relevant scattering mechanisms are crucial for topological and quantum magnonics. Cu$_2$OSeO$_3$ with its ultra-low damping~\cite{stasinopoulos2017low} is an excellent candidate for probing such nonlinear excitations.

\subsection*{Magnons in skyrmion lattices enhancing quantum circuits}

The combination of ultra-low Gilbert damping (as low as $10^{-4}$~at few Kelvin temperature)~\cite{stasinopoulos2017low} and multiferroicity~\cite{okamura2013microwave} makes Cu$_2$OSeO$_3$ a compelling material for studying the coupling between skyrmion dynamics, cavity photons, and superconducting qubits. In spherical Cu$_2$OSeO$_3$ crystals, distinct magnon modes can be excited either by dynamic magnetic fields via dipolar coupling or through ME coupling, as illustrated in Fig.~\ref{fig:Ping2}(a) and (b). The magnon-photon hybridization induced by ME coupling enables access to the quadrupole modes.

In a Cu$_2$OSeO$_3$ sphere, skyrmion lattices align with the direction of the applied magnetic field, effectively eliminating the influence of magnetic anisotropy. This freedom in orientation allows for the excitation and coupling of variable magnon modes with microwave cavities. By positioning the Cu$_2$OSeO$_3$ sphere at different locations within a microwave cavity (Fig.~\ref{fig:Ping2}(c)), one can exploit the spatial distribution of electric and magnetic fields in a cavity to selectively couple to different magnon modes. Additionally, spatially variant modes such as the sextupole and octupole modes (compare Fig.~\ref{fig:Ping1}(c)) are expected to couple differently with the cavity fields depending on their symmetry. With an additional superconducting qubit, it becomes possible to investigate coupling between a topological magnon and the qubit under various coupling regimes. The emergence of topological chiral edge modes in skyrmion lattices, with their unidirectional propagation and protection from decoherence, further supports the realization of long-lived hybrid quantum systems and directional coupling between a series of superconducting qubits.

\subsection*{Time quasicrystal in helimagnet}

In a helimagnet, i.e., a chiral magnet with helical order, an oscillating magnetic field induces a coherent screw-like motion of the helical spin texture, reminiscent of an Archimedean screw~\cite{del2021archimedean}. Under strong driving conditions, the system enters a nonlinear dynamical regime in which the uniform helical motion becomes unstable. This instability gives rise to intricate spatiotemporal patterns. In both the helical and skyrmion lattice phases, such nonlinear excitation can lead to the formation of a time quasicrystal—a state where the magnetization exhibits oscillations that are incommensurate with both the intrinsic modulation vector of the spin texture and the frequency of the external field~\cite{del2021archimedean}. The emergence of this time quasicrystal order opens a pathway toward exploiting nonlinear spin dynamics in skyrmion lattices for constructing quantum memory architectures.
\begin{figure*}[h!]
	\centering
	\includegraphics[width=0.95\linewidth]{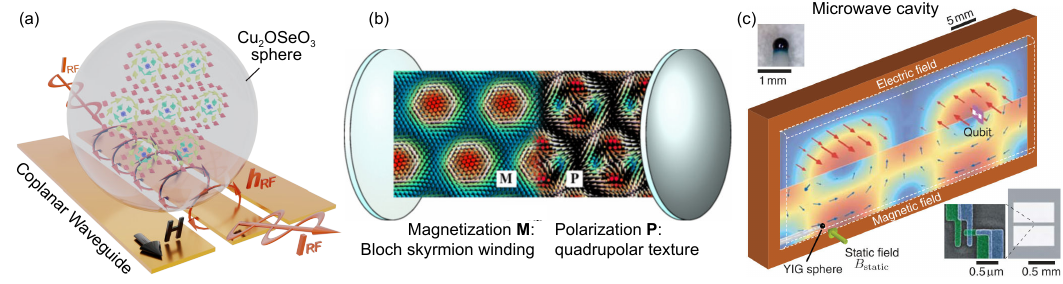}
	\caption{\textbf{Coherent coupling between magnons in skyrmion lattices hosted in a Cu$_2$OSeO$_3$ sphere with photons and qubits.} (a) Proposed configuration for investigating coherent long-lifetime magnon modes with finite wavevectors in a skyrmion lattice phase in a Cu$_2$OSeO$_3$ sphere. (b) Schematic illustration of magnon-photon coupling in a skyrmion lattice phase in the multiferroic chiral magnet Cu$_2$OSeO$_3$. The sample might be placed in a microwave cavity as sketched in (c). Reprinted from Ref.~\cite{hirosawa2022magnetoelectric}. CC BY 4.0. (c) Microwave cavity incorporating a sphere-shaped magnet with magnons (here YIG) and a superconducting qubit. The electric field (red arrows) is shown in the upper part and the magnetic field (blue arrows) is shown in the lower part. A Cu$_2$OSeO$_3$ sphere hosting a skyrmion lattice can replace the YIG sphere. The cavity fields can then couple the superconducting qubit and the magnons in the skyrmion lattice phase. From Ref.~\cite{tabuchi2015coherent}. Reprinted with permission from AAAS.}
	\label{fig:Ping2}
\end{figure*}

\section*{Challenges}

When aiming at the functionalization of topological magnon modes at microwave frequencies, several key challenges remain. In particular, the coherent excitation of magnons and the controlled coupling of multipole skyrmion dynamic modes with cavity photons and superconducting qubits are not yet fully established. To activate the corresponding minibands, short-wavelength magnons must be excited through magneto-optical or magneto-electric coupling. However, the mechanisms enabling their coherent excitation have not yet been systematically demonstrated or experimentally verified.

As lattice constants in skyrmion lattices typically have a characteristic length below 100 nanometers, imaging short-wavelength (high-wavevector) magnons in these systems is technically demanding. Optical techniques in laboratory settings are constrained by the diffraction limit to typically larger length scales. Synchrotron-based approaches, such as scanning transmission X-ray microscopy, offer a spatial resolution around 20 nm and are suitable for local imaging of high-wavevector magnons. However, there have been no experimental demonstrations to date, largely due to the weak magnon signal in skyrmion phases at cryogenic temperatures requiring the further improvement of the signal-to-noise ratio in low-temperature experiments. 

Furthermore, it is important to identify skyrmion-hosting materials that combine low magnetic damping with room-temperature stability. Quantum sensing methods such as nitrogen vacancy center magnetometry could become a viable tool here, provided that the operational frequencies and magnetic field orientations can be controlled with high precision.

Minimizing magnetic losses is essential for implementing skyrmion-based quantum computing. Phase coherence of the magnon modes is critical for enabling interaction between magnon minibands in topological magnon band structures and solid-state qubits. Progress in developing room-temperature, single-crystalline materials with spherical geometry and low damping could open the path to quantum information processing based on bulk skyrmion dynamics. Similarly, the realization of a time quasicrystal in skyrmion lattices and helical phases relies on magnon-magnon interactions and requires sufficient energy injection into the system. Low damping gives also access to non-linear effects~\cite{tengdin2022imaging} which control the arrangement and/or number of skyrmions in a device similar to magnon-modified spin structures in the canted antiferromagnet hematite~\cite{chen2025deterministic}. Achieving very low damping at room temperature in a skyrmion-hosting material, preferably an insulator, remains hence a fundamental challenge in this pursuit with intriguing prospects. 

\section*{Acknowledgments}
PC acknowledges the support from a government grant managed by the ANR as part of the France 2030 investment plan from PEPR SPIN ANR-22-EXSP 0002 (CHIREX) and the from ANR under contract no. ANR-22-CE30-0014 (DeMIuRGe). DG acknowledges support by the SNSF via grant 197360.

\endgroup

\newpage

\section{X-ray Ptychography for imaging skyrmions and their dynamics}
\begingroup
    \let\section\subsection
    \let\subsection\subsubsection
    \let\subsubsection\paragraph
    \let\paragraph\subparagraph

Simone Finizio$^1$, and Christopher H. Marrows$^2$
\vspace{0.5cm}

\textit{$^1$ Paul Scherrer Institut, 5232 Villigen PSI, Switzerland\\
$^2$ School of Physics \& Astronomy, University of Leeds, Leeds LS2 9JT, United Kingdom}\\


\section*{Introduction}
Skyrmions are the prototypical topologically non-trivial spin textures. Their structure and dynamics have been studied intensively for the past decade or so \cite{Marrows2021} in a variety of different ferromagnetic materials, including bulk crystals, epitaxial films, and multilayers. In addition to the fascinating physics of real space non-trivial spin texture topology and the associated emergent electrodynamics associated with the Berry phase they induce in transport electron wavefunctions, they are candidates for a variety of potential spintronic technologies owing to their small size, topological stability, and ability to be nucleated, manipulated, and detected electrically \cite{Marrows2021}. Over the past few years, they have been joined by a variety of other spin textures with distinct topological properties that have been proposed, and in some cases observed, for the first time. These include skyrmionium, chiral bobbers, antiskyrmions, merons, and hopfions \cite{gobel2021skyrmions}. In addition to ferromagnets, the range of magnetic materials being considered by researchers is being extended to antiferromagnets, ferrimagnets, and their synthetic counterparts including synthetic antiferromagnets (SAFs). 

Imaging the structure and dynamics of these different objects presents a variety of challenges. First is the need for higher spatial resolution in the future. Skyrmionium is a simple example of what are known as skyrmion bags, composite objects in which a variable number of inner skyrmions are surrounded by a much larger, and possibly irregularly shaped, outer skyrmion. There is a huge amount of spatial detail to resolve in these systems. Meanwhile, typical skyrmion sizes in systems studied up to now are mostly in the range of several tens to hundreds of nanometers. Skyrmions with sizes not much above 10~nm have been observed in the compensated ferrimagnet GdCo \cite{Caretta2018}, at the resolution limit of the X-ray holography technique that was used to image them. Smaller skyrmions can be expected in true antiferromagnets. Few nanometre-scale skyrmions are known to be possible in pristine magnetic monolayers \cite{Romming2015}. There is also the prospect of multiferroic skyrmions on the same scale, since nm-sized polar skyrmions have been observed in (PbTiO$_3$)$_n$/(SrTiO$_3$)$_n$ ferroelectric superlattices \cite{Das2019}: can similar things be achieved with related multiferroic materials? Imaging techniques with single-digit nanometre spatial resolution are required for further progress in all of these areas. 

\section*{Relevance, Vision and Challenges}
Within the quiver of high-resolution magnetic imaging techniques, real space X-ray microscopy has been one of the ``workhorses'' of the skyrmionics community \cite{Marrows2021}. However, a major limitation of real space imaging techniques such as scanning transmission X-ray microscopy (STXM) is that achieving single-digit nanometre resolution is extremely challenging. Therefore, different X-ray magnetic microscopy approaches have to be explored, in order to meet the demand for such high resolution whilst keeping the elemental sensitivity, the possibility to probe bulk information, and the capability of performing time-resolved imaging that make X-ray microscopy complementary to other high-resolution techniques such as NV magnetometry and TEM.

In parallel to the evolving requirements of the user community, many synchrotron lightsources have either recently undergone an upgrade or are planning an upgrade to diffraction limited storage rings (DLSRs). By reducing the electron beam's horizontal emittance down to the one of the X-ray beam, DLSRs offer a significant increase in the coherent photon flux. Taking the Swiss Light Source as example, a 40$\times$ increase in the coherent flux at the Fe/Co/Ni L\textsubscript{2-3} edges will be delivered if compared to the old source. This significant improvement will allow for the routine use of ``photon-hungry'' X-ray coherent diffractive imaging (CDI) techniques, heavily relying on coherent illumination. In contrast to direct imaging techniques, CDI techniques record, in far-field conditions, an image of the scattered X-ray light produced by the interaction of the coherent beam with the sample. This information is then used to reconstruct an image of the sample where the spatial resolution is limited by the maximum scattering angle of the X-ray light that is recorded, allowing one to go beyond the limits imposed by the optics. 

Among the various CDI techniques, X-ray ptychography has recently emerged as a replacement for STXM imaging in quasistatic conditions. Thanks to the recent extension to lower energies of 2D counting detectors with high dynamic ranges such as the eiger-lgad \cite{Baruffaldi2025}, soft X-ray ptychography at sub-10~nm imaging resolution is becoming routinely accessible, which will allow for the imaging of ultrasmall topological features whilst keeping all of the advantages of quasistatic X-ray microscopy. In addition, CDI techniques give access to phase contrast information, which allows for the imaging of $\upmu$m thick thin film systems otherwise inaccessible at soft X-ray energies, as magnetic phase contrast is present well before the absorption edge \cite{Neethirajan2024}.

\begin{figure*}[h!]
    \centering
    \includegraphics[width=0.95\textwidth]{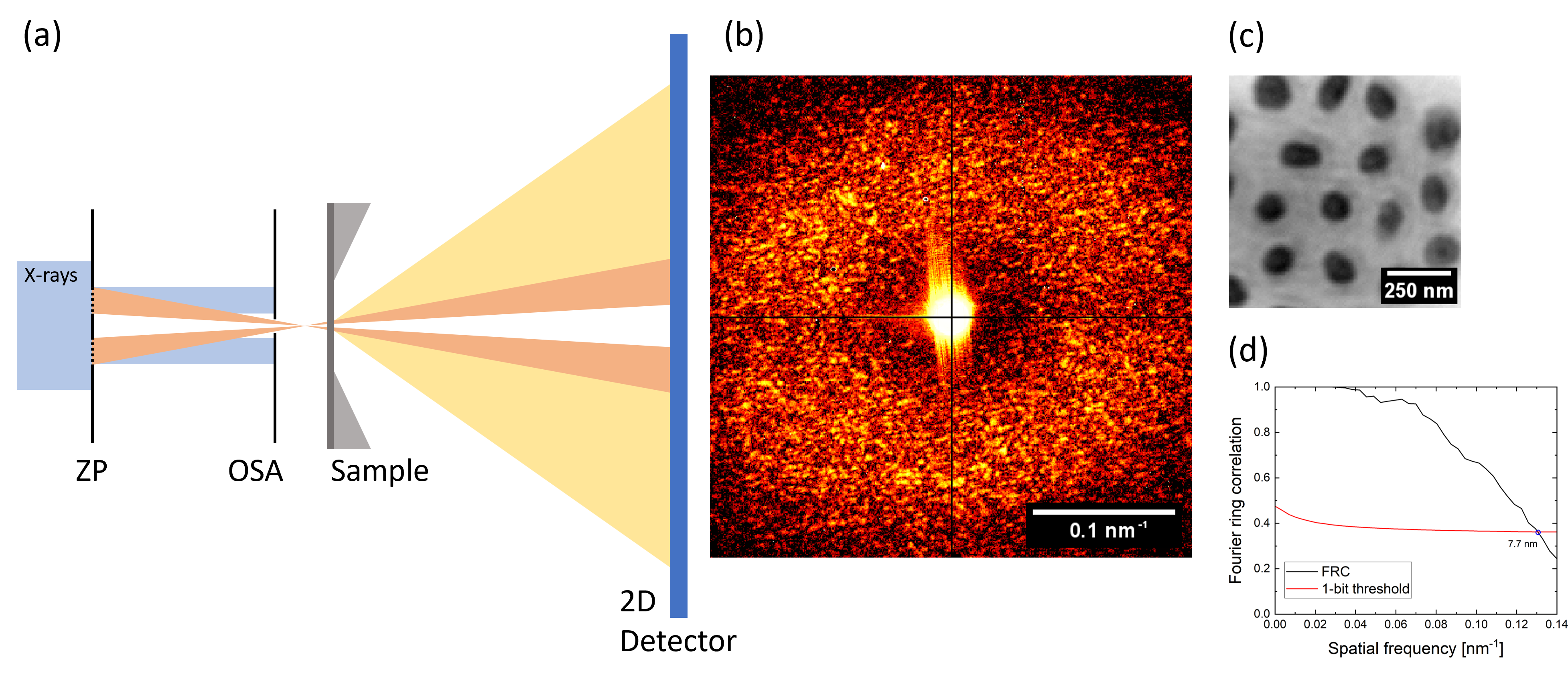}
    \caption{(a) Sketch of the operating principle of X-ray ptychography imaging, where an X-ray transparent sample is scanned with a slightly defocused X-ray beam, and the transmitted beam intensity is recorded in far field conditions by a 2D pixelated X-ray detector. A zoneplate (ZP) combined with an order selecting aperture (OSA) is used to focus the X-ray beam. (b) Example of a frame recorded by the 2D detector, depicted in logarithmic scale, showing the unscattered beam in the center and the coherently scattered X-ray beam in the regions corresponding to higher spatial frequencies. (c) Example of a reconstructed ptychography image (amplitude contrast) of the skyrmion state in a [Pt/CoFeB/MgO]$_{15}$ stack~\cite{kern2022deterministic}. Image courtesy of Lisa-Marie Kern. (d) Corresponding Fourier ring correlation calculation, showing that spatial resolutions well beyond the limits of the optics can be obtained.}
    \label{fig:Finizio}
\end{figure*}

Enhanced temporal resolution is needed to accompany this improved spatial resolution. We have previously observed, through pump-probe imaging, the electrical-current driven nucleation of a skyrmion in a Pt/CoB/Ir multilayer, a process occurring in response to a current pulse with a duration of a few nanoseconds, with a resolution of 200~ps \cite{Marrows2021}. Nevertheless, preliminary quasistatic data indicates that nucleation in response to current pulses as short as a few tens of picoseconds is possible, requiring vastly higher time resolution to track the details of the spin dynamics. Other results indicate the need for the ability to study skyrmion motion at enhanced temporal resolution as well. For instance, SAF skyrmions have been observed to move at very high velocities of hundreds of ms$^{-1}$ on the basis of quasistatic imaging before and after the application of a current pulse \cite{pham2024fast}. Unravelling the details of this very fast motion involves imaging the skyrmion structure whilst it is in flight: this may, for instance, shed light on the question of whether the SAF skyrmions do indeed display the predicted inertia that is associated with the internal degree of freedom arising from the indirect interlayer exchange coupling \cite{Panigrahy2022}.

While ptychography will enable access to single-digit nanometre resolutions and to phase contrast information, the operation of 2D detectors in pump-probe regime is very challenging, hindering the possibility to perform time-resolved ptychography. While there is some progress also on this issue, e.g. from the TimePix project \cite{Llopart2022}, other non synchrotron-based ptychography opportunities have to be considered. For this, free electron lasers (FELs) are a promising avenue, as they offer extremely short pulses of X-ray light (in the fs range) at repetition rates (usually 50-100~Hz) that are compatible with the current frame rates of 2D detectors. However,  FELs exhibit some specific challenges for time-resolved ptychography that still have to be fully overcome. In particular, the two main challenges are to be found in the timing and pointing instabilities of a FEL source. For the timing instabilities, a possible solution could be to employ a seeded source. For the pointing instabilities, instead, a possible solution could be to employ single-shot ptychographic imaging, using position refining phase recovery algorithms \cite{Pancaldi2024}. However, some challenges, such as e.g. guaranteeing that samples can survive the high flux of a FEL source, still need to be addressed.

Some of these topological textures are three-dimensional and so require imaging from multiple directions or by multiple modalities for a full understanding to emerge. This includes chiral bobbers, which are bound to a surface \cite{gobel2021skyrmions}; skyrmion cocoons, which are embedded within the material away from surfaces \cite{grelier2022threedimensional}; and weak/partial skyrmions, which only occupy certain layers within a multilayer stack \cite{mandru2020coexistence}. Hopfions are a family of particularly complex three-dimensional structures that may co-exist with skyrmion bags: the full complexity of the three-dimensional spin texture needs to be revealed in order to confidently determine the hopfion number of the texture~\cite{gobel2021skyrmions}.

By integrating soft X-ray ptychography with laminography \cite{Witte2020}, 3D high-resolution imaging will also be possible. Here, it will be possible to acquire 3D magnetic images with sub-10~nm voxel sizes, however requiring the acquisition of a large number of high-resolution 2D projections. However, there are still some challenges to be resolved, the main one being the integration of an out-of-plane magnetic field with the laminography setup, requiring one to take the miniaturization of the setup and the stability of the magnetic field upon rotating the sample into consideration \cite{Witte2020}.

\section*{Conclusion}

As we have described here, remarkable progress is being made on the cutting-edge synchrotron techniques needed to address the combined challenges of ultrahigh spatial and temporal resolution, combined with the move from 2D to 3D imaging. Thus, the study of magnetic skyrmions and related spin textures with ptychography a bright future ahead of it.

\endgroup

\newpage

\section{Current-driven antiskyrmions at room temperature towards skyrmionics}
\begingroup
    \let\section\subsection
    \let\subsection\subsubsection
    \let\subsubsection\paragraph
    \let\paragraph\subparagraph
Kosuke Karube$^1$, and Xiuzhen Yu$^1$
\vspace{0.5cm}

\noindent
\textit{$^1$ RIKEN Center for Emergent Matter Science (CEMS), Wako, 351-0198, Japan}

\section*{Introduction}
Magnetic skyrmions - vortexlike spin textures characterized by integer topological numbers - have attracted significant attention as research platform for studying a wide range of emergent electromagnetic phenomena in contemporary condensed matter physics~\cite{back20202020}. Skyrmions act as topologically protected particles and can be manipulated using ultra-low current densities ($\approx \SI{1e6}{Am^{-2}}$), which are 5-6 orders of magnitude lower than those required to drive ferromagnetic domain walls. Such remarkable properties make skyrmions promising candidates for high-density low-power-consumption magnetic memory devices. For practical application of skyrmions in next-generation spintronics technologies — collectively referred to as ``skyrmionics'' —, it is essential to stabilize and control skyrmions at temperatures above room temperature.

Thus far, the considerable progress has been made in identifying suitable materials hosting skyrmions. The first phase of skyrmion research focused on bulk chiral magnets~\cite{muehlbauer2009skyrmion, yu2010real}, where Bloch-type skyrmions arise from the competition between ferromagnetic exchange interaction and Dzyaloshinskii–Moriya interaction (DMI). Following it, substantial advancements have been achieved in the study of Néel-type skyrmions in thin-film heterostructures with interfacial DMI, as well as in bulk polar magnets. In parallel, significant progress has been made in understanding of skyrmion dynamics under electric currents through numerical studies on the basis of Landau-Lifshitz-Gilbert and Thiele equations ~\cite{iwasaki2013universal, fert2013skyrmions}. Experimentally, a single robust skyrmion was successfully driven by pulsed current at room temperature, and the current-dependent skyrmion Hall motion was observed ~\cite{peng2021dynamic}. These findings highlight the key roles of spin-transfer torque and the pinning effect in skyrmion motions.

Antiskyrmions — antiparticles of skyrmions with opposite topological charge — also hold promise for future device applications. Unlike skyrmions with either Bloch- or Néel-type twisted spins, antiskyrmions exhibit spin textures composed of both Bloch and Néel-type twists, resulting in rotationally anisotropic spin configurations. Experimental progress of antiskyrmions lagged behind that of skyrmions, primarily due to the scarcity of suitable host materials. The first example of antiskyrmions was found in tetragonal Heusler alloys $\rm Mn_{1.4}Pt_{0.9}Pd_{0.1}Sn$ and $\rm Mn_{1.4}PtSn$ with $D_{2d}$ symmetry in 2017~\cite{nayak2017magnetic}. Subsequently, the present authors' group discovered stable antiskyrmions in Pd-doped schreibersite $\rm (Fe_{0.63}Ni_{0.3}Pd_{0.07})_3P$ (FNPP) with $S_4$ symmetry [Fig.~\ref{fig:Karube}(a)]~\cite{karube2021room}. In these magnets, antiskyrmions are stabilized even above room temperature via complex interplay among the exchange interaction, anisotropic DMI, uniaxial magnetic anisotropy and magnetic dipolar interaction. Due to the dominant role of dipolar interaction, antiskyrmions tend to form square shapes [Fig.~\ref{fig:Karube}(b)], and can transform into non-topological bubbles and elliptically deformed Bloch-type skyrmions by tuning external parameters such as magnetic fields~\cite{karube2021room, peng2020controlled}. Moreover, antiskyrmions and skyrmions (and non-topological bubbles) with distinct topological charges can coexist under appropriate conditions, which is useful for designing confined antiskyrmion racetrack memory devises~\cite{gobel2021skyrmion}. To realize the antiskyrmion-based devices, it is essential to demonstrate the electric manipulation of antiskyrmions in confined geometries and to understand their dynamics.

\section*{Relevance And Vision}

\begin{figure*}[h!]
    \centering
    \includegraphics[width=0.95\textwidth]{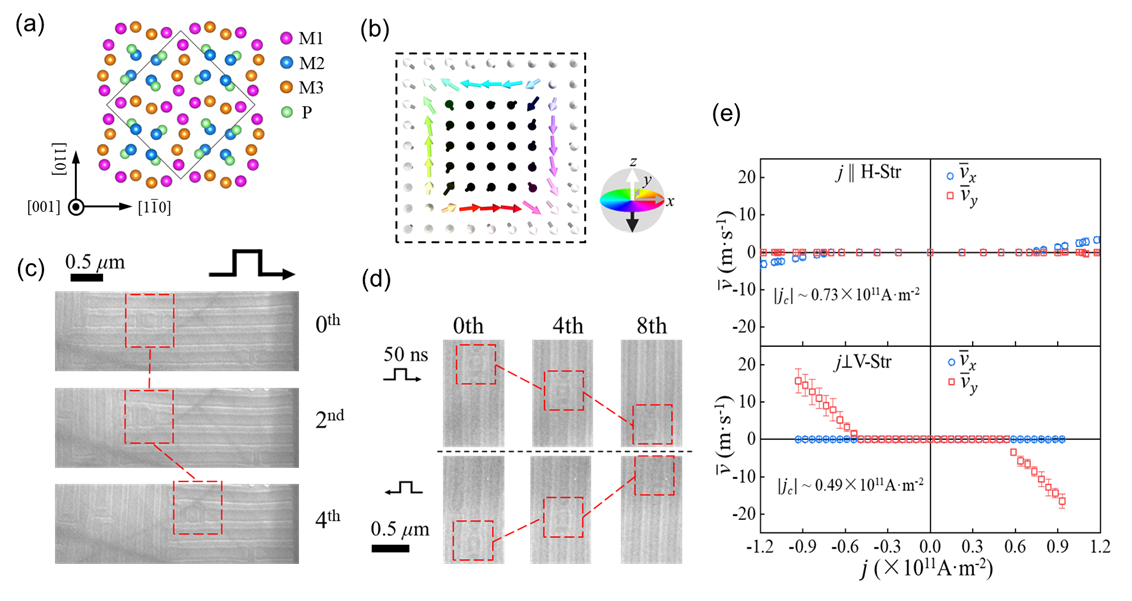}
    \caption{(a) Crystal structure of M3P [$\rm (Fe_{0.63}Ni_{0.3}Pd_{0.07})_3P$] with $I\bar{4}$ space group along the [001] axis. The M1, M2 and M3 denote three inequivalent transition metal sites for Fe, Ni or Pd. (b) Schematic of the spin structure of an antiskyrmion. (c, d) LTEM images observed after sequential 50 ns current-pulse applications showing the current-driven antiskyrmion motion (marked by the red rectangular dashed line) in (c) horizontal stripes (H-Str) and (d) vertical stripes (V-Str) at room temperature and at $\mu_0H$ = 50 mT. (e) The corresponding average antiskyrmion velocity  versus current density  in (c, d). Reprinted from ref. \cite{guang2024confined}. Copyright 2024, Springer Nature.}
    \label{fig:Karube}
\end{figure*}

The authors' group has recently succeeded in electrically driving antiskyrmions at room temperature~\cite{guang2024confined}. We designed microdevices of FNPP, and performed in situ observation using Lorentz transmission electron microscopy (LTEM) under current application. In FNPP, square-shaped antiskyrmions form at room temperature upon application of magnetic field and persist as a robust metastable state even after the field is reduced. By adjusting the magnetic field and angle of the thin plate, we created a single metastable antiskyrmion embedded in horizontal and vertical magnetic stripes (called H-Str and V-Str, respectively) [Fig.~\ref{fig:Karube}(c, d)] and investigated their current-driven dynamics. This approach enables the study of antiskyrmion motion confined to one-dimensional stripes, effectively mimicking the geometry of a racetrack memory device while utilizing a standard LTEM sample.

Upon repeated excitation of 50 ns pulsed current, the antiskyrmion confined in the H-Str moves horizontally, parallel to the current [Fig.~\ref{fig:Karube}(c)]. On the other hand, the antiskyrmion confined in the V-Str moves vertically, perpendicular to the current [Fig.~\ref{fig:Karube}(d)]. When the current is applied in the opposite direction, the antiskyrmion motion is also reversed. Figure~\ref{fig:Karube}(e) shows the average velocity of antiskyrmion as a function of a current density $j$ for the $j \parallel $H-Str and $j \perp $V-Str geometries. The antiskyrmion begins to move above the threshold current density $|j_c| \approx \SI{0.73e-11}{Am^{-2}}$ for $j \parallel$ H-Str and above $|j_c| \approx \SI{0.49e-11}{Am^{-2}}$ for $j \perp $ V-Str, respectively. In both configurations, the antiskyrmion velocity increases linearly with the current density. Importantly, the antiskyrmion velocities are much higher for $j \perp $ V-Str when the current flow is perpendicular to the stripes than those for $j \parallel$ H-Str.

Micromagnetic simulations using the Thiele equation explain the observed antiskyrmion dynamics confined in the stripe background in terms of spin transfer torque. These findings provide a new understanding of antiskyrmion dynamics in confined geometries and enhance the potential for antiskyrmion applications in racetrack memory devices.

\section*{Challenges}

Although the Heusler alloys and FNPP are suitable materials for studying stability and current-driven dynamics of antiskyrmions at room temperature, and are therefore promising for skyrmionics applications, there are several challenges to overcome. First, the antiskyrmion diameter is on the order of 100 nm and increases with crystal thickness due to the dominant role of dipolar interactions. Reducing the antiskyrmion size to around 10 nm or even smaller is a key challenge for realizing high-density skyrmionics devices. One potential solution is to enhance the spin–orbit interaction, or to exploit the Ruderman–Kittel–Kasuya–Yosida (RKKY) interaction. The latter has recently been demonstrated for skyrmions in Gd-based frustrated magnets, although these magnets exhibit very low magnetic ordering temperatures. Second, the threshold current density for antiskyrmion motion is on the order of $10^{10}-10^{11}\, {Am^{-2}}$, much higher than the ideal value for manipulating skyrmions due to the strong pinning effect. The key to lowering the critical current density may be to reduce crystallographic site disorder.

The realization of high-performance skyrmionics devices also requires high-frequency (GHz to subTHz) manipulations of antiskyrmions, such as creation, drive, detection and annihilation, by electrical and optical methods. Recently, ultrafast dynamics of skyrmions at room temperature using a ns pulsed laser excitation was demonstrated in a chiral magnet~\cite{shimojima2021nano}. Similar experiments should be attempted with antiskyrmion materials to develop ultrafast control of antiskyrmion dynamics.

In conclusion, recent progress in antiskyrmion research from materials development to understanding current-driven dynamics in confined geometries demonstrates the potential applications of antiskyrmions. In the future, further research development in fundamental science and device applications of antiskyrmion is expected to overcome the hurdles mentioned above and move even closer to the realization of skyrmionics.

\section*{Acknowledgments}
We are grateful to Dr. Yasujiro Taguchi for his insightful discussions and invaluable help in preparing this manuscript. This work is supported by JST CREST (Grant No. JPMJCR20T1) and the RIKEN TRIP initiative (Many-body Electron Systems).

\endgroup

\newpage

\section{Skyrmion Hall Effect and Guided Motion}
\begingroup
    \let\section\subsection
    \let\subsection\subsubsection
    \let\subsubsection\paragraph
    \let\paragraph\subparagraph
Lisa-Marie Kern$^1$, and Kai Litzius$^2$
\vspace{0.5cm}

\noindent
\textit{$^1$ Department of Materials Science and Engineering, Massachusetts Institute of Technology, Cambridge, Massachusetts 02139, USA\\
$^2$ Experimental Physics V, Center for Electronic Correlations and Magnetism, University of Augsburg, 86159 Augsburg, Germany}

\section*{Introduction}
Magnetic skyrmions are valued for their topological stability, nanoscale size, low depinning currents, and unique electrodynamic effects, enabling advanced memory and logic devices. Their topology, however, causes a transverse motion due to the Magnus force from a real-space Berry phase, leading to the skyrmion Hall effect (SkHE) (Fig.~\ref{fig:Kern1}).~\cite{iwasaki2013universal} The dynamics are described by the Thiele equation, balancing gyroscopic, dissipative, and external forces to predict steady-state motion.~\cite{tomasello2014strategy, jiang2017direct} For spin-orbit torque (SOT) driven skyrmions, the corresponding Thiele formalism reads:
\begin{equation}
    \mathbf{G} \times \mathbf{v} - \alpha_G \mathcal{D} \cdot \mathbf{v} + B \hat{\mathcal{R}} \cdot \mathbf{j} = 0.
\end{equation}

Here, $\mathbf{G}$ denotes the \textit{gyrocoupling vector}, while $\mathbf{v} = [v_x, v_y]$ represents the skyrmion velocity, the \textit{dissipative tensor} $\mathcal{D}$ captures the effect of dissipative forces acting on the moving skyrmion, $\alpha_G$ corresponds to the \textit{Gilbert damping}, and $B$ represents the \textit{spin Hall effect (SHE)} due to current $\mathbf{j}$. The matrix $\hat{\mathcal{R}}$ denotes an \textit{in-plane rotation matrix}, accounting for the skyrmion’s orientation relative to the current direction.~\cite{tomasello2014strategy}
This results in the final solutions for Néel and Bloch skyrmions:

\begin{equation}
    \begin{array}{c@{\hskip 1.0cm}c}
    \mathrm{N\acute{e}el} \left\{
    \begin{aligned}
    v_x &= \frac{\alpha_G \mathcal{D} B}{1 + \alpha_G^2 \mathcal{D}^2} j, \\
    v_y &= \frac{B}{1 + \alpha_G^2 \mathcal{D}^2} \mathbf{j}.
    \end{aligned}
    \right.
    &
    \mathrm{Bloch} \left\{
    \begin{aligned}
    v_x &= \frac{B}{1 + \alpha_G^2 \mathcal{D}^2} j, \\
    v_y &= \frac{\alpha_G \mathcal{D} B}{1 + \alpha_G^2 \mathcal{D}^2} j.
    \end{aligned}
    \right.
    \end{array}
\end{equation}

As directly visible from these equations, the skyrmion Hall effect is inverted for Néel and Bloch configurations.~\cite{tomasello2014strategy} Assuming that the domain wall width $\Delta$ is much smaller than the skyrmion radius $R$, the skyrmion Hall angle simplifies to:~\cite{litzius2017skyrmion}

\begin{equation}
    \begin{array}{c@{\hskip 1.cm}c}
    \begin{aligned}
    \tan\Theta_{\mathrm{SkH, N\acute{e}el}} &= \pm \frac{1}{\alpha_G \mathcal{D}} \approx \pm \frac{2 \Delta}{\alpha_G R},
    \end{aligned}
    \\
    \\
    \begin{aligned}
    \tan\Theta_{\mathrm{SkH, Bloch}} &= \pm \frac{\alpha_G \mathcal{D}}{1} \approx \pm \frac{\alpha_G R}{2 \Delta}.
    \end{aligned}
    \end{array}
\end{equation}

Instead of moving strictly along the driving current, skyrmions deflect at an angle $\Theta_{\mathrm{SkH}}$, analogous to the classical Hall effect for charged particles. In a less naive approach, this characteristic skyrmion Hall angle additionally depends on the gyrotropic force and deformations of the skyrmion shape~\cite{bo2024suppression}, i.e. the skyrmion's speed, as well as disorder~\cite{kern2022deterministic}. Importantly, the direction of this deflection is determined by the sign of the topological charge, making it a direct manifestation of the skyrmion’s internal structure (Fig.~\ref{fig:Kern1}c and d). Note however, that the presence of a non-trivial spin texture does not necessarily require a net topological charge. 

\begin{figure*}[h!]
    \centering
    \includegraphics[width=0.95\linewidth]{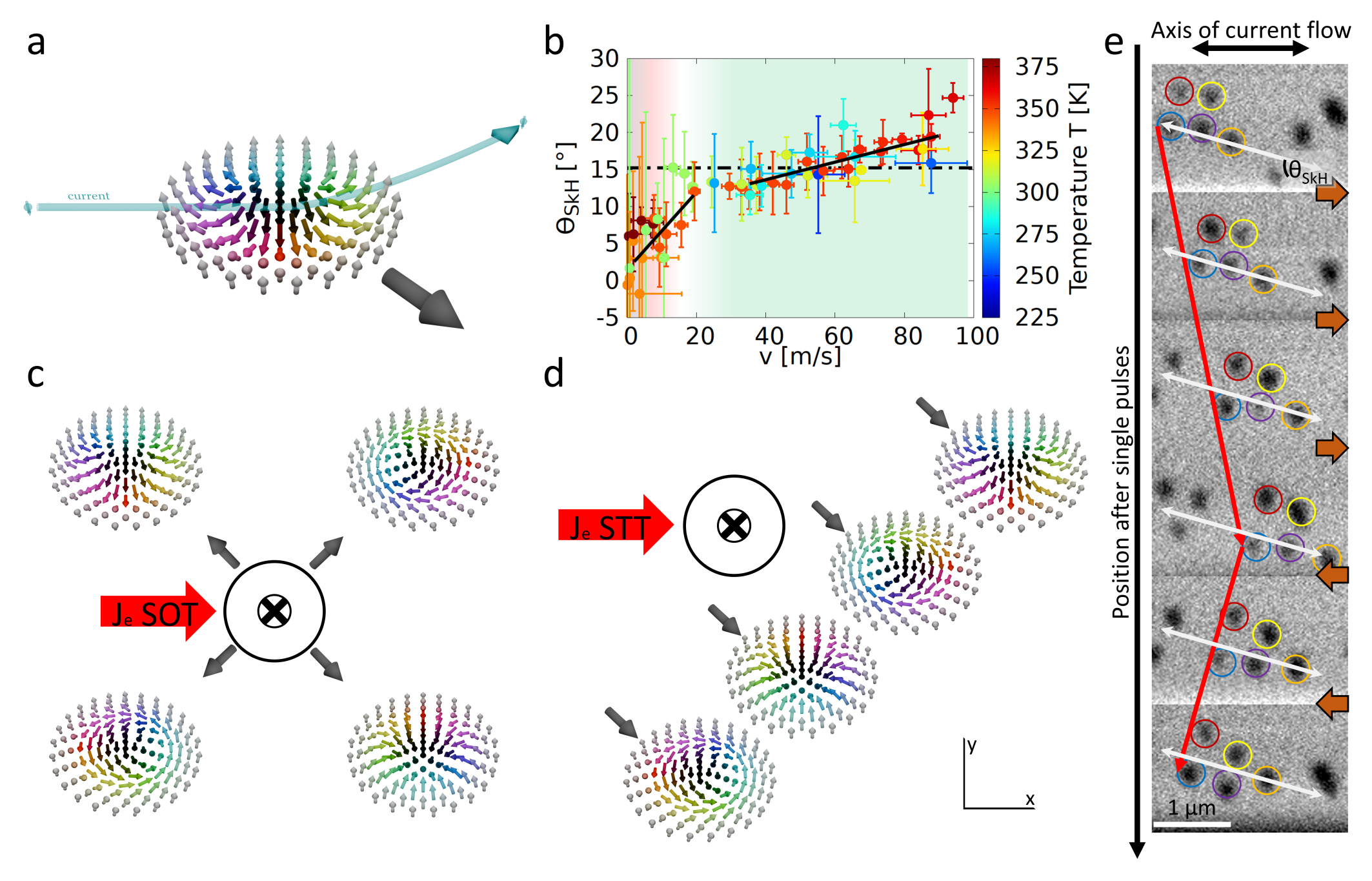}
    \caption{Skyrmion Hall effect in various configurations. a) Spin Transfer Torque (STT) current flowing through a skyrmion, causing it to be diverted and in turn push the skyrmion off of its axial trajectory. b) Experimental observation of a current-dependent skyrmion Hall angle in spin–orbit torque (SOT) geometry, while it appears to be unaffected by temperature. c+d) Skyrmion Hall effect for c) SOT and d) STT drive and different spin textures: While STT consistently drives the skyrmions in the same direction, SOT-induced motion strongly depends on chirality and skyrmion type. Shown are down magnetized skyrmions and an electron current $J_\mathrm{e}$, grey arrows show the displacement direction of the individual skyrmion types from an original position (indicated down-magnetized circle). e) Experimental image sequence (scanning transmission X-ray microscopy) directly visualizing the skyrmion Hall effect. Data taken from K. Litzius, PhD thesis, Johannes Gutenberg-University Mainz, 2018.}
    \label{fig:Kern1}
\end{figure*}

\section*{Relevance and Vision} 
The skyrmion Hall effect has been demonstrated across a range of experimental and theoretical platforms, including ferro-, ferri-, antiferro-, and altermagnetic systems, providing broad insights into its origin and dynamics~\cite{litzius2017skyrmion,jiang2017direct,woo2018current,pham2024fast,Jin2024}. High-resolution real-space imaging techniques such as transmission X-ray microscopy, Lorentz transmission electron microscopy (LTEM), magneto-optical Kerr effect (MOKE) imaging, and scanning probe microscopy (SPM) have directly visualized the lateral drift of skyrmions (Fig.~\ref{fig:Kern1}e).~\cite{Yang2024_Fundamentals_applications} These methods enable quantitative extraction of the Hall angle and its dependence on current density, skyrmion size, and temperature.~\cite{litzius2020role} Electrical transport measurements show indirect signatures of the SkHE via asymmetric Hall voltages, often attributed to the topological Hall effect (THE) superimposed with emergent electrodynamics. Micromagnetic simulations incorporating STT and SOT support experiments and predict nontrivial behaviors like angle suppression in confined geometries or due to edge repulsion.\\

The SkHE is a defining feature of skyrmionic spin textures and a clear indicator of their topological nature. At the same time, its tendency to deflect skyrmions from a straight path poses a serious challenge for applications that rely on precise, directional control of magnetic textures. This dual role of the SkHE, as both a fundamental signature and a practical hurdle, has fueled a shift in research perspectives. Recent efforts aim to harness or suppress it in a controlled way, turning what was once seen as a drawback into an opportunity.

\section*{Challenges} 

As research advances towards practical skyrmion-based technologies, controlling the skyrmion Hall effect is expected to become a defining objective. Future efforts are converging around three primary classes of action (Fig.~\ref{fig:Kern2}):\\

\begin{figure*}[h!]
    \centering
    \includegraphics[width=0.95\linewidth]{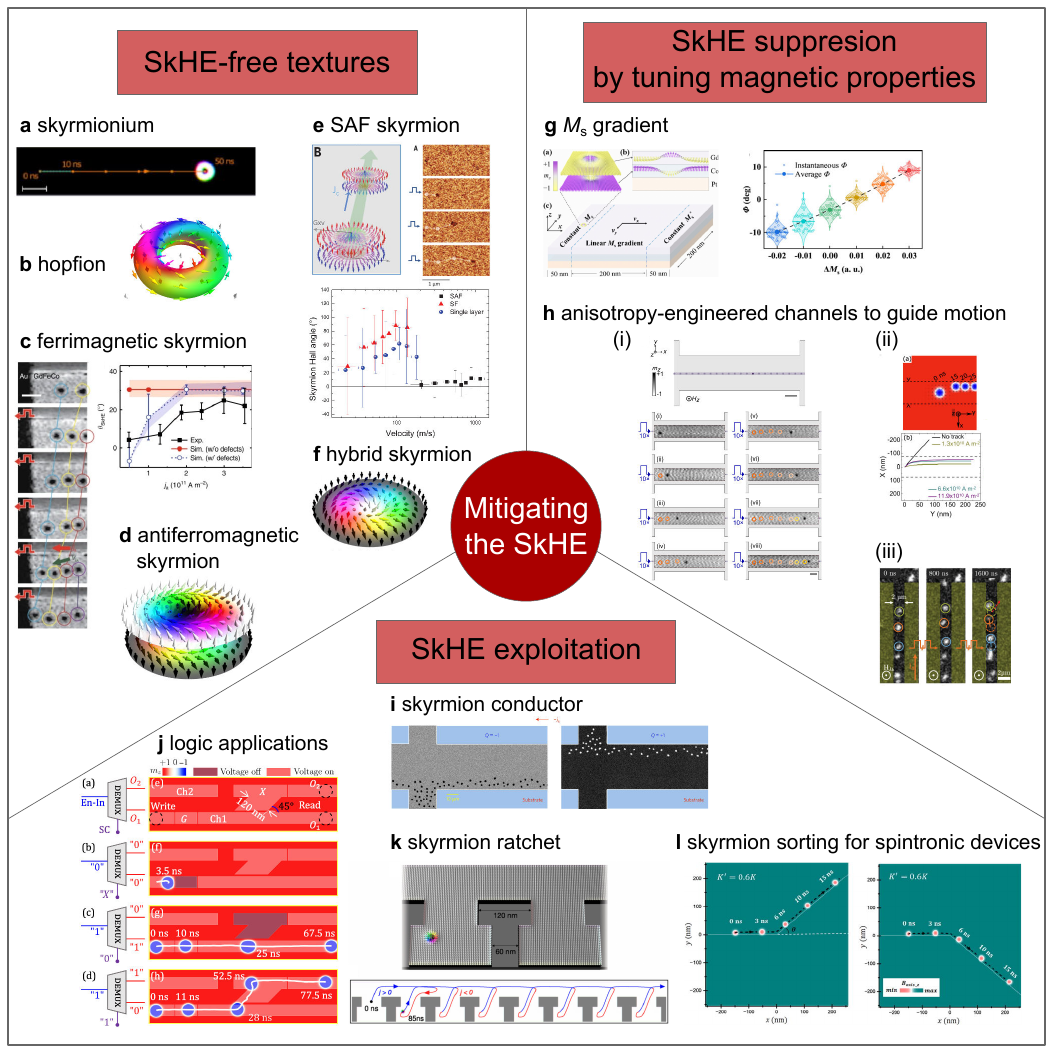}
    \captionsetup{font=tiny,skip=-1pt,justification=raggedright,singlelinecheck=false}
    \caption{Mitigation of the skyrmion Hall effect via SkHE-free textures including a) skyrmionium, b) hopfion, c) ferrimagnetic skyrmion, d) antiferromagnetic skyrmion, e) SAF skyrmion and f) hybrid skyrmion; or via SkHE-suppression tuning g) the saturation magnetization or h) the magnetic anisotropy. Alternatively, the SkHE can be harnessed in application concepts including i) a skyrmion conductor, j) logic gates, k) a skyrmion ratchet or l) skyrmion sorting channels. a) is reprinted with permission from~[\citenum{Yang2024_Fundamentals_applications}]. b), d) and f) are reprinted with permission from~[\citenum{gobel2021skyrmion}]. c) is reprinted with permission from~[\citenum{woo2018current}]. e) is reprinted with permission from~[\citenum{pham2024fast}]. Copyright 2024, The American Association for the Advancement of Science. g) is reprinted with permission from~[\citenum{bo2024suppression}]. h) (i) is reprinted with permission from~[\citenum{kern2022deterministic}], (ii) from~[\citenum{juge2021helium}]. Copyright 2021 American Chemical Society., and (iii) from~[\citenum{ahrens2023skyrmions}]. i) is reprinted with permission from~[\citenum{jiang2017direct}]. Copyright 2017 Springer Nature. j) is reprinted from~[\citenum{he2024guided}]. Copyright 2025 American Physical Society. k) is reprinted from~[\citenum{gobel2021skyrmion}]. l) is reproduced from~[\citenum{ma2025deterministic}], with the permission of AIP Publishing. a)-d), f)-g), h (i) and (iii), and k) are licensed under CC BY 4.0.}
    \label{fig:Kern2}
\end{figure*}

The first strategy aims to eliminate the Skyrmion Hall effect through the design of spin textures that intrinsically compensate for the transverse Magnus force through coupled spin textures. Several such textures have been proposed and realized across a growing range of material systems. A skyrmionium, for example, consists of two concentric skyrmions with opposite polarity, forming a ring-like structure (Fig.\ref{fig:Kern2}a) that is theoretically free of SkHE due to internal force cancellation.~\cite{Yang2024_Fundamentals_applications} Its three-dimensional counterpart, the hopfion (Fig.\ref{fig:Kern2}b), is likewise predicted to exhibit no SkHE. In ferri- and antiferromagnets, sublattices host skyrmions of opposite polarity, forming ferrimagnetic (Fig.\ref{fig:Kern2}c) or antiferromagnetic skyrmions (Fig.\ref{fig:Kern2}d). Ferrimagnetic skyrmions exhibit a reduced but finite skyrmion Hall angle under current,~\cite{woo2018current} while antiferromagnetic skyrmions are predicted to fully cancel the SkHE. A promising platform is synthetic antiferromagnets (SAFs), where two ferromagnetic layers, antiferromagnetically coupled across a spacer, each host a skyrmion of opposite polarity, thus yielding a SAF skyrmion with no net topological charge (Fig.\ref{fig:Kern2}e).~\cite{pham2024fast} Similarly, bimerons, composed of two sub-merons with opposite polarity, exhibit no net SkHE due to the internal compensation of Hall effects. Hybrid skyrmions in engineered synthetic bilayers (Fig.\ref{fig:Kern2}f) may also suppress the SkHE through helicity tuning.~\cite{chang2025suppressed}  All these more complex textures share the key characteristic that under current-driven motion, the opposing Magnus forces from each sub-skyrmion cancel out, intrinsically suppressing the SkHE. This not only eliminates transverse deflection but can also enable significantly faster skyrmion motion.\\

The second class of action focuses on the suppression of any undesired deviations from the current axis caused by the skyrmion Hall effect. Geometric confinement like notch-guiding or edge repulsion were among the first strategies to suppress transverse deflection. Additionally, tailored current pulse shaping can dynamically steer motion and minimize lateral drift. Recent approaches increasingly aim for tailored magnetic properties and material engineering to reliably suppress the SkHE. Locally modifying magnetic properties, including the saturation magnetization~$\mathrm{M}_\mathrm{S}$ or the perpendicular magnetic anisotropy (PMA), can steer skyrmion motion with respect to velocity and skyrmion Hall angle. Even a $1\%$ variation in $\mathrm{M}_\mathrm{S}$ can effectively suppress the skyrmion Hall effect in synthetic ferrimagnets (Fig.~\ref{fig:Kern2}g)~\cite{bo2024suppression}. Local (helium) ion irradiation can modify magnetic multilayers and their interfaces by inducing atomic displacements and intermixing, primarily reducing PMA. The nanometer spatial precision of focused ion beams enables the engineering of tailored anisotropy landscapes, creating individual pinning sites and guiding channels (Fig.~\ref{fig:Kern2}h) without altering the topography of the material.~\cite{juge2021helium,kern2022deterministic} In these channels, skyrmions can be selectively addressed and guided along straight paths over micrometer distances with high fidelity and precision, effectively suppressing the SkHE.~\cite{kern2022deterministic} Alternatively, skyrmions can be driven along a straight path by irradiating everything around the channel and leaving the channel itself intact.~\cite{ahrens2023skyrmions}\\

The third approach, instead of working against the nature of these systems, leverages the skyrmion Hall effect as an inherent feature to be harnessed in spintronic applications. Although skyrmion-boundary interactions are typically associated with data loss, an accumulation of skyrmions on either side of a conductor could also be employed for functional purposes (Fig.~\ref{fig:Kern2}i).~\cite{jiang2017direct} While various device concepts have been proposed to functionalize individual skyrmions, ranging from racetrack memories to reservoir computing, the focus is here on application concepts requiring the presence of the skyrmion Hall effect: In voltage-controlled strain channels, repulsive or attractive forces could be used to incorporate the transverse motion of a skyrmion to distribute data in logic circuits or more involved demultiplexer gates (Fig.~\ref{fig:Kern2}j).~\cite{he2024guided} Using a skyrmion ratchet mechanism in an asymmetric racetrack, the SkHE could convert an alternating current (AC) into directed skyrmion motion, potentially leading to more efficient and compact data storage solutions (Fig.~\ref{fig:Kern2}k).~\cite{gobel2021skyrmion} Alternatively, skyrmion guiding channels which were originally proposed to suppress the SkHE~\cite{ma2025deterministic} could be extended to enable skyrmion sorting by exploiting the SkHE at intersections (Fig.~\ref{fig:Kern2}l).\\

These strategies to eliminate, suppress, or harness the skyrmion Hall effect attempt to reframe it as a new driver for spintronic innovation. Its challenges foster advances in materials research as well as in theoretical predictions for steering, sorting and reconfiguring magnetic skyrmions. While many concepts remain to be experimentally validated, this growing design space lays the groundwork for robust, scalable skyrmion technologies, including applications in logic, memory, and neuromorphic computing.

\section*{Acknowledgements}
We thank Moritz Kamm for preparing the skyrmion renders in Fig.~\ref{fig:Kern1}. L.-M.K. acknowledges financial support from the Deutsche Forschungsgemeinschaft (DFG, German Research Foundation) - project number 546268067.
\endgroup

\newpage

\section{Electrical manipulation of skyrmions in chiral magnets}
\begingroup
    \let\section\subsection
    \let\subsection\subsubsection
    \let\subsubsection\paragraph
    \let\paragraph\subparagraph
Yizhou Liu$^1$, and Haifeng Du$^1$
\vspace{0.5cm}

\noindent
\textit{$^1$ Anhui Province Key Laboratory of Low-Energy Quantum Materials and Devices, High Magnetic Field Laboratory, HFIPS, Chinese Academy of Sciences, Hefei 230031, China
}

\section*{Introduction}
Magnetic skyrmions represent a class of topological spin textures characterized by swirling arrangements of spins that wrap around a unit sphere an integer number of times. These nanoscale spin textures exhibit particle-like dynamics and remarkable properties (e.g., low current density-driven motion), making them interesting for both next-generation information technologies and fundamental research~\cite{nagaosa2013topological}. Magnetic skyrmions were first observed experimentally in chiral magnets~\cite{muehlbauer2009skyrmion}, materials with broken inversion symmetry that support the Dzyaloshinskii-Moriya interaction (DMI). Due to their single crystalline nature, simple magnetic model, and the ease of generating single/crystalized skyrmions, chiral magnets (like MnSi and FeGe) have served as a pivotal material system for studying skyrmion dynamics.

The particle-like dynamics of skyrmions make them suitable to serve as information carriers with complex operations, opening new avenues for memories, logic, and neuromorphic computing systems. The electrical manipulation of skyrmions represents a crucial step toward the realization of practical skyrmion-based devices. The significance of electrical manipulation extends beyond mere current-driven motion. It encompasses the entire lifecycle of skyrmions in devices: from controlled nucleation through current-induced motion, to directional steering along defined paths, to reliable detection, and finally to deterministic annihilation when required. Investigating these processes is essential for understanding the fundamental properties of skyrmion and developing skyrmionic devices.

\section*{Status}

Recent years have witnessed remarkable progress in the electrical manipulation of skyrmions in chiral magnets, establishing a foundation for further device applications. Studies have demonstrated motion, creation, and annihilation of single magnetic skyrmions through electrical means in various chiral magnetic systems.

In chiral magnets, the current-driven single skyrmion motion is mostly induced by the spin transfer torque (STT) and is described by the Thiele equation

\begin{equation}
    \mathbf{G}\times (\mathbf{v}_s - \mathbf{v}_d) + \mathcal{D}(\beta\mathbf{v}_s - \alpha\mathbf{v}_d) + \mathbf{F} = 0.
    \label{eq:liu1}
\end{equation}

The first term describes the gyrotropic effect, where $\mathbf{G} = (0, 0, 4\pi Q)$ ($Q$ is the topological charge of skyrmion). $\mathbf{v}_d$ and $\mathbf{v}_s$ are velocities of skyrmion and conduction electrons, respectively. $\mathbf{v}_s$ is defined in terms of the driving current density $j$ as $\mathbf{v}_s = -(gP\mu_B/2eM_s)j$, where $g$ is the Landé g-factor, $\mu_B$ is the Bohr magneton, $e$ is the elementary charge, $M_s$ is the saturation magnetization, and $P$ is the spin polarization of the conduction electrons. The second term describes the dissipative forces with the Gilbert damping coefficient $\alpha$, non-adiabatic STT parameter $\beta$, and the dissipative tensor $\mathcal{D}$, which is determined by the shape of skyrmion. The third term $\mathbf{F}$ accounts for the forces exerted on skyrmion related to external fields, geometrical confinement, and impurities. From Eq.~\eqref{eq:liu1}, the skyrmion velocity along the current direction in the flow regime (assuming the current is along the $x$-direction), the skyrmion Hall angle, and the depinning transition can be obtained. These theoretical predictions have been confirmed in recent experimental works on skyrmion motion under nanosecond current pulses in FeGe~\cite{yu2020motion, song2024steady} and $\rm Co_9Zn_9Mn_2$~\cite{peng2021dynamic}.

\begin{figure*}[h!]
\centering
\includegraphics[width=0.95\textwidth]{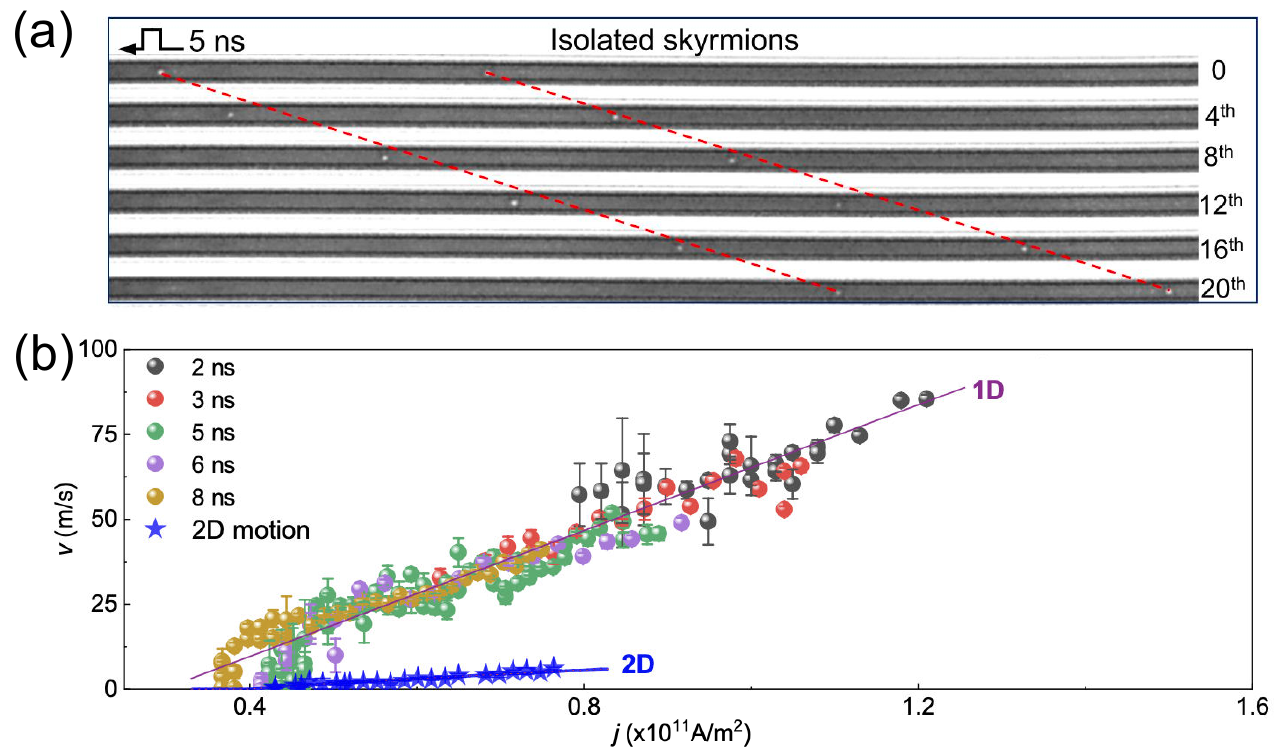}
\caption{(a) Snapshots of the current-driven motion of an 80 nm skyrmion in a 100-nm width FeGe stripe. A series of current pulses with $\SI{8.47e10}{Am^{-2}}$ amplitude and 5 ns duration were employed to drive the skyrmion4. (b) The corresponding current density-velocity plot for various current pulse duration. Star symbols show the current density-velocity relation of the skyrmion motion in a 2D FeGe system~\cite{song2024steady}. }
\label{fig:Liu1}
\end{figure*}

On the other hand, it is noticed that in a confined nanostripe with width comparable to the size of skyrmion, the transverse motion of skyrmion is suppressed, resulting in a modified skyrmion velocity $v_x \approx -\frac{\beta}{\alpha}v_s$. For $\beta > \alpha$, the skyrmion velocity gets enhanced. Such enhancement has been observed in an experiment that aims to build racetrack-type devices by fabricating a 100-nm confined FeGe nanostripe (the skyrmion size is 80 nm, \ref{fig:Liu1}a)~\cite{song2024steady}. Comparing with the two-dimensional (2D) current-driven skyrmion motion in FeGe, a great enhancement of the skyrmion velocity (about 8 times) is achieved due to the high $\frac{\beta}{\alpha}$ value of FeGe (Fig.~\ref{fig:Liu1}b). It can be seen that the enhanced skyrmion driving efficiency (defined as the slope of the current density-velocity curve) is even comparable to that by spin orbit torque (SOT) in synthetic antiferromagnet/ferrimagnet~\cite{pham2024fast, mallick2024driving}. The device also demonstrated that the skyrmion Hall effect might not be a significant problem in device application with optimized geometry. In a similar manner, interstitial skyrmions, i.e., skyrmions confined within helical stripes, show similar velocity enhancement although they are confined by soft boundaries (the spin helices) rather than hard boundaries of the sample~\cite{song2022experimental}.

Other than skyrmions, the exploration of antiskyrmions, topological spin structures with opposite topological charge compared to conventional skyrmions, has expanded the repertoire of manipulable objects in chiral magnets. Antiskyrmion mostly resides in materials with D2d or S4 symmetry in which the DMI is anisotropic. Experiments have demonstrated that antiskyrmions can also be controlled electrically~\cite{he2024experimental, guang2024confined}, exhibiting distinct motion characteristics that could be exploited for specialized applications. So far, antiskyrmion dynamics has only been observed when enclosed by spin helices. As the antiskyrmion dynamics also fits Eq. (1), the antiskyrmion velocity follows a similar relationship $v_x \approx -\frac{\beta}{\alpha}v_s$ when confined in helical stripes. Moreover, when the current is applied perpendicular to the boundary rather than along the helical direction, the antiskyrmion velocity is $v_x \propto -(4\pi Q/\alpha)v_s$ (assuming the current is along the $y$-direction). Therefore, an enhancement of the antiskyrmion velocity is expected as $\alpha$ is usually much smaller than unity. This scenario has been confirmed in a recent experimental work (Fig.~\ref{fig:Karube}c-e)~\cite{guang2024confined}.

In addition to current-driven motion, the electrical creation and annihilation of skyrmions have also been realized experimentally in chiral magnet\cite{wang2022electrical}. Current pulses applied through a nanostructured notch can induce local heating and spin torques that facilitate skyrmion nucleation at predetermined positions. Similarly, applying current pulses in the opposite direction can annihilate existing skyrmions, providing a mechanism for information writing and erasing operations.

\section*{Challenges}
\begin{figure}[h!]
\centering
\includegraphics[width=0.95\linewidth]{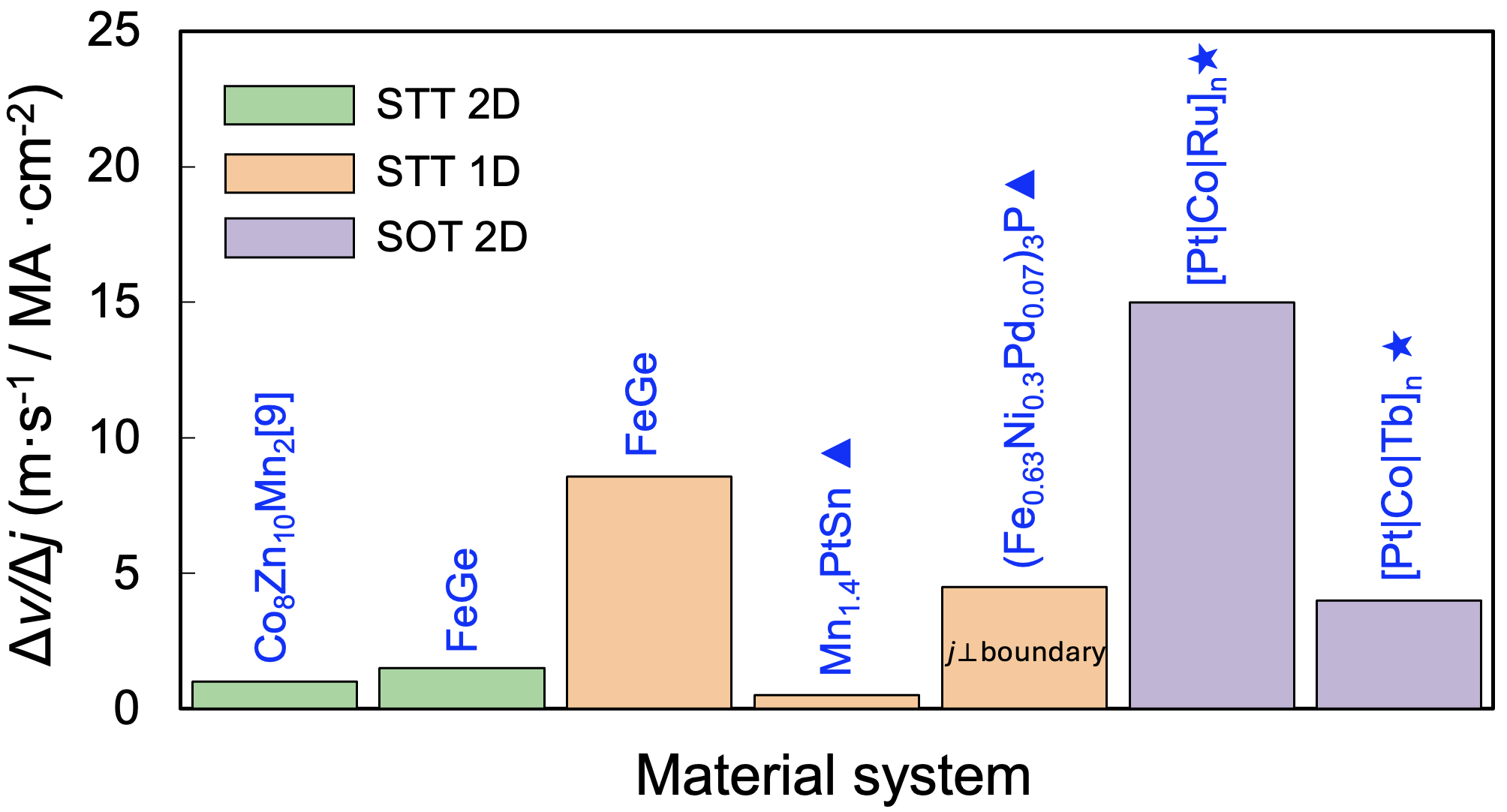}
\caption{The current-driven efficiency (slope of the current density-velocity relation in the flow regime, ) for skyrmion, antiskyrmion (triangle symbol), and synthetic anti(ferro)magnetic skyrmion (star symbol) dynamics in various material systems. STT and SOT stands for spin transfer torque and spin orbital torque, respectively. For (Fe0.63Ni0.3Pd0.07)3P, the current is applied perpendicular to the soft boundary made by spin helix. For [Pt/Co/Ru]n, synthetic antimagnetic skyrmion is studied. For [Pt/Co/Tb]n, synthetic ferrimagnetic skyrmion is studied. The values are estimated based on the data shown in Refs.~\cite{pham2024fast, mallick2024driving, song2022experimental, he2024experimental, guang2024confined, wang2022electrical}.}
\label{fig:Liu2}
\end{figure}

Despite the impressive advances in skyrmion manipulation in chiral magnets, several critical challenges remain in achieving fast and precise electrical control. For the skyrmion motion, finding materials that have large $\frac{\beta}{\alpha}$ values at room temperature is crucial for employing conventional STT driving scheme in chiral magnets. While the skyrmion velocity is enhanced in the one-dimensional nanostripe, its physical origin, the nonadiabaticity  in STT, has been controversial for decades and is still under debate. This impedes further searches for materials with large $\frac{\beta}{\alpha}$ value. In addition, while the current-driven efficiency in nanostripes is comparable to that driven by SOT in synthetic antiferromagnet/ferrimagnet (Fig.~\ref{fig:Liu2}), how to enhance the skyrmion velocity while maintaining its stability is also a challenge. Indeed, theoretical study has once suggested that typical Bloch skyrmions in combination with SOT could lead to even higher current-driven effeiciency (but the applied current density availabe is limited)~\cite{tomasello2014strategy}. Therefore, it is also worth to explore this kind of driving scheme in chiral magnet with SOT driving.

Another critical issue is the enhancement of detection sensitivity for individual skyrmions. While various magnetoresistance effects can detect the presence of skyrmions in chiral magnets, achieving a high on/off ratio for reliable reading operations remains difficult, especially for smaller skyrmions. New detection schemes, possibly combining magnetic tunnel junctions (MTJs) need to be developed and optimized, but the compatibility between chiral magnet and MTJ materials poses a challenge. Furthermore, MTJs could also be employed for skyrmion creation and annihilation due to their high efficiency. From this perspective, chiral magnetic thin film should be employed for both SOT driving and MTJ detection, which sets challenge for material growth and magnetic imaging.

One advantage of employing skyrmions as information carriers is their nanometer sizes that could lead to higher device density. Finding materials that host skyrmions with few-nanometer sizes at room-temperature has long been an important topic to pursue. The recent report of few-nanomter sized skyrmions in centrosymmetric materials like $\rm Gd_2PdSi_3$ provides new hints along this direction.

\section*{Concluding Remarks}

One ultimate goal of research on electrical manipulation of skyrmions is the development of transformative technologies that leverage their unique properties. Due to the well-ordered skyrmion crystal phase and the ease of creating diverse topological spin textures, chiral magnets have been a versitle platform to study skyrmion physics and validate schemes for skyrmion manipulation. The recent emergence of three-dimensional topological spin textures like magnetic hopfions also adds a new flavor to the study of chiral magnets. Their sophisticated structures could lead to novel physics due to the rich degrees of freedom in three dimensions. The possibility of harnessing these three-dimensional topological spin textures for practical devices presents an exciting frontier that has yet to be explored experimentally.

\section*{Acknowledgements}

We acknowledge the support from the National Key R\&D Program of China (Grant No. 2022YFA1403603), the CAS project for Young Scientists in Basic Research (YSBR-084), and the National Natural Science Foundation of China (Grant No. 12241406).

\endgroup

\newpage

\section{Skyrmion Magnonics}
\begingroup
    \let\section\subsection
    \let\subsection\subsubsection
    \let\subsubsection\paragraph
    \let\paragraph\subparagraph
Martin Lonsky$^1$, and Axel Hoffmann$^2$
\vspace{0.5cm}

\noindent
\textit{$^1$ Institute of Physics, Goethe University Frankfurt, 60438 Frankfurt, Germany\\
$^2$ Materials Research Laboratory and Department of Materials Science and Engineering, The Grainger College of Engineering, University of Illinois Urbana-Champaign, Urbana, Illinois 61801, USA}

\section*{Introduction} \label{intro}
Skyrmions are topologically distinct chiral spin textures that exist in a wide range of materials and are considered for applications in future data storage, unconventional computing and information processing. While many research works focus on the interaction with electric currents ({\em e.g.}, for racetrack memory applications), the coupling with high-frequency spin excitations -- so-called spin waves, or magnons -- opens new avenues for fast, versatile, and energy-efficient devices. In other words, \textit{skyrmion magnonics}, the combination of magnonics and skyrmionics, gives rise to novel physical phenomena and fruitful synergies that are worth exploring. 

As shown in Fig.~\ref{fig:Lonsky1}, key research directions  are rooted in the complex interaction between skyrmions and magnons: First, individual skyrmions exhibit distinct spin eigen-excitations when subjected to radiofrequency magnetic fields or spin torques \cite{garst2017collective, lonsky2020dynamic, Lonsky2020coupled}. These modes correspond to magnon-skyrmion bound states, which represent one solution of the magnon-skyrmion scattering problem. Additionally, in analogy to the case of domain walls \cite{Yu_2021}, a translational motion of skyrmions can be induced by spin waves, enabling control without electric currents. Furthermore, magnons propagate through periodic skyrmion lattices or arrays, whereby they interact with the spin textures in complex ways. The system can be regarded as a magnonic crystal and enables band engineering and harnessing nonreciprocal effects. Recently, magnon-skyrmion hybrid systems with the prospect for coherent information processing and systems with nonlinear magnon-skyrmion interactions have been proposed as additional relevant platforms \cite{Pan_2024}. 

The phenomena mentioned above exhibit dynamic properties that can be tuned by a variety of parameters such as the choice of materials, geometry, disorder, and skyrmion size. Consequently, precise control of these parameters is required for meaningful research and applications. Key challenges include material optimization ({\em e.g.}, reducing Gilbert damping), time- and spatially resolved characterization capability of magnons, and integration with existing technology. Here, the progress and open questions in these areas are highlighted. 

\begin{figure*}[h!]
  \centering
  \includegraphics[width=0.95\textwidth]{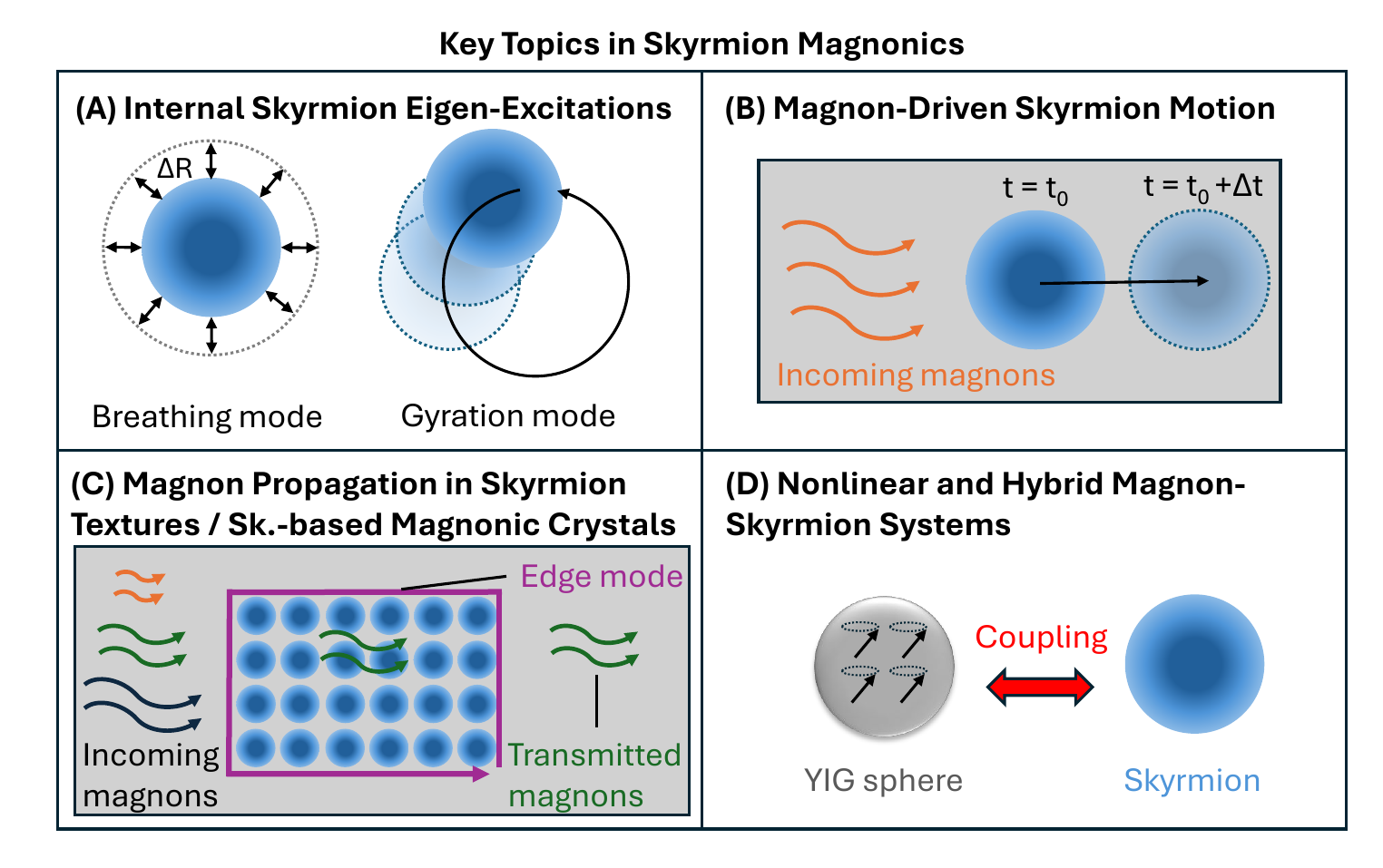}
  \caption{Overview of key topics in skyrmion magnonics. (a) Resonant skyrmion dynamics: breathing mode with oscillating skyrmion radius R and (counterclockwise) gyration mode with rotating skyrmion motion. (b) Magnons can drive translational skyrmion motion. (c) Skyrmion-based magnonic crystal: transmission of selected magnons (spin-wave filtering) and excitation of topological edge states. (d) Example hybrid system consisting of an yttrium iron garnet (YIG) sphere coupled to a skyrmion, based on \cite{Pan_2024}.} 
  \label{fig:Lonsky1}%
\end{figure*}%

\section*{Key Topics, Challenges and Advances} \label{current}

\subsection*{Skyrmion internal dynamics} 
The eigen-excitations of skyrmions ({\em i.e.}, spin-wave resonances, typically in the GHz frequency range) in various materials and geometries have been extensively studied \cite{garst2017collective, lonsky2020dynamic}. These include breathing and gyration modes which can be excited by microwave magnetic fields, electric fields, spin torques, laser pulses, and other means. While most studies have relied on micromagnetic simulations, experimental investigation has recently become possible using techniques such as microwave spectroscopy on carefully fabricated materials and devices. Possible applications include unconventional computing and electrical skyrmion detection based on unique dynamic fingerprints. 

Pioneering experiments were conducted on single-crystal Cu$_2$OSeO$_3$, where propagating skyrmion excitations were measured with high precision \cite{Seki_2020}. The damping parameter was shown to be as low as $\alpha \approx 10^{-4}$ at $T=5\,$K. However, thin-film systems and room temperature functionality are preferred with regards to future applications. Only recently, experimental evidence of resonant skyrmion dynamics was reported for Ir/Fe/Co/Pt \cite{satywali2021microwave} and [Pt/FeCoB/AlO$_{\mathrm{x}}$] \cite{Srivastava_2023} multilayers with low Gilbert damping ($\alpha \leq 0.05$) at room temperature.

Current challenges include the limited range of suitable materials, especially for low-damping, room-temperature thin-film systems, and resolving high-frequency ($>50\,$GHz) or spatially localized modes of nanoscale skyrmions. Meanwhile, synthetic antiferromagnets have emerged with distinct advantages, as their interlayer exchange coupling enables zero-field stable skyrmions and reduced stray fields. Their intermediate eigenmode frequencies -- between those of ferromagnets and conventional antiferromagnets -- also facilitate experimental detection.

\subsection*{Magnon-driven skyrmion motion} 
Magnons can efficiently drive skyrmion motion via momentum transfer, as demonstrated numerically for nanowires where magnon currents induce skyrmion propagation with velocities scaling as $v\propto J/\alpha$ (where $J$ denotes the magnon current density) \cite{Zhang_2017}. 
This phenomenon is governed by three mechanisms. First, there is skew scattering of magnons interacting with skyrmions. Second, in confined geometries, such as nanowires, interactions between skyrmions and boundaries promote directional motion along edges. Third, at high magnon currents the resonant excitation of internal skyrmion breathing and additional magnon modes leads to deformation and eventually annihilation of the skyrmion, thus limiting the maximum velocity. 
This approach offers energy-efficient, current-free manipulation of skyrmions for spintronic devices, though challenges such as relatively high damping in many room-temperature skyrmion hosts persist. Furthermore, the skyrmion size needs to be reliably controlled, as it plays a role in the driving force, which increases as a function of the diameter due to the enhanced skyrmion surface that interacts with the spin wave. Lastly, pinning forces of skyrmions at defects have to be overcome, implying the need of sufficiently low defect densities in materials and large enough magnon currents without destabilizing skyrmions. Detecting magnon-induced skyrmion motion is not trivial, particularly distinguishing it from dynamics driven by thermal gradients.
At this time, an experimental realization of magnon-driven skyrmion motion has not yet been reported, but may be expected in the near future, given the progress being made in material optimization and experimental technique developments.  

\subsection*{Magnon propagation in skyrmion textures} 
Skyrmion-based magnonic crystals (MCs) represent a promising platform for the dynamic control of spin waves, offering tunable band structures and topological magnon modes \cite{Chen_2021}. These systems leverage the periodic arrangement of skyrmions to create magnonic bandgaps, which can be dynamically modulated via external fields, or currents, enabling functionalities such as on-demand spin wave transmission or rejection. Such dynamically reconfigurable MCs stand in stark contrast to geometrically patterned, static MCs with fixed bandgaps, as they offer more flexibility and facilitate miniaturization without the constraints of lithography. 

Recent theoretical and numerical advances involve the discovered potential of skyrmion-based MCs in topological magnonics. The interplay between skyrmion chirality and spin-wave dynamics generates topologically protected magnon bands, supporting unidirectional edge states that are robust against scattering. 
It was shown that these chiral edge modes, analogous to quantum Hall edge states, arise from finite Chern numbers in the magnon spectrum, enabling low-dissipation spin-wave guides \cite{Chen_2021}. In principle, it is possible to deliberately switch topologically nontrivial magnon edge states on and off, or change their directionality, for instance by using external magnetic fields, and thus enhance the tunability of skyrmion-based MCs \cite{Diaz_2020}. 
The experimental detection of topological magnon modes could be realized by using ferromagnetic resonance in combination with spatially resolved imaging techniques, such as micro-focused Brillouin light scattering \cite{Sebastian_2015} or nitrogen vacancy-center based techniques \cite{Casola_2018}.  

Despite the rapid theoretical and numerical advances, practical challenges remain in stabilizing skyrmion lattices at room temperature and reducing Gilbert damping of thin-film skyrmion hosts to enhance spin wave propagation for practical implementations. Precise control of skyrmion periodicity is required to achieve usable bandgaps, but defects and pinning disrupt long-range order. In a first step, artificial pinning arrays or geometric confinement ({\em e.g.}, nanostructured tracks) could be used in proof-of-concept experiments to enhance lattice regularity, though at the cost of reduced reconfigurability \cite{szulc2025multifunctional}. Thus far, there only exist a few preliminary experimental works, but no comprehensive reports on skyrmion-based MCs \cite{Chen_2021}.

\subsection*{Nonlinear and hybrid magnon-skyrmion systems} 
Further recent advances are based on the unique interplay between magnons and skyrmion textures, where skyrmions act as a versatile medium for spin wave manipulation. For instance, nonlinear magnon-skyrmion scattering generates a magnonic frequency comb with mode spacing determined by the skyrmion breathing frequency, enabling tunable frequency conversion and comb-based magnonic devices \cite{Wang_2021}. Moreover, the interference and delayed relaxation of spin waves in skyrmion crystals can be exploited for reservoir computing \cite{Lee_2022}. 
Magnon-skyrmion hybrid systems enable strong coupling between magnons and skyrmion dynamics, which can be harnessed for coherent information processing and nonreciprocal signal control \cite{Li_2022}. One approach is the strong coupling of a skyrmion with micromagnets such as yttrium iron garnet (YIG) spheres, which exhibit long-lived spin wave modes \cite{Pan_2024}. The purpose is to initiate magnon-mediated coherent and dissipative couplings between skyrmion dynamics. Ultimately, a nonreciprocal interaction between distant skyrmions can be achieved and harnessed for coherent information processing.
To summarize, hybrid platforms hold promise for advancing spintronics, novel computing paradigms, and the study of topological excitations.

\section*{Concluding Remarks} \label{conclusion}

Skyrmion magnonics is a promising platform for potential information processing and computing applications. Nevertheless, critical challenges span all its subfields, and a significant portion of conducted research to this point has been solely of numerical nature. 
Material limitations -- particularly high damping and necessity of low temperatures for many skyrmion hosts  -- constrain applications. 
While synthetic antiferromagnets and metallic multilayers show promise, achieving sufficiently low damping in thin-film systems remains difficult. Nanoscale control of skyrmion lattices is essential, too, as defects disrupt bandgap engineering and magnon-driven motion. Another task is given by integrating skyrmion-based devices with conventional magnonic circuits. Addressing these challenges through advances in materials, characterization, and device integration will likely unlock the potential of skyrmion magnonics for novel technologies.

\section*{Acknowledgements}

The manuscript preparation was partially supported
by the NSF through the University of Illinois Urbana-Champaign Materials Research Science and Engineering Center Grant No. DMR-1720633 and DMR-2309037.



\clearpage

\endgroup

\newpage

\section{Optical transformation and control of
skyrmionic spin textures}
\begingroup
    \let\section\subsection
    \let\subsection\subsubsection
    \let\subsubsection\paragraph
    \let\paragraph\subparagraph
Daniel Steil$^1$, and Stefan Mathias$^{1,2}$
\vspace{0.5cm}

\noindent
\textit{$^1$  $\rm 1^{st}$ Institute of Physics, University of Göttingen, 37077 Göttingen, Germany}\\
\textit{$^2$ International Center for Advanced Studies of Energy Conversion (ICASEC), University of Göttingen, 37077 Göttingen, Germany} 

\section*{Introduction}
Skyrmionic spin textures have attracted great interest from a fundamental scientific perspective, but particularly also in the context of technological application in spintronics and magnonics. Potential applications range from usage in racetrack memory, nanoscale spin wave emitters, utilization in spin wave logic to neuromorphic computing schemes. To make such applications possible, suitable control schemes are needed. Next to electrical or microwave manipulation, ultrashort light pulses provide the intriguing opportunity to achieve contact-free manipulation and readout with ultimate speed. Essential operations for utilizing light manipulation of skyrmions include the creation and annihilation of such spin textures, achieving time-domain optical readout and controlling their dynamical properties, such as gyration and breathing modes.
Recent experiments have demonstrated different aspects of these operations. For instance, Je et al.~\cite{je2018creation} and Büttner et al.~\cite{buttner2021observation}  optically created (and annihilated) such textures, as also depicted in Fig.~\ref{fig:Steil1}a. Gerlinger et al. ~\cite{gerlinger2021application} and Zhu et al.~\cite{zhu2024ultrafast} showed toggle switching between single domain and skyrmionic textures using a combination of optical pulses and magnetic fields. Ogawa et al. were the first to demonstrate optical readout via the magnetooptical Faraday effect~\cite{ogawa2015ultrafast}, as shown in Fig.~\ref{fig:Steil1}b. In Fe/Gd multilayers, we and our co-authors could show all three essential operations in one system: optical creation and annihilation of a mixed lattice of bubbles and skyrmions, detection of different spin textures via the magnetooptical Kerr effect~\cite{titze2024laserinduced, titze2025pathways}, and control of optically excited breathing modes in amplitude and phase~\cite{titze2024alloptical}, see Fig.~\ref{fig:Steil1}c.

\begin{figure*}[h!]
    \centering
    \includegraphics[width=0.95\textwidth]{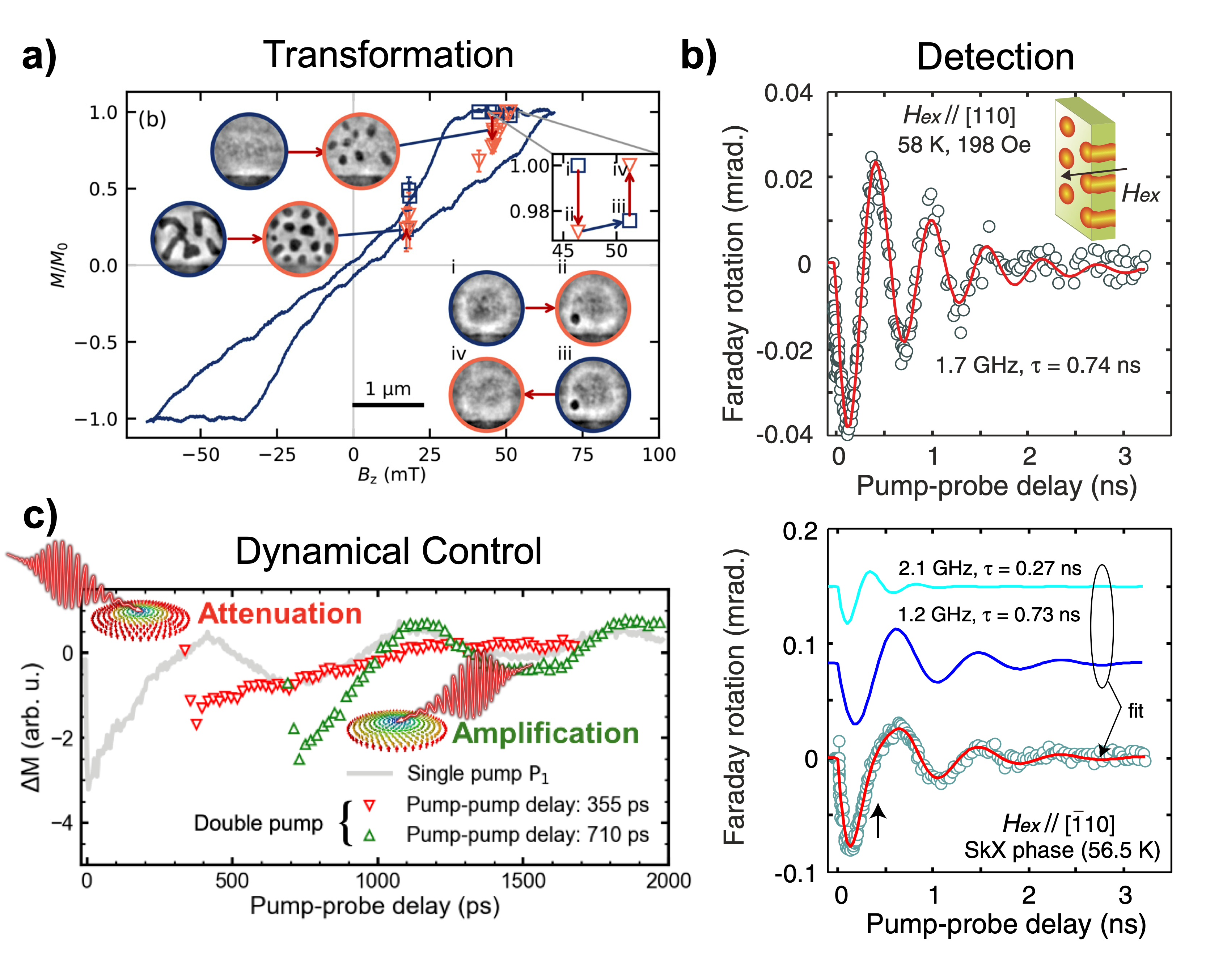}
    \caption{Transformation: Overlay of soft x-ray holography images highlighting nucleation of skyrmions from the saturated state and from stripe domains by laser excitation (left top; blue: initial state, orange: final state) as well as deterministic creation or annihilation of a single skyrmion using light excitation together with magnetic field control (right bottom) onto a MOKE hysteresis loop of the Pt/CoFeB/MgO sample used. The square inset shows the magnetic field values used for creation/annihilation of the skyrmion. Adapted from~\cite{gerlinger2021application} under a CC BY 4.0 license. (b) Detection: Coherent spin signals in transient Faraday rotation corresponding to the breathing mode (top) and gyrational modes (bottom) of skyrmions excited by the inverse Faraday effect. Adapted from \cite{ogawa2015ultrafast} under a CC BY 4.0 license, Copyright © 2015, The Author(s); (c) Dynamical Control: Demonstration of attenuation and amplification of already breathing bubbles and skyrmions depending on the delay between two exciting optical pulses. Adapted from \cite{titze2024alloptical} under a CC BY 4.0 license.}
    \label{fig:Steil1}
\end{figure*}

\section*{Relevance and Vision}
Optical detection and manipulation of skyrmionic textures enables contact-free studies of both nucleation and annihilation of such spin objects, as well as researching their spatio-temporal dynamics. The use of ultrashort laser pulses provides unprecedented temporal resolution to study the transformation of spin textures and offers the fastest and potentially most energy efficient way of manipulating them. Ultrashort light pulses also enable access to metastable spin textures that cannot be obtained by quasi-equilibrium methods. Different classes of samples can be addressed by different light interaction channels. Spin textures in metallic or semiconducting films can be manipulated by changing materials properties like the magnetic anisotropy by light absorption~\cite{titze2024laserinduced, titze2024alloptical, padmanabhan2019optically, kalin2022optically}, while in insulators absorption-free interactions like the inverse Faraday effect can be used~\cite{ogawa2015ultrafast}. Microscopy or near-field methods may facilitate single site detection and manipulation without utilizing complex samples designs that are often needed for methods relying on electric currents, or electric and microwave fields. Light excitation may even allow to create reprogrammable magnonic crystal structures on the fly.

Applying light as such a tool has already been discussed for various technical applications. Gerlinger et al.~\cite{gerlinger2021application} recently proposed three application concepts of optically created skyrmions: all-optical writing and deleting, e.g., for random access data storage and memory applications, a skyrmion reshuffler for probabilistic computing, and a racetrack memory system based on optical and spin-orbit torque manipulation of skyrmions. These devices primarily rely on the presence of skyrmionic textures. On the other hand, coupling between skyrmions may lead to the creation of magnonic crystals allowing for tuneable eigenmodes of the systems as discussed in~\cite{lonsky2020dynamic}. These in turn can be used to create devices like spin wave filters, sensors, spin wave phase shifters, resonators, or parts of logic gates, see, e.g.,~\cite{chumak2017magnonic}.

To unlock this potential, further detailed studies on the interaction of light with skyrmionic textures are needed. Concerning nucleation and annihilation, an in-depth understanding of relevant mechanisms~\cite{je2018creation, buttner2021observation}, both thermal and non-thermal, is required for fundamental science and applications. This understanding will enable minimum energy expenditure and fastest write and delete speeds, as well as long term stability of written textures. Potential advantages of using optomagnetic effects like the inverse Faraday effect are largely unexplored experimentally.  Here, the ultimate goal is the creation of arbitrary spin object structures at will.

For optical detection and dynamical optical control schemes, it is necessary to systematically analyze how dynamics are influenced by different stabilizing mechanisms (dipolar, bulk \& interface Dzyaloshinskii-Moriya interaction [DMI]) or different types of skyrmionic textures (Bloch, Néel, hybrid, 3D spin objects like chiral bobbers~\cite{zheng2018experimental}  or higher order skyrmions~\cite{hassan2024dipolar}, see also Fig.~\ref{fig:Steil2}a). For applications, technologically relevant systems need to move into the focus of research (i.e., ambient T, thin films). The few investigated systems so far comprise mostly insulators/semiconductors at cryogenic temperatures~\cite{ogawa2015ultrafast, padmanabhan2019optically, kalin2022optically}. Developing robust capabilities to optically tailor skyrmion lattices will allow to study interactions between localized skyrmionic textures and pave the way towards full tuneability of magnon dispersions and band structures (see Fig.~\ref{fig:Steil2}c).

\section*{Challenges}
Technologically relevant skyrmion-host materials for optical manipulation need to be grown by standard, easily tunable thin film methods and should exhibit (meta)stable skyrmionics textures at ambient temperature. Currently many systems are either single crystalline materials (e.g. FeGe), or complex alloys and compounds, which exhibit skyrmionic textures only at low temperatures~\cite{tokura2021magnetic, lonsky2020dynamic}.

\begin{figure*}[h!]
    \centering
    \includegraphics[width=0.95\textwidth]{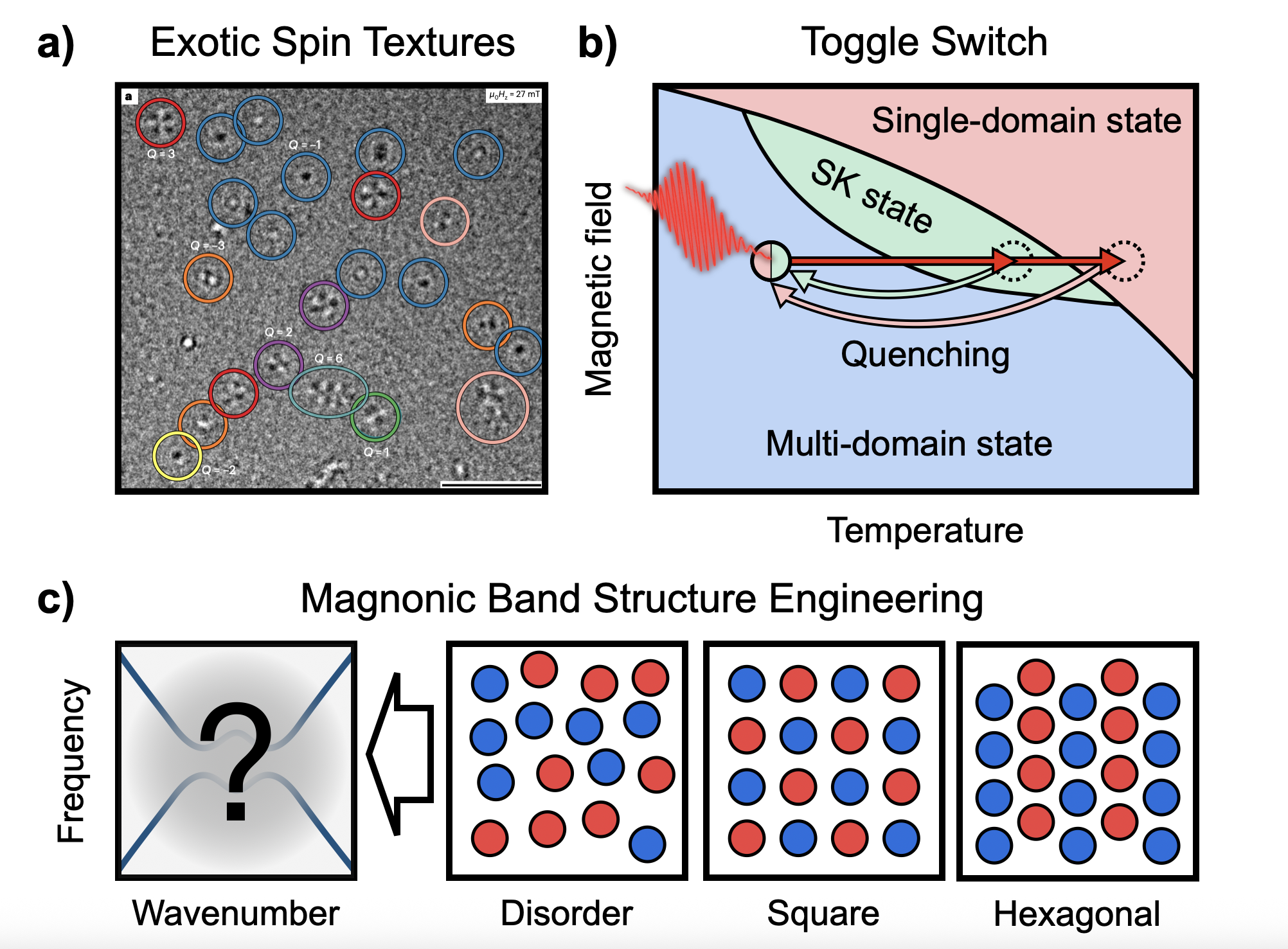}
    \caption{a) Exotic textures: Overview of different skyrmionics spin textures in a [Co/Ni]$_{10}$ multilayer sample ranging from simple $Q=-1$ skyrmions, $Q=1$ antiskyrmions to high order skyrmionic textures. Lorentz-TEM image from ~\cite{hassan2024dipolar} under a CC BY 4.0 license. Scale bar is 1 µm. b) Toggle switching: Idealized magnetic phase diagram for toggle switching between a skyrmion state and a single domain state without magnetic field change. Here the idea is to use heating to either drive the system into the skyrmion state or the single domain state, which remains as a local metastable state after a thermal quench. Alternatively, the system could be reset by heating over the Curie temperature. c) Magnonic band structure engineering can be achieved by creating different lattice arrangements including different types of topological spin objects. Three examples are given with lattice ordering including different spin objects in red and blue (e.g., clockwise vs. counterclockwise skyrmions or skyrmions and antiskyrmions), highlighting the vast possibilities.}
    \label{fig:Steil2}
\end{figure*}

Ideally, optical creation and annihilation of skyrmionic textures should occur by a purely optical toggle switching mechanism without additional control parameters as in Refs.~\cite{gerlinger2021application, zhu2024ultrafast}. This would provide a two-state system comprised of a skyrmionic state and an alternate spin state, as schematically shown in Fig.~\ref{fig:Steil2}b. Realizing advanced logic operations and advanced dynamical control strategies then requires optically created skyrmion lattices with arbitrary spin object positions. To achieve this, robust solutions to deterministically nucleate and annihilate topological spin textures on freely defined positions (see Fig.~\ref{fig:Steil2}c) or suitable sample patterning methods to determine nucleation sites~\cite{kern2022tailoring} need to be developed. Predefined pinning sites may help prevent unwanted rearrangement of written structures. Notably, the recent demonstration of the controlled nucleation of higher order skyrmionic textures in thin films~\cite{kern2025controlled}, may pave the way for such studies in pump-probe experiments.

Exploring coherent excitations of skyrmionic textures using ultrashort laser pulses is often hindered by the fact that many thin film systems  exhibit strong Gilbert damping due to the use of heavy elements in their composition, in particular in systems utilizing interfacial DMI for skyrmion creation. This effect also prevents long distance propagation of spin waves, suppressing interactions between skyrmionic textures. To overcome these limitations, systems with low damping of spin excitations and long spin-wave propagation are needed. Moreover, the few systems studied so far ~\cite{ogawa2015ultrafast, titze2024laserinduced, padmanabhan2019optically, kalin2022optically} have relatively low-frequency modes in the range of a few GHz, restricting experimental analysis, e.g., of mode dispersions, in optical experiments due to the limited length of optical delay lines. Systems with higher frequency modes (on the order of 10 GHz or higher) will enable more sophisticated studies and potentially high-speed applications. Finally, exploring dynamics of skyrmionic textures with arbitrary topological charge either requires creating homogeneous lattices of such objects or developing time-resolved experimental setups with sub-µm spatial resolution to study single spin objects. Such setups would also allow for the exploration of nonlocal effects like propagation of skyrmion excitations.

\section*{Acknowledgments:}
We thank Tim Titze for support in figure preparation and for critical reading of the manuscript.

\endgroup

\newpage

\section{Ultimate Speed of Topological Switching}
\begingroup
    \let\section\subsection
    \let\subsection\subsubsection
    \let\subsubsection\paragraph
    \let\paragraph\subparagraph
Bastian Pfau$^1$, and Felix Büttner$^{2,3}$
\vspace{0.5cm}

\noindent
$^1$ Max Born Institute for Nonlinear Optics and Short Pulse Spectroscopy, 12489 Berlin, Germany\\
$^2$ Experimental Physics V, Center for Electronic Correlations and Magnetism, University of Augsburg, 86159 Augsburg, Germany\\
$^3$ Helmholtz-Zentrum Berlin für Materialien und Energie GmbH, 14109 Berlin, Germany

\section*{Introduction}

Magnetization dynamics span an extraordinary range of temporal and spatial scales, from the rare thermally activated switching events of bits in magnetic storage devices to the ultrafast all-optical switching on femtosecond timescales. This breadth is also relevant for topological spin textures, such as skyrmions, where the underlying processes can be broadly classified into three regimes (Fig.~\ref{fig:Pfau1}). In the thermodynamic regime, slow fluctuations govern the lifetime and stability of skyrmions~\cite{wild2017entropy}, ultimately limiting data retention and device reliability. On nanosecond timescales, the dynamics can be effectively described using micromagnetic models that capture the collective precessional motion of spins subject to an effective field. Skyrmion creation and motion via spin–orbit torques (SOT) is described in this way. Finally, in the ultrafast atomistic regime, impulsive excitations—such as laser pulses or ultrashort field transients—drive spin dynamics beyond the macrospin approximation, involving spin-flip scattering and transient modifications of electronic and magnetic order. The research of skyrmion nucleation within these regimes has revealed a rich landscape of switching phenomena, with characteristic timescales ranging from seconds to sub-nanoseconds depending on the excitation mechanism and materials system. Recent advances in time-resolved X-ray and electron microscopy have begun to reveal aspects of these processes and their characteristic timescales, offering valuable insights while highlighting the need for further development to fully capture ultrafast topological switching dynamics.
\begin{figure*}[h!]
    \centering
    \includegraphics[width=0.95\textwidth]{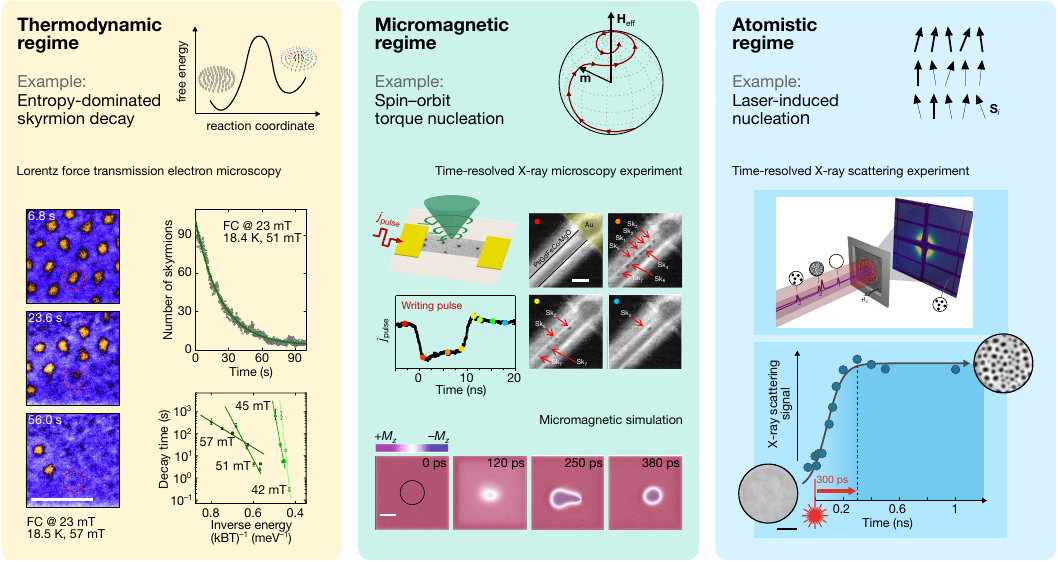}
    \caption{General regimes of magnetization dynamics and examples of topological transitions. Thermodynamic regime: Thermally activated decay of skyrmions in the chiral B20 material $\rm Fe_{0.5}Co_{0.5}Si$. The skyrmion lifetime generally obeys the Arrhenius law with higher energy barriers at lower fields. However, an attempt frequency that dramatically increases due to entropic effects substantially reduces the lifetime of the skyrmions at lower fields. Adapted from Ref.~\cite{wild2017entropy}. Micromagnetic regime: Time-resolved imaging of the SOT-driven creation of a skyrmion in a ferrimagnetic racetrack. Several skyrmions are nucleated transiently. A current pulse of opposite polarity later erases the persisting skyrmions. Adapted from Ref.~\cite{woo2018deterministic}. Atomistic regime: Ultrafast creation of a topological skyrmion state from a uniform ferromagnetic state by a femtosecond laser pulse. An X-ray scattering measurement at European XFEL revealed a creation time of 300 ps. Adapted from Ref.~\cite{buttner2021observation}.}
    \label{fig:Pfau1}
\end{figure*}
Within these regimes, a range of approaches has been developed to achieve controlled skyrmion nucleation down to sub-nanosecond timescales. SOT generated by fast current pulses in chiral multilayers, such as CoFeB-based stacks, has enabled deterministic creation with pulse durations down to 0.75 ns, limited primarily by the electrical excitation itself~\cite{buttner2017fieldfree, woo2018deterministic}. Spatially precise nucleation has been demonstrated by combining SOT with engineered pinning sites produced via focused ion beam irradiation~\cite{kern2022deterministic}. Beyond current-driven mechanisms, localized spin torques and Joule heating beneath vertical nanocontacts provide nanosecond switching in both metallic systems and two-dimensional magnets~\cite{finizio2019deterministic, powalla2022single}. Surface acoustic waves have also been used to trigger skyrmion formation within similar temporal windows~\cite{yokouchi2020creation}.

Ultrafast optical excitation has emerged as a broadly applicable approach to induce skyrmion nucleation across various material classes, including bulk chiral magnets~\cite{berruto2018laserinduced}, metallic multilayers~\cite{buttner2021observation, je2018creation}, and two-dimensional van der Waals magnets~\cite{khela2023laserinduced}. Time-resolved X-ray scattering at free-electron laser facilities has demonstrated that nucleation can proceed within just 300 picoseconds~\cite{buttner2021observation}, highlighting the potential for extremely rapid switching. Beyond speed, optical methods offer the capability to flexibly control skyrmion density, enabling the creation of individual isolated skyrmions as well as extended skyrmion lattices.

Finally, electric-field-driven switching offers a promising route to efficient control without charge flow, with recent demonstrations achieving deterministic creation and deletion~\cite{hsu2017electricfielddriven, bhattacharya2020creation}. The timescales of this switching mechanism remain unexplored so far.

\section*{Relevance And Vision}

Understanding and controlling skyrmion nucleation on ultrafast timescales is essential for both fundamental scientific research and the development of future spintronic technologies. From a basic perspective, exploring how topological spin textures emerge within hundreds of picoseconds provides a unique opportunity to probe the interplay of exchange interactions, Dzyaloshinskii–Moriya interactions, and thermal fluctuations during nonequilibrium phase transitions. Such studies may clarify whether topology itself imposes intrinsic speed limits on magnetic switching. In applied contexts, the ability to deterministically create and delete skyrmions at sub-nanosecond speeds is a prerequisite for integrating skyrmion-based devices into GHz-frequency logic and memory architectures. Beyond conventional memory applications, ultrafast skyrmion control could also enable reconfigurable magnonic circuits and hybrid optoelectronic components that combine electrical and optical stimuli. Establishing reliable, scalable methods to achieve this performance will be a key milestone in demonstrating the technological viability of topological spin textures.

\section*{Challenges And Opportunities}

Electrically driven approaches, including spin–orbit torques from current pulses and electric-field-induced switching, are particularly attractive for integration with conventional CMOS architectures. These mechanisms offer deterministic control and can be embedded directly into device layouts. However, their dynamic limits remain poorly explored. While the generation of electrical signals with sub-100-ps duration is well established in RF electronics, such ultrafast stimuli have yet to be exploited in skyrmion systems. This represents an untapped opportunity to probe the true temporal limits of electrically driven topological switching, even though the time scale of the driver will be ultimately limited to picoseconds.

\begin{figure*}[h!]
    \centering
    \includegraphics[width=0.95\textwidth]{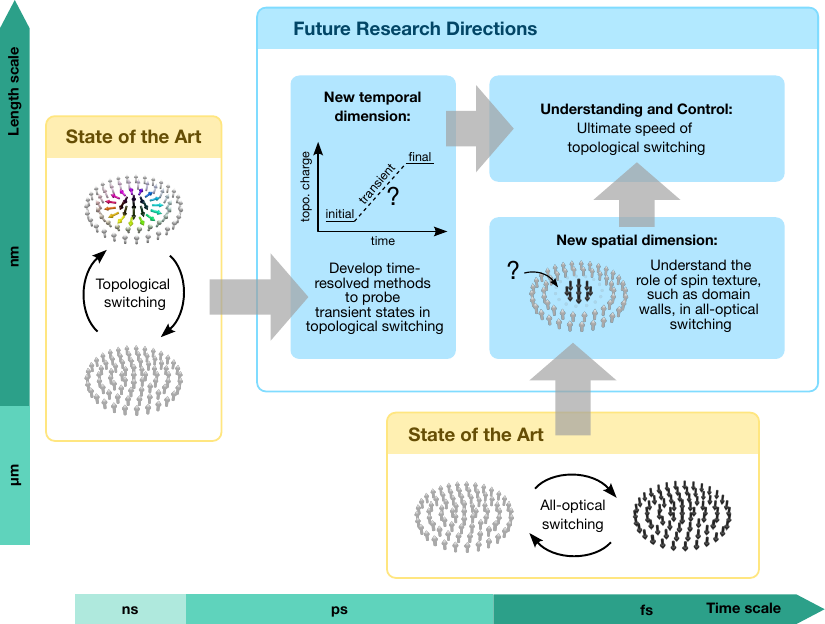}
    \caption{Perspectives for research on ultrafast topological switching motivated by the state-of-the-art in nanoscale topological switching and ultrafast all-optical switching.}
    \label{fig:Pfau2}
\end{figure*}

In contrast, optical methods offer direct access to the fastest dynamical processes. All-optical switching (AOS) has demonstrated sub-picosecond magnetization reversal in ferrimagnetic rare-earth transition-metal alloys after femtosecond photo-excitation (see Fig.~\ref{fig:Pfau2}). Yet, the switched areas typically span micrometers, and the spin structure, topology, and formation dynamics of the surrounding domain walls have received limited attention so far. This is remarkable as domain walls are not passive boundaries, but instead are known to determine, through their chiral (Bloch or Néel) structure, almost any spintronic behavior, from spin texture motion to topology. Furthermore, domain walls themselves can host topological defects, such as vertical Bloch lines, which may introduce stochasticity, limit the scalability or stability of final switched states, or, conversely, reveal hidden aspects of the transformation trajectory~\cite{metternich2025defects}. Understanding the nature and evolution of the domain boundary, especially following ultrafast excitation, is therefore of both fundamental and applied relevance, particularly in topological switching scenarios where the final magnetic state is inherently structured.

In contrast to AOS in ferrimagnetic alloys, laser-induced skyrmion nucleation in ferromagnets has enabled true nanoscale control and the creation of topologically nontrivial states. However, unlike AOS, skyrmion nucleation times observed so far fall in the regime of hundreds of picoseconds to nanoseconds. These nucleation times are likely limited by slow heat dissipation in the membrane samples rather than by fundamental mechanisms. Whether topological switching can occur on femtosecond timescales remains an open question. Addressing this question will require developing and exploring alternative material systems offering more efficient energy dissipation pathways, while still being compatible with detection schemes. In addition, antiferromagnets and synthetic antiferromagnetically coupled multilayers exhibit intrinsically faster magnetization dynamics as switching can, in principle, proceed without angular momentum dissipation.

On the methodical side, the transient states leading to skyrmion formation remain largely inaccessible. A technique to directly detect topological charge in a time-resolved measurement does not exist, and imaging with both vector sensitivity and sub-picosecond resolution is still beyond reach. As a result, the ultrafast dynamics and possible transient topological states during any topological switching trajectory remain largely uncharted.

As illustrated in Fig.~\ref{fig:Pfau2}, bridging the temporal and spatial extremes by combining the speed of optical switching with the nanoscale control and topological richness of skyrmions defines a compelling long-term objective. Addressing this challenge will require advances in both metrology and materials. Simultaneously, technological hurdles remain: energy efficiency, endurance, and compatibility with device scaling will all be critical for translating these phenomena into real-world applications. Progress in these areas will define whether skyrmionics can deliver on its promise as a platform for ultrafast, low-power spintronic and hybrid opto-spintronic technologies.

\endgroup

\newpage

\section{Electrical detection of skymions in magnetic tunnel junctions and their efficient manipulation by gate voltage and current}
\begingroup
    \let\section\subsection
    \let\subsection\subsubsection
    \let\subsubsection\paragraph
    \let\paragraph\subparagraph
Olivier Boulle$^1$, and Hélène Béa$^{1,2}$\\
\vspace{0.5cm}

\textit{$^1$ Univ. Grenoble Alpes, CNRS, CEA, SPINTEC, 38054 Grenoble, France}\\
\textit{$^2$ Institut Universitaire de France (IUF), 75000 Paris, France}

\section*{Introduction}
The nanometer size of magnetic skyrmions, their topological protection and efficient electrical manipulation have opened promising technological outlook for memory and logics devices based on skyrmion manipulation~\cite{fert2017magnetic, zhang2020skyrmion, vakili2021skyrmionics}. This includes racetrack memory, skyrmion logic gates, exploiting the natural repulsion between skyrmions, as well as non-conventional computing schemes, including skyrmion synapses and neurons, Brownian computing or reservoir computing. Skyrmion based technologies require several key elements: nanometer scale skyrmions at room temperature, low power nucleation for write operations, fast and controlled propagation in tracks, electrical detection for the readout operation, low power manipulation for logics. Important advances were done over the last ten years in this direction, in particular with the demonstration of room temperature skyrmions in ultrathin films~\cite{fert2017magnetic, zhang2020skyrmion, vakili2021skyrmionics, jiang2015blowing, boulle2016room}, their efficient current induced motion and electrical detection using the anomalous Hall effect (AHE)~\cite{fert2017magnetic, zhang2020skyrmion, vakili2021skyrmionics}.

\begin{figure*}[h!]
    \centering
    \includegraphics[width=0.95\textwidth]{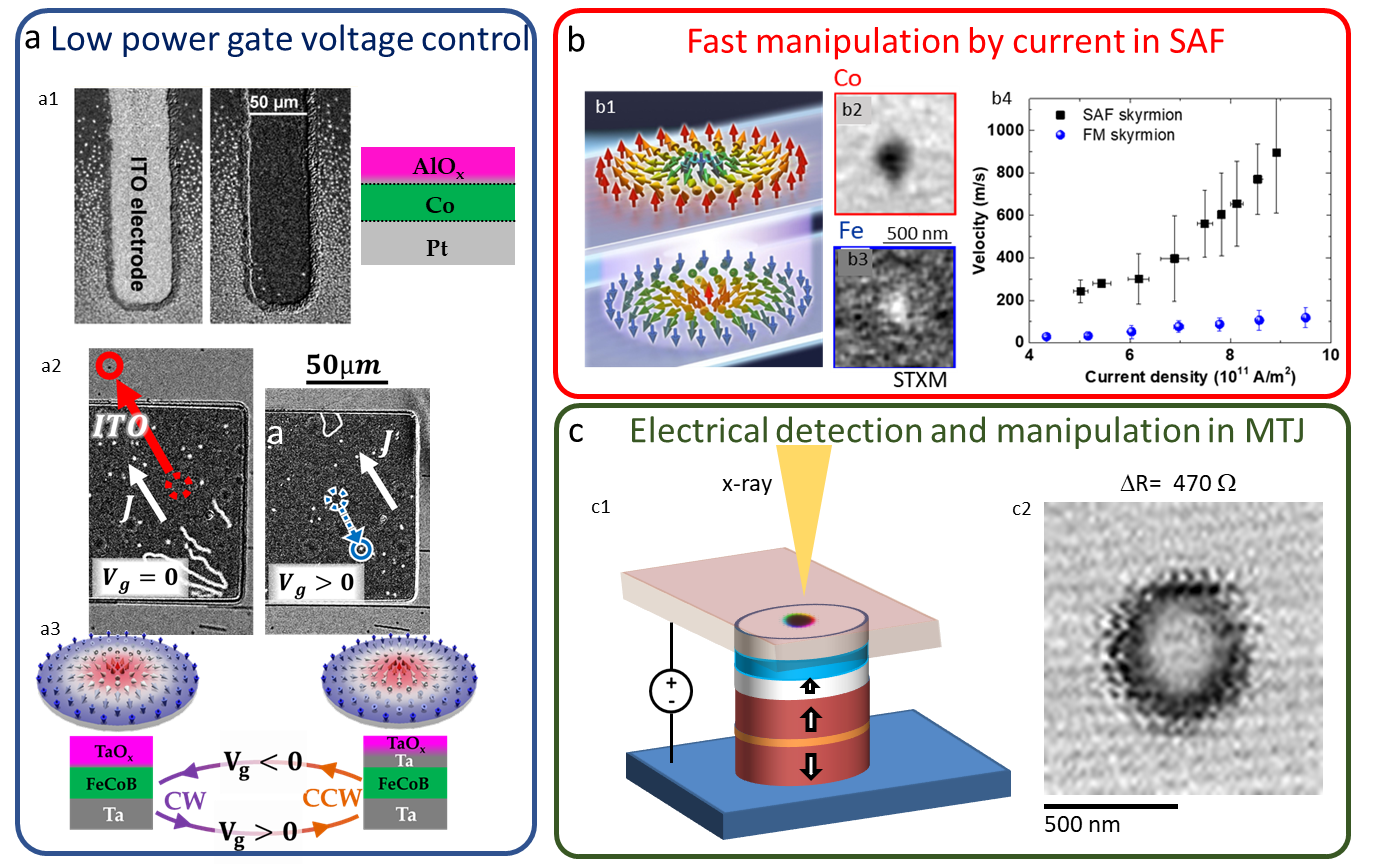}
    \caption{(a) a1 Nucleation and annihilation of skyrmions in Pt/Co/AlOx samples using a gate voltage applied through a transparent Indium tin oxide (ITO) electrode, measured by p-MOKE (Adapted with permission from ~\cite{schott2017skyrmion}. Copyright 2017 American Chemical Society). a2-a3 Inversion of skyrmion chirality by a gate voltage under ITO electrode, measured by p-MOKE, in Ta/FeCoB/TaOx stacks (adapted from ~\cite{fillion2022gate}). Oxygen migration due to the gate explains the inversion of chirality. (b). b1 - Schematic representation of the spin texture of a magnetic skyrmion in a synthetic antiferromagnet Credit B. Bourgeois – O. Boulle. b2-b3. Scanning transmission x-ray microscopy (STXM) image of a skyrmion in a Pt/Co/NiFe/Ru/Pt/Co/Ru ultrathin film. b2 is acquired at the Co L3 edge, and b3 at the Fe L3 edge. from Ref.~\cite{juge2022skyrmions} Copyright Springer Nature. b4 Skyrmion velocity vs current density measured in a Pt/Co/Ru/Pt/Co/Ru thin films (black square dots) and in Pt/Co/Ru thin films adapted from Ref ~\cite{pham2024fast}. (c) c1 Schematic illustration of a MTJ hosting a skyrmion in the Free layer. The skyrmion is observed via STXM while measuring the MTJ resistance. c2 XMCD-STXM image of a skyrmion in a 500 nm MTJ dot (Fe L3 edge). The image was acquired after the application of a 10 ns voltage pulse on the MTJ from a uniform magnetic state (parallel resistance state). This led to the skyrmion nucleation and to an increase of the MTJ resistance of 470 $\Omega$. Adapted with permission from Ref.~\cite{urrestarazu2024electrical}. Copyright American Chemical Society.}
    \label{fig:Boulle1}
\end{figure*}

\section*{Low Energy Manipulation of Skyrmion Stability, Chirality and Dynamics via Gate Voltage}

The nucleation of skyrmions in thin films with perpendicular magnetic anisotropy is done by overcoming an energy barrier arising from their topology thanks to thermal fluctuations. Nucleation using various external stimuli such as magnetic field, local current injection, strain, laser illumination~\cite{fert2017magnetic, zhang2020skyrmion, vakili2021skyrmionics, je2018creation, juge2022skyrmions} was reported. Skyrmions can also be nucleated and manipulated individually using gate voltage with low power~\cite{schott2017skyrmion} (see Fig.~\ref{fig:Boulle1}a), which is of particular interest for logics application.

The nucleation and annihilation energy barriers are strongly dependent on magnetic parameters, such as the magnetic anisotropy, the Dzyaloshinskii-Moriya interaction (DMI) and the magnetic field. The use of gate voltages to tune either magnetic anisotropy, via voltage controlled magnetic anisotropy (VCMA), or DMI by more than 100\% (VCDMI)~\cite{srivastava2018large} is a dynamic, low power, reversible and local method that allows the control of the skyrmion stability~\cite{schott2017skyrmion, srivastava2018large, ameziane2023solid},. The control of DMI amplitude by a gate voltage also allowed the demonstration of the cancellation of the undesired transverse skyrmion Hall angle ~\cite{dai2023electric}.

Additionally, in a system with weak DMI (Ta/FeCoB/TaOx trilayers), a gate voltage of only few volts allows the chirality of skyrmions and domain walls to be reversed~\cite{fillion2022gate}, thanks to an inversion of the DMI coefficient sign via oxygen ion migration (see Fig.~\ref{fig:Boulle1}a). The inversion of a Néel skyrmion chirality is allowed by topology, if the transient Bloch skyrmion does not annihilate. Hence, fine optimizations of VCMA and VCDMI are necessary to reach such skyrmion chirality inversion.

This dynamic and local manipulation of interfacial magnetic properties that are PMA and DMI establishes an additional degree of control to engineer programmable skyrmion-based memory or logic devices and lead for instance to the demonstration of skyrmion transistor based on the modulation of magnetic anisotropy via VCMA~\cite{yang2023magnetic}.

\section*{Fast Manipulation of Antiferromagnetic Skyrmions with Currents}

While magnetic skyrmions in ultrathin films are particularly appealing for applications, in particular due to their industry compatible sputtering deposition and smaller operating current, they suffer from several limitations. Firstly, they are stabilized by the stray field energy, which limits their minimal size to typically several tens of nanometers. Secondly, their dynamics in tracks are perturbed by the skyrmion Hall effect, a motion transverse to the current. This effect, resulting from a gyrotropic force, is a consequence of its topology, and is an important issue for devices as it can lead to the skyrmion annihilation on the track edge. The skyrmion Hall effect also slows down the skyrmion motion since part of the longitudinal driving force from the current is transferred to a transverse motion.

These issues can be solved by considering antiferromagnetic (AF) skyrmions. They are composed of skyrmions in different sub-lattices, which are antiferromagnetically coupled (see Fig.~\ref{fig:Boulle1}b). The AF coupled skyrmions experience opposite gyrotropic forces, which overall compensate, leading to a cancellation of the skyrmion Hall effect. A convenient class of material to stabilize antiferromagnetic skyrmions are synthetic antiferromagnets (SAF), composed of ultrathin ferromagnetic layers which are AF coupled. Using x-ray microscopy, we have demonstrated that AF skyrmions can be stabilized at room temperature and zero magnetic field in SAFs composed of (Pt/Co/Ru/Pt/Co/NiFe) stacks~\cite{juge2022skyrmions}.
Skyrmions in Pt/Co/Ru based SAF can be moved up to 900 m/s along the current direction~\cite{pham2024fast} (Fig.~\ref{fig:Boulle1}b) , i.e around 9 times faster than in single Pt/Co/Ru layer, where significant Hall deviation is observed ($\approx 60°$). Analytical models and micromagnetic simulations show that this large enhancement of the velocity is explained by the cancellation of the skyrmion Hall effect in the SAF. Large current induced velocity with reduced skyrmion Hall effect was also reported in rare earth transition metal ferrimagnetic stacks, such as Pt/CoGd~\cite{quessab2022zero}. Faster motion, up to 10 km/s, is expected in true antiferromagnetic materials~\cite{tremsina2022atomistic}, where topological spin textures were reported recently~\cite{jani2021antiferromagnetic}. Non-colinear antiferromagnets~\cite{higo2022thin} as well as altermagnets~\cite{song2025altermagnets} and 2D antiferromagnets seems particularly promising, owing to their spin polarized transport and the possibility to detect spin textures via AHE, tunnel magnetoresistance (TMR)~\cite{chen2023octupole, qin2023room} and magnetic imaging (MOKE and/or XMCD). This opens a path for fast ($\approx$ 100 fs) and low energy ($\approx$ 0.1 fJ) memory elements based on the full electrical AF skyrmion manipulation and detection (see Fig.~\ref{fig:Boulle2}b).

\section*{Electrical Nucleation and Detection of Skyrmions in Magnetic Tunnel Junctions}

An important challenges for skyrmion based applications is also the electrical nucleation, manipulation and  detection of  skyrmions in magnetic tunnel junctions (MTJ). This was recently demonstrated in our group using \textit{operando} scanning transmission x-ray microscopy (STXM). To favor the skyrmion nucleation, the FeCoB free layer was optimized to be close to the in-plane/out-of-plane reorientation transition. A skyrmion was nucleated in the MTJ FeCoB free layer using a short (10 ns) voltage pulse, as observed via STXM (see Fig.~\ref{fig:Boulle1}c). This led to an increase of the junction resistance of about 470 $\Omega$. The electrical nucleation and manipulation can be explained by the VCMA effect. Similar results were recently reported by several other groups~\cite{chen2024allelectrical, li2022experimental, guang2023electrical}. This opens promising outlooks for 2 terminal memory elements using additional electrically tunable skyrmionic states to encode the information, excite/detect microwave dynamics or achieve pattern recognition. This includes skyrmion based multistate memory, skyrmionic memristors where skyrmions are excited with low power via VCMA (see Fig.~\ref{fig:Boulle2}a), skyrmionic oscillators or microwave detector, or skyrmionic based reservoir computing.

\begin{figure*}[h!]
    \centering
    \includegraphics[width=0.95\linewidth]{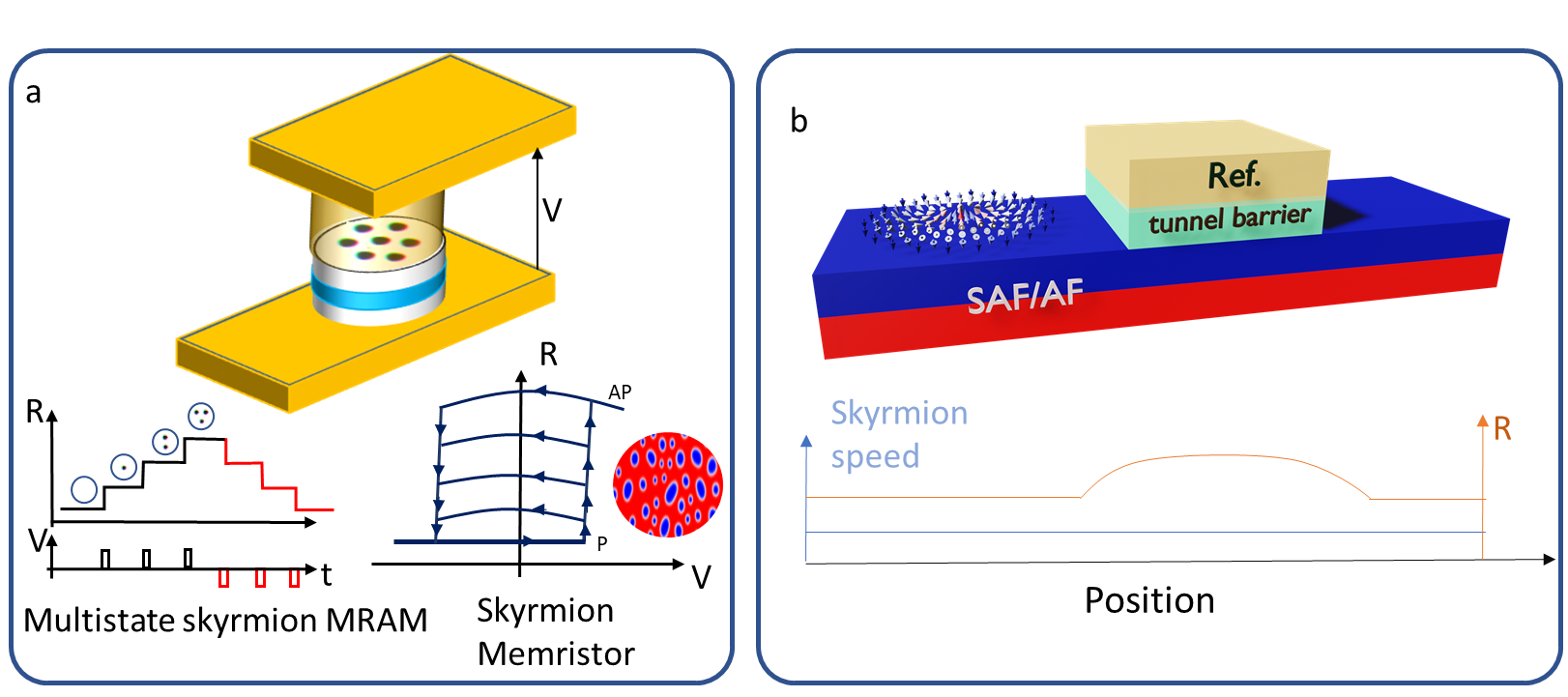}
    \caption{\textbf{ Skyrmion device outlook.} (a) Skyrmions in a two terminal MTJ for low power multistate skyrmion MRAM or skyrmion memristor. Skyrmions can be nucleated electrically with low power using VCMA. (b) Memory elements based on the fast manipulation of AF skyrmions in tracks. The AF skyrmion motion is detected electrically using an AF MTJ.}
    \label{fig:Boulle2}
\end{figure*}

\section*{Challenges}

Several key challenges still need to be addressed for skyrmion based technologies. Regarding racetrack application, the skyrmion maximal velocity in AF is limited by the maximal spin wave velocity in the material as well as skyrmion deformation~\cite{tremsina2022atomistic}. In addition, the skyrmion velocity is expected to be proportional to its size, such that nanometer size skyrmion would move slower. Another major issue is that skyrmions are prone to pin on material disorder, in particular in sputtered ultrathin thin films, which limits the reliability of the device. Proper engineering of novel AF materials is needed to overcome these issues.

The demonstration of three terminal devices  where skyrmions in a AF track are detected electrically using an MTJ is also an important milestone  for fast memory applications. This would require challenging AF track and MTJ material engineering  as well as device design to achieve seamless motion of the skyrmions in the track with high TMR detection.

While many logic concepts were proposed, with promising efficient logic-in-memory computing schemes~\cite{sisodia2022robust, sisodia2022programmable}, an experimental demonstration of such concepts is still lacking. This requires challenging control of the skyrmion position, synchronization and interactions. This could be addressed via local modification of the material properties using ion irradiation~\cite{juge2021helium}, heat~\cite{albisetti2016nanopatterning} or local voltage control ~\cite{fillion2022gate}. Another path is to consider low energy non-conventional computing schemes which are error and speed tolerant and where the interaction with disorder is exploited to generate novel functionalities. This would include integrated neuromorphic computing devices, such as skyrmion synapses or neurons ~\cite{dacamarasantaclaragomes2025neuromorphic}, or reservoir computing schemes, where the non-linear interactions can be exploited for pattern recognition.

\section*{Acknowledgements}
This work was supported by the French ANR via Contracts No. ELECSPIN ANR-16-CE24-0018, ADMIS ANR-, the DARPA TEE program through Grant MIPR No. HR0011831554

\endgroup

\newpage

\section{Efficient skyrmion generation in magnetic devices}
\begingroup
    \let\section\subsection
    \let\subsection\subsubsection
    \let\subsubsection\paragraph
    \let\paragraph\subparagraph
Ale\v{s} Hrabec$^{1,2}$, and Stanislas Rohart$^3$
\vspace{0.5cm}

\noindent
$^1$ Laboratory for Mesoscopic Systems, Department of Materials, ETH Zurich, 8093 Zurich, Switzerland\\
$^2$ Paul Scherrer Institut, 5232 Villigen PSI, Switzerland \\
$^3$ Université Paris-Saclay, CNRS, Laboratoire de Physique des Solides, 91405 Orsay, France\\

\section*{Introduction}
The controlled nucleation of skyrmions—meaning their reliable creation at a precise location and time—is a key challenge in the development of skyrmion-based technologies. This applies both to racetrack-like devices, where itinerant skyrmions must be injected at one location and moved along a track, as well as to quasistatic architectures such as multistate memories, where information is stored as the number of skyrmions present in a confined area. In both cases, the ability to generate individual skyrmions with low energy and high reproducibility is essential for writing data and enabling functional operation. From a physical point of view, skyrmion nucleation is a nontrivial process since it requires a topological transition. This can happen smoothly at the track edge, while nucleation within the film requires the formation of a transitory magnetic singularity. This adds complexity to the process and impacts not only the energy but also the time required for skyrmion nucleation.

\section*{Relevance and Vision}
Early demonstrations of skyrmion nucleation relied on the application of out-of-plane magnetic fields. This approach provided a means to generate skyrmions for their further investigation, but is neither local nor energy-efficient, making it unsuitable for device integration. The research community has since explored three main avenues: current-, electric field-, and laser-induced nucleation. Each comes with its own physical mechanisms, benefits and limitations.

In current-based nucleation experiments [Fig.~\ref{fig:Hrabec}(a)], skyrmions have been successfully generated using local current injection through point contacts~\cite{hrabec2017current, finizio2019deterministic, juge2022skyrmions}, artificial defects produced by helium irradiation ~\cite{kern2022deterministic}, or lithographically defined notches ~\cite{buttner2017fieldfree, dacamarasantaclaragomes2025neuromorphic}. These techniques allow spatial localization and reproducible creation of skyrmions, but at a large energy expense (about 100 pJ in the best cases~\cite{hrabec2017current, buttner2017fieldfree}). However, the underlying mechanisms remain difficult to separate experimentally. Two effects are typically cited: Joule heating and spin torques (either spin-transfer or spin-orbit torques). In geometries where the current is concentrated, the thermal contribution may dominate, facilitating nucleation due to magnetic fluctuations. While this helps the nucleation process, it also brings some stochasticity. For practical devices, improving repeatability requires an athermal process such as spin transfer torque. A promising method to favor this mechanism is to tilt the magnetization away from the easy axis (e.g. using an external magnetic field), which enhances the efficiency of the spin-orbit torque so that it can overcome the thermal contribution~\cite{akhtar2019currentinduced}.

In contrast, electric field-induced nucleation [Fig.~\ref{fig:Hrabec}(b)] modifies the magnetic energy landscape via voltage control of magnetic anisotropy (VCMA), and does so without any net charge current. This approach drastically reduces energy consumption and is naturally compatible with magnetic tunnel junctions (MTJ). The formation or annihilation of skyrmions in ultrathin films by applying a gate voltage has been demonstrated in Ref.~\cite{schott2017skyrmion}, and, more recently, the nucleation and detection of single skyrmions was demonstrated by combining VCMA control with tunnel magnetoresistance readout in an MTJ device~\cite{urrestarazu2024electrical, chen2024allelectrical}. While early implementations required about 30 pJ per nucleation event, further optimization has led to sub-100 fJ write energy~\cite{chen2024allelectrical}. This is a significant milestone that positions VCMA as the most promising low-energy solution. Moreover, scaling arguments~\cite{chen2024allelectrical} suggest that this energy could be reduced even further by increasing the resistance-area product and decreasing device dimensions, giving electric field control a strong advantage over current-based schemes.

Ultrafast optical pulses provide another route for skyrmion generation [Fig.~\ref{fig:Hrabec}(c)]. By rapidly heating and quenching a magnetic system, laser excitation can drive topological transitions at picosecond timescales~\cite{juge2022skyrmions, finazzi2013laserinduced, buttner2021observation}. However, integrating this approach into scalable electronic devices remains difficult due to the need for incorporation ultrafast photonic elements in standard electronic architectures.

\section*{Challenges}
Among these approaches, electric field control emerges as the leading candidate for integration into nonvolatile memory technologies. The recent realizations of low-energy nucleation in realistic MTJ devices mark a substantial step forward. However, one important limitation remains: so far, VCMA-based nucleation has only been shown to give static configurations, where the skyrmion is created and remains in place. For itinerant skyrmions, the injection into a track and subsequent motion along it is yet to be demonstrated, ideally relying on the same electric field-based principle. This integration is not straightforward and will require careful design of three-terminal geometries, where a gate electrode can locally enable the nucleation in a well-defined region, from which the skyrmion can then be moved along the track using conventional current-induced motion.

Another important aspect is the time taken for nucleation. Simulations, particularly those presented in~\cite{finizio2019deterministic, buttner2017fieldfree, urrestarazu2024electrical}, show that the nucleation process, whether induced by current or electric field, generally proceeds in three stages: initial formation of a reversed bubble, transition into a topological skyrmion, and final relaxation. Each stage introduces a delay, making the process inherently slow. Experimentally, the few studies with time resolution confirm that nucleation takes several nanoseconds~\cite{finizio2019deterministic}. Optical pulses offer an alternative by enabling local demagnetization at ultrafast timescales, potentially allowing the skyrmion to form directly with the appropriate topology and thus reducing the overall nucleation time. Although integration is currently challenging, this approach may overcome the speed limitations of electrical methods. Generally, developing an understanding of how magnetization damping and other magnetic parameters affect nucleation dynamics could help to identify faster, yet reliable, writing strategies.

\begin{figure*}[h!]
    \centering
    \includegraphics[width=0.95\textwidth]{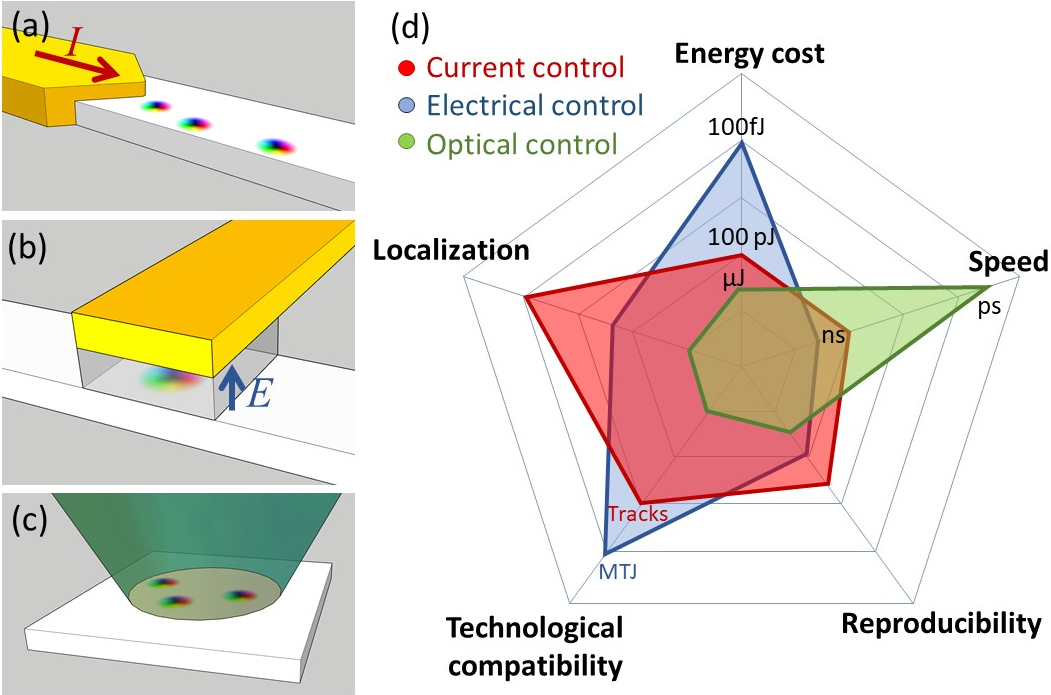}
    \caption{ Illustration of key strategies and challenges for skyrmion generation. (a) Current-induced nucleation via local current concentration near a point contact. (b) Electric-field-induced nucleation relying on voltage-controlled magnetic anisotropy. (c) Ultrafast laser-induced nucleation through thermal excitation and topological transition. (d) Qualitative comparison of the three stimuli based on five key criteria: energy cost, speed, reproducibility, technological compatibility, and spatial localization.}
    \label{fig:Hrabec}
\end{figure*}

Integration and localization of the nucleation process are equally critical. Point contacts and notches allow for spatial selectivity, but may reduce device endurance due to repeated stress and heating. In electric field-based approaches applied to itinerant skyrmion devices, the challenge is to reduce the size of the control electrodes and integrate them in compact geometries. Another promising route is the use of ion irradiation to locally tailor magnetic properties and facilitate nucleation in predefined regions. The promising preliminary realization of this concept reported in Ref.~\cite{kern2022deterministic} calls for more systematic investigation.

Moving toward more advanced materials, antiferromagnetic and ferrimagnetic systems offer potential advantages in terms of speed, reduced stray fields, and robustness to external perturbations. However, electric field-induced nucleation remains to be demonstrated in such materials, marking an important direction for future exploration. Similarly, while laser-induced nucleation of skyrmions has been partially explored in synthetic antiferromagnets~\cite{juge2022skyrmions}, more work is needed to understand how the energy and nucleation time scale in these systems. More broadly, given that dynamics is faster in antiferromagnets, it remains an open question whether nucleation could similarly benefit from these enhanced dynamics.

Combining multiple stimuli to exploit their complementary strengths, depicted in Fig.~\ref{fig:Hrabec}(d), one could provide a promising strategy to enhance the efficiency and control of skyrmion generation. For example, electric fields can be used to define locally a low-energy state, while a short current or optical pulse would supply the activation energy needed to trigger the nucleation. Such hybrid schemes are attractive because they balance energy efficiency, speed, and control. They also open the way to multifunctional devices, where the same structure can be used for memory, logic, or neuromorphic operations depending on the excitation protocol.

Beyond the stimuli themselves, a key metric for practical use is reproducibility. A nucleation process that is not deterministic will lead to data corruption or device failure. Achieving ``skyrmions-on-demand'' is therefore a major objective. So far, near-perfect reproducibility under controlled conditions has rarely been achieved (e.g. in Ref.~\cite{finizio2019deterministic}) and this remains one of the main bottlenecks for deployment.

\section*{Concluding Remarks}
Overall, the control of skyrmion nucleation has reached a reasonable level of maturity, with several experimental platforms demonstrating low-energy, localized, and detectable skyrmion generation. The challenge now lies in extending the fast, reliable and low-energy skyrmion generation to fully integrated systems. With continued efforts on material optimization, creating hybrid excitation schemes, and the development of device architecture, the field is well positioned to push skyrmion nucleation toward realistic performance for technological applications.

\section*{Acknowledgements}
This work was supported by France 2030 (plan project Chirex PEPR SPIN ANR-22- EXSP-0002).

\endgroup

\newpage

\section{Unconventional computing using magnetic skyrmions}
\begingroup
    \let\section\subsection
    \let\subsection\subsubsection
    \let\subsubsection\paragraph
    \let\paragraph\subparagraph
Hidekazu Kurebayashi $^{1,2,3,4}$, and Christian Back $^{5,6}$

\noindent
\textit{$^1$ Department of Electronic and Electrical Engineering, University College London, London, WC1E 7JE, UK\\
$^2$ London Centre for Nanotechnology, University College London, 17-19 Gordon
Street, London, WC1H 0AH, United Kingdom\\
$^3$ WPI Advanced Institute for Materials Research, Tohoku University, 2-1-1,
Katahira, Sendai 980-8577, Japan\\
$^4$ Center for Science and Innovation in Spintronics, Tohoku University, 2-1-1,
Katahira, Sendai, 980-8577 Japan\\
$^5$ Department of Physics, School of Natural Sciences, Technical University of Munich, Munich, Germany\\
$^6$ Center for Quantum Engineering (ZQE), Technical University of Munich, Munich, Germany}

\section*{Introduction}

The relentless growth of data-driven technologies has exposed critical limitations in conventional von Neumann computing architectures, particularly their energy inefficiency and scalability constraints. As Moore's Law approaches fundamental physical barriers, researchers are increasingly exploring unconventional computing paradigms that leverage novel physical phenomena. Spin-based systems, with their inherent non-volatility, time non-locality, and rich nonlinear dynamics, offer a transformative pathway by combining magnetism that provides a diverse palette of ground states and interactions and spintronics to electrically control them efficiently. Among these, magnetic skyrmions~\cite{fert2017magnetic}, topologically defined nanoscale spin textures, have emerged as a promising platform for low-power, high-density computing - for example, magnetic skyrmions with sizes as small as 1.9 nm have been reported in $\rm GdRu_2Si_2$ \cite{khanh2020nanometric} for potential high-density skyrmion data storing and processing.

In the early stages of magnetic skyrmion research, one of the main foci was their potential application in racetrack memory devices, and research was driven by their high stability and the prospect of low-power operation through spin-transfer and spin-orbit torques~\cite{fert2017magnetic}.

More recently, scientists have begun to explore further functionalities of skyrmion dynamics, such as analog-like behaviour of skyrmion creation and annihilation to mimic biological synapse potentiation and depression \cite{song2020skyrmionbased} and the thermal diffusion of skyrmion displacement for probabilistic computing \cite{zazvorka2019thermal}. Continuing on these early works of using magnetic skyrmions as an essential building block of unconventional computing, the community's long-term goal of developing commercially viable skyrmion technologies requires the exploration and understanding of their particle-like dynamics, topological stability, and efficient current-driven motion that might in the future redefine the look of information storage, logic, and neuromorphic systems at ambient temperature.

Here we showcase recent developments of integrating magnetic skyrmions as key elements of unconventional computing architectures. We focus on reservoir computing (RC) that is one of many recurrent neural networks, particularly suited for computing small-scale, low-complexity tasks with time-series data due to its low-cost and rapid training properties \cite{nakajima2021reservoir}.

\begin{figure}[h!]
\centering
\includegraphics[width=\linewidth]{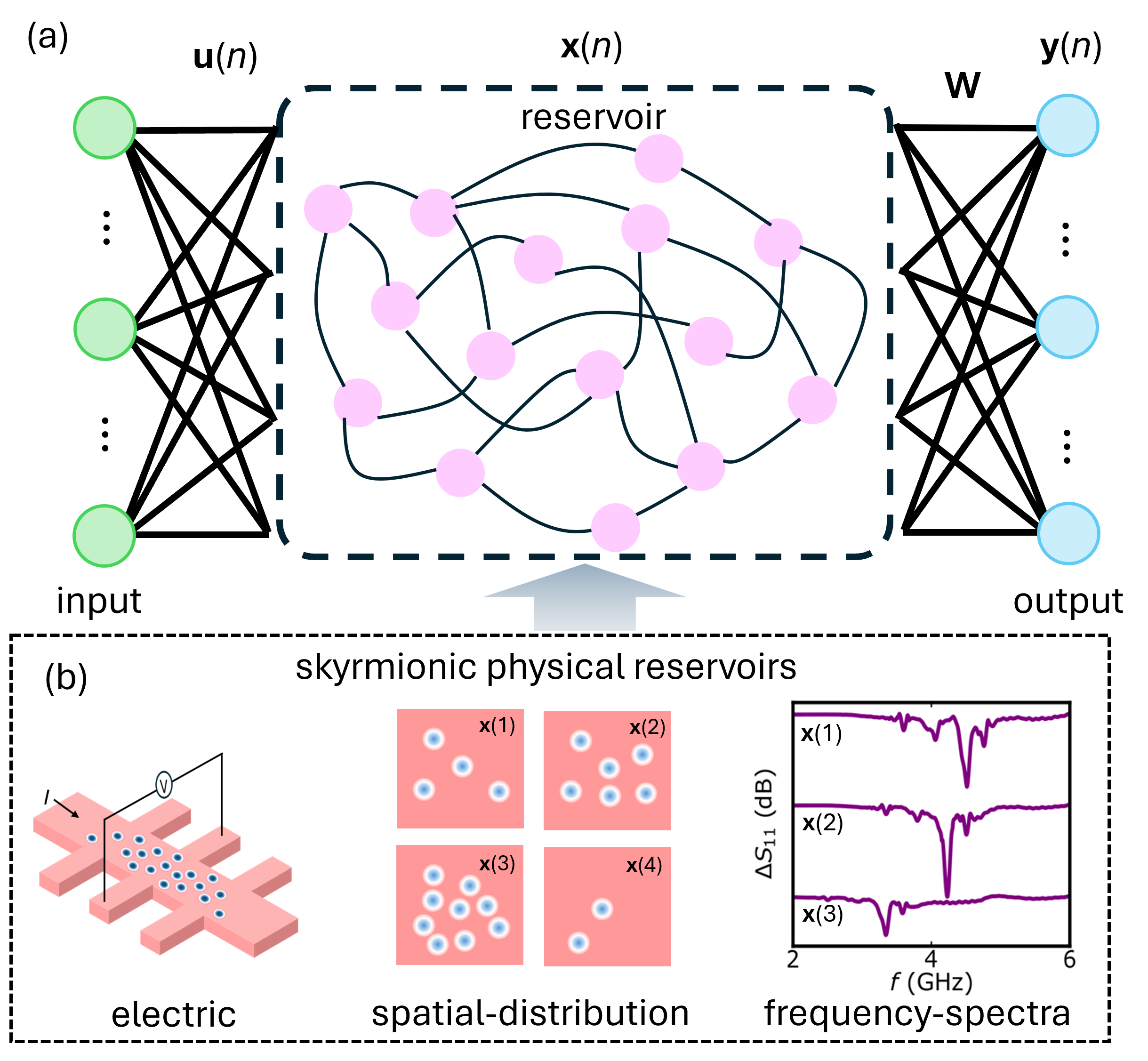}
\caption{(a) Schematic of reservoir computing. Green dots represent input data ($\mathbf{u}(n)$ with $n$ time steps) that are fed into the reservoir component with time-dependent states $\mathbf{x}(n)$  in the middle. The output layer $\mathbf{y}(n)$ is connected to the reservoir with the weight function $\mathbf{W}$. (b) A variety of responses in magnetic skyrmions can be exploited as a physical reservoir, e.g. electric voltages in a Hall bar device, spatial distribution and GHz skyrmion dynamics.}
\label{fig:Back1}
\end{figure}

\section*{Reservoir Computing Using Magnetic Skyrmions}

A basic RC architecture is illustrated in Fig.~\ref{fig:Back1}(a). Input data ($\mathbf{u}(n)$) with $n$ time-steps are fed into the ``reservoir'' component ($\mathbf{
x}(n)$) possessing nonlinear and short-term memory dynamics that is fixed during the training period of machine learning, whereas the weights ($\mathbf{W}$) of the reservoir's output layer ($\mathbf{y}(n)$) are only trained with the targeted data set by relatively simple linear and/or ridge regression. The operation scheme of fixing the reservoir dynamics during the machine learning processes is particularly suitable for the use of dynamical responses from physical systems in the framework of neuromorphic hardware, so-called physical reservoir computing, because unlike many other neuromorphic architecture candidates, we do not have to physically define individual neurons as well as their synaptic connections. Several physical systems, including spintronic, optical and memoristive devices, display inherently rich and distinct nonlinearity and short-term memory capacity and have been exploited as the reservoir element \cite{nakajima2021reservoir}. A key requirement of developing a physical reservoir is high-dimensional mapping, namely, a scheme to generate multi-dimensional responses by a single input excitation. This is particularly effective to process the input data within the high dimensional feature space to enhance the computational capabilities.

\begin{figure*}[h!]
\centering
\includegraphics[width=\textwidth]{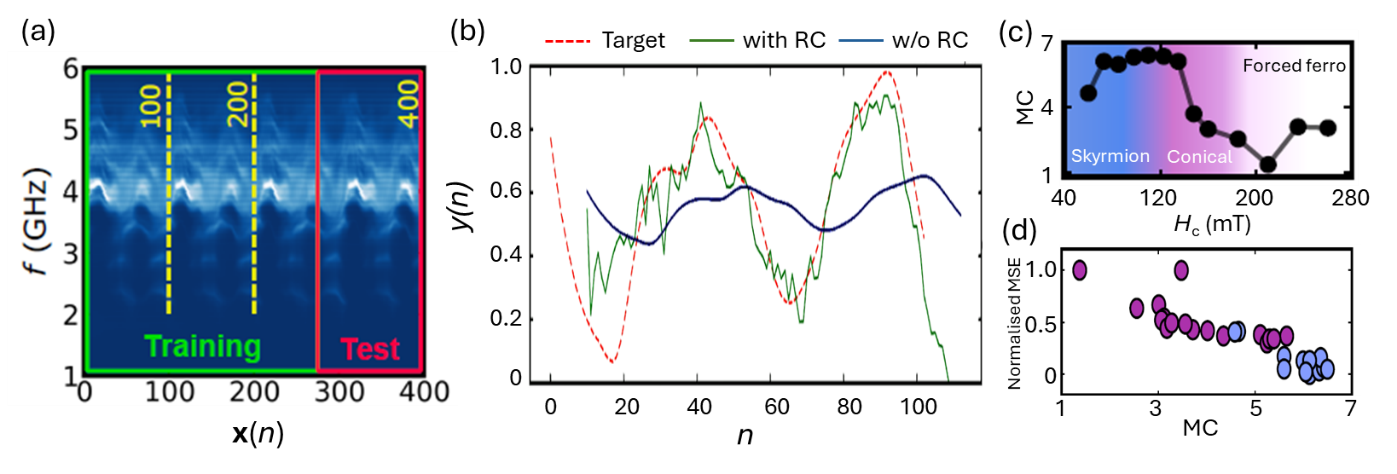}
\caption{(a) Spinwave spectra of magnetic skyrmions in {$\rm Cu_2OSeO_3$}. 70\% of the data were used for training and the rest of 30\% were for testing. (b) An example of applying the skyrmion reservoir for MG time-series prediction tasks. The red dotted curve is the target function and the green(blue) solid curve us the predicted values for ten step ahead with(without) the skyrmion reservoir. (c) Memory capacity (MC) metric varies by magnetic field that can control magnetic phases. (d) The correlation between normalised mean squared error (MSE) and MC.}
\label{fig:Back2}
\end{figure*}

Among many possible physical systems to be used for the reservoir, magnetic skyrmions have been intensively studied (see e.g. a recent review article \cite{lee2023perspective}), leveraging skyrmions’ inherent nonlinearity and distinct memory capacities by using the skyrmion population and skyrmion dynamics (Fig.~\ref{fig:Back1}(b)). Yokouchi et al. utilised the nonlinear and memory-dependent dynamics of magnetic skyrmions in response to quasi-static magnetic fields for physical reservoir computing \cite{yokouchi2022pattern}. The waveform of periodic magnetic field excitations, such as square and sine forms, can modulate the skyrmion population profile detected by electric voltages via the anomalous Hall effect (AHE) in magnetic multilayers. In order to demonstrate handwritten digit recognition tasks, they map two-dimensional pixel greyscales onto a one-dimensional magnetic field oscillation strength, together with high-dimensional mapping of the reservoir by multiplexing with varied dc applied fields. They successfully achieved an accuracy of $94.7 \pm 0.3\%$ in the handwritten digit recognition tasks.

Sun et al. demonstrated a similar skyrmion reservoir with electric voltages as its input and the same AHE readout as an output~\cite{sun2023experimental}. Strain applied via a piezoelectric substrate can efficiently control the skyrmion density of the magnetic multilayers grown on top, displaying nonlinear voltage dependence of the anomalous Hall resistance suitable for reservoir computing. In their waveform recognition demonstration, high dimensional mapping has been achieved by taking multiple AHE voltage outputs within one periodic excitation. The skyrmion phase has consistently shown better waveform recognition than the magnetisation-saturation phase, indicating the skyrmion's rich responses and memory properties over other phase. Good performance for a Mackey-Glass (MG) chaotic time-series prediction task has been demonstrated using the strain-mediated skyrmion reservoir.

Lee et al. exploited rich magnetic phases of skyrmion hosting chiral magnets (e.g. the chiral magnet $\rm Cu_2OSeO_3$) as physical reservoirs that can reconfigure the reservoir properties via phase changes and optimise the performance across the diverse set of computing tasks~\cite{lee2024taskadaptive}. This task-adaptive approach is not readily available in physical reservoir computing due to the relatively narrow response properties of physical devices once fabricated. This is a drawback, compared to software-based machine learning where model performance can be tailored and optimised via hyperparameter tuning. Lee et al. exploited magnetic field and temperature as a phase control, and used spin-wave spectra as a high dimensional mapping technique where one magnetic-field state can expand its dimension into the frequency spectral space. They mapped input data onto the magnetic field profile and performed reservoir computing tasks using spin-wave spectra of magnetic skyrmions as shown in Fig.~\ref{fig:Back2}(a). Using the skyrmion's strong memory properties, the skyrmion reservoir perform strongly on future prediction tasks benchmarked by chaotic MG time series (Fig.~\ref{fig:Back2}(b)). The memory capacity (MC) metric of skyrmion reservoirs has been quantified and is found to excel other magnetic phases, all controlled by magnetic field ($H_c$). The correlation between the computational performance rated by normalised mean squared error (MSE) and MC has been revealed by Fig.~\ref{fig:Back2}(d) where the negative correlation has been identified. Apart from this, the signal transformation requires strong nonlinearity which is best performed in the conical phase. Hence, by choosing the magnetic phases, it is possible to tune the reservoir properties on demand.

Raab et al. \cite{raab2022brownian} proposed the use of the thermally activated diffusive motion of skyrmions as a computational resource. The magnetic-field-biasing on skyrmion positions in confined geometries displays nonlinearity and memory effect monitored by magneto-optical Kerr microscopy for logic operations. The method was further adopted for more real-world problems, i.e. hand gesture recognition recorded by a Range-Doppler radar \cite{beneke2024gesture}.

\section*{Challenges And Outlook}

One of the most promising applications for neuromorphic computing hardware, including magnetic skyrmions, is edge computing where small-sized computing problems are solved on demand at the timescale matching to data generation. To this goals, we highlight two main challenges to be tackled and addressed, in order to develop viable neuromorphic hardware relevant to real applications. The first is to integrate materials hosting magnetic skryrmions at temperatures well above room temperature on industry-ready substrates, AND to electrically excite and measure the dynamical response for computational operation. Another challenge is to identify real-life problems where skyrmion applications stand out against any rivalling physical systems. The timescale of short-term memories should be tuned with that of a real-life problem to maximise the computational performance, as demonstrated by Beneke et al. \cite{beneke2024gesture}. We envisage that further demonstrations of challenging real-like problems are beneficial to shed light on a potential role of skyrmion reservoir computing for the landscape of unconventional computing.
\endgroup

\newpage

\section{Neuromorphic hardware solutions with multiple non-collinear magnetic textures}
\begingroup
    \let\section\subsection
    \let\subsection\subsubsection
    \let\subsubsection\paragraph
    \let\paragraph\subparagraph
Nicolas Reyren$^1$, and Vincent Cros$^1$
\vspace{0.5cm}

\noindent
\textit{$^1$ Laboratoire Albert Fert, CNRS, Thales, Université Paris-Saclay, 91767 Palaiseau, France}

\section*{Status}

Encoding information into different type of magnetic textures or their combination could bring an original advantage to foreseen neuromorphic spintronics.

In experiments, only few non-trivial topological textures with particle-like properties were observed, among which skyrmions or skyrmion bubbles received the strongest attention these last twenty years~\cite{back20202020}. In fact, currently only a few examples of systems hosting different particle-like textures are demonstrated. 

A first family of such structures are built around modulation of the skyrmion tubes. For instance, the tubes can be arranged in ``skyrmion bundles'' with a tuneable topological number~\cite{tang2021magnetic}. A skyrmion tube can be inserted into another skyrmion tube of larger diameter and opposite polarity, resulting in the formation of a so-called ``target skyrmion'' which is topologically trivial~\cite{zheng2017direct}. This process can be iterated, producing n$\pi$-skyrmions. Theoretically, soliton solutions with a higher number of internal ``oscillations'' have also been predicted~\cite{borisov2002vortices}. Additionally, skyrmion tubes can be ``zipped'' through the film thickness to form together Y-shaped structures, displaying one skyrmion at one interface, and two at the other~\cite{seki2022direct}. In principle, this process can also be iterated at will, providing the material is sufficiently thick. 

Instead of multiplying skyrmion tubes, they can be truncated, either from one surface or from both, leading to the formation of bobbers (with a topological charge of 1/2)~\cite{rybakov2015new, zheng2018experimental} or dipole strings also called torons (with zero topological charge)~\cite{leonov2021surface, seki2022direct}, respectively. These textures host one and two Bloch points at their extremity. In systems lacking of ``magnetic continuity'' along the thickness, such as metallic multilayers, analogous textures called ``skyrmionic cocoons''~\cite{grelier2022threedimensional} can be stabilized. In this case, Bloch points are absent~\cite{grelier2022threedimensional}, and a single topological number cannot be used to define the topology of the set of 2D planes constituting the multilayers.

Finally another type of particle-like textures, that have recently attracted considerable interest is the hopfion~\cite{rybakov2022magnetic}, which has likely been observed experimentally.

\begin{figure*}[h!]
    \centering
    \includegraphics[width=0.95\textwidth]{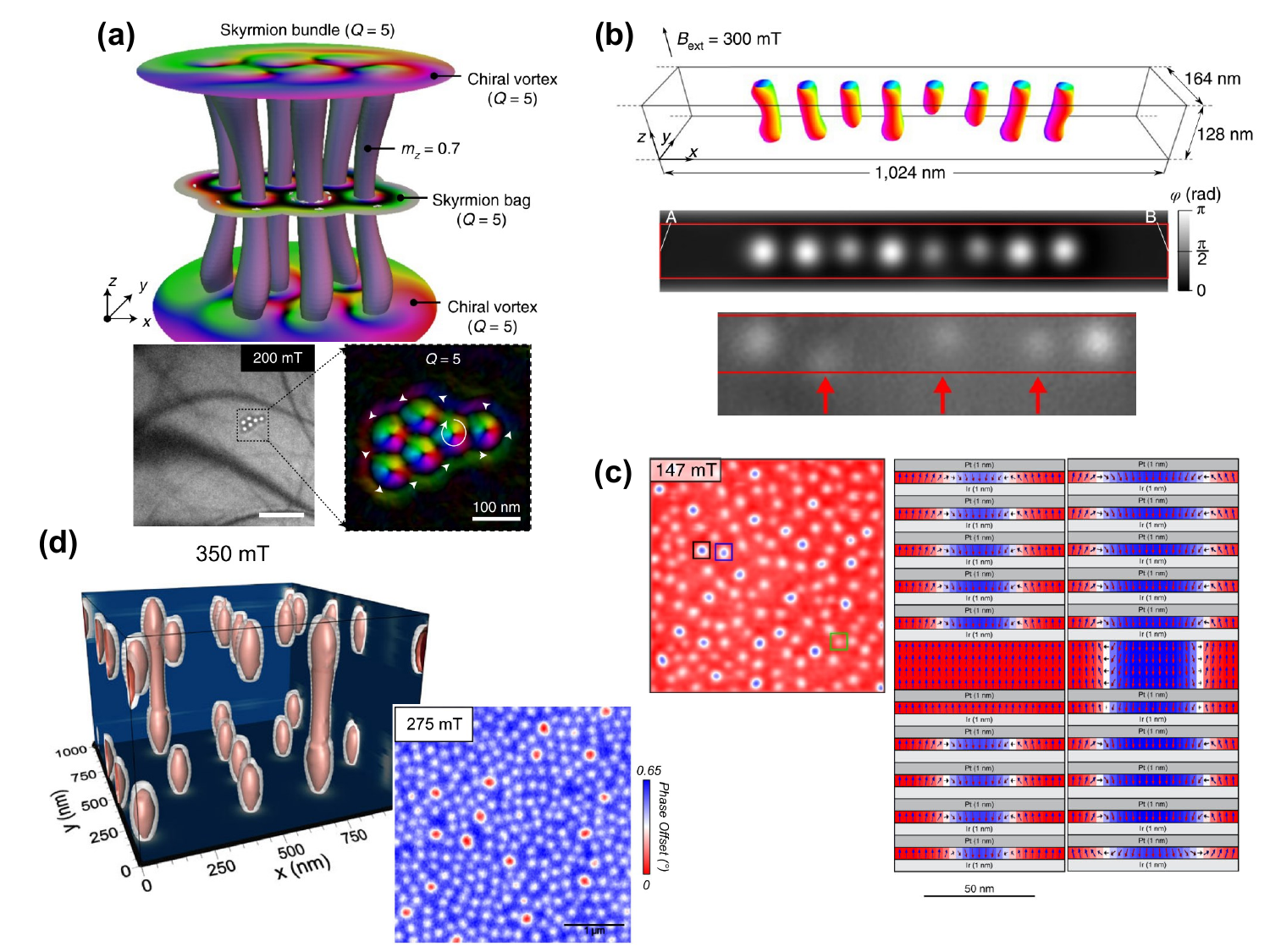}
    \caption{Collection of three dimensional magnetic textures in bulk and multilayer systems. (a) A skyrmion bundle with a topological charge $Q = 5$ as it could be observed in FeGe plates using Lorentz-TEM. \cite{tang2021magnetic} (b) From top to bottom: simulation of the magnetic configuration of skyrmion tubes and chiral bobbers in FeGe, simulation of the phase shift that would be obtained with such textures in an electron beam image; actual magnetic phase shit image in an FeGe wedge-shaped stripe (arrows indicate bobbers). \cite{zheng2018experimental} (c) Magnetic force microscopy of two distinct skyrmion phases (incomplete and tubular) in magnetic multilayers and the corresponding simulated configuration. \cite{mandru2020coexistence} (d) Magnetic force microscopy of coexisting skyrmion tubes and skyrmionic cocoons in magnetic multilayers and the corresponding simulated configuration. \cite{grelier2022threedimensional}.}
    \label{fig:Cros}
\end{figure*}

What is the interest of combining different types of textures? Combining different magnetic topological textures---such as skyrmions, skyrmioniums, and cocoons---enables richer neuromorphic functionalities by leveraging their distinct dynamics, stability, and interactions. One key advantage is that, under certain conditions, different information channels can be transported within a single physical track, allowing for signal multiplexing, parallel computation, and significant circuit footprint reduction~\cite{chen2020skyrmionic}. Employing different types of magnetic quasiparticles could enable the hardware to dynamically adapt to various tasks, mirroring the brain's ability to reallocate resources. Another advantage is that their diverse response times can be leveraged for time-based processing and learning, paving the way for more efficient, brain-inspired computing.

However, to date, only a limited number of studies have demonstrated the coexistence of multiple particle-like textures in extended films or tracks~\cite{zheng2018experimental, tang2021magnetic, grelier2022threedimensional, grelier2023xray}.

\section*{Current and Future Challenges}

Implementing neuromorphic hardware based on magnetic topological textures faces several important challenges. Many textures require external magnetic fields for stabilization---often different for each type---so ensuring their stability without external fields, or within practical ranges, is essential. The effects of confinement in nanostructures are not yet fully understood and could limit the applicability of certain textures. Electrically nucleating diverse textures within a single material system remains difficult, especially when aiming for low energy consumption. Ensuring reliable operation also requires addressing the influence of local material inhomogeneities and pinning centers, which can interfere with deterministic behaviour. Another challenge is reducing the energy needed to control the motion of different textures. On the detection side, improving the electrical readout signal and enabling clear discrimination between textures are key for practical applications. It is also important to enable independent voltage control over the properties of each texture type, and to ensure that all operations occur at voltages compatible with standard CMOS electronics. Finally, integration into chip fabrication processes, particularly for complex multilayer stacks or epitaxial systems, remains a major hurdle for scalable implementation.

\section*{Advances in Science and Technology to Meet These Thallenges}

Several scientific and technological advances are needed to tackle these challenges. Materials science will be crucial, including the development of novel magnetic multilayers, such as emerging 2D materials/heterostructures that offer robust, room-temperature stabilization of textures with minimal pinning effects. This also involves interfacial engineering, tuning the Dzyaloshinskii–Moriya interaction (DMI), voltage-controlled anisotropy, to further optimize these materials for field-free operation. Nanofabrication advances such as 3D patterning/nanomagnetism~\cite{gubbiotti20252025} and localized strain will improve control over confinement effects and/or guiding of particles. On the control side, spin–orbit torque and voltage-controlled magnetic anisotropy breakthroughs are crucial for low-energy, deterministic nucleation and manipulation of multiple textures. Energy efficiency could also drastically improve by using reconfigurable devices that recycle textures instead of destroying them, inspired by biological systems. Parallel progress in algorithms and software will be essential for modelling quasiparticle interactions, optimizing architectures, and developing self-learning rules.

\section*{Concluding Remarks}

Combining multiple magnetic textures holds great promise for energy-efficient, adaptable neuromorphic computing. While skyrmions have been widely studied, new textures offer richer functionality but face challenges in stability, control, and integration. Advances in materials, fabrication, control methods, and algorithms are key to unlocking scalable, brain-inspired hardware solutions.

\section*{Acknowledgments}

Authors acknowledge Julie Grollier for inspiring discussions. Financial support from a France 2030 government grant managed by the French national research agency (ANR) PEPR SPIN CHIREX (ANR-22-EXSP-0002), SPINMAT (ANR-22-EXSP-0007), by the European Research Council advanced grant GrenaDyn (n° 101020684), and by the EU project SkyANN (n° 101135729) is acknowledged.

\endgroup

\newpage

\section{Microwave Technology with Skyrmions}
\begingroup
    \let\section\subsection
    \let\subsection\subsubsection
    \let\subsubsection\paragraph
    \let\paragraph\subparagraph
Riccardo Tomasello$^1$, and Giovanni Finocchio$^2$
\vspace{0.5cm}

\noindent
\textit{$^1$ Department of Electrical and Information Engineering,
Politecnico di Bari, 70125 Bari, Italy\\
$^2$ Department of Mathematical and Computer Sciences, Physical Sciences and Earth Sciences, University of Messina, Messina 98166,Italy}

\section*{Status}
Microwave technology plays a crucial role in modern society for global connectivity, monitoring, as well as medical diagnostics. It also provides critical components for applications in medicine, security, and industrial processing.

Spintronics offers a promising route to miniaturize microwave devices by exploiting the spin degree of freedom of electrons, enabling compact and energy-efficient designs of both oscillators~\cite{zeng2013spin}, and detectors~\cite{finocchio2021perspectives}. These two device types represent complementary functionalities. Spintronic oscillators convert a dc current into a high-frequency magnetization precession when the spin-transfer torque (STT) balances the damping losses, achieving steady-state dynamics from MHz up to tens of GHz in ferromagnets. Conversely, spintronic detectors rely on the ferromagnetic resonant response which generates a rectified dc voltage via the spin-torque diode (STD) effect. The device at the core of this microwave technology is magnetic tunnel junction (MTJ).

Spintronic microwave technology can additionally benefit from the properties of magnetic skyrmions, which can enable highly tunable and low-power consumption devices by exploiting the excitation of their resonant modes (gyration, breathing). STT skyrmion-based oscillators and detectors have been predicted in both MTJs supplied with a localized perpendicularly polarized spin current \cite{zhou2015dynamically, carpentieri2015topological, zhang2015current, finocchio2015skyrmion}, and spin-hall nano-oscillators (Fig.~\ref{fig:Tomasello1}(a))~\cite{giordano2016spinhall}. In ~\cite{zhou2015dynamically, carpentieri2015topological}, the dynamically-stabilized skyrmions, named topological dynamical solitons, can exhibit magnetization dynamics in the GHz range (Fig.~\ref{fig:Tomasello1}(b)). Similarly in ~\cite{giordano2016spinhall}, the dc current excites a dynamical soliton together with a gyration which gives rise to the propagation of spin-waves in the form of a spiral mode, thus combining skyrmionics with magnonics. Whereas, in ~\cite{zhang2015current}, the skyrmion is a metastable state of the MTJ free layer and the dc STT drives it out of the nano-contact region, thus generating a gyration of the skyrmion (Fig.~\ref{fig:Tomasello1}(c)). Indeed, a different proposal was shown in~\cite{garcia2016skyrmion}, where the dc STT is applied uniformly in all the MTJ free layer, but the spin-current is polarized by a vortex-like configuration of the pinned layer magnetization (Fig.~\ref{fig:Tomasello1}(d)-(e)).

\begin{figure*}[h!]
    \centering
    \includegraphics[width=0.95\linewidth]{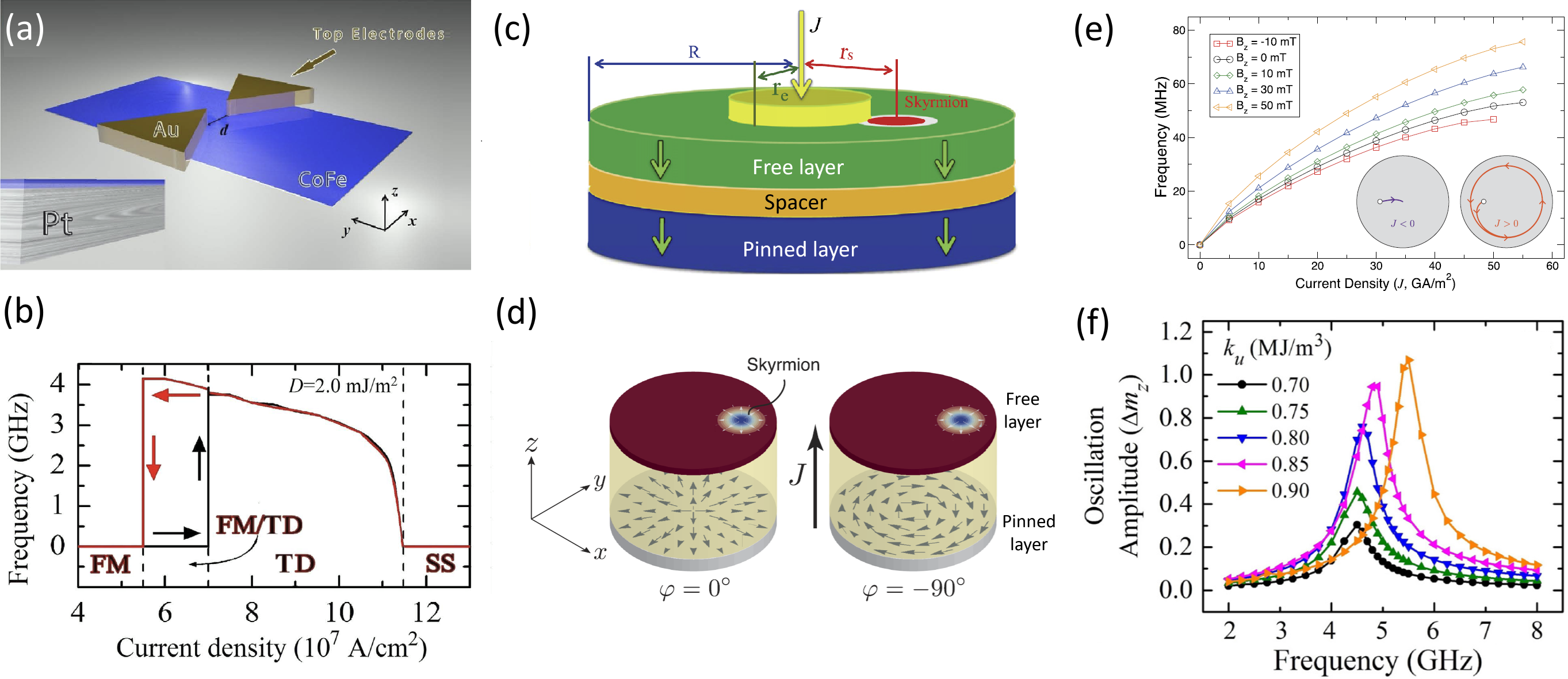}
    \caption{a) Sketch of a spin-hall oscillator where a topological dynamical soliton can be stabilized, reproduced with permission from Ref.~\cite{giordano2016spinhall}. (b) Frequency of a topological dynamical soliton as a function of the injected dc current density in a nano-contact MTJ with a perpendicular magnetization of the pinned layer, reproduced with permission from Ref.~\cite{carpentieri2015topological}. (c) Sketch of a nano-contact MTJ with a perpendicular magnetization of the pinned layer where the skyrmion gyration generates the magnetization oscillations, reproduced with permission from Ref.~\cite{zhang2015current}. (d) Sketch of an MTJ with a pinned layer magnetization in the form of a vortex, where the injected current is applied uniformly throughout the entire section, reproduced with permission from Ref.~\cite{garcia2016skyrmion}. (e) Frequency of the skyrmion gyration as a function of the injected current density in the MTJ sketched in (e), reproduced with permission from Ref.~\cite{garcia2016skyrmion}. (f) Linear resonant response of a skyrmion in a nano-contact MTJ with a perpendicular magnetization of the pinned layer, reproduced with permission from Ref.~\cite{finocchio2015skyrmion}.}
    \label{fig:Tomasello1}
\end{figure*}

On the other hand, when a metastable skyrmion is excited by an ac perpendicularly polarized STT in MTJs, it expands (shrinks) within the semi-period where the spin-current polarization is parallel (antiparallel) to its core. Therefore, a steady-state breathing mode is achieved which is responsible of the periodic variation of the magnetization z-component of the free layer, leading to the STD effect. The amplitude of such oscillation is larger as the ac STT frequency approaches the skyrmion resonance frequency, thus showing a typical resonant response (Fig.~\ref{fig:Tomasello1}(f)).

The room-temperature experimental characterization of internal modes of skyrmions in magnetic multilayers with interfacial Dzyaloshinskii–Moriya Interaction (DMI) has been performed via the application of an ac field, as shown in Fig.~\ref{fig:Tomasello2}(a)~\cite{satywali2021microwave}. Three main modes where identified (Fig.~\ref{fig:Tomasello2}(b)): a low-frequency (LF) mode corresponding to a skyrmion gyration localized in the ring around the skyrmion; a high-frequency (HF) mode localized in the area between neighboring skyrmions; a ferromagnetic (FM) mode in the absence of skyrmions and corresponding to the uniform precession of the magnetization.

Later, room temperature oscillatory skyrmion dynamics were measured experimentally in MTJs. Many efforts have been put to optimize the MTJ stack for hosting a single skyrmion at room temperature with a large enough tunnelling magnetoresistance (TMR) and small resistance area product (RA). Thanks to the strategy of combining an optimized MTJ stack with a skyrmion-hosting magnetic multilayer introduced in ~\cite{guang2023electrical}(Fig.~\ref{fig:Tomasello2}(c)), nowadays we are able to achieve MTJ hosting skyrmions with a 100\% TMR~\cite{zhao2024electrical} More importantly, the first experimental evidence of a microwave detector based on a single skyrmion was obtained in an MTJ <300 nm, with the first indirect observation of a room-temperature skyrmion breathing mode~\cite{Liu2025} (see region A in Fig.~\ref{fig:Tomasello2}(d)).

\begin{figure}[h!]
    \centering
    \includegraphics[width=0.95\linewidth]{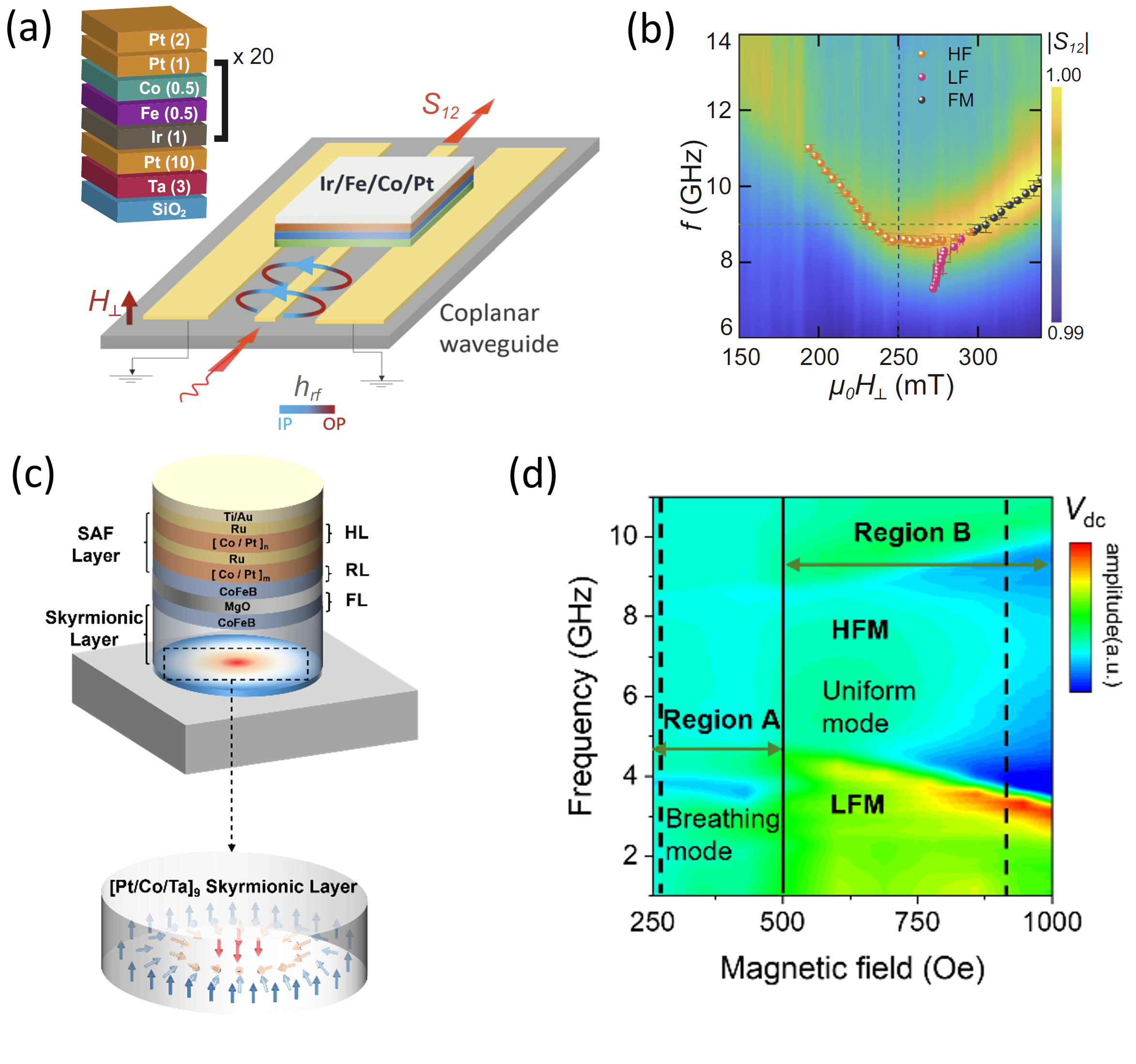}
    \caption{Sketch of the experimental setup to characterize the microwave
response of skyrmions at room temperature via the application of an ac
field, reproduced with permission from Ref.~\cite{satywali2021microwave}. (b) Phase diagram of the
excited modes in the system sketched in (a), reproduced with permission
from Ref.~\cite{satywali2021microwave}. (c) Sketch of the MTJ including a skyrmion-hosting magnetic
multilayer (skyrmionic layer) where the first experimental skyrmion-based
STD effect was measured, reproduced with permission from Ref.~\cite{Liu2025}. (d)
Phase diagram of the excited modes in the MTJ sketched in (c), reproduced
with permission from Ref.~\cite{Liu2025}. The high-frequency mode (HFM) and the
low-frequency mode (LFM) are related to the uniform precession of the free
layer and pinned layer, respectively.}
    \label{fig:Tomasello2}
\end{figure}

\section*{Challenges}

Numerous challenges must be addressed to enable the practical implementation of effective skyrmion-based microwave technologies. While skyrmion-based oscillators have remained theoretical so far, recent experimental demonstrations of MTJs hosting skyrmions ~\cite{Liu2025} pave the way toward their realization. For devices where the oscillations rely on skyrmion gyration, the frequency is low in the MHz regime and the motion is likely to be hindered by pinning effects caused by intrinsic defects in the thin free layer, similarly to what is observed in current-driven skyrmion motion in racetrack memory applications. More importantly, both MTJ geometries that support skyrmion gyration face a fundamental limitation: the electrical detection of the oscillation remains inherently challenging. In Fig.~\ref{fig:Tomasello1}(c), the skyrmion moves out of the nano-contact region, with no contribution to the TMR. In Fig.~\ref{fig:Tomasello1}(d), the skyrmion moves as a rigid object, resulting in no variation of the TMR. Moreover, the pinned layer magnetization in this configuration is non-uniform, implying that any TMR signal would originate solely from the in-plane spin components of the skyrmion domain wall. All these challenges suggest that the only viable solution toward the practical implementation of the skyrmion-based microwave oscillator lies in exploiting the topological dynamical soliton. This oscillatory regime could be achieved in the MTJ configuration shown in Fig.~\ref{fig:Tomasello2}(c), following an ad hoc optimization of the external magnetic field to operate at the edge of the skyrmion stability region.

Skyrmion-based microwave detectors are a step ahead of their oscillator counterparts, as a first proof-of-concept has already been successfully demonstrated ~\cite{Liu2025}. However, the controlled electrical nucleation of skyrmions remains a critical requirement. To date, skyrmions in MTJs have been nucleated via relaxation processes induced by variations in the external magnetic field. Although voltage-controlled magnetic anisotropy (VCMA)-based nucleation has been demonstrated, it typically requires a thicker insulating barrier, which hinders current flow and thereby compromises the functionality of skyrmion-based microwave detectors.

The output voltage observed in the first proof-of-concept device remains very low ($\mu$V)~\cite{Liu2025}. One potential strategy to enhance the signal involves improving the quality of the insulating interface, thereby increasing the TMR while maintaining a sufficiently low RA. Additionally, exploiting non-linear resonance phenomena, such as parametric resonance, could further improve the performance. This behavior may be enabled through the combined use of STT and VCMA within the same skyrmion-based MTJ, an approach already proposed in theoretical studies. Such hybrid mechanism has been already successfully employed in MTJs with uniform free layers to enhance detector efficiency > 200 kV/W.

To enable compact integration, the size of skyrmion-based microwave detectors must be scaled down below 100 nm. However, as device dimensions reduce, the detrimental magnetostatic effects arising from sample boundaries become increasingly significant. To counteract this and maintain skyrmion stability, an enhancement of the DMI is required.

A more comprehensive characterization of the detector parameters—including noise-equivalent power (NEP) and conversion efficiency—is essential to assess the competitiveness of skyrmion-based detectors against alternative technologies.

Finally, theoretical predictions~\cite{finocchio2015skyrmion} have suggested that skyrmion-based microwave detectors can achieve superior sensitivity when a nano-contact MTJ geometry is employed. This enhancement arises from the localized electrical readout of the skyrmion breathing mode, which enables full exploitation of the TMR response. Consequently, efforts toward the design and optimization of nano-contact skyrmion-based MTJs would be highly beneficial for the development of high-performance detectors

\section*{Conclusion}

In summary, skyrmion-based microwave technology promises the development of highly compact detectors featuring enhanced sensitivity, improved frequency selectivity, and mutual synchronization. An intriguing perspective involves the use of multiple MTJs sharing a common free layer, with a single skyrmion excited beneath each junction. To ensure that each skyrmion remains confined beneath its respective MTJ, thus minimizing the effects of Brownian motion, artificial pinning sites can be introduced. In this configuration, magnetostatic coupling may serve a twofold purpose: synchronizing the breathing modes of the individual skyrmions and reinforcing their collective response, thereby significantly increasing the generated dc output voltage.

Skyrmions in MTJs offer prospects beyond microwave detection, enabling the integration of sensing and computing within a single device. At the core of skyrmion-based reservoir computing lies the ability to detect skyrmion dynamics—for example, by placing an MTJ over a pinned skyrmion and using rectification under in-plane excitation. Similarly, skyrmion detectors can measure GHz dynamics in task-adaptive systems. In neuromorphic architectures, their ability to respond to microwave signals could enable MTJ-based networks where each element acts as both oscillator and detector, supporting many-to-many microwave communication. Furthermore, the experimental detection of quantized helicity excitations is key to advancing skyrmion-based quantum computing, where dynamic skyrmion readout may play a central role.

\section*{Acknowledgements}

This work was supported by the project PRIN 20222N9A73 ``SKYrmion-based magnetic tunnel junction to design a temperature SENSor – SkySens", funded by the Italian Ministry of Research, and by the project number 101070287—SWAN-on-chip—HORIZON-CL4-2021-DIGITAL EMERGING-01. RT and GF are with the PETASPIN team and thank the support of the PETASPIN association (\url{http://www.petaspin.com/}).
\endgroup

\newpage

\section{From Electrical Detection of Skyrmions to Functional Devices}
\begingroup
    \let\section\subsection
    \let\subsection\subsubsection
    \let\subsubsection\paragraph
    \let\paragraph\subparagraph
Le Zhao$^1$, and Wanjun Jiang$^{2,3}$
\vspace{0.5cm}

\noindent
\textit{$^1$ Institute of Applied Physics, TU Wien, Wiedner Hauptstraße 8-10, Vienna, 1040, Austria\\
$^2$ State Key Laboratory of Low-Dimensional Quantum Physics and Department of Physics, Tsinghua University, Beijing 100084, China\\
$^3$ Frontier Science Center for Quantum Information, Tsinghua University, Beijing 100084, China}

\section*{Introduction}

Magnetic skyrmions are nanoscale, topologically stable spin textures that can be efficiently driven by ultra-low current densities. These features make them attractive for the next-generation spintronic devices. Toward functional implementations, achieving reliable and efficient detection of individual skyrmions, while maintaining their mobility and compatibility with device-level integration, remain central challenges.

Imaging methods such as Lorentz transmission electron microscopy~\cite{yu2010real} and magneto-optical Kerr effect microscopy~\cite{jiang2017direct} have provided valuable insights into skyrmion studies. However, their limited spatial or temporal resolution, along with the requirement for bulky equipment, makes them incompatible with device-level integration. By contrast, direct electrical readout techniques offer the opportunity of high-speed detection and better scalability, which are increasingly explored as practical solutions for skyrmion detection. The anomalous Hall effect (AHE)~\cite{maccariello2018electrical} and anomalous Nernst effect (ANE)~\cite{wang2020thermal, fernandezscarioni2021thermoelectric} have proven effective for detecting skyrmions. In particular, electrical readout by the AHE method has been implemented into spintronic computing paradigms~\cite{song2020skyrmionbased, dacamarasantaclaragomes2025neuromorphic}, owing to its simplicity and compatibility. However, the amplitudes of both AHE and ANE signals are limited by the magnetization. Consequently, the output voltages linearly scale with the total number and size of skyrmions, or equivalently, with the net magnetization plus an additional contribution from topological charges. As device shrinks in dimensions, the amplitudes of signals also decrease, leading to the reduced signal-to-noise ratio and increased readout error, which can introduce considerable difficulties for high-density integration. Moreover, these methods do not resolve the exact position of individual skyrmions, which may hinder device concepts relying on well-defined skyrmion trajectories or controlled skyrmion-skyrmion interactions.

To overcome this scaling limitation, tunnel magnetoresistance (TMR) has emerged as a promising alternative. In perpendicular device geometries, the local tunneling conductance depends on the relative orientation of magnetizations in two magnetic layers separated by an insulating barrier. Meanwhile, the overall resistance difference between the parallel and antiparallel states scales inversely with the junction's cross-sectional area. This scaling-up behavior is inherently favorable for the integration of the high-density nanodevices, while enabling enhanced spatial resolution that can be useful for detecting nanoscale skyrmions. Recent experiments have demonstrated the electrical detection of skyrmions in MTJs~\cite{guang2023electrical, chen2024allelectrical, urrestarazu2024electrical} (Fig.~\ref{fig:Jiang1}(A-G)). Specifically, the (all-)electrical manipulation of skyrmions via vertical electric fields was implemented and simultaneously facilitated by the voltage-controlled magnetic anisotropy (VCMA) effect. These advances consolidate the feasibility of electrical readout and control of skyrmions, opening new avenues for developing low-power skyrmion-based device architectures. Furthermore, device platforms enabling tunneling-based detection of mobile skyrmions have been also realized~\cite{zhao2024electrical}, marking a key step toward real-time, scalable detection schemes that are demanded with current-driven skyrmion dynamics for building the skyrmion racetrack memory.

\begin{figure*}[h!]
    \centering
     \includegraphics[width = 0.55\textwidth]{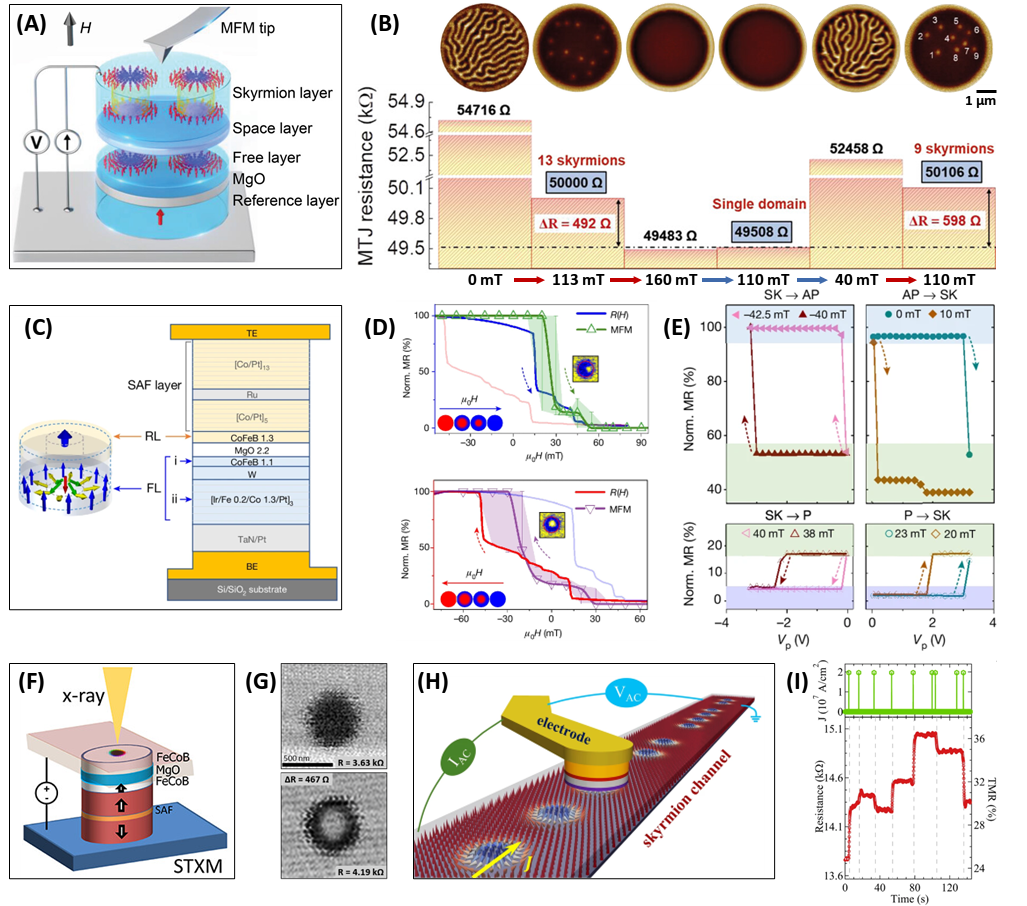}    \captionsetup{font=tiny,skip=-1pt,justification=raggedright,singlelinecheck=false}
    \caption{Skyrmion detection using MTJs. (A) Schematic illustration of skyrmion detection in a MTJ, with in-situ magnetic force microscopy (MFM) being used to verify the presence of skyrmions. (B) Tunneling magnetoresistance (TMR) measured under different magnetic fields, alongside the corresponding MFM images. (C) Multilayer stack structure for a skyrmionic MTJ. (D) Hysteresis loops of TMR as a function of magnetic field, accompanied by in-situ MFM images showing the nucleation of a single skyrmion. (E) Voltage-pulse-induced switching from the skyrmion state to antiparallel or parallel states. (F) Schematic representation of the MTJ pillar with in-situ scanning transmission X-ray microscopy (STXM) measurements. In this configuration, skyrmions form in the top single CoFeB layer. (G) STXM images showing the generation of a single skyrmion, with a corresponding TMR change of $\Delta R = 467~\Omega$. Note that in (A),(C) and (F), skyrmions are confined to the MTJ pillars and cannot propagate. (H) Illustration of detecting mobile skyrmions using an MTJ. (I) Changes in TMR that are measured during the application of current pulses along the skyrmion channel. (A-B) Adapted under the terms of the CC-BY license from~\cite{guang2023electrical}. Copyright 2022 by the Wiley-VCH GmbH. (C-E) Adapted with permission under the terms of the CC-BY-NC license from~\cite{chen2024allelectrical}. Copyright 2024 by Springer Nature. (F-G) Adapted with permission from~\cite{urrestarazu2024electrical}. Copyright 2024 by the American Chemical Society. (H-I) Adapted under the terms of the CC-BY license from~\cite{zhao2024electrical}. Copyright 2024 by Springer Nature.}
    \label{fig:Jiang1}
\end{figure*}

\section*{Relevance And Vision}

Skyrmions have been proposed as active information carriers in a variety of spintronic devices, including racetrack memory~\cite{yu2017roomtemperature}, reconfigurable logic gates~\cite{yan2021skyrmionbased}, and neuromorphic computing schemes~\cite{lee2023handwritten} (Fig.~\ref{fig:Jiang2}(A-D)). In these applications, device functionalities are manifested not only through the presence of skyrmions, but also their precise locations within the device. For instance, the information encoded in the spatial sequence of skyrmions requires the localized detection at predefined regions for an accurate retrieval. Likewise, the product of logical operations, including those in non-von Neumann computing schemes~\cite{raab2022brownian, lee2023perspective}, also demands a high signal-to-noise ratio at the targeted output sites to ensure correct signal interpretation. Moreover, as many device architectures are expected to rely on the current-driven skyrmion motion, adapting MTJ-based detection accordingly will be critical for their practical applications.

Owing to the top-down nature of conventional MTJ fabrication processes, the junction must be positioned above the relatively larger racetrack, where skyrmions are expected to move freely. As a high TMR ratio requires exceptionally well-defined interfaces, typically based on CoFeB/MgO/CoFeB stacks, the multilayer structure must be carefully designed and deposited. Specifically, the magnetic multilayers that host skyrmions should be located underneath the MgO barrier, while the reference layer should be above. These constraints place tight constraints on film growth.

Building upon these considerations, a prototypical device platform has been developed, enabling the simultaneous TMR detection of mobile skyrmions while maintaining compatibility with current-driven skyrmion operations (Fig.~\ref{fig:Jiang1}(H-I)). The skyrmion-hosting multilayer consists of ten repetitions of [Pt/Co/Ta]$_{10}$, within which skyrmions with diameters on the order of 400 nm can be observed. [Pt/Co/Ta]$_{10}$ multilayer is ferromagnetically coupled to the CoFeB free layer via a 1 nm Ta spacer. Another CoFeB layer that is on top of the MgO barrier, is pinned by a synthetic antiferromagnet (SAF) composed of Co/[Pt/Co]$_{2}$/Ru/[Co/Pt]$_{3}$. The stack is patterned into a typical three-terminal spin-torque device featuring a narrow channel and a tunneling junction on top. During the junction definition, the etching is stopped immediately after reaching the MgO layer, ensuring that the skyrmion-hosting layer beneath remains continuous. Through systematic optimization of magnetic multilayers and the device fabrication process, a TMR ratio of 103\% can be achieved at room temperature, allowing high-fidelity local detection of magnetic skyrmions. Upon applying nanosecond current pulses along the narrow channel, skyrmions can be driven into motion with the help of spin-orbit torque. As skyrmions pass beneath the junction area, the position and shape of the local spin textures could modulate the associate values of TMR, producing prominent changes in the voltage readout. Through the proposed methodology combining material stack design and fabrication strategies, a broad range of previously envisioned skyrmion device architectures can be experimentally implemented.

 \begin{figure*}[h!]
     \centering
     \includegraphics[width = 0.55\textwidth]{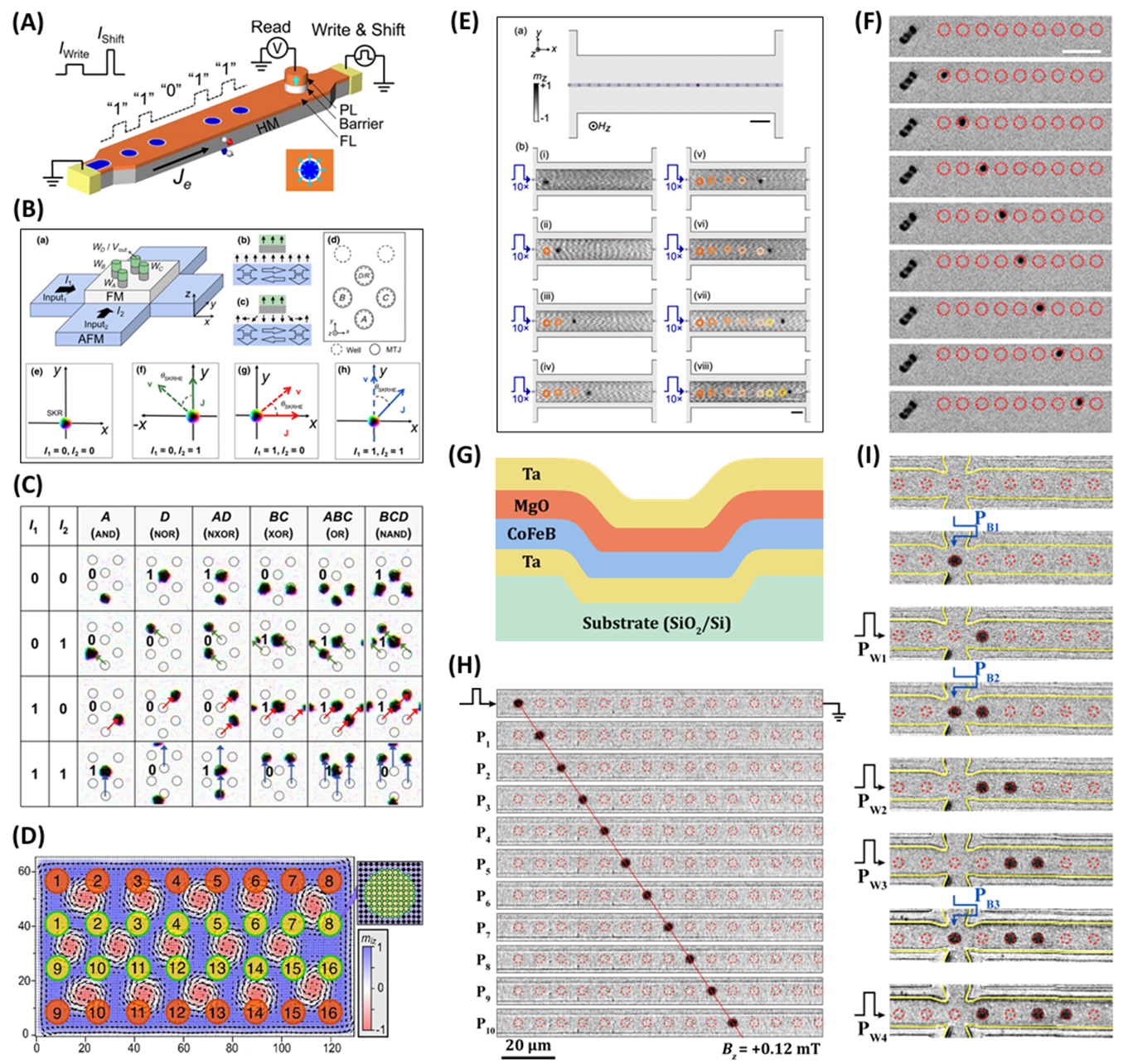}     \captionsetup{font=tiny,skip=-1pt,justification=raggedright,singlelinecheck=false}
     \caption{Advances in application-oriented skyrmion studies. (A) Conceptual sketches of skyrmion racetrack memory devices. (B) Design of a reconfigurable skyrmion Boolean logic device composed of five pinning wells and four MTJs for local writing and detection. (C) Simulated truth tables corresponding to different logic functions. (D) Illustration of neuromorphic computing schemes based on skyrmion lattices. Input and output nodes are marked in red and yellow, respectively, and can be implemented using MTJs in practical devices. (E) Confinement of skyrmion motion into a narrow channel irradiated with He+ ion. (F) Deterministic shifting of a skyrmion achieved by defining periodic pinning sites via laser irradiation. (G) Cross-sectional schematic of the material stack deposited on a substrate with pre-defined indentations. (I) Deterministic motion of skyrmion along a narrow channel with periodic indentations, where the step size of each shift is strictly defined by the geometric confinement. (J) Generation of the skyrmion data string ``1101'' within a cross-shaped device fabricated on an indented substrate. (A) Reprinted with permission from~\cite{yu2017roomtemperature}. Copyright 2017 by the American Chemical Society. (B-C) Reprinted with permission from~\cite{yan2021skyrmionbased}. Copyright 2021 by the American Physical Society. (D) Adapted under the terms of the CC-BY license from~\cite{lee2023handwritten}. Copyright 2023 by Springer Nature. (E) Reprinted under the terms of the CC-BY-NC-ND license from~\cite{kern2022deterministic}. Copyright 2022 by the American Chemical Society. (F) Adapted with permission from~\cite{he2023allelectrical}. Copyright 2023 by the American Chemical Society. (G-J) Adapted under the terms of the CC-BY license from~\cite{zhao2024realization}. Copyright 2024 by Science China Press.}
     \label{fig:Jiang2}
 \end{figure*}

\section*{Challenges}

Despite these advances, several challenges must be addressed before skyrmion-based functionalities can be widely integrated into practical devices.

One key issue is the thermally induced stochastic behavior of skyrmions~\cite{desplat2018thermal, troncoso2014brownian, zazvorka2019thermal}, which can lead to uncertainty of their position. This fact can compromise deterministic device operations, especially for memory and logic applications. To enable reliable computing elements, guided~\cite{kern2022deterministic} or site-selective~\cite{he2023allelectrical} methods for skyrmion displacement are needed (Fig.~\ref{fig:Jiang2}(E-F)). In a recent study~\cite{zhao2024realization}, engineered morphological features on the substrate were implemented to serve as effective energy landscapes for skyrmion pinning and routing, offering a promising strategy for achieving deterministic motion of skyrmion (Fig.~\ref{fig:Jiang2}(G-I)). Remarkably, this bottom-up technique does not interfere with subsequent material deposition or device fabrication processes, making it highly attractive for MTJ-based detection schemes, and the subsequent device applications.

Another major challenge lies on the large current density and overall energy consumption required for driving skyrmions. While spin-orbit torque could enable the manipulation of skyrmions, the current threshold in practical implementations remains relatively high, posing a serious limitation for the energy-constrained systems. Material optimization of parameters such as damping, effective spin/orbital Hall angle and net magnetization, together with improved interface engineering, will be essential for obtaining the increased skyrmion velocity with the reduced energy consumption.

In addition to the driving efficiency, the tolerance of skyrmion to variations in temperature and magnetic fields remains another critical issue for device reliability. Aside from thermal fluctuations, the intrinsic temperature dependence of magnetic properties can alter skyrmion size and mobility, while their size and density are also influenced by the external magnetic field. Identifying a suitable material system that remains robust under varying temperatures and strong magnetic fields would greatly broaden the application potential of skyrmion-based devices.

From a system-level perspective, achieving fully electrical control over device initialization, data input and readout is critical for practical implementation. Achieving compact and multifunctional devices that combine writing, controlled motion and high-fidelity readout. Skyrmions can offer unique dynamics and distinct functionalities, compared to other magnetic solitons such as chiral domain walls and trivial magnetic bubbles. To fully exploit these properties, novel device concepts, tailored to topological features, skyrmion-skyrmion interactions, and current-driven behaviors will be required.

\section*{Concluding remarks}

The convergence of real-time tunneling magnetoresistance detection, substrate-mediated precise positioning, and current-driven skyrmion motion offer a promising pathway toward functional skyrmion-based devices. With the continued advances in material stability, energy efficiency, and device integration, skyrmion technologies are expected to evolve from laboratory demonstrations to practical spintronic applications.

\section*{Acknowledgements}

We thank funding support from the European Research Council (ERC) under the Horizon 2020 Program (Grant Agreement ID: 101001290, 3DNANOMAG), the National Natural Science Foundation under the distinguished Young Scholar program (NSFC Grant No. 12225409), the NSFC Basic Science Center Project (NSFC Grant No. 52388201), the NSFC general program (Grant Nos. 52271181, 12404138), the National Key R\&D Program of China (Grant No. 2022YFA1405100), the Innovation Program for Quantum Science and Technology (Grant No. 2023ZD0300500) and Beijing Natural Science Foundation (Grant No. Z240006).

\endgroup

\newpage
%


\begin{thebibliography}{441}%
\makeatletter
\providecommand \@ifxundefined [1]{%
 \@ifx{#1\undefined}
}%
\providecommand \@ifnum [1]{%
 \ifnum #1\expandafter \@firstoftwo
 \else \expandafter \@secondoftwo
 \fi
}%
\providecommand \@ifx [1]{%
 \ifx #1\expandafter \@firstoftwo
 \else \expandafter \@secondoftwo
 \fi
}%
\providecommand \natexlab [1]{#1}%
\providecommand \enquote  [1]{``#1''}%
\providecommand \bibnamefont  [1]{#1}%
\providecommand \bibfnamefont [1]{#1}%
\providecommand \citenamefont [1]{#1}%
\providecommand \href@noop [0]{\@secondoftwo}%
\providecommand \href [0]{\begingroup \@sanitize@url \@href}%
\providecommand \@href[1]{\@@startlink{#1}\@@href}%
\providecommand \@@href[1]{\endgroup#1\@@endlink}%
\providecommand \@sanitize@url [0]{\catcode `\\12\catcode `\$12\catcode
  `\&12\catcode `\#12\catcode `\^12\catcode `\_12\catcode `\%12\relax}%
\providecommand \@@startlink[1]{}%
\providecommand \@@endlink[0]{}%
\providecommand \url  [0]{\begingroup\@sanitize@url \@url }%
\providecommand \@url [1]{\endgroup\@href {#1}{\urlprefix }}%
\providecommand \urlprefix  [0]{URL }%
\providecommand \Eprint [0]{\href }%
\providecommand \doibase [0]{https://doi.org/}%
\providecommand \selectlanguage [0]{\@gobble}%
\providecommand \bibinfo  [0]{\@secondoftwo}%
\providecommand \bibfield  [0]{\@secondoftwo}%
\providecommand \translation [1]{[#1]}%
\providecommand \BibitemOpen [0]{}%
\providecommand \bibitemStop [0]{}%
\providecommand \bibitemNoStop [0]{.\EOS\space}%
\providecommand \EOS [0]{\spacefactor3000\relax}%
\providecommand \BibitemShut  [1]{\csname bibitem#1\endcsname}%
\let\auto@bib@innerbib\@empty
\bibitem [{\citenamefont {Skyrme}(1962)}]{skyrme1962unified}%
  \BibitemOpen
  \bibfield  {author} {\bibinfo {author} {\bibfnamefont {T.~H.~R.}\
  \bibnamefont {Skyrme}},\ }\bibfield  {title} {\bibinfo {title} {A unified
  field theory of mesons and baryons},\ }\href@noop {} {\bibfield  {journal}
  {\bibinfo  {journal} {Nucl. Phys.}\ }\textbf {\bibinfo {volume} {31}},\
  \bibinfo {pages} {556} (\bibinfo {year} {1962})}\BibitemShut {NoStop}%
\bibitem [{\citenamefont {Manton}\ and\ \citenamefont
  {Sutcliffe}(2004)}]{manton2004topological}%
  \BibitemOpen
  \bibfield  {author} {\bibinfo {author} {\bibfnamefont {N.}~\bibnamefont
  {Manton}}\ and\ \bibinfo {author} {\bibfnamefont {P.}~\bibnamefont
  {Sutcliffe}},\ }\href@noop {} {\emph {\bibinfo {title} {Topological
  solitons}}}\ (\bibinfo  {publisher} {Cambridge University Press},\ \bibinfo
  {year} {2004})\BibitemShut {NoStop}%
\bibitem [{\citenamefont {Bogdanov}\ and\ \citenamefont
  {Yablonskii}(1989)}]{Bogdanov1989}%
  \BibitemOpen
  \bibfield  {author} {\bibinfo {author} {\bibfnamefont {A.~N.}\ \bibnamefont
  {Bogdanov}}\ and\ \bibinfo {author} {\bibfnamefont {D.}~\bibnamefont
  {Yablonskii}},\ }\bibfield  {title} {\bibinfo {title} {Thermodynamically
  stable “vortices” in magnetically ordered crystals. the mixed state of
  magnets},\ }\href@noop {} {\bibfield  {journal} {\bibinfo  {journal} {Zh.
  Eksp. Teor. Fiz}\ }\textbf {\bibinfo {volume} {95}},\ \bibinfo {pages} {178}
  (\bibinfo {year} {1989})}\BibitemShut {NoStop}%
\bibitem [{\citenamefont {Pokrovsky}(1979)}]{pokrovsky1979properties}%
  \BibitemOpen
  \bibfield  {author} {\bibinfo {author} {\bibfnamefont {V.}~\bibnamefont
  {Pokrovsky}},\ }\bibfield  {title} {\bibinfo {title} {Properties of ordered,
  continuously degenerate systems},\ }\href@noop {} {\bibfield  {journal}
  {\bibinfo  {journal} {Adv. Phys.}\ }\textbf {\bibinfo {volume} {28}},\
  \bibinfo {pages} {595} (\bibinfo {year} {1979})}\BibitemShut {NoStop}%
\bibitem [{\citenamefont {Roessler}\ \emph {et~al.}(2006)\citenamefont
  {Roessler}, \citenamefont {Bogdanov},\ and\ \citenamefont
  {Pfleiderer}}]{roessler2006spontaneous}%
  \BibitemOpen
  \bibfield  {author} {\bibinfo {author} {\bibfnamefont {U.~K.}\ \bibnamefont
  {Roessler}}, \bibinfo {author} {\bibfnamefont {A.}~\bibnamefont {Bogdanov}},\
  and\ \bibinfo {author} {\bibfnamefont {C.}~\bibnamefont {Pfleiderer}},\
  }\bibfield  {title} {\bibinfo {title} {Spontaneous skyrmion ground states in
  magnetic metals},\ }\href@noop {} {\bibfield  {journal} {\bibinfo  {journal}
  {Nature}\ }\textbf {\bibinfo {volume} {442}},\ \bibinfo {pages} {797}
  (\bibinfo {year} {2006})}\BibitemShut {NoStop}%
\bibitem [{\citenamefont {Smalyukh}\ \emph {et~al.}(2010)\citenamefont
  {Smalyukh}, \citenamefont {Lansac}, \citenamefont {Clark},\ and\
  \citenamefont {Trivedi}}]{smalyukh2010threedimensional}%
  \BibitemOpen
  \bibfield  {author} {\bibinfo {author} {\bibfnamefont {I.~I.}\ \bibnamefont
  {Smalyukh}}, \bibinfo {author} {\bibfnamefont {Y.}~\bibnamefont {Lansac}},
  \bibinfo {author} {\bibfnamefont {N.~A.}\ \bibnamefont {Clark}},\ and\
  \bibinfo {author} {\bibfnamefont {R.~P.}\ \bibnamefont {Trivedi}},\
  }\bibfield  {title} {\bibinfo {title} {Three-dimensional structure and
  multistable optical switching of triple-twisted particle-like excitations in
  anisotropic fluids},\ }\href {https://doi.org/10.1038/nmat2592} {\bibfield
  {journal} {\bibinfo  {journal} {Nat. Mater.}\ }\textbf {\bibinfo {volume}
  {9}},\ \bibinfo {pages} {139} (\bibinfo {year} {2010})}\BibitemShut {NoStop}%
\bibitem [{\citenamefont {Das}\ \emph {et~al.}(2019)\citenamefont {Das},
  \citenamefont {Tang}, \citenamefont {Hong}, \citenamefont {Gon\c{c}alves},
  \citenamefont {McCarter}, \citenamefont {Klewe}, \citenamefont {Nguyen},
  \citenamefont {G\'{o}mez-Ortiz}, \citenamefont {Shafer}, \citenamefont
  {Arenholz}, \citenamefont {Stoica}, \citenamefont {Hsu}, \citenamefont
  {Wang}, \citenamefont {Ophus}, \citenamefont {Liu}, \citenamefont {Nelson},
  \citenamefont {Saremi}, \citenamefont {Prasad}, \citenamefont {Mei},
  \citenamefont {Schlom}, \citenamefont {\'I{\~n}iguez}, \citenamefont
  {Garc\'{\i}a-Fern\'{a}ndez}, \citenamefont {Muller}, \citenamefont {Chen},
  \citenamefont {Junquera}, \citenamefont {Martin},\ and\ \citenamefont
  {Ramesh}}]{Das2019}%
  \BibitemOpen
  \bibfield  {author} {\bibinfo {author} {\bibfnamefont {S.}~\bibnamefont
  {Das}}, \bibinfo {author} {\bibfnamefont {Y.~L.}\ \bibnamefont {Tang}},
  \bibinfo {author} {\bibfnamefont {Z.}~\bibnamefont {Hong}}, \bibinfo {author}
  {\bibfnamefont {M.~A.~P.}\ \bibnamefont {Gon\c{c}alves}}, \bibinfo {author}
  {\bibfnamefont {M.~R.}\ \bibnamefont {McCarter}}, \bibinfo {author}
  {\bibfnamefont {C.}~\bibnamefont {Klewe}}, \bibinfo {author} {\bibfnamefont
  {K.~X.}\ \bibnamefont {Nguyen}}, \bibinfo {author} {\bibfnamefont
  {F.}~\bibnamefont {G\'{o}mez-Ortiz}}, \bibinfo {author} {\bibfnamefont
  {P.}~\bibnamefont {Shafer}}, \bibinfo {author} {\bibfnamefont
  {E.}~\bibnamefont {Arenholz}}, \bibinfo {author} {\bibfnamefont {V.~A.}\
  \bibnamefont {Stoica}}, \bibinfo {author} {\bibfnamefont {S.-L.}\
  \bibnamefont {Hsu}}, \bibinfo {author} {\bibfnamefont {B.}~\bibnamefont
  {Wang}}, \bibinfo {author} {\bibfnamefont {C.}~\bibnamefont {Ophus}},
  \bibinfo {author} {\bibfnamefont {J.~F.}\ \bibnamefont {Liu}}, \bibinfo
  {author} {\bibfnamefont {C.~T.}\ \bibnamefont {Nelson}}, \bibinfo {author}
  {\bibfnamefont {S.}~\bibnamefont {Saremi}}, \bibinfo {author} {\bibfnamefont
  {B.}~\bibnamefont {Prasad}}, \bibinfo {author} {\bibfnamefont {A.~B.}\
  \bibnamefont {Mei}}, \bibinfo {author} {\bibfnamefont {D.~G.}\ \bibnamefont
  {Schlom}}, \bibinfo {author} {\bibfnamefont {J.}~\bibnamefont
  {\'I{\~n}iguez}}, \bibinfo {author} {\bibfnamefont {P.}~\bibnamefont
  {Garc\'{\i}a-Fern\'{a}ndez}}, \bibinfo {author} {\bibfnamefont {D.~A.}\
  \bibnamefont {Muller}}, \bibinfo {author} {\bibfnamefont {L.~Q.}\
  \bibnamefont {Chen}}, \bibinfo {author} {\bibfnamefont {J.}~\bibnamefont
  {Junquera}}, \bibinfo {author} {\bibfnamefont {L.~W.}\ \bibnamefont
  {Martin}},\ and\ \bibinfo {author} {\bibfnamefont {R.}~\bibnamefont
  {Ramesh}},\ }\bibfield  {title} {\bibinfo {title} {Observation of
  room-temperature polar skyrmions},\ }\href
  {https://doi.org/https://doi.org/10.1038/s41586-019-1092-8} {\bibfield
  {journal} {\bibinfo  {journal} {Nature}\ }\textbf {\bibinfo {volume} {568}},\
  \bibinfo {pages} {368–372} (\bibinfo {year} {2019})}\BibitemShut {NoStop}%
\bibitem [{\citenamefont {M\"uhlbauer}\ \emph {et~al.}(2009)\citenamefont
  {M\"uhlbauer}, \citenamefont {Binz}, \citenamefont {Jonietz}, \citenamefont
  {Pfleiderer}, \citenamefont {Rosch}, \citenamefont {Neubauer}, \citenamefont
  {Georgii},\ and\ \citenamefont {Böni}}]{muehlbauer2009skyrmion}%
  \BibitemOpen
  \bibfield  {author} {\bibinfo {author} {\bibfnamefont {S.}~\bibnamefont
  {M\"uhlbauer}}, \bibinfo {author} {\bibfnamefont {B.}~\bibnamefont {Binz}},
  \bibinfo {author} {\bibfnamefont {F.}~\bibnamefont {Jonietz}}, \bibinfo
  {author} {\bibfnamefont {C.}~\bibnamefont {Pfleiderer}}, \bibinfo {author}
  {\bibfnamefont {A.}~\bibnamefont {Rosch}}, \bibinfo {author} {\bibfnamefont
  {A.}~\bibnamefont {Neubauer}}, \bibinfo {author} {\bibfnamefont
  {R.}~\bibnamefont {Georgii}},\ and\ \bibinfo {author} {\bibfnamefont
  {P.}~\bibnamefont {Böni}},\ }\bibfield  {title} {\bibinfo {title} {Skyrmion
  lattice in a chiral magnet},\ }\href@noop {} {\bibfield  {journal} {\bibinfo
  {journal} {Science}\ }\textbf {\bibinfo {volume} {323}},\ \bibinfo {pages}
  {915} (\bibinfo {year} {2009})}\BibitemShut {NoStop}%
\bibitem [{\citenamefont {Yu}\ \emph {et~al.}(2010)\citenamefont {Yu},
  \citenamefont {Onose}, \citenamefont {Kanazawa}, \citenamefont {Park},
  \citenamefont {Han}, \citenamefont {Matsui}, \citenamefont {Nagaosa},\ and\
  \citenamefont {Tokura}}]{yu2010real}%
  \BibitemOpen
  \bibfield  {author} {\bibinfo {author} {\bibfnamefont {X.~Z.}\ \bibnamefont
  {Yu}}, \bibinfo {author} {\bibfnamefont {Y.}~\bibnamefont {Onose}}, \bibinfo
  {author} {\bibfnamefont {N.}~\bibnamefont {Kanazawa}}, \bibinfo {author}
  {\bibfnamefont {J.~H.}\ \bibnamefont {Park}}, \bibinfo {author}
  {\bibfnamefont {J.~H.}\ \bibnamefont {Han}}, \bibinfo {author} {\bibfnamefont
  {Y.}~\bibnamefont {Matsui}}, \bibinfo {author} {\bibfnamefont
  {N.}~\bibnamefont {Nagaosa}},\ and\ \bibinfo {author} {\bibfnamefont
  {Y.}~\bibnamefont {Tokura}},\ }\bibfield  {title} {\bibinfo {title}
  {Real-space observation of a two-dimensional skyrmion crystal},\ }\href
  {https://doi.org/10.1038/nature09124} {\bibfield  {journal} {\bibinfo
  {journal} {Nature}\ }\textbf {\bibinfo {volume} {465}},\ \bibinfo {pages}
  {901} (\bibinfo {year} {2010})}\BibitemShut {NoStop}%
\bibitem [{\citenamefont {Heinze}\ \emph {et~al.}(2011)\citenamefont {Heinze},
  \citenamefont {Von~Bergmann}, \citenamefont {Menzel}, \citenamefont {Brede},
  \citenamefont {Kubetzka}, \citenamefont {Wiesendanger}, \citenamefont
  {Bihlmayer},\ and\ \citenamefont {Bl{\"u}gel}}]{heinze2011spontaneous}%
  \BibitemOpen
  \bibfield  {author} {\bibinfo {author} {\bibfnamefont {S.}~\bibnamefont
  {Heinze}}, \bibinfo {author} {\bibfnamefont {K.}~\bibnamefont
  {Von~Bergmann}}, \bibinfo {author} {\bibfnamefont {M.}~\bibnamefont
  {Menzel}}, \bibinfo {author} {\bibfnamefont {J.}~\bibnamefont {Brede}},
  \bibinfo {author} {\bibfnamefont {A.}~\bibnamefont {Kubetzka}}, \bibinfo
  {author} {\bibfnamefont {R.}~\bibnamefont {Wiesendanger}}, \bibinfo {author}
  {\bibfnamefont {G.}~\bibnamefont {Bihlmayer}},\ and\ \bibinfo {author}
  {\bibfnamefont {S.}~\bibnamefont {Bl{\"u}gel}},\ }\bibfield  {title}
  {\bibinfo {title} {Spontaneous atomic-scale magnetic skyrmion lattice in two
  dimensions},\ }\href@noop {} {\bibfield  {journal} {\bibinfo  {journal} {Nat.
  Phys.}\ }\textbf {\bibinfo {volume} {7}},\ \bibinfo {pages} {713} (\bibinfo
  {year} {2011})}\BibitemShut {NoStop}%
\bibitem [{\citenamefont {Jiang}\ \emph {et~al.}(2015)\citenamefont {Jiang},
  \citenamefont {Upadhyaya}, \citenamefont {Zhang}, \citenamefont {Yu},
  \citenamefont {Jungfleisch}, \citenamefont {Fradin}, \citenamefont {Pearson},
  \citenamefont {Tserkovnyak}, \citenamefont {Wang}, \citenamefont {Heinonen},
  \citenamefont {{te Velthuis}},\ and\ \citenamefont
  {Hoffmann}}]{jiang2015blowing}%
  \BibitemOpen
  \bibfield  {author} {\bibinfo {author} {\bibfnamefont {W.}~\bibnamefont
  {Jiang}}, \bibinfo {author} {\bibfnamefont {P.}~\bibnamefont {Upadhyaya}},
  \bibinfo {author} {\bibfnamefont {W.}~\bibnamefont {Zhang}}, \bibinfo
  {author} {\bibfnamefont {G.}~\bibnamefont {Yu}}, \bibinfo {author}
  {\bibfnamefont {M.~B.}\ \bibnamefont {Jungfleisch}}, \bibinfo {author}
  {\bibfnamefont {F.~Y.}\ \bibnamefont {Fradin}}, \bibinfo {author}
  {\bibfnamefont {J.~E.}\ \bibnamefont {Pearson}}, \bibinfo {author}
  {\bibfnamefont {Y.}~\bibnamefont {Tserkovnyak}}, \bibinfo {author}
  {\bibfnamefont {K.~L.}\ \bibnamefont {Wang}}, \bibinfo {author}
  {\bibfnamefont {O.}~\bibnamefont {Heinonen}}, \bibinfo {author}
  {\bibfnamefont {S.~G.~E.}\ \bibnamefont {{te Velthuis}}},\ and\ \bibinfo
  {author} {\bibfnamefont {A.}~\bibnamefont {Hoffmann}},\ }\bibfield  {title}
  {\bibinfo {title} {Blowing magnetic skyrmion bubbles},\ }\href
  {https://doi.org/10.1126/science.aaa1442} {\bibfield  {journal} {\bibinfo
  {journal} {Science}\ }\textbf {\bibinfo {volume} {349}},\ \bibinfo {pages}
  {283} (\bibinfo {year} {2015})}\BibitemShut {NoStop}%
\bibitem [{\citenamefont {Boulle}\ \emph {et~al.}(2016)\citenamefont {Boulle},
  \citenamefont {Vogel}, \citenamefont {Yang}, \citenamefont {Pizzini},
  \citenamefont {de~Souza~Chaves}, \citenamefont {Locatelli}, \citenamefont
  {Mente{\c{s}}}, \citenamefont {Sala}, \citenamefont {Buda-Prejbeanu},
  \citenamefont {Klein} \emph {et~al.}}]{boulle2016room}%
  \BibitemOpen
  \bibfield  {author} {\bibinfo {author} {\bibfnamefont {O.}~\bibnamefont
  {Boulle}}, \bibinfo {author} {\bibfnamefont {J.}~\bibnamefont {Vogel}},
  \bibinfo {author} {\bibfnamefont {H.}~\bibnamefont {Yang}}, \bibinfo {author}
  {\bibfnamefont {S.}~\bibnamefont {Pizzini}}, \bibinfo {author} {\bibfnamefont
  {D.}~\bibnamefont {de~Souza~Chaves}}, \bibinfo {author} {\bibfnamefont
  {A.}~\bibnamefont {Locatelli}}, \bibinfo {author} {\bibfnamefont {T.~O.}\
  \bibnamefont {Mente{\c{s}}}}, \bibinfo {author} {\bibfnamefont
  {A.}~\bibnamefont {Sala}}, \bibinfo {author} {\bibfnamefont {L.~D.}\
  \bibnamefont {Buda-Prejbeanu}}, \bibinfo {author} {\bibfnamefont
  {O.}~\bibnamefont {Klein}}, \emph {et~al.},\ }\bibfield  {title} {\bibinfo
  {title} {Room-temperature chiral magnetic skyrmions in ultrathin magnetic
  nanostructures},\ }\href@noop {} {\bibfield  {journal} {\bibinfo  {journal}
  {Nat. Nanotechnol.}\ }\textbf {\bibinfo {volume} {11}},\ \bibinfo {pages}
  {449} (\bibinfo {year} {2016})}\BibitemShut {NoStop}%
\bibitem [{\citenamefont {{Moreau-Luchaire}}\ \emph {et~al.}(2016)\citenamefont
  {{Moreau-Luchaire}}, \citenamefont {Moutafis}, \citenamefont {Reyren},
  \citenamefont {Sampaio}, \citenamefont {Vaz}, \citenamefont {Van~Horne},
  \citenamefont {Bouzehouane}, \citenamefont {Garcia}, \citenamefont
  {Deranlot}, \citenamefont {Warnicke}, \citenamefont {Wohlh{\"u}ter},
  \citenamefont {George}, \citenamefont {Weigand}, \citenamefont {Raabe},
  \citenamefont {Cros},\ and\ \citenamefont
  {Fert}}]{moreau-luchaire2016additive}%
  \BibitemOpen
  \bibfield  {author} {\bibinfo {author} {\bibfnamefont {C.}~\bibnamefont
  {{Moreau-Luchaire}}}, \bibinfo {author} {\bibfnamefont {C.}~\bibnamefont
  {Moutafis}}, \bibinfo {author} {\bibfnamefont {N.}~\bibnamefont {Reyren}},
  \bibinfo {author} {\bibfnamefont {J.}~\bibnamefont {Sampaio}}, \bibinfo
  {author} {\bibfnamefont {C.~a.~F.}\ \bibnamefont {Vaz}}, \bibinfo {author}
  {\bibfnamefont {N.}~\bibnamefont {Van~Horne}}, \bibinfo {author}
  {\bibfnamefont {K.}~\bibnamefont {Bouzehouane}}, \bibinfo {author}
  {\bibfnamefont {K.}~\bibnamefont {Garcia}}, \bibinfo {author} {\bibfnamefont
  {C.}~\bibnamefont {Deranlot}}, \bibinfo {author} {\bibfnamefont
  {P.}~\bibnamefont {Warnicke}}, \bibinfo {author} {\bibfnamefont
  {P.}~\bibnamefont {Wohlh{\"u}ter}}, \bibinfo {author} {\bibfnamefont {J.-M.}\
  \bibnamefont {George}}, \bibinfo {author} {\bibfnamefont {M.}~\bibnamefont
  {Weigand}}, \bibinfo {author} {\bibfnamefont {J.}~\bibnamefont {Raabe}},
  \bibinfo {author} {\bibfnamefont {V.}~\bibnamefont {Cros}},\ and\ \bibinfo
  {author} {\bibfnamefont {A.}~\bibnamefont {Fert}},\ }\bibfield  {title}
  {\bibinfo {title} {Additive interfacial chiral interaction in multilayers for
  stabilization of small individual skyrmions at room temperature},\ }\href
  {https://doi.org/10.1038/nnano.2015.313} {\bibfield  {journal} {\bibinfo
  {journal} {Nature Nanotech}\ }\textbf {\bibinfo {volume} {11}},\ \bibinfo
  {pages} {444} (\bibinfo {year} {2016})}\BibitemShut {NoStop}%
\bibitem [{\citenamefont {Qin}\ \emph {et~al.}(2018)\citenamefont {Qin},
  \citenamefont {Wang}, \citenamefont {Zhu}, \citenamefont {Jin}, \citenamefont
  {Fu}, \citenamefont {Liu},\ and\ \citenamefont {Cao}}]{qin2018stabilization}%
  \BibitemOpen
  \bibfield  {author} {\bibinfo {author} {\bibfnamefont {Z.}~\bibnamefont
  {Qin}}, \bibinfo {author} {\bibfnamefont {Y.}~\bibnamefont {Wang}}, \bibinfo
  {author} {\bibfnamefont {S.}~\bibnamefont {Zhu}}, \bibinfo {author}
  {\bibfnamefont {C.}~\bibnamefont {Jin}}, \bibinfo {author} {\bibfnamefont
  {J.}~\bibnamefont {Fu}}, \bibinfo {author} {\bibfnamefont {Q.}~\bibnamefont
  {Liu}},\ and\ \bibinfo {author} {\bibfnamefont {J.}~\bibnamefont {Cao}},\
  }\bibfield  {title} {\bibinfo {title} {Stabilization and reversal of skyrmion
  lattice in ta/cofeb/mgo multilayers},\ }\href@noop {} {\bibfield  {journal}
  {\bibinfo  {journal} {ACS Appl. Mater. Interfaces}\ }\textbf {\bibinfo
  {volume} {10}},\ \bibinfo {pages} {36556} (\bibinfo {year}
  {2018})}\BibitemShut {NoStop}%
\bibitem [{\citenamefont {Zhang}\ \emph
  {et~al.}(2018{\natexlab{a}})\citenamefont {Zhang}, \citenamefont {Zhang},
  \citenamefont {Wen}, \citenamefont {Chudnovsky},\ and\ \citenamefont
  {Zhang}}]{zhang2018creation}%
  \BibitemOpen
  \bibfield  {author} {\bibinfo {author} {\bibfnamefont {S.}~\bibnamefont
  {Zhang}}, \bibinfo {author} {\bibfnamefont {J.}~\bibnamefont {Zhang}},
  \bibinfo {author} {\bibfnamefont {Y.}~\bibnamefont {Wen}}, \bibinfo {author}
  {\bibfnamefont {E.~M.}\ \bibnamefont {Chudnovsky}},\ and\ \bibinfo {author}
  {\bibfnamefont {X.}~\bibnamefont {Zhang}},\ }\bibfield  {title} {\bibinfo
  {title} {Creation of a thermally assisted skyrmion lattice in pt/co/ta
  multilayer films},\ }\href@noop {} {\bibfield  {journal} {\bibinfo  {journal}
  {Appl. Phys. Lett.}\ }\textbf {\bibinfo {volume} {113}} (\bibinfo {year}
  {2018}{\natexlab{a}})}\BibitemShut {NoStop}%
\bibitem [{\citenamefont {Wang}\ \emph
  {et~al.}(2019{\natexlab{a}})\citenamefont {Wang}, \citenamefont {Liu},
  \citenamefont {Mehmood}, \citenamefont {Han}, \citenamefont {Wang},
  \citenamefont {Xu}, \citenamefont {Feng}, \citenamefont {Hou}, \citenamefont
  {Peng}, \citenamefont {Gao} \emph {et~al.}}]{wang2019construction}%
  \BibitemOpen
  \bibfield  {author} {\bibinfo {author} {\bibfnamefont {L.}~\bibnamefont
  {Wang}}, \bibinfo {author} {\bibfnamefont {C.}~\bibnamefont {Liu}}, \bibinfo
  {author} {\bibfnamefont {N.}~\bibnamefont {Mehmood}}, \bibinfo {author}
  {\bibfnamefont {G.}~\bibnamefont {Han}}, \bibinfo {author} {\bibfnamefont
  {Y.}~\bibnamefont {Wang}}, \bibinfo {author} {\bibfnamefont {X.}~\bibnamefont
  {Xu}}, \bibinfo {author} {\bibfnamefont {C.}~\bibnamefont {Feng}}, \bibinfo
  {author} {\bibfnamefont {Z.}~\bibnamefont {Hou}}, \bibinfo {author}
  {\bibfnamefont {Y.}~\bibnamefont {Peng}}, \bibinfo {author} {\bibfnamefont
  {X.}~\bibnamefont {Gao}}, \emph {et~al.},\ }\bibfield  {title} {\bibinfo
  {title} {Construction of a room-temperature pt/co/ta multilayer film with
  ultrahigh-density skyrmions for memory application},\ }\href@noop {}
  {\bibfield  {journal} {\bibinfo  {journal} {ACS Appl. Mater. Interfaces}\
  }\textbf {\bibinfo {volume} {11}},\ \bibinfo {pages} {12098} (\bibinfo {year}
  {2019}{\natexlab{a}})}\BibitemShut {NoStop}%
\bibitem [{\citenamefont {Jaiswal}\ \emph {et~al.}(2017)\citenamefont
  {Jaiswal}, \citenamefont {Litzius}, \citenamefont {Lemesh}, \citenamefont
  {B{\"u}ttner}, \citenamefont {Finizio}, \citenamefont {Raabe}, \citenamefont
  {Weigand}, \citenamefont {Lee}, \citenamefont {Langer}, \citenamefont {Ocker}
  \emph {et~al.}}]{jaiswal2017investigation}%
  \BibitemOpen
  \bibfield  {author} {\bibinfo {author} {\bibfnamefont {S.}~\bibnamefont
  {Jaiswal}}, \bibinfo {author} {\bibfnamefont {K.}~\bibnamefont {Litzius}},
  \bibinfo {author} {\bibfnamefont {I.}~\bibnamefont {Lemesh}}, \bibinfo
  {author} {\bibfnamefont {F.}~\bibnamefont {B{\"u}ttner}}, \bibinfo {author}
  {\bibfnamefont {S.}~\bibnamefont {Finizio}}, \bibinfo {author} {\bibfnamefont
  {J.}~\bibnamefont {Raabe}}, \bibinfo {author} {\bibfnamefont
  {M.}~\bibnamefont {Weigand}}, \bibinfo {author} {\bibfnamefont
  {K.}~\bibnamefont {Lee}}, \bibinfo {author} {\bibfnamefont {J.}~\bibnamefont
  {Langer}}, \bibinfo {author} {\bibfnamefont {B.}~\bibnamefont {Ocker}}, \emph
  {et~al.},\ }\bibfield  {title} {\bibinfo {title} {Investigation of the
  dzyaloshinskii-moriya interaction and room temperature skyrmions in
  w/cofeb/mgo thin films and microwires},\ }\href@noop {} {\bibfield  {journal}
  {\bibinfo  {journal} {Appl. Phys. Lett.}\ }\textbf {\bibinfo {volume} {111}}
  (\bibinfo {year} {2017})}\BibitemShut {NoStop}%
\bibitem [{\citenamefont {Montoya}\ \emph {et~al.}(2017)\citenamefont
  {Montoya}, \citenamefont {Couture}, \citenamefont {Chess}, \citenamefont
  {Lee}, \citenamefont {Kent}, \citenamefont {Henze}, \citenamefont {Sinha},
  \citenamefont {Im}, \citenamefont {Kevan}, \citenamefont {Fischer} \emph
  {et~al.}}]{montoya2017tailoring}%
  \BibitemOpen
  \bibfield  {author} {\bibinfo {author} {\bibfnamefont {S.}~\bibnamefont
  {Montoya}}, \bibinfo {author} {\bibfnamefont {S.}~\bibnamefont {Couture}},
  \bibinfo {author} {\bibfnamefont {J.}~\bibnamefont {Chess}}, \bibinfo
  {author} {\bibfnamefont {J.}~\bibnamefont {Lee}}, \bibinfo {author}
  {\bibfnamefont {N.}~\bibnamefont {Kent}}, \bibinfo {author} {\bibfnamefont
  {D.}~\bibnamefont {Henze}}, \bibinfo {author} {\bibfnamefont
  {S.}~\bibnamefont {Sinha}}, \bibinfo {author} {\bibfnamefont {M.-Y.}\
  \bibnamefont {Im}}, \bibinfo {author} {\bibfnamefont {S.}~\bibnamefont
  {Kevan}}, \bibinfo {author} {\bibfnamefont {P.}~\bibnamefont {Fischer}},
  \emph {et~al.},\ }\bibfield  {title} {\bibinfo {title} {Tailoring magnetic
  energies to form dipole skyrmions and skyrmion lattices},\ }\href@noop {}
  {\bibfield  {journal} {\bibinfo  {journal} {Phys. Rev. B}\ }\textbf {\bibinfo
  {volume} {95}},\ \bibinfo {pages} {024415} (\bibinfo {year}
  {2017})}\BibitemShut {NoStop}%
\bibitem [{\citenamefont {Okubo}\ \emph {et~al.}(2012)\citenamefont {Okubo},
  \citenamefont {Chung},\ and\ \citenamefont
  {Kawamura}}]{Okubo_PhysRevLett.108.017206}%
  \BibitemOpen
  \bibfield  {author} {\bibinfo {author} {\bibfnamefont {T.}~\bibnamefont
  {Okubo}}, \bibinfo {author} {\bibfnamefont {S.}~\bibnamefont {Chung}},\ and\
  \bibinfo {author} {\bibfnamefont {H.}~\bibnamefont {Kawamura}},\ }\bibfield
  {title} {\bibinfo {title} {Multiple-$q$ states and the skyrmion lattice of
  the triangular-lattice {Heisenberg} antiferromagnet under magnetic fields},\
  }\href {https://doi.org/10.1103/PhysRevLett.108.017206} {\bibfield  {journal}
  {\bibinfo  {journal} {Phys. Rev. Lett.}\ }\textbf {\bibinfo {volume} {108}},\
  \bibinfo {pages} {017206} (\bibinfo {year} {2012})}\BibitemShut {NoStop}%
\bibitem [{\citenamefont {Kurumaji}\ \emph {et~al.}(2019)\citenamefont
  {Kurumaji}, \citenamefont {Nakajima}, \citenamefont {Hirschberger},
  \citenamefont {Kikkawa}, \citenamefont {Yamasaki}, \citenamefont {Sagayama},
  \citenamefont {Nakao}, \citenamefont {Taguchi}, \citenamefont {Arima},\ and\
  \citenamefont {Tokura}}]{kurumaji2019skyrmion}%
  \BibitemOpen
  \bibfield  {author} {\bibinfo {author} {\bibfnamefont {T.}~\bibnamefont
  {Kurumaji}}, \bibinfo {author} {\bibfnamefont {T.}~\bibnamefont {Nakajima}},
  \bibinfo {author} {\bibfnamefont {M.}~\bibnamefont {Hirschberger}}, \bibinfo
  {author} {\bibfnamefont {A.}~\bibnamefont {Kikkawa}}, \bibinfo {author}
  {\bibfnamefont {Y.}~\bibnamefont {Yamasaki}}, \bibinfo {author}
  {\bibfnamefont {H.}~\bibnamefont {Sagayama}}, \bibinfo {author}
  {\bibfnamefont {H.}~\bibnamefont {Nakao}}, \bibinfo {author} {\bibfnamefont
  {Y.}~\bibnamefont {Taguchi}}, \bibinfo {author} {\bibfnamefont {T.-h.}\
  \bibnamefont {Arima}},\ and\ \bibinfo {author} {\bibfnamefont
  {Y.}~\bibnamefont {Tokura}},\ }\bibfield  {title} {\bibinfo {title} {Skyrmion
  lattice with a giant topological {{Hall}} effect in a frustrated
  triangular-lattice magnet},\ }\href {https://doi.org/10.1126/science.aau0968}
  {\bibfield  {journal} {\bibinfo  {journal} {Science}\ }\textbf {\bibinfo
  {volume} {365}},\ \bibinfo {pages} {914} (\bibinfo {year}
  {2019})}\BibitemShut {NoStop}%
\bibitem [{\citenamefont {Milde}\ \emph {et~al.}(2013)\citenamefont {Milde},
  \citenamefont {K{\"o}hler}, \citenamefont {Seidel}, \citenamefont {Eng},
  \citenamefont {Bauer}, \citenamefont {Chacon}, \citenamefont {Kindervater},
  \citenamefont {M{\"u}hlbauer}, \citenamefont {Pfleiderer}, \citenamefont
  {Buhrandt} \emph {et~al.}}]{milde2013unwinding}%
  \BibitemOpen
  \bibfield  {author} {\bibinfo {author} {\bibfnamefont {P.}~\bibnamefont
  {Milde}}, \bibinfo {author} {\bibfnamefont {D.}~\bibnamefont {K{\"o}hler}},
  \bibinfo {author} {\bibfnamefont {J.}~\bibnamefont {Seidel}}, \bibinfo
  {author} {\bibfnamefont {L.}~\bibnamefont {Eng}}, \bibinfo {author}
  {\bibfnamefont {A.}~\bibnamefont {Bauer}}, \bibinfo {author} {\bibfnamefont
  {A.}~\bibnamefont {Chacon}}, \bibinfo {author} {\bibfnamefont
  {J.}~\bibnamefont {Kindervater}}, \bibinfo {author} {\bibfnamefont
  {S.}~\bibnamefont {M{\"u}hlbauer}}, \bibinfo {author} {\bibfnamefont
  {C.}~\bibnamefont {Pfleiderer}}, \bibinfo {author} {\bibfnamefont
  {S.}~\bibnamefont {Buhrandt}}, \emph {et~al.},\ }\bibfield  {title} {\bibinfo
  {title} {Unwinding of a skyrmion lattice by magnetic monopoles},\ }\href@noop
  {} {\bibfield  {journal} {\bibinfo  {journal} {Science}\ }\textbf {\bibinfo
  {volume} {340}},\ \bibinfo {pages} {1076} (\bibinfo {year}
  {2013})}\BibitemShut {NoStop}%
\bibitem [{\citenamefont {Woo}\ \emph {et~al.}(2018{\natexlab{a}})\citenamefont
  {Woo}, \citenamefont {Song}, \citenamefont {Zhang}, \citenamefont {Zhou},
  \citenamefont {Ezawa}, \citenamefont {Liu}, \citenamefont {Finizio},
  \citenamefont {Raabe}, \citenamefont {Lee}, \citenamefont {Kim} \emph
  {et~al.}}]{woo2018current}%
  \BibitemOpen
  \bibfield  {author} {\bibinfo {author} {\bibfnamefont {S.}~\bibnamefont
  {Woo}}, \bibinfo {author} {\bibfnamefont {K.~M.}\ \bibnamefont {Song}},
  \bibinfo {author} {\bibfnamefont {X.}~\bibnamefont {Zhang}}, \bibinfo
  {author} {\bibfnamefont {Y.}~\bibnamefont {Zhou}}, \bibinfo {author}
  {\bibfnamefont {M.}~\bibnamefont {Ezawa}}, \bibinfo {author} {\bibfnamefont
  {X.}~\bibnamefont {Liu}}, \bibinfo {author} {\bibfnamefont {S.}~\bibnamefont
  {Finizio}}, \bibinfo {author} {\bibfnamefont {J.}~\bibnamefont {Raabe}},
  \bibinfo {author} {\bibfnamefont {N.~J.}\ \bibnamefont {Lee}}, \bibinfo
  {author} {\bibfnamefont {S.-I.}\ \bibnamefont {Kim}}, \emph {et~al.},\
  }\bibfield  {title} {\bibinfo {title} {Current-driven dynamics and inhibition
  of the skyrmion {H}all effect of ferrimagnetic skyrmions in gdfeco films},\
  }\href@noop {} {\bibfield  {journal} {\bibinfo  {journal} {Nat. Commun.}\
  }\textbf {\bibinfo {volume} {9}},\ \bibinfo {pages} {959} (\bibinfo {year}
  {2018}{\natexlab{a}})}\BibitemShut {NoStop}%
\bibitem [{\citenamefont {Li}\ \emph {et~al.}(2019)\citenamefont {Li},
  \citenamefont {Bykova}, \citenamefont {Zhang}, \citenamefont {Yu},
  \citenamefont {Tomasello}, \citenamefont {Carpentieri}, \citenamefont {Liu},
  \citenamefont {Guang}, \citenamefont {Gr{\"a}fe}, \citenamefont {Weigand}
  \emph {et~al.}}]{li2019anatomy}%
  \BibitemOpen
  \bibfield  {author} {\bibinfo {author} {\bibfnamefont {W.}~\bibnamefont
  {Li}}, \bibinfo {author} {\bibfnamefont {I.}~\bibnamefont {Bykova}}, \bibinfo
  {author} {\bibfnamefont {S.}~\bibnamefont {Zhang}}, \bibinfo {author}
  {\bibfnamefont {G.}~\bibnamefont {Yu}}, \bibinfo {author} {\bibfnamefont
  {R.}~\bibnamefont {Tomasello}}, \bibinfo {author} {\bibfnamefont
  {M.}~\bibnamefont {Carpentieri}}, \bibinfo {author} {\bibfnamefont
  {Y.}~\bibnamefont {Liu}}, \bibinfo {author} {\bibfnamefont {Y.}~\bibnamefont
  {Guang}}, \bibinfo {author} {\bibfnamefont {J.}~\bibnamefont {Gr{\"a}fe}},
  \bibinfo {author} {\bibfnamefont {M.}~\bibnamefont {Weigand}}, \emph
  {et~al.},\ }\bibfield  {title} {\bibinfo {title} {Anatomy of skyrmionic
  textures in magnetic multilayers},\ }\href@noop {} {\bibfield  {journal}
  {\bibinfo  {journal} {Adv. Mater.}\ }\textbf {\bibinfo {volume} {31}},\
  \bibinfo {pages} {1807683} (\bibinfo {year} {2019})}\BibitemShut {NoStop}%
\bibitem [{\citenamefont {Turnbull}\ \emph {et~al.}(2020)\citenamefont
  {Turnbull}, \citenamefont {Birch}, \citenamefont {Laurenson}, \citenamefont
  {Bukin}, \citenamefont {Burgos-Parra}, \citenamefont {Popescu}, \citenamefont
  {Wilson}, \citenamefont {Stefan{\v c}i{\v c}}, \citenamefont {Balakrishnan},
  \citenamefont {Ogrin} \emph {et~al.}}]{turnbull2020tilted}%
  \BibitemOpen
  \bibfield  {author} {\bibinfo {author} {\bibfnamefont {L.~A.}\ \bibnamefont
  {Turnbull}}, \bibinfo {author} {\bibfnamefont {M.~T.}\ \bibnamefont {Birch}},
  \bibinfo {author} {\bibfnamefont {A.}~\bibnamefont {Laurenson}}, \bibinfo
  {author} {\bibfnamefont {N.}~\bibnamefont {Bukin}}, \bibinfo {author}
  {\bibfnamefont {E.~O.}\ \bibnamefont {Burgos-Parra}}, \bibinfo {author}
  {\bibfnamefont {H.}~\bibnamefont {Popescu}}, \bibinfo {author} {\bibfnamefont
  {M.~N.}\ \bibnamefont {Wilson}}, \bibinfo {author} {\bibfnamefont
  {A.}~\bibnamefont {Stefan{\v c}i{\v c}}}, \bibinfo {author} {\bibfnamefont
  {G.}~\bibnamefont {Balakrishnan}}, \bibinfo {author} {\bibfnamefont {F.~Y.}\
  \bibnamefont {Ogrin}}, \emph {et~al.},\ }\bibfield  {title} {\bibinfo {title}
  {Tilted x-ray holography of magnetic bubbles in mnniga lamellae},\
  }\href@noop {} {\bibfield  {journal} {\bibinfo  {journal} {ACS Nano}\
  }\textbf {\bibinfo {volume} {15}},\ \bibinfo {pages} {387} (\bibinfo {year}
  {2020})}\BibitemShut {NoStop}%
\bibitem [{\citenamefont {Grelier}\ \emph {et~al.}(2023)\citenamefont
  {Grelier}, \citenamefont {Godel}, \citenamefont {Vecchiola}, \citenamefont
  {Collin}, \citenamefont {Bouzehouane}, \citenamefont {Cros}, \citenamefont
  {Reyren}, \citenamefont {Battistelli}, \citenamefont {Popescu}, \citenamefont
  {L{\'e}veill{\'e}}, \citenamefont {Jaouen},\ and\ \citenamefont
  {B{\"u}ttner}}]{grelier2023xray}%
  \BibitemOpen
  \bibfield  {author} {\bibinfo {author} {\bibfnamefont {M.}~\bibnamefont
  {Grelier}}, \bibinfo {author} {\bibfnamefont {F.}~\bibnamefont {Godel}},
  \bibinfo {author} {\bibfnamefont {A.}~\bibnamefont {Vecchiola}}, \bibinfo
  {author} {\bibfnamefont {S.}~\bibnamefont {Collin}}, \bibinfo {author}
  {\bibfnamefont {K.}~\bibnamefont {Bouzehouane}}, \bibinfo {author}
  {\bibfnamefont {V.}~\bibnamefont {Cros}}, \bibinfo {author} {\bibfnamefont
  {N.}~\bibnamefont {Reyren}}, \bibinfo {author} {\bibfnamefont
  {R.}~\bibnamefont {Battistelli}}, \bibinfo {author} {\bibfnamefont
  {H.}~\bibnamefont {Popescu}}, \bibinfo {author} {\bibfnamefont
  {C.}~\bibnamefont {L{\'e}veill{\'e}}}, \bibinfo {author} {\bibfnamefont
  {N.}~\bibnamefont {Jaouen}},\ and\ \bibinfo {author} {\bibfnamefont
  {F.}~\bibnamefont {B{\"u}ttner}},\ }\bibfield  {title} {\bibinfo {title}
  {X-ray holography of skyrmionic cocoons in aperiodic magnetic multilayers},\
  }\href {https://doi.org/10.1103/PhysRevB.107.L220405} {\bibfield  {journal}
  {\bibinfo  {journal} {Phys. Rev. B}\ }\textbf {\bibinfo {volume} {107}},\
  \bibinfo {pages} {L220405} (\bibinfo {year} {2023})}\BibitemShut {NoStop}%
\bibitem [{\citenamefont {Donnelly}\ \emph {et~al.}(2017)\citenamefont
  {Donnelly}, \citenamefont {Guizar-Sicairos}, \citenamefont {Scagnoli},
  \citenamefont {Gliga}, \citenamefont {Holler}, \citenamefont {Raabe},\ and\
  \citenamefont {Heyderman}}]{donnelly2017three}%
  \BibitemOpen
  \bibfield  {author} {\bibinfo {author} {\bibfnamefont {C.}~\bibnamefont
  {Donnelly}}, \bibinfo {author} {\bibfnamefont {M.}~\bibnamefont
  {Guizar-Sicairos}}, \bibinfo {author} {\bibfnamefont {V.}~\bibnamefont
  {Scagnoli}}, \bibinfo {author} {\bibfnamefont {S.}~\bibnamefont {Gliga}},
  \bibinfo {author} {\bibfnamefont {M.}~\bibnamefont {Holler}}, \bibinfo
  {author} {\bibfnamefont {J.}~\bibnamefont {Raabe}},\ and\ \bibinfo {author}
  {\bibfnamefont {L.~J.}\ \bibnamefont {Heyderman}},\ }\bibfield  {title}
  {\bibinfo {title} {Three-dimensional magnetization structures revealed with
  x-ray vector nanotomography},\ }\href@noop {} {\bibfield  {journal} {\bibinfo
   {journal} {Nature}\ }\textbf {\bibinfo {volume} {547}},\ \bibinfo {pages}
  {328} (\bibinfo {year} {2017})}\BibitemShut {NoStop}%
\bibitem [{\citenamefont {Donnelly}\ \emph {et~al.}(2016)\citenamefont
  {Donnelly}, \citenamefont {Scagnoli}, \citenamefont {Guizar-Sicairos},
  \citenamefont {Holler}, \citenamefont {Wilhelm}, \citenamefont {Guillou},
  \citenamefont {Rogalev}, \citenamefont {Detlefs}, \citenamefont {Menzel},
  \citenamefont {Raabe} \emph {et~al.}}]{donnelly2016high}%
  \BibitemOpen
  \bibfield  {author} {\bibinfo {author} {\bibfnamefont {C.}~\bibnamefont
  {Donnelly}}, \bibinfo {author} {\bibfnamefont {V.}~\bibnamefont {Scagnoli}},
  \bibinfo {author} {\bibfnamefont {M.}~\bibnamefont {Guizar-Sicairos}},
  \bibinfo {author} {\bibfnamefont {M.}~\bibnamefont {Holler}}, \bibinfo
  {author} {\bibfnamefont {F.}~\bibnamefont {Wilhelm}}, \bibinfo {author}
  {\bibfnamefont {F.}~\bibnamefont {Guillou}}, \bibinfo {author} {\bibfnamefont
  {A.}~\bibnamefont {Rogalev}}, \bibinfo {author} {\bibfnamefont
  {C.}~\bibnamefont {Detlefs}}, \bibinfo {author} {\bibfnamefont
  {A.}~\bibnamefont {Menzel}}, \bibinfo {author} {\bibfnamefont
  {J.}~\bibnamefont {Raabe}}, \emph {et~al.},\ }\bibfield  {title} {\bibinfo
  {title} {High-resolution hard x-ray magnetic imaging with dichroic
  ptychography},\ }\href@noop {} {\bibfield  {journal} {\bibinfo  {journal}
  {Phys. Rev. B}\ }\textbf {\bibinfo {volume} {94}},\ \bibinfo {pages} {064421}
  (\bibinfo {year} {2016})}\BibitemShut {NoStop}%
\bibitem [{\citenamefont {McVitie}\ \emph {et~al.}(2018)\citenamefont
  {McVitie}, \citenamefont {Hughes}, \citenamefont {Fallon}, \citenamefont
  {McFadzean}, \citenamefont {McGrouther}, \citenamefont {Krajnak},
  \citenamefont {Legrand}, \citenamefont {Maccariello}, \citenamefont {Collin},
  \citenamefont {Garcia} \emph {et~al.}}]{mcvitie2018transmission}%
  \BibitemOpen
  \bibfield  {author} {\bibinfo {author} {\bibfnamefont {S.}~\bibnamefont
  {McVitie}}, \bibinfo {author} {\bibfnamefont {S.}~\bibnamefont {Hughes}},
  \bibinfo {author} {\bibfnamefont {K.}~\bibnamefont {Fallon}}, \bibinfo
  {author} {\bibfnamefont {S.}~\bibnamefont {McFadzean}}, \bibinfo {author}
  {\bibfnamefont {D.}~\bibnamefont {McGrouther}}, \bibinfo {author}
  {\bibfnamefont {M.}~\bibnamefont {Krajnak}}, \bibinfo {author} {\bibfnamefont
  {W.}~\bibnamefont {Legrand}}, \bibinfo {author} {\bibfnamefont
  {D.}~\bibnamefont {Maccariello}}, \bibinfo {author} {\bibfnamefont
  {S.}~\bibnamefont {Collin}}, \bibinfo {author} {\bibfnamefont
  {K.}~\bibnamefont {Garcia}}, \emph {et~al.},\ }\bibfield  {title} {\bibinfo
  {title} {A transmission electron microscope study of n{\'e}el skyrmion
  magnetic textures in multilayer thin film systems with large interfacial
  chiral interaction},\ }\href@noop {} {\bibfield  {journal} {\bibinfo
  {journal} {Sci. Rep.}\ }\textbf {\bibinfo {volume} {8}},\ \bibinfo {pages}
  {5703} (\bibinfo {year} {2018})}\BibitemShut {NoStop}%
\bibitem [{\citenamefont {Denneulin}\ \emph {et~al.}(2021)\citenamefont
  {Denneulin}, \citenamefont {Caron}, \citenamefont {Hoffmann}, \citenamefont
  {Lin}, \citenamefont {Tan}, \citenamefont {Kov{\'a}cs}, \citenamefont
  {Bl{\"u}gel},\ and\ \citenamefont {Dunin-Borkowski}}]{denneulin2021off}%
  \BibitemOpen
  \bibfield  {author} {\bibinfo {author} {\bibfnamefont {T.}~\bibnamefont
  {Denneulin}}, \bibinfo {author} {\bibfnamefont {J.}~\bibnamefont {Caron}},
  \bibinfo {author} {\bibfnamefont {M.}~\bibnamefont {Hoffmann}}, \bibinfo
  {author} {\bibfnamefont {M.}~\bibnamefont {Lin}}, \bibinfo {author}
  {\bibfnamefont {H.}~\bibnamefont {Tan}}, \bibinfo {author} {\bibfnamefont
  {A.}~\bibnamefont {Kov{\'a}cs}}, \bibinfo {author} {\bibfnamefont
  {S.}~\bibnamefont {Bl{\"u}gel}},\ and\ \bibinfo {author} {\bibfnamefont
  {R.}~\bibnamefont {Dunin-Borkowski}},\ }\bibfield  {title} {\bibinfo {title}
  {Off-axis electron holography of n{\'e}el-type skyrmions in multilayers of
  heavy metals and ferromagnets},\ }\href@noop {} {\bibfield  {journal}
  {\bibinfo  {journal} {Ultramicroscopy}\ }\textbf {\bibinfo {volume} {220}},\
  \bibinfo {pages} {113155} (\bibinfo {year} {2021})}\BibitemShut {NoStop}%
\bibitem [{\citenamefont {Langner}\ \emph {et~al.}(2014)\citenamefont
  {Langner}, \citenamefont {Roy}, \citenamefont {Mishra}, \citenamefont {Lee},
  \citenamefont {Shi}, \citenamefont {Hossain}, \citenamefont {Chuang},
  \citenamefont {Seki}, \citenamefont {Tokura}, \citenamefont {Kevan} \emph
  {et~al.}}]{langner2014coupled}%
  \BibitemOpen
  \bibfield  {author} {\bibinfo {author} {\bibfnamefont {M.}~\bibnamefont
  {Langner}}, \bibinfo {author} {\bibfnamefont {S.}~\bibnamefont {Roy}},
  \bibinfo {author} {\bibfnamefont {S.}~\bibnamefont {Mishra}}, \bibinfo
  {author} {\bibfnamefont {J.}~\bibnamefont {Lee}}, \bibinfo {author}
  {\bibfnamefont {X.}~\bibnamefont {Shi}}, \bibinfo {author} {\bibfnamefont
  {M.}~\bibnamefont {Hossain}}, \bibinfo {author} {\bibfnamefont {Y.-D.}\
  \bibnamefont {Chuang}}, \bibinfo {author} {\bibfnamefont {S.}~\bibnamefont
  {Seki}}, \bibinfo {author} {\bibfnamefont {Y.}~\bibnamefont {Tokura}},
  \bibinfo {author} {\bibfnamefont {S.}~\bibnamefont {Kevan}}, \emph {et~al.},\
  }\bibfield  {title} {\bibinfo {title} {Coupled skyrmion sublattices in cu 2
  oseo 3},\ }\href@noop {} {\bibfield  {journal} {\bibinfo  {journal} {Phys.
  Rev. Lett.}\ }\textbf {\bibinfo {volume} {112}},\ \bibinfo {pages} {167202}
  (\bibinfo {year} {2014})}\BibitemShut {NoStop}%
\bibitem [{\citenamefont {Nagaosa}\ and\ \citenamefont
  {Tokura}(2013)}]{nagaosa2013topological}%
  \BibitemOpen
  \bibfield  {author} {\bibinfo {author} {\bibfnamefont {N.}~\bibnamefont
  {Nagaosa}}\ and\ \bibinfo {author} {\bibfnamefont {Y.}~\bibnamefont
  {Tokura}},\ }\bibfield  {title} {\bibinfo {title} {Topological properties and
  dynamics of magnetic skyrmions},\ }\href
  {https://doi.org/10.1038/nnano.2013.243} {\bibfield  {journal} {\bibinfo
  {journal} {Nature Nanotech}\ }\textbf {\bibinfo {volume} {8}},\ \bibinfo
  {pages} {899} (\bibinfo {year} {2013})}\BibitemShut {NoStop}%
\bibitem [{\citenamefont {Neubauer}\ \emph {et~al.}(2009)\citenamefont
  {Neubauer}, \citenamefont {Pfleiderer}, \citenamefont {Binz}, \citenamefont
  {Rosch}, \citenamefont {Ritz}, \citenamefont {Niklowitz},\ and\ \citenamefont
  {B{\"o}ni}}]{neubauer2009topological}%
  \BibitemOpen
  \bibfield  {author} {\bibinfo {author} {\bibfnamefont {A.}~\bibnamefont
  {Neubauer}}, \bibinfo {author} {\bibfnamefont {C.}~\bibnamefont
  {Pfleiderer}}, \bibinfo {author} {\bibfnamefont {B.}~\bibnamefont {Binz}},
  \bibinfo {author} {\bibfnamefont {A.}~\bibnamefont {Rosch}}, \bibinfo
  {author} {\bibfnamefont {R.}~\bibnamefont {Ritz}}, \bibinfo {author}
  {\bibfnamefont {P.~G.}\ \bibnamefont {Niklowitz}},\ and\ \bibinfo {author}
  {\bibfnamefont {P.}~\bibnamefont {B{\"o}ni}},\ }\bibfield  {title} {\bibinfo
  {title} {Topological {{Hall Effect}} in the \${{A}}\$ {{Phase}} of
  {{MnSi}}},\ }\href {https://doi.org/10.1103/PhysRevLett.102.186602}
  {\bibfield  {journal} {\bibinfo  {journal} {Phys. Rev. Lett.}\ }\textbf
  {\bibinfo {volume} {102}},\ \bibinfo {pages} {186602} (\bibinfo {year}
  {2009})}\BibitemShut {NoStop}%
\bibitem [{\citenamefont {Shiomi}\ \emph {et~al.}(2013)\citenamefont {Shiomi},
  \citenamefont {Kanazawa}, \citenamefont {Shibata}, \citenamefont {Onose},\
  and\ \citenamefont {Tokura}}]{shiomi2013topological}%
  \BibitemOpen
  \bibfield  {author} {\bibinfo {author} {\bibfnamefont {Y.}~\bibnamefont
  {Shiomi}}, \bibinfo {author} {\bibfnamefont {N.}~\bibnamefont {Kanazawa}},
  \bibinfo {author} {\bibfnamefont {K.}~\bibnamefont {Shibata}}, \bibinfo
  {author} {\bibfnamefont {Y.}~\bibnamefont {Onose}},\ and\ \bibinfo {author}
  {\bibfnamefont {Y.}~\bibnamefont {Tokura}},\ }\bibfield  {title} {\bibinfo
  {title} {Topological nernst effect in a three-dimensional skyrmion-lattice
  phase},\ }\href@noop {} {\bibfield  {journal} {\bibinfo  {journal} {Phys.
  Rev. B—Condensed Matter and Materials Physics}\ }\textbf {\bibinfo {volume}
  {88}},\ \bibinfo {pages} {064409} (\bibinfo {year} {2013})}\BibitemShut
  {NoStop}%
\bibitem [{\citenamefont {Koshibae}\ \emph {et~al.}(2015)\citenamefont
  {Koshibae}, \citenamefont {Kaneko}, \citenamefont {Iwasaki}, \citenamefont
  {Kawasaki}, \citenamefont {Tokura},\ and\ \citenamefont
  {Nagaosa}}]{koshibae2015memory}%
  \BibitemOpen
  \bibfield  {author} {\bibinfo {author} {\bibfnamefont {W.}~\bibnamefont
  {Koshibae}}, \bibinfo {author} {\bibfnamefont {Y.}~\bibnamefont {Kaneko}},
  \bibinfo {author} {\bibfnamefont {J.}~\bibnamefont {Iwasaki}}, \bibinfo
  {author} {\bibfnamefont {M.}~\bibnamefont {Kawasaki}}, \bibinfo {author}
  {\bibfnamefont {Y.}~\bibnamefont {Tokura}},\ and\ \bibinfo {author}
  {\bibfnamefont {N.}~\bibnamefont {Nagaosa}},\ }\bibfield  {title} {\bibinfo
  {title} {Memory functions of magnetic skyrmions},\ }\href@noop {} {\bibfield
  {journal} {\bibinfo  {journal} {Jpn. J. Appl. Phys.}\ }\textbf {\bibinfo
  {volume} {54}},\ \bibinfo {pages} {053001} (\bibinfo {year}
  {2015})}\BibitemShut {NoStop}%
\bibitem [{\citenamefont {Penthorn}\ \emph {et~al.}(2019)\citenamefont
  {Penthorn}, \citenamefont {Hao}, \citenamefont {Wang}, \citenamefont {Huai},\
  and\ \citenamefont {Jiang}}]{penthorn2019experimental}%
  \BibitemOpen
  \bibfield  {author} {\bibinfo {author} {\bibfnamefont {N.}~\bibnamefont
  {Penthorn}}, \bibinfo {author} {\bibfnamefont {X.}~\bibnamefont {Hao}},
  \bibinfo {author} {\bibfnamefont {Z.}~\bibnamefont {Wang}}, \bibinfo {author}
  {\bibfnamefont {Y.}~\bibnamefont {Huai}},\ and\ \bibinfo {author}
  {\bibfnamefont {H.}~\bibnamefont {Jiang}},\ }\bibfield  {title} {\bibinfo
  {title} {Experimental observation of single skyrmion signatures in a magnetic
  tunnel junction},\ }\href@noop {} {\bibfield  {journal} {\bibinfo  {journal}
  {Phys. Rev. Lett.}\ }\textbf {\bibinfo {volume} {122}},\ \bibinfo {pages}
  {257201} (\bibinfo {year} {2019})}\BibitemShut {NoStop}%
\bibitem [{\citenamefont {Guang}\ \emph {et~al.}(2023)\citenamefont {Guang},
  \citenamefont {Zhang}, \citenamefont {Zhang}, \citenamefont {Wang},
  \citenamefont {Zhao}, \citenamefont {Tomasello}, \citenamefont {Zhang},
  \citenamefont {He}, \citenamefont {Li}, \citenamefont {Liu} \emph
  {et~al.}}]{guang2023electrical}%
  \BibitemOpen
  \bibfield  {author} {\bibinfo {author} {\bibfnamefont {Y.}~\bibnamefont
  {Guang}}, \bibinfo {author} {\bibfnamefont {L.}~\bibnamefont {Zhang}},
  \bibinfo {author} {\bibfnamefont {J.}~\bibnamefont {Zhang}}, \bibinfo
  {author} {\bibfnamefont {Y.}~\bibnamefont {Wang}}, \bibinfo {author}
  {\bibfnamefont {Y.}~\bibnamefont {Zhao}}, \bibinfo {author} {\bibfnamefont
  {R.}~\bibnamefont {Tomasello}}, \bibinfo {author} {\bibfnamefont
  {S.}~\bibnamefont {Zhang}}, \bibinfo {author} {\bibfnamefont
  {B.}~\bibnamefont {He}}, \bibinfo {author} {\bibfnamefont {J.}~\bibnamefont
  {Li}}, \bibinfo {author} {\bibfnamefont {Y.}~\bibnamefont {Liu}}, \emph
  {et~al.},\ }\bibfield  {title} {\bibinfo {title} {Electrical detection of
  magnetic skyrmions in a magnetic tunnel junction},\ }\href@noop {} {\bibfield
   {journal} {\bibinfo  {journal} {Adv. Electron. Mater.}\ }\textbf {\bibinfo
  {volume} {9}},\ \bibinfo {pages} {2200570} (\bibinfo {year}
  {2023})}\BibitemShut {NoStop}%
\bibitem [{\citenamefont {Parkin}\ \emph {et~al.}(2008)\citenamefont {Parkin},
  \citenamefont {Hayashi},\ and\ \citenamefont {Thomas}}]{parkin2008magnetic}%
  \BibitemOpen
  \bibfield  {author} {\bibinfo {author} {\bibfnamefont {S.~S.~P.}\
  \bibnamefont {Parkin}}, \bibinfo {author} {\bibfnamefont {M.}~\bibnamefont
  {Hayashi}},\ and\ \bibinfo {author} {\bibfnamefont {L.}~\bibnamefont
  {Thomas}},\ }\bibfield  {title} {\bibinfo {title} {Magnetic {{Domain-Wall
  Racetrack Memory}}},\ }\href {https://doi.org/10.1126/science.1145799}
  {\bibfield  {journal} {\bibinfo  {journal} {Science}\ }\textbf {\bibinfo
  {volume} {320}},\ \bibinfo {pages} {190} (\bibinfo {year}
  {2008})}\BibitemShut {NoStop}%
\bibitem [{\citenamefont {Fert}\ \emph {et~al.}(2013)\citenamefont {Fert},
  \citenamefont {Cros},\ and\ \citenamefont {Sampaio}}]{fert2013skyrmions}%
  \BibitemOpen
  \bibfield  {author} {\bibinfo {author} {\bibfnamefont {A.}~\bibnamefont
  {Fert}}, \bibinfo {author} {\bibfnamefont {V.}~\bibnamefont {Cros}},\ and\
  \bibinfo {author} {\bibfnamefont {J.}~\bibnamefont {Sampaio}},\ }\bibfield
  {title} {\bibinfo {title} {Skyrmions on the track},\ }\href
  {https://doi.org/10.1038/nnano.2013.29} {\bibfield  {journal} {\bibinfo
  {journal} {Nature Nanotech}\ }\textbf {\bibinfo {volume} {8}},\ \bibinfo
  {pages} {152} (\bibinfo {year} {2013})}\BibitemShut {NoStop}%
\bibitem [{\citenamefont {Zhao}\ \emph
  {et~al.}(2024{\natexlab{a}})\citenamefont {Zhao}, \citenamefont {Hua},
  \citenamefont {Song}, \citenamefont {Yu},\ and\ \citenamefont
  {Jiang}}]{zhao2024realization}%
  \BibitemOpen
  \bibfield  {author} {\bibinfo {author} {\bibfnamefont {L.}~\bibnamefont
  {Zhao}}, \bibinfo {author} {\bibfnamefont {C.}~\bibnamefont {Hua}}, \bibinfo
  {author} {\bibfnamefont {C.}~\bibnamefont {Song}}, \bibinfo {author}
  {\bibfnamefont {W.}~\bibnamefont {Yu}},\ and\ \bibinfo {author}
  {\bibfnamefont {W.}~\bibnamefont {Jiang}},\ }\bibfield  {title} {\bibinfo
  {title} {Realization of skyrmion shift register},\ }\href
  {https://doi.org/10.1016/j.scib.2024.05.035} {\bibfield  {journal} {\bibinfo
  {journal} {Sci. Bull.}\ }\textbf {\bibinfo {volume} {69}},\ \bibinfo {pages}
  {2370} (\bibinfo {year} {2024}{\natexlab{a}})}\BibitemShut {NoStop}%
\bibitem [{\citenamefont {Litzius}\ \emph {et~al.}(2017)\citenamefont
  {Litzius}, \citenamefont {Lemesh}, \citenamefont {Kr{\"u}ger}, \citenamefont
  {Bassirian}, \citenamefont {Caretta}, \citenamefont {Richter}, \citenamefont
  {B{\"u}ttner}, \citenamefont {Sato}, \citenamefont {Tretiakov}, \citenamefont
  {F{\"o}rster} \emph {et~al.}}]{litzius2017skyrmion}%
  \BibitemOpen
  \bibfield  {author} {\bibinfo {author} {\bibfnamefont {K.}~\bibnamefont
  {Litzius}}, \bibinfo {author} {\bibfnamefont {I.}~\bibnamefont {Lemesh}},
  \bibinfo {author} {\bibfnamefont {B.}~\bibnamefont {Kr{\"u}ger}}, \bibinfo
  {author} {\bibfnamefont {P.}~\bibnamefont {Bassirian}}, \bibinfo {author}
  {\bibfnamefont {L.}~\bibnamefont {Caretta}}, \bibinfo {author} {\bibfnamefont
  {K.}~\bibnamefont {Richter}}, \bibinfo {author} {\bibfnamefont
  {F.}~\bibnamefont {B{\"u}ttner}}, \bibinfo {author} {\bibfnamefont
  {K.}~\bibnamefont {Sato}}, \bibinfo {author} {\bibfnamefont {O.~A.}\
  \bibnamefont {Tretiakov}}, \bibinfo {author} {\bibfnamefont {J.}~\bibnamefont
  {F{\"o}rster}}, \emph {et~al.},\ }\bibfield  {title} {\bibinfo {title}
  {Skyrmion {H}all effect revealed by direct time-resolved x-ray microscopy},\
  }\href@noop {} {\bibfield  {journal} {\bibinfo  {journal} {Nat. Phys.}\
  }\textbf {\bibinfo {volume} {13}},\ \bibinfo {pages} {170} (\bibinfo {year}
  {2017})}\BibitemShut {NoStop}%
\bibitem [{\citenamefont {Jiang}\ \emph {et~al.}(2017)\citenamefont {Jiang},
  \citenamefont {Zhang}, \citenamefont {Yu}, \citenamefont {Zhang},
  \citenamefont {Wang}, \citenamefont {Benjamin~Jungfleisch}, \citenamefont
  {Pearson}, \citenamefont {Cheng}, \citenamefont {Heinonen}, \citenamefont
  {Wang} \emph {et~al.}}]{jiang2017direct}%
  \BibitemOpen
  \bibfield  {author} {\bibinfo {author} {\bibfnamefont {W.}~\bibnamefont
  {Jiang}}, \bibinfo {author} {\bibfnamefont {X.}~\bibnamefont {Zhang}},
  \bibinfo {author} {\bibfnamefont {G.}~\bibnamefont {Yu}}, \bibinfo {author}
  {\bibfnamefont {W.}~\bibnamefont {Zhang}}, \bibinfo {author} {\bibfnamefont
  {X.}~\bibnamefont {Wang}}, \bibinfo {author} {\bibfnamefont {M.}~\bibnamefont
  {Benjamin~Jungfleisch}}, \bibinfo {author} {\bibfnamefont {J.~E.}\
  \bibnamefont {Pearson}}, \bibinfo {author} {\bibfnamefont {X.}~\bibnamefont
  {Cheng}}, \bibinfo {author} {\bibfnamefont {O.}~\bibnamefont {Heinonen}},
  \bibinfo {author} {\bibfnamefont {K.~L.}\ \bibnamefont {Wang}}, \emph
  {et~al.},\ }\bibfield  {title} {\bibinfo {title} {Direct observation of the
  skyrmion {H}all effect},\ }\href@noop {} {\bibfield  {journal} {\bibinfo
  {journal} {Nat. Phys.}\ }\textbf {\bibinfo {volume} {13}},\ \bibinfo {pages}
  {162} (\bibinfo {year} {2017})}\BibitemShut {NoStop}%
\bibitem [{\citenamefont {Tomasello}\ \emph {et~al.}(2014)\citenamefont
  {Tomasello}, \citenamefont {Martinez}, \citenamefont {Zivieri}, \citenamefont
  {Torres}, \citenamefont {Carpentieri},\ and\ \citenamefont
  {Finocchio}}]{tomasello2014strategy}%
  \BibitemOpen
  \bibfield  {author} {\bibinfo {author} {\bibfnamefont {R.}~\bibnamefont
  {Tomasello}}, \bibinfo {author} {\bibfnamefont {E.}~\bibnamefont {Martinez}},
  \bibinfo {author} {\bibfnamefont {R.}~\bibnamefont {Zivieri}}, \bibinfo
  {author} {\bibfnamefont {L.}~\bibnamefont {Torres}}, \bibinfo {author}
  {\bibfnamefont {M.}~\bibnamefont {Carpentieri}},\ and\ \bibinfo {author}
  {\bibfnamefont {G.}~\bibnamefont {Finocchio}},\ }\bibfield  {title} {\bibinfo
  {title} {A strategy for the design of skyrmion racetrack memories},\ }\href
  {https://doi.org/10.1038/srep06784} {\bibfield  {journal} {\bibinfo
  {journal} {Sci Rep}\ }\textbf {\bibinfo {volume} {4}},\ \bibinfo {pages}
  {6784} (\bibinfo {year} {2014})}\BibitemShut {NoStop}%
\bibitem [{\citenamefont {Legrand}\ \emph {et~al.}(2020)\citenamefont
  {Legrand}, \citenamefont {Maccariello}, \citenamefont {Ajejas}, \citenamefont
  {Collin}, \citenamefont {Vecchiola}, \citenamefont {Bouzehouane},
  \citenamefont {Reyren}, \citenamefont {Cros},\ and\ \citenamefont
  {Fert}}]{legrand2020room}%
  \BibitemOpen
  \bibfield  {author} {\bibinfo {author} {\bibfnamefont {W.}~\bibnamefont
  {Legrand}}, \bibinfo {author} {\bibfnamefont {D.}~\bibnamefont
  {Maccariello}}, \bibinfo {author} {\bibfnamefont {F.}~\bibnamefont {Ajejas}},
  \bibinfo {author} {\bibfnamefont {S.}~\bibnamefont {Collin}}, \bibinfo
  {author} {\bibfnamefont {A.}~\bibnamefont {Vecchiola}}, \bibinfo {author}
  {\bibfnamefont {K.}~\bibnamefont {Bouzehouane}}, \bibinfo {author}
  {\bibfnamefont {N.}~\bibnamefont {Reyren}}, \bibinfo {author} {\bibfnamefont
  {V.}~\bibnamefont {Cros}},\ and\ \bibinfo {author} {\bibfnamefont
  {A.}~\bibnamefont {Fert}},\ }\bibfield  {title} {\bibinfo {title}
  {Room-temperature stabilization of antiferromagnetic skyrmions in synthetic
  antiferromagnets},\ }\href@noop {} {\bibfield  {journal} {\bibinfo  {journal}
  {Nat. Mater.}\ }\textbf {\bibinfo {volume} {19}},\ \bibinfo {pages} {34}
  (\bibinfo {year} {2020})}\BibitemShut {NoStop}%
\bibitem [{\citenamefont {Dohi}\ \emph {et~al.}(2019)\citenamefont {Dohi},
  \citenamefont {DuttaGupta}, \citenamefont {Fukami},\ and\ \citenamefont
  {Ohno}}]{dohi2019formation}%
  \BibitemOpen
  \bibfield  {author} {\bibinfo {author} {\bibfnamefont {T.}~\bibnamefont
  {Dohi}}, \bibinfo {author} {\bibfnamefont {S.}~\bibnamefont {DuttaGupta}},
  \bibinfo {author} {\bibfnamefont {S.}~\bibnamefont {Fukami}},\ and\ \bibinfo
  {author} {\bibfnamefont {H.}~\bibnamefont {Ohno}},\ }\bibfield  {title}
  {\bibinfo {title} {Formation and current-induced motion of synthetic
  antiferromagnetic skyrmion bubbles},\ }\href@noop {} {\bibfield  {journal}
  {\bibinfo  {journal} {Nat. Commun.}\ }\textbf {\bibinfo {volume} {10}},\
  \bibinfo {pages} {5153} (\bibinfo {year} {2019})}\BibitemShut {NoStop}%
\bibitem [{\citenamefont {Pham}\ \emph {et~al.}(2024)\citenamefont {Pham},
  \citenamefont {Sisodia}, \citenamefont {Di~Manici}, \citenamefont
  {{Urrestarazu-Larra{\~n}aga}}, \citenamefont {Bairagi}, \citenamefont
  {{Pelloux-Prayer}}, \citenamefont {Guedas}, \citenamefont {{Buda-Prejbeanu}},
  \citenamefont {Auffret}, \citenamefont {Locatelli}, \citenamefont {Mente{\c
  s}}, \citenamefont {Pizzini}, \citenamefont {Kumar}, \citenamefont {Finco},
  \citenamefont {Jacques}, \citenamefont {Gaudin},\ and\ \citenamefont
  {Boulle}}]{pham2024fast}%
  \BibitemOpen
  \bibfield  {author} {\bibinfo {author} {\bibfnamefont {V.~T.}\ \bibnamefont
  {Pham}}, \bibinfo {author} {\bibfnamefont {N.}~\bibnamefont {Sisodia}},
  \bibinfo {author} {\bibfnamefont {I.}~\bibnamefont {Di~Manici}}, \bibinfo
  {author} {\bibfnamefont {J.}~\bibnamefont {{Urrestarazu-Larra{\~n}aga}}},
  \bibinfo {author} {\bibfnamefont {K.}~\bibnamefont {Bairagi}}, \bibinfo
  {author} {\bibfnamefont {J.}~\bibnamefont {{Pelloux-Prayer}}}, \bibinfo
  {author} {\bibfnamefont {R.}~\bibnamefont {Guedas}}, \bibinfo {author}
  {\bibfnamefont {L.~D.}\ \bibnamefont {{Buda-Prejbeanu}}}, \bibinfo {author}
  {\bibfnamefont {S.}~\bibnamefont {Auffret}}, \bibinfo {author} {\bibfnamefont
  {A.}~\bibnamefont {Locatelli}}, \bibinfo {author} {\bibfnamefont {T.~O.}\
  \bibnamefont {Mente{\c s}}}, \bibinfo {author} {\bibfnamefont
  {S.}~\bibnamefont {Pizzini}}, \bibinfo {author} {\bibfnamefont
  {P.}~\bibnamefont {Kumar}}, \bibinfo {author} {\bibfnamefont
  {A.}~\bibnamefont {Finco}}, \bibinfo {author} {\bibfnamefont
  {V.}~\bibnamefont {Jacques}}, \bibinfo {author} {\bibfnamefont
  {G.}~\bibnamefont {Gaudin}},\ and\ \bibinfo {author} {\bibfnamefont
  {O.}~\bibnamefont {Boulle}},\ }\bibfield  {title} {\bibinfo {title} {Fast
  current-induced skyrmion motion in synthetic antiferromagnets},\ }\href
  {https://doi.org/10.1126/science.add5751} {\bibfield  {journal} {\bibinfo
  {journal} {Science}\ }\textbf {\bibinfo {volume} {384}},\ \bibinfo {pages}
  {307} (\bibinfo {year} {2024})}\BibitemShut {NoStop}%
\bibitem [{\citenamefont {Finocchio}\ \emph {et~al.}(2015)\citenamefont
  {Finocchio}, \citenamefont {Ricci}, \citenamefont {Tomasello}, \citenamefont
  {Giordano}, \citenamefont {Lanuzza}, \citenamefont {Puliafito}, \citenamefont
  {Burrascano}, \citenamefont {Azzerboni},\ and\ \citenamefont
  {Carpentieri}}]{finocchio2015skyrmion}%
  \BibitemOpen
  \bibfield  {author} {\bibinfo {author} {\bibfnamefont {G.}~\bibnamefont
  {Finocchio}}, \bibinfo {author} {\bibfnamefont {M.}~\bibnamefont {Ricci}},
  \bibinfo {author} {\bibfnamefont {R.}~\bibnamefont {Tomasello}}, \bibinfo
  {author} {\bibfnamefont {A.}~\bibnamefont {Giordano}}, \bibinfo {author}
  {\bibfnamefont {M.}~\bibnamefont {Lanuzza}}, \bibinfo {author} {\bibfnamefont
  {V.}~\bibnamefont {Puliafito}}, \bibinfo {author} {\bibfnamefont
  {P.}~\bibnamefont {Burrascano}}, \bibinfo {author} {\bibfnamefont
  {B.}~\bibnamefont {Azzerboni}},\ and\ \bibinfo {author} {\bibfnamefont
  {M.}~\bibnamefont {Carpentieri}},\ }\bibfield  {title} {\bibinfo {title}
  {Skyrmion based microwave detectors and harvesting},\ }\href
  {https://doi.org/10.1063/1.4938539} {\bibfield  {journal} {\bibinfo
  {journal} {Appl. Phys. Lett.}\ }\textbf {\bibinfo {volume} {107}},\ \bibinfo
  {pages} {262401} (\bibinfo {year} {2015})}\BibitemShut {NoStop}%
\bibitem [{\citenamefont {Song}\ \emph {et~al.}(2020)\citenamefont {Song},
  \citenamefont {Jeong}, \citenamefont {Pan}, \citenamefont {Zhang},
  \citenamefont {Xia}, \citenamefont {Cha}, \citenamefont {Park}, \citenamefont
  {Kim}, \citenamefont {Finizio}, \citenamefont {Raabe}, \citenamefont {Chang},
  \citenamefont {Zhou}, \citenamefont {Zhao}, \citenamefont {Kang},
  \citenamefont {Ju},\ and\ \citenamefont {Woo}}]{song2020skyrmionbased}%
  \BibitemOpen
  \bibfield  {author} {\bibinfo {author} {\bibfnamefont {K.~M.}\ \bibnamefont
  {Song}}, \bibinfo {author} {\bibfnamefont {J.-S.}\ \bibnamefont {Jeong}},
  \bibinfo {author} {\bibfnamefont {B.}~\bibnamefont {Pan}}, \bibinfo {author}
  {\bibfnamefont {X.}~\bibnamefont {Zhang}}, \bibinfo {author} {\bibfnamefont
  {J.}~\bibnamefont {Xia}}, \bibinfo {author} {\bibfnamefont {S.}~\bibnamefont
  {Cha}}, \bibinfo {author} {\bibfnamefont {T.-E.}\ \bibnamefont {Park}},
  \bibinfo {author} {\bibfnamefont {K.}~\bibnamefont {Kim}}, \bibinfo {author}
  {\bibfnamefont {S.}~\bibnamefont {Finizio}}, \bibinfo {author} {\bibfnamefont
  {J.}~\bibnamefont {Raabe}}, \bibinfo {author} {\bibfnamefont
  {J.}~\bibnamefont {Chang}}, \bibinfo {author} {\bibfnamefont
  {Y.}~\bibnamefont {Zhou}}, \bibinfo {author} {\bibfnamefont {W.}~\bibnamefont
  {Zhao}}, \bibinfo {author} {\bibfnamefont {W.}~\bibnamefont {Kang}}, \bibinfo
  {author} {\bibfnamefont {H.}~\bibnamefont {Ju}},\ and\ \bibinfo {author}
  {\bibfnamefont {S.}~\bibnamefont {Woo}},\ }\bibfield  {title} {\bibinfo
  {title} {Skyrmion-based artificial synapses for neuromorphic computing},\
  }\href {https://doi.org/10.1038/s41928-020-0385-0} {\bibfield  {journal}
  {\bibinfo  {journal} {Nat. Electron.}\ }\textbf {\bibinfo {volume} {3}},\
  \bibinfo {pages} {148} (\bibinfo {year} {2020})}\BibitemShut {NoStop}%
\bibitem [{\citenamefont {Pinna}\ \emph {et~al.}(2020)\citenamefont {Pinna},
  \citenamefont {Bourianoff},\ and\ \citenamefont
  {{Everschor-Sitte}}}]{pinna2020reservoir}%
  \BibitemOpen
  \bibfield  {author} {\bibinfo {author} {\bibfnamefont {D.}~\bibnamefont
  {Pinna}}, \bibinfo {author} {\bibfnamefont {G.}~\bibnamefont {Bourianoff}},\
  and\ \bibinfo {author} {\bibfnamefont {K.}~\bibnamefont
  {{Everschor-Sitte}}},\ }\bibfield  {title} {\bibinfo {title} {Reservoir
  {{Computing}} with {{Random Skyrmion Textures}}},\ }\href
  {https://doi.org/10.1103/PhysRevApplied.14.054020} {\bibfield  {journal}
  {\bibinfo  {journal} {Phys. Rev. Appl.}\ }\textbf {\bibinfo {volume} {14}},\
  \bibinfo {pages} {054020} (\bibinfo {year} {2020})}\BibitemShut {NoStop}%
\bibitem [{\citenamefont {Lee}\ \emph {et~al.}(2024)\citenamefont {Lee},
  \citenamefont {Wei}, \citenamefont {Stenning}, \citenamefont {Gartside},
  \citenamefont {Prestwood}, \citenamefont {Seki}, \citenamefont {Aqeel},
  \citenamefont {Karube}, \citenamefont {Kanazawa}, \citenamefont {Taguchi},
  \citenamefont {Back}, \citenamefont {Tokura}, \citenamefont {Branford},\ and\
  \citenamefont {Kurebayashi}}]{lee2024taskadaptive}%
  \BibitemOpen
  \bibfield  {author} {\bibinfo {author} {\bibfnamefont {O.}~\bibnamefont
  {Lee}}, \bibinfo {author} {\bibfnamefont {T.}~\bibnamefont {Wei}}, \bibinfo
  {author} {\bibfnamefont {K.~D.}\ \bibnamefont {Stenning}}, \bibinfo {author}
  {\bibfnamefont {J.~C.}\ \bibnamefont {Gartside}}, \bibinfo {author}
  {\bibfnamefont {D.}~\bibnamefont {Prestwood}}, \bibinfo {author}
  {\bibfnamefont {S.}~\bibnamefont {Seki}}, \bibinfo {author} {\bibfnamefont
  {A.}~\bibnamefont {Aqeel}}, \bibinfo {author} {\bibfnamefont
  {K.}~\bibnamefont {Karube}}, \bibinfo {author} {\bibfnamefont
  {N.}~\bibnamefont {Kanazawa}}, \bibinfo {author} {\bibfnamefont
  {Y.}~\bibnamefont {Taguchi}}, \bibinfo {author} {\bibfnamefont
  {C.}~\bibnamefont {Back}}, \bibinfo {author} {\bibfnamefont {Y.}~\bibnamefont
  {Tokura}}, \bibinfo {author} {\bibfnamefont {W.~R.}\ \bibnamefont
  {Branford}},\ and\ \bibinfo {author} {\bibfnamefont {H.}~\bibnamefont
  {Kurebayashi}},\ }\bibfield  {title} {\bibinfo {title} {Task-adaptive
  physical reservoir computing},\ }\href
  {https://doi.org/10.1038/s41563-023-01698-8} {\bibfield  {journal} {\bibinfo
  {journal} {Nat. Mater.}\ }\textbf {\bibinfo {volume} {23}},\ \bibinfo {pages}
  {79} (\bibinfo {year} {2024})}\BibitemShut {NoStop}%
\bibitem [{\citenamefont {Lee}\ and\ \citenamefont
  {Mochizuki}(2023)}]{lee2023handwritten}%
  \BibitemOpen
  \bibfield  {author} {\bibinfo {author} {\bibfnamefont {M.-K.}\ \bibnamefont
  {Lee}}\ and\ \bibinfo {author} {\bibfnamefont {M.}~\bibnamefont
  {Mochizuki}},\ }\bibfield  {title} {\bibinfo {title} {Handwritten digit
  recognition by spin waves in a {{Skyrmion}} reservoir},\ }\href
  {https://doi.org/10.1038/s41598-023-46677-w} {\bibfield  {journal} {\bibinfo
  {journal} {Sci Rep}\ }\textbf {\bibinfo {volume} {13}},\ \bibinfo {pages}
  {19423} (\bibinfo {year} {2023})}\BibitemShut {NoStop}%
\bibitem [{\citenamefont {{da C{\^a}mara Santa Clara Gomes}}\ \emph
  {et~al.}(2025)\citenamefont {{da C{\^a}mara Santa Clara Gomes}},
  \citenamefont {Sassi}, \citenamefont {{Sanz-Hern{\'a}ndez}}, \citenamefont
  {Krishnia}, \citenamefont {Collin}, \citenamefont {Martin}, \citenamefont
  {Seneor}, \citenamefont {Cros}, \citenamefont {Grollier},\ and\ \citenamefont
  {Reyren}}]{dacamarasantaclaragomes2025neuromorphic}%
  \BibitemOpen
  \bibfield  {author} {\bibinfo {author} {\bibfnamefont {T.}~\bibnamefont {{da
  C{\^a}mara Santa Clara Gomes}}}, \bibinfo {author} {\bibfnamefont
  {Y.}~\bibnamefont {Sassi}}, \bibinfo {author} {\bibfnamefont
  {D.}~\bibnamefont {{Sanz-Hern{\'a}ndez}}}, \bibinfo {author} {\bibfnamefont
  {S.}~\bibnamefont {Krishnia}}, \bibinfo {author} {\bibfnamefont
  {S.}~\bibnamefont {Collin}}, \bibinfo {author} {\bibfnamefont {M.-B.}\
  \bibnamefont {Martin}}, \bibinfo {author} {\bibfnamefont {P.}~\bibnamefont
  {Seneor}}, \bibinfo {author} {\bibfnamefont {V.}~\bibnamefont {Cros}},
  \bibinfo {author} {\bibfnamefont {J.}~\bibnamefont {Grollier}},\ and\
  \bibinfo {author} {\bibfnamefont {N.}~\bibnamefont {Reyren}},\ }\bibfield
  {title} {\bibinfo {title} {Neuromorphic weighted sums with magnetic
  skyrmions},\ }\href {https://doi.org/10.1038/s41928-024-01303-z} {\bibfield
  {journal} {\bibinfo  {journal} {Nat. Electron.}\ }\textbf {\bibinfo {volume}
  {8}},\ \bibinfo {pages} {204} (\bibinfo {year} {2025})}\BibitemShut {NoStop}%
\bibitem [{\citenamefont {Nayak}\ \emph {et~al.}(2017)\citenamefont {Nayak},
  \citenamefont {Kumar}, \citenamefont {Ma}, \citenamefont {Werner},
  \citenamefont {Pippel}, \citenamefont {Sahoo}, \citenamefont {Damay},
  \citenamefont {R{\"o}{\ss}ler}, \citenamefont {Felser},\ and\ \citenamefont
  {Parkin}}]{nayak2017magnetic}%
  \BibitemOpen
  \bibfield  {author} {\bibinfo {author} {\bibfnamefont {A.~K.}\ \bibnamefont
  {Nayak}}, \bibinfo {author} {\bibfnamefont {V.}~\bibnamefont {Kumar}},
  \bibinfo {author} {\bibfnamefont {T.}~\bibnamefont {Ma}}, \bibinfo {author}
  {\bibfnamefont {P.}~\bibnamefont {Werner}}, \bibinfo {author} {\bibfnamefont
  {E.}~\bibnamefont {Pippel}}, \bibinfo {author} {\bibfnamefont
  {R.}~\bibnamefont {Sahoo}}, \bibinfo {author} {\bibfnamefont
  {F.}~\bibnamefont {Damay}}, \bibinfo {author} {\bibfnamefont {U.~K.}\
  \bibnamefont {R{\"o}{\ss}ler}}, \bibinfo {author} {\bibfnamefont
  {C.}~\bibnamefont {Felser}},\ and\ \bibinfo {author} {\bibfnamefont
  {S.~S.~P.}\ \bibnamefont {Parkin}},\ }\bibfield  {title} {\bibinfo {title}
  {Magnetic antiskyrmions above room temperature in tetragonal {{Heusler}}
  materials},\ }\href {https://doi.org/10.1038/nature23466} {\bibfield
  {journal} {\bibinfo  {journal} {Nature}\ }\textbf {\bibinfo {volume} {548}},\
  \bibinfo {pages} {561} (\bibinfo {year} {2017})}\BibitemShut {NoStop}%
\bibitem [{\citenamefont {Jena}\ \emph
  {et~al.}(2020{\natexlab{a}})\citenamefont {Jena}, \citenamefont {G{\"o}bel},
  \citenamefont {Ma}, \citenamefont {Kumar}, \citenamefont {Saha},
  \citenamefont {Mertig}, \citenamefont {Felser},\ and\ \citenamefont
  {Parkin}}]{jena2020elliptical}%
  \BibitemOpen
  \bibfield  {author} {\bibinfo {author} {\bibfnamefont {J.}~\bibnamefont
  {Jena}}, \bibinfo {author} {\bibfnamefont {B.}~\bibnamefont {G{\"o}bel}},
  \bibinfo {author} {\bibfnamefont {T.}~\bibnamefont {Ma}}, \bibinfo {author}
  {\bibfnamefont {V.}~\bibnamefont {Kumar}}, \bibinfo {author} {\bibfnamefont
  {R.}~\bibnamefont {Saha}}, \bibinfo {author} {\bibfnamefont {I.}~\bibnamefont
  {Mertig}}, \bibinfo {author} {\bibfnamefont {C.}~\bibnamefont {Felser}},\
  and\ \bibinfo {author} {\bibfnamefont {S.~S.~P.}\ \bibnamefont {Parkin}},\
  }\bibfield  {title} {\bibinfo {title} {Elliptical {{Bloch}} skyrmion chiral
  twins in an antiskyrmion system},\ }\href
  {https://doi.org/10.1038/s41467-020-14925-6} {\bibfield  {journal} {\bibinfo
  {journal} {Nat. Commun.}\ }\textbf {\bibinfo {volume} {11}},\ \bibinfo
  {pages} {1115} (\bibinfo {year} {2020}{\natexlab{a}})}\BibitemShut {NoStop}%
\bibitem [{\citenamefont {Hoffmann}\ \emph {et~al.}(2021)\citenamefont
  {Hoffmann}, \citenamefont {M{\"u}ller}, \citenamefont {Melcher},\ and\
  \citenamefont {Bl{\"u}gel}}]{hoffmann2021skyrmion}%
  \BibitemOpen
  \bibfield  {author} {\bibinfo {author} {\bibfnamefont {M.}~\bibnamefont
  {Hoffmann}}, \bibinfo {author} {\bibfnamefont {G.~P.}\ \bibnamefont
  {M{\"u}ller}}, \bibinfo {author} {\bibfnamefont {C.}~\bibnamefont
  {Melcher}},\ and\ \bibinfo {author} {\bibfnamefont {S.}~\bibnamefont
  {Bl{\"u}gel}},\ }\bibfield  {title} {\bibinfo {title} {Skyrmion-antiskyrmion
  racetrack memory in rank-one dmi materials},\ }\href@noop {} {\bibfield
  {journal} {\bibinfo  {journal} {Front. Phys.}\ }\textbf {\bibinfo {volume}
  {9}},\ \bibinfo {pages} {769873} (\bibinfo {year} {2021})}\BibitemShut
  {NoStop}%
\bibitem [{\citenamefont {Heigl}\ \emph {et~al.}(2021)\citenamefont {Heigl},
  \citenamefont {Koraltan}, \citenamefont {Va{\v{n}}atka}, \citenamefont
  {Kraft}, \citenamefont {Abert}, \citenamefont {Vogler}, \citenamefont
  {Semisalova}, \citenamefont {Che}, \citenamefont {Ullrich}, \citenamefont
  {Schmidt} \emph {et~al.}}]{heigl2021dipolar}%
  \BibitemOpen
  \bibfield  {author} {\bibinfo {author} {\bibfnamefont {M.}~\bibnamefont
  {Heigl}}, \bibinfo {author} {\bibfnamefont {S.}~\bibnamefont {Koraltan}},
  \bibinfo {author} {\bibfnamefont {M.}~\bibnamefont {Va{\v{n}}atka}}, \bibinfo
  {author} {\bibfnamefont {R.}~\bibnamefont {Kraft}}, \bibinfo {author}
  {\bibfnamefont {C.}~\bibnamefont {Abert}}, \bibinfo {author} {\bibfnamefont
  {C.}~\bibnamefont {Vogler}}, \bibinfo {author} {\bibfnamefont
  {A.}~\bibnamefont {Semisalova}}, \bibinfo {author} {\bibfnamefont
  {P.}~\bibnamefont {Che}}, \bibinfo {author} {\bibfnamefont {A.}~\bibnamefont
  {Ullrich}}, \bibinfo {author} {\bibfnamefont {T.}~\bibnamefont {Schmidt}},
  \emph {et~al.},\ }\bibfield  {title} {\bibinfo {title} {Dipolar-stabilized
  first and second-order antiskyrmions in ferrimagnetic multilayers},\
  }\href@noop {} {\bibfield  {journal} {\bibinfo  {journal} {Nat. Commun.}\
  }\textbf {\bibinfo {volume} {12}},\ \bibinfo {pages} {2611} (\bibinfo {year}
  {2021})}\BibitemShut {NoStop}%
\bibitem [{\citenamefont {Battye}\ and\ \citenamefont
  {Sutcliffe}(1998)}]{battye1998knots}%
  \BibitemOpen
  \bibfield  {author} {\bibinfo {author} {\bibfnamefont {R.~A.}\ \bibnamefont
  {Battye}}\ and\ \bibinfo {author} {\bibfnamefont {P.~M.}\ \bibnamefont
  {Sutcliffe}},\ }\bibfield  {title} {\bibinfo {title} {Knots as stable soliton
  solutions in a three-dimensional classical field theory},\ }\href
  {https://doi.org/10.1103/PhysRevLett.81.4798} {\bibfield  {journal} {\bibinfo
   {journal} {Phys. Rev. Lett.}\ }\textbf {\bibinfo {volume} {81}},\ \bibinfo
  {pages} {4798} (\bibinfo {year} {1998})}\BibitemShut {NoStop}%
\bibitem [{\citenamefont {Kent}\ \emph {et~al.}(2021)\citenamefont {Kent},
  \citenamefont {Reynolds}, \citenamefont {Raftrey}, \citenamefont {Campbell},
  \citenamefont {Virasawmy}, \citenamefont {Dhuey}, \citenamefont {Chopdekar},
  \citenamefont {{Hierro-Rodriguez}}, \citenamefont {Sorrentino}, \citenamefont
  {Pereiro}, \citenamefont {Ferrer}, \citenamefont {Hellman}, \citenamefont
  {Sutcliffe},\ and\ \citenamefont {Fischer}}]{kent2021creation}%
  \BibitemOpen
  \bibfield  {author} {\bibinfo {author} {\bibfnamefont {N.}~\bibnamefont
  {Kent}}, \bibinfo {author} {\bibfnamefont {N.}~\bibnamefont {Reynolds}},
  \bibinfo {author} {\bibfnamefont {D.}~\bibnamefont {Raftrey}}, \bibinfo
  {author} {\bibfnamefont {I.~T.~G.}\ \bibnamefont {Campbell}}, \bibinfo
  {author} {\bibfnamefont {S.}~\bibnamefont {Virasawmy}}, \bibinfo {author}
  {\bibfnamefont {S.}~\bibnamefont {Dhuey}}, \bibinfo {author} {\bibfnamefont
  {R.~V.}\ \bibnamefont {Chopdekar}}, \bibinfo {author} {\bibfnamefont
  {A.}~\bibnamefont {{Hierro-Rodriguez}}}, \bibinfo {author} {\bibfnamefont
  {A.}~\bibnamefont {Sorrentino}}, \bibinfo {author} {\bibfnamefont
  {E.}~\bibnamefont {Pereiro}}, \bibinfo {author} {\bibfnamefont
  {S.}~\bibnamefont {Ferrer}}, \bibinfo {author} {\bibfnamefont
  {F.}~\bibnamefont {Hellman}}, \bibinfo {author} {\bibfnamefont
  {P.}~\bibnamefont {Sutcliffe}},\ and\ \bibinfo {author} {\bibfnamefont
  {P.}~\bibnamefont {Fischer}},\ }\bibfield  {title} {\bibinfo {title}
  {Creation and observation of {{Hopfions}} in magnetic multilayer systems},\
  }\href {https://doi.org/10.1038/s41467-021-21846-5} {\bibfield  {journal}
  {\bibinfo  {journal} {Nat. Commun.}\ }\textbf {\bibinfo {volume} {12}},\
  \bibinfo {pages} {1562} (\bibinfo {year} {2021})}\BibitemShut {NoStop}%
\bibitem [{\citenamefont {Rybakov}\ \emph {et~al.}(2022)\citenamefont
  {Rybakov}, \citenamefont {Kiselev}, \citenamefont {Borisov}, \citenamefont
  {D{\"o}ring}, \citenamefont {Melcher},\ and\ \citenamefont
  {Bl{\"u}gel}}]{rybakov2022magnetic}%
  \BibitemOpen
  \bibfield  {author} {\bibinfo {author} {\bibfnamefont {F.~N.}\ \bibnamefont
  {Rybakov}}, \bibinfo {author} {\bibfnamefont {N.~S.}\ \bibnamefont
  {Kiselev}}, \bibinfo {author} {\bibfnamefont {A.~B.}\ \bibnamefont
  {Borisov}}, \bibinfo {author} {\bibfnamefont {L.}~\bibnamefont {D{\"o}ring}},
  \bibinfo {author} {\bibfnamefont {C.}~\bibnamefont {Melcher}},\ and\ \bibinfo
  {author} {\bibfnamefont {S.}~\bibnamefont {Bl{\"u}gel}},\ }\bibfield  {title}
  {\bibinfo {title} {Magnetic hopfions in solids},\ }\href
  {https://doi.org/10.1063/5.0099942} {\bibfield  {journal} {\bibinfo
  {journal} {APL Mater.}\ }\textbf {\bibinfo {volume} {10}},\ \bibinfo {pages}
  {111113} (\bibinfo {year} {2022})}\BibitemShut {NoStop}%
\bibitem [{\citenamefont {Zheng}\ \emph {et~al.}(2023)\citenamefont {Zheng},
  \citenamefont {Kiselev}, \citenamefont {Rybakov}, \citenamefont {Yang},
  \citenamefont {Shi}, \citenamefont {Bl{\"u}gel},\ and\ \citenamefont
  {{Dunin-Borkowski}}}]{zheng2023hopfion}%
  \BibitemOpen
  \bibfield  {author} {\bibinfo {author} {\bibfnamefont {F.}~\bibnamefont
  {Zheng}}, \bibinfo {author} {\bibfnamefont {N.~S.}\ \bibnamefont {Kiselev}},
  \bibinfo {author} {\bibfnamefont {F.~N.}\ \bibnamefont {Rybakov}}, \bibinfo
  {author} {\bibfnamefont {L.}~\bibnamefont {Yang}}, \bibinfo {author}
  {\bibfnamefont {W.}~\bibnamefont {Shi}}, \bibinfo {author} {\bibfnamefont
  {S.}~\bibnamefont {Bl{\"u}gel}},\ and\ \bibinfo {author} {\bibfnamefont
  {R.~E.}\ \bibnamefont {{Dunin-Borkowski}}},\ }\bibfield  {title} {\bibinfo
  {title} {Hopfion rings in a cubic chiral magnet},\ }\href
  {https://doi.org/10.1038/s41586-023-06658-5} {\bibfield  {journal} {\bibinfo
  {journal} {Nature}\ }\textbf {\bibinfo {volume} {623}},\ \bibinfo {pages}
  {718} (\bibinfo {year} {2023})}\BibitemShut {NoStop}%
\bibitem [{\citenamefont {Ba{\'c}ani}\ \emph {et~al.}(2019)\citenamefont
  {Ba{\'c}ani}, \citenamefont {Marioni}, \citenamefont {Schwenk},\ and\
  \citenamefont {Hug}}]{bacani2019how}%
  \BibitemOpen
  \bibfield  {author} {\bibinfo {author} {\bibfnamefont {M.}~\bibnamefont
  {Ba{\'c}ani}}, \bibinfo {author} {\bibfnamefont {M.~A.}\ \bibnamefont
  {Marioni}}, \bibinfo {author} {\bibfnamefont {J.}~\bibnamefont {Schwenk}},\
  and\ \bibinfo {author} {\bibfnamefont {H.~J.}\ \bibnamefont {Hug}},\
  }\bibfield  {title} {\bibinfo {title} {How to measure the local
  {{Dzyaloshinskii-Moriya Interaction}} in {{Skyrmion Thin-Film
  Multilayers}}},\ }\href {https://doi.org/10.1038/s41598-019-39501-x}
  {\bibfield  {journal} {\bibinfo  {journal} {Sci Rep}\ }\textbf {\bibinfo
  {volume} {9}},\ \bibinfo {pages} {3114} (\bibinfo {year} {2019})}\BibitemShut
  {NoStop}%
\bibitem [{\citenamefont {Abert}(2019)}]{abert2019micromagnetics}%
  \BibitemOpen
  \bibfield  {author} {\bibinfo {author} {\bibfnamefont {C.}~\bibnamefont
  {Abert}},\ }\bibfield  {title} {\bibinfo {title} {Micromagnetics and
  spintronics: {Models} and numerical methods},\ }\href@noop {} {\bibfield
  {journal} {\bibinfo  {journal} {Eur. Phys. J. B}\ }\textbf {\bibinfo {volume}
  {92}},\ \bibinfo {pages} {1} (\bibinfo {year} {2019})}\BibitemShut {NoStop}%
\bibitem [{\citenamefont {Donahue}\ and\ \citenamefont
  {Porter}(1999)}]{donahue1999oommf}%
  \BibitemOpen
  \bibfield  {author} {\bibinfo {author} {\bibfnamefont {M.~J.}\ \bibnamefont
  {Donahue}}\ and\ \bibinfo {author} {\bibfnamefont {D.~G.}\ \bibnamefont
  {Porter}},\ }\href {https://doi.org/10.6028/NIST.IR.6376} {\emph {\bibinfo
  {title} {{{OOMMF}} User's Guide, Version 1.0}}},\ \bibinfo {type} {Tech.
  Rep.}\ \bibinfo {number} {NIST IR 6376}\ (\bibinfo  {institution} {{National
  Institute of Standards and Technology}},\ \bibinfo {address} {Gaithersburg,
  MD},\ \bibinfo {year} {1999})\BibitemShut {NoStop}%
\bibitem [{\citenamefont {Vansteenkiste}\ \emph {et~al.}(2014)\citenamefont
  {Vansteenkiste}, \citenamefont {Leliaert}, \citenamefont {Dvornik},
  \citenamefont {Helsen}, \citenamefont {{Garcia-Sanchez}},\ and\ \citenamefont
  {Van~Waeyenberge}}]{vansteenkiste2014design}%
  \BibitemOpen
  \bibfield  {author} {\bibinfo {author} {\bibfnamefont {A.}~\bibnamefont
  {Vansteenkiste}}, \bibinfo {author} {\bibfnamefont {J.}~\bibnamefont
  {Leliaert}}, \bibinfo {author} {\bibfnamefont {M.}~\bibnamefont {Dvornik}},
  \bibinfo {author} {\bibfnamefont {M.}~\bibnamefont {Helsen}}, \bibinfo
  {author} {\bibfnamefont {F.}~\bibnamefont {{Garcia-Sanchez}}},\ and\ \bibinfo
  {author} {\bibfnamefont {B.}~\bibnamefont {Van~Waeyenberge}},\ }\bibfield
  {title} {\bibinfo {title} {The design and verification of {{MuMax3}}},\
  }\href {https://doi.org/10.1063/1.4899186} {\bibfield  {journal} {\bibinfo
  {journal} {AIP Adv.}\ }\textbf {\bibinfo {volume} {4}},\ \bibinfo {pages}
  {107133} (\bibinfo {year} {2014})}\BibitemShut {NoStop}%
\bibitem [{\citenamefont {Ducevic}\ \emph {et~al.}(2025)\citenamefont
  {Ducevic}, \citenamefont {Bruckner}, \citenamefont {Abert}, \citenamefont
  {Vogler},\ and\ \citenamefont {Suess}}]{bruckner2023magnum}%
  \BibitemOpen
  \bibfield  {author} {\bibinfo {author} {\bibfnamefont {A.}~\bibnamefont
  {Ducevic}}, \bibinfo {author} {\bibfnamefont {F.}~\bibnamefont {Bruckner}},
  \bibinfo {author} {\bibfnamefont {C.}~\bibnamefont {Abert}}, \bibinfo
  {author} {\bibfnamefont {C.}~\bibnamefont {Vogler}},\ and\ \bibinfo {author}
  {\bibfnamefont {D.}~\bibnamefont {Suess}},\ }\bibfield  {title} {\bibinfo
  {title} {Micromagnetic simulations with periodic strayfield calculation of
  soft magnetic composite-materials},\ }\bibfield  {journal} {\bibinfo
  {journal} {Sci. Rep.}\ }\textbf {\bibinfo {volume} {15}},\ \href
  {https://doi.org/10.1038/s41598-025-01881-8} {10.1038/s41598-025-01881-8}
  (\bibinfo {year} {2025})\BibitemShut {NoStop}%
\bibitem [{\citenamefont {Abert}\ \emph {et~al.}(2025)\citenamefont {Abert},
  \citenamefont {Bruckner}, \citenamefont {Voronov}, \citenamefont {Lang},
  \citenamefont {Pathak}, \citenamefont {Holt}, \citenamefont {Kraft},
  \citenamefont {Allayarov}, \citenamefont {Flauger}, \citenamefont {Koraltan},
  \citenamefont {Schrefl}, \citenamefont {Chumak}, \citenamefont {Fangohr},\
  and\ \citenamefont {Suess}}]{abert2025neuralmag}%
  \BibitemOpen
  \bibfield  {author} {\bibinfo {author} {\bibfnamefont {C.}~\bibnamefont
  {Abert}}, \bibinfo {author} {\bibfnamefont {F.}~\bibnamefont {Bruckner}},
  \bibinfo {author} {\bibfnamefont {A.}~\bibnamefont {Voronov}}, \bibinfo
  {author} {\bibfnamefont {M.}~\bibnamefont {Lang}}, \bibinfo {author}
  {\bibfnamefont {S.~A.}\ \bibnamefont {Pathak}}, \bibinfo {author}
  {\bibfnamefont {S.}~\bibnamefont {Holt}}, \bibinfo {author} {\bibfnamefont
  {R.}~\bibnamefont {Kraft}}, \bibinfo {author} {\bibfnamefont
  {R.}~\bibnamefont {Allayarov}}, \bibinfo {author} {\bibfnamefont
  {P.}~\bibnamefont {Flauger}}, \bibinfo {author} {\bibfnamefont
  {S.}~\bibnamefont {Koraltan}}, \bibinfo {author} {\bibfnamefont
  {T.}~\bibnamefont {Schrefl}}, \bibinfo {author} {\bibfnamefont
  {A.}~\bibnamefont {Chumak}}, \bibinfo {author} {\bibfnamefont
  {H.}~\bibnamefont {Fangohr}},\ and\ \bibinfo {author} {\bibfnamefont
  {D.}~\bibnamefont {Suess}},\ }\bibfield  {title} {\bibinfo {title}
  {{{NeuralMag}}: An open-source nodal finite-difference code for inverse
  micromagnetics},\ }\href {https://doi.org/10.1038/s41524-025-01688-1}
  {\bibfield  {journal} {\bibinfo  {journal} {npj Comput Mater}\ }\textbf
  {\bibinfo {volume} {11}},\ \bibinfo {pages} {193} (\bibinfo {year}
  {2025})}\BibitemShut {NoStop}%
\bibitem [{\citenamefont {Rohart}\ and\ \citenamefont
  {Thiaville}(2013)}]{rohart2013skyrmion}%
  \BibitemOpen
  \bibfield  {author} {\bibinfo {author} {\bibfnamefont {S.}~\bibnamefont
  {Rohart}}\ and\ \bibinfo {author} {\bibfnamefont {A.}~\bibnamefont
  {Thiaville}},\ }\bibfield  {title} {\bibinfo {title} {Skyrmion confinement in
  ultrathin film nanostructures in the presence of {{Dzyaloshinskii-Moriya}}
  interaction},\ }\href {https://doi.org/10.1103/PhysRevB.88.184422} {\bibfield
   {journal} {\bibinfo  {journal} {Phys. Rev. B}\ }\textbf {\bibinfo {volume}
  {88}},\ \bibinfo {pages} {184422} (\bibinfo {year} {2013})}\BibitemShut
  {NoStop}%
\bibitem [{\citenamefont {Suess}\ \emph {et~al.}(2019)\citenamefont {Suess},
  \citenamefont {Vogler}, \citenamefont {Bruckner}, \citenamefont
  {Heistracher}, \citenamefont {Slanovc},\ and\ \citenamefont
  {Abert}}]{suess2019spin}%
  \BibitemOpen
  \bibfield  {author} {\bibinfo {author} {\bibfnamefont {D.}~\bibnamefont
  {Suess}}, \bibinfo {author} {\bibfnamefont {C.}~\bibnamefont {Vogler}},
  \bibinfo {author} {\bibfnamefont {F.}~\bibnamefont {Bruckner}}, \bibinfo
  {author} {\bibfnamefont {P.}~\bibnamefont {Heistracher}}, \bibinfo {author}
  {\bibfnamefont {F.}~\bibnamefont {Slanovc}},\ and\ \bibinfo {author}
  {\bibfnamefont {C.}~\bibnamefont {Abert}},\ }\bibfield  {title} {\bibinfo
  {title} {Spin {{Torque Efficiency}} and {{Analytic Error Rate Estimates}} of
  {{Skyrmion Racetrack Memory}}},\ }\href
  {https://doi.org/10.1038/s41598-019-41062-y} {\bibfield  {journal} {\bibinfo
  {journal} {Sci Rep}\ }\textbf {\bibinfo {volume} {9}},\ \bibinfo {pages}
  {4827} (\bibinfo {year} {2019})}\BibitemShut {NoStop}%
\bibitem [{\citenamefont {Bruckner}\ \emph {et~al.}(2017)\citenamefont
  {Bruckner}, \citenamefont {Abert}, \citenamefont {Vogler}, \citenamefont
  {Bruckner},\ and\ \citenamefont {Suess}}]{bruckner2017large}%
  \BibitemOpen
  \bibfield  {author} {\bibinfo {author} {\bibfnamefont {F.}~\bibnamefont
  {Bruckner}}, \bibinfo {author} {\bibfnamefont {C.}~\bibnamefont {Abert}},
  \bibinfo {author} {\bibfnamefont {C.}~\bibnamefont {Vogler}}, \bibinfo
  {author} {\bibfnamefont {F.}~\bibnamefont {Bruckner}},\ and\ \bibinfo
  {author} {\bibfnamefont {D.}~\bibnamefont {Suess}},\ }\bibfield  {title}
  {\bibinfo {title} {Large-scale micromagnetic simulations using the adjoint
  method for sensitivity analysis},\ }\href@noop {} {\bibfield  {journal}
  {\bibinfo  {journal} {J. Magn. Magn. Mater.}\ }\textbf {\bibinfo {volume}
  {428}},\ \bibinfo {pages} {296} (\bibinfo {year} {2017})}\BibitemShut
  {NoStop}%
\bibitem [{\citenamefont {Wang}\ \emph
  {et~al.}(2021{\natexlab{a}})\citenamefont {Wang}, \citenamefont {Chumak},\
  and\ \citenamefont {Pirro}}]{wang2021inverse}%
  \BibitemOpen
  \bibfield  {author} {\bibinfo {author} {\bibfnamefont {Q.}~\bibnamefont
  {Wang}}, \bibinfo {author} {\bibfnamefont {A.~V.}\ \bibnamefont {Chumak}},\
  and\ \bibinfo {author} {\bibfnamefont {P.}~\bibnamefont {Pirro}},\ }\bibfield
   {title} {\bibinfo {title} {Inverse-design magnonic devices},\ }\href
  {https://doi.org/10.1038/s41467-021-22897-4} {\bibfield  {journal} {\bibinfo
  {journal} {Nat. Commun.}\ }\textbf {\bibinfo {volume} {12}},\ \bibinfo
  {pages} {2636} (\bibinfo {year} {2021}{\natexlab{a}})}\BibitemShut {NoStop}%
\bibitem [{\citenamefont {Papp}\ \emph {et~al.}(2021)\citenamefont {Papp},
  \citenamefont {Porod},\ and\ \citenamefont {Csaba}}]{papp2021nanoscale}%
  \BibitemOpen
  \bibfield  {author} {\bibinfo {author} {\bibfnamefont {{\'A}.}~\bibnamefont
  {Papp}}, \bibinfo {author} {\bibfnamefont {W.}~\bibnamefont {Porod}},\ and\
  \bibinfo {author} {\bibfnamefont {G.}~\bibnamefont {Csaba}},\ }\bibfield
  {title} {\bibinfo {title} {Nanoscale neural network using non-linear
  spin-wave interference},\ }\href {https://doi.org/10.1038/s41467-021-26711-z}
  {\bibfield  {journal} {\bibinfo  {journal} {Nat. Commun.}\ }\textbf {\bibinfo
  {volume} {12}},\ \bibinfo {pages} {6422} (\bibinfo {year}
  {2021})}\BibitemShut {NoStop}%
\bibitem [{\citenamefont {Voronov}\ \emph {et~al.}(2025)\citenamefont
  {Voronov}, \citenamefont {Abert}, \citenamefont
  {Br\{{\textbackslash}''u\}ckner},\ and\ \citenamefont
  {Suess}}]{voronov2025inverse}%
  \BibitemOpen
  \bibfield  {author} {\bibinfo {author} {\bibfnamefont {A.~A.}\ \bibnamefont
  {Voronov}}, \bibinfo {author} {\bibfnamefont {C.}~\bibnamefont {Abert}},
  \bibinfo {author} {\bibfnamefont {F.}~\bibnamefont
  {Br\{{\textbackslash}''u\}ckner}},\ and\ \bibinfo {author} {\bibfnamefont
  {D.}~\bibnamefont {Suess}},\ }\bibfield  {title} {\bibinfo {title}
  {Inverse-design topology optimization of magnonic devices using level-set
  method},\ }\href@noop {} {\bibfield  {journal} {\bibinfo  {journal} {npj
  Spintronics}\ }\textbf {\bibinfo {volume} {3}},\ \bibinfo {pages} {1}
  (\bibinfo {year} {2025})}\BibitemShut {NoStop}%
\bibitem [{\citenamefont {Suess}\ \emph {et~al.}(2024)\citenamefont {Suess},
  \citenamefont {GUETTINGER},\ and\ \citenamefont {Satz}}]{suess2024devices}%
  \BibitemOpen
  \bibfield  {author} {\bibinfo {author} {\bibfnamefont {D.}~\bibnamefont
  {Suess}}, \bibinfo {author} {\bibfnamefont {J.}~\bibnamefont {GUETTINGER}},\
  and\ \bibinfo {author} {\bibfnamefont {A.}~\bibnamefont {Satz}},\ }\href@noop
  {} {\bibinfo {title} {Devices for determining a number of rotations of a
  magnetic field}} (\bibinfo {year} {2024})\BibitemShut {NoStop}%
\bibitem [{\citenamefont {Raab}\ \emph {et~al.}(2022)\citenamefont {Raab},
  \citenamefont {Brems}, \citenamefont {Beneke}, \citenamefont {Dohi},
  \citenamefont {Roth{\"o}rl}, \citenamefont {Kammerbauer}, \citenamefont
  {Mentink},\ and\ \citenamefont {Kl{\"a}ui}}]{raab2022brownian}%
  \BibitemOpen
  \bibfield  {author} {\bibinfo {author} {\bibfnamefont {K.}~\bibnamefont
  {Raab}}, \bibinfo {author} {\bibfnamefont {M.~A.}\ \bibnamefont {Brems}},
  \bibinfo {author} {\bibfnamefont {G.}~\bibnamefont {Beneke}}, \bibinfo
  {author} {\bibfnamefont {T.}~\bibnamefont {Dohi}}, \bibinfo {author}
  {\bibfnamefont {J.}~\bibnamefont {Roth{\"o}rl}}, \bibinfo {author}
  {\bibfnamefont {F.}~\bibnamefont {Kammerbauer}}, \bibinfo {author}
  {\bibfnamefont {J.~H.}\ \bibnamefont {Mentink}},\ and\ \bibinfo {author}
  {\bibfnamefont {M.}~\bibnamefont {Kl{\"a}ui}},\ }\bibfield  {title} {\bibinfo
  {title} {Brownian reservoir computing realized using geometrically confined
  skyrmion dynamics},\ }\href {https://doi.org/10.1038/s41467-022-34309-2}
  {\bibfield  {journal} {\bibinfo  {journal} {Nat. Commun.}\ }\textbf {\bibinfo
  {volume} {13}},\ \bibinfo {pages} {6982} (\bibinfo {year}
  {2022})}\BibitemShut {NoStop}%
\bibitem [{\citenamefont {Gubbiotti}\ \emph {et~al.}(2025)\citenamefont
  {Gubbiotti}, \citenamefont {Barman}, \citenamefont {Ladak}, \citenamefont
  {Bran}, \citenamefont {Grundler}, \citenamefont {Huth}, \citenamefont
  {Plank}, \citenamefont {Schmidt}, \citenamefont {{van Dijken}}, \citenamefont
  {Streubel}, \citenamefont {Dobrovoloskiy}, \citenamefont {Scagnoli},
  \citenamefont {Heyderman}, \citenamefont {Donnelly}, \citenamefont {Hellwig},
  \citenamefont {Fallarino}, \citenamefont {Jungfleisch}, \citenamefont
  {Farhan}, \citenamefont {Maccaferri}, \citenamefont {Vavassori},
  \citenamefont {Fischer}, \citenamefont {Tomasello}, \citenamefont
  {Finocchio}, \citenamefont {Cl{\'e}rac}, \citenamefont {Sessoli},
  \citenamefont {Makarov}, \citenamefont {Sheka}, \citenamefont {Krawczyk},
  \citenamefont {Gallardo}, \citenamefont {Landeros}, \citenamefont
  {{d'Aquino}}, \citenamefont {Hertel}, \citenamefont {Pirro}, \citenamefont
  {Ciubotaru}, \citenamefont {Becherer}, \citenamefont {Gartside},
  \citenamefont {Ono}, \citenamefont {Bortolotti},\ and\ \citenamefont
  {{Fern{\'a}ndez-Pacheco}}}]{gubbiotti20252025}%
  \BibitemOpen
  \bibfield  {author} {\bibinfo {author} {\bibfnamefont {G.}~\bibnamefont
  {Gubbiotti}}, \bibinfo {author} {\bibfnamefont {A.}~\bibnamefont {Barman}},
  \bibinfo {author} {\bibfnamefont {S.}~\bibnamefont {Ladak}}, \bibinfo
  {author} {\bibfnamefont {C.}~\bibnamefont {Bran}}, \bibinfo {author}
  {\bibfnamefont {D.}~\bibnamefont {Grundler}}, \bibinfo {author}
  {\bibfnamefont {M.}~\bibnamefont {Huth}}, \bibinfo {author} {\bibfnamefont
  {H.}~\bibnamefont {Plank}}, \bibinfo {author} {\bibfnamefont
  {G.}~\bibnamefont {Schmidt}}, \bibinfo {author} {\bibfnamefont
  {S.}~\bibnamefont {{van Dijken}}}, \bibinfo {author} {\bibfnamefont
  {R.}~\bibnamefont {Streubel}}, \bibinfo {author} {\bibfnamefont
  {O.}~\bibnamefont {Dobrovoloskiy}}, \bibinfo {author} {\bibfnamefont
  {V.}~\bibnamefont {Scagnoli}}, \bibinfo {author} {\bibfnamefont
  {L.}~\bibnamefont {Heyderman}}, \bibinfo {author} {\bibfnamefont
  {C.}~\bibnamefont {Donnelly}}, \bibinfo {author} {\bibfnamefont
  {O.}~\bibnamefont {Hellwig}}, \bibinfo {author} {\bibfnamefont
  {L.}~\bibnamefont {Fallarino}}, \bibinfo {author} {\bibfnamefont {M.~B.}\
  \bibnamefont {Jungfleisch}}, \bibinfo {author} {\bibfnamefont
  {A.}~\bibnamefont {Farhan}}, \bibinfo {author} {\bibfnamefont
  {N.}~\bibnamefont {Maccaferri}}, \bibinfo {author} {\bibfnamefont
  {P.}~\bibnamefont {Vavassori}}, \bibinfo {author} {\bibfnamefont
  {P.}~\bibnamefont {Fischer}}, \bibinfo {author} {\bibfnamefont
  {R.}~\bibnamefont {Tomasello}}, \bibinfo {author} {\bibfnamefont
  {G.}~\bibnamefont {Finocchio}}, \bibinfo {author} {\bibfnamefont
  {R.}~\bibnamefont {Cl{\'e}rac}}, \bibinfo {author} {\bibfnamefont
  {R.}~\bibnamefont {Sessoli}}, \bibinfo {author} {\bibfnamefont
  {D.}~\bibnamefont {Makarov}}, \bibinfo {author} {\bibfnamefont {D.~D.}\
  \bibnamefont {Sheka}}, \bibinfo {author} {\bibfnamefont {M.}~\bibnamefont
  {Krawczyk}}, \bibinfo {author} {\bibfnamefont {R.}~\bibnamefont {Gallardo}},
  \bibinfo {author} {\bibfnamefont {P.}~\bibnamefont {Landeros}}, \bibinfo
  {author} {\bibfnamefont {M.}~\bibnamefont {{d'Aquino}}}, \bibinfo {author}
  {\bibfnamefont {R.}~\bibnamefont {Hertel}}, \bibinfo {author} {\bibfnamefont
  {P.}~\bibnamefont {Pirro}}, \bibinfo {author} {\bibfnamefont
  {F.}~\bibnamefont {Ciubotaru}}, \bibinfo {author} {\bibfnamefont
  {M.}~\bibnamefont {Becherer}}, \bibinfo {author} {\bibfnamefont
  {J.}~\bibnamefont {Gartside}}, \bibinfo {author} {\bibfnamefont
  {T.}~\bibnamefont {Ono}}, \bibinfo {author} {\bibfnamefont {P.}~\bibnamefont
  {Bortolotti}},\ and\ \bibinfo {author} {\bibfnamefont {A.}~\bibnamefont
  {{Fern{\'a}ndez-Pacheco}}},\ }\bibfield  {title} {\bibinfo {title} {2025
  roadmap on {{3D}} nanomagnetism},\ }\href
  {https://doi.org/10.1088/1361-648X/ad9655} {\bibfield  {journal} {\bibinfo
  {journal} {J. Phys.: Condens. Matter}\ }\textbf {\bibinfo {volume} {37}},\
  \bibinfo {pages} {143502} (\bibinfo {year} {2025})}\BibitemShut {NoStop}%
\bibitem [{\citenamefont {Faddeev}\ and\ \citenamefont
  {Niemi}(1997)}]{Faddeev1997}%
  \BibitemOpen
  \bibfield  {author} {\bibinfo {author} {\bibfnamefont {L.}~\bibnamefont
  {Faddeev}}\ and\ \bibinfo {author} {\bibfnamefont {A.~J.}\ \bibnamefont
  {Niemi}},\ }\bibfield  {title} {\bibinfo {title} {Stable knot-like structures
  in classical field theory},\ }\href {https://doi.org/10.1038/387058a0}
  {\bibfield  {journal} {\bibinfo  {journal} {Nature}\ }\textbf {\bibinfo
  {volume} {387}},\ \bibinfo {pages} {58} (\bibinfo {year} {1997})}\BibitemShut
  {NoStop}%
\bibitem [{\citenamefont {Whitehead}(1947)}]{whitehead1947}%
  \BibitemOpen
  \bibfield  {author} {\bibinfo {author} {\bibfnamefont {J.}~\bibnamefont
  {Whitehead}},\ }\bibfield  {title} {\bibinfo {title} {An expression of hopf's
  invariant as an integral},\ }\href {https://doi.org/10.1073/pnas.33.5.117}
  {\bibfield  {journal} {\bibinfo  {journal} {Proc. Natl. Acad. Sci. U.S.A.}\
  }\textbf {\bibinfo {volume} {33}},\ \bibinfo {pages} {117} (\bibinfo {year}
  {1947})}\BibitemShut {NoStop}%
\bibitem [{\citenamefont {Christensen}\ \emph {et~al.}(2024)\citenamefont
  {Christensen}, \citenamefont {Staub}, \citenamefont {Devidas}, \citenamefont
  {Kalisky}, \citenamefont {Nowack}, \citenamefont {Webb}, \citenamefont
  {Andersen}, \citenamefont {Huck}, \citenamefont {Broadway}, \citenamefont
  {Wagner}, \citenamefont {Maletinsky}, \citenamefont {van~der Sar},
  \citenamefont {Du}, \citenamefont {Yacoby}, \citenamefont {Collomb},
  \citenamefont {Bending}, \citenamefont {Oral}, \citenamefont {Hug},
  \citenamefont {Mandru}, \citenamefont {Neu}, \citenamefont {Schumacher},
  \citenamefont {Sievers}, \citenamefont {Saito}, \citenamefont
  {Khajetoorians}, \citenamefont {Hauptmann}, \citenamefont {Baumann},
  \citenamefont {Eichler}, \citenamefont {Degen}, \citenamefont {McCord},
  \citenamefont {Vogel}, \citenamefont {Fiebig}, \citenamefont {Fischer},
  \citenamefont {Hierro-Rodriguez}, \citenamefont {Finizio}, \citenamefont
  {Dhesi}, \citenamefont {Donnelly}, \citenamefont {Büttner}, \citenamefont
  {Kfir}, \citenamefont {Hu}, \citenamefont {Zayko}, \citenamefont {Eisebitt},
  \citenamefont {Pfau}, \citenamefont {Frömter}, \citenamefont {Kläui},
  \citenamefont {Yasin}, \citenamefont {McMorran}, \citenamefont {Seki},
  \citenamefont {Yu}, \citenamefont {Lubk}, \citenamefont {Wolf}, \citenamefont
  {Pryds}, \citenamefont {Makarov},\ and\ \citenamefont
  {Poggio}}]{Christensen_2024}%
  \BibitemOpen
  \bibfield  {author} {\bibinfo {author} {\bibfnamefont {D.~V.}\ \bibnamefont
  {Christensen}}, \bibinfo {author} {\bibfnamefont {U.}~\bibnamefont {Staub}},
  \bibinfo {author} {\bibfnamefont {T.~R.}\ \bibnamefont {Devidas}}, \bibinfo
  {author} {\bibfnamefont {B.}~\bibnamefont {Kalisky}}, \bibinfo {author}
  {\bibfnamefont {K.~C.}\ \bibnamefont {Nowack}}, \bibinfo {author}
  {\bibfnamefont {J.~L.}\ \bibnamefont {Webb}}, \bibinfo {author}
  {\bibfnamefont {U.~L.}\ \bibnamefont {Andersen}}, \bibinfo {author}
  {\bibfnamefont {A.}~\bibnamefont {Huck}}, \bibinfo {author} {\bibfnamefont
  {D.~A.}\ \bibnamefont {Broadway}}, \bibinfo {author} {\bibfnamefont
  {K.}~\bibnamefont {Wagner}}, \bibinfo {author} {\bibfnamefont
  {P.}~\bibnamefont {Maletinsky}}, \bibinfo {author} {\bibfnamefont
  {T.}~\bibnamefont {van~der Sar}}, \bibinfo {author} {\bibfnamefont {C.~R.}\
  \bibnamefont {Du}}, \bibinfo {author} {\bibfnamefont {A.}~\bibnamefont
  {Yacoby}}, \bibinfo {author} {\bibfnamefont {D.}~\bibnamefont {Collomb}},
  \bibinfo {author} {\bibfnamefont {S.}~\bibnamefont {Bending}}, \bibinfo
  {author} {\bibfnamefont {A.}~\bibnamefont {Oral}}, \bibinfo {author}
  {\bibfnamefont {H.~J.}\ \bibnamefont {Hug}}, \bibinfo {author} {\bibfnamefont
  {A.-O.}\ \bibnamefont {Mandru}}, \bibinfo {author} {\bibfnamefont
  {V.}~\bibnamefont {Neu}}, \bibinfo {author} {\bibfnamefont {H.~W.}\
  \bibnamefont {Schumacher}}, \bibinfo {author} {\bibfnamefont
  {S.}~\bibnamefont {Sievers}}, \bibinfo {author} {\bibfnamefont
  {H.}~\bibnamefont {Saito}}, \bibinfo {author} {\bibfnamefont {A.~A.}\
  \bibnamefont {Khajetoorians}}, \bibinfo {author} {\bibfnamefont
  {N.}~\bibnamefont {Hauptmann}}, \bibinfo {author} {\bibfnamefont
  {S.}~\bibnamefont {Baumann}}, \bibinfo {author} {\bibfnamefont
  {A.}~\bibnamefont {Eichler}}, \bibinfo {author} {\bibfnamefont {C.~L.}\
  \bibnamefont {Degen}}, \bibinfo {author} {\bibfnamefont {J.}~\bibnamefont
  {McCord}}, \bibinfo {author} {\bibfnamefont {M.}~\bibnamefont {Vogel}},
  \bibinfo {author} {\bibfnamefont {M.}~\bibnamefont {Fiebig}}, \bibinfo
  {author} {\bibfnamefont {P.}~\bibnamefont {Fischer}}, \bibinfo {author}
  {\bibfnamefont {A.}~\bibnamefont {Hierro-Rodriguez}}, \bibinfo {author}
  {\bibfnamefont {S.}~\bibnamefont {Finizio}}, \bibinfo {author} {\bibfnamefont
  {S.~S.}\ \bibnamefont {Dhesi}}, \bibinfo {author} {\bibfnamefont
  {C.}~\bibnamefont {Donnelly}}, \bibinfo {author} {\bibfnamefont
  {F.}~\bibnamefont {Büttner}}, \bibinfo {author} {\bibfnamefont
  {O.}~\bibnamefont {Kfir}}, \bibinfo {author} {\bibfnamefont {W.}~\bibnamefont
  {Hu}}, \bibinfo {author} {\bibfnamefont {S.}~\bibnamefont {Zayko}}, \bibinfo
  {author} {\bibfnamefont {S.}~\bibnamefont {Eisebitt}}, \bibinfo {author}
  {\bibfnamefont {B.}~\bibnamefont {Pfau}}, \bibinfo {author} {\bibfnamefont
  {R.}~\bibnamefont {Frömter}}, \bibinfo {author} {\bibfnamefont
  {M.}~\bibnamefont {Kläui}}, \bibinfo {author} {\bibfnamefont {F.~S.}\
  \bibnamefont {Yasin}}, \bibinfo {author} {\bibfnamefont {B.~J.}\ \bibnamefont
  {McMorran}}, \bibinfo {author} {\bibfnamefont {S.}~\bibnamefont {Seki}},
  \bibinfo {author} {\bibfnamefont {X.}~\bibnamefont {Yu}}, \bibinfo {author}
  {\bibfnamefont {A.}~\bibnamefont {Lubk}}, \bibinfo {author} {\bibfnamefont
  {D.}~\bibnamefont {Wolf}}, \bibinfo {author} {\bibfnamefont {N.}~\bibnamefont
  {Pryds}}, \bibinfo {author} {\bibfnamefont {D.}~\bibnamefont {Makarov}},\
  and\ \bibinfo {author} {\bibfnamefont {M.}~\bibnamefont {Poggio}},\
  }\bibfield  {title} {\bibinfo {title} {2024 roadmap on magnetic microscopy
  techniques and their applications in materials science},\ }\href
  {https://doi.org/10.1088/2515-7639/ad31b5} {\bibfield  {journal} {\bibinfo
  {journal} {J. Phys.: Mater.}\ }\textbf {\bibinfo {volume} {7}},\ \bibinfo
  {pages} {032501} (\bibinfo {year} {2024})}\BibitemShut {NoStop}%
\bibitem [{\citenamefont {Azhar}\ \emph {et~al.}(2022)\citenamefont {Azhar},
  \citenamefont {Kravchuk},\ and\ \citenamefont {Garst}}]{Azhar2022}%
  \BibitemOpen
  \bibfield  {author} {\bibinfo {author} {\bibfnamefont {M.}~\bibnamefont
  {Azhar}}, \bibinfo {author} {\bibfnamefont {V.~P.}\ \bibnamefont
  {Kravchuk}},\ and\ \bibinfo {author} {\bibfnamefont {M.}~\bibnamefont
  {Garst}},\ }\bibfield  {title} {\bibinfo {title} {Screw dislocations in
  chiral magnets},\ }\href {https://doi.org/10.1103/PhysRevLett.128.157204}
  {\bibfield  {journal} {\bibinfo  {journal} {Phys. Rev. Lett.}\ }\textbf
  {\bibinfo {volume} {128}},\ \bibinfo {pages} {157204} (\bibinfo {year}
  {2022})}\BibitemShut {NoStop}%
\bibitem [{\citenamefont {Azhar}\ \emph {et~al.}(2024)\citenamefont {Azhar},
  \citenamefont {Shaju}, \citenamefont {Knapman}, \citenamefont {Pignedoli},\
  and\ \citenamefont {Everschor-Sitte}}]{azhar2024}%
  \BibitemOpen
  \bibfield  {author} {\bibinfo {author} {\bibfnamefont {M.}~\bibnamefont
  {Azhar}}, \bibinfo {author} {\bibfnamefont {S.~C.}\ \bibnamefont {Shaju}},
  \bibinfo {author} {\bibfnamefont {R.}~\bibnamefont {Knapman}}, \bibinfo
  {author} {\bibfnamefont {A.}~\bibnamefont {Pignedoli}},\ and\ \bibinfo
  {author} {\bibfnamefont {K.}~\bibnamefont {Everschor-Sitte}},\ }\bibfield
  {title} {\bibinfo {title} {3d magnetic textures with mixed topology:
  Unlocking the tunable hopf index},\ }\href
  {https://arxiv.org/abs/2411.06929v1} {\bibfield  {journal} {\bibinfo
  {journal} {arXiv:2411.06929}\ } (\bibinfo {year} {2024})}\BibitemShut
  {NoStop}%
\bibitem [{\citenamefont {Wu}\ \emph {et~al.}(2025)\citenamefont {Wu},
  \citenamefont {Mata-Cervera}, \citenamefont {Wang}, \citenamefont {Zhu},
  \citenamefont {Qiu},\ and\ \citenamefont {Shen}}]{wu2025photonic}%
  \BibitemOpen
  \bibfield  {author} {\bibinfo {author} {\bibfnamefont {H.}~\bibnamefont
  {Wu}}, \bibinfo {author} {\bibfnamefont {N.}~\bibnamefont {Mata-Cervera}},
  \bibinfo {author} {\bibfnamefont {H.}~\bibnamefont {Wang}}, \bibinfo {author}
  {\bibfnamefont {Z.}~\bibnamefont {Zhu}}, \bibinfo {author} {\bibfnamefont
  {C.}~\bibnamefont {Qiu}},\ and\ \bibinfo {author} {\bibfnamefont
  {Y.}~\bibnamefont {Shen}},\ }\bibfield  {title} {\bibinfo {title} {Photonic
  torons with 3d topology transitions and tunable spin monopoles},\ }\href@noop
  {} {\bibfield  {journal} {\bibinfo  {journal} {Phys. Rev. Lett.}\ }\textbf
  {\bibinfo {volume} {135}},\ \bibinfo {pages} {063802} (\bibinfo {year}
  {2025})}\BibitemShut {NoStop}%
\bibitem [{\citenamefont {Knapman}\ \emph {et~al.}(2025)\citenamefont
  {Knapman}, \citenamefont {Azhar}, \citenamefont {Pignedoli}, \citenamefont
  {Gallard}, \citenamefont {Hertel}, \citenamefont {Leliaert},\ and\
  \citenamefont {Everschor-Sitte}}]{Knapman2025}%
  \BibitemOpen
  \bibfield  {author} {\bibinfo {author} {\bibfnamefont {R.}~\bibnamefont
  {Knapman}}, \bibinfo {author} {\bibfnamefont {M.}~\bibnamefont {Azhar}},
  \bibinfo {author} {\bibfnamefont {A.}~\bibnamefont {Pignedoli}}, \bibinfo
  {author} {\bibfnamefont {L.}~\bibnamefont {Gallard}}, \bibinfo {author}
  {\bibfnamefont {R.}~\bibnamefont {Hertel}}, \bibinfo {author} {\bibfnamefont
  {J.}~\bibnamefont {Leliaert}},\ and\ \bibinfo {author} {\bibfnamefont
  {K.}~\bibnamefont {Everschor-Sitte}},\ }\bibfield  {title} {\bibinfo {title}
  {Numerical calculation of the hopf index for three-dimensional magnetic
  textures},\ }\href {https://doi.org/10.1103/PhysRevB.111.134408} {\bibfield
  {journal} {\bibinfo  {journal} {Phys. Rev. B}\ }\textbf {\bibinfo {volume}
  {111}},\ \bibinfo {pages} {134408} (\bibinfo {year} {2025})}\BibitemShut
  {NoStop}%
\bibitem [{\citenamefont {Yu}\ \emph {et~al.}(2023)\citenamefont {Yu},
  \citenamefont {Liu}, \citenamefont {Iakoubovskii}, \citenamefont {Nakajima},
  \citenamefont {Kanazawa}, \citenamefont {Nagaosa},\ and\ \citenamefont
  {Tokura}}]{Yu2023}%
  \BibitemOpen
  \bibfield  {author} {\bibinfo {author} {\bibfnamefont {X.}~\bibnamefont
  {Yu}}, \bibinfo {author} {\bibfnamefont {Y.}~\bibnamefont {Liu}}, \bibinfo
  {author} {\bibfnamefont {K.~V.}\ \bibnamefont {Iakoubovskii}}, \bibinfo
  {author} {\bibfnamefont {K.}~\bibnamefont {Nakajima}}, \bibinfo {author}
  {\bibfnamefont {N.}~\bibnamefont {Kanazawa}}, \bibinfo {author}
  {\bibfnamefont {N.}~\bibnamefont {Nagaosa}},\ and\ \bibinfo {author}
  {\bibfnamefont {Y.}~\bibnamefont {Tokura}},\ }\bibfield  {title} {\bibinfo
  {title} {Realization and current-driven dynamics of fractional hopfions and
  their ensembles in a helimagnet fege},\ }\href
  {https://doi.org/https://doi.org/10.1002/adma.202210646} {\bibfield
  {journal} {\bibinfo  {journal} {Adv. Mater.}\ }\textbf {\bibinfo {volume}
  {35}},\ \bibinfo {pages} {2210646} (\bibinfo {year} {2023})}\BibitemShut
  {NoStop}%
\bibitem [{\citenamefont {Desplat}\ \emph {et~al.}(2018)\citenamefont
  {Desplat}, \citenamefont {Suess}, \citenamefont {Kim},\ and\ \citenamefont
  {Stamps}}]{desplat2018thermal}%
  \BibitemOpen
  \bibfield  {author} {\bibinfo {author} {\bibfnamefont {L.}~\bibnamefont
  {Desplat}}, \bibinfo {author} {\bibfnamefont {D.}~\bibnamefont {Suess}},
  \bibinfo {author} {\bibfnamefont {J.-V.}\ \bibnamefont {Kim}},\ and\ \bibinfo
  {author} {\bibfnamefont {R.~L.}\ \bibnamefont {Stamps}},\ }\bibfield  {title}
  {\bibinfo {title} {Thermal stability of metastable magnetic skyrmions:
  Entropic narrowing and significance of internal eigenmodes},\ }\href
  {https://doi.org/10.1103/PhysRevB.98.134407} {\bibfield  {journal} {\bibinfo
  {journal} {Phys. Rev. B}\ }\textbf {\bibinfo {volume} {98}},\ \bibinfo
  {pages} {134407} (\bibinfo {year} {2018})}\BibitemShut {NoStop}%
\bibitem [{\citenamefont {Desplat}\ \emph {et~al.}(2019)\citenamefont
  {Desplat}, \citenamefont {Kim},\ and\ \citenamefont
  {Stamps}}]{desplat2019paths}%
  \BibitemOpen
  \bibfield  {author} {\bibinfo {author} {\bibfnamefont {L.}~\bibnamefont
  {Desplat}}, \bibinfo {author} {\bibfnamefont {J.-V.}\ \bibnamefont {Kim}},\
  and\ \bibinfo {author} {\bibfnamefont {R.~L.}\ \bibnamefont {Stamps}},\
  }\bibfield  {title} {\bibinfo {title} {Paths to annihilation of first-and
  second-order (anti) skyrmions via (anti) meron nucleation on the frustrated
  square lattice},\ }\href {https://doi.org/10.1103/PhysRevB.99.174409}
  {\bibfield  {journal} {\bibinfo  {journal} {Phys. Rev. B}\ }\textbf {\bibinfo
  {volume} {99}},\ \bibinfo {pages} {174409} (\bibinfo {year}
  {2019})}\BibitemShut {NoStop}%
\bibitem [{\citenamefont {Desplat}\ \emph {et~al.}(2020)\citenamefont
  {Desplat}, \citenamefont {Vogler}, \citenamefont {Kim}, \citenamefont
  {Stamps},\ and\ \citenamefont {Suess}}]{desplat2020path}%
  \BibitemOpen
  \bibfield  {author} {\bibinfo {author} {\bibfnamefont {L.}~\bibnamefont
  {Desplat}}, \bibinfo {author} {\bibfnamefont {C.}~\bibnamefont {Vogler}},
  \bibinfo {author} {\bibfnamefont {J.-V.}\ \bibnamefont {Kim}}, \bibinfo
  {author} {\bibfnamefont {R.~L.}\ \bibnamefont {Stamps}},\ and\ \bibinfo
  {author} {\bibfnamefont {D.}~\bibnamefont {Suess}},\ }\bibfield  {title}
  {\bibinfo {title} {Path sampling for lifetimes of metastable magnetic
  skyrmions and direct comparison with {Kramers}' method},\ }\href
  {https://doi.org/10.1103/PhysRevB.101.060403} {\bibfield  {journal} {\bibinfo
   {journal} {Phys. Rev. B}\ }\textbf {\bibinfo {volume} {101}},\ \bibinfo
  {pages} {060403(R)} (\bibinfo {year} {2020})}\BibitemShut {NoStop}%
\bibitem [{\citenamefont {Bessarab}\ \emph {et~al.}(2018)\citenamefont
  {Bessarab}, \citenamefont {M{\"u}ller}, \citenamefont {Lobanov},
  \citenamefont {Rybakov}, \citenamefont {Kiselev}, \citenamefont
  {J{\'o}nsson}, \citenamefont {Uzdin}, \citenamefont {Bl{\"u}gel},
  \citenamefont {Bergqvist},\ and\ \citenamefont
  {Delin}}]{bessarab2018lifetime}%
  \BibitemOpen
  \bibfield  {author} {\bibinfo {author} {\bibfnamefont {P.~F.}\ \bibnamefont
  {Bessarab}}, \bibinfo {author} {\bibfnamefont {G.~P.}\ \bibnamefont
  {M{\"u}ller}}, \bibinfo {author} {\bibfnamefont {I.~S.}\ \bibnamefont
  {Lobanov}}, \bibinfo {author} {\bibfnamefont {F.~N.}\ \bibnamefont
  {Rybakov}}, \bibinfo {author} {\bibfnamefont {N.~S.}\ \bibnamefont
  {Kiselev}}, \bibinfo {author} {\bibfnamefont {H.}~\bibnamefont
  {J{\'o}nsson}}, \bibinfo {author} {\bibfnamefont {V.~M.}\ \bibnamefont
  {Uzdin}}, \bibinfo {author} {\bibfnamefont {S.}~\bibnamefont {Bl{\"u}gel}},
  \bibinfo {author} {\bibfnamefont {L.}~\bibnamefont {Bergqvist}},\ and\
  \bibinfo {author} {\bibfnamefont {A.}~\bibnamefont {Delin}},\ }\bibfield
  {title} {\bibinfo {title} {Lifetime of racetrack skyrmions},\ }\href
  {https://doi.org/10.1038/s41598-018-21623-3} {\bibfield  {journal} {\bibinfo
  {journal} {Sci. Rep.}\ }\textbf {\bibinfo {volume} {8}},\ \bibinfo {pages}
  {1} (\bibinfo {year} {2018})}\BibitemShut {NoStop}%
\bibitem [{\citenamefont {von Malottki}\ \emph {et~al.}(2019)\citenamefont {von
  Malottki}, \citenamefont {Bessarab}, \citenamefont {Haldar}, \citenamefont
  {Delin},\ and\ \citenamefont {Heinze}}]{von2019skyrmion}%
  \BibitemOpen
  \bibfield  {author} {\bibinfo {author} {\bibfnamefont {S.}~\bibnamefont {von
  Malottki}}, \bibinfo {author} {\bibfnamefont {P.~F.}\ \bibnamefont
  {Bessarab}}, \bibinfo {author} {\bibfnamefont {S.}~\bibnamefont {Haldar}},
  \bibinfo {author} {\bibfnamefont {A.}~\bibnamefont {Delin}},\ and\ \bibinfo
  {author} {\bibfnamefont {S.}~\bibnamefont {Heinze}},\ }\bibfield  {title}
  {\bibinfo {title} {Skyrmion lifetime in ultrathin films},\ }\href
  {https://doi.org/10.1103/PhysRevB.99.060409} {\bibfield  {journal} {\bibinfo
  {journal} {Phys. Rev. B}\ }\textbf {\bibinfo {volume} {99}},\ \bibinfo
  {pages} {060409(R)} (\bibinfo {year} {2019})}\BibitemShut {NoStop}%
\bibitem [{\citenamefont {Goerzen}\ \emph {et~al.}(2022)\citenamefont
  {Goerzen}, \citenamefont {von Malottki}, \citenamefont {Kwiatkowski},
  \citenamefont {Bessarab},\ and\ \citenamefont
  {Heinze}}]{goerzen2022atomistic}%
  \BibitemOpen
  \bibfield  {author} {\bibinfo {author} {\bibfnamefont {M.~A.}\ \bibnamefont
  {Goerzen}}, \bibinfo {author} {\bibfnamefont {S.}~\bibnamefont {von
  Malottki}}, \bibinfo {author} {\bibfnamefont {G.~J.}\ \bibnamefont
  {Kwiatkowski}}, \bibinfo {author} {\bibfnamefont {P.~F.}\ \bibnamefont
  {Bessarab}},\ and\ \bibinfo {author} {\bibfnamefont {S.}~\bibnamefont
  {Heinze}},\ }\bibfield  {title} {\bibinfo {title} {Atomistic spin simulations
  of electric-field-assisted nucleation and annihilation of magnetic skyrmions
  in {Pd/Fe/Ir(111)}},\ }\href {https://doi.org/10.1103/PhysRevB.105.214435}
  {\bibfield  {journal} {\bibinfo  {journal} {Phys. Rev. B}\ }\textbf {\bibinfo
  {volume} {105}},\ \bibinfo {pages} {214435} (\bibinfo {year}
  {2022})}\BibitemShut {NoStop}%
\bibitem [{\citenamefont {Muckel}\ \emph {et~al.}(2021)\citenamefont {Muckel},
  \citenamefont {von Malottki}, \citenamefont {Holl}, \citenamefont {Pestka},
  \citenamefont {Pratzer}, \citenamefont {Bessarab}, \citenamefont {Heinze},\
  and\ \citenamefont {Morgenstern}}]{muckel2021experimental}%
  \BibitemOpen
  \bibfield  {author} {\bibinfo {author} {\bibfnamefont {F.}~\bibnamefont
  {Muckel}}, \bibinfo {author} {\bibfnamefont {S.}~\bibnamefont {von
  Malottki}}, \bibinfo {author} {\bibfnamefont {C.}~\bibnamefont {Holl}},
  \bibinfo {author} {\bibfnamefont {B.}~\bibnamefont {Pestka}}, \bibinfo
  {author} {\bibfnamefont {M.}~\bibnamefont {Pratzer}}, \bibinfo {author}
  {\bibfnamefont {P.~F.}\ \bibnamefont {Bessarab}}, \bibinfo {author}
  {\bibfnamefont {S.}~\bibnamefont {Heinze}},\ and\ \bibinfo {author}
  {\bibfnamefont {M.}~\bibnamefont {Morgenstern}},\ }\bibfield  {title}
  {\bibinfo {title} {Experimental identification of two distinct skyrmion
  collapse mechanisms},\ }\href {https://doi.org/10.1038/s41567-020-01101-2}
  {\bibfield  {journal} {\bibinfo  {journal} {Nat. Phys.}\ }\textbf {\bibinfo
  {volume} {17}},\ \bibinfo {pages} {395} (\bibinfo {year} {2021})}\BibitemShut
  {NoStop}%
\bibitem [{\citenamefont {Wild}\ \emph {et~al.}(2017)\citenamefont {Wild},
  \citenamefont {Meier}, \citenamefont {P{\"o}llath}, \citenamefont
  {Kronseder}, \citenamefont {Bauer}, \citenamefont {Chacon}, \citenamefont
  {Halder}, \citenamefont {Schowalter}, \citenamefont {Rosenauer},
  \citenamefont {Zweck} \emph {et~al.}}]{wild2017entropy}%
  \BibitemOpen
  \bibfield  {author} {\bibinfo {author} {\bibfnamefont {J.}~\bibnamefont
  {Wild}}, \bibinfo {author} {\bibfnamefont {T.~N.}\ \bibnamefont {Meier}},
  \bibinfo {author} {\bibfnamefont {S.}~\bibnamefont {P{\"o}llath}}, \bibinfo
  {author} {\bibfnamefont {M.}~\bibnamefont {Kronseder}}, \bibinfo {author}
  {\bibfnamefont {A.}~\bibnamefont {Bauer}}, \bibinfo {author} {\bibfnamefont
  {A.}~\bibnamefont {Chacon}}, \bibinfo {author} {\bibfnamefont
  {M.}~\bibnamefont {Halder}}, \bibinfo {author} {\bibfnamefont
  {M.}~\bibnamefont {Schowalter}}, \bibinfo {author} {\bibfnamefont
  {A.}~\bibnamefont {Rosenauer}}, \bibinfo {author} {\bibfnamefont
  {J.}~\bibnamefont {Zweck}}, \emph {et~al.},\ }\bibfield  {title} {\bibinfo
  {title} {Entropy-limited topological protection of skyrmions},\ }\href
  {https://doi.org/10.1126/sciadv.1701704} {\bibfield  {journal} {\bibinfo
  {journal} {Sci. Adv.}\ }\textbf {\bibinfo {volume} {3}},\ \bibinfo {pages}
  {e1701704} (\bibinfo {year} {2017})}\BibitemShut {NoStop}%
\bibitem [{\citenamefont {Prychynenko}\ \emph {et~al.}(2018)\citenamefont
  {Prychynenko}, \citenamefont {Sitte}, \citenamefont {Litzius}, \citenamefont
  {Kr\"uger}, \citenamefont {Bourianoff}, \citenamefont {Kl\"aui},
  \citenamefont {Sinova},\ and\ \citenamefont
  {Everschor-Sitte}}]{prychynenko2018magnetic}%
  \BibitemOpen
  \bibfield  {author} {\bibinfo {author} {\bibfnamefont {D.}~\bibnamefont
  {Prychynenko}}, \bibinfo {author} {\bibfnamefont {M.}~\bibnamefont {Sitte}},
  \bibinfo {author} {\bibfnamefont {K.}~\bibnamefont {Litzius}}, \bibinfo
  {author} {\bibfnamefont {B.}~\bibnamefont {Kr\"uger}}, \bibinfo {author}
  {\bibfnamefont {G.}~\bibnamefont {Bourianoff}}, \bibinfo {author}
  {\bibfnamefont {M.}~\bibnamefont {Kl\"aui}}, \bibinfo {author} {\bibfnamefont
  {J.}~\bibnamefont {Sinova}},\ and\ \bibinfo {author} {\bibfnamefont
  {K.}~\bibnamefont {Everschor-Sitte}},\ }\bibfield  {title} {\bibinfo {title}
  {Magnetic skyrmion as a nonlinear resistive element: A potential building
  block for reservoir computing},\ }\href
  {https://doi.org/10.1103/PhysRevApplied.9.014034} {\bibfield  {journal}
  {\bibinfo  {journal} {Phys. Rev. Appl.}\ }\textbf {\bibinfo {volume} {9}},\
  \bibinfo {pages} {014034} (\bibinfo {year} {2018})}\BibitemShut {NoStop}%
\bibitem [{\citenamefont {Legrand}\ \emph {et~al.}(2017)\citenamefont
  {Legrand}, \citenamefont {Maccariello}, \citenamefont {Reyren}, \citenamefont
  {Garcia}, \citenamefont {Moutafis}, \citenamefont {Moreau-Luchaire},
  \citenamefont {Collin}, \citenamefont {Bouzehouane}, \citenamefont {Cros},\
  and\ \citenamefont {Fert}}]{legrand2017room}%
  \BibitemOpen
  \bibfield  {author} {\bibinfo {author} {\bibfnamefont {W.}~\bibnamefont
  {Legrand}}, \bibinfo {author} {\bibfnamefont {D.}~\bibnamefont
  {Maccariello}}, \bibinfo {author} {\bibfnamefont {N.}~\bibnamefont {Reyren}},
  \bibinfo {author} {\bibfnamefont {K.}~\bibnamefont {Garcia}}, \bibinfo
  {author} {\bibfnamefont {C.}~\bibnamefont {Moutafis}}, \bibinfo {author}
  {\bibfnamefont {C.}~\bibnamefont {Moreau-Luchaire}}, \bibinfo {author}
  {\bibfnamefont {S.}~\bibnamefont {Collin}}, \bibinfo {author} {\bibfnamefont
  {K.}~\bibnamefont {Bouzehouane}}, \bibinfo {author} {\bibfnamefont
  {V.}~\bibnamefont {Cros}},\ and\ \bibinfo {author} {\bibfnamefont
  {A.}~\bibnamefont {Fert}},\ }\bibfield  {title} {\bibinfo {title}
  {Room-temperature current-induced generation and motion of sub-100 nm
  skyrmions},\ }\href {https://doi.org/10.1021/acs.nanolett.7b00649} {\bibfield
   {journal} {\bibinfo  {journal} {Nano Lett.}\ }\textbf {\bibinfo {volume}
  {17}},\ \bibinfo {pages} {2703} (\bibinfo {year} {2017})}\BibitemShut
  {NoStop}%
\bibitem [{\citenamefont {Schott}\ \emph {et~al.}(2017)\citenamefont {Schott},
  \citenamefont {{Bernand-Mantel}}, \citenamefont {Ranno}, \citenamefont
  {Pizzini}, \citenamefont {Vogel}, \citenamefont {B{\'e}a}, \citenamefont
  {Baraduc}, \citenamefont {Auffret}, \citenamefont {Gaudin},\ and\
  \citenamefont {Givord}}]{schott2017skyrmion}%
  \BibitemOpen
  \bibfield  {author} {\bibinfo {author} {\bibfnamefont {M.}~\bibnamefont
  {Schott}}, \bibinfo {author} {\bibfnamefont {A.}~\bibnamefont
  {{Bernand-Mantel}}}, \bibinfo {author} {\bibfnamefont {L.}~\bibnamefont
  {Ranno}}, \bibinfo {author} {\bibfnamefont {S.}~\bibnamefont {Pizzini}},
  \bibinfo {author} {\bibfnamefont {J.}~\bibnamefont {Vogel}}, \bibinfo
  {author} {\bibfnamefont {H.}~\bibnamefont {B{\'e}a}}, \bibinfo {author}
  {\bibfnamefont {C.}~\bibnamefont {Baraduc}}, \bibinfo {author} {\bibfnamefont
  {S.}~\bibnamefont {Auffret}}, \bibinfo {author} {\bibfnamefont
  {G.}~\bibnamefont {Gaudin}},\ and\ \bibinfo {author} {\bibfnamefont
  {D.}~\bibnamefont {Givord}},\ }\bibfield  {title} {\bibinfo {title} {The
  {{Skyrmion Switch}}: {{Turning Magnetic Skyrmion Bubbles}} on and off with an
  {{Electric Field}}},\ }\href {https://doi.org/10.1021/acs.nanolett.7b00328}
  {\bibfield  {journal} {\bibinfo  {journal} {Nano Lett.}\ }\textbf {\bibinfo
  {volume} {17}},\ \bibinfo {pages} {3006} (\bibinfo {year}
  {2017})}\BibitemShut {NoStop}%
\bibitem [{\citenamefont {Troncoso}\ and\ \citenamefont
  {N{\'u}{\~n}ez}(2014)}]{troncoso2014brownian}%
  \BibitemOpen
  \bibfield  {author} {\bibinfo {author} {\bibfnamefont {R.~E.}\ \bibnamefont
  {Troncoso}}\ and\ \bibinfo {author} {\bibfnamefont {{\'A}.~S.}\ \bibnamefont
  {N{\'u}{\~n}ez}},\ }\bibfield  {title} {\bibinfo {title} {Brownian motion of
  massive skyrmions in magnetic thin films},\ }\href
  {https://doi.org/10.1016/j.aop.2014.10.007} {\bibfield  {journal} {\bibinfo
  {journal} {Annals of Physics}\ }\textbf {\bibinfo {volume} {351}},\ \bibinfo
  {pages} {850} (\bibinfo {year} {2014})}\BibitemShut {NoStop}%
\bibitem [{\citenamefont {Ritzmann}\ \emph {et~al.}(2018)\citenamefont
  {Ritzmann}, \citenamefont {{von Malottki}}, \citenamefont {Kim},
  \citenamefont {Heinze}, \citenamefont {Sinova},\ and\ \citenamefont
  {Dup{\'e}}}]{ritzmann2018trochoidal}%
  \BibitemOpen
  \bibfield  {author} {\bibinfo {author} {\bibfnamefont {U.}~\bibnamefont
  {Ritzmann}}, \bibinfo {author} {\bibfnamefont {S.}~\bibnamefont {{von
  Malottki}}}, \bibinfo {author} {\bibfnamefont {J.-V.}\ \bibnamefont {Kim}},
  \bibinfo {author} {\bibfnamefont {S.}~\bibnamefont {Heinze}}, \bibinfo
  {author} {\bibfnamefont {J.}~\bibnamefont {Sinova}},\ and\ \bibinfo {author}
  {\bibfnamefont {B.}~\bibnamefont {Dup{\'e}}},\ }\bibfield  {title} {\bibinfo
  {title} {{Trochoidal motion and pair generation in skyrmion and antiskyrmion
  dynamics under spin--orbit torques}},\ }\href
  {https://doi.org/10.1038/s41928-018-0114-0} {\bibfield  {journal} {\bibinfo
  {journal} {Nat. Electron.}\ }\textbf {\bibinfo {volume} {1}},\ \bibinfo
  {pages} {451 } (\bibinfo {year} {2018})}\BibitemShut {NoStop}%
\bibitem [{\citenamefont {Shimizu}\ \emph {et~al.}(2022)\citenamefont
  {Shimizu}, \citenamefont {Okumura}, \citenamefont {Kato},\ and\ \citenamefont
  {Motome}}]{Shimizu2022}%
  \BibitemOpen
  \bibfield  {author} {\bibinfo {author} {\bibfnamefont {K.}~\bibnamefont
  {Shimizu}}, \bibinfo {author} {\bibfnamefont {S.}~\bibnamefont {Okumura}},
  \bibinfo {author} {\bibfnamefont {Y.}~\bibnamefont {Kato}},\ and\ \bibinfo
  {author} {\bibfnamefont {Y.}~\bibnamefont {Motome}},\ }\bibfield  {title}
  {\bibinfo {title} {Phase degree of freedom and topology in multiple-$q$ spin
  textures},\ }\href {https://doi.org/10.1103/PhysRevB.105.224405} {\bibfield
  {journal} {\bibinfo  {journal} {Phys. Rev. B}\ }\textbf {\bibinfo {volume}
  {105}},\ \bibinfo {pages} {224405} (\bibinfo {year} {2022})}\BibitemShut
  {NoStop}%
\bibitem [{\citenamefont {Hayami}\ \emph {et~al.}(2021)\citenamefont {Hayami},
  \citenamefont {Okubo},\ and\ \citenamefont {Motome}}]{Hayami2021}%
  \BibitemOpen
  \bibfield  {author} {\bibinfo {author} {\bibfnamefont {S.}~\bibnamefont
  {Hayami}}, \bibinfo {author} {\bibfnamefont {T.}~\bibnamefont {Okubo}},\ and\
  \bibinfo {author} {\bibfnamefont {Y.}~\bibnamefont {Motome}},\ }\bibfield
  {title} {\bibinfo {title} {Phase shift in skyrmion crystals},\ }\href
  {https://doi.org/10.1038/s41467-021-27083-0} {\bibfield  {journal} {\bibinfo
  {journal} {Nat. Commun.}\ }\textbf {\bibinfo {volume} {12}},\ \bibinfo
  {pages} {6927} (\bibinfo {year} {2021})}\BibitemShut {NoStop}%
\bibitem [{\citenamefont {G{\"o}bel}\ \emph {et~al.}(2017)\citenamefont
  {G{\"o}bel}, \citenamefont {Mook}, \citenamefont {Henk},\ and\ \citenamefont
  {Mertig}}]{gobel2017antiferromagnetic}%
  \BibitemOpen
  \bibfield  {author} {\bibinfo {author} {\bibfnamefont {B.}~\bibnamefont
  {G{\"o}bel}}, \bibinfo {author} {\bibfnamefont {A.}~\bibnamefont {Mook}},
  \bibinfo {author} {\bibfnamefont {J.}~\bibnamefont {Henk}},\ and\ \bibinfo
  {author} {\bibfnamefont {I.}~\bibnamefont {Mertig}},\ }\bibfield  {title}
  {\bibinfo {title} {Antiferromagnetic skyrmion crystals: {{Generation}},
  topological {{Hall}}, and topological spin {{Hall}} effect},\ }\href
  {https://doi.org/10.1103/PhysRevB.96.060406} {\bibfield  {journal} {\bibinfo
  {journal} {Phys. Rev. B}\ }\textbf {\bibinfo {volume} {96}},\ \bibinfo
  {pages} {060406} (\bibinfo {year} {2017})}\BibitemShut {NoStop}%
\bibitem [{\citenamefont {Hamamoto}\ \emph {et~al.}(2015)\citenamefont
  {Hamamoto}, \citenamefont {Ezawa},\ and\ \citenamefont
  {Nagaosa}}]{Nagaosa2015}%
  \BibitemOpen
  \bibfield  {author} {\bibinfo {author} {\bibfnamefont {K.}~\bibnamefont
  {Hamamoto}}, \bibinfo {author} {\bibfnamefont {M.}~\bibnamefont {Ezawa}},\
  and\ \bibinfo {author} {\bibfnamefont {N.}~\bibnamefont {Nagaosa}},\
  }\bibfield  {title} {\bibinfo {title} {Quantized topological hall effect in
  skyrmion crystal},\ }\href {https://doi.org/10.1103/PhysRevB.92.115417}
  {\bibfield  {journal} {\bibinfo  {journal} {Phys. Rev. B}\ }\textbf {\bibinfo
  {volume} {92}},\ \bibinfo {pages} {115417} (\bibinfo {year}
  {2015})}\BibitemShut {NoStop}%
\bibitem [{\citenamefont {Lux}\ \emph {et~al.}(2024)\citenamefont {Lux},
  \citenamefont {Ghosh}, \citenamefont {Prass}, \citenamefont {Prodan},\ and\
  \citenamefont {Mokrousov}}]{Lux2024}%
  \BibitemOpen
  \bibfield  {author} {\bibinfo {author} {\bibfnamefont {F.~R.}\ \bibnamefont
  {Lux}}, \bibinfo {author} {\bibfnamefont {S.}~\bibnamefont {Ghosh}}, \bibinfo
  {author} {\bibfnamefont {P.}~\bibnamefont {Prass}}, \bibinfo {author}
  {\bibfnamefont {E.}~\bibnamefont {Prodan}},\ and\ \bibinfo {author}
  {\bibfnamefont {Y.}~\bibnamefont {Mokrousov}},\ }\bibfield  {title} {\bibinfo
  {title} {Unified topological characterization of electronic states in spin
  textures from noncommutative $k$-theory},\ }\href
  {https://doi.org/10.1103/PhysRevResearch.6.013102} {\bibfield  {journal}
  {\bibinfo  {journal} {Phys. Rev. Res.}\ }\textbf {\bibinfo {volume} {6}},\
  \bibinfo {pages} {013102} (\bibinfo {year} {2024})}\BibitemShut {NoStop}%
\bibitem [{\citenamefont {Prass}\ \emph {et~al.}(2024)\citenamefont {Prass},
  \citenamefont {Lux}, \citenamefont {Prodan}, \citenamefont {van Straten},\
  and\ \citenamefont {Mokrousov}}]{PascalSciPost}%
  \BibitemOpen
  \bibfield  {author} {\bibinfo {author} {\bibfnamefont {P.}~\bibnamefont
  {Prass}}, \bibinfo {author} {\bibfnamefont {F.~R.}\ \bibnamefont {Lux}},
  \bibinfo {author} {\bibfnamefont {E.}~\bibnamefont {Prodan}}, \bibinfo
  {author} {\bibfnamefont {D.}~\bibnamefont {van Straten}},\ and\ \bibinfo
  {author} {\bibfnamefont {Y.}~\bibnamefont {Mokrousov}},\ }\bibfield  {title}
  {\bibinfo {title} {{A $C^\ast$-algebraic view on the interaction of real- and
  reciprocal space topology in skyrmion crystals}},\ }\href
  {https://doi.org/10.21468/SciPostPhysCore.7.4.080} {\bibfield  {journal}
  {\bibinfo  {journal} {SciPost Phys. Core}\ }\textbf {\bibinfo {volume} {7}},\
  \bibinfo {pages} {080} (\bibinfo {year} {2024})}\BibitemShut {NoStop}%
\bibitem [{\citenamefont {Lux}\ \emph {et~al.}(2018)\citenamefont {Lux},
  \citenamefont {Freimuth}, \citenamefont {Bl{\"u}gel},\ and\ \citenamefont
  {Mokrousov}}]{Lux2018}%
  \BibitemOpen
  \bibfield  {author} {\bibinfo {author} {\bibfnamefont {F.~R.}\ \bibnamefont
  {Lux}}, \bibinfo {author} {\bibfnamefont {F.}~\bibnamefont {Freimuth}},
  \bibinfo {author} {\bibfnamefont {S.}~\bibnamefont {Bl{\"u}gel}},\ and\
  \bibinfo {author} {\bibfnamefont {Y.}~\bibnamefont {Mokrousov}},\ }\bibfield
  {title} {\bibinfo {title} {Engineering chiral and topological orbital
  magnetism of domain walls and skyrmions},\ }\href
  {https://doi.org/10.1038/s42005-018-0055-y} {\bibfield  {journal} {\bibinfo
  {journal} {Commun. Phys.}\ }\textbf {\bibinfo {volume} {1}},\ \bibinfo
  {pages} {60} (\bibinfo {year} {2018})}\BibitemShut {NoStop}%
\bibitem [{\citenamefont {Paddison}\ \emph {et~al.}(2022)\citenamefont
  {Paddison}, \citenamefont {Rai}, \citenamefont {May}, \citenamefont {Calder},
  \citenamefont {Stone}, \citenamefont {Frontzek},\ and\ \citenamefont
  {Christianson}}]{Paddison2022}%
  \BibitemOpen
  \bibfield  {author} {\bibinfo {author} {\bibfnamefont {J.~A.~M.}\
  \bibnamefont {Paddison}}, \bibinfo {author} {\bibfnamefont {B.~K.}\
  \bibnamefont {Rai}}, \bibinfo {author} {\bibfnamefont {A.~F.}\ \bibnamefont
  {May}}, \bibinfo {author} {\bibfnamefont {S.}~\bibnamefont {Calder}},
  \bibinfo {author} {\bibfnamefont {M.~B.}\ \bibnamefont {Stone}}, \bibinfo
  {author} {\bibfnamefont {M.~D.}\ \bibnamefont {Frontzek}},\ and\ \bibinfo
  {author} {\bibfnamefont {A.~D.}\ \bibnamefont {Christianson}},\ }\bibfield
  {title} {\bibinfo {title} {Magnetic interactions of the centrosymmetric
  skyrmion material ${\mathrm{gd}}_{2}{\mathrm{pdsi}}_{3}$},\ }\href
  {https://doi.org/10.1103/PhysRevLett.129.137202} {\bibfield  {journal}
  {\bibinfo  {journal} {Phys. Rev. Lett.}\ }\textbf {\bibinfo {volume} {129}},\
  \bibinfo {pages} {137202} (\bibinfo {year} {2022})}\BibitemShut {NoStop}%
\bibitem [{\citenamefont {Khanh}\ \emph {et~al.}(2020)\citenamefont {Khanh},
  \citenamefont {Nakajima}, \citenamefont {Yu}, \citenamefont {Gao},
  \citenamefont {Shibata}, \citenamefont {Hirschberger}, \citenamefont
  {Yamasaki}, \citenamefont {Sagayama}, \citenamefont {Nakao}, \citenamefont
  {Peng}, \citenamefont {Nakajima}, \citenamefont {Takagi}, \citenamefont
  {Arima}, \citenamefont {Tokura},\ and\ \citenamefont
  {Seki}}]{khanh2020nanometric}%
  \BibitemOpen
  \bibfield  {author} {\bibinfo {author} {\bibfnamefont {N.~D.}\ \bibnamefont
  {Khanh}}, \bibinfo {author} {\bibfnamefont {T.}~\bibnamefont {Nakajima}},
  \bibinfo {author} {\bibfnamefont {X.}~\bibnamefont {Yu}}, \bibinfo {author}
  {\bibfnamefont {S.}~\bibnamefont {Gao}}, \bibinfo {author} {\bibfnamefont
  {K.}~\bibnamefont {Shibata}}, \bibinfo {author} {\bibfnamefont
  {M.}~\bibnamefont {Hirschberger}}, \bibinfo {author} {\bibfnamefont
  {Y.}~\bibnamefont {Yamasaki}}, \bibinfo {author} {\bibfnamefont
  {H.}~\bibnamefont {Sagayama}}, \bibinfo {author} {\bibfnamefont
  {H.}~\bibnamefont {Nakao}}, \bibinfo {author} {\bibfnamefont
  {L.}~\bibnamefont {Peng}}, \bibinfo {author} {\bibfnamefont {K.}~\bibnamefont
  {Nakajima}}, \bibinfo {author} {\bibfnamefont {R.}~\bibnamefont {Takagi}},
  \bibinfo {author} {\bibfnamefont {T.-h.}\ \bibnamefont {Arima}}, \bibinfo
  {author} {\bibfnamefont {Y.}~\bibnamefont {Tokura}},\ and\ \bibinfo {author}
  {\bibfnamefont {S.}~\bibnamefont {Seki}},\ }\bibfield  {title} {\bibinfo
  {title} {Nanometric square skyrmion lattice in a centrosymmetric tetragonal
  magnet},\ }\href {https://doi.org/10.1038/s41565-020-0684-7} {\bibfield
  {journal} {\bibinfo  {journal} {Nat. Nanotechnol.}\ }\textbf {\bibinfo
  {volume} {15}},\ \bibinfo {pages} {444} (\bibinfo {year} {2020})}\BibitemShut
  {NoStop}%
\bibitem [{\citenamefont {Bouaziz}\ \emph {et~al.}(2022)\citenamefont
  {Bouaziz}, \citenamefont {Mendive-Tapia}, \citenamefont {Bl\"ugel},\ and\
  \citenamefont {Staunton}}]{Bouaziz2022}%
  \BibitemOpen
  \bibfield  {author} {\bibinfo {author} {\bibfnamefont {J.}~\bibnamefont
  {Bouaziz}}, \bibinfo {author} {\bibfnamefont {E.}~\bibnamefont
  {Mendive-Tapia}}, \bibinfo {author} {\bibfnamefont {S.}~\bibnamefont
  {Bl\"ugel}},\ and\ \bibinfo {author} {\bibfnamefont {J.~B.}\ \bibnamefont
  {Staunton}},\ }\bibfield  {title} {\bibinfo {title} {Fermi-surface origin of
  skyrmion lattices in centrosymmetric rare-earth intermetallics},\ }\href
  {https://doi.org/10.1103/PhysRevLett.128.157206} {\bibfield  {journal}
  {\bibinfo  {journal} {Phys. Rev. Lett.}\ }\textbf {\bibinfo {volume} {128}},\
  \bibinfo {pages} {157206} (\bibinfo {year} {2022})}\BibitemShut {NoStop}%
\bibitem [{\citenamefont {Chen}\ \emph
  {et~al.}(2025{\natexlab{a}})\citenamefont {Chen}, \citenamefont {Nomoto},
  \citenamefont {Hirschberger},\ and\ \citenamefont
  {Arita}}]{chen2025topological}%
  \BibitemOpen
  \bibfield  {author} {\bibinfo {author} {\bibfnamefont {H.-Y.}\ \bibnamefont
  {Chen}}, \bibinfo {author} {\bibfnamefont {T.}~\bibnamefont {Nomoto}},
  \bibinfo {author} {\bibfnamefont {M.}~\bibnamefont {Hirschberger}},\ and\
  \bibinfo {author} {\bibfnamefont {R.}~\bibnamefont {Arita}},\ }\bibfield
  {title} {\bibinfo {title} {Topological hall effect of skyrmions from first
  principles},\ }\href {https://doi.org/10.1103/PhysRevX.15.011054} {\bibfield
  {journal} {\bibinfo  {journal} {Phys. Rev. X}\ }\textbf {\bibinfo {volume}
  {15}},\ \bibinfo {pages} {011054} (\bibinfo {year}
  {2025}{\natexlab{a}})}\BibitemShut {NoStop}%
\bibitem [{\citenamefont {Petrovi\ifmmode~\acute{c}\else \'{c}\fi{}}\ \emph
  {et~al.}(2025)\citenamefont {Petrovi\ifmmode~\acute{c}\else \'{c}\fi{}},
  \citenamefont {Psaroudaki}, \citenamefont {Fischer}, \citenamefont {Garst},\
  and\ \citenamefont {Panagopoulos}}]{Petrovic2025}%
  \BibitemOpen
  \bibfield  {author} {\bibinfo {author} {\bibfnamefont {A.~P.}\ \bibnamefont
  {Petrovi\ifmmode~\acute{c}\else \'{c}\fi{}}}, \bibinfo {author}
  {\bibfnamefont {C.}~\bibnamefont {Psaroudaki}}, \bibinfo {author}
  {\bibfnamefont {P.}~\bibnamefont {Fischer}}, \bibinfo {author} {\bibfnamefont
  {M.}~\bibnamefont {Garst}},\ and\ \bibinfo {author} {\bibfnamefont
  {C.}~\bibnamefont {Panagopoulos}},\ }\bibfield  {title} {\bibinfo {title}
  {Colloquium: Quantum properties and functionalities of magnetic skyrmions},\
  }\href {https://doi.org/10.1103/RevModPhys.97.031001} {\bibfield  {journal}
  {\bibinfo  {journal} {Rev. Mod. Phys.}\ }\textbf {\bibinfo {volume} {97}},\
  \bibinfo {pages} {031001} (\bibinfo {year} {2025})}\BibitemShut {NoStop}%
\bibitem [{\citenamefont {Papanicolaou}\ and\ \citenamefont
  {Tomaras}(1991)}]{Papanicolaou1991}%
  \BibitemOpen
  \bibfield  {author} {\bibinfo {author} {\bibfnamefont {N.}~\bibnamefont
  {Papanicolaou}}\ and\ \bibinfo {author} {\bibfnamefont {T.}~\bibnamefont
  {Tomaras}},\ }\bibfield  {title} {\bibinfo {title} {Dynamics of magnetic
  vortices},\ }\href
  {https://doi.org/https://doi.org/10.1016/0550-3213(91)90410-Y} {\bibfield
  {journal} {\bibinfo  {journal} {Nucl. Phys. B}\ }\textbf {\bibinfo {volume}
  {360}},\ \bibinfo {pages} {425} (\bibinfo {year} {1991})}\BibitemShut
  {NoStop}%
\bibitem [{\citenamefont {Ochoa}\ and\ \citenamefont
  {Tserkovnyak}(2019)}]{Ochoa2019}%
  \BibitemOpen
  \bibfield  {author} {\bibinfo {author} {\bibfnamefont {H.}~\bibnamefont
  {Ochoa}}\ and\ \bibinfo {author} {\bibfnamefont {Y.}~\bibnamefont
  {Tserkovnyak}},\ }\bibfield  {title} {\bibinfo {title} {Quantum
  skyrmionics},\ }\href {https://doi.org/10.1142/S0217979219300056} {\bibfield
  {journal} {\bibinfo  {journal} {International Journal of Modern Physics B}\
  }\textbf {\bibinfo {volume} {33}},\ \bibinfo {pages} {1930005} (\bibinfo
  {year} {2019})},\ \Eprint
  {https://arxiv.org/abs/https://doi.org/10.1142/S0217979219300056}
  {https://doi.org/10.1142/S0217979219300056} \BibitemShut {NoStop}%
\bibitem [{\citenamefont {Sorn}\ \emph {et~al.}(2025)\citenamefont {Sorn},
  \citenamefont {Schmalian},\ and\ \citenamefont {Garst}}]{Sorn2024}%
  \BibitemOpen
  \bibfield  {author} {\bibinfo {author} {\bibfnamefont {S.}~\bibnamefont
  {Sorn}}, \bibinfo {author} {\bibfnamefont {J.}~\bibnamefont {Schmalian}},\
  and\ \bibinfo {author} {\bibfnamefont {M.}~\bibnamefont {Garst}},\ }\bibfield
   {title} {\bibinfo {title} {Topological dipoles of quantum skyrmions},\
  }\href@noop {} {\bibfield  {journal} {\bibinfo  {journal} {Physical Review
  X}\ }\textbf {\bibinfo {volume} {15}},\ \bibinfo {pages} {041037} (\bibinfo
  {year} {2025})}\BibitemShut {NoStop}%
\bibitem [{\citenamefont {Arovas}(2020)}]{Arovas}%
  \BibitemOpen
  \bibfield  {author} {\bibinfo {author} {\bibfnamefont {D.}~\bibnamefont
  {Arovas}},\ }\href@noop {} {\bibinfo {title} {Lecture notes on quantum hall
  effect (a work in progress)}} (\bibinfo {year} {2020}),\ \bibinfo {note}
  {unpublished lecture notes}\BibitemShut {NoStop}%
\bibitem [{\citenamefont {Takashima}\ \emph {et~al.}(2016)\citenamefont
  {Takashima}, \citenamefont {Ishizuka},\ and\ \citenamefont
  {Balents}}]{Balents2016}%
  \BibitemOpen
  \bibfield  {author} {\bibinfo {author} {\bibfnamefont {R.}~\bibnamefont
  {Takashima}}, \bibinfo {author} {\bibfnamefont {H.}~\bibnamefont
  {Ishizuka}},\ and\ \bibinfo {author} {\bibfnamefont {L.}~\bibnamefont
  {Balents}},\ }\bibfield  {title} {\bibinfo {title} {Quantum skyrmions in
  two-dimensional chiral magnets},\ }\href
  {https://doi.org/10.1103/PhysRevB.94.134415} {\bibfield  {journal} {\bibinfo
  {journal} {Phys. Rev. B}\ }\textbf {\bibinfo {volume} {94}},\ \bibinfo
  {pages} {134415} (\bibinfo {year} {2016})}\BibitemShut {NoStop}%
\bibitem [{\citenamefont {Bhowmick}\ \emph {et~al.}(2025)\citenamefont
  {Bhowmick}, \citenamefont {Haller}, \citenamefont {Kathyat}, \citenamefont
  {Schmidt},\ and\ \citenamefont {Sengupta}}]{Haller2025}%
  \BibitemOpen
  \bibfield  {author} {\bibinfo {author} {\bibfnamefont {D.}~\bibnamefont
  {Bhowmick}}, \bibinfo {author} {\bibfnamefont {A.}~\bibnamefont {Haller}},
  \bibinfo {author} {\bibfnamefont {D.~S.}\ \bibnamefont {Kathyat}}, \bibinfo
  {author} {\bibfnamefont {T.~L.}\ \bibnamefont {Schmidt}},\ and\ \bibinfo
  {author} {\bibfnamefont {P.}~\bibnamefont {Sengupta}},\ }\bibfield  {title}
  {\bibinfo {title} {Quantum skyrmion liquid},\ }\href
  {https://doi.org/10.1103/PhysRevB.111.134410} {\bibfield  {journal} {\bibinfo
   {journal} {Phys. Rev. B}\ }\textbf {\bibinfo {volume} {111}},\ \bibinfo
  {pages} {134410} (\bibinfo {year} {2025})}\BibitemShut {NoStop}%
\bibitem [{\citenamefont {Gromov}\ and\ \citenamefont
  {Radzihovsky}(2024)}]{Gromov2024}%
  \BibitemOpen
  \bibfield  {author} {\bibinfo {author} {\bibfnamefont {A.}~\bibnamefont
  {Gromov}}\ and\ \bibinfo {author} {\bibfnamefont {L.}~\bibnamefont
  {Radzihovsky}},\ }\bibfield  {title} {\bibinfo {title} {Colloquium: Fracton
  matter},\ }\href {https://doi.org/10.1103/RevModPhys.96.011001} {\bibfield
  {journal} {\bibinfo  {journal} {Rev. Mod. Phys.}\ }\textbf {\bibinfo {volume}
  {96}},\ \bibinfo {pages} {011001} (\bibinfo {year} {2024})}\BibitemShut
  {NoStop}%
\bibitem [{\citenamefont {Sala}\ \emph {et~al.}(2020)\citenamefont {Sala},
  \citenamefont {Rakovszky}, \citenamefont {Verresen}, \citenamefont {Knap},\
  and\ \citenamefont {Pollmann}}]{Pollmann2020}%
  \BibitemOpen
  \bibfield  {author} {\bibinfo {author} {\bibfnamefont {P.}~\bibnamefont
  {Sala}}, \bibinfo {author} {\bibfnamefont {T.}~\bibnamefont {Rakovszky}},
  \bibinfo {author} {\bibfnamefont {R.}~\bibnamefont {Verresen}}, \bibinfo
  {author} {\bibfnamefont {M.}~\bibnamefont {Knap}},\ and\ \bibinfo {author}
  {\bibfnamefont {F.}~\bibnamefont {Pollmann}},\ }\bibfield  {title} {\bibinfo
  {title} {Ergodicity breaking arising from hilbert space fragmentation in
  dipole-conserving hamiltonians},\ }\href
  {https://doi.org/10.1103/PhysRevX.10.011047} {\bibfield  {journal} {\bibinfo
  {journal} {Phys. Rev. X}\ }\textbf {\bibinfo {volume} {10}},\ \bibinfo
  {pages} {011047} (\bibinfo {year} {2020})}\BibitemShut {NoStop}%
\bibitem [{\citenamefont {Yang}\ \emph {et~al.}(2012)\citenamefont {Yang},
  \citenamefont {Hu}, \citenamefont {Papi\ifmmode~\acute{c}\else \'{c}\fi{}},\
  and\ \citenamefont {Haldane}}]{Haldane2012}%
  \BibitemOpen
  \bibfield  {author} {\bibinfo {author} {\bibfnamefont {B.}~\bibnamefont
  {Yang}}, \bibinfo {author} {\bibfnamefont {Z.-X.}\ \bibnamefont {Hu}},
  \bibinfo {author} {\bibfnamefont {Z.}~\bibnamefont
  {Papi\ifmmode~\acute{c}\else \'{c}\fi{}}},\ and\ \bibinfo {author}
  {\bibfnamefont {F.~D.~M.}\ \bibnamefont {Haldane}},\ }\bibfield  {title}
  {\bibinfo {title} {Model wave functions for the collective modes and the
  magnetoroton theory of the fractional quantum hall effect},\ }\href
  {https://doi.org/10.1103/PhysRevLett.108.256807} {\bibfield  {journal}
  {\bibinfo  {journal} {Phys. Rev. Lett.}\ }\textbf {\bibinfo {volume} {108}},\
  \bibinfo {pages} {256807} (\bibinfo {year} {2012})}\BibitemShut {NoStop}%
\bibitem [{\citenamefont {Pfleiderer}\ \emph {et~al.}(2001)\citenamefont
  {Pfleiderer}, \citenamefont {Julian},\ and\ \citenamefont
  {Lonzarich}}]{Pfleiderer2001}%
  \BibitemOpen
  \bibfield  {author} {\bibinfo {author} {\bibfnamefont {C.}~\bibnamefont
  {Pfleiderer}}, \bibinfo {author} {\bibfnamefont {S.~R.}\ \bibnamefont
  {Julian}},\ and\ \bibinfo {author} {\bibfnamefont {G.~G.}\ \bibnamefont
  {Lonzarich}},\ }\bibfield  {title} {\bibinfo {title} {Non-fermi-liquid nature
  of the normal state of itinerant-electron ferromagnets},\ }\href
  {https://doi.org/10.1038/35106527} {\bibfield  {journal} {\bibinfo  {journal}
  {Nature}\ }\textbf {\bibinfo {volume} {414}},\ \bibinfo {pages} {427}
  (\bibinfo {year} {2001})}\BibitemShut {NoStop}%
\bibitem [{\citenamefont {Pedrazzini}\ \emph {et~al.}(2007)\citenamefont
  {Pedrazzini}, \citenamefont {Wilhelm}, \citenamefont {Jaccard}, \citenamefont
  {Jarlborg}, \citenamefont {Schmidt}, \citenamefont {Hanfland}, \citenamefont
  {Akselrud}, \citenamefont {Yuan}, \citenamefont {Schwarz}, \citenamefont
  {Grin},\ and\ \citenamefont {Steglich}}]{Pedrazzini2007}%
  \BibitemOpen
  \bibfield  {author} {\bibinfo {author} {\bibfnamefont {P.}~\bibnamefont
  {Pedrazzini}}, \bibinfo {author} {\bibfnamefont {H.}~\bibnamefont {Wilhelm}},
  \bibinfo {author} {\bibfnamefont {D.}~\bibnamefont {Jaccard}}, \bibinfo
  {author} {\bibfnamefont {T.}~\bibnamefont {Jarlborg}}, \bibinfo {author}
  {\bibfnamefont {M.}~\bibnamefont {Schmidt}}, \bibinfo {author} {\bibfnamefont
  {M.}~\bibnamefont {Hanfland}}, \bibinfo {author} {\bibfnamefont
  {L.}~\bibnamefont {Akselrud}}, \bibinfo {author} {\bibfnamefont {H.~Q.}\
  \bibnamefont {Yuan}}, \bibinfo {author} {\bibfnamefont {U.}~\bibnamefont
  {Schwarz}}, \bibinfo {author} {\bibfnamefont {Y.}~\bibnamefont {Grin}},\ and\
  \bibinfo {author} {\bibfnamefont {F.}~\bibnamefont {Steglich}},\ }\bibfield
  {title} {\bibinfo {title} {Metallic state in cubic fege beyond its quantum
  phase transition},\ }\href {https://doi.org/10.1103/PhysRevLett.98.047204}
  {\bibfield  {journal} {\bibinfo  {journal} {Phys. Rev. Lett.}\ }\textbf
  {\bibinfo {volume} {98}},\ \bibinfo {pages} {047204} (\bibinfo {year}
  {2007})}\BibitemShut {NoStop}%
\bibitem [{\citenamefont {Hug}(2025)}]{Hug2021chapter}%
  \BibitemOpen
  \bibfield  {author} {\bibinfo {author} {\bibfnamefont {H.~J.}\ \bibnamefont
  {Hug}},\ }\bibfield  {title} {\bibinfo {title} {Mapping the magnetic field of
  skyrmions and spin spirals byscanning probe microscopy},\ }in\ \href
  {https://doi.org/10.1016/b978-0-12-820815-1.00016-x} {\emph {\bibinfo
  {booktitle} {Magnetic Skyrmions and Their Applications}}},\ \bibinfo {editor}
  {edited by\ \bibinfo {editor} {\bibfnamefont {G.}~\bibnamefont {Finocchio}}\
  and\ \bibinfo {editor} {\bibfnamefont {C.}~\bibnamefont {Panagopoulos}}}\
  (\bibinfo  {publisher} {Woodhead Publishing Series in Electronic and Optical
  Materials},\ \bibinfo {year} {2025})\ \bibinfo {edition} {1st}\ ed.,\ pp.\
  \bibinfo {pages} {99--139}\BibitemShut {NoStop}%
\bibitem [{\citenamefont {Jaafar}\ \emph {et~al.}(2011)\citenamefont {Jaafar},
  \citenamefont {Iglesias-Freire}, \citenamefont {Serrano-Ram{\'o}n},
  \citenamefont {Ibarra}, \citenamefont {de~Teresa},\ and\ \citenamefont
  {Asenjo}}]{jaafar2011distinguishing}%
  \BibitemOpen
  \bibfield  {author} {\bibinfo {author} {\bibfnamefont {M.}~\bibnamefont
  {Jaafar}}, \bibinfo {author} {\bibfnamefont {O.}~\bibnamefont
  {Iglesias-Freire}}, \bibinfo {author} {\bibfnamefont {L.}~\bibnamefont
  {Serrano-Ram{\'o}n}}, \bibinfo {author} {\bibfnamefont {M.~R.}\ \bibnamefont
  {Ibarra}}, \bibinfo {author} {\bibfnamefont {J.~M.}\ \bibnamefont
  {de~Teresa}},\ and\ \bibinfo {author} {\bibfnamefont {A.}~\bibnamefont
  {Asenjo}},\ }\bibfield  {title} {\bibinfo {title} {Distinguishing magnetic
  and electrostatic interactions by a kelvin probe force microscopy--magnetic
  force microscopy combination},\ }\href@noop {} {\bibfield  {journal}
  {\bibinfo  {journal} {Beilstein J. Nanotechnol.}\ }\textbf {\bibinfo {volume}
  {2}},\ \bibinfo {pages} {552} (\bibinfo {year} {2011})}\BibitemShut {NoStop}%
\bibitem [{\citenamefont {Soumyanarayanan}\ \emph {et~al.}(2017)\citenamefont
  {Soumyanarayanan}, \citenamefont {Raju}, \citenamefont {Gonzalez~Oyarce},
  \citenamefont {Tan}, \citenamefont {Im}, \citenamefont {Petrovi{\'c}},
  \citenamefont {Ho}, \citenamefont {Khoo}, \citenamefont {Tran}, \citenamefont
  {Gan} \emph {et~al.}}]{soumyanarayanan2017tunable}%
  \BibitemOpen
  \bibfield  {author} {\bibinfo {author} {\bibfnamefont {A.}~\bibnamefont
  {Soumyanarayanan}}, \bibinfo {author} {\bibfnamefont {M.}~\bibnamefont
  {Raju}}, \bibinfo {author} {\bibfnamefont {A.}~\bibnamefont
  {Gonzalez~Oyarce}}, \bibinfo {author} {\bibfnamefont {A.~K.}\ \bibnamefont
  {Tan}}, \bibinfo {author} {\bibfnamefont {M.-Y.}\ \bibnamefont {Im}},
  \bibinfo {author} {\bibfnamefont {A.~P.}\ \bibnamefont {Petrovi{\'c}}},
  \bibinfo {author} {\bibfnamefont {P.}~\bibnamefont {Ho}}, \bibinfo {author}
  {\bibfnamefont {K.}~\bibnamefont {Khoo}}, \bibinfo {author} {\bibfnamefont
  {M.}~\bibnamefont {Tran}}, \bibinfo {author} {\bibfnamefont {C.}~\bibnamefont
  {Gan}}, \emph {et~al.},\ }\bibfield  {title} {\bibinfo {title} {Tunable
  room-temperature magnetic skyrmions in ir/fe/co/pt multilayers},\ }\href@noop
  {} {\bibfield  {journal} {\bibinfo  {journal} {Nat. Mater.}\ }\textbf
  {\bibinfo {volume} {16}},\ \bibinfo {pages} {898} (\bibinfo {year}
  {2017})}\BibitemShut {NoStop}%
\bibitem [{\citenamefont {Mandru}\ \emph {et~al.}(2020)\citenamefont {Mandru},
  \citenamefont {Y{\i}ld{\i}r{\i}m}, \citenamefont {Tomasello}, \citenamefont
  {Heistracher}, \citenamefont {Penedo}, \citenamefont {Giordano},
  \citenamefont {Suess}, \citenamefont {Finocchio},\ and\ \citenamefont
  {Hug}}]{mandru2020coexistence}%
  \BibitemOpen
  \bibfield  {author} {\bibinfo {author} {\bibfnamefont {A.-O.}\ \bibnamefont
  {Mandru}}, \bibinfo {author} {\bibfnamefont {O.}~\bibnamefont
  {Y{\i}ld{\i}r{\i}m}}, \bibinfo {author} {\bibfnamefont {R.}~\bibnamefont
  {Tomasello}}, \bibinfo {author} {\bibfnamefont {P.}~\bibnamefont
  {Heistracher}}, \bibinfo {author} {\bibfnamefont {M.}~\bibnamefont {Penedo}},
  \bibinfo {author} {\bibfnamefont {A.}~\bibnamefont {Giordano}}, \bibinfo
  {author} {\bibfnamefont {D.}~\bibnamefont {Suess}}, \bibinfo {author}
  {\bibfnamefont {G.}~\bibnamefont {Finocchio}},\ and\ \bibinfo {author}
  {\bibfnamefont {H.~J.}\ \bibnamefont {Hug}},\ }\bibfield  {title} {\bibinfo
  {title} {Coexistence of distinct skyrmion phases observed in hybrid
  ferromagnetic/ferrimagnetic multilayers},\ }\href@noop {} {\bibfield
  {journal} {\bibinfo  {journal} {Nat. Commun.}\ }\textbf {\bibinfo {volume}
  {11}},\ \bibinfo {pages} {6365} (\bibinfo {year} {2020})}\BibitemShut
  {NoStop}%
\bibitem [{\citenamefont {Grelier}\ \emph {et~al.}(2022)\citenamefont
  {Grelier}, \citenamefont {Godel}, \citenamefont {Vecchiola}, \citenamefont
  {Collin}, \citenamefont {Bouzehouane}, \citenamefont {Fert}, \citenamefont
  {Cros},\ and\ \citenamefont {Reyren}}]{grelier2022threedimensional}%
  \BibitemOpen
  \bibfield  {author} {\bibinfo {author} {\bibfnamefont {M.}~\bibnamefont
  {Grelier}}, \bibinfo {author} {\bibfnamefont {F.}~\bibnamefont {Godel}},
  \bibinfo {author} {\bibfnamefont {A.}~\bibnamefont {Vecchiola}}, \bibinfo
  {author} {\bibfnamefont {S.}~\bibnamefont {Collin}}, \bibinfo {author}
  {\bibfnamefont {K.}~\bibnamefont {Bouzehouane}}, \bibinfo {author}
  {\bibfnamefont {A.}~\bibnamefont {Fert}}, \bibinfo {author} {\bibfnamefont
  {V.}~\bibnamefont {Cros}},\ and\ \bibinfo {author} {\bibfnamefont
  {N.}~\bibnamefont {Reyren}},\ }\bibfield  {title} {\bibinfo {title}
  {Three-dimensional skyrmionic cocoons in magnetic multilayers},\ }\href
  {https://doi.org/10.1038/s41467-022-34370-x} {\bibfield  {journal} {\bibinfo
  {journal} {Nat. Commun.}\ }\textbf {\bibinfo {volume} {13}},\ \bibinfo
  {pages} {6843} (\bibinfo {year} {2022})}\BibitemShut {NoStop}%
\bibitem [{\citenamefont {Liu}\ \emph {et~al.}(2025{\natexlab{a}})\citenamefont
  {Liu}, \citenamefont {Tomasello}, \citenamefont {Wu}, \citenamefont {Fang},
  \citenamefont {Chen}, \citenamefont {Zheng}, \citenamefont {Zhang},
  \citenamefont {Darwin}, \citenamefont {Hug}, \citenamefont {Carpentieri}
  \emph {et~al.}}]{Liu2025}%
  \BibitemOpen
  \bibfield  {author} {\bibinfo {author} {\bibfnamefont {S.}~\bibnamefont
  {Liu}}, \bibinfo {author} {\bibfnamefont {R.}~\bibnamefont {Tomasello}},
  \bibinfo {author} {\bibfnamefont {Y.}~\bibnamefont {Wu}}, \bibinfo {author}
  {\bibfnamefont {B.}~\bibnamefont {Fang}}, \bibinfo {author} {\bibfnamefont
  {A.}~\bibnamefont {Chen}}, \bibinfo {author} {\bibfnamefont {D.}~\bibnamefont
  {Zheng}}, \bibinfo {author} {\bibfnamefont {B.}~\bibnamefont {Zhang}},
  \bibinfo {author} {\bibfnamefont {E.}~\bibnamefont {Darwin}}, \bibinfo
  {author} {\bibfnamefont {H.~J.}\ \bibnamefont {Hug}}, \bibinfo {author}
  {\bibfnamefont {M.}~\bibnamefont {Carpentieri}}, \emph {et~al.},\ }\bibfield
  {title} {\bibinfo {title} {Topological skyrmion-based spin-torque-diode
  effect in magnetic tunnel junctions},\ }\href@noop {} {\bibfield  {journal}
  {\bibinfo  {journal} {Adv. Electron. Mater.}\ ,\ \bibinfo {pages} {e00130}}
  (\bibinfo {year} {2025}{\natexlab{a}})}\BibitemShut {NoStop}%
\bibitem [{\citenamefont {Maccariello}\ \emph {et~al.}(2018)\citenamefont
  {Maccariello}, \citenamefont {Legrand}, \citenamefont {Reyren}, \citenamefont
  {Garcia}, \citenamefont {Bouzehouane}, \citenamefont {Collin}, \citenamefont
  {Cros},\ and\ \citenamefont {Fert}}]{maccariello2018electrical}%
  \BibitemOpen
  \bibfield  {author} {\bibinfo {author} {\bibfnamefont {D.}~\bibnamefont
  {Maccariello}}, \bibinfo {author} {\bibfnamefont {W.}~\bibnamefont
  {Legrand}}, \bibinfo {author} {\bibfnamefont {N.}~\bibnamefont {Reyren}},
  \bibinfo {author} {\bibfnamefont {K.}~\bibnamefont {Garcia}}, \bibinfo
  {author} {\bibfnamefont {K.}~\bibnamefont {Bouzehouane}}, \bibinfo {author}
  {\bibfnamefont {S.}~\bibnamefont {Collin}}, \bibinfo {author} {\bibfnamefont
  {V.}~\bibnamefont {Cros}},\ and\ \bibinfo {author} {\bibfnamefont
  {A.}~\bibnamefont {Fert}},\ }\bibfield  {title} {\bibinfo {title} {Electrical
  detection of single magnetic skyrmions in metallic multilayers at room
  temperature},\ }\href {https://doi.org/10.1038/s41565-017-0044-4} {\bibfield
  {journal} {\bibinfo  {journal} {Nature Nanotech}\ }\textbf {\bibinfo {volume}
  {13}},\ \bibinfo {pages} {233} (\bibinfo {year} {2018})}\BibitemShut
  {NoStop}%
\bibitem [{\citenamefont {Hrabec}\ \emph {et~al.}(2017)\citenamefont {Hrabec},
  \citenamefont {Sampaio}, \citenamefont {Belmeguenai}, \citenamefont {Gross},
  \citenamefont {Weil}, \citenamefont {Ch{\'e}rif}, \citenamefont
  {Stashkevich}, \citenamefont {Jacques}, \citenamefont {Thiaville},\ and\
  \citenamefont {Rohart}}]{hrabec2017current}%
  \BibitemOpen
  \bibfield  {author} {\bibinfo {author} {\bibfnamefont {A.}~\bibnamefont
  {Hrabec}}, \bibinfo {author} {\bibfnamefont {J.}~\bibnamefont {Sampaio}},
  \bibinfo {author} {\bibfnamefont {M.}~\bibnamefont {Belmeguenai}}, \bibinfo
  {author} {\bibfnamefont {I.}~\bibnamefont {Gross}}, \bibinfo {author}
  {\bibfnamefont {R.}~\bibnamefont {Weil}}, \bibinfo {author} {\bibfnamefont
  {S.~M.}\ \bibnamefont {Ch{\'e}rif}}, \bibinfo {author} {\bibfnamefont
  {A.}~\bibnamefont {Stashkevich}}, \bibinfo {author} {\bibfnamefont
  {V.}~\bibnamefont {Jacques}}, \bibinfo {author} {\bibfnamefont
  {A.}~\bibnamefont {Thiaville}},\ and\ \bibinfo {author} {\bibfnamefont
  {S.}~\bibnamefont {Rohart}},\ }\bibfield  {title} {\bibinfo {title}
  {Current-induced skyrmion generation and dynamics in symmetric bilayers},\
  }\href@noop {} {\bibfield  {journal} {\bibinfo  {journal} {Nat. Commun.}\
  }\textbf {\bibinfo {volume} {8}},\ \bibinfo {pages} {15765} (\bibinfo {year}
  {2017})}\BibitemShut {NoStop}%
\bibitem [{\citenamefont {Raju}\ \emph {et~al.}(2021)\citenamefont {Raju},
  \citenamefont {Petrović}, \citenamefont {Yagil}, \citenamefont {Denisov},
  \citenamefont {Duong}, \citenamefont {Göbel}, \citenamefont {Şaşıoğlu},
  \citenamefont {Auslaender}, \citenamefont {Mertig}, \citenamefont
  {Rozhansky},\ and\ \citenamefont {Panagopoulos}}]{Raju2021}%
  \BibitemOpen
  \bibfield  {author} {\bibinfo {author} {\bibfnamefont {M.}~\bibnamefont
  {Raju}}, \bibinfo {author} {\bibfnamefont {A.~P.}\ \bibnamefont {Petrović}},
  \bibinfo {author} {\bibfnamefont {A.}~\bibnamefont {Yagil}}, \bibinfo
  {author} {\bibfnamefont {K.~S.}\ \bibnamefont {Denisov}}, \bibinfo {author}
  {\bibfnamefont {N.~K.}\ \bibnamefont {Duong}}, \bibinfo {author}
  {\bibfnamefont {B.}~\bibnamefont {Göbel}}, \bibinfo {author} {\bibfnamefont
  {E.}~\bibnamefont {Şaşıoğlu}}, \bibinfo {author} {\bibfnamefont {O.~M.}\
  \bibnamefont {Auslaender}}, \bibinfo {author} {\bibfnamefont
  {I.}~\bibnamefont {Mertig}}, \bibinfo {author} {\bibfnamefont {I.~V.}\
  \bibnamefont {Rozhansky}},\ and\ \bibinfo {author} {\bibfnamefont
  {C.}~\bibnamefont {Panagopoulos}},\ }\bibfield  {title} {\bibinfo {title}
  {{Colossal topological Hall effect at the transition between isolated and
  lattice-phase interfacial skyrmions}},\ }\href
  {https://doi.org/10.1038/s41467-021-22976-6} {\bibfield  {journal} {\bibinfo
  {journal} {Nat. Commun.}\ }\textbf {\bibinfo {volume} {12}},\ \bibinfo
  {pages} {2758} (\bibinfo {year} {2021})},\ \Eprint
  {https://arxiv.org/abs/2105.08245} {2105.08245} \BibitemShut {NoStop}%
\bibitem [{\citenamefont {Feng}\ \emph {et~al.}(2022)\citenamefont {Feng},
  \citenamefont {Vaghefi}, \citenamefont {Vranjkovic}, \citenamefont {Penedo},
  \citenamefont {Kappenberger}, \citenamefont {Schwenk}, \citenamefont {Zhao},
  \citenamefont {Mandru},\ and\ \citenamefont {Hug}}]{Feng2022}%
  \BibitemOpen
  \bibfield  {author} {\bibinfo {author} {\bibfnamefont {Y.}~\bibnamefont
  {Feng}}, \bibinfo {author} {\bibfnamefont {P.~M.}\ \bibnamefont {Vaghefi}},
  \bibinfo {author} {\bibfnamefont {S.}~\bibnamefont {Vranjkovic}}, \bibinfo
  {author} {\bibfnamefont {M.}~\bibnamefont {Penedo}}, \bibinfo {author}
  {\bibfnamefont {P.}~\bibnamefont {Kappenberger}}, \bibinfo {author}
  {\bibfnamefont {J.}~\bibnamefont {Schwenk}}, \bibinfo {author} {\bibfnamefont
  {X.}~\bibnamefont {Zhao}}, \bibinfo {author} {\bibfnamefont {A.-O.}\
  \bibnamefont {Mandru}},\ and\ \bibinfo {author} {\bibfnamefont
  {H.}~\bibnamefont {Hug}},\ }\bibfield  {title} {\bibinfo {title} {{Magnetic
  force microscopy contrast formation and field sensitivity}},\ }\href
  {https://doi.org/10.1016/j.jmmm.2022.169073} {\bibfield  {journal} {\bibinfo
  {journal} {J. Magn. Magn. Mater.}\ }\textbf {\bibinfo {volume} {551}},\
  \bibinfo {pages} {169073} (\bibinfo {year} {2022})}\BibitemShut {NoStop}%
\bibitem [{\citenamefont {Meyer}\ \emph {et~al.}(2021)\citenamefont {Meyer},
  \citenamefont {Bennewitz},\ and\ \citenamefont {Hug}}]{meyer2021scanning}%
  \BibitemOpen
  \bibfield  {author} {\bibinfo {author} {\bibfnamefont {E.}~\bibnamefont
  {Meyer}}, \bibinfo {author} {\bibfnamefont {R.}~\bibnamefont {Bennewitz}},\
  and\ \bibinfo {author} {\bibfnamefont {H.~J.}\ \bibnamefont {Hug}},\
  }\href@noop {} {\emph {\bibinfo {title} {Scanning probe microscopy: the lab
  on a tip}}}\ (\bibinfo  {publisher} {Springer},\ \bibinfo {year}
  {2021})\BibitemShut {NoStop}%
\bibitem [{\citenamefont {Chen}\ \emph {et~al.}(2024)\citenamefont {Chen},
  \citenamefont {Lourembam}, \citenamefont {Ho}, \citenamefont {Toh},
  \citenamefont {Huang}, \citenamefont {Chen}, \citenamefont {Tan},
  \citenamefont {Yap}, \citenamefont {Lim}, \citenamefont {Tan}, \citenamefont
  {Suraj}, \citenamefont {Sim}, \citenamefont {Toh}, \citenamefont {Lim},
  \citenamefont {Lim}, \citenamefont {Zhou}, \citenamefont {Chung},
  \citenamefont {Lim},\ and\ \citenamefont
  {Soumyanarayanan}}]{chen2024allelectrical}%
  \BibitemOpen
  \bibfield  {author} {\bibinfo {author} {\bibfnamefont {S.}~\bibnamefont
  {Chen}}, \bibinfo {author} {\bibfnamefont {J.}~\bibnamefont {Lourembam}},
  \bibinfo {author} {\bibfnamefont {P.}~\bibnamefont {Ho}}, \bibinfo {author}
  {\bibfnamefont {A.~K.~J.}\ \bibnamefont {Toh}}, \bibinfo {author}
  {\bibfnamefont {J.}~\bibnamefont {Huang}}, \bibinfo {author} {\bibfnamefont
  {X.}~\bibnamefont {Chen}}, \bibinfo {author} {\bibfnamefont {H.~K.}\
  \bibnamefont {Tan}}, \bibinfo {author} {\bibfnamefont {S.~L.~K.}\
  \bibnamefont {Yap}}, \bibinfo {author} {\bibfnamefont {R.~J.~J.}\
  \bibnamefont {Lim}}, \bibinfo {author} {\bibfnamefont {H.~R.}\ \bibnamefont
  {Tan}}, \bibinfo {author} {\bibfnamefont {T.~S.}\ \bibnamefont {Suraj}},
  \bibinfo {author} {\bibfnamefont {M.~I.}\ \bibnamefont {Sim}}, \bibinfo
  {author} {\bibfnamefont {Y.~T.}\ \bibnamefont {Toh}}, \bibinfo {author}
  {\bibfnamefont {I.}~\bibnamefont {Lim}}, \bibinfo {author} {\bibfnamefont
  {N.~C.~B.}\ \bibnamefont {Lim}}, \bibinfo {author} {\bibfnamefont
  {J.}~\bibnamefont {Zhou}}, \bibinfo {author} {\bibfnamefont {H.~J.}\
  \bibnamefont {Chung}}, \bibinfo {author} {\bibfnamefont {S.~T.}\ \bibnamefont
  {Lim}},\ and\ \bibinfo {author} {\bibfnamefont {A.}~\bibnamefont
  {Soumyanarayanan}},\ }\bibfield  {title} {\bibinfo {title} {All-electrical
  skyrmionic magnetic tunnel junction},\ }\href
  {https://doi.org/10.1038/s41586-024-07131-7} {\bibfield  {journal} {\bibinfo
  {journal} {Nature}\ }\textbf {\bibinfo {volume} {627}},\ \bibinfo {pages}
  {522} (\bibinfo {year} {2024})}\BibitemShut {NoStop}%
\bibitem [{\citenamefont {Casiraghi}\ \emph {et~al.}(2019)\citenamefont
  {Casiraghi}, \citenamefont {Corte-Le{\'o}n}, \citenamefont {Vafaee},
  \citenamefont {Garcia-Sanchez}, \citenamefont {Durin}, \citenamefont
  {Pasquale}, \citenamefont {Jakob}, \citenamefont {Kl{\"a}ui},\ and\
  \citenamefont {Kazakova}}]{casiraghi2019individual}%
  \BibitemOpen
  \bibfield  {author} {\bibinfo {author} {\bibfnamefont {A.}~\bibnamefont
  {Casiraghi}}, \bibinfo {author} {\bibfnamefont {H.}~\bibnamefont
  {Corte-Le{\'o}n}}, \bibinfo {author} {\bibfnamefont {M.}~\bibnamefont
  {Vafaee}}, \bibinfo {author} {\bibfnamefont {F.}~\bibnamefont
  {Garcia-Sanchez}}, \bibinfo {author} {\bibfnamefont {G.}~\bibnamefont
  {Durin}}, \bibinfo {author} {\bibfnamefont {M.}~\bibnamefont {Pasquale}},
  \bibinfo {author} {\bibfnamefont {G.}~\bibnamefont {Jakob}}, \bibinfo
  {author} {\bibfnamefont {M.}~\bibnamefont {Kl{\"a}ui}},\ and\ \bibinfo
  {author} {\bibfnamefont {O.}~\bibnamefont {Kazakova}},\ }\bibfield  {title}
  {\bibinfo {title} {Individual skyrmion manipulation by local magnetic field
  gradients},\ }\href@noop {} {\bibfield  {journal} {\bibinfo  {journal}
  {Commun. Phys.}\ }\textbf {\bibinfo {volume} {2}},\ \bibinfo {pages} {145}
  (\bibinfo {year} {2019})}\BibitemShut {NoStop}%
\bibitem [{\citenamefont {Meng}\ \emph {et~al.}(2019)\citenamefont {Meng},
  \citenamefont {Ahmed}, \citenamefont {Ba{\'c}ani}, \citenamefont {Mandru},
  \citenamefont {Zhao}, \citenamefont {Bagu{\'e}s}, \citenamefont {Esser},
  \citenamefont {Flores}, \citenamefont {McComb}, \citenamefont {Hug} \emph
  {et~al.}}]{meng2019observation}%
  \BibitemOpen
  \bibfield  {author} {\bibinfo {author} {\bibfnamefont {K.-Y.}\ \bibnamefont
  {Meng}}, \bibinfo {author} {\bibfnamefont {A.~S.}\ \bibnamefont {Ahmed}},
  \bibinfo {author} {\bibfnamefont {M.}~\bibnamefont {Ba{\'c}ani}}, \bibinfo
  {author} {\bibfnamefont {A.-O.}\ \bibnamefont {Mandru}}, \bibinfo {author}
  {\bibfnamefont {X.}~\bibnamefont {Zhao}}, \bibinfo {author} {\bibfnamefont
  {N.}~\bibnamefont {Bagu{\'e}s}}, \bibinfo {author} {\bibfnamefont {B.~D.}\
  \bibnamefont {Esser}}, \bibinfo {author} {\bibfnamefont {J.}~\bibnamefont
  {Flores}}, \bibinfo {author} {\bibfnamefont {D.~W.}\ \bibnamefont {McComb}},
  \bibinfo {author} {\bibfnamefont {H.~J.}\ \bibnamefont {Hug}}, \emph
  {et~al.},\ }\bibfield  {title} {\bibinfo {title} {Observation of nanoscale
  skyrmions in sriro3/srruo3 bilayers},\ }\href@noop {} {\bibfield  {journal}
  {\bibinfo  {journal} {Nano Lett.}\ }\textbf {\bibinfo {volume} {19}},\
  \bibinfo {pages} {3169} (\bibinfo {year} {2019})}\BibitemShut {NoStop}%
\bibitem [{\citenamefont {Raju}\ \emph {et~al.}(2019)\citenamefont {Raju},
  \citenamefont {Yagil}, \citenamefont {Soumyanarayanan}, \citenamefont {Tan},
  \citenamefont {Almoalem}, \citenamefont {Ma}, \citenamefont {Auslaender},\
  and\ \citenamefont {Panagopoulos}}]{raju2019evolution}%
  \BibitemOpen
  \bibfield  {author} {\bibinfo {author} {\bibfnamefont {M.}~\bibnamefont
  {Raju}}, \bibinfo {author} {\bibfnamefont {A.}~\bibnamefont {Yagil}},
  \bibinfo {author} {\bibfnamefont {A.}~\bibnamefont {Soumyanarayanan}},
  \bibinfo {author} {\bibfnamefont {A.~K.}\ \bibnamefont {Tan}}, \bibinfo
  {author} {\bibfnamefont {A.}~\bibnamefont {Almoalem}}, \bibinfo {author}
  {\bibfnamefont {F.}~\bibnamefont {Ma}}, \bibinfo {author} {\bibfnamefont
  {O.}~\bibnamefont {Auslaender}},\ and\ \bibinfo {author} {\bibfnamefont
  {C.}~\bibnamefont {Panagopoulos}},\ }\bibfield  {title} {\bibinfo {title}
  {The evolution of skyrmions in ir/fe/co/pt multilayers and their topological
  hall signature},\ }\href@noop {} {\bibfield  {journal} {\bibinfo  {journal}
  {Nat. Commun.}\ }\textbf {\bibinfo {volume} {10}},\ \bibinfo {pages} {696}
  (\bibinfo {year} {2019})}\BibitemShut {NoStop}%
\bibitem [{\citenamefont {Zhao}\ \emph {et~al.}(2018)\citenamefont {Zhao},
  \citenamefont {Schwenk}, \citenamefont {Mandru}, \citenamefont {Penedo},
  \citenamefont {Ba{\'c}ani}, \citenamefont {Marioni},\ and\ \citenamefont
  {Hug}}]{zhao2018magnetic}%
  \BibitemOpen
  \bibfield  {author} {\bibinfo {author} {\bibfnamefont {X.}~\bibnamefont
  {Zhao}}, \bibinfo {author} {\bibfnamefont {J.}~\bibnamefont {Schwenk}},
  \bibinfo {author} {\bibfnamefont {A.}~\bibnamefont {Mandru}}, \bibinfo
  {author} {\bibfnamefont {M.}~\bibnamefont {Penedo}}, \bibinfo {author}
  {\bibfnamefont {M.}~\bibnamefont {Ba{\'c}ani}}, \bibinfo {author}
  {\bibfnamefont {M.}~\bibnamefont {Marioni}},\ and\ \bibinfo {author}
  {\bibfnamefont {H.}~\bibnamefont {Hug}},\ }\bibfield  {title} {\bibinfo
  {title} {Magnetic force microscopy with frequency-modulated capacitive
  tip--sample distance control},\ }\href@noop {} {\bibfield  {journal}
  {\bibinfo  {journal} {New J. Phys.}\ }\textbf {\bibinfo {volume} {20}},\
  \bibinfo {pages} {013018} (\bibinfo {year} {2018})}\BibitemShut {NoStop}%
\bibitem [{\citenamefont {Koraltan}\ \emph {et~al.}(2025)\citenamefont
  {Koraltan}, \citenamefont {Sunny}, \citenamefont {Karaman}, \citenamefont
  {Peremadathil-Pradeep}, \citenamefont {Darwin}, \citenamefont {B{\"u}ttner},
  \citenamefont {Suess}, \citenamefont {Hug},\ and\ \citenamefont
  {Albrecht}}]{koraltan2025signatures}%
  \BibitemOpen
  \bibfield  {author} {\bibinfo {author} {\bibfnamefont {S.}~\bibnamefont
  {Koraltan}}, \bibinfo {author} {\bibfnamefont {J.}~\bibnamefont {Sunny}},
  \bibinfo {author} {\bibfnamefont {T.}~\bibnamefont {Karaman}}, \bibinfo
  {author} {\bibfnamefont {R.}~\bibnamefont {Peremadathil-Pradeep}}, \bibinfo
  {author} {\bibfnamefont {E.}~\bibnamefont {Darwin}}, \bibinfo {author}
  {\bibfnamefont {F.}~\bibnamefont {B{\"u}ttner}}, \bibinfo {author}
  {\bibfnamefont {D.}~\bibnamefont {Suess}}, \bibinfo {author} {\bibfnamefont
  {H.~J.}\ \bibnamefont {Hug}},\ and\ \bibinfo {author} {\bibfnamefont
  {M.}~\bibnamefont {Albrecht}},\ }\bibfield  {title} {\bibinfo {title}
  {Signatures of higher order skyrmionic textures revealed by magnetic force
  microscopy},\ }\href@noop {} {\bibfield  {journal} {\bibinfo  {journal}
  {arXiv preprint arXiv:2501.04499}\ } (\bibinfo {year} {2025})}\BibitemShut
  {NoStop}%
\bibitem [{\citenamefont {Sim}\ \emph {et~al.}(2025)\citenamefont {Sim},
  \citenamefont {Thian}, \citenamefont {Maddu}, \citenamefont {Chen},
  \citenamefont {Tan}, \citenamefont {Li}, \citenamefont {Ho},\ and\
  \citenamefont {Soumyanarayanan}}]{Sim2025}%
  \BibitemOpen
  \bibfield  {author} {\bibinfo {author} {\bibfnamefont {M.~I.}\ \bibnamefont
  {Sim}}, \bibinfo {author} {\bibfnamefont {D.}~\bibnamefont {Thian}}, \bibinfo
  {author} {\bibfnamefont {R.}~\bibnamefont {Maddu}}, \bibinfo {author}
  {\bibfnamefont {X.}~\bibnamefont {Chen}}, \bibinfo {author} {\bibfnamefont
  {H.~K.}\ \bibnamefont {Tan}}, \bibinfo {author} {\bibfnamefont
  {C.}~\bibnamefont {Li}}, \bibinfo {author} {\bibfnamefont {P.}~\bibnamefont
  {Ho}},\ and\ \bibinfo {author} {\bibfnamefont {A.}~\bibnamefont
  {Soumyanarayanan}},\ }\bibfield  {title} {\bibinfo {title} {{Zero Field
  Antiferromagnetically Coupled Skyrmions and their Field‐Driven Uncoupling
  in Composite Chiral Multilayers}},\ }\bibfield  {journal} {\bibinfo
  {journal} {Adv. Funct. Mater.}\ }\href
  {https://doi.org/10.1002/adfm.202416927} {10.1002/adfm.202416927} (\bibinfo
  {year} {2025}),\ \Eprint {https://arxiv.org/abs/2501.03588} {2501.03588}
  \BibitemShut {NoStop}%
\bibitem [{\citenamefont {Dugato}\ \emph {et~al.}(2025)\citenamefont {Dugato},
  \citenamefont {Jalil}, \citenamefont {Cardias}, \citenamefont {Albuquerque},
  \citenamefont {Costa}, \citenamefont {Almeida}, \citenamefont {Fallon},
  \citenamefont {Kov{\'a}cs}, \citenamefont {McVitie}, \citenamefont
  {Dunin-Borkowski} \emph {et~al.}}]{dugato2025curved}%
  \BibitemOpen
  \bibfield  {author} {\bibinfo {author} {\bibfnamefont {D.~A.}\ \bibnamefont
  {Dugato}}, \bibinfo {author} {\bibfnamefont {W.~B.}\ \bibnamefont {Jalil}},
  \bibinfo {author} {\bibfnamefont {R.}~\bibnamefont {Cardias}}, \bibinfo
  {author} {\bibfnamefont {M.}~\bibnamefont {Albuquerque}}, \bibinfo {author}
  {\bibfnamefont {M.}~\bibnamefont {Costa}}, \bibinfo {author} {\bibfnamefont
  {T.~P.}\ \bibnamefont {Almeida}}, \bibinfo {author} {\bibfnamefont
  {K.}~\bibnamefont {Fallon}}, \bibinfo {author} {\bibfnamefont
  {A.}~\bibnamefont {Kov{\'a}cs}}, \bibinfo {author} {\bibfnamefont
  {S.}~\bibnamefont {McVitie}}, \bibinfo {author} {\bibfnamefont {R.~E.}\
  \bibnamefont {Dunin-Borkowski}}, \emph {et~al.},\ }\bibfield  {title}
  {\bibinfo {title} {Curved nanomagnets: An archetype for the skyrmionic states
  at ambient conditions},\ }\href@noop {} {\bibfield  {journal} {\bibinfo
  {journal} {Nano Lett.}\ } (\bibinfo {year} {2025})}\BibitemShut {NoStop}%
\bibitem [{\citenamefont {{Fern{\'a}ndez-Pacheco}}\ \emph
  {et~al.}(2017)\citenamefont {{Fern{\'a}ndez-Pacheco}}, \citenamefont
  {Streubel}, \citenamefont {Fruchart}, \citenamefont {Hertel}, \citenamefont
  {Fischer},\ and\ \citenamefont
  {Cowburn}}]{fernandez-pacheco2017threedimensional}%
  \BibitemOpen
  \bibfield  {author} {\bibinfo {author} {\bibfnamefont {A.}~\bibnamefont
  {{Fern{\'a}ndez-Pacheco}}}, \bibinfo {author} {\bibfnamefont
  {R.}~\bibnamefont {Streubel}}, \bibinfo {author} {\bibfnamefont
  {O.}~\bibnamefont {Fruchart}}, \bibinfo {author} {\bibfnamefont
  {R.}~\bibnamefont {Hertel}}, \bibinfo {author} {\bibfnamefont
  {P.}~\bibnamefont {Fischer}},\ and\ \bibinfo {author} {\bibfnamefont {R.~P.}\
  \bibnamefont {Cowburn}},\ }\bibfield  {title} {\bibinfo {title}
  {Three-dimensional nanomagnetism},\ }\href
  {https://doi.org/10.1038/ncomms15756} {\bibfield  {journal} {\bibinfo
  {journal} {Nat. Commun.}\ }\textbf {\bibinfo {volume} {8}},\ \bibinfo {pages}
  {15756} (\bibinfo {year} {2017})}\BibitemShut {NoStop}%
\bibitem [{\citenamefont {Bhattacharya}\ \emph {et~al.}(2025)\citenamefont
  {Bhattacharya}, \citenamefont {Langton}, \citenamefont {Rajib}, \citenamefont
  {Marlowe}, \citenamefont {Chen}, \citenamefont {Al~Misba}, \citenamefont
  {Atulasimha}, \citenamefont {Zhang}, \citenamefont {Yin},\ and\ \citenamefont
  {Liu}}]{bhattacharya2025selfassembled}%
  \BibitemOpen
  \bibfield  {author} {\bibinfo {author} {\bibfnamefont {D.}~\bibnamefont
  {Bhattacharya}}, \bibinfo {author} {\bibfnamefont {C.}~\bibnamefont
  {Langton}}, \bibinfo {author} {\bibfnamefont {M.~M.}\ \bibnamefont {Rajib}},
  \bibinfo {author} {\bibfnamefont {E.}~\bibnamefont {Marlowe}}, \bibinfo
  {author} {\bibfnamefont {Z.}~\bibnamefont {Chen}}, \bibinfo {author}
  {\bibfnamefont {W.}~\bibnamefont {Al~Misba}}, \bibinfo {author}
  {\bibfnamefont {J.}~\bibnamefont {Atulasimha}}, \bibinfo {author}
  {\bibfnamefont {X.}~\bibnamefont {Zhang}}, \bibinfo {author} {\bibfnamefont
  {G.}~\bibnamefont {Yin}},\ and\ \bibinfo {author} {\bibfnamefont
  {K.}~\bibnamefont {Liu}},\ }\bibfield  {title} {\bibinfo {title}
  {Self-assembled {{3D Interconnected Magnetic Nanowire Networks}} for
  {{Neuromorphic Computing}}},\ }\href {https://doi.org/10.1021/acsami.4c22620}
  {\bibfield  {journal} {\bibinfo  {journal} {ACS Appl. Mater. Interfaces}\
  }\textbf {\bibinfo {volume} {17}},\ \bibinfo {pages} {20087} (\bibinfo {year}
  {2025})}\BibitemShut {NoStop}%
\bibitem [{\citenamefont {Pip}\ \emph {et~al.}(2020)\citenamefont {Pip},
  \citenamefont {Donnelly}, \citenamefont {D{\"o}beli}, \citenamefont
  {Gunderson}, \citenamefont {Heyderman},\ and\ \citenamefont
  {Philippe}}]{pip2020electroless}%
  \BibitemOpen
  \bibfield  {author} {\bibinfo {author} {\bibfnamefont {P.}~\bibnamefont
  {Pip}}, \bibinfo {author} {\bibfnamefont {C.}~\bibnamefont {Donnelly}},
  \bibinfo {author} {\bibfnamefont {M.}~\bibnamefont {D{\"o}beli}}, \bibinfo
  {author} {\bibfnamefont {C.}~\bibnamefont {Gunderson}}, \bibinfo {author}
  {\bibfnamefont {L.~J.}\ \bibnamefont {Heyderman}},\ and\ \bibinfo {author}
  {\bibfnamefont {L.}~\bibnamefont {Philippe}},\ }\bibfield  {title} {\bibinfo
  {title} {Electroless {{Deposition}} of {{Ni}}--{{Fe Alloys}} on {{Scaffolds}}
  for {{3D Nanomagnetism}}},\ }\href {https://doi.org/10.1002/smll.202004099}
  {\bibfield  {journal} {\bibinfo  {journal} {Small}\ }\textbf {\bibinfo
  {volume} {16}},\ \bibinfo {pages} {2004099} (\bibinfo {year}
  {2020})}\BibitemShut {NoStop}%
\bibitem [{\citenamefont {Xu}\ \emph {et~al.}(2025)\citenamefont {Xu},
  \citenamefont {Deenen}, \citenamefont {Guo}, \citenamefont
  {Morales-Fern{\'a}ndez}, \citenamefont {Wintz}, \citenamefont {Zhakina},
  \citenamefont {Weigand}, \citenamefont {Donnelly},\ and\ \citenamefont
  {Grundler}}]{xu2025geometry}%
  \BibitemOpen
  \bibfield  {author} {\bibinfo {author} {\bibfnamefont {M.}~\bibnamefont
  {Xu}}, \bibinfo {author} {\bibfnamefont {A.~J.}\ \bibnamefont {Deenen}},
  \bibinfo {author} {\bibfnamefont {H.}~\bibnamefont {Guo}}, \bibinfo {author}
  {\bibfnamefont {P.}~\bibnamefont {Morales-Fern{\'a}ndez}}, \bibinfo {author}
  {\bibfnamefont {S.}~\bibnamefont {Wintz}}, \bibinfo {author} {\bibfnamefont
  {E.}~\bibnamefont {Zhakina}}, \bibinfo {author} {\bibfnamefont
  {M.}~\bibnamefont {Weigand}}, \bibinfo {author} {\bibfnamefont
  {C.}~\bibnamefont {Donnelly}},\ and\ \bibinfo {author} {\bibfnamefont
  {D.}~\bibnamefont {Grundler}},\ }\bibfield  {title} {\bibinfo {title}
  {Geometry-induced spin chirality in a non-chiral ferromagnet at zero field},\
  }\href@noop {} {\bibfield  {journal} {\bibinfo  {journal} {Nat.
  Nanotechnol.}\ ,\ \bibinfo {pages} {1}} (\bibinfo {year} {2025})}\BibitemShut
  {NoStop}%
\bibitem [{\citenamefont {Farinha}\ \emph {et~al.}(2025)\citenamefont
  {Farinha}, \citenamefont {Yang}, \citenamefont {Yoon}, \citenamefont {Pal},\
  and\ \citenamefont {Parkin}}]{farinha2025interplay}%
  \BibitemOpen
  \bibfield  {author} {\bibinfo {author} {\bibfnamefont {A.~M.~A.}\
  \bibnamefont {Farinha}}, \bibinfo {author} {\bibfnamefont {S.-H.}\
  \bibnamefont {Yang}}, \bibinfo {author} {\bibfnamefont {J.}~\bibnamefont
  {Yoon}}, \bibinfo {author} {\bibfnamefont {B.}~\bibnamefont {Pal}},\ and\
  \bibinfo {author} {\bibfnamefont {S.~S.~P.}\ \bibnamefont {Parkin}},\
  }\bibfield  {title} {\bibinfo {title} {Interplay of geometrical and spin
  chiralities in {{3D}} twisted magnetic ribbons},\ }\href
  {https://doi.org/10.1038/s41586-024-08582-8} {\bibfield  {journal} {\bibinfo
  {journal} {Nature}\ }\textbf {\bibinfo {volume} {639}},\ \bibinfo {pages}
  {67} (\bibinfo {year} {2025})}\BibitemShut {NoStop}%
\bibitem [{\citenamefont {Skoric}\ \emph {et~al.}(2020)\citenamefont {Skoric},
  \citenamefont {{Sanz-Hern{\'a}ndez}}, \citenamefont {Meng}, \citenamefont
  {Donnelly}, \citenamefont {{Merino-Aceituno}},\ and\ \citenamefont
  {{Fern{\'a}ndez-Pacheco}}}]{skoric2020layerbylayer}%
  \BibitemOpen
  \bibfield  {author} {\bibinfo {author} {\bibfnamefont {L.}~\bibnamefont
  {Skoric}}, \bibinfo {author} {\bibfnamefont {D.}~\bibnamefont
  {{Sanz-Hern{\'a}ndez}}}, \bibinfo {author} {\bibfnamefont {F.}~\bibnamefont
  {Meng}}, \bibinfo {author} {\bibfnamefont {C.}~\bibnamefont {Donnelly}},
  \bibinfo {author} {\bibfnamefont {S.}~\bibnamefont {{Merino-Aceituno}}},\
  and\ \bibinfo {author} {\bibfnamefont {A.}~\bibnamefont
  {{Fern{\'a}ndez-Pacheco}}},\ }\bibfield  {title} {\bibinfo {title}
  {Layer-by-{{Layer Growth}} of {{Complex-Shaped Three-Dimensional
  Nanostructures}} with {{Focused Electron Beams}}},\ }\href
  {https://doi.org/10.1021/acs.nanolett.9b03565} {\bibfield  {journal}
  {\bibinfo  {journal} {Nano Lett.}\ }\textbf {\bibinfo {volume} {20}},\
  \bibinfo {pages} {184} (\bibinfo {year} {2020})}\BibitemShut {NoStop}%
\bibitem [{\citenamefont {Phatak}\ \emph {et~al.}(2014)\citenamefont {Phatak},
  \citenamefont {Liu}, \citenamefont {Gulsoy}, \citenamefont {Schmidt},
  \citenamefont {{Franke-Schubert}},\ and\ \citenamefont
  {{Petford-Long}}}]{phatak2014visualization}%
  \BibitemOpen
  \bibfield  {author} {\bibinfo {author} {\bibfnamefont {C.}~\bibnamefont
  {Phatak}}, \bibinfo {author} {\bibfnamefont {Y.}~\bibnamefont {Liu}},
  \bibinfo {author} {\bibfnamefont {E.~B.}\ \bibnamefont {Gulsoy}}, \bibinfo
  {author} {\bibfnamefont {D.}~\bibnamefont {Schmidt}}, \bibinfo {author}
  {\bibfnamefont {E.}~\bibnamefont {{Franke-Schubert}}},\ and\ \bibinfo
  {author} {\bibfnamefont {A.}~\bibnamefont {{Petford-Long}}},\ }\bibfield
  {title} {\bibinfo {title} {Visualization of the {{Magnetic Structure}} of
  {{Sculpted Three-Dimensional Cobalt Nanospirals}}},\ }\href
  {https://doi.org/10.1021/nl404071u} {\bibfield  {journal} {\bibinfo
  {journal} {Nano Lett.}\ }\textbf {\bibinfo {volume} {14}},\ \bibinfo {pages}
  {759} (\bibinfo {year} {2014})}\BibitemShut {NoStop}%
\bibitem [{\citenamefont {Llandro}\ \emph {et~al.}(2020)\citenamefont
  {Llandro}, \citenamefont {Love}, \citenamefont {Kov{\'a}cs}, \citenamefont
  {Caron}, \citenamefont {Vyas}, \citenamefont {K{\'a}kay}, \citenamefont
  {Salikhov}, \citenamefont {Lenz}, \citenamefont {Fassbender}, \citenamefont
  {Scherer}, \citenamefont {Cimorra}, \citenamefont {Steiner}, \citenamefont
  {Barnes}, \citenamefont {{Dunin-Borkowski}}, \citenamefont {Fukami},\ and\
  \citenamefont {Ohno}}]{llandro2020visualizing}%
  \BibitemOpen
  \bibfield  {author} {\bibinfo {author} {\bibfnamefont {J.}~\bibnamefont
  {Llandro}}, \bibinfo {author} {\bibfnamefont {D.~M.}\ \bibnamefont {Love}},
  \bibinfo {author} {\bibfnamefont {A.}~\bibnamefont {Kov{\'a}cs}}, \bibinfo
  {author} {\bibfnamefont {J.}~\bibnamefont {Caron}}, \bibinfo {author}
  {\bibfnamefont {K.~N.}\ \bibnamefont {Vyas}}, \bibinfo {author}
  {\bibfnamefont {A.}~\bibnamefont {K{\'a}kay}}, \bibinfo {author}
  {\bibfnamefont {R.}~\bibnamefont {Salikhov}}, \bibinfo {author}
  {\bibfnamefont {K.}~\bibnamefont {Lenz}}, \bibinfo {author} {\bibfnamefont
  {J.}~\bibnamefont {Fassbender}}, \bibinfo {author} {\bibfnamefont {M.~R.~J.}\
  \bibnamefont {Scherer}}, \bibinfo {author} {\bibfnamefont {C.}~\bibnamefont
  {Cimorra}}, \bibinfo {author} {\bibfnamefont {U.}~\bibnamefont {Steiner}},
  \bibinfo {author} {\bibfnamefont {C.~H.~W.}\ \bibnamefont {Barnes}}, \bibinfo
  {author} {\bibfnamefont {R.~E.}\ \bibnamefont {{Dunin-Borkowski}}}, \bibinfo
  {author} {\bibfnamefont {S.}~\bibnamefont {Fukami}},\ and\ \bibinfo {author}
  {\bibfnamefont {H.}~\bibnamefont {Ohno}},\ }\bibfield  {title} {\bibinfo
  {title} {Visualizing {{Magnetic Structure}} in {{3D Nanoscale Ni}}--{{Fe
  Gyroid Networks}}},\ }\href {https://doi.org/10.1021/acs.nanolett.0c00578}
  {\bibfield  {journal} {\bibinfo  {journal} {Nano Lett.}\ }\textbf {\bibinfo
  {volume} {20}},\ \bibinfo {pages} {3642} (\bibinfo {year}
  {2020})}\BibitemShut {NoStop}%
\bibitem [{\citenamefont {Skoric}\ \emph {et~al.}(2022)\citenamefont {Skoric},
  \citenamefont {Donnelly}, \citenamefont {{Hierro-Rodriguez}}, \citenamefont
  {Cascales~Sandoval}, \citenamefont {{Ruiz-G{\'o}mez}}, \citenamefont
  {Foerster}, \citenamefont {Ni{\~n}o}, \citenamefont {Belkhou}, \citenamefont
  {Abert}, \citenamefont {Suess},\ and\ \citenamefont
  {{Fern{\'a}ndez-Pacheco}}}]{skoric2022domain}%
  \BibitemOpen
  \bibfield  {author} {\bibinfo {author} {\bibfnamefont {L.}~\bibnamefont
  {Skoric}}, \bibinfo {author} {\bibfnamefont {C.}~\bibnamefont {Donnelly}},
  \bibinfo {author} {\bibfnamefont {A.}~\bibnamefont {{Hierro-Rodriguez}}},
  \bibinfo {author} {\bibfnamefont {M.~A.}\ \bibnamefont {Cascales~Sandoval}},
  \bibinfo {author} {\bibfnamefont {S.}~\bibnamefont {{Ruiz-G{\'o}mez}}},
  \bibinfo {author} {\bibfnamefont {M.}~\bibnamefont {Foerster}}, \bibinfo
  {author} {\bibfnamefont {M.~A.}\ \bibnamefont {Ni{\~n}o}}, \bibinfo {author}
  {\bibfnamefont {R.}~\bibnamefont {Belkhou}}, \bibinfo {author} {\bibfnamefont
  {C.}~\bibnamefont {Abert}}, \bibinfo {author} {\bibfnamefont
  {D.}~\bibnamefont {Suess}},\ and\ \bibinfo {author} {\bibfnamefont
  {A.}~\bibnamefont {{Fern{\'a}ndez-Pacheco}}},\ }\bibfield  {title} {\bibinfo
  {title} {Domain {{Wall Automotion}} in {{Three-Dimensional Magnetic Helical
  Interconnectors}}},\ }\href {https://doi.org/10.1021/acsnano.1c10345}
  {\bibfield  {journal} {\bibinfo  {journal} {ACS Nano}\ }\textbf {\bibinfo
  {volume} {16}},\ \bibinfo {pages} {8860} (\bibinfo {year}
  {2022})}\BibitemShut {NoStop}%
\bibitem [{\citenamefont {Volkov}\ \emph {et~al.}(2023)\citenamefont {Volkov},
  \citenamefont {Wolf}, \citenamefont {Pylypovskyi}, \citenamefont {K{\'a}kay},
  \citenamefont {Sheka}, \citenamefont {B{\"u}chner}, \citenamefont
  {Fassbender}, \citenamefont {Lubk},\ and\ \citenamefont
  {Makarov}}]{volkov2023chirality}%
  \BibitemOpen
  \bibfield  {author} {\bibinfo {author} {\bibfnamefont {O.~M.}\ \bibnamefont
  {Volkov}}, \bibinfo {author} {\bibfnamefont {D.}~\bibnamefont {Wolf}},
  \bibinfo {author} {\bibfnamefont {O.~V.}\ \bibnamefont {Pylypovskyi}},
  \bibinfo {author} {\bibfnamefont {A.}~\bibnamefont {K{\'a}kay}}, \bibinfo
  {author} {\bibfnamefont {D.~D.}\ \bibnamefont {Sheka}}, \bibinfo {author}
  {\bibfnamefont {B.}~\bibnamefont {B{\"u}chner}}, \bibinfo {author}
  {\bibfnamefont {J.}~\bibnamefont {Fassbender}}, \bibinfo {author}
  {\bibfnamefont {A.}~\bibnamefont {Lubk}},\ and\ \bibinfo {author}
  {\bibfnamefont {D.}~\bibnamefont {Makarov}},\ }\bibfield  {title} {\bibinfo
  {title} {Chirality coupling in topological magnetic textures with multiple
  magnetochiral parameters},\ }\href
  {https://doi.org/10.1038/s41467-023-37081-z} {\bibfield  {journal} {\bibinfo
  {journal} {Nat. Commun.}\ }\textbf {\bibinfo {volume} {14}},\ \bibinfo
  {pages} {1491} (\bibinfo {year} {2023})}\BibitemShut {NoStop}%
\bibitem [{\citenamefont {Volkov}\ \emph {et~al.}(2024)\citenamefont {Volkov},
  \citenamefont {Pylypovskyi}, \citenamefont {Porrati}, \citenamefont
  {Kronast}, \citenamefont {{Fernandez-Roldan}}, \citenamefont {K{\'a}kay},
  \citenamefont {Kuprava}, \citenamefont {Barth}, \citenamefont {Rybakov},
  \citenamefont {Eriksson}, \citenamefont {{Lamb-Camarena}}, \citenamefont
  {Makushko}, \citenamefont {Mawass}, \citenamefont {Shakeel}, \citenamefont
  {Dobrovolskiy}, \citenamefont {Huth},\ and\ \citenamefont
  {Makarov}}]{volkov2024threedimensional}%
  \BibitemOpen
  \bibfield  {author} {\bibinfo {author} {\bibfnamefont {O.~M.}\ \bibnamefont
  {Volkov}}, \bibinfo {author} {\bibfnamefont {O.~V.}\ \bibnamefont
  {Pylypovskyi}}, \bibinfo {author} {\bibfnamefont {F.}~\bibnamefont
  {Porrati}}, \bibinfo {author} {\bibfnamefont {F.}~\bibnamefont {Kronast}},
  \bibinfo {author} {\bibfnamefont {J.~A.}\ \bibnamefont {{Fernandez-Roldan}}},
  \bibinfo {author} {\bibfnamefont {A.}~\bibnamefont {K{\'a}kay}}, \bibinfo
  {author} {\bibfnamefont {A.}~\bibnamefont {Kuprava}}, \bibinfo {author}
  {\bibfnamefont {S.}~\bibnamefont {Barth}}, \bibinfo {author} {\bibfnamefont
  {F.~N.}\ \bibnamefont {Rybakov}}, \bibinfo {author} {\bibfnamefont
  {O.}~\bibnamefont {Eriksson}}, \bibinfo {author} {\bibfnamefont
  {S.}~\bibnamefont {{Lamb-Camarena}}}, \bibinfo {author} {\bibfnamefont
  {P.}~\bibnamefont {Makushko}}, \bibinfo {author} {\bibfnamefont {M.-A.}\
  \bibnamefont {Mawass}}, \bibinfo {author} {\bibfnamefont {S.}~\bibnamefont
  {Shakeel}}, \bibinfo {author} {\bibfnamefont {O.~V.}\ \bibnamefont
  {Dobrovolskiy}}, \bibinfo {author} {\bibfnamefont {M.}~\bibnamefont {Huth}},\
  and\ \bibinfo {author} {\bibfnamefont {D.}~\bibnamefont {Makarov}},\
  }\bibfield  {title} {\bibinfo {title} {Three-dimensional magnetic
  nanotextures with high-order vorticity in soft magnetic wireframes},\ }\href
  {https://doi.org/10.1038/s41467-024-46403-8} {\bibfield  {journal} {\bibinfo
  {journal} {Nat. Commun.}\ }\textbf {\bibinfo {volume} {15}},\ \bibinfo
  {pages} {2193} (\bibinfo {year} {2024})}\BibitemShut {NoStop}%
\bibitem [{\citenamefont {Kravchuk}\ \emph {et~al.}(2018)\citenamefont
  {Kravchuk}, \citenamefont {Sheka}, \citenamefont {K{\'a}kay}, \citenamefont
  {Volkov}, \citenamefont {R{\"o}{\ss}ler}, \citenamefont {{van den Brink}},
  \citenamefont {Makarov},\ and\ \citenamefont
  {Gaididei}}]{kravchuk2018multiplet}%
  \BibitemOpen
  \bibfield  {author} {\bibinfo {author} {\bibfnamefont {V.~P.}\ \bibnamefont
  {Kravchuk}}, \bibinfo {author} {\bibfnamefont {D.~D.}\ \bibnamefont {Sheka}},
  \bibinfo {author} {\bibfnamefont {A.}~\bibnamefont {K{\'a}kay}}, \bibinfo
  {author} {\bibfnamefont {O.~M.}\ \bibnamefont {Volkov}}, \bibinfo {author}
  {\bibfnamefont {U.~K.}\ \bibnamefont {R{\"o}{\ss}ler}}, \bibinfo {author}
  {\bibfnamefont {J.}~\bibnamefont {{van den Brink}}}, \bibinfo {author}
  {\bibfnamefont {D.}~\bibnamefont {Makarov}},\ and\ \bibinfo {author}
  {\bibfnamefont {Y.}~\bibnamefont {Gaididei}},\ }\bibfield  {title} {\bibinfo
  {title} {Multiplet of {{Skyrmion States}} on a {{Curvilinear Defect}}:
  {{Reconfigurable Skyrmion Lattices}}},\ }\href
  {https://doi.org/10.1103/PhysRevLett.120.067201} {\bibfield  {journal}
  {\bibinfo  {journal} {Phys. Rev. Lett.}\ }\textbf {\bibinfo {volume} {120}},\
  \bibinfo {pages} {067201} (\bibinfo {year} {2018})}\BibitemShut {NoStop}%
\bibitem [{\citenamefont {Korniienko}\ \emph {et~al.}(2020)\citenamefont
  {Korniienko}, \citenamefont {K{\'a}kay}, \citenamefont {Sheka},\ and\
  \citenamefont {Kravchuk}}]{korniienko2020effect}%
  \BibitemOpen
  \bibfield  {author} {\bibinfo {author} {\bibfnamefont {A.}~\bibnamefont
  {Korniienko}}, \bibinfo {author} {\bibfnamefont {A.}~\bibnamefont
  {K{\'a}kay}}, \bibinfo {author} {\bibfnamefont {D.~D.}\ \bibnamefont
  {Sheka}},\ and\ \bibinfo {author} {\bibfnamefont {V.~P.}\ \bibnamefont
  {Kravchuk}},\ }\bibfield  {title} {\bibinfo {title} {Effect of curvature on
  the eigenstates of magnetic skyrmions},\ }\href@noop {} {\bibfield  {journal}
  {\bibinfo  {journal} {Phys. Rev. B}\ }\textbf {\bibinfo {volume} {102}},\
  \bibinfo {pages} {014432} (\bibinfo {year} {2020})}\BibitemShut {NoStop}%
\bibitem [{\citenamefont {Fullerton}\ \emph {et~al.}(2025)\citenamefont
  {Fullerton}, \citenamefont {Leo}, \citenamefont {Jurczyk}, \citenamefont
  {Donnelly}, \citenamefont {{Sanz-Hern{\'a}ndez}}, \citenamefont {Skoric},
  \citenamefont {Mille}, \citenamefont {Stanescu}, \citenamefont {MacLaren},
  \citenamefont {Belkhou}, \citenamefont {{Hierro-Rodriguez}},\ and\
  \citenamefont {{Fern{\'a}ndez-Pacheco}}}]{fullertonfractional}%
  \BibitemOpen
  \bibfield  {author} {\bibinfo {author} {\bibfnamefont {J.}~\bibnamefont
  {Fullerton}}, \bibinfo {author} {\bibfnamefont {N.}~\bibnamefont {Leo}},
  \bibinfo {author} {\bibfnamefont {J.}~\bibnamefont {Jurczyk}}, \bibinfo
  {author} {\bibfnamefont {C.}~\bibnamefont {Donnelly}}, \bibinfo {author}
  {\bibfnamefont {D.}~\bibnamefont {{Sanz-Hern{\'a}ndez}}}, \bibinfo {author}
  {\bibfnamefont {L.}~\bibnamefont {Skoric}}, \bibinfo {author} {\bibfnamefont
  {N.}~\bibnamefont {Mille}}, \bibinfo {author} {\bibfnamefont
  {S.}~\bibnamefont {Stanescu}}, \bibinfo {author} {\bibfnamefont {D.~A.}\
  \bibnamefont {MacLaren}}, \bibinfo {author} {\bibfnamefont {R.}~\bibnamefont
  {Belkhou}}, \bibinfo {author} {\bibfnamefont {A.}~\bibnamefont
  {{Hierro-Rodriguez}}},\ and\ \bibinfo {author} {\bibfnamefont
  {A.}~\bibnamefont {{Fern{\'a}ndez-Pacheco}}},\ }\bibfield  {title} {\bibinfo
  {title} {Fractional {{Skyrmion Tubes}} in {{Chiral-Interfaced 3D Magnetic
  Nanowires}}},\ }\href {https://doi.org/10.1002/adfm.202501615} {\bibfield
  {journal} {\bibinfo  {journal} {Adv. Funct. Mater.}\ ,\ \bibinfo {pages}
  {2501615}} (\bibinfo {year} {2025})}\BibitemShut {NoStop}%
\bibitem [{\citenamefont {Sanz-Hern{\'a}ndez}\ \emph
  {et~al.}(2020)\citenamefont {Sanz-Hern{\'a}ndez}, \citenamefont
  {Hierro-Rodriguez}, \citenamefont {Donnelly}, \citenamefont {Pablo-Navarro},
  \citenamefont {Sorrentino}, \citenamefont {Pereiro}, \citenamefont
  {Mag{\'e}n}, \citenamefont {McVitie}, \citenamefont {de~Teresa},
  \citenamefont {Ferrer} \emph {et~al.}}]{sanz2020artificial}%
  \BibitemOpen
  \bibfield  {author} {\bibinfo {author} {\bibfnamefont {D.}~\bibnamefont
  {Sanz-Hern{\'a}ndez}}, \bibinfo {author} {\bibfnamefont {A.}~\bibnamefont
  {Hierro-Rodriguez}}, \bibinfo {author} {\bibfnamefont {C.}~\bibnamefont
  {Donnelly}}, \bibinfo {author} {\bibfnamefont {J.}~\bibnamefont
  {Pablo-Navarro}}, \bibinfo {author} {\bibfnamefont {A.}~\bibnamefont
  {Sorrentino}}, \bibinfo {author} {\bibfnamefont {E.}~\bibnamefont {Pereiro}},
  \bibinfo {author} {\bibfnamefont {C.}~\bibnamefont {Mag{\'e}n}}, \bibinfo
  {author} {\bibfnamefont {S.}~\bibnamefont {McVitie}}, \bibinfo {author}
  {\bibfnamefont {J.~M.}\ \bibnamefont {de~Teresa}}, \bibinfo {author}
  {\bibfnamefont {S.}~\bibnamefont {Ferrer}}, \emph {et~al.},\ }\bibfield
  {title} {\bibinfo {title} {Artificial double-helix for geometrical control of
  magnetic chirality},\ }\href@noop {} {\bibfield  {journal} {\bibinfo
  {journal} {ACS Nano}\ }\textbf {\bibinfo {volume} {14}},\ \bibinfo {pages}
  {8084} (\bibinfo {year} {2020})}\BibitemShut {NoStop}%
\bibitem [{\citenamefont {Donnelly}\ \emph {et~al.}(2022)\citenamefont
  {Donnelly}, \citenamefont {Hierro-Rodr{\'\i}guez}, \citenamefont {Abert},
  \citenamefont {Witte}, \citenamefont {Skoric}, \citenamefont
  {Sanz-Hern{\'a}ndez}, \citenamefont {Finizio}, \citenamefont {Meng},
  \citenamefont {McVitie}, \citenamefont {Raabe} \emph
  {et~al.}}]{donnelly2022complex}%
  \BibitemOpen
  \bibfield  {author} {\bibinfo {author} {\bibfnamefont {C.}~\bibnamefont
  {Donnelly}}, \bibinfo {author} {\bibfnamefont {A.}~\bibnamefont
  {Hierro-Rodr{\'\i}guez}}, \bibinfo {author} {\bibfnamefont {C.}~\bibnamefont
  {Abert}}, \bibinfo {author} {\bibfnamefont {K.}~\bibnamefont {Witte}},
  \bibinfo {author} {\bibfnamefont {L.}~\bibnamefont {Skoric}}, \bibinfo
  {author} {\bibfnamefont {D.}~\bibnamefont {Sanz-Hern{\'a}ndez}}, \bibinfo
  {author} {\bibfnamefont {S.}~\bibnamefont {Finizio}}, \bibinfo {author}
  {\bibfnamefont {F.}~\bibnamefont {Meng}}, \bibinfo {author} {\bibfnamefont
  {S.}~\bibnamefont {McVitie}}, \bibinfo {author} {\bibfnamefont
  {J.}~\bibnamefont {Raabe}}, \emph {et~al.},\ }\bibfield  {title} {\bibinfo
  {title} {Complex free-space magnetic field textures induced by
  three-dimensional magnetic nanostructures},\ }\href@noop {} {\bibfield
  {journal} {\bibinfo  {journal} {Nat. Nanotechnol.}\ }\textbf {\bibinfo
  {volume} {17}},\ \bibinfo {pages} {136} (\bibinfo {year} {2022})}\BibitemShut
  {NoStop}%
\bibitem [{\citenamefont {Fullerton}\ and\ \citenamefont
  {Phatak}(2025)}]{fullerton2025design}%
  \BibitemOpen
  \bibfield  {author} {\bibinfo {author} {\bibfnamefont {J.}~\bibnamefont
  {Fullerton}}\ and\ \bibinfo {author} {\bibfnamefont {C.}~\bibnamefont
  {Phatak}},\ }\bibfield  {title} {\bibinfo {title} {Design and control of
  three-dimensional topological magnetic fields using interwoven helical
  nanostructures},\ }\href@noop {} {\bibfield  {journal} {\bibinfo  {journal}
  {Nano Lett.}\ }\textbf {\bibinfo {volume} {25}},\ \bibinfo {pages} {5148}
  (\bibinfo {year} {2025})}\BibitemShut {NoStop}%
\bibitem [{\citenamefont {Da~Col}\ \emph {et~al.}(2014)\citenamefont {Da~Col},
  \citenamefont {Jamet}, \citenamefont {Rougemaille}, \citenamefont
  {Locatelli}, \citenamefont {Mentes}, \citenamefont {Burgos}, \citenamefont
  {Afid}, \citenamefont {Darques}, \citenamefont {Cagnon}, \citenamefont
  {Toussaint},\ and\ \citenamefont {Fruchart}}]{dacol2014observation}%
  \BibitemOpen
  \bibfield  {author} {\bibinfo {author} {\bibfnamefont {S.}~\bibnamefont
  {Da~Col}}, \bibinfo {author} {\bibfnamefont {S.}~\bibnamefont {Jamet}},
  \bibinfo {author} {\bibfnamefont {N.}~\bibnamefont {Rougemaille}}, \bibinfo
  {author} {\bibfnamefont {A.}~\bibnamefont {Locatelli}}, \bibinfo {author}
  {\bibfnamefont {T.~O.}\ \bibnamefont {Mentes}}, \bibinfo {author}
  {\bibfnamefont {B.~S.}\ \bibnamefont {Burgos}}, \bibinfo {author}
  {\bibfnamefont {R.}~\bibnamefont {Afid}}, \bibinfo {author} {\bibfnamefont
  {M.}~\bibnamefont {Darques}}, \bibinfo {author} {\bibfnamefont
  {L.}~\bibnamefont {Cagnon}}, \bibinfo {author} {\bibfnamefont {J.~C.}\
  \bibnamefont {Toussaint}},\ and\ \bibinfo {author} {\bibfnamefont
  {O.}~\bibnamefont {Fruchart}},\ }\bibfield  {title} {\bibinfo {title}
  {Observation of {{Bloch-point}} domain walls in cylindrical magnetic
  nanowires},\ }\href {https://doi.org/10.1103/PhysRevB.89.180405} {\bibfield
  {journal} {\bibinfo  {journal} {Phys. Rev. B}\ }\textbf {\bibinfo {volume}
  {89}},\ \bibinfo {pages} {180405} (\bibinfo {year} {2014})}\BibitemShut
  {NoStop}%
\bibitem [{\citenamefont {{Ruiz-G{\'o}mez}}\ \emph {et~al.}(2025)\citenamefont
  {{Ruiz-G{\'o}mez}}, \citenamefont {Abert}, \citenamefont
  {{Morales-Fern{\'a}ndez}}, \citenamefont {{Fern{\'a}ndez-Gonz{\'a}lez}},
  \citenamefont {Koraltan}, \citenamefont {Danesi}, \citenamefont {Suess},
  \citenamefont {Varela}, \citenamefont {{S{\'a}nchez-Santolino}},
  \citenamefont {Bagu{\'e}s}, \citenamefont {Foerster}, \citenamefont
  {Ni{\~n}o}, \citenamefont {Mandziak}, \citenamefont {{Wilgocka-{\'S}l{\k
  e}zak}}, \citenamefont {Nita}, \citenamefont {Koenig}, \citenamefont
  {Seifert}, \citenamefont {{Hierro-Rodriguez}}, \citenamefont
  {{Fern{\'a}ndez-Pacheco}},\ and\ \citenamefont
  {Donnelly}}]{ruiz-gomez2025tailoring}%
  \BibitemOpen
  \bibfield  {author} {\bibinfo {author} {\bibfnamefont {S.}~\bibnamefont
  {{Ruiz-G{\'o}mez}}}, \bibinfo {author} {\bibfnamefont {C.}~\bibnamefont
  {Abert}}, \bibinfo {author} {\bibfnamefont {P.}~\bibnamefont
  {{Morales-Fern{\'a}ndez}}}, \bibinfo {author} {\bibfnamefont
  {C.}~\bibnamefont {{Fern{\'a}ndez-Gonz{\'a}lez}}}, \bibinfo {author}
  {\bibfnamefont {S.}~\bibnamefont {Koraltan}}, \bibinfo {author}
  {\bibfnamefont {L.}~\bibnamefont {Danesi}}, \bibinfo {author} {\bibfnamefont
  {D.}~\bibnamefont {Suess}}, \bibinfo {author} {\bibfnamefont
  {M.}~\bibnamefont {Varela}}, \bibinfo {author} {\bibfnamefont
  {G.}~\bibnamefont {{S{\'a}nchez-Santolino}}}, \bibinfo {author}
  {\bibfnamefont {N.}~\bibnamefont {Bagu{\'e}s}}, \bibinfo {author}
  {\bibfnamefont {M.}~\bibnamefont {Foerster}}, \bibinfo {author}
  {\bibfnamefont {M.~{\'A}.}\ \bibnamefont {Ni{\~n}o}}, \bibinfo {author}
  {\bibfnamefont {A.}~\bibnamefont {Mandziak}}, \bibinfo {author}
  {\bibfnamefont {D.}~\bibnamefont {{Wilgocka-{\'S}l{\k e}zak}}}, \bibinfo
  {author} {\bibfnamefont {P.}~\bibnamefont {Nita}}, \bibinfo {author}
  {\bibfnamefont {M.}~\bibnamefont {Koenig}}, \bibinfo {author} {\bibfnamefont
  {S.}~\bibnamefont {Seifert}}, \bibinfo {author} {\bibfnamefont
  {A.}~\bibnamefont {{Hierro-Rodriguez}}}, \bibinfo {author} {\bibfnamefont
  {A.}~\bibnamefont {{Fern{\'a}ndez-Pacheco}}},\ and\ \bibinfo {author}
  {\bibfnamefont {C.}~\bibnamefont {Donnelly}},\ }\bibfield  {title} {\bibinfo
  {title} {Tailoring the energy landscape of a bloch point domain wall with
  curvature},\ }\href {https://doi.org/10.1038/s41467-025-62705-x} {\bibfield
  {journal} {\bibinfo  {journal} {Nat. Commun.}\ }\textbf {\bibinfo {volume}
  {16}},\ \bibinfo {pages} {7422} (\bibinfo {year} {2025})}\BibitemShut
  {NoStop}%
\bibitem [{\citenamefont {{Castillo-Sep{\'u}lveda}}\ \emph
  {et~al.}(2021)\citenamefont {{Castillo-Sep{\'u}lveda}}, \citenamefont
  {Cacilhas}, \citenamefont {{Carvalho-Santos}}, \citenamefont {Corona},\ and\
  \citenamefont {Altbir}}]{castillo-sepulveda2021magnetic}%
  \BibitemOpen
  \bibfield  {author} {\bibinfo {author} {\bibfnamefont {S.}~\bibnamefont
  {{Castillo-Sep{\'u}lveda}}}, \bibinfo {author} {\bibfnamefont
  {R.}~\bibnamefont {Cacilhas}}, \bibinfo {author} {\bibfnamefont {V.~L.}\
  \bibnamefont {{Carvalho-Santos}}}, \bibinfo {author} {\bibfnamefont {R.~M.}\
  \bibnamefont {Corona}},\ and\ \bibinfo {author} {\bibfnamefont
  {D.}~\bibnamefont {Altbir}},\ }\bibfield  {title} {\bibinfo {title} {Magnetic
  hopfions in toroidal nanostructures driven by an {{Oersted}} magnetic
  field},\ }\href {https://doi.org/10.1103/PhysRevB.104.184406} {\bibfield
  {journal} {\bibinfo  {journal} {Phys. Rev. B}\ }\textbf {\bibinfo {volume}
  {104}},\ \bibinfo {pages} {184406} (\bibinfo {year} {2021})}\BibitemShut
  {NoStop}%
\bibitem [{\citenamefont {Wang}\ \emph
  {et~al.}(2019{\natexlab{b}})\citenamefont {Wang}, \citenamefont
  {Qaiumzadeh},\ and\ \citenamefont {Brataas}}]{wang2019current}%
  \BibitemOpen
  \bibfield  {author} {\bibinfo {author} {\bibfnamefont {X.}~\bibnamefont
  {Wang}}, \bibinfo {author} {\bibfnamefont {A.}~\bibnamefont {Qaiumzadeh}},\
  and\ \bibinfo {author} {\bibfnamefont {A.}~\bibnamefont {Brataas}},\
  }\bibfield  {title} {\bibinfo {title} {Current-driven dynamics of magnetic
  hopfions},\ }\href@noop {} {\bibfield  {journal} {\bibinfo  {journal} {Phys.
  Rev. Lett.}\ }\textbf {\bibinfo {volume} {123}},\ \bibinfo {pages} {147203}
  (\bibinfo {year} {2019}{\natexlab{b}})}\BibitemShut {NoStop}%
\bibitem [{\citenamefont {Khodzhaev}\ and\ \citenamefont
  {Turgut}(2022)}]{khodzhaev2022hopfion}%
  \BibitemOpen
  \bibfield  {author} {\bibinfo {author} {\bibfnamefont {Z.}~\bibnamefont
  {Khodzhaev}}\ and\ \bibinfo {author} {\bibfnamefont {E.}~\bibnamefont
  {Turgut}},\ }\bibfield  {title} {\bibinfo {title} {Hopfion dynamics in chiral
  magnets},\ }\href@noop {} {\bibfield  {journal} {\bibinfo  {journal} {J.
  Phys.: Condens. Matter}\ }\textbf {\bibinfo {volume} {34}},\ \bibinfo {pages}
  {225805} (\bibinfo {year} {2022})}\BibitemShut {NoStop}%
\bibitem [{\citenamefont {Bo}\ \emph {et~al.}(2021)\citenamefont {Bo},
  \citenamefont {Ji}, \citenamefont {Hu}, \citenamefont {Zhao}, \citenamefont
  {Li}, \citenamefont {Zhang},\ and\ \citenamefont {Zhang}}]{bo2021spin}%
  \BibitemOpen
  \bibfield  {author} {\bibinfo {author} {\bibfnamefont {L.}~\bibnamefont
  {Bo}}, \bibinfo {author} {\bibfnamefont {L.}~\bibnamefont {Ji}}, \bibinfo
  {author} {\bibfnamefont {C.}~\bibnamefont {Hu}}, \bibinfo {author}
  {\bibfnamefont {R.}~\bibnamefont {Zhao}}, \bibinfo {author} {\bibfnamefont
  {Y.}~\bibnamefont {Li}}, \bibinfo {author} {\bibfnamefont {J.}~\bibnamefont
  {Zhang}},\ and\ \bibinfo {author} {\bibfnamefont {X.}~\bibnamefont {Zhang}},\
  }\bibfield  {title} {\bibinfo {title} {Spin excitation spectrum of a magnetic
  hopfion},\ }\href@noop {} {\bibfield  {journal} {\bibinfo  {journal} {Appl.
  Phys. Lett.}\ }\textbf {\bibinfo {volume} {119}} (\bibinfo {year}
  {2021})}\BibitemShut {NoStop}%
\bibitem [{\citenamefont {Kosterlitz}\ and\ \citenamefont
  {Thouless}(1973)}]{kosterlitz1973ordering}%
  \BibitemOpen
  \bibfield  {author} {\bibinfo {author} {\bibfnamefont {J.~M.}\ \bibnamefont
  {Kosterlitz}}\ and\ \bibinfo {author} {\bibfnamefont {D.~J.}\ \bibnamefont
  {Thouless}},\ }\bibfield  {title} {\bibinfo {title} {Ordering, metastability
  and phase transitions in two-dimensional systems},\ }\href@noop {} {\bibfield
   {journal} {\bibinfo  {journal} {J. Phys. C: Solid State Phys.}\ }\textbf
  {\bibinfo {volume} {6}},\ \bibinfo {pages} {1181} (\bibinfo {year}
  {1973})}\BibitemShut {NoStop}%
\bibitem [{\citenamefont {Nelson}\ and\ \citenamefont
  {Halperin}(1979)}]{nelson1979dislocation}%
  \BibitemOpen
  \bibfield  {author} {\bibinfo {author} {\bibfnamefont {D.~R.}\ \bibnamefont
  {Nelson}}\ and\ \bibinfo {author} {\bibfnamefont {B.}~\bibnamefont
  {Halperin}},\ }\bibfield  {title} {\bibinfo {title} {Dislocation-mediated
  melting in two dimensions},\ }\href@noop {} {\bibfield  {journal} {\bibinfo
  {journal} {Phys. Rev. B}\ }\textbf {\bibinfo {volume} {19}},\ \bibinfo
  {pages} {2457} (\bibinfo {year} {1979})}\BibitemShut {NoStop}%
\bibitem [{\citenamefont {Huang}\ \emph {et~al.}(2020)\citenamefont {Huang},
  \citenamefont {Sch{\"o}nenberger}, \citenamefont {Cantoni}, \citenamefont
  {Heinen}, \citenamefont {Magrez}, \citenamefont {Rosch}, \citenamefont
  {Carbone},\ and\ \citenamefont {R{\o}nnow}}]{huang2020melting}%
  \BibitemOpen
  \bibfield  {author} {\bibinfo {author} {\bibfnamefont {P.}~\bibnamefont
  {Huang}}, \bibinfo {author} {\bibfnamefont {T.}~\bibnamefont
  {Sch{\"o}nenberger}}, \bibinfo {author} {\bibfnamefont {M.}~\bibnamefont
  {Cantoni}}, \bibinfo {author} {\bibfnamefont {L.}~\bibnamefont {Heinen}},
  \bibinfo {author} {\bibfnamefont {A.}~\bibnamefont {Magrez}}, \bibinfo
  {author} {\bibfnamefont {A.}~\bibnamefont {Rosch}}, \bibinfo {author}
  {\bibfnamefont {F.}~\bibnamefont {Carbone}},\ and\ \bibinfo {author}
  {\bibfnamefont {H.~M.}\ \bibnamefont {R{\o}nnow}},\ }\bibfield  {title}
  {\bibinfo {title} {Melting of a skyrmion lattice to a skyrmion liquid via a
  hexatic phase},\ }\href@noop {} {\bibfield  {journal} {\bibinfo  {journal}
  {Nat. Nanotechnol.}\ }\textbf {\bibinfo {volume} {15}},\ \bibinfo {pages}
  {761} (\bibinfo {year} {2020})}\BibitemShut {NoStop}%
\bibitem [{\citenamefont {Gruber}\ \emph {et~al.}(2025)\citenamefont {Gruber},
  \citenamefont {Roth{\"o}rl}, \citenamefont {Fr{\"o}hlich}, \citenamefont
  {Brems}, \citenamefont {Kammerbauer}, \citenamefont {Syskaki}, \citenamefont
  {Jefremovas}, \citenamefont {Krishnia}, \citenamefont {Sudb{\o}},
  \citenamefont {Virnau} \emph {et~al.}}]{gruber2025imaging}%
  \BibitemOpen
  \bibfield  {author} {\bibinfo {author} {\bibfnamefont {R.}~\bibnamefont
  {Gruber}}, \bibinfo {author} {\bibfnamefont {J.}~\bibnamefont {Roth{\"o}rl}},
  \bibinfo {author} {\bibfnamefont {S.~M.}\ \bibnamefont {Fr{\"o}hlich}},
  \bibinfo {author} {\bibfnamefont {M.~A.}\ \bibnamefont {Brems}}, \bibinfo
  {author} {\bibfnamefont {F.}~\bibnamefont {Kammerbauer}}, \bibinfo {author}
  {\bibfnamefont {M.-A.}\ \bibnamefont {Syskaki}}, \bibinfo {author}
  {\bibfnamefont {E.~M.}\ \bibnamefont {Jefremovas}}, \bibinfo {author}
  {\bibfnamefont {S.}~\bibnamefont {Krishnia}}, \bibinfo {author}
  {\bibfnamefont {A.}~\bibnamefont {Sudb{\o}}}, \bibinfo {author}
  {\bibfnamefont {P.}~\bibnamefont {Virnau}}, \emph {et~al.},\ }\bibfield
  {title} {\bibinfo {title} {Imaging topological defect dynamics mediating 2d
  skyrmion lattice melting},\ }\href@noop {} {\bibfield  {journal} {\bibinfo
  {journal} {arXiv preprint arXiv:2501.13151}\ } (\bibinfo {year}
  {2025})}\BibitemShut {NoStop}%
\bibitem [{\citenamefont {Z{\'a}zvorka}\ \emph {et~al.}(2020)\citenamefont
  {Z{\'a}zvorka}, \citenamefont {Dittrich}, \citenamefont {Ge}, \citenamefont
  {Kerber}, \citenamefont {Raab}, \citenamefont {Winkler}, \citenamefont
  {Litzius}, \citenamefont {Veis}, \citenamefont {Virnau},\ and\ \citenamefont
  {Kl{\"a}ui}}]{zazvorka2020skyrmion}%
  \BibitemOpen
  \bibfield  {author} {\bibinfo {author} {\bibfnamefont {J.}~\bibnamefont
  {Z{\'a}zvorka}}, \bibinfo {author} {\bibfnamefont {F.}~\bibnamefont
  {Dittrich}}, \bibinfo {author} {\bibfnamefont {Y.}~\bibnamefont {Ge}},
  \bibinfo {author} {\bibfnamefont {N.}~\bibnamefont {Kerber}}, \bibinfo
  {author} {\bibfnamefont {K.}~\bibnamefont {Raab}}, \bibinfo {author}
  {\bibfnamefont {T.}~\bibnamefont {Winkler}}, \bibinfo {author} {\bibfnamefont
  {K.}~\bibnamefont {Litzius}}, \bibinfo {author} {\bibfnamefont
  {M.}~\bibnamefont {Veis}}, \bibinfo {author} {\bibfnamefont {P.}~\bibnamefont
  {Virnau}},\ and\ \bibinfo {author} {\bibfnamefont {M.}~\bibnamefont
  {Kl{\"a}ui}},\ }\bibfield  {title} {\bibinfo {title} {Skyrmion {{Lattice
  Phases}} in {{Thin Film Multilayer}}},\ }\href
  {https://doi.org/10.1002/adfm.202004037} {\bibfield  {journal} {\bibinfo
  {journal} {Adv. Funct. Mater.}\ }\textbf {\bibinfo {volume} {30}},\ \bibinfo
  {pages} {2004037} (\bibinfo {year} {2020})}\BibitemShut {NoStop}%
\bibitem [{\citenamefont {Gruber}\ \emph {et~al.}(2023)\citenamefont {Gruber},
  \citenamefont {Brems}, \citenamefont {Roth{\"o}rl}, \citenamefont {Sparmann},
  \citenamefont {Schmitt}, \citenamefont {Kononenko}, \citenamefont
  {Kammerbauer}, \citenamefont {Syskaki}, \citenamefont {Farago}, \citenamefont
  {Virnau} \emph {et~al.}}]{gruber2023300}%
  \BibitemOpen
  \bibfield  {author} {\bibinfo {author} {\bibfnamefont {R.}~\bibnamefont
  {Gruber}}, \bibinfo {author} {\bibfnamefont {M.~A.}\ \bibnamefont {Brems}},
  \bibinfo {author} {\bibfnamefont {J.}~\bibnamefont {Roth{\"o}rl}}, \bibinfo
  {author} {\bibfnamefont {T.}~\bibnamefont {Sparmann}}, \bibinfo {author}
  {\bibfnamefont {M.}~\bibnamefont {Schmitt}}, \bibinfo {author} {\bibfnamefont
  {I.}~\bibnamefont {Kononenko}}, \bibinfo {author} {\bibfnamefont
  {F.}~\bibnamefont {Kammerbauer}}, \bibinfo {author} {\bibfnamefont {M.-A.}\
  \bibnamefont {Syskaki}}, \bibinfo {author} {\bibfnamefont {O.}~\bibnamefont
  {Farago}}, \bibinfo {author} {\bibfnamefont {P.}~\bibnamefont {Virnau}},
  \emph {et~al.},\ }\bibfield  {title} {\bibinfo {title} {300-times-increased
  diffusive skyrmion dynamics and effective pinning reduction by periodic field
  excitation},\ }\href@noop {} {\bibfield  {journal} {\bibinfo  {journal} {Adv.
  Mater.}\ }\textbf {\bibinfo {volume} {35}},\ \bibinfo {pages} {2208922}
  (\bibinfo {year} {2023})}\BibitemShut {NoStop}%
\bibitem [{\citenamefont {Kl{\"a}ui}(2020)}]{klaui2020freezing}%
  \BibitemOpen
  \bibfield  {author} {\bibinfo {author} {\bibfnamefont {M.}~\bibnamefont
  {Kl{\"a}ui}},\ }\bibfield  {title} {\bibinfo {title} {Freezing and melting
  skyrmions in {{2D}}},\ }\href {https://doi.org/10.1038/s41565-020-0726-1}
  {\bibfield  {journal} {\bibinfo  {journal} {Nat. Nanotechnol.}\ }\textbf
  {\bibinfo {volume} {15}},\ \bibinfo {pages} {726} (\bibinfo {year}
  {2020})}\BibitemShut {NoStop}%
\bibitem [{\citenamefont {Gruber}\ \emph {et~al.}(2022)\citenamefont {Gruber},
  \citenamefont {Z{\'a}zvorka}, \citenamefont {Brems}, \citenamefont
  {Rodrigues}, \citenamefont {Dohi}, \citenamefont {Kerber}, \citenamefont
  {Seng}, \citenamefont {Vafaee}, \citenamefont {{Everschor-Sitte}},
  \citenamefont {Virnau},\ and\ \citenamefont
  {Kl{\"a}ui}}]{gruber2022skyrmion}%
  \BibitemOpen
  \bibfield  {author} {\bibinfo {author} {\bibfnamefont {R.}~\bibnamefont
  {Gruber}}, \bibinfo {author} {\bibfnamefont {J.}~\bibnamefont
  {Z{\'a}zvorka}}, \bibinfo {author} {\bibfnamefont {M.~A.}\ \bibnamefont
  {Brems}}, \bibinfo {author} {\bibfnamefont {D.~R.}\ \bibnamefont
  {Rodrigues}}, \bibinfo {author} {\bibfnamefont {T.}~\bibnamefont {Dohi}},
  \bibinfo {author} {\bibfnamefont {N.}~\bibnamefont {Kerber}}, \bibinfo
  {author} {\bibfnamefont {B.}~\bibnamefont {Seng}}, \bibinfo {author}
  {\bibfnamefont {M.}~\bibnamefont {Vafaee}}, \bibinfo {author} {\bibfnamefont
  {K.}~\bibnamefont {{Everschor-Sitte}}}, \bibinfo {author} {\bibfnamefont
  {P.}~\bibnamefont {Virnau}},\ and\ \bibinfo {author} {\bibfnamefont
  {M.}~\bibnamefont {Kl{\"a}ui}},\ }\bibfield  {title} {\bibinfo {title}
  {Skyrmion pinning energetics in thin film systems},\ }\href
  {https://doi.org/10.1038/s41467-022-30743-4} {\bibfield  {journal} {\bibinfo
  {journal} {Nat. Commun.}\ }\textbf {\bibinfo {volume} {13}},\ \bibinfo
  {pages} {3144} (\bibinfo {year} {2022})}\BibitemShut {NoStop}%
\bibitem [{\citenamefont {Reichhardt}\ \emph {et~al.}(2022)\citenamefont
  {Reichhardt}, \citenamefont {Reichhardt},\ and\ \citenamefont
  {Milo{\v{s}}evi{\'c}}}]{reichhardt2022statics}%
  \BibitemOpen
  \bibfield  {author} {\bibinfo {author} {\bibfnamefont {C.}~\bibnamefont
  {Reichhardt}}, \bibinfo {author} {\bibfnamefont {C.~J.~O.}\ \bibnamefont
  {Reichhardt}},\ and\ \bibinfo {author} {\bibfnamefont {M.}~\bibnamefont
  {Milo{\v{s}}evi{\'c}}},\ }\bibfield  {title} {\bibinfo {title} {Statics and
  dynamics of skyrmions interacting with disorder and nanostructures},\
  }\href@noop {} {\bibfield  {journal} {\bibinfo  {journal} {Rev. Mod. Phys.}\
  }\textbf {\bibinfo {volume} {94}},\ \bibinfo {pages} {035005} (\bibinfo
  {year} {2022})}\BibitemShut {NoStop}%
\bibitem [{\citenamefont {Kern}\ \emph
  {et~al.}(2022{\natexlab{a}})\citenamefont {Kern}, \citenamefont {Pfau},
  \citenamefont {Deinhart}, \citenamefont {Schneider}, \citenamefont {Klose},
  \citenamefont {Gerlinger}, \citenamefont {Wittrock}, \citenamefont {Engel},
  \citenamefont {Will}, \citenamefont {G{\"u}nther}, \citenamefont
  {Liefferink}, \citenamefont {Mentink}, \citenamefont {Wintz}, \citenamefont
  {Weigand}, \citenamefont {Huang}, \citenamefont {Battistelli}, \citenamefont
  {Metternich}, \citenamefont {B{\"u}ttner}, \citenamefont {H{\"o}flich},\ and\
  \citenamefont {Eisebitt}}]{kern2022deterministic}%
  \BibitemOpen
  \bibfield  {author} {\bibinfo {author} {\bibfnamefont {L.-M.}\ \bibnamefont
  {Kern}}, \bibinfo {author} {\bibfnamefont {B.}~\bibnamefont {Pfau}}, \bibinfo
  {author} {\bibfnamefont {V.}~\bibnamefont {Deinhart}}, \bibinfo {author}
  {\bibfnamefont {M.}~\bibnamefont {Schneider}}, \bibinfo {author}
  {\bibfnamefont {C.}~\bibnamefont {Klose}}, \bibinfo {author} {\bibfnamefont
  {K.}~\bibnamefont {Gerlinger}}, \bibinfo {author} {\bibfnamefont
  {S.}~\bibnamefont {Wittrock}}, \bibinfo {author} {\bibfnamefont
  {D.}~\bibnamefont {Engel}}, \bibinfo {author} {\bibfnamefont
  {I.}~\bibnamefont {Will}}, \bibinfo {author} {\bibfnamefont {C.~M.}\
  \bibnamefont {G{\"u}nther}}, \bibinfo {author} {\bibfnamefont
  {R.}~\bibnamefont {Liefferink}}, \bibinfo {author} {\bibfnamefont {J.~H.}\
  \bibnamefont {Mentink}}, \bibinfo {author} {\bibfnamefont {S.}~\bibnamefont
  {Wintz}}, \bibinfo {author} {\bibfnamefont {M.}~\bibnamefont {Weigand}},
  \bibinfo {author} {\bibfnamefont {M.-J.}\ \bibnamefont {Huang}}, \bibinfo
  {author} {\bibfnamefont {R.}~\bibnamefont {Battistelli}}, \bibinfo {author}
  {\bibfnamefont {D.}~\bibnamefont {Metternich}}, \bibinfo {author}
  {\bibfnamefont {F.}~\bibnamefont {B{\"u}ttner}}, \bibinfo {author}
  {\bibfnamefont {K.}~\bibnamefont {H{\"o}flich}},\ and\ \bibinfo {author}
  {\bibfnamefont {S.}~\bibnamefont {Eisebitt}},\ }\bibfield  {title} {\bibinfo
  {title} {Deterministic {{Generation}} and {{Guided Motion}} of {{Magnetic
  Skyrmions}} by {{Focused He}}+-{{Ion Irradiation}}},\ }\href
  {https://doi.org/10.1021/acs.nanolett.2c00670} {\bibfield  {journal}
  {\bibinfo  {journal} {Nano Lett.}\ }\textbf {\bibinfo {volume} {22}},\
  \bibinfo {pages} {4028} (\bibinfo {year} {2022}{\natexlab{a}})}\BibitemShut
  {NoStop}%
\bibitem [{\citenamefont {Lee}\ \emph {et~al.}(2023)\citenamefont {Lee},
  \citenamefont {Msiska}, \citenamefont {Brems}, \citenamefont {Kl{\"a}ui},
  \citenamefont {Kurebayashi},\ and\ \citenamefont
  {{Everschor-Sitte}}}]{lee2023perspective}%
  \BibitemOpen
  \bibfield  {author} {\bibinfo {author} {\bibfnamefont {O.}~\bibnamefont
  {Lee}}, \bibinfo {author} {\bibfnamefont {R.}~\bibnamefont {Msiska}},
  \bibinfo {author} {\bibfnamefont {M.~A.}\ \bibnamefont {Brems}}, \bibinfo
  {author} {\bibfnamefont {M.}~\bibnamefont {Kl{\"a}ui}}, \bibinfo {author}
  {\bibfnamefont {H.}~\bibnamefont {Kurebayashi}},\ and\ \bibinfo {author}
  {\bibfnamefont {K.}~\bibnamefont {{Everschor-Sitte}}},\ }\bibfield  {title}
  {\bibinfo {title} {Perspective on unconventional computing using magnetic
  skyrmions},\ }\href {https://doi.org/10.1063/5.0148469} {\bibfield  {journal}
  {\bibinfo  {journal} {Appl. Phys. Lett.}\ }\textbf {\bibinfo {volume}
  {122}},\ \bibinfo {pages} {260501} (\bibinfo {year} {2023})}\BibitemShut
  {NoStop}%
\bibitem [{\citenamefont {Brems}\ \emph {et~al.}(2025)\citenamefont {Brems},
  \citenamefont {Sparmann}, \citenamefont {Fr{\"o}hlich}, \citenamefont {Dany},
  \citenamefont {Roth{\"o}rl}, \citenamefont {Kammerbauer}, \citenamefont
  {Jefremovas}, \citenamefont {Farago}, \citenamefont {Kl{\"a}ui},\ and\
  \citenamefont {Virnau}}]{brems2025realizing}%
  \BibitemOpen
  \bibfield  {author} {\bibinfo {author} {\bibfnamefont {M.~A.}\ \bibnamefont
  {Brems}}, \bibinfo {author} {\bibfnamefont {T.}~\bibnamefont {Sparmann}},
  \bibinfo {author} {\bibfnamefont {S.~M.}\ \bibnamefont {Fr{\"o}hlich}},
  \bibinfo {author} {\bibfnamefont {L.-C.}\ \bibnamefont {Dany}}, \bibinfo
  {author} {\bibfnamefont {J.}~\bibnamefont {Roth{\"o}rl}}, \bibinfo {author}
  {\bibfnamefont {F.}~\bibnamefont {Kammerbauer}}, \bibinfo {author}
  {\bibfnamefont {E.~M.}\ \bibnamefont {Jefremovas}}, \bibinfo {author}
  {\bibfnamefont {O.}~\bibnamefont {Farago}}, \bibinfo {author} {\bibfnamefont
  {M.}~\bibnamefont {Kl{\"a}ui}},\ and\ \bibinfo {author} {\bibfnamefont
  {P.}~\bibnamefont {Virnau}},\ }\bibfield  {title} {\bibinfo {title}
  {Realizing quantitative quasiparticle modeling of skyrmion dynamics in
  arbitrary potentials},\ }\href@noop {} {\bibfield  {journal} {\bibinfo
  {journal} {Phys. Rev. Lett.}\ }\textbf {\bibinfo {volume} {134}},\ \bibinfo
  {pages} {046701} (\bibinfo {year} {2025})}\BibitemShut {NoStop}%
\bibitem [{\citenamefont {Kosevich}\ \emph {et~al.}(1990)\citenamefont
  {Kosevich}, \citenamefont {Ivanov},\ and\ \citenamefont
  {Kovalev}}]{kosevich1990magnetic}%
  \BibitemOpen
  \bibfield  {author} {\bibinfo {author} {\bibfnamefont {A.~M.}\ \bibnamefont
  {Kosevich}}, \bibinfo {author} {\bibfnamefont {B.~A.}\ \bibnamefont
  {Ivanov}},\ and\ \bibinfo {author} {\bibfnamefont {A.~S.}\ \bibnamefont
  {Kovalev}},\ }\bibfield  {title} {\bibinfo {title} {Magnetic {{Solitons}}},\
  }\href {https://doi.org/10.1016/0370-1573(90)90130-T} {\bibfield  {journal}
  {\bibinfo  {journal} {Phys. Rep.}\ }\textbf {\bibinfo {volume} {194}},\
  \bibinfo {pages} {117} (\bibinfo {year} {1990})}\BibitemShut {NoStop}%
\bibitem [{\citenamefont {Barker}\ and\ \citenamefont
  {Tretiakov}(2016)}]{barker2016static}%
  \BibitemOpen
  \bibfield  {author} {\bibinfo {author} {\bibfnamefont {J.}~\bibnamefont
  {Barker}}\ and\ \bibinfo {author} {\bibfnamefont {O.~A.}\ \bibnamefont
  {Tretiakov}},\ }\bibfield  {title} {\bibinfo {title} {Static and {{Dynamical
  Properties}} of {{Antiferromagnetic Skyrmions}} in the {{Presence}} of
  {{Applied Current}} and {{Temperature}}},\ }\href
  {https://doi.org/10.1103/PhysRevLett.116.147203} {\bibfield  {journal}
  {\bibinfo  {journal} {Phys. Rev. Lett.}\ }\textbf {\bibinfo {volume} {116}},\
  \bibinfo {pages} {147203} (\bibinfo {year} {2016})}\BibitemShut {NoStop}%
\bibitem [{\citenamefont {Shen}\ \emph {et~al.}(2020)\citenamefont {Shen},
  \citenamefont {Xia}, \citenamefont {Zhang}, \citenamefont {Ezawa},
  \citenamefont {Tretiakov}, \citenamefont {Liu}, \citenamefont {Zhao},\ and\
  \citenamefont {Zhou}}]{shen2020currentinduced}%
  \BibitemOpen
  \bibfield  {author} {\bibinfo {author} {\bibfnamefont {L.}~\bibnamefont
  {Shen}}, \bibinfo {author} {\bibfnamefont {J.}~\bibnamefont {Xia}}, \bibinfo
  {author} {\bibfnamefont {X.}~\bibnamefont {Zhang}}, \bibinfo {author}
  {\bibfnamefont {M.}~\bibnamefont {Ezawa}}, \bibinfo {author} {\bibfnamefont
  {O.~A.}\ \bibnamefont {Tretiakov}}, \bibinfo {author} {\bibfnamefont
  {X.}~\bibnamefont {Liu}}, \bibinfo {author} {\bibfnamefont {G.}~\bibnamefont
  {Zhao}},\ and\ \bibinfo {author} {\bibfnamefont {Y.}~\bibnamefont {Zhou}},\
  }\bibfield  {title} {\bibinfo {title} {Current-{{Induced Dynamics}} and
  {{Chaos}} of {{Antiferromagnetic Bimerons}}},\ }\href
  {https://doi.org/10.1103/PhysRevLett.124.037202} {\bibfield  {journal}
  {\bibinfo  {journal} {Phys. Rev. Lett.}\ }\textbf {\bibinfo {volume} {124}},\
  \bibinfo {pages} {037202} (\bibinfo {year} {2020})}\BibitemShut {NoStop}%
\bibitem [{\citenamefont {Chmiel}\ \emph {et~al.}(2018)\citenamefont {Chmiel},
  \citenamefont {Waterfield~Price}, \citenamefont {Johnson}, \citenamefont
  {Lamirand}, \citenamefont {Schad}, \citenamefont {{van der Laan}},
  \citenamefont {Harris}, \citenamefont {Irwin}, \citenamefont {Rzchowski},
  \citenamefont {Eom},\ and\ \citenamefont {Radaelli}}]{chmiel2018observation}%
  \BibitemOpen
  \bibfield  {author} {\bibinfo {author} {\bibfnamefont {F.~P.}\ \bibnamefont
  {Chmiel}}, \bibinfo {author} {\bibfnamefont {N.}~\bibnamefont
  {Waterfield~Price}}, \bibinfo {author} {\bibfnamefont {R.~D.}\ \bibnamefont
  {Johnson}}, \bibinfo {author} {\bibfnamefont {A.~D.}\ \bibnamefont
  {Lamirand}}, \bibinfo {author} {\bibfnamefont {J.}~\bibnamefont {Schad}},
  \bibinfo {author} {\bibfnamefont {G.}~\bibnamefont {{van der Laan}}},
  \bibinfo {author} {\bibfnamefont {D.~T.}\ \bibnamefont {Harris}}, \bibinfo
  {author} {\bibfnamefont {J.}~\bibnamefont {Irwin}}, \bibinfo {author}
  {\bibfnamefont {M.~S.}\ \bibnamefont {Rzchowski}}, \bibinfo {author}
  {\bibfnamefont {C.-B.}\ \bibnamefont {Eom}},\ and\ \bibinfo {author}
  {\bibfnamefont {P.~G.}\ \bibnamefont {Radaelli}},\ }\bibfield  {title}
  {\bibinfo {title} {Observation of magnetic vortex pairs at room temperature
  in a planar {$\alpha$}-{{Fe2O3}}/{{Co}} heterostructure},\ }\href
  {https://doi.org/10.1038/s41563-018-0101-x} {\bibfield  {journal} {\bibinfo
  {journal} {Nat. Mater.}\ }\textbf {\bibinfo {volume} {17}},\ \bibinfo {pages}
  {581} (\bibinfo {year} {2018})}\BibitemShut {NoStop}%
\bibitem [{\citenamefont {Jani}\ \emph {et~al.}(2021)\citenamefont {Jani},
  \citenamefont {Lin}, \citenamefont {Chen}, \citenamefont {Harrison},
  \citenamefont {Maccherozzi}, \citenamefont {Schad}, \citenamefont {Prakash},
  \citenamefont {Eom}, \citenamefont {Ariando}, \citenamefont {Venkatesan},\
  and\ \citenamefont {Radaelli}}]{jani2021antiferromagnetic}%
  \BibitemOpen
  \bibfield  {author} {\bibinfo {author} {\bibfnamefont {H.}~\bibnamefont
  {Jani}}, \bibinfo {author} {\bibfnamefont {J.-C.}\ \bibnamefont {Lin}},
  \bibinfo {author} {\bibfnamefont {J.}~\bibnamefont {Chen}}, \bibinfo {author}
  {\bibfnamefont {J.}~\bibnamefont {Harrison}}, \bibinfo {author}
  {\bibfnamefont {F.}~\bibnamefont {Maccherozzi}}, \bibinfo {author}
  {\bibfnamefont {J.}~\bibnamefont {Schad}}, \bibinfo {author} {\bibfnamefont
  {S.}~\bibnamefont {Prakash}}, \bibinfo {author} {\bibfnamefont {C.-B.}\
  \bibnamefont {Eom}}, \bibinfo {author} {\bibfnamefont {A.}~\bibnamefont
  {Ariando}}, \bibinfo {author} {\bibfnamefont {T.}~\bibnamefont
  {Venkatesan}},\ and\ \bibinfo {author} {\bibfnamefont {P.~G.}\ \bibnamefont
  {Radaelli}},\ }\bibfield  {title} {\bibinfo {title} {Antiferromagnetic
  half-skyrmions and bimerons at room temperature},\ }\href
  {https://doi.org/10.1038/s41586-021-03219-6} {\bibfield  {journal} {\bibinfo
  {journal} {Nature}\ }\textbf {\bibinfo {volume} {590}},\ \bibinfo {pages}
  {74} (\bibinfo {year} {2021})}\BibitemShut {NoStop}%
\bibitem [{\citenamefont {Amin}\ \emph {et~al.}(2023)\citenamefont {Amin},
  \citenamefont {Poole}, \citenamefont {Reimers}, \citenamefont {Barton},
  \citenamefont {Dal~Din}, \citenamefont {Maccherozzi}, \citenamefont {Dhesi},
  \citenamefont {Nov{\'a}k}, \citenamefont {Krizek}, \citenamefont {Chauhan},
  \citenamefont {Campion}, \citenamefont {Rushforth}, \citenamefont
  {Jungwirth}, \citenamefont {Tretiakov}, \citenamefont {Edmonds},\ and\
  \citenamefont {Wadley}}]{amin2023antiferromagnetic}%
  \BibitemOpen
  \bibfield  {author} {\bibinfo {author} {\bibfnamefont {O.~J.}\ \bibnamefont
  {Amin}}, \bibinfo {author} {\bibfnamefont {S.~F.}\ \bibnamefont {Poole}},
  \bibinfo {author} {\bibfnamefont {S.}~\bibnamefont {Reimers}}, \bibinfo
  {author} {\bibfnamefont {L.~X.}\ \bibnamefont {Barton}}, \bibinfo {author}
  {\bibfnamefont {A.}~\bibnamefont {Dal~Din}}, \bibinfo {author} {\bibfnamefont
  {F.}~\bibnamefont {Maccherozzi}}, \bibinfo {author} {\bibfnamefont {S.~S.}\
  \bibnamefont {Dhesi}}, \bibinfo {author} {\bibfnamefont {V.}~\bibnamefont
  {Nov{\'a}k}}, \bibinfo {author} {\bibfnamefont {F.}~\bibnamefont {Krizek}},
  \bibinfo {author} {\bibfnamefont {J.~S.}\ \bibnamefont {Chauhan}}, \bibinfo
  {author} {\bibfnamefont {R.~P.}\ \bibnamefont {Campion}}, \bibinfo {author}
  {\bibfnamefont {A.~W.}\ \bibnamefont {Rushforth}}, \bibinfo {author}
  {\bibfnamefont {T.}~\bibnamefont {Jungwirth}}, \bibinfo {author}
  {\bibfnamefont {O.~A.}\ \bibnamefont {Tretiakov}}, \bibinfo {author}
  {\bibfnamefont {K.~W.}\ \bibnamefont {Edmonds}},\ and\ \bibinfo {author}
  {\bibfnamefont {P.}~\bibnamefont {Wadley}},\ }\bibfield  {title} {\bibinfo
  {title} {Antiferromagnetic half-skyrmions electrically generated and
  controlled at room temperature},\ }\href
  {https://doi.org/10.1038/s41565-023-01386-3} {\bibfield  {journal} {\bibinfo
  {journal} {Nat. Nanotechnol.}\ }\textbf {\bibinfo {volume} {18}},\ \bibinfo
  {pages} {849} (\bibinfo {year} {2023})}\BibitemShut {NoStop}%
\bibitem [{\citenamefont {Jani}\ \emph {et~al.}(2024)\citenamefont {Jani},
  \citenamefont {Harrison}, \citenamefont {Hooda}, \citenamefont {Prakash},
  \citenamefont {Nandi}, \citenamefont {Hu}, \citenamefont {Zeng},
  \citenamefont {Lin}, \citenamefont {Godfrey}, \citenamefont {ji~Omar},
  \citenamefont {Butcher}, \citenamefont {Raabe}, \citenamefont {Finizio},
  \citenamefont {Thean}, \citenamefont {Ariando},\ and\ \citenamefont
  {Radaelli}}]{jani2024spatially}%
  \BibitemOpen
  \bibfield  {author} {\bibinfo {author} {\bibfnamefont {H.}~\bibnamefont
  {Jani}}, \bibinfo {author} {\bibfnamefont {J.}~\bibnamefont {Harrison}},
  \bibinfo {author} {\bibfnamefont {S.}~\bibnamefont {Hooda}}, \bibinfo
  {author} {\bibfnamefont {S.}~\bibnamefont {Prakash}}, \bibinfo {author}
  {\bibfnamefont {P.}~\bibnamefont {Nandi}}, \bibinfo {author} {\bibfnamefont
  {J.}~\bibnamefont {Hu}}, \bibinfo {author} {\bibfnamefont {Z.}~\bibnamefont
  {Zeng}}, \bibinfo {author} {\bibfnamefont {J.-C.}\ \bibnamefont {Lin}},
  \bibinfo {author} {\bibfnamefont {C.}~\bibnamefont {Godfrey}}, \bibinfo
  {author} {\bibfnamefont {G.}~\bibnamefont {ji~Omar}}, \bibinfo {author}
  {\bibfnamefont {T.~A.}\ \bibnamefont {Butcher}}, \bibinfo {author}
  {\bibfnamefont {J.}~\bibnamefont {Raabe}}, \bibinfo {author} {\bibfnamefont
  {S.}~\bibnamefont {Finizio}}, \bibinfo {author} {\bibfnamefont {A.~V.-Y.}\
  \bibnamefont {Thean}}, \bibinfo {author} {\bibfnamefont {A.}~\bibnamefont
  {Ariando}},\ and\ \bibinfo {author} {\bibfnamefont {P.~G.}\ \bibnamefont
  {Radaelli}},\ }\bibfield  {title} {\bibinfo {title} {Spatially reconfigurable
  antiferromagnetic states in topologically rich free-standing nanomembranes},\
  }\href {https://doi.org/10.1038/s41563-024-01806-2} {\bibfield  {journal}
  {\bibinfo  {journal} {Nat. Mater.}\ }\textbf {\bibinfo {volume} {23}},\
  \bibinfo {pages} {619} (\bibinfo {year} {2024})}\BibitemShut {NoStop}%
\bibitem [{\citenamefont {Harrison}\ \emph {et~al.}(2025)\citenamefont
  {Harrison}, \citenamefont {Hu}, \citenamefont {Godfrey}, \citenamefont {Lin},
  \citenamefont {Butcher}, \citenamefont {Raabe}, \citenamefont {Finizio},
  \citenamefont {Jani},\ and\ \citenamefont {Radaelli}}]{harrison2025}%
  \BibitemOpen
  \bibfield  {author} {\bibinfo {author} {\bibfnamefont {J.}~\bibnamefont
  {Harrison}}, \bibinfo {author} {\bibfnamefont {J.}~\bibnamefont {Hu}},
  \bibinfo {author} {\bibfnamefont {C.}~\bibnamefont {Godfrey}}, \bibinfo
  {author} {\bibfnamefont {J.-C.}\ \bibnamefont {Lin}}, \bibinfo {author}
  {\bibfnamefont {T.~A.}\ \bibnamefont {Butcher}}, \bibinfo {author}
  {\bibfnamefont {J.}~\bibnamefont {Raabe}}, \bibinfo {author} {\bibfnamefont
  {S.}~\bibnamefont {Finizio}}, \bibinfo {author} {\bibfnamefont
  {H.}~\bibnamefont {Jani}},\ and\ \bibinfo {author} {\bibfnamefont {P.~G.}\
  \bibnamefont {Radaelli}},\ }\bibfield  {title} {\bibinfo {title} {Room
  temperature control of axial and basal antiferromagnetic anisotropies using
  strain},\ }\href@noop {} {\bibfield  {journal} {\bibinfo  {journal} {ACS
  nano}\ } (\bibinfo {year} {2025})}\BibitemShut {NoStop}%
\bibitem [{\citenamefont {Harrison}\ \emph {et~al.}(2024)\citenamefont
  {Harrison}, \citenamefont {Jani}, \citenamefont {Hu}, \citenamefont {Lal},
  \citenamefont {Lin}, \citenamefont {Popescu}, \citenamefont {Brown},
  \citenamefont {Jaouen}, \citenamefont {Ariando},\ and\ \citenamefont
  {Radaelli}}]{harrison2024holographic}%
  \BibitemOpen
  \bibfield  {author} {\bibinfo {author} {\bibfnamefont {J.}~\bibnamefont
  {Harrison}}, \bibinfo {author} {\bibfnamefont {H.}~\bibnamefont {Jani}},
  \bibinfo {author} {\bibfnamefont {J.}~\bibnamefont {Hu}}, \bibinfo {author}
  {\bibfnamefont {M.}~\bibnamefont {Lal}}, \bibinfo {author} {\bibfnamefont
  {J.-C.}\ \bibnamefont {Lin}}, \bibinfo {author} {\bibfnamefont
  {H.}~\bibnamefont {Popescu}}, \bibinfo {author} {\bibfnamefont
  {J.}~\bibnamefont {Brown}}, \bibinfo {author} {\bibfnamefont
  {N.}~\bibnamefont {Jaouen}}, \bibinfo {author} {\bibfnamefont
  {A.}~\bibnamefont {Ariando}},\ and\ \bibinfo {author} {\bibfnamefont {P.~G.}\
  \bibnamefont {Radaelli}},\ }\bibfield  {title} {\bibinfo {title} {Holographic
  imaging of antiferromagnetic domains with in-situ magnetic field},\ }\href
  {https://doi.org/10.1364/OE.508005} {\bibfield  {journal} {\bibinfo
  {journal} {Opt. Express}\ }\textbf {\bibinfo {volume} {32}},\ \bibinfo
  {pages} {5885} (\bibinfo {year} {2024})}\BibitemShut {NoStop}%
\bibitem [{\citenamefont {Tan}\ \emph {et~al.}(2024)\citenamefont {Tan},
  \citenamefont {Jani}, \citenamefont {H{\"o}gen}, \citenamefont {Stefan},
  \citenamefont {Castelnovo}, \citenamefont {Braund}, \citenamefont {Geim},
  \citenamefont {Mechnich}, \citenamefont {Feuer}, \citenamefont {Knowles},
  \citenamefont {Ariando}, \citenamefont {Radaelli},\ and\ \citenamefont
  {Atat{\"u}re}}]{tan2024revealing}%
  \BibitemOpen
  \bibfield  {author} {\bibinfo {author} {\bibfnamefont {A.~K.~C.}\
  \bibnamefont {Tan}}, \bibinfo {author} {\bibfnamefont {H.}~\bibnamefont
  {Jani}}, \bibinfo {author} {\bibfnamefont {M.}~\bibnamefont {H{\"o}gen}},
  \bibinfo {author} {\bibfnamefont {L.}~\bibnamefont {Stefan}}, \bibinfo
  {author} {\bibfnamefont {C.}~\bibnamefont {Castelnovo}}, \bibinfo {author}
  {\bibfnamefont {D.}~\bibnamefont {Braund}}, \bibinfo {author} {\bibfnamefont
  {A.}~\bibnamefont {Geim}}, \bibinfo {author} {\bibfnamefont {A.}~\bibnamefont
  {Mechnich}}, \bibinfo {author} {\bibfnamefont {M.~S.~G.}\ \bibnamefont
  {Feuer}}, \bibinfo {author} {\bibfnamefont {H.~S.}\ \bibnamefont {Knowles}},
  \bibinfo {author} {\bibfnamefont {A.}~\bibnamefont {Ariando}}, \bibinfo
  {author} {\bibfnamefont {P.~G.}\ \bibnamefont {Radaelli}},\ and\ \bibinfo
  {author} {\bibfnamefont {M.}~\bibnamefont {Atat{\"u}re}},\ }\bibfield
  {title} {\bibinfo {title} {Revealing emergent magnetic charge in an
  antiferromagnet with diamond quantum magnetometry},\ }\href
  {https://doi.org/10.1038/s41563-023-01737-4} {\bibfield  {journal} {\bibinfo
  {journal} {Nat. Mater.}\ }\textbf {\bibinfo {volume} {23}},\ \bibinfo {pages}
  {205} (\bibinfo {year} {2024})}\BibitemShut {NoStop}%
\bibitem [{\citenamefont {Amin}\ \emph {et~al.}(2024)\citenamefont {Amin},
  \citenamefont {Dal~Din}, \citenamefont {Golias}, \citenamefont {Niu},
  \citenamefont {Zakharov}, \citenamefont {Fromage}, \citenamefont {Fields},
  \citenamefont {Heywood}, \citenamefont {Cousins}, \citenamefont
  {Maccherozzi}, \citenamefont {Krempask{\'y}}, \citenamefont {Dil},
  \citenamefont {Kriegner}, \citenamefont {Kiraly}, \citenamefont {Campion},
  \citenamefont {Rushforth}, \citenamefont {Edmonds}, \citenamefont {Dhesi},
  \citenamefont {{\v S}mejkal}, \citenamefont {Jungwirth},\ and\ \citenamefont
  {Wadley}}]{amin2024nanoscale}%
  \BibitemOpen
  \bibfield  {author} {\bibinfo {author} {\bibfnamefont {O.~J.}\ \bibnamefont
  {Amin}}, \bibinfo {author} {\bibfnamefont {A.}~\bibnamefont {Dal~Din}},
  \bibinfo {author} {\bibfnamefont {E.}~\bibnamefont {Golias}}, \bibinfo
  {author} {\bibfnamefont {Y.}~\bibnamefont {Niu}}, \bibinfo {author}
  {\bibfnamefont {A.}~\bibnamefont {Zakharov}}, \bibinfo {author}
  {\bibfnamefont {S.~C.}\ \bibnamefont {Fromage}}, \bibinfo {author}
  {\bibfnamefont {C.~J.~B.}\ \bibnamefont {Fields}}, \bibinfo {author}
  {\bibfnamefont {S.~L.}\ \bibnamefont {Heywood}}, \bibinfo {author}
  {\bibfnamefont {R.~B.}\ \bibnamefont {Cousins}}, \bibinfo {author}
  {\bibfnamefont {F.}~\bibnamefont {Maccherozzi}}, \bibinfo {author}
  {\bibfnamefont {J.}~\bibnamefont {Krempask{\'y}}}, \bibinfo {author}
  {\bibfnamefont {J.~H.}\ \bibnamefont {Dil}}, \bibinfo {author} {\bibfnamefont
  {D.}~\bibnamefont {Kriegner}}, \bibinfo {author} {\bibfnamefont
  {B.}~\bibnamefont {Kiraly}}, \bibinfo {author} {\bibfnamefont {R.~P.}\
  \bibnamefont {Campion}}, \bibinfo {author} {\bibfnamefont {A.~W.}\
  \bibnamefont {Rushforth}}, \bibinfo {author} {\bibfnamefont {K.~W.}\
  \bibnamefont {Edmonds}}, \bibinfo {author} {\bibfnamefont {S.~S.}\
  \bibnamefont {Dhesi}}, \bibinfo {author} {\bibfnamefont {L.}~\bibnamefont
  {{\v S}mejkal}}, \bibinfo {author} {\bibfnamefont {T.}~\bibnamefont
  {Jungwirth}},\ and\ \bibinfo {author} {\bibfnamefont {P.}~\bibnamefont
  {Wadley}},\ }\bibfield  {title} {\bibinfo {title} {Nanoscale imaging and
  control of altermagnetism in {{MnTe}}},\ }\href
  {https://doi.org/10.1038/s41586-024-08234-x} {\bibfield  {journal} {\bibinfo
  {journal} {Nature}\ }\textbf {\bibinfo {volume} {636}},\ \bibinfo {pages}
  {348} (\bibinfo {year} {2024})}\BibitemShut {NoStop}%
\bibitem [{\citenamefont {Ovcharov}\ \emph {et~al.}(2022)\citenamefont
  {Ovcharov}, \citenamefont {Galkina}, \citenamefont {Ivanov},\ and\
  \citenamefont {Khymyn}}]{ovcharov2022spin}%
  \BibitemOpen
  \bibfield  {author} {\bibinfo {author} {\bibfnamefont {R.}~\bibnamefont
  {Ovcharov}}, \bibinfo {author} {\bibfnamefont {E.}~\bibnamefont {Galkina}},
  \bibinfo {author} {\bibfnamefont {B.}~\bibnamefont {Ivanov}},\ and\ \bibinfo
  {author} {\bibfnamefont {R.}~\bibnamefont {Khymyn}},\ }\bibfield  {title}
  {\bibinfo {title} {Spin {{Hall Nano-Oscillator Based}} on an
  {{Antiferromagnetic Domain Wall}}},\ }\href
  {https://doi.org/10.1103/PhysRevApplied.18.024047} {\bibfield  {journal}
  {\bibinfo  {journal} {Phys. Rev. Appl.}\ }\textbf {\bibinfo {volume} {18}},\
  \bibinfo {pages} {024047} (\bibinfo {year} {2022})}\BibitemShut {NoStop}%
\bibitem [{\citenamefont {Finco}\ \emph {et~al.}(2021)\citenamefont {Finco},
  \citenamefont {Haykal}, \citenamefont {Tanos}, \citenamefont {Fabre},
  \citenamefont {Chouaieb}, \citenamefont {Akhtar}, \citenamefont
  {{Robert-Philip}}, \citenamefont {Legrand}, \citenamefont {Ajejas},
  \citenamefont {Bouzehouane}, \citenamefont {Reyren}, \citenamefont
  {Devolder}, \citenamefont {Adam}, \citenamefont {Kim}, \citenamefont {Cros},\
  and\ \citenamefont {Jacques}}]{finco2021imaging}%
  \BibitemOpen
  \bibfield  {author} {\bibinfo {author} {\bibfnamefont {A.}~\bibnamefont
  {Finco}}, \bibinfo {author} {\bibfnamefont {A.}~\bibnamefont {Haykal}},
  \bibinfo {author} {\bibfnamefont {R.}~\bibnamefont {Tanos}}, \bibinfo
  {author} {\bibfnamefont {F.}~\bibnamefont {Fabre}}, \bibinfo {author}
  {\bibfnamefont {S.}~\bibnamefont {Chouaieb}}, \bibinfo {author}
  {\bibfnamefont {W.}~\bibnamefont {Akhtar}}, \bibinfo {author} {\bibfnamefont
  {I.}~\bibnamefont {{Robert-Philip}}}, \bibinfo {author} {\bibfnamefont
  {W.}~\bibnamefont {Legrand}}, \bibinfo {author} {\bibfnamefont
  {F.}~\bibnamefont {Ajejas}}, \bibinfo {author} {\bibfnamefont
  {K.}~\bibnamefont {Bouzehouane}}, \bibinfo {author} {\bibfnamefont
  {N.}~\bibnamefont {Reyren}}, \bibinfo {author} {\bibfnamefont
  {T.}~\bibnamefont {Devolder}}, \bibinfo {author} {\bibfnamefont {J.-P.}\
  \bibnamefont {Adam}}, \bibinfo {author} {\bibfnamefont {J.-V.}\ \bibnamefont
  {Kim}}, \bibinfo {author} {\bibfnamefont {V.}~\bibnamefont {Cros}},\ and\
  \bibinfo {author} {\bibfnamefont {V.}~\bibnamefont {Jacques}},\ }\bibfield
  {title} {\bibinfo {title} {Imaging non-collinear antiferromagnetic textures
  via single spin relaxometry},\ }\href
  {https://doi.org/10.1038/s41467-021-20995-x} {\bibfield  {journal} {\bibinfo
  {journal} {Nat. Commun.}\ }\textbf {\bibinfo {volume} {12}},\ \bibinfo
  {pages} {767} (\bibinfo {year} {2021})}\BibitemShut {NoStop}%
\bibitem [{\citenamefont {Harrison}\ \emph {et~al.}(2022)\citenamefont
  {Harrison}, \citenamefont {Jani},\ and\ \citenamefont
  {Radaelli}}]{harrison2022route}%
  \BibitemOpen
  \bibfield  {author} {\bibinfo {author} {\bibfnamefont {J.}~\bibnamefont
  {Harrison}}, \bibinfo {author} {\bibfnamefont {H.}~\bibnamefont {Jani}},\
  and\ \bibinfo {author} {\bibfnamefont {P.~G.}\ \bibnamefont {Radaelli}},\
  }\bibfield  {title} {\bibinfo {title} {Route towards stable homochiral
  topological textures in a-type antiferromagnets},\ }\href
  {https://doi.org/10.1103/PhysRevB.105.224424} {\bibfield  {journal} {\bibinfo
   {journal} {Phys. Rev. B}\ }\textbf {\bibinfo {volume} {105}},\ \bibinfo
  {pages} {224424} (\bibinfo {year} {2022})}\BibitemShut {NoStop}%
\bibitem [{\citenamefont {El~Kanj}\ \emph {et~al.}(2023)\citenamefont
  {El~Kanj}, \citenamefont {Gomonay}, \citenamefont {Boventer}, \citenamefont
  {Bortolotti}, \citenamefont {Cros}, \citenamefont {Anane},\ and\
  \citenamefont {Lebrun}}]{elkanj2023antiferromagnetic}%
  \BibitemOpen
  \bibfield  {author} {\bibinfo {author} {\bibfnamefont {A.}~\bibnamefont
  {El~Kanj}}, \bibinfo {author} {\bibfnamefont {O.}~\bibnamefont {Gomonay}},
  \bibinfo {author} {\bibfnamefont {I.}~\bibnamefont {Boventer}}, \bibinfo
  {author} {\bibfnamefont {P.}~\bibnamefont {Bortolotti}}, \bibinfo {author}
  {\bibfnamefont {V.}~\bibnamefont {Cros}}, \bibinfo {author} {\bibfnamefont
  {A.}~\bibnamefont {Anane}},\ and\ \bibinfo {author} {\bibfnamefont
  {R.}~\bibnamefont {Lebrun}},\ }\bibfield  {title} {\bibinfo {title}
  {Antiferromagnetic magnon spintronic based on nonreciprocal and
  nondegenerated ultra-fast spin-waves in the canted antiferromagnet
  {$\alpha$}-{{Fe2O3}}},\ }\href {https://doi.org/10.1126/sciadv.adh1601}
  {\bibfield  {journal} {\bibinfo  {journal} {Sci. Adv.}\ }\textbf {\bibinfo
  {volume} {9}},\ \bibinfo {pages} {eadh1601} (\bibinfo {year}
  {2023})}\BibitemShut {NoStop}%
\bibitem [{\citenamefont {Bogdanov}\ \emph {et~al.}(2002)\citenamefont
  {Bogdanov}, \citenamefont {R{\"o}{\ss}ler}, \citenamefont {Wolf},\ and\
  \citenamefont {M{\"u}ller}}]{bogdanov2002magnetic}%
  \BibitemOpen
  \bibfield  {author} {\bibinfo {author} {\bibfnamefont {A.~N.}\ \bibnamefont
  {Bogdanov}}, \bibinfo {author} {\bibfnamefont {U.~K.}\ \bibnamefont
  {R{\"o}{\ss}ler}}, \bibinfo {author} {\bibfnamefont {M.}~\bibnamefont
  {Wolf}},\ and\ \bibinfo {author} {\bibfnamefont {K.-H.}\ \bibnamefont
  {M{\"u}ller}},\ }\bibfield  {title} {\bibinfo {title} {Magnetic structures
  and reorientation transitions in noncentrosymmetric uniaxial
  antiferromagnets},\ }\href {https://doi.org/10.1103/PhysRevB.66.214410}
  {\bibfield  {journal} {\bibinfo  {journal} {Phys. Rev. B}\ }\textbf {\bibinfo
  {volume} {66}},\ \bibinfo {pages} {214410} (\bibinfo {year}
  {2002})}\BibitemShut {NoStop}%
\bibitem [{\citenamefont {Meshcheriakova}\ \emph {et~al.}(2014)\citenamefont
  {Meshcheriakova}, \citenamefont {Chadov}, \citenamefont {Nayak},
  \citenamefont {R{\"o}{\ss}ler}, \citenamefont {K{\"u}bler}, \citenamefont
  {Andr{\'e}}, \citenamefont {Tsirlin}, \citenamefont {Kiss}, \citenamefont
  {Hausdorf}, \citenamefont {Kalache}, \citenamefont {Schnelle}, \citenamefont
  {Nicklas},\ and\ \citenamefont {Felser}}]{meshcheriakova2014large}%
  \BibitemOpen
  \bibfield  {author} {\bibinfo {author} {\bibfnamefont {O.}~\bibnamefont
  {Meshcheriakova}}, \bibinfo {author} {\bibfnamefont {S.}~\bibnamefont
  {Chadov}}, \bibinfo {author} {\bibfnamefont {A.~K.}\ \bibnamefont {Nayak}},
  \bibinfo {author} {\bibfnamefont {U.~K.}\ \bibnamefont {R{\"o}{\ss}ler}},
  \bibinfo {author} {\bibfnamefont {J.}~\bibnamefont {K{\"u}bler}}, \bibinfo
  {author} {\bibfnamefont {G.}~\bibnamefont {Andr{\'e}}}, \bibinfo {author}
  {\bibfnamefont {A.~A.}\ \bibnamefont {Tsirlin}}, \bibinfo {author}
  {\bibfnamefont {J.}~\bibnamefont {Kiss}}, \bibinfo {author} {\bibfnamefont
  {S.}~\bibnamefont {Hausdorf}}, \bibinfo {author} {\bibfnamefont
  {A.}~\bibnamefont {Kalache}}, \bibinfo {author} {\bibfnamefont
  {W.}~\bibnamefont {Schnelle}}, \bibinfo {author} {\bibfnamefont
  {M.}~\bibnamefont {Nicklas}},\ and\ \bibinfo {author} {\bibfnamefont
  {C.}~\bibnamefont {Felser}},\ }\bibfield  {title} {\bibinfo {title} {Large
  {{Noncollinearity}} and {{Spin Reorientation}} in the {{Novel}}
  \$\{{\textbackslash}mathrm\{\vphantom{\}\}}{{Mn}}\vphantom\{\}\vphantom\{\}\_\{2\}{\textbackslash}mathrm\{\vphantom\}{{RhSn}}\vphantom\{\}\$
  {{Heusler Magnet}}},\ }\href {https://doi.org/10.1103/PhysRevLett.113.087203}
  {\bibfield  {journal} {\bibinfo  {journal} {Phys. Rev. Lett.}\ }\textbf
  {\bibinfo {volume} {113}},\ \bibinfo {pages} {087203} (\bibinfo {year}
  {2014})}\BibitemShut {NoStop}%
\bibitem [{\citenamefont {Jena}\ \emph
  {et~al.}(2020{\natexlab{b}})\citenamefont {Jena}, \citenamefont {Stinshoff},
  \citenamefont {Saha}, \citenamefont {Srivastava}, \citenamefont {Ma},
  \citenamefont {Deniz}, \citenamefont {Werner}, \citenamefont {Felser},\ and\
  \citenamefont {Parkin}}]{jena2020observation}%
  \BibitemOpen
  \bibfield  {author} {\bibinfo {author} {\bibfnamefont {J.}~\bibnamefont
  {Jena}}, \bibinfo {author} {\bibfnamefont {R.}~\bibnamefont {Stinshoff}},
  \bibinfo {author} {\bibfnamefont {R.}~\bibnamefont {Saha}}, \bibinfo {author}
  {\bibfnamefont {A.~K.}\ \bibnamefont {Srivastava}}, \bibinfo {author}
  {\bibfnamefont {T.}~\bibnamefont {Ma}}, \bibinfo {author} {\bibfnamefont
  {H.}~\bibnamefont {Deniz}}, \bibinfo {author} {\bibfnamefont
  {P.}~\bibnamefont {Werner}}, \bibinfo {author} {\bibfnamefont
  {C.}~\bibnamefont {Felser}},\ and\ \bibinfo {author} {\bibfnamefont
  {S.~S.~P.}\ \bibnamefont {Parkin}},\ }\bibfield  {title} {\bibinfo {title}
  {Observation of {{Magnetic Antiskyrmions}} in the {{Low Magnetization
  Ferrimagnet Mn2Rh0}}.{{95Ir0}}.{{05Sn}}},\ }\href
  {https://doi.org/10.1021/acs.nanolett.9b02973} {\bibfield  {journal}
  {\bibinfo  {journal} {Nano Lett.}\ }\textbf {\bibinfo {volume} {20}},\
  \bibinfo {pages} {59} (\bibinfo {year} {2020}{\natexlab{b}})}\BibitemShut
  {NoStop}%
\bibitem [{\citenamefont {Jena}\ \emph {et~al.}(2024)\citenamefont {Jena},
  \citenamefont {Koraltan}, \citenamefont {Bruckner}, \citenamefont {Holst},
  \citenamefont {Tangi}, \citenamefont {Abert}, \citenamefont {Felser},
  \citenamefont {Suess},\ and\ \citenamefont {Parkin}}]{jena2024topological}%
  \BibitemOpen
  \bibfield  {author} {\bibinfo {author} {\bibfnamefont {J.}~\bibnamefont
  {Jena}}, \bibinfo {author} {\bibfnamefont {S.}~\bibnamefont {Koraltan}},
  \bibinfo {author} {\bibfnamefont {F.}~\bibnamefont {Bruckner}}, \bibinfo
  {author} {\bibfnamefont {K.}~\bibnamefont {Holst}}, \bibinfo {author}
  {\bibfnamefont {M.}~\bibnamefont {Tangi}}, \bibinfo {author} {\bibfnamefont
  {C.}~\bibnamefont {Abert}}, \bibinfo {author} {\bibfnamefont
  {C.}~\bibnamefont {Felser}}, \bibinfo {author} {\bibfnamefont
  {D.}~\bibnamefont {Suess}},\ and\ \bibinfo {author} {\bibfnamefont {S.~S.}\
  \bibnamefont {Parkin}},\ }\bibfield  {title} {\bibinfo {title} {Topological
  {{Phase Transformation}} and {{Collapse Dynamics}} of {{Spin Textures}} in a
  {{Non-Centrosymmetric D2d System}}},\ }\href
  {https://doi.org/10.1002/adfm.202403358} {\bibfield  {journal} {\bibinfo
  {journal} {Adv. Funct. Mater.}\ }\textbf {\bibinfo {volume} {34}},\ \bibinfo
  {pages} {2403358} (\bibinfo {year} {2024})}\BibitemShut {NoStop}%
\bibitem [{\citenamefont {Saha}\ \emph {et~al.}(2019)\citenamefont {Saha},
  \citenamefont {Srivastava}, \citenamefont {Ma}, \citenamefont {Jena},
  \citenamefont {Werner}, \citenamefont {Kumar}, \citenamefont {Felser},\ and\
  \citenamefont {Parkin}}]{saha2019intrinsic}%
  \BibitemOpen
  \bibfield  {author} {\bibinfo {author} {\bibfnamefont {R.}~\bibnamefont
  {Saha}}, \bibinfo {author} {\bibfnamefont {A.~K.}\ \bibnamefont
  {Srivastava}}, \bibinfo {author} {\bibfnamefont {T.}~\bibnamefont {Ma}},
  \bibinfo {author} {\bibfnamefont {J.}~\bibnamefont {Jena}}, \bibinfo {author}
  {\bibfnamefont {P.}~\bibnamefont {Werner}}, \bibinfo {author} {\bibfnamefont
  {V.}~\bibnamefont {Kumar}}, \bibinfo {author} {\bibfnamefont
  {C.}~\bibnamefont {Felser}},\ and\ \bibinfo {author} {\bibfnamefont
  {S.~S.~P.}\ \bibnamefont {Parkin}},\ }\bibfield  {title} {\bibinfo {title}
  {Intrinsic stability of magnetic anti-skyrmions in the tetragonal inverse
  {{Heusler}} compound {{Mn1}}.{{4Pt0}}.{{9Pd0}}.{{1Sn}}},\ }\href
  {https://doi.org/10.1038/s41467-019-13323-x} {\bibfield  {journal} {\bibinfo
  {journal} {Nat. Commun.}\ }\textbf {\bibinfo {volume} {10}},\ \bibinfo
  {pages} {5305} (\bibinfo {year} {2019})}\BibitemShut {NoStop}%
\bibitem [{\citenamefont {Ma}\ \emph {et~al.}(2020)\citenamefont {Ma},
  \citenamefont {Sharma}, \citenamefont {Saha}, \citenamefont {Srivastava},
  \citenamefont {Werner}, \citenamefont {Vir}, \citenamefont {Kumar},
  \citenamefont {Felser},\ and\ \citenamefont {Parkin}}]{ma2020tunable}%
  \BibitemOpen
  \bibfield  {author} {\bibinfo {author} {\bibfnamefont {T.}~\bibnamefont
  {Ma}}, \bibinfo {author} {\bibfnamefont {A.~K.}\ \bibnamefont {Sharma}},
  \bibinfo {author} {\bibfnamefont {R.}~\bibnamefont {Saha}}, \bibinfo {author}
  {\bibfnamefont {A.~K.}\ \bibnamefont {Srivastava}}, \bibinfo {author}
  {\bibfnamefont {P.}~\bibnamefont {Werner}}, \bibinfo {author} {\bibfnamefont
  {P.}~\bibnamefont {Vir}}, \bibinfo {author} {\bibfnamefont {V.}~\bibnamefont
  {Kumar}}, \bibinfo {author} {\bibfnamefont {C.}~\bibnamefont {Felser}},\ and\
  \bibinfo {author} {\bibfnamefont {S.~S.~P.}\ \bibnamefont {Parkin}},\
  }\bibfield  {title} {\bibinfo {title} {Tunable {{Magnetic Antiskyrmion Size}}
  and {{Helical Period}} from {{Nanometers}} to {{Micrometers}} in a {{D2d
  Heusler Compound}}},\ }\href {https://doi.org/10.1002/adma.202002043}
  {\bibfield  {journal} {\bibinfo  {journal} {Adv. Mater.}\ }\textbf {\bibinfo
  {volume} {32}},\ \bibinfo {pages} {2002043} (\bibinfo {year}
  {2020})}\BibitemShut {NoStop}%
\bibitem [{\citenamefont {Yu}\ \emph {et~al.}(2011)\citenamefont {Yu},
  \citenamefont {Kanazawa}, \citenamefont {Onose}, \citenamefont {Kimoto},
  \citenamefont {Zhang}, \citenamefont {Ishiwata}, \citenamefont {Matsui},\
  and\ \citenamefont {Tokura}}]{yu2011near}%
  \BibitemOpen
  \bibfield  {author} {\bibinfo {author} {\bibfnamefont {X.}~\bibnamefont
  {Yu}}, \bibinfo {author} {\bibfnamefont {N.}~\bibnamefont {Kanazawa}},
  \bibinfo {author} {\bibfnamefont {Y.}~\bibnamefont {Onose}}, \bibinfo
  {author} {\bibfnamefont {K.}~\bibnamefont {Kimoto}}, \bibinfo {author}
  {\bibfnamefont {W.}~\bibnamefont {Zhang}}, \bibinfo {author} {\bibfnamefont
  {S.}~\bibnamefont {Ishiwata}}, \bibinfo {author} {\bibfnamefont
  {Y.}~\bibnamefont {Matsui}},\ and\ \bibinfo {author} {\bibfnamefont
  {Y.}~\bibnamefont {Tokura}},\ }\bibfield  {title} {\bibinfo {title} {Near
  room-temperature formation of a skyrmion crystal in thin-films of the
  helimagnet fege},\ }\href@noop {} {\bibfield  {journal} {\bibinfo  {journal}
  {Nat. Mater.}\ }\textbf {\bibinfo {volume} {10}},\ \bibinfo {pages} {106}
  (\bibinfo {year} {2011})}\BibitemShut {NoStop}%
\bibitem [{\citenamefont {Jena}\ \emph {et~al.}(2022)\citenamefont {Jena},
  \citenamefont {G{\"o}bel}, \citenamefont {Hirosawa}, \citenamefont
  {D{\'i}az}, \citenamefont {Wolf}, \citenamefont {Hinokihara}, \citenamefont
  {Kumar}, \citenamefont {Mertig}, \citenamefont {Felser}, \citenamefont
  {Lubk}, \citenamefont {Loss},\ and\ \citenamefont
  {Parkin}}]{jena2022observation}%
  \BibitemOpen
  \bibfield  {author} {\bibinfo {author} {\bibfnamefont {J.}~\bibnamefont
  {Jena}}, \bibinfo {author} {\bibfnamefont {B.}~\bibnamefont {G{\"o}bel}},
  \bibinfo {author} {\bibfnamefont {T.}~\bibnamefont {Hirosawa}}, \bibinfo
  {author} {\bibfnamefont {S.~A.}\ \bibnamefont {D{\'i}az}}, \bibinfo {author}
  {\bibfnamefont {D.}~\bibnamefont {Wolf}}, \bibinfo {author} {\bibfnamefont
  {T.}~\bibnamefont {Hinokihara}}, \bibinfo {author} {\bibfnamefont
  {V.}~\bibnamefont {Kumar}}, \bibinfo {author} {\bibfnamefont
  {I.}~\bibnamefont {Mertig}}, \bibinfo {author} {\bibfnamefont
  {C.}~\bibnamefont {Felser}}, \bibinfo {author} {\bibfnamefont
  {A.}~\bibnamefont {Lubk}}, \bibinfo {author} {\bibfnamefont {D.}~\bibnamefont
  {Loss}},\ and\ \bibinfo {author} {\bibfnamefont {S.~S.~P.}\ \bibnamefont
  {Parkin}},\ }\bibfield  {title} {\bibinfo {title} {Observation of fractional
  spin textures in a {{Heusler}} material},\ }\href
  {https://doi.org/10.1038/s41467-022-29991-1} {\bibfield  {journal} {\bibinfo
  {journal} {Nat. Commun.}\ }\textbf {\bibinfo {volume} {13}},\ \bibinfo
  {pages} {2348} (\bibinfo {year} {2022})}\BibitemShut {NoStop}%
\bibitem [{\citenamefont {Hirosawa}\ \emph {et~al.}(2020)\citenamefont
  {Hirosawa}, \citenamefont {D{\'i}az}, \citenamefont {Klinovaja},\ and\
  \citenamefont {Loss}}]{hirosawa2020magnonic}%
  \BibitemOpen
  \bibfield  {author} {\bibinfo {author} {\bibfnamefont {T.}~\bibnamefont
  {Hirosawa}}, \bibinfo {author} {\bibfnamefont {S.~A.}\ \bibnamefont
  {D{\'i}az}}, \bibinfo {author} {\bibfnamefont {J.}~\bibnamefont
  {Klinovaja}},\ and\ \bibinfo {author} {\bibfnamefont {D.}~\bibnamefont
  {Loss}},\ }\bibfield  {title} {\bibinfo {title} {Magnonic {{Quadrupole
  Topological Insulator}} in {{Antiskyrmion Crystals}}},\ }\href
  {https://doi.org/10.1103/PhysRevLett.125.207204} {\bibfield  {journal}
  {\bibinfo  {journal} {Phys. Rev. Lett.}\ }\textbf {\bibinfo {volume} {125}},\
  \bibinfo {pages} {207204} (\bibinfo {year} {2020})}\BibitemShut {NoStop}%
\bibitem [{\citenamefont {Karube}\ \emph {et~al.}(2021)\citenamefont {Karube},
  \citenamefont {Peng}, \citenamefont {Masell}, \citenamefont {Yu},
  \citenamefont {Kagawa}, \citenamefont {Tokura},\ and\ \citenamefont
  {Taguchi}}]{karube2021room}%
  \BibitemOpen
  \bibfield  {author} {\bibinfo {author} {\bibfnamefont {K.}~\bibnamefont
  {Karube}}, \bibinfo {author} {\bibfnamefont {L.}~\bibnamefont {Peng}},
  \bibinfo {author} {\bibfnamefont {J.}~\bibnamefont {Masell}}, \bibinfo
  {author} {\bibfnamefont {X.}~\bibnamefont {Yu}}, \bibinfo {author}
  {\bibfnamefont {F.}~\bibnamefont {Kagawa}}, \bibinfo {author} {\bibfnamefont
  {Y.}~\bibnamefont {Tokura}},\ and\ \bibinfo {author} {\bibfnamefont
  {Y.}~\bibnamefont {Taguchi}},\ }\bibfield  {title} {\bibinfo {title}
  {Room-temperature antiskyrmions and sawtooth surface textures in a
  non-centrosymmetric magnet with s 4 symmetry},\ }\href@noop {} {\bibfield
  {journal} {\bibinfo  {journal} {Nat. Mater.}\ }\textbf {\bibinfo {volume}
  {20}},\ \bibinfo {pages} {335} (\bibinfo {year} {2021})}\BibitemShut
  {NoStop}%
\bibitem [{\citenamefont {Tang}\ \emph {et~al.}(2021)\citenamefont {Tang},
  \citenamefont {Wu}, \citenamefont {Wang}, \citenamefont {Kong}, \citenamefont
  {Lv}, \citenamefont {Wei}, \citenamefont {Zang}, \citenamefont {Tian},\ and\
  \citenamefont {Du}}]{tang2021magnetic}%
  \BibitemOpen
  \bibfield  {author} {\bibinfo {author} {\bibfnamefont {J.}~\bibnamefont
  {Tang}}, \bibinfo {author} {\bibfnamefont {Y.}~\bibnamefont {Wu}}, \bibinfo
  {author} {\bibfnamefont {W.}~\bibnamefont {Wang}}, \bibinfo {author}
  {\bibfnamefont {L.}~\bibnamefont {Kong}}, \bibinfo {author} {\bibfnamefont
  {B.}~\bibnamefont {Lv}}, \bibinfo {author} {\bibfnamefont {W.}~\bibnamefont
  {Wei}}, \bibinfo {author} {\bibfnamefont {J.}~\bibnamefont {Zang}}, \bibinfo
  {author} {\bibfnamefont {M.}~\bibnamefont {Tian}},\ and\ \bibinfo {author}
  {\bibfnamefont {H.}~\bibnamefont {Du}},\ }\bibfield  {title} {\bibinfo
  {title} {Magnetic skyrmion bundles and their current-driven dynamics},\
  }\href {https://doi.org/10.1038/s41565-021-00954-9} {\bibfield  {journal}
  {\bibinfo  {journal} {Nat. Nanotechnol.}\ }\textbf {\bibinfo {volume} {16}},\
  \bibinfo {pages} {1086} (\bibinfo {year} {2021})}\BibitemShut {NoStop}%
\bibitem [{\citenamefont {Hassan}\ \emph {et~al.}(2024)\citenamefont {Hassan},
  \citenamefont {Koraltan}, \citenamefont {Ullrich}, \citenamefont {Bruckner},
  \citenamefont {Serha}, \citenamefont {Levchenko}, \citenamefont {Varvaro},
  \citenamefont {Kiselev}, \citenamefont {Heigl}, \citenamefont {Abert},
  \citenamefont {Suess},\ and\ \citenamefont {Albrecht}}]{hassan2024dipolar}%
  \BibitemOpen
  \bibfield  {author} {\bibinfo {author} {\bibfnamefont {M.}~\bibnamefont
  {Hassan}}, \bibinfo {author} {\bibfnamefont {S.}~\bibnamefont {Koraltan}},
  \bibinfo {author} {\bibfnamefont {A.}~\bibnamefont {Ullrich}}, \bibinfo
  {author} {\bibfnamefont {F.}~\bibnamefont {Bruckner}}, \bibinfo {author}
  {\bibfnamefont {R.~O.}\ \bibnamefont {Serha}}, \bibinfo {author}
  {\bibfnamefont {K.~V.}\ \bibnamefont {Levchenko}}, \bibinfo {author}
  {\bibfnamefont {G.}~\bibnamefont {Varvaro}}, \bibinfo {author} {\bibfnamefont
  {N.~S.}\ \bibnamefont {Kiselev}}, \bibinfo {author} {\bibfnamefont
  {M.}~\bibnamefont {Heigl}}, \bibinfo {author} {\bibfnamefont
  {C.}~\bibnamefont {Abert}}, \bibinfo {author} {\bibfnamefont
  {D.}~\bibnamefont {Suess}},\ and\ \bibinfo {author} {\bibfnamefont
  {M.}~\bibnamefont {Albrecht}},\ }\bibfield  {title} {\bibinfo {title}
  {Dipolar skyrmions and antiskyrmions of arbitrary topological charge at room
  temperature},\ }\href {https://doi.org/10.1038/s41567-023-02358-z} {\bibfield
   {journal} {\bibinfo  {journal} {Nat. Phys.}\ }\textbf {\bibinfo {volume}
  {20}},\ \bibinfo {pages} {615} (\bibinfo {year} {2024})}\BibitemShut
  {NoStop}%
\bibitem [{\citenamefont {Sharma}\ \emph {et~al.}(2021)\citenamefont {Sharma},
  \citenamefont {Jena}, \citenamefont {Rana}, \citenamefont {Markou},
  \citenamefont {Meyerheim}, \citenamefont {Mohseni}, \citenamefont
  {Srivastava}, \citenamefont {Kostanoskiy}, \citenamefont {Felser},\ and\
  \citenamefont {Parkin}}]{sharma2021nanoscale}%
  \BibitemOpen
  \bibfield  {author} {\bibinfo {author} {\bibfnamefont {A.~K.}\ \bibnamefont
  {Sharma}}, \bibinfo {author} {\bibfnamefont {J.}~\bibnamefont {Jena}},
  \bibinfo {author} {\bibfnamefont {K.~G.}\ \bibnamefont {Rana}}, \bibinfo
  {author} {\bibfnamefont {A.}~\bibnamefont {Markou}}, \bibinfo {author}
  {\bibfnamefont {H.~L.}\ \bibnamefont {Meyerheim}}, \bibinfo {author}
  {\bibfnamefont {K.}~\bibnamefont {Mohseni}}, \bibinfo {author} {\bibfnamefont
  {A.~K.}\ \bibnamefont {Srivastava}}, \bibinfo {author} {\bibfnamefont
  {I.}~\bibnamefont {Kostanoskiy}}, \bibinfo {author} {\bibfnamefont
  {C.}~\bibnamefont {Felser}},\ and\ \bibinfo {author} {\bibfnamefont
  {S.~S.~P.}\ \bibnamefont {Parkin}},\ }\bibfield  {title} {\bibinfo {title}
  {Nanoscale {{Noncollinear Spin Textures}} in {{Thin Films}} of a {{D2d
  Heusler Compound}}},\ }\href {https://doi.org/10.1002/adma.202101323}
  {\bibfield  {journal} {\bibinfo  {journal} {Adv. Mater.}\ }\textbf {\bibinfo
  {volume} {33}},\ \bibinfo {pages} {2101323} (\bibinfo {year}
  {2021})}\BibitemShut {NoStop}%
\bibitem [{\citenamefont {Jena}\ \emph
  {et~al.}(2020{\natexlab{c}})\citenamefont {Jena}, \citenamefont {G{\"o}bel},
  \citenamefont {Kumar}, \citenamefont {Mertig}, \citenamefont {Felser},\ and\
  \citenamefont {Parkin}}]{jena2020evolution}%
  \BibitemOpen
  \bibfield  {author} {\bibinfo {author} {\bibfnamefont {J.}~\bibnamefont
  {Jena}}, \bibinfo {author} {\bibfnamefont {B.}~\bibnamefont {G{\"o}bel}},
  \bibinfo {author} {\bibfnamefont {V.}~\bibnamefont {Kumar}}, \bibinfo
  {author} {\bibfnamefont {I.}~\bibnamefont {Mertig}}, \bibinfo {author}
  {\bibfnamefont {C.}~\bibnamefont {Felser}},\ and\ \bibinfo {author}
  {\bibfnamefont {S.}~\bibnamefont {Parkin}},\ }\bibfield  {title} {\bibinfo
  {title} {Evolution and competition between chiral spin textures in
  nanostripes with {{D2d}} symmetry},\ }\href
  {https://doi.org/10.1126/sciadv.abc0723} {\bibfield  {journal} {\bibinfo
  {journal} {Sci. Adv.}\ }\textbf {\bibinfo {volume} {6}},\ \bibinfo {pages}
  {eabc0723} (\bibinfo {year} {2020}{\natexlab{c}})}\BibitemShut {NoStop}%
\bibitem [{\citenamefont {Yang}\ \emph {et~al.}(2015)\citenamefont {Yang},
  \citenamefont {Ryu},\ and\ \citenamefont {Parkin}}]{yang2015domainwall}%
  \BibitemOpen
  \bibfield  {author} {\bibinfo {author} {\bibfnamefont {S.-H.}\ \bibnamefont
  {Yang}}, \bibinfo {author} {\bibfnamefont {K.-S.}\ \bibnamefont {Ryu}},\ and\
  \bibinfo {author} {\bibfnamefont {S.}~\bibnamefont {Parkin}},\ }\bibfield
  {title} {\bibinfo {title} {Domain-wall velocities of up to 750 m s-1 driven
  by exchange-coupling torque in synthetic antiferromagnets},\ }\href
  {https://doi.org/10.1038/nnano.2014.324} {\bibfield  {journal} {\bibinfo
  {journal} {Nature Nanotech}\ }\textbf {\bibinfo {volume} {10}},\ \bibinfo
  {pages} {221} (\bibinfo {year} {2015})}\BibitemShut {NoStop}%
\bibitem [{\citenamefont {Jeon}\ \emph {et~al.}(2024)\citenamefont {Jeon},
  \citenamefont {Migliorini}, \citenamefont {Yoon}, \citenamefont {Jeong},\
  and\ \citenamefont {Parkin}}]{jeon2024multicore}%
  \BibitemOpen
  \bibfield  {author} {\bibinfo {author} {\bibfnamefont {J.-C.}\ \bibnamefont
  {Jeon}}, \bibinfo {author} {\bibfnamefont {A.}~\bibnamefont {Migliorini}},
  \bibinfo {author} {\bibfnamefont {J.}~\bibnamefont {Yoon}}, \bibinfo {author}
  {\bibfnamefont {J.}~\bibnamefont {Jeong}},\ and\ \bibinfo {author}
  {\bibfnamefont {S.~S.~P.}\ \bibnamefont {Parkin}},\ }\bibfield  {title}
  {\bibinfo {title} {Multicore memristor from electrically readable nanoscopic
  racetracks},\ }\href {https://doi.org/10.1126/science.adh3419} {\bibfield
  {journal} {\bibinfo  {journal} {Science}\ }\textbf {\bibinfo {volume}
  {386}},\ \bibinfo {pages} {315} (\bibinfo {year} {2024})}\BibitemShut
  {NoStop}%
\bibitem [{\citenamefont {Junquera}\ \emph {et~al.}(2023)\citenamefont
  {Junquera}, \citenamefont {Nahas}, \citenamefont {Prokhorenko}, \citenamefont
  {Bellaiche}, \citenamefont {\'I\~niguez}, \citenamefont {Schlom},
  \citenamefont {Chen}, \citenamefont {Salahuddin}, \citenamefont {Muller},
  \citenamefont {Martin},\ and\ \citenamefont {Ramesh}}]{Junquera-23}%
  \BibitemOpen
  \bibfield  {author} {\bibinfo {author} {\bibfnamefont {J.}~\bibnamefont
  {Junquera}}, \bibinfo {author} {\bibfnamefont {Y.}~\bibnamefont {Nahas}},
  \bibinfo {author} {\bibfnamefont {S.}~\bibnamefont {Prokhorenko}}, \bibinfo
  {author} {\bibfnamefont {L.}~\bibnamefont {Bellaiche}}, \bibinfo {author}
  {\bibfnamefont {J.}~\bibnamefont {\'I\~niguez}}, \bibinfo {author}
  {\bibfnamefont {D.~G.}\ \bibnamefont {Schlom}}, \bibinfo {author}
  {\bibfnamefont {L.-Q.}\ \bibnamefont {Chen}}, \bibinfo {author}
  {\bibfnamefont {S.}~\bibnamefont {Salahuddin}}, \bibinfo {author}
  {\bibfnamefont {D.~A.}\ \bibnamefont {Muller}}, \bibinfo {author}
  {\bibfnamefont {L.~W.}\ \bibnamefont {Martin}},\ and\ \bibinfo {author}
  {\bibfnamefont {R.}~\bibnamefont {Ramesh}},\ }\bibfield  {title} {\bibinfo
  {title} {Topological phases in polar oxide nanostructures},\ }\href
  {https://doi.org/10.1103/RevModPhys.95.025001} {\bibfield  {journal}
  {\bibinfo  {journal} {Rev. Mod. Phys.}\ }\textbf {\bibinfo {volume} {95}},\
  \bibinfo {pages} {025001} (\bibinfo {year} {2023})}\BibitemShut {NoStop}%
\bibitem [{\citenamefont {S{\'a}nchez-Santolino}\ \emph
  {et~al.}(2024)\citenamefont {S{\'a}nchez-Santolino}, \citenamefont {Rouco},
  \citenamefont {Puebla}, \citenamefont {Aramberri}, \citenamefont {Zamora},
  \citenamefont {Cabero}, \citenamefont {Cuellar}, \citenamefont {Munuera},
  \citenamefont {Mompean}, \citenamefont {Garcia-Hernandez}, \citenamefont
  {Castellanos-Gomez}, \citenamefont {{\'I}{\~{n}}iguez}, \citenamefont
  {Leon},\ and\ \citenamefont {Santamaria}}]{Santolino-24}%
  \BibitemOpen
  \bibfield  {author} {\bibinfo {author} {\bibfnamefont {G.}~\bibnamefont
  {S{\'a}nchez-Santolino}}, \bibinfo {author} {\bibfnamefont {V.}~\bibnamefont
  {Rouco}}, \bibinfo {author} {\bibfnamefont {S.}~\bibnamefont {Puebla}},
  \bibinfo {author} {\bibfnamefont {H.}~\bibnamefont {Aramberri}}, \bibinfo
  {author} {\bibfnamefont {V.}~\bibnamefont {Zamora}}, \bibinfo {author}
  {\bibfnamefont {M.}~\bibnamefont {Cabero}}, \bibinfo {author} {\bibfnamefont
  {F.~A.}\ \bibnamefont {Cuellar}}, \bibinfo {author} {\bibfnamefont
  {C.}~\bibnamefont {Munuera}}, \bibinfo {author} {\bibfnamefont
  {F.}~\bibnamefont {Mompean}}, \bibinfo {author} {\bibfnamefont
  {M.}~\bibnamefont {Garcia-Hernandez}}, \bibinfo {author} {\bibfnamefont
  {A.}~\bibnamefont {Castellanos-Gomez}}, \bibinfo {author} {\bibfnamefont
  {J.}~\bibnamefont {{\'I}{\~{n}}iguez}}, \bibinfo {author} {\bibfnamefont
  {C.}~\bibnamefont {Leon}},\ and\ \bibinfo {author} {\bibfnamefont
  {J.}~\bibnamefont {Santamaria}},\ }\bibfield  {title} {\bibinfo {title} {A
  2$\mathrm{D}$ ferroelectric vortex pattern in twisted
  $\mathrm{B}$a$\mathrm{T}$i$\mathrm{O}_{3}$ freestanding layers},\ }\href
  {https://doi.org/10.1038/s41586-023-06978-6} {\bibfield  {journal} {\bibinfo
  {journal} {Nature}\ }\textbf {\bibinfo {volume} {626}},\ \bibinfo {pages}
  {529} (\bibinfo {year} {2024})}\BibitemShut {NoStop}%
\bibitem [{\citenamefont {Bennett}\ \emph
  {et~al.}(2023{\natexlab{a}})\citenamefont {Bennett}, \citenamefont
  {Chaudhary}, \citenamefont {Slager}, \citenamefont {Bousquet},\ and\
  \citenamefont {Ghosez}}]{Bennett-23}%
  \BibitemOpen
  \bibfield  {author} {\bibinfo {author} {\bibfnamefont {D.}~\bibnamefont
  {Bennett}}, \bibinfo {author} {\bibfnamefont {G.}~\bibnamefont {Chaudhary}},
  \bibinfo {author} {\bibfnamefont {R.-J.}\ \bibnamefont {Slager}}, \bibinfo
  {author} {\bibfnamefont {E.}~\bibnamefont {Bousquet}},\ and\ \bibinfo
  {author} {\bibfnamefont {P.}~\bibnamefont {Ghosez}},\ }\bibfield  {title}
  {\bibinfo {title} {Polar meron-antimeron networks in strained and twisted
  bilayers},\ }\href {https://doi.org/10.1038/s41467-023-37337-8} {\bibfield
  {journal} {\bibinfo  {journal} {Nat. Commun.}\ }\textbf {\bibinfo {volume}
  {14}},\ \bibinfo {pages} {1629} (\bibinfo {year}
  {2023}{\natexlab{a}})}\BibitemShut {NoStop}%
\bibitem [{\citenamefont {Pan}\ \emph {et~al.}(2025)\citenamefont {Pan},
  \citenamefont {Li}, \citenamefont {Yang}, \citenamefont {Niu}, \citenamefont
  {Bian}, \citenamefont {Liu}, \citenamefont {Chen}, \citenamefont {Dong},
  \citenamefont {Wang}, \citenamefont {Zhou}, \citenamefont {Zhou},
  \citenamefont {Luo}, \citenamefont {Chu}, \citenamefont {Lin}, \citenamefont
  {Li},\ and\ \citenamefont {Liu}}]{Pan-25}%
  \BibitemOpen
  \bibfield  {author} {\bibinfo {author} {\bibfnamefont {E.}~\bibnamefont
  {Pan}}, \bibinfo {author} {\bibfnamefont {Z.}~\bibnamefont {Li}}, \bibinfo
  {author} {\bibfnamefont {F.}~\bibnamefont {Yang}}, \bibinfo {author}
  {\bibfnamefont {K.}~\bibnamefont {Niu}}, \bibinfo {author} {\bibfnamefont
  {R.}~\bibnamefont {Bian}}, \bibinfo {author} {\bibfnamefont {Q.}~\bibnamefont
  {Liu}}, \bibinfo {author} {\bibfnamefont {J.}~\bibnamefont {Chen}}, \bibinfo
  {author} {\bibfnamefont {B.}~\bibnamefont {Dong}}, \bibinfo {author}
  {\bibfnamefont {R.}~\bibnamefont {Wang}}, \bibinfo {author} {\bibfnamefont
  {T.}~\bibnamefont {Zhou}}, \bibinfo {author} {\bibfnamefont {A.}~\bibnamefont
  {Zhou}}, \bibinfo {author} {\bibfnamefont {X.}~\bibnamefont {Luo}}, \bibinfo
  {author} {\bibfnamefont {J.}~\bibnamefont {Chu}}, \bibinfo {author}
  {\bibfnamefont {J.}~\bibnamefont {Lin}}, \bibinfo {author} {\bibfnamefont
  {W.}~\bibnamefont {Li}},\ and\ \bibinfo {author} {\bibfnamefont
  {F.}~\bibnamefont {Liu}},\ }\bibfield  {title} {\bibinfo {title} {Observation
  and manipulation of two-dimensional topological polar texture confined in
  moir{\'e} interface},\ }\href {https://doi.org/10.1038/s41467-025-58105-w}
  {\bibfield  {journal} {\bibinfo  {journal} {Nat. Commun.}\ }\textbf {\bibinfo
  {volume} {16}},\ \bibinfo {pages} {3026} (\bibinfo {year}
  {2025})}\BibitemShut {NoStop}%
\bibitem [{\citenamefont {Liu}\ \emph {et~al.}(2025{\natexlab{b}})\citenamefont
  {Liu}, \citenamefont {Zhang}, \citenamefont {Shapovalov}, \citenamefont
  {Niu}, \citenamefont {Cairney}, \citenamefont {Liao}, \citenamefont
  {Roleder}, \citenamefont {Majchrowski}, \citenamefont {Arbiol}, \citenamefont
  {Ghosez},\ and\ \citenamefont {Catalan}}]{Liu-25}%
  \BibitemOpen
  \bibfield  {author} {\bibinfo {author} {\bibfnamefont {Y.}~\bibnamefont
  {Liu}}, \bibinfo {author} {\bibfnamefont {H.}~\bibnamefont {Zhang}}, \bibinfo
  {author} {\bibfnamefont {K.}~\bibnamefont {Shapovalov}}, \bibinfo {author}
  {\bibfnamefont {R.}~\bibnamefont {Niu}}, \bibinfo {author} {\bibfnamefont
  {J.~M.}\ \bibnamefont {Cairney}}, \bibinfo {author} {\bibfnamefont
  {X.}~\bibnamefont {Liao}}, \bibinfo {author} {\bibfnamefont {K.}~\bibnamefont
  {Roleder}}, \bibinfo {author} {\bibfnamefont {A.}~\bibnamefont
  {Majchrowski}}, \bibinfo {author} {\bibfnamefont {J.}~\bibnamefont {Arbiol}},
  \bibinfo {author} {\bibfnamefont {P.}~\bibnamefont {Ghosez}},\ and\ \bibinfo
  {author} {\bibfnamefont {G.}~\bibnamefont {Catalan}},\ }\bibfield  {title}
  {\bibinfo {title} {Vortices and antivortices in antiferroelectric
  $\mathrm{P}$b$\mathrm{Z}$r$\mathrm{O}_{3}$},\ }\bibfield  {journal} {\bibinfo
   {journal} {Nat. Mater.}\ }\href {https://doi.org/10.1038/s41563-025-02245-3}
  {10.1038/s41563-025-02245-3} (\bibinfo {year}
  {2025}{\natexlab{b}})\BibitemShut {NoStop}%
\bibitem [{\citenamefont {Bennett}\ \emph
  {et~al.}(2023{\natexlab{b}})\citenamefont {Bennett}, \citenamefont
  {Jankowski}, \citenamefont {Chaudhary}, \citenamefont {Kaxiras},\ and\
  \citenamefont {Slager}}]{Bennett-23.2}%
  \BibitemOpen
  \bibfield  {author} {\bibinfo {author} {\bibfnamefont {D.}~\bibnamefont
  {Bennett}}, \bibinfo {author} {\bibfnamefont {W.~J.}\ \bibnamefont
  {Jankowski}}, \bibinfo {author} {\bibfnamefont {G.}~\bibnamefont
  {Chaudhary}}, \bibinfo {author} {\bibfnamefont {E.}~\bibnamefont {Kaxiras}},\
  and\ \bibinfo {author} {\bibfnamefont {R.-J.}\ \bibnamefont {Slager}},\
  }\bibfield  {title} {\bibinfo {title} {Theory of polarization textures in
  crystal supercells},\ }\href
  {https://doi.org/10.1103/PhysRevResearch.5.033216} {\bibfield  {journal}
  {\bibinfo  {journal} {Phys. Rev. Res.}\ }\textbf {\bibinfo {volume} {5}},\
  \bibinfo {pages} {033216} (\bibinfo {year} {2023}{\natexlab{b}})}\BibitemShut
  {NoStop}%
\bibitem [{\citenamefont {Bennett}\ and\ \citenamefont
  {Ghosez}(2024)}]{Bennett-24}%
  \BibitemOpen
  \bibfield  {author} {\bibinfo {author} {\bibfnamefont {D.}~\bibnamefont
  {Bennett}}\ and\ \bibinfo {author} {\bibfnamefont {P.}~\bibnamefont
  {Ghosez}},\ }\bibfield  {title} {\bibinfo {title} {Asymmetric dynamical
  charges in two-dimensional ferroelectrics},\ }\href
  {https://doi.org/10.1103/PhysRevB.110.L041101} {\bibfield  {journal}
  {\bibinfo  {journal} {Phys. Rev. B}\ }\textbf {\bibinfo {volume} {110}},\
  \bibinfo {pages} {L041101} (\bibinfo {year} {2024})}\BibitemShut {NoStop}%
\bibitem [{\citenamefont {Aramberri}\ and\ \citenamefont {\'I\~niguez
  Gonz\'alez}(2024)}]{Aramberri-24}%
  \BibitemOpen
  \bibfield  {author} {\bibinfo {author} {\bibfnamefont {H.}~\bibnamefont
  {Aramberri}}\ and\ \bibinfo {author} {\bibfnamefont {J.}~\bibnamefont
  {\'I\~niguez Gonz\'alez}},\ }\bibfield  {title} {\bibinfo {title} {Brownian
  electric bubble quasiparticles},\ }\href
  {https://doi.org/10.1103/PhysRevLett.132.136801} {\bibfield  {journal}
  {\bibinfo  {journal} {Phys. Rev. Lett.}\ }\textbf {\bibinfo {volume} {132}},\
  \bibinfo {pages} {136801} (\bibinfo {year} {2024})}\BibitemShut {NoStop}%
\bibitem [{\citenamefont {Gao}\ \emph {et~al.}(2024)\citenamefont {Gao},
  \citenamefont {Prokhorenko}, \citenamefont {Nahas},\ and\ \citenamefont
  {Bellaiche}}]{Gao-24}%
  \BibitemOpen
  \bibfield  {author} {\bibinfo {author} {\bibfnamefont {L.}~\bibnamefont
  {Gao}}, \bibinfo {author} {\bibfnamefont {S.}~\bibnamefont {Prokhorenko}},
  \bibinfo {author} {\bibfnamefont {Y.}~\bibnamefont {Nahas}},\ and\ \bibinfo
  {author} {\bibfnamefont {L.}~\bibnamefont {Bellaiche}},\ }\bibfield  {title}
  {\bibinfo {title} {Dynamical control of topology in polar skyrmions via
  twisted light},\ }\href {https://doi.org/10.1103/PhysRevLett.132.026902}
  {\bibfield  {journal} {\bibinfo  {journal} {Phys. Rev. Lett.}\ }\textbf
  {\bibinfo {volume} {132}},\ \bibinfo {pages} {026902} (\bibinfo {year}
  {2024})}\BibitemShut {NoStop}%
\bibitem [{\citenamefont {Zatterin}\ \emph {et~al.}(2024)\citenamefont
  {Zatterin}, \citenamefont {Ondrejkovic}, \citenamefont {Bastogne},
  \citenamefont {Lichtensteiger}, \citenamefont {Tovaglieri}, \citenamefont
  {Chaney}, \citenamefont {Sasani}, \citenamefont {Sch\"ulli}, \citenamefont
  {Bosak}, \citenamefont {Leake}, \citenamefont {Zubko}, \citenamefont
  {Ghosez}, \citenamefont {Hlinka}, \citenamefont {Triscone},\ and\
  \citenamefont {Hadjimichael}}]{Zatterin-24}%
  \BibitemOpen
  \bibfield  {author} {\bibinfo {author} {\bibfnamefont {E.}~\bibnamefont
  {Zatterin}}, \bibinfo {author} {\bibfnamefont {P.}~\bibnamefont
  {Ondrejkovic}}, \bibinfo {author} {\bibfnamefont {L.}~\bibnamefont
  {Bastogne}}, \bibinfo {author} {\bibfnamefont {C.}~\bibnamefont
  {Lichtensteiger}}, \bibinfo {author} {\bibfnamefont {L.}~\bibnamefont
  {Tovaglieri}}, \bibinfo {author} {\bibfnamefont {D.~A.}\ \bibnamefont
  {Chaney}}, \bibinfo {author} {\bibfnamefont {A.}~\bibnamefont {Sasani}},
  \bibinfo {author} {\bibfnamefont {T.}~\bibnamefont {Sch\"ulli}}, \bibinfo
  {author} {\bibfnamefont {A.}~\bibnamefont {Bosak}}, \bibinfo {author}
  {\bibfnamefont {S.}~\bibnamefont {Leake}}, \bibinfo {author} {\bibfnamefont
  {P.}~\bibnamefont {Zubko}}, \bibinfo {author} {\bibfnamefont
  {P.}~\bibnamefont {Ghosez}}, \bibinfo {author} {\bibfnamefont
  {J.}~\bibnamefont {Hlinka}}, \bibinfo {author} {\bibfnamefont {J.-M.}\
  \bibnamefont {Triscone}},\ and\ \bibinfo {author} {\bibfnamefont
  {M.}~\bibnamefont {Hadjimichael}},\ }\bibfield  {title} {\bibinfo {title}
  {Assessing the ubiquity of $\mathrm{B}$loch domain walls in ferroelectric
  lead titanate superlattices},\ }\href
  {https://doi.org/10.1103/PhysRevX.14.041052} {\bibfield  {journal} {\bibinfo
  {journal} {Phys. Rev. X}\ }\textbf {\bibinfo {volume} {14}},\ \bibinfo
  {pages} {041052} (\bibinfo {year} {2024})}\BibitemShut {NoStop}%
\bibitem [{\citenamefont {Ma}\ \emph {et~al.}(2025{\natexlab{a}})\citenamefont
  {Ma}, \citenamefont {Chen}, \citenamefont {He}, \citenamefont {Yu},
  \citenamefont {Prokhorenko}, \citenamefont {Wen}, \citenamefont {Zhong},
  \citenamefont {{\'I}{\~{n}}iguez-Gonz{\'a}lez}, \citenamefont {Bellaiche},
  \citenamefont {Wu},\ and\ \citenamefont {Yang}}]{Ma-25}%
  \BibitemOpen
  \bibfield  {author} {\bibinfo {author} {\bibfnamefont {X.}~\bibnamefont
  {Ma}}, \bibinfo {author} {\bibfnamefont {H.}~\bibnamefont {Chen}}, \bibinfo
  {author} {\bibfnamefont {R.}~\bibnamefont {He}}, \bibinfo {author}
  {\bibfnamefont {Z.}~\bibnamefont {Yu}}, \bibinfo {author} {\bibfnamefont
  {S.}~\bibnamefont {Prokhorenko}}, \bibinfo {author} {\bibfnamefont
  {Z.}~\bibnamefont {Wen}}, \bibinfo {author} {\bibfnamefont {Z.}~\bibnamefont
  {Zhong}}, \bibinfo {author} {\bibfnamefont {J.}~\bibnamefont
  {{\'I}{\~{n}}iguez-Gonz{\'a}lez}}, \bibinfo {author} {\bibfnamefont
  {L.}~\bibnamefont {Bellaiche}}, \bibinfo {author} {\bibfnamefont
  {D.}~\bibnamefont {Wu}},\ and\ \bibinfo {author} {\bibfnamefont
  {Y.}~\bibnamefont {Yang}},\ }\bibfield  {title} {\bibinfo {title} {Active
  learning of effective hamiltonian for super-large-scale atomic structures},\
  }\href {https://doi.org/10.1038/s41524-025-01563-z} {\bibfield  {journal}
  {\bibinfo  {journal} {npj Comput. Mater.}\ }\textbf {\bibinfo {volume}
  {11}},\ \bibinfo {pages} {70} (\bibinfo {year}
  {2025}{\natexlab{a}})}\BibitemShut {NoStop}%
\bibitem [{\citenamefont {Govinden}\ \emph {et~al.}(2023)\citenamefont
  {Govinden}, \citenamefont {Prokhorenko}, \citenamefont {Zhang}, \citenamefont
  {Rijal}, \citenamefont {Nahas}, \citenamefont {Bellaiche},\ and\
  \citenamefont {Valanoor}}]{Govinden-23}%
  \BibitemOpen
  \bibfield  {author} {\bibinfo {author} {\bibfnamefont {V.}~\bibnamefont
  {Govinden}}, \bibinfo {author} {\bibfnamefont {S.}~\bibnamefont
  {Prokhorenko}}, \bibinfo {author} {\bibfnamefont {Q.}~\bibnamefont {Zhang}},
  \bibinfo {author} {\bibfnamefont {S.}~\bibnamefont {Rijal}}, \bibinfo
  {author} {\bibfnamefont {Y.}~\bibnamefont {Nahas}}, \bibinfo {author}
  {\bibfnamefont {L.}~\bibnamefont {Bellaiche}},\ and\ \bibinfo {author}
  {\bibfnamefont {N.}~\bibnamefont {Valanoor}},\ }\bibfield  {title} {\bibinfo
  {title} {Spherical ferroelectric solitons},\ }\href
  {https://doi.org/10.1038/s41563-023-01527-y} {\bibfield  {journal} {\bibinfo
  {journal} {Nat. Mater.}\ }\textbf {\bibinfo {volume} {22}},\ \bibinfo {pages}
  {553} (\bibinfo {year} {2023})}\BibitemShut {NoStop}%
\bibitem [{\citenamefont {Lukyanchuk}\ \emph {et~al.}(2025)\citenamefont
  {Lukyanchuk}, \citenamefont {Razumnaya}, \citenamefont {Kondovych},
  \citenamefont {Tikhonov}, \citenamefont {Khesin},\ and\ \citenamefont
  {Vinokur}}]{Lukyanchuk-25}%
  \BibitemOpen
  \bibfield  {author} {\bibinfo {author} {\bibfnamefont {I.~A.}\ \bibnamefont
  {Lukyanchuk}}, \bibinfo {author} {\bibfnamefont {A.~G.}\ \bibnamefont
  {Razumnaya}}, \bibinfo {author} {\bibfnamefont {S.}~\bibnamefont
  {Kondovych}}, \bibinfo {author} {\bibfnamefont {Y.~A.}\ \bibnamefont
  {Tikhonov}}, \bibinfo {author} {\bibfnamefont {B.}~\bibnamefont {Khesin}},\
  and\ \bibinfo {author} {\bibfnamefont {V.~M.}\ \bibnamefont {Vinokur}},\
  }\bibfield  {title} {\bibinfo {title} {Topological foundations of
  ferroelectricity},\ }\href
  {https://doi.org/https://doi.org/10.1016/j.physrep.2025.01.002} {\bibfield
  {journal} {\bibinfo  {journal} {Phys. Rep.}\ }\textbf {\bibinfo {volume}
  {1110}},\ \bibinfo {pages} {1} (\bibinfo {year} {2025})}\BibitemShut
  {NoStop}%
\bibitem [{\citenamefont {Back}\ \emph {et~al.}(2020)\citenamefont {Back},
  \citenamefont {Cros}, \citenamefont {Ebert}, \citenamefont {Everschor-Sitte},
  \citenamefont {Fert}, \citenamefont {Garst}, \citenamefont {Ma},
  \citenamefont {Mankovsky}, \citenamefont {Monchesky}, \citenamefont
  {Mostovoy} \emph {et~al.}}]{back20202020}%
  \BibitemOpen
  \bibfield  {author} {\bibinfo {author} {\bibfnamefont {C.}~\bibnamefont
  {Back}}, \bibinfo {author} {\bibfnamefont {V.}~\bibnamefont {Cros}}, \bibinfo
  {author} {\bibfnamefont {H.}~\bibnamefont {Ebert}}, \bibinfo {author}
  {\bibfnamefont {K.}~\bibnamefont {Everschor-Sitte}}, \bibinfo {author}
  {\bibfnamefont {A.}~\bibnamefont {Fert}}, \bibinfo {author} {\bibfnamefont
  {M.}~\bibnamefont {Garst}}, \bibinfo {author} {\bibfnamefont
  {T.}~\bibnamefont {Ma}}, \bibinfo {author} {\bibfnamefont {S.}~\bibnamefont
  {Mankovsky}}, \bibinfo {author} {\bibfnamefont {T.}~\bibnamefont
  {Monchesky}}, \bibinfo {author} {\bibfnamefont {M.}~\bibnamefont {Mostovoy}},
  \emph {et~al.},\ }\bibfield  {title} {\bibinfo {title} {The 2020 skyrmionics
  roadmap},\ }\href@noop {} {\bibfield  {journal} {\bibinfo  {journal} {J.
  Phys. D: Appl. Phys.}\ }\textbf {\bibinfo {volume} {53}},\ \bibinfo {pages}
  {363001} (\bibinfo {year} {2020})}\BibitemShut {NoStop}%
\bibitem [{\citenamefont {Foster}\ \emph {et~al.}(2019)\citenamefont {Foster},
  \citenamefont {Kind}, \citenamefont {Ackerman}, \citenamefont {Tai},
  \citenamefont {Dennis},\ and\ \citenamefont {Smalyukh}}]{foster2019two}%
  \BibitemOpen
  \bibfield  {author} {\bibinfo {author} {\bibfnamefont {D.}~\bibnamefont
  {Foster}}, \bibinfo {author} {\bibfnamefont {C.}~\bibnamefont {Kind}},
  \bibinfo {author} {\bibfnamefont {P.~J.}\ \bibnamefont {Ackerman}}, \bibinfo
  {author} {\bibfnamefont {J.-S.~B.}\ \bibnamefont {Tai}}, \bibinfo {author}
  {\bibfnamefont {M.~R.}\ \bibnamefont {Dennis}},\ and\ \bibinfo {author}
  {\bibfnamefont {I.~I.}\ \bibnamefont {Smalyukh}},\ }\bibfield  {title}
  {\bibinfo {title} {Two-dimensional skyrmion bags in liquid crystals and
  ferromagnets},\ }\href@noop {} {\bibfield  {journal} {\bibinfo  {journal}
  {Nat. Phys.}\ }\textbf {\bibinfo {volume} {15}},\ \bibinfo {pages} {655}
  (\bibinfo {year} {2019})}\BibitemShut {NoStop}%
\bibitem [{\citenamefont {Zhang}\ \emph
  {et~al.}(2024{\natexlab{a}})\citenamefont {Zhang}, \citenamefont {Tang},
  \citenamefont {Wu}, \citenamefont {Shi}, \citenamefont {Xu}, \citenamefont
  {Wang}, \citenamefont {Tian},\ and\ \citenamefont {Du}}]{zhang2024stable}%
  \BibitemOpen
  \bibfield  {author} {\bibinfo {author} {\bibfnamefont {Y.}~\bibnamefont
  {Zhang}}, \bibinfo {author} {\bibfnamefont {J.}~\bibnamefont {Tang}},
  \bibinfo {author} {\bibfnamefont {Y.}~\bibnamefont {Wu}}, \bibinfo {author}
  {\bibfnamefont {M.}~\bibnamefont {Shi}}, \bibinfo {author} {\bibfnamefont
  {X.}~\bibnamefont {Xu}}, \bibinfo {author} {\bibfnamefont {S.}~\bibnamefont
  {Wang}}, \bibinfo {author} {\bibfnamefont {M.}~\bibnamefont {Tian}},\ and\
  \bibinfo {author} {\bibfnamefont {H.}~\bibnamefont {Du}},\ }\bibfield
  {title} {\bibinfo {title} {Stable skyrmion bundles at room temperature and
  zero magnetic field in a chiral magnet},\ }\href@noop {} {\bibfield
  {journal} {\bibinfo  {journal} {Nat. Commun.}\ }\textbf {\bibinfo {volume}
  {15}},\ \bibinfo {pages} {3391} (\bibinfo {year}
  {2024}{\natexlab{a}})}\BibitemShut {NoStop}%
\bibitem [{\citenamefont {Kern}\ \emph {et~al.}(2025)\citenamefont {Kern},
  \citenamefont {Kuchkin}, \citenamefont {Deinhart}, \citenamefont {Klose},
  \citenamefont {Sidiropoulos}, \citenamefont {Auer}, \citenamefont {Gaebel},
  \citenamefont {Gerlinger}, \citenamefont {Battistelli}, \citenamefont
  {Wittrock} \emph {et~al.}}]{kern2025controlled}%
  \BibitemOpen
  \bibfield  {author} {\bibinfo {author} {\bibfnamefont {L.-M.}\ \bibnamefont
  {Kern}}, \bibinfo {author} {\bibfnamefont {V.~M.}\ \bibnamefont {Kuchkin}},
  \bibinfo {author} {\bibfnamefont {V.}~\bibnamefont {Deinhart}}, \bibinfo
  {author} {\bibfnamefont {C.}~\bibnamefont {Klose}}, \bibinfo {author}
  {\bibfnamefont {T.}~\bibnamefont {Sidiropoulos}}, \bibinfo {author}
  {\bibfnamefont {M.}~\bibnamefont {Auer}}, \bibinfo {author} {\bibfnamefont
  {S.}~\bibnamefont {Gaebel}}, \bibinfo {author} {\bibfnamefont
  {K.}~\bibnamefont {Gerlinger}}, \bibinfo {author} {\bibfnamefont
  {R.}~\bibnamefont {Battistelli}}, \bibinfo {author} {\bibfnamefont
  {S.}~\bibnamefont {Wittrock}}, \emph {et~al.},\ }\bibfield  {title} {\bibinfo
  {title} {Controlled formation of skyrmion bags},\ }\href@noop {} {\bibfield
  {journal} {\bibinfo  {journal} {Adv. Mater.}\ ,\ \bibinfo {pages} {2501250}}
  (\bibinfo {year} {2025})}\BibitemShut {NoStop}%
\bibitem [{\citenamefont {Rosencwaig}\ \emph {et~al.}(1972)\citenamefont
  {Rosencwaig}, \citenamefont {Tabor},\ and\ \citenamefont
  {Nelson}}]{rosencwaig1972new}%
  \BibitemOpen
  \bibfield  {author} {\bibinfo {author} {\bibfnamefont {A.}~\bibnamefont
  {Rosencwaig}}, \bibinfo {author} {\bibfnamefont {W.~J.}\ \bibnamefont
  {Tabor}},\ and\ \bibinfo {author} {\bibfnamefont {T.}~\bibnamefont
  {Nelson}},\ }\bibfield  {title} {\bibinfo {title} {New domain-wall
  configuration for magnetic bubbles},\ }\href@noop {} {\bibfield  {journal}
  {\bibinfo  {journal} {Phys. Rev. Lett.}\ }\textbf {\bibinfo {volume} {29}},\
  \bibinfo {pages} {946} (\bibinfo {year} {1972})}\BibitemShut {NoStop}%
\bibitem [{\citenamefont {Grundy}(1977)}]{grundy1977magnetic}%
  \BibitemOpen
  \bibfield  {author} {\bibinfo {author} {\bibfnamefont {P.}~\bibnamefont
  {Grundy}},\ }\bibfield  {title} {\bibinfo {title} {Magnetic bubbles and their
  observation in the electron microscope},\ }\href@noop {} {\bibfield
  {journal} {\bibinfo  {journal} {Contemp. Phys.}\ }\textbf {\bibinfo {volume}
  {18}},\ \bibinfo {pages} {47} (\bibinfo {year} {1977})}\BibitemShut {NoStop}%
\bibitem [{\citenamefont {Slonczewski}\ \emph {et~al.}(1973)\citenamefont
  {Slonczewski}, \citenamefont {Malozemoff},\ and\ \citenamefont
  {Voegeli}}]{slonczewski1973statics}%
  \BibitemOpen
  \bibfield  {author} {\bibinfo {author} {\bibfnamefont {J.}~\bibnamefont
  {Slonczewski}}, \bibinfo {author} {\bibfnamefont {A.}~\bibnamefont
  {Malozemoff}},\ and\ \bibinfo {author} {\bibfnamefont {O.}~\bibnamefont
  {Voegeli}},\ }\bibfield  {title} {\bibinfo {title} {Statics and dynamics of
  bubbles containing bloch lines},\ }in\ \href@noop {} {\emph {\bibinfo
  {booktitle} {AIP Conf. Proc.}}},\ Vol.~\bibinfo {volume} {10}\ (\bibinfo
  {organization} {American Institute of Physics},\ \bibinfo {year} {1973})\
  pp.\ \bibinfo {pages} {458--477}\BibitemShut {NoStop}%
\bibitem [{\citenamefont {Finizio}\ \emph {et~al.}(2019)\citenamefont
  {Finizio}, \citenamefont {Zeissler}, \citenamefont {Wintz}, \citenamefont
  {Mayr}, \citenamefont {We{\ss}els}, \citenamefont {Huxtable}, \citenamefont
  {Burnell}, \citenamefont {Marrows},\ and\ \citenamefont
  {Raabe}}]{finizio2019deterministic}%
  \BibitemOpen
  \bibfield  {author} {\bibinfo {author} {\bibfnamefont {S.}~\bibnamefont
  {Finizio}}, \bibinfo {author} {\bibfnamefont {K.}~\bibnamefont {Zeissler}},
  \bibinfo {author} {\bibfnamefont {S.}~\bibnamefont {Wintz}}, \bibinfo
  {author} {\bibfnamefont {S.}~\bibnamefont {Mayr}}, \bibinfo {author}
  {\bibfnamefont {T.}~\bibnamefont {We{\ss}els}}, \bibinfo {author}
  {\bibfnamefont {A.~J.}\ \bibnamefont {Huxtable}}, \bibinfo {author}
  {\bibfnamefont {G.}~\bibnamefont {Burnell}}, \bibinfo {author} {\bibfnamefont
  {C.~H.}\ \bibnamefont {Marrows}},\ and\ \bibinfo {author} {\bibfnamefont
  {J.}~\bibnamefont {Raabe}},\ }\bibfield  {title} {\bibinfo {title}
  {Deterministic {{Field-Free Skyrmion Nucleation}} at a {{Nanoengineered
  Injector Device}}},\ }\href {https://doi.org/10.1021/acs.nanolett.9b02840}
  {\bibfield  {journal} {\bibinfo  {journal} {Nano Lett.}\ }\textbf {\bibinfo
  {volume} {19}},\ \bibinfo {pages} {7246} (\bibinfo {year}
  {2019})}\BibitemShut {NoStop}%
\bibitem [{\citenamefont {Yang}\ \emph {et~al.}(2022)\citenamefont {Yang},
  \citenamefont {Ju}, \citenamefont {Kim}, \citenamefont {Kim}, \citenamefont
  {An}, \citenamefont {Moon}, \citenamefont {Park},\ and\ \citenamefont
  {Hwang}}]{yang2022magnetic}%
  \BibitemOpen
  \bibfield  {author} {\bibinfo {author} {\bibfnamefont {S.}~\bibnamefont
  {Yang}}, \bibinfo {author} {\bibfnamefont {T.-S.}\ \bibnamefont {Ju}},
  \bibinfo {author} {\bibfnamefont {C.}~\bibnamefont {Kim}}, \bibinfo {author}
  {\bibfnamefont {H.-J.}\ \bibnamefont {Kim}}, \bibinfo {author} {\bibfnamefont
  {K.}~\bibnamefont {An}}, \bibinfo {author} {\bibfnamefont {K.-W.}\
  \bibnamefont {Moon}}, \bibinfo {author} {\bibfnamefont {S.}~\bibnamefont
  {Park}},\ and\ \bibinfo {author} {\bibfnamefont {C.}~\bibnamefont {Hwang}},\
  }\bibfield  {title} {\bibinfo {title} {Magnetic field magnitudes needed for
  skyrmion generation in a general perpendicularly magnetized film},\
  }\href@noop {} {\bibfield  {journal} {\bibinfo  {journal} {Nano Lett.}\
  }\textbf {\bibinfo {volume} {22}},\ \bibinfo {pages} {8430} (\bibinfo {year}
  {2022})}\BibitemShut {NoStop}%
\bibitem [{\citenamefont {B{\"u}ttner}\ \emph {et~al.}(2017)\citenamefont
  {B{\"u}ttner}, \citenamefont {Lemesh}, \citenamefont {Schneider},
  \citenamefont {Pfau}, \citenamefont {G{\"u}nther}, \citenamefont {Hessing},
  \citenamefont {Geilhufe}, \citenamefont {Caretta}, \citenamefont {Engel},
  \citenamefont {Kr{\"u}ger}, \citenamefont {Viefhaus}, \citenamefont
  {Eisebitt},\ and\ \citenamefont {Beach}}]{buttner2017fieldfree}%
  \BibitemOpen
  \bibfield  {author} {\bibinfo {author} {\bibfnamefont {F.}~\bibnamefont
  {B{\"u}ttner}}, \bibinfo {author} {\bibfnamefont {I.}~\bibnamefont {Lemesh}},
  \bibinfo {author} {\bibfnamefont {M.}~\bibnamefont {Schneider}}, \bibinfo
  {author} {\bibfnamefont {B.}~\bibnamefont {Pfau}}, \bibinfo {author}
  {\bibfnamefont {C.~M.}\ \bibnamefont {G{\"u}nther}}, \bibinfo {author}
  {\bibfnamefont {P.}~\bibnamefont {Hessing}}, \bibinfo {author} {\bibfnamefont
  {J.}~\bibnamefont {Geilhufe}}, \bibinfo {author} {\bibfnamefont
  {L.}~\bibnamefont {Caretta}}, \bibinfo {author} {\bibfnamefont
  {D.}~\bibnamefont {Engel}}, \bibinfo {author} {\bibfnamefont
  {B.}~\bibnamefont {Kr{\"u}ger}}, \bibinfo {author} {\bibfnamefont
  {J.}~\bibnamefont {Viefhaus}}, \bibinfo {author} {\bibfnamefont
  {S.}~\bibnamefont {Eisebitt}},\ and\ \bibinfo {author} {\bibfnamefont
  {G.~S.~D.}\ \bibnamefont {Beach}},\ }\bibfield  {title} {\bibinfo {title}
  {Field-free deterministic ultrafast creation of magnetic skyrmions by
  spin--orbit torques},\ }\href {https://doi.org/10.1038/nnano.2017.178}
  {\bibfield  {journal} {\bibinfo  {journal} {Nature Nanotech}\ }\textbf
  {\bibinfo {volume} {12}},\ \bibinfo {pages} {1040} (\bibinfo {year}
  {2017})}\BibitemShut {NoStop}%
\bibitem [{\citenamefont {Koraltan}\ \emph {et~al.}(2023)\citenamefont
  {Koraltan}, \citenamefont {Abert}, \citenamefont {Bruckner}, \citenamefont
  {Heigl}, \citenamefont {Albrecht},\ and\ \citenamefont
  {Suess}}]{koraltan2023generation}%
  \BibitemOpen
  \bibfield  {author} {\bibinfo {author} {\bibfnamefont {S.}~\bibnamefont
  {Koraltan}}, \bibinfo {author} {\bibfnamefont {C.}~\bibnamefont {Abert}},
  \bibinfo {author} {\bibfnamefont {F.}~\bibnamefont {Bruckner}}, \bibinfo
  {author} {\bibfnamefont {M.}~\bibnamefont {Heigl}}, \bibinfo {author}
  {\bibfnamefont {M.}~\bibnamefont {Albrecht}},\ and\ \bibinfo {author}
  {\bibfnamefont {D.}~\bibnamefont {Suess}},\ }\bibfield  {title} {\bibinfo
  {title} {Generation and annihilation of skyrmions and antiskyrmions in
  magnetic heterostructures},\ }\href@noop {} {\bibfield  {journal} {\bibinfo
  {journal} {Phys. Rev. B}\ }\textbf {\bibinfo {volume} {108}},\ \bibinfo
  {pages} {134401} (\bibinfo {year} {2023})}\BibitemShut {NoStop}%
\bibitem [{\citenamefont {Takagi}\ \emph {et~al.}(2022)\citenamefont {Takagi},
  \citenamefont {Matsuyama}, \citenamefont {Ukleev}, \citenamefont {Yu},
  \citenamefont {White}, \citenamefont {Francoual}, \citenamefont {Mardegan},
  \citenamefont {Hayami}, \citenamefont {Saito}, \citenamefont {Kaneko},
  \citenamefont {Ohishi}, \citenamefont {{\=O}nuki}, \citenamefont {Arima},
  \citenamefont {Tokura}, \citenamefont {Nakajima},\ and\ \citenamefont
  {Seki}}]{takagi2022square}%
  \BibitemOpen
  \bibfield  {author} {\bibinfo {author} {\bibfnamefont {R.}~\bibnamefont
  {Takagi}}, \bibinfo {author} {\bibfnamefont {N.}~\bibnamefont {Matsuyama}},
  \bibinfo {author} {\bibfnamefont {V.}~\bibnamefont {Ukleev}}, \bibinfo
  {author} {\bibfnamefont {L.}~\bibnamefont {Yu}}, \bibinfo {author}
  {\bibfnamefont {J.~S.}\ \bibnamefont {White}}, \bibinfo {author}
  {\bibfnamefont {S.}~\bibnamefont {Francoual}}, \bibinfo {author}
  {\bibfnamefont {J.~R.~L.}\ \bibnamefont {Mardegan}}, \bibinfo {author}
  {\bibfnamefont {S.}~\bibnamefont {Hayami}}, \bibinfo {author} {\bibfnamefont
  {H.}~\bibnamefont {Saito}}, \bibinfo {author} {\bibfnamefont
  {K.}~\bibnamefont {Kaneko}}, \bibinfo {author} {\bibfnamefont
  {K.}~\bibnamefont {Ohishi}}, \bibinfo {author} {\bibfnamefont
  {Y.}~\bibnamefont {{\=O}nuki}}, \bibinfo {author} {\bibfnamefont {T.-h.}\
  \bibnamefont {Arima}}, \bibinfo {author} {\bibfnamefont {Y.}~\bibnamefont
  {Tokura}}, \bibinfo {author} {\bibfnamefont {T.}~\bibnamefont {Nakajima}},\
  and\ \bibinfo {author} {\bibfnamefont {S.}~\bibnamefont {Seki}},\ }\bibfield
  {title} {\bibinfo {title} {Square and rhombic lattices of magnetic skyrmions
  in a centrosymmetric binary compound},\ }\href
  {https://doi.org/10.1038/s41467-022-29131-9} {\bibfield  {journal} {\bibinfo
  {journal} {Nat. Commun.}\ }\textbf {\bibinfo {volume} {13}},\ \bibinfo
  {pages} {1472} (\bibinfo {year} {2022})}\BibitemShut {NoStop}%
\bibitem [{\citenamefont {Yoshimochi}\ \emph {et~al.}(2024)\citenamefont
  {Yoshimochi}, \citenamefont {Takagi}, \citenamefont {Ju}, \citenamefont
  {Khanh}, \citenamefont {Saito}, \citenamefont {Sagayama}, \citenamefont
  {Nakao}, \citenamefont {Itoh}, \citenamefont {Tokura}, \citenamefont {Arima},
  \citenamefont {Hayami}, \citenamefont {Nakajima},\ and\ \citenamefont
  {Seki}}]{yoshimochi2024multi}%
  \BibitemOpen
  \bibfield  {author} {\bibinfo {author} {\bibfnamefont {H.}~\bibnamefont
  {Yoshimochi}}, \bibinfo {author} {\bibfnamefont {R.}~\bibnamefont {Takagi}},
  \bibinfo {author} {\bibfnamefont {J.}~\bibnamefont {Ju}}, \bibinfo {author}
  {\bibfnamefont {N.}~\bibnamefont {Khanh}}, \bibinfo {author} {\bibfnamefont
  {H.}~\bibnamefont {Saito}}, \bibinfo {author} {\bibfnamefont
  {H.}~\bibnamefont {Sagayama}}, \bibinfo {author} {\bibfnamefont
  {H.}~\bibnamefont {Nakao}}, \bibinfo {author} {\bibfnamefont
  {S.}~\bibnamefont {Itoh}}, \bibinfo {author} {\bibfnamefont {Y.}~\bibnamefont
  {Tokura}}, \bibinfo {author} {\bibfnamefont {T.}~\bibnamefont {Arima}},
  \bibinfo {author} {\bibfnamefont {S.}~\bibnamefont {Hayami}}, \bibinfo
  {author} {\bibfnamefont {T.}~\bibnamefont {Nakajima}},\ and\ \bibinfo
  {author} {\bibfnamefont {S.}~\bibnamefont {Seki}},\ }\bibfield  {title}
  {\bibinfo {title} {Multistep topological transitions among meron and skyrmion
  crystals in a centrosymmetric magnet},\ }\href
  {https://doi.org/https://doi.org/10.1038/s41567-024-02445-9} {\bibfield
  {journal} {\bibinfo  {journal} {Nat. Phys.}\ }\textbf {\bibinfo {volume}
  {20}},\ \bibinfo {pages} {1001} (\bibinfo {year} {2024})}\BibitemShut
  {NoStop}%
\bibitem [{\citenamefont {Ishiwata}\ \emph {et~al.}(2020)\citenamefont
  {Ishiwata}, \citenamefont {Nakajima}, \citenamefont {Kim}, \citenamefont
  {Inosov}, \citenamefont {Kanazawa}, \citenamefont {White}, \citenamefont
  {Gavilano}, \citenamefont {Georgii}, \citenamefont {Seemann}, \citenamefont
  {Brandl}, \citenamefont {Manuel}, \citenamefont {Khalyavin}, \citenamefont
  {Seki}, \citenamefont {Tokunaga}, \citenamefont {Kinoshita}, \citenamefont
  {Long}, \citenamefont {Kaneko}, \citenamefont {Taguchi}, \citenamefont
  {Arima}, \citenamefont {Keimer},\ and\ \citenamefont
  {Tokura}}]{Ishiwata_PhysRevB.101.134406}%
  \BibitemOpen
  \bibfield  {author} {\bibinfo {author} {\bibfnamefont {S.}~\bibnamefont
  {Ishiwata}}, \bibinfo {author} {\bibfnamefont {T.}~\bibnamefont {Nakajima}},
  \bibinfo {author} {\bibfnamefont {J.-H.}\ \bibnamefont {Kim}}, \bibinfo
  {author} {\bibfnamefont {D.~S.}\ \bibnamefont {Inosov}}, \bibinfo {author}
  {\bibfnamefont {N.}~\bibnamefont {Kanazawa}}, \bibinfo {author}
  {\bibfnamefont {J.~S.}\ \bibnamefont {White}}, \bibinfo {author}
  {\bibfnamefont {J.~L.}\ \bibnamefont {Gavilano}}, \bibinfo {author}
  {\bibfnamefont {R.}~\bibnamefont {Georgii}}, \bibinfo {author} {\bibfnamefont
  {K.~M.}\ \bibnamefont {Seemann}}, \bibinfo {author} {\bibfnamefont
  {G.}~\bibnamefont {Brandl}}, \bibinfo {author} {\bibfnamefont
  {P.}~\bibnamefont {Manuel}}, \bibinfo {author} {\bibfnamefont {D.~D.}\
  \bibnamefont {Khalyavin}}, \bibinfo {author} {\bibfnamefont {S.}~\bibnamefont
  {Seki}}, \bibinfo {author} {\bibfnamefont {Y.}~\bibnamefont {Tokunaga}},
  \bibinfo {author} {\bibfnamefont {M.}~\bibnamefont {Kinoshita}}, \bibinfo
  {author} {\bibfnamefont {Y.~W.}\ \bibnamefont {Long}}, \bibinfo {author}
  {\bibfnamefont {Y.}~\bibnamefont {Kaneko}}, \bibinfo {author} {\bibfnamefont
  {Y.}~\bibnamefont {Taguchi}}, \bibinfo {author} {\bibfnamefont
  {T.}~\bibnamefont {Arima}}, \bibinfo {author} {\bibfnamefont
  {B.}~\bibnamefont {Keimer}},\ and\ \bibinfo {author} {\bibfnamefont
  {Y.}~\bibnamefont {Tokura}},\ }\bibfield  {title} {\bibinfo {title}
  {{Emergent topological spin structures in the centrosymmetric cubic
  perovskite ${\mathrm{SrFeO}}_{3}$}},\ }\href
  {https://doi.org/10.1103/PhysRevB.101.134406} {\bibfield  {journal} {\bibinfo
   {journal} {Phys. Rev. B}\ }\textbf {\bibinfo {volume} {101}},\ \bibinfo
  {pages} {134406} (\bibinfo {year} {2020})}\BibitemShut {NoStop}%
\bibitem [{\citenamefont {Hayami}\ and\ \citenamefont
  {Motome}(2021)}]{hayami2021topological}%
  \BibitemOpen
  \bibfield  {author} {\bibinfo {author} {\bibfnamefont {S.}~\bibnamefont
  {Hayami}}\ and\ \bibinfo {author} {\bibfnamefont {Y.}~\bibnamefont
  {Motome}},\ }\bibfield  {title} {\bibinfo {title} {Topological spin crystals
  by itinerant frustration},\ }\href {https://doi.org/10.1088/1361-648x/ac1a30}
  {\bibfield  {journal} {\bibinfo  {journal} {J. Phys.: Condens. Matter}\
  }\textbf {\bibinfo {volume} {33}},\ \bibinfo {pages} {443001} (\bibinfo
  {year} {2021})}\BibitemShut {NoStop}%
\bibitem [{\citenamefont {Dong}\ \emph {et~al.}(2025)\citenamefont {Dong},
  \citenamefont {Kinoshita}, \citenamefont {Ochi}, \citenamefont {Nakachi},
  \citenamefont {Higashinaka}, \citenamefont {Hayami}, \citenamefont {Wan},
  \citenamefont {Arai}, \citenamefont {Huh}, \citenamefont {Hashimoto},
  \citenamefont {Lu}, \citenamefont {Tokunaga}, \citenamefont {Aoki},
  \citenamefont {Matsuda},\ and\ \citenamefont {Kondo}}]{dong2025pseudogap}%
  \BibitemOpen
  \bibfield  {author} {\bibinfo {author} {\bibfnamefont {Y.}~\bibnamefont
  {Dong}}, \bibinfo {author} {\bibfnamefont {Y.}~\bibnamefont {Kinoshita}},
  \bibinfo {author} {\bibfnamefont {M.}~\bibnamefont {Ochi}}, \bibinfo {author}
  {\bibfnamefont {R.}~\bibnamefont {Nakachi}}, \bibinfo {author} {\bibfnamefont
  {R.}~\bibnamefont {Higashinaka}}, \bibinfo {author} {\bibfnamefont
  {S.}~\bibnamefont {Hayami}}, \bibinfo {author} {\bibfnamefont
  {Y.}~\bibnamefont {Wan}}, \bibinfo {author} {\bibfnamefont {Y.}~\bibnamefont
  {Arai}}, \bibinfo {author} {\bibfnamefont {S.}~\bibnamefont {Huh}}, \bibinfo
  {author} {\bibfnamefont {M.}~\bibnamefont {Hashimoto}}, \bibinfo {author}
  {\bibfnamefont {D.}~\bibnamefont {Lu}}, \bibinfo {author} {\bibfnamefont
  {M.}~\bibnamefont {Tokunaga}}, \bibinfo {author} {\bibfnamefont
  {Y.}~\bibnamefont {Aoki}}, \bibinfo {author} {\bibfnamefont {T.~D.}\
  \bibnamefont {Matsuda}},\ and\ \bibinfo {author} {\bibfnamefont
  {T.}~\bibnamefont {Kondo}},\ }\bibfield  {title} {\bibinfo {title} {Pseudogap
  and fermi arc induced by fermi surface nesting in a centrosymmetric skyrmion
  magnet},\ }\href {https://doi.org/10.1126/science.adj7710} {\bibfield
  {journal} {\bibinfo  {journal} {Science}\ }\textbf {\bibinfo {volume}
  {388}},\ \bibinfo {pages} {624} (\bibinfo {year} {2025})}\BibitemShut
  {NoStop}%
\bibitem [{\citenamefont {Hayami}\ and\ \citenamefont
  {Yambe}(2024)}]{hayami2024stabilization}%
  \BibitemOpen
  \bibfield  {author} {\bibinfo {author} {\bibfnamefont {S.}~\bibnamefont
  {Hayami}}\ and\ \bibinfo {author} {\bibfnamefont {R.}~\bibnamefont {Yambe}},\
  }\bibfield  {title} {\bibinfo {title} {Stabilization mechanisms of magnetic
  skyrmion crystal and multiple-$q$ states based on momentum-resolved spin
  interactions},\ }\href
  {https://doi.org/https://doi.org/10.1016/j.mtquan.2024.100010} {\bibfield
  {journal} {\bibinfo  {journal} {Mater. Today Quantum}\ }\textbf {\bibinfo
  {volume} {3}},\ \bibinfo {pages} {100010} (\bibinfo {year}
  {2024})}\BibitemShut {NoStop}%
\bibitem [{\citenamefont {Miyazaki}(2023)}]{miyazaki2023machine}%
  \BibitemOpen
  \bibfield  {author} {\bibinfo {author} {\bibfnamefont {Y.}~\bibnamefont
  {Miyazaki}},\ }\bibfield  {title} {\bibinfo {title} {Equivariant neural
  networks for spin dynamics simulations of itinerant magnets},\ }\href
  {https://doi.org/10.1088/2632-2153/acffa2} {\bibfield  {journal} {\bibinfo
  {journal} {Mach. Learn.: Sci. Technol.}\ }\textbf {\bibinfo {volume} {4}},\
  \bibinfo {pages} {045006} (\bibinfo {year} {2023})}\BibitemShut {NoStop}%
\bibitem [{\citenamefont {Sharma}\ \emph {et~al.}(2023)\citenamefont {Sharma},
  \citenamefont {Wang},\ and\ \citenamefont {Batista}}]{sharma2023machine}%
  \BibitemOpen
  \bibfield  {author} {\bibinfo {author} {\bibfnamefont {V.}~\bibnamefont
  {Sharma}}, \bibinfo {author} {\bibfnamefont {Z.}~\bibnamefont {Wang}},\ and\
  \bibinfo {author} {\bibfnamefont {C.~D.}\ \bibnamefont {Batista}},\
  }\bibfield  {title} {\bibinfo {title} {Machine learning assisted derivation
  of minimal low-energy models for metallic magnets},\ }\href@noop {}
  {\bibfield  {journal} {\bibinfo  {journal} {npj Comput. Mater.}\ }\textbf
  {\bibinfo {volume} {9}},\ \bibinfo {pages} {192} (\bibinfo {year}
  {2023})}\BibitemShut {NoStop}%
\bibitem [{\citenamefont {Nomoto}\ and\ \citenamefont
  {Arita}(2023)}]{Nomoto2023}%
  \BibitemOpen
  \bibfield  {author} {\bibinfo {author} {\bibfnamefont {T.}~\bibnamefont
  {Nomoto}}\ and\ \bibinfo {author} {\bibfnamefont {R.}~\bibnamefont {Arita}},\
  }\bibfield  {title} {\bibinfo {title} {{Ab initio exploration of short-pitch
  skyrmion materials: Role of orbital frustration}},\ }\bibfield  {journal}
  {\bibinfo  {journal} {J. Appl. Phys.}\ }\textbf {\bibinfo {volume} {133}},\
  \href {https://doi.org/10.1063/5.0141628} {10.1063/5.0141628} (\bibinfo
  {year} {2023}),\ \bibinfo {note} {150901},\ \Eprint
  {https://arxiv.org/abs/https://pubs.aip.org/aip/jap/article-pdf/doi/10.1063/5.0141628/16824864/150901\_1\_5.0141628.pdf}
  {https://pubs.aip.org/aip/jap/article-pdf/doi/10.1063/5.0141628/16824864/150901\_1\_5.0141628.pdf}
  \BibitemShut {NoStop}%
\bibitem [{\citenamefont {Mertelj}\ \emph {et~al.}(2013)\citenamefont
  {Mertelj}, \citenamefont {Lisjak}, \citenamefont {Drofenik},\ and\
  \citenamefont {{\v C}opi{\v c}}}]{mertelj2013ferromagnetism}%
  \BibitemOpen
  \bibfield  {author} {\bibinfo {author} {\bibfnamefont {A.}~\bibnamefont
  {Mertelj}}, \bibinfo {author} {\bibfnamefont {D.}~\bibnamefont {Lisjak}},
  \bibinfo {author} {\bibfnamefont {M.}~\bibnamefont {Drofenik}},\ and\
  \bibinfo {author} {\bibfnamefont {M.}~\bibnamefont {{\v C}opi{\v c}}},\
  }\bibfield  {title} {\bibinfo {title} {Ferromagnetism in suspensions of
  magnetic platelets in liquid crystal},\ }\href
  {https://doi.org/10.1038/nature12863} {\bibfield  {journal} {\bibinfo
  {journal} {Nature}\ }\textbf {\bibinfo {volume} {504}},\ \bibinfo {pages}
  {237} (\bibinfo {year} {2013})}\BibitemShut {NoStop}%
\bibitem [{\citenamefont {Mundoor}\ \emph {et~al.}(2016)\citenamefont
  {Mundoor}, \citenamefont {Senyuk},\ and\ \citenamefont
  {Smalyukh}}]{mundoor2016triclinic}%
  \BibitemOpen
  \bibfield  {author} {\bibinfo {author} {\bibfnamefont {H.}~\bibnamefont
  {Mundoor}}, \bibinfo {author} {\bibfnamefont {B.}~\bibnamefont {Senyuk}},\
  and\ \bibinfo {author} {\bibfnamefont {I.~I.}\ \bibnamefont {Smalyukh}},\
  }\bibfield  {title} {\bibinfo {title} {Triclinic nematic colloidal crystals
  from competing elastic and electrostatic interactions},\ }\href
  {https://doi.org/10.1126/science.aaf0801} {\bibfield  {journal} {\bibinfo
  {journal} {Science}\ }\textbf {\bibinfo {volume} {352}},\ \bibinfo {pages}
  {69} (\bibinfo {year} {2016})}\BibitemShut {NoStop}%
\bibitem [{\citenamefont {Smalyukh}(2020)}]{smalyukh2020review}%
  \BibitemOpen
  \bibfield  {author} {\bibinfo {author} {\bibfnamefont {I.~I.}\ \bibnamefont
  {Smalyukh}},\ }\bibfield  {title} {\bibinfo {title} {Review: Knots and other
  new topological effects in liquid crystals and colloids},\ }\href
  {https://doi.org/10.1088/1361-6633/abaa39} {\bibfield  {journal} {\bibinfo
  {journal} {Rep. Prog. Phys.}\ }\textbf {\bibinfo {volume} {83}},\ \bibinfo
  {pages} {106601} (\bibinfo {year} {2020})}\BibitemShut {NoStop}%
\bibitem [{\citenamefont {Fowler}\ and\ \citenamefont
  {Dierking}(2017)}]{fowler2017kibble}%
  \BibitemOpen
  \bibfield  {author} {\bibinfo {author} {\bibfnamefont {N.}~\bibnamefont
  {Fowler}}\ and\ \bibinfo {author} {\bibfnamefont {D.~I.}\ \bibnamefont
  {Dierking}},\ }\bibfield  {title} {\bibinfo {title} {Kibble--{{Zurek
  Scaling}} during {{Defect Formation}} in a {{Nematic Liquid Crystal}}},\
  }\href {https://doi.org/10.1002/cphc.201700023} {\bibfield  {journal}
  {\bibinfo  {journal} {ChemPhysChem}\ }\textbf {\bibinfo {volume} {18}},\
  \bibinfo {pages} {812} (\bibinfo {year} {2017})}\BibitemShut {NoStop}%
\bibitem [{\citenamefont {Ackerman}\ \emph {et~al.}(2017)\citenamefont
  {Ackerman}, \citenamefont {Boyle},\ and\ \citenamefont
  {Smalyukh}}]{ackerman2017squirming}%
  \BibitemOpen
  \bibfield  {author} {\bibinfo {author} {\bibfnamefont {P.~J.}\ \bibnamefont
  {Ackerman}}, \bibinfo {author} {\bibfnamefont {T.}~\bibnamefont {Boyle}},\
  and\ \bibinfo {author} {\bibfnamefont {I.~I.}\ \bibnamefont {Smalyukh}},\
  }\bibfield  {title} {\bibinfo {title} {Squirming motion of baby skyrmions in
  nematic fluids},\ }\href {https://doi.org/10.1038/s41467-017-00659-5}
  {\bibfield  {journal} {\bibinfo  {journal} {Nat. Commun.}\ }\textbf {\bibinfo
  {volume} {8}},\ \bibinfo {pages} {673} (\bibinfo {year} {2017})}\BibitemShut
  {NoStop}%
\bibitem [{\citenamefont {Wu}\ and\ \citenamefont
  {Smalyukh}(2022)}]{wu2022hopfions}%
  \BibitemOpen
  \bibfield  {author} {\bibinfo {author} {\bibfnamefont {J.-S.}\ \bibnamefont
  {Wu}}\ and\ \bibinfo {author} {\bibfnamefont {I.~I.}\ \bibnamefont
  {Smalyukh}},\ }\bibfield  {title} {\bibinfo {title} {Hopfions, heliknotons,
  skyrmions, torons and both abelian and nonabelian vortices in chiral liquid
  crystals},\ }\href {https://doi.org/10.1080/21680396.2022.2040058} {\bibfield
   {journal} {\bibinfo  {journal} {Liq. Cryst. Rev.}\ }\textbf {\bibinfo
  {volume} {10}},\ \bibinfo {pages} {34} (\bibinfo {year} {2022})}\BibitemShut
  {NoStop}%
\bibitem [{\citenamefont {Donnelly}\ \emph {et~al.}(2021)\citenamefont
  {Donnelly}, \citenamefont {Metlov}, \citenamefont {Scagnoli}, \citenamefont
  {{Guizar-Sicairos}}, \citenamefont {Holler}, \citenamefont {Bingham},
  \citenamefont {Raabe}, \citenamefont {Heyderman}, \citenamefont {Cooper},\
  and\ \citenamefont {Gliga}}]{donnelly2021experimental}%
  \BibitemOpen
  \bibfield  {author} {\bibinfo {author} {\bibfnamefont {C.}~\bibnamefont
  {Donnelly}}, \bibinfo {author} {\bibfnamefont {K.~L.}\ \bibnamefont
  {Metlov}}, \bibinfo {author} {\bibfnamefont {V.}~\bibnamefont {Scagnoli}},
  \bibinfo {author} {\bibfnamefont {M.}~\bibnamefont {{Guizar-Sicairos}}},
  \bibinfo {author} {\bibfnamefont {M.}~\bibnamefont {Holler}}, \bibinfo
  {author} {\bibfnamefont {N.~S.}\ \bibnamefont {Bingham}}, \bibinfo {author}
  {\bibfnamefont {J.}~\bibnamefont {Raabe}}, \bibinfo {author} {\bibfnamefont
  {L.~J.}\ \bibnamefont {Heyderman}}, \bibinfo {author} {\bibfnamefont {N.~R.}\
  \bibnamefont {Cooper}},\ and\ \bibinfo {author} {\bibfnamefont
  {S.}~\bibnamefont {Gliga}},\ }\bibfield  {title} {\bibinfo {title}
  {Experimental observation of vortex rings in a bulk magnet},\ }\href
  {https://doi.org/10.1038/s41567-020-01057-3} {\bibfield  {journal} {\bibinfo
  {journal} {Nat. Phys.}\ }\textbf {\bibinfo {volume} {17}},\ \bibinfo {pages}
  {316} (\bibinfo {year} {2021})}\BibitemShut {NoStop}%
\bibitem [{\citenamefont {Sohn}\ \emph
  {et~al.}(2019{\natexlab{a}})\citenamefont {Sohn}, \citenamefont {Liu},\ and\
  \citenamefont {Smalyukh}}]{sohn2019schools}%
  \BibitemOpen
  \bibfield  {author} {\bibinfo {author} {\bibfnamefont {H.~R.~O.}\
  \bibnamefont {Sohn}}, \bibinfo {author} {\bibfnamefont {C.~D.}\ \bibnamefont
  {Liu}},\ and\ \bibinfo {author} {\bibfnamefont {I.~I.}\ \bibnamefont
  {Smalyukh}},\ }\bibfield  {title} {\bibinfo {title} {Schools of skyrmions
  with electrically tunable elastic interactions},\ }\href
  {https://doi.org/10.1038/s41467-019-12723-3} {\bibfield  {journal} {\bibinfo
  {journal} {Nat. Commun.}\ }\textbf {\bibinfo {volume} {10}},\ \bibinfo
  {pages} {4744} (\bibinfo {year} {2019}{\natexlab{a}})}\BibitemShut {NoStop}%
\bibitem [{\citenamefont {Sohn}\ and\ \citenamefont
  {Smalyukh}(2020)}]{sohn2020electrically}%
  \BibitemOpen
  \bibfield  {author} {\bibinfo {author} {\bibfnamefont {H.~R.~O.}\
  \bibnamefont {Sohn}}\ and\ \bibinfo {author} {\bibfnamefont {I.~I.}\
  \bibnamefont {Smalyukh}},\ }\bibfield  {title} {\bibinfo {title}
  {Electrically powered motions of toron crystallites in chiral liquid
  crystals},\ }\href {https://doi.org/10.1073/pnas.1922198117} {\bibfield
  {journal} {\bibinfo  {journal} {Proc. Natl. Acad. Sci. U.S.A.}\ }\textbf
  {\bibinfo {volume} {117}},\ \bibinfo {pages} {6437} (\bibinfo {year}
  {2020})}\BibitemShut {NoStop}%
\bibitem [{\citenamefont {Tai}\ and\ \citenamefont
  {Smalyukh}(2019)}]{tai2019threedimensional}%
  \BibitemOpen
  \bibfield  {author} {\bibinfo {author} {\bibfnamefont {J.-S.~B.}\
  \bibnamefont {Tai}}\ and\ \bibinfo {author} {\bibfnamefont {I.~I.}\
  \bibnamefont {Smalyukh}},\ }\bibfield  {title} {\bibinfo {title}
  {Three-dimensional crystals of adaptive knots},\ }\href
  {https://doi.org/10.1126/science.aay1638} {\bibfield  {journal} {\bibinfo
  {journal} {Science}\ }\textbf {\bibinfo {volume} {365}},\ \bibinfo {pages}
  {1449} (\bibinfo {year} {2019})}\BibitemShut {NoStop}%
\bibitem [{\citenamefont {Sohn}\ \emph
  {et~al.}(2019{\natexlab{b}})\citenamefont {Sohn}, \citenamefont {Liu},
  \citenamefont {Wang},\ and\ \citenamefont
  {Smalyukh}}]{sohn2019lightcontrolled}%
  \BibitemOpen
  \bibfield  {author} {\bibinfo {author} {\bibfnamefont {H.~R.~O.}\
  \bibnamefont {Sohn}}, \bibinfo {author} {\bibfnamefont {C.~D.}\ \bibnamefont
  {Liu}}, \bibinfo {author} {\bibfnamefont {Y.}~\bibnamefont {Wang}},\ and\
  \bibinfo {author} {\bibfnamefont {I.~I.}\ \bibnamefont {Smalyukh}},\
  }\bibfield  {title} {\bibinfo {title} {Light-controlled skyrmions and torons
  as reconfigurable particles},\ }\href {https://doi.org/10.1364/OE.27.029055}
  {\bibfield  {journal} {\bibinfo  {journal} {Opt. Express}\ }\textbf {\bibinfo
  {volume} {27}},\ \bibinfo {pages} {29055} (\bibinfo {year}
  {2019}{\natexlab{b}})}\BibitemShut {NoStop}%
\bibitem [{\citenamefont {Bogdanov}\ and\ \citenamefont
  {Hubert}(1994)}]{bogdanov1994thermodynamically}%
  \BibitemOpen
  \bibfield  {author} {\bibinfo {author} {\bibfnamefont {A.}~\bibnamefont
  {Bogdanov}}\ and\ \bibinfo {author} {\bibfnamefont {A.}~\bibnamefont
  {Hubert}},\ }\bibfield  {title} {\bibinfo {title} {Thermodynamically stable
  magnetic vortex states in magnetic crystals},\ }\href@noop {} {\bibfield
  {journal} {\bibinfo  {journal} {J. Magn. Magn. Mater.}\ }\textbf {\bibinfo
  {volume} {138}},\ \bibinfo {pages} {255} (\bibinfo {year}
  {1994})}\BibitemShut {NoStop}%
\bibitem [{\citenamefont {Cheong}\ and\ \citenamefont
  {Xu}(2022)}]{cheong2022magnetic}%
  \BibitemOpen
  \bibfield  {author} {\bibinfo {author} {\bibfnamefont {S.-W.}\ \bibnamefont
  {Cheong}}\ and\ \bibinfo {author} {\bibfnamefont {X.}~\bibnamefont {Xu}},\
  }\bibfield  {title} {\bibinfo {title} {Magnetic chirality},\ }\href@noop {}
  {\bibfield  {journal} {\bibinfo  {journal} {npj Quantum Mater.}\ }\textbf
  {\bibinfo {volume} {7}},\ \bibinfo {pages} {40} (\bibinfo {year}
  {2022})}\BibitemShut {NoStop}%
\bibitem [{\citenamefont {Togawa}\ \emph {et~al.}(2012)\citenamefont {Togawa},
  \citenamefont {Koyama}, \citenamefont {Takayanagi}, \citenamefont {Mori},
  \citenamefont {Kousaka}, \citenamefont {Akimitsu}, \citenamefont {Nishihara},
  \citenamefont {Inoue}, \citenamefont {Ovchinnikov},\ and\ \citenamefont
  {Kishine}}]{togawa2012chiral}%
  \BibitemOpen
  \bibfield  {author} {\bibinfo {author} {\bibfnamefont {Y.}~\bibnamefont
  {Togawa}}, \bibinfo {author} {\bibfnamefont {T.}~\bibnamefont {Koyama}},
  \bibinfo {author} {\bibfnamefont {K.}~\bibnamefont {Takayanagi}}, \bibinfo
  {author} {\bibfnamefont {S.}~\bibnamefont {Mori}}, \bibinfo {author}
  {\bibfnamefont {Y.}~\bibnamefont {Kousaka}}, \bibinfo {author} {\bibfnamefont
  {J.}~\bibnamefont {Akimitsu}}, \bibinfo {author} {\bibfnamefont
  {S.}~\bibnamefont {Nishihara}}, \bibinfo {author} {\bibfnamefont
  {K.}~\bibnamefont {Inoue}}, \bibinfo {author} {\bibfnamefont
  {A.}~\bibnamefont {Ovchinnikov}},\ and\ \bibinfo {author} {\bibfnamefont
  {J.-i.}\ \bibnamefont {Kishine}},\ }\bibfield  {title} {\bibinfo {title}
  {Chiral magnetic soliton lattice on a chiral helimagnet},\ }\href@noop {}
  {\bibfield  {journal} {\bibinfo  {journal} {Phys. Rev. Lett.}\ }\textbf
  {\bibinfo {volume} {108}},\ \bibinfo {pages} {107202} (\bibinfo {year}
  {2012})}\BibitemShut {NoStop}%
\bibitem [{\citenamefont {Leonov}\ and\ \citenamefont
  {Mostovoy}(2015)}]{leonov2015multiply}%
  \BibitemOpen
  \bibfield  {author} {\bibinfo {author} {\bibfnamefont {A.}~\bibnamefont
  {Leonov}}\ and\ \bibinfo {author} {\bibfnamefont {M.}~\bibnamefont
  {Mostovoy}},\ }\bibfield  {title} {\bibinfo {title} {Multiply periodic states
  and isolated skyrmions in an anisotropic frustrated magnet},\ }\href@noop {}
  {\bibfield  {journal} {\bibinfo  {journal} {Nat. Commun.}\ }\textbf {\bibinfo
  {volume} {6}},\ \bibinfo {pages} {8275} (\bibinfo {year} {2015})}\BibitemShut
  {NoStop}%
\bibitem [{\citenamefont {Karube}\ \emph {et~al.}(2018)\citenamefont {Karube},
  \citenamefont {White}, \citenamefont {Morikawa}, \citenamefont {Dewhurst},
  \citenamefont {Cubitt}, \citenamefont {Kikkawa}, \citenamefont {Yu},
  \citenamefont {Tokunaga}, \citenamefont {Arima}, \citenamefont {R{\o}nnow}
  \emph {et~al.}}]{karube2018disordered}%
  \BibitemOpen
  \bibfield  {author} {\bibinfo {author} {\bibfnamefont {K.}~\bibnamefont
  {Karube}}, \bibinfo {author} {\bibfnamefont {J.~S.}\ \bibnamefont {White}},
  \bibinfo {author} {\bibfnamefont {D.}~\bibnamefont {Morikawa}}, \bibinfo
  {author} {\bibfnamefont {C.~D.}\ \bibnamefont {Dewhurst}}, \bibinfo {author}
  {\bibfnamefont {R.}~\bibnamefont {Cubitt}}, \bibinfo {author} {\bibfnamefont
  {A.}~\bibnamefont {Kikkawa}}, \bibinfo {author} {\bibfnamefont
  {X.}~\bibnamefont {Yu}}, \bibinfo {author} {\bibfnamefont {Y.}~\bibnamefont
  {Tokunaga}}, \bibinfo {author} {\bibfnamefont {T.-h.}\ \bibnamefont {Arima}},
  \bibinfo {author} {\bibfnamefont {H.~M.}\ \bibnamefont {R{\o}nnow}}, \emph
  {et~al.},\ }\bibfield  {title} {\bibinfo {title} {Disordered skyrmion phase
  stabilized by magnetic frustration in a chiral magnet},\ }\href@noop {}
  {\bibfield  {journal} {\bibinfo  {journal} {Sci. Adv.}\ }\textbf {\bibinfo
  {volume} {4}},\ \bibinfo {pages} {eaar7043} (\bibinfo {year}
  {2018})}\BibitemShut {NoStop}%
\bibitem [{\citenamefont {Chacon}\ \emph {et~al.}(2018)\citenamefont {Chacon},
  \citenamefont {Heinen}, \citenamefont {Halder}, \citenamefont {Bauer},
  \citenamefont {Simeth}, \citenamefont {M{\"u}hlbauer}, \citenamefont
  {Berger}, \citenamefont {Garst}, \citenamefont {Rosch},\ and\ \citenamefont
  {Pfleiderer}}]{chacon2018observation}%
  \BibitemOpen
  \bibfield  {author} {\bibinfo {author} {\bibfnamefont {A.}~\bibnamefont
  {Chacon}}, \bibinfo {author} {\bibfnamefont {L.}~\bibnamefont {Heinen}},
  \bibinfo {author} {\bibfnamefont {M.}~\bibnamefont {Halder}}, \bibinfo
  {author} {\bibfnamefont {A.}~\bibnamefont {Bauer}}, \bibinfo {author}
  {\bibfnamefont {W.}~\bibnamefont {Simeth}}, \bibinfo {author} {\bibfnamefont
  {S.}~\bibnamefont {M{\"u}hlbauer}}, \bibinfo {author} {\bibfnamefont
  {H.}~\bibnamefont {Berger}}, \bibinfo {author} {\bibfnamefont
  {M.}~\bibnamefont {Garst}}, \bibinfo {author} {\bibfnamefont
  {A.}~\bibnamefont {Rosch}},\ and\ \bibinfo {author} {\bibfnamefont
  {C.}~\bibnamefont {Pfleiderer}},\ }\bibfield  {title} {\bibinfo {title}
  {Observation of two independent skyrmion phases in a chiral magnetic
  material},\ }\href@noop {} {\bibfield  {journal} {\bibinfo  {journal} {Nat.
  Phys.}\ }\textbf {\bibinfo {volume} {14}},\ \bibinfo {pages} {936} (\bibinfo
  {year} {2018})}\BibitemShut {NoStop}%
\bibitem [{\citenamefont {Bannenberg}\ \emph {et~al.}(2019)\citenamefont
  {Bannenberg}, \citenamefont {Wilhelm}, \citenamefont {Cubitt}, \citenamefont
  {Labh}, \citenamefont {Schmidt}, \citenamefont {Leli{\`e}vre-Berna},
  \citenamefont {Pappas}, \citenamefont {Mostovoy},\ and\ \citenamefont
  {Leonov}}]{bannenberg2019multiple}%
  \BibitemOpen
  \bibfield  {author} {\bibinfo {author} {\bibfnamefont {L.~J.}\ \bibnamefont
  {Bannenberg}}, \bibinfo {author} {\bibfnamefont {H.}~\bibnamefont {Wilhelm}},
  \bibinfo {author} {\bibfnamefont {R.}~\bibnamefont {Cubitt}}, \bibinfo
  {author} {\bibfnamefont {A.}~\bibnamefont {Labh}}, \bibinfo {author}
  {\bibfnamefont {M.~P.}\ \bibnamefont {Schmidt}}, \bibinfo {author}
  {\bibfnamefont {E.}~\bibnamefont {Leli{\`e}vre-Berna}}, \bibinfo {author}
  {\bibfnamefont {C.}~\bibnamefont {Pappas}}, \bibinfo {author} {\bibfnamefont
  {M.}~\bibnamefont {Mostovoy}},\ and\ \bibinfo {author} {\bibfnamefont
  {A.~O.}\ \bibnamefont {Leonov}},\ }\bibfield  {title} {\bibinfo {title}
  {Multiple low-temperature skyrmionic states in a bulk chiral magnet},\
  }\href@noop {} {\bibfield  {journal} {\bibinfo  {journal} {npj Quantum
  Mater.}\ }\textbf {\bibinfo {volume} {4}},\ \bibinfo {pages} {11} (\bibinfo
  {year} {2019})}\BibitemShut {NoStop}%
\bibitem [{\citenamefont {Ukleev}\ \emph {et~al.}(2021)\citenamefont {Ukleev},
  \citenamefont {Karube}, \citenamefont {Derlet}, \citenamefont {Wang},
  \citenamefont {Luetkens}, \citenamefont {Morikawa}, \citenamefont {Kikkawa},
  \citenamefont {Mangin-Thro}, \citenamefont {Wildes}, \citenamefont {Yamasaki}
  \emph {et~al.}}]{ukleev2021frustration}%
  \BibitemOpen
  \bibfield  {author} {\bibinfo {author} {\bibfnamefont {V.}~\bibnamefont
  {Ukleev}}, \bibinfo {author} {\bibfnamefont {K.}~\bibnamefont {Karube}},
  \bibinfo {author} {\bibfnamefont {P.}~\bibnamefont {Derlet}}, \bibinfo
  {author} {\bibfnamefont {C.}~\bibnamefont {Wang}}, \bibinfo {author}
  {\bibfnamefont {H.}~\bibnamefont {Luetkens}}, \bibinfo {author}
  {\bibfnamefont {D.}~\bibnamefont {Morikawa}}, \bibinfo {author}
  {\bibfnamefont {A.}~\bibnamefont {Kikkawa}}, \bibinfo {author} {\bibfnamefont
  {L.}~\bibnamefont {Mangin-Thro}}, \bibinfo {author} {\bibfnamefont
  {A.}~\bibnamefont {Wildes}}, \bibinfo {author} {\bibfnamefont
  {Y.}~\bibnamefont {Yamasaki}}, \emph {et~al.},\ }\bibfield  {title} {\bibinfo
  {title} {Frustration-driven magnetic fluctuations as the origin of the
  low-temperature skyrmion phase in {Co$_7$Zn$_7$Mn$_6$}},\ }\href@noop {}
  {\bibfield  {journal} {\bibinfo  {journal} {npj Quantum Mater.}\ }\textbf
  {\bibinfo {volume} {6}},\ \bibinfo {pages} {40} (\bibinfo {year}
  {2021})}\BibitemShut {NoStop}%
\bibitem [{\citenamefont {Baral}\ \emph {et~al.}(2023)\citenamefont {Baral},
  \citenamefont {Utesov}, \citenamefont {Luo}, \citenamefont {Radu},
  \citenamefont {Magrez}, \citenamefont {White},\ and\ \citenamefont
  {Ukleev}}]{baral2023direct}%
  \BibitemOpen
  \bibfield  {author} {\bibinfo {author} {\bibfnamefont {P.~R.}\ \bibnamefont
  {Baral}}, \bibinfo {author} {\bibfnamefont {O.~I.}\ \bibnamefont {Utesov}},
  \bibinfo {author} {\bibfnamefont {C.}~\bibnamefont {Luo}}, \bibinfo {author}
  {\bibfnamefont {F.}~\bibnamefont {Radu}}, \bibinfo {author} {\bibfnamefont
  {A.}~\bibnamefont {Magrez}}, \bibinfo {author} {\bibfnamefont {J.~S.}\
  \bibnamefont {White}},\ and\ \bibinfo {author} {\bibfnamefont
  {V.}~\bibnamefont {Ukleev}},\ }\bibfield  {title} {\bibinfo {title} {Direct
  observation of exchange anisotropy in the helimagnetic insulator
  {Cu$_2$OSeO$_3$}},\ }\href@noop {} {\bibfield  {journal} {\bibinfo  {journal}
  {Phys. Rev. Res.}\ }\textbf {\bibinfo {volume} {5}},\ \bibinfo {pages}
  {L032019} (\bibinfo {year} {2023})}\BibitemShut {NoStop}%
\bibitem [{\citenamefont {Ukleev}\ \emph {et~al.}(2024)\citenamefont {Ukleev},
  \citenamefont {Utesov}, \citenamefont {Luo}, \citenamefont {Radu},
  \citenamefont {Wintz}, \citenamefont {Weigand}, \citenamefont {Finizio},
  \citenamefont {Winter}, \citenamefont {Tahn}, \citenamefont {Rellinghaus}
  \emph {et~al.}}]{ukleev2024competing}%
  \BibitemOpen
  \bibfield  {author} {\bibinfo {author} {\bibfnamefont {V.}~\bibnamefont
  {Ukleev}}, \bibinfo {author} {\bibfnamefont {O.~I.}\ \bibnamefont {Utesov}},
  \bibinfo {author} {\bibfnamefont {C.}~\bibnamefont {Luo}}, \bibinfo {author}
  {\bibfnamefont {F.}~\bibnamefont {Radu}}, \bibinfo {author} {\bibfnamefont
  {S.}~\bibnamefont {Wintz}}, \bibinfo {author} {\bibfnamefont
  {M.}~\bibnamefont {Weigand}}, \bibinfo {author} {\bibfnamefont
  {S.}~\bibnamefont {Finizio}}, \bibinfo {author} {\bibfnamefont
  {M.}~\bibnamefont {Winter}}, \bibinfo {author} {\bibfnamefont
  {A.}~\bibnamefont {Tahn}}, \bibinfo {author} {\bibfnamefont {B.}~\bibnamefont
  {Rellinghaus}}, \emph {et~al.},\ }\bibfield  {title} {\bibinfo {title}
  {Competing anisotropies in the chiral cubic magnet {Co$_8$Zn$_8$Mn$_4$}
  unveiled by resonant x-ray magnetic scattering},\ }\href@noop {} {\bibfield
  {journal} {\bibinfo  {journal} {Phys. Rev. B}\ }\textbf {\bibinfo {volume}
  {109}},\ \bibinfo {pages} {184415} (\bibinfo {year} {2024})}\BibitemShut
  {NoStop}%
\bibitem [{\citenamefont {Moody}\ \emph {et~al.}(2021)\citenamefont {Moody},
  \citenamefont {Nielsen}, \citenamefont {Wilson}, \citenamefont {Venero},
  \citenamefont {{\v{S}}tefan{\v{c}}i{\v{c}}}, \citenamefont {Balakrishnan},\
  and\ \citenamefont {Hatton}}]{moody2021experimental}%
  \BibitemOpen
  \bibfield  {author} {\bibinfo {author} {\bibfnamefont {S.}~\bibnamefont
  {Moody}}, \bibinfo {author} {\bibfnamefont {P.}~\bibnamefont {Nielsen}},
  \bibinfo {author} {\bibfnamefont {M.}~\bibnamefont {Wilson}}, \bibinfo
  {author} {\bibfnamefont {D.~A.}\ \bibnamefont {Venero}}, \bibinfo {author}
  {\bibfnamefont {A.}~\bibnamefont {{\v{S}}tefan{\v{c}}i{\v{c}}}}, \bibinfo
  {author} {\bibfnamefont {G.}~\bibnamefont {Balakrishnan}},\ and\ \bibinfo
  {author} {\bibfnamefont {P.}~\bibnamefont {Hatton}},\ }\bibfield  {title}
  {\bibinfo {title} {Experimental evidence of a change of exchange anisotropy
  sign with temperature in {Zn-substituted Cu$_2$OSeO$_3$}},\ }\href@noop {}
  {\bibfield  {journal} {\bibinfo  {journal} {Phys. Rev. Res.}\ }\textbf
  {\bibinfo {volume} {3}},\ \bibinfo {pages} {043149} (\bibinfo {year}
  {2021})}\BibitemShut {NoStop}%
\bibitem [{\citenamefont {Nagase}\ \emph {et~al.}(2021)\citenamefont {Nagase},
  \citenamefont {So}, \citenamefont {Yasui}, \citenamefont {Ishida},
  \citenamefont {Yoshida}, \citenamefont {Tanaka}, \citenamefont {Saitoh},
  \citenamefont {Ikarashi}, \citenamefont {Kawaguchi}, \citenamefont {Kuwahara}
  \emph {et~al.}}]{nagase2021observation}%
  \BibitemOpen
  \bibfield  {author} {\bibinfo {author} {\bibfnamefont {T.}~\bibnamefont
  {Nagase}}, \bibinfo {author} {\bibfnamefont {Y.-G.}\ \bibnamefont {So}},
  \bibinfo {author} {\bibfnamefont {H.}~\bibnamefont {Yasui}}, \bibinfo
  {author} {\bibfnamefont {T.}~\bibnamefont {Ishida}}, \bibinfo {author}
  {\bibfnamefont {H.~K.}\ \bibnamefont {Yoshida}}, \bibinfo {author}
  {\bibfnamefont {Y.}~\bibnamefont {Tanaka}}, \bibinfo {author} {\bibfnamefont
  {K.}~\bibnamefont {Saitoh}}, \bibinfo {author} {\bibfnamefont
  {N.}~\bibnamefont {Ikarashi}}, \bibinfo {author} {\bibfnamefont
  {Y.}~\bibnamefont {Kawaguchi}}, \bibinfo {author} {\bibfnamefont
  {M.}~\bibnamefont {Kuwahara}}, \emph {et~al.},\ }\bibfield  {title} {\bibinfo
  {title} {Observation of domain wall bimerons in chiral magnets},\ }\href@noop
  {} {\bibfield  {journal} {\bibinfo  {journal} {Nat. Commun.}\ }\textbf
  {\bibinfo {volume} {12}},\ \bibinfo {pages} {3490} (\bibinfo {year}
  {2021})}\BibitemShut {NoStop}%
\bibitem [{\citenamefont {Yu}\ \emph {et~al.}(2018)\citenamefont {Yu},
  \citenamefont {Koshibae}, \citenamefont {Tokunaga}, \citenamefont {Shibata},
  \citenamefont {Taguchi}, \citenamefont {Nagaosa},\ and\ \citenamefont
  {Tokura}}]{yu2018transformation}%
  \BibitemOpen
  \bibfield  {author} {\bibinfo {author} {\bibfnamefont {X.}~\bibnamefont
  {Yu}}, \bibinfo {author} {\bibfnamefont {W.}~\bibnamefont {Koshibae}},
  \bibinfo {author} {\bibfnamefont {Y.}~\bibnamefont {Tokunaga}}, \bibinfo
  {author} {\bibfnamefont {K.}~\bibnamefont {Shibata}}, \bibinfo {author}
  {\bibfnamefont {Y.}~\bibnamefont {Taguchi}}, \bibinfo {author} {\bibfnamefont
  {N.}~\bibnamefont {Nagaosa}},\ and\ \bibinfo {author} {\bibfnamefont
  {Y.}~\bibnamefont {Tokura}},\ }\bibfield  {title} {\bibinfo {title}
  {Transformation between meron and skyrmion topological spin textures in a
  chiral magnet},\ }\href@noop {} {\bibfield  {journal} {\bibinfo  {journal}
  {Nature}\ }\textbf {\bibinfo {volume} {564}},\ \bibinfo {pages} {95}
  (\bibinfo {year} {2018})}\BibitemShut {NoStop}%
\bibitem [{\citenamefont {White}\ \emph {et~al.}(2025)\citenamefont {White},
  \citenamefont {Ukleev}, \citenamefont {Yu}, \citenamefont {Tokura},
  \citenamefont {Taguchi},\ and\ \citenamefont {Karube}}]{white2025anisotropy}%
  \BibitemOpen
  \bibfield  {author} {\bibinfo {author} {\bibfnamefont {J.~S.}\ \bibnamefont
  {White}}, \bibinfo {author} {\bibfnamefont {V.}~\bibnamefont {Ukleev}},
  \bibinfo {author} {\bibfnamefont {L.}~\bibnamefont {Yu}}, \bibinfo {author}
  {\bibfnamefont {Y.}~\bibnamefont {Tokura}}, \bibinfo {author} {\bibfnamefont
  {Y.}~\bibnamefont {Taguchi}},\ and\ \bibinfo {author} {\bibfnamefont
  {K.}~\bibnamefont {Karube}},\ }\bibfield  {title} {\bibinfo {title}
  {Anisotropy-dependent decay of room temperature metastable skyrmions and a
  nascent double-$q$ spin texture in co$_8$zn$_9$mn$_3$},\ }\href@noop {}
  {\bibfield  {journal} {\bibinfo  {journal} {Adv. Mater.}\ ,\ \bibinfo {pages}
  {2501146}} (\bibinfo {year} {2025})}\BibitemShut {NoStop}%
\bibitem [{\citenamefont {Psaroudaki}\ and\ \citenamefont
  {Panagopoulos}(2021)}]{psaroudaki2021skyrmion}%
  \BibitemOpen
  \bibfield  {author} {\bibinfo {author} {\bibfnamefont {C.}~\bibnamefont
  {Psaroudaki}}\ and\ \bibinfo {author} {\bibfnamefont {C.}~\bibnamefont
  {Panagopoulos}},\ }\bibfield  {title} {\bibinfo {title} {Skyrmion qubits: A
  new class of quantum logic elements based on nanoscale magnetization},\
  }\href@noop {} {\bibfield  {journal} {\bibinfo  {journal} {Phys. Rev. Lett.}\
  }\textbf {\bibinfo {volume} {127}},\ \bibinfo {pages} {067201} (\bibinfo
  {year} {2021})}\BibitemShut {NoStop}%
\bibitem [{\citenamefont {Ado}\ \emph {et~al.}(2021)\citenamefont {Ado},
  \citenamefont {Tchernyshyov},\ and\ \citenamefont
  {Titov}}]{ado2021noncollinear}%
  \BibitemOpen
  \bibfield  {author} {\bibinfo {author} {\bibfnamefont {I.}~\bibnamefont
  {Ado}}, \bibinfo {author} {\bibfnamefont {O.}~\bibnamefont {Tchernyshyov}},\
  and\ \bibinfo {author} {\bibfnamefont {M.}~\bibnamefont {Titov}},\ }\bibfield
   {title} {\bibinfo {title} {Noncollinear ground state from a four-spin chiral
  exchange in a tetrahedral magnet},\ }\href@noop {} {\bibfield  {journal}
  {\bibinfo  {journal} {Phys. Rev. Lett.}\ }\textbf {\bibinfo {volume} {127}},\
  \bibinfo {pages} {127204} (\bibinfo {year} {2021})}\BibitemShut {NoStop}%
\bibitem [{\citenamefont {Rybakov}\ \emph {et~al.}(2021)\citenamefont
  {Rybakov}, \citenamefont {Pervishko}, \citenamefont {Eriksson},\ and\
  \citenamefont {Babaev}}]{rybakov2021antichiral}%
  \BibitemOpen
  \bibfield  {author} {\bibinfo {author} {\bibfnamefont {F.~N.}\ \bibnamefont
  {Rybakov}}, \bibinfo {author} {\bibfnamefont {A.}~\bibnamefont {Pervishko}},
  \bibinfo {author} {\bibfnamefont {O.}~\bibnamefont {Eriksson}},\ and\
  \bibinfo {author} {\bibfnamefont {E.}~\bibnamefont {Babaev}},\ }\bibfield
  {title} {\bibinfo {title} {Antichiral ferromagnetism},\ }\href@noop {}
  {\bibfield  {journal} {\bibinfo  {journal} {Phys. Rev. B}\ }\textbf {\bibinfo
  {volume} {104}},\ \bibinfo {pages} {L020406} (\bibinfo {year}
  {2021})}\BibitemShut {NoStop}%
\bibitem [{\citenamefont {Yasin}\ \emph {et~al.}(2024)\citenamefont {Yasin},
  \citenamefont {Masell}, \citenamefont {Takahashi}, \citenamefont {Akashi},
  \citenamefont {Baba}, \citenamefont {Karube}, \citenamefont {Shindo},
  \citenamefont {Arima}, \citenamefont {Taguchi}, \citenamefont {Tokura},
  \citenamefont {Tanigaki},\ and\ \citenamefont {Yu}}]{Yasin2024}%
  \BibitemOpen
  \bibfield  {author} {\bibinfo {author} {\bibfnamefont {F.~S.}\ \bibnamefont
  {Yasin}}, \bibinfo {author} {\bibfnamefont {J.}~\bibnamefont {Masell}},
  \bibinfo {author} {\bibfnamefont {Y.}~\bibnamefont {Takahashi}}, \bibinfo
  {author} {\bibfnamefont {T.}~\bibnamefont {Akashi}}, \bibinfo {author}
  {\bibfnamefont {N.}~\bibnamefont {Baba}}, \bibinfo {author} {\bibfnamefont
  {K.}~\bibnamefont {Karube}}, \bibinfo {author} {\bibfnamefont
  {D.}~\bibnamefont {Shindo}}, \bibinfo {author} {\bibfnamefont
  {T.}~\bibnamefont {Arima}}, \bibinfo {author} {\bibfnamefont
  {Y.}~\bibnamefont {Taguchi}}, \bibinfo {author} {\bibfnamefont
  {Y.}~\bibnamefont {Tokura}}, \bibinfo {author} {\bibfnamefont
  {T.}~\bibnamefont {Tanigaki}},\ and\ \bibinfo {author} {\bibfnamefont
  {X.}~\bibnamefont {Yu}},\ }\bibfield  {title} {\bibinfo {title} {Bloch point
  quadrupole constituting hybrid topological strings revealed with electron
  holographic vector field tomography},\ }\href
  {https://doi.org/10.1002/adma.202311737} {\bibfield  {journal} {\bibinfo
  {journal} {Adv. Mater.}\ }\textbf {\bibinfo {volume} {36}},\ \bibinfo {pages}
  {2311737} (\bibinfo {year} {2024})}\BibitemShut {NoStop}%
\bibitem [{\citenamefont {Guang}\ \emph {et~al.}(2024)\citenamefont {Guang},
  \citenamefont {Zhang}, \citenamefont {Liu}, \citenamefont {Peng},
  \citenamefont {Yasin}, \citenamefont {Karube}, \citenamefont {Nakamura},
  \citenamefont {Nagaosa}, \citenamefont {Taguchi}, \citenamefont {Mochizuki}
  \emph {et~al.}}]{guang2024confined}%
  \BibitemOpen
  \bibfield  {author} {\bibinfo {author} {\bibfnamefont {Y.}~\bibnamefont
  {Guang}}, \bibinfo {author} {\bibfnamefont {X.}~\bibnamefont {Zhang}},
  \bibinfo {author} {\bibfnamefont {Y.}~\bibnamefont {Liu}}, \bibinfo {author}
  {\bibfnamefont {L.}~\bibnamefont {Peng}}, \bibinfo {author} {\bibfnamefont
  {F.~S.}\ \bibnamefont {Yasin}}, \bibinfo {author} {\bibfnamefont
  {K.}~\bibnamefont {Karube}}, \bibinfo {author} {\bibfnamefont
  {D.}~\bibnamefont {Nakamura}}, \bibinfo {author} {\bibfnamefont
  {N.}~\bibnamefont {Nagaosa}}, \bibinfo {author} {\bibfnamefont
  {Y.}~\bibnamefont {Taguchi}}, \bibinfo {author} {\bibfnamefont
  {M.}~\bibnamefont {Mochizuki}}, \emph {et~al.},\ }\bibfield  {title}
  {\bibinfo {title} {Confined antiskyrmion motion driven by electric current
  excitations},\ }\href@noop {} {\bibfield  {journal} {\bibinfo  {journal}
  {Nat. Commun.}\ }\textbf {\bibinfo {volume} {15}},\ \bibinfo {pages} {7701}
  (\bibinfo {year} {2024})}\BibitemShut {NoStop}%
\bibitem [{\citenamefont {Yasin}\ \emph {et~al.}(2023)\citenamefont {Yasin},
  \citenamefont {Masell}, \citenamefont {Karube}, \citenamefont {Shindo},
  \citenamefont {Taguchi}, \citenamefont {Tokura},\ and\ \citenamefont
  {Yu}}]{Yasin2023}%
  \BibitemOpen
  \bibfield  {author} {\bibinfo {author} {\bibfnamefont {F.~S.}\ \bibnamefont
  {Yasin}}, \bibinfo {author} {\bibfnamefont {J.}~\bibnamefont {Masell}},
  \bibinfo {author} {\bibfnamefont {K.}~\bibnamefont {Karube}}, \bibinfo
  {author} {\bibfnamefont {D.}~\bibnamefont {Shindo}}, \bibinfo {author}
  {\bibfnamefont {Y.}~\bibnamefont {Taguchi}}, \bibinfo {author} {\bibfnamefont
  {Y.}~\bibnamefont {Tokura}},\ and\ \bibinfo {author} {\bibfnamefont
  {X.}~\bibnamefont {Yu}},\ }\bibfield  {title} {\bibinfo {title} {Heat
  current-driven topological spin texture transformations and helical q-vector
  switching},\ }\href {https://doi.org/10.1038/s41467-023-42846-7} {\bibfield
  {journal} {\bibinfo  {journal} {Nat. Commun.}\ }\textbf {\bibinfo {volume}
  {14}},\ \bibinfo {pages} {7094} (\bibinfo {year} {2023})}\BibitemShut
  {NoStop}%
\bibitem [{\citenamefont {Camosi}\ \emph {et~al.}(2018)\citenamefont {Camosi},
  \citenamefont {Rougemaille}, \citenamefont {Fruchart}, \citenamefont
  {Vogel},\ and\ \citenamefont {Rohart}}]{Camosi2018}%
  \BibitemOpen
  \bibfield  {author} {\bibinfo {author} {\bibfnamefont {L.}~\bibnamefont
  {Camosi}}, \bibinfo {author} {\bibfnamefont {N.}~\bibnamefont {Rougemaille}},
  \bibinfo {author} {\bibfnamefont {O.}~\bibnamefont {Fruchart}}, \bibinfo
  {author} {\bibfnamefont {J.}~\bibnamefont {Vogel}},\ and\ \bibinfo {author}
  {\bibfnamefont {S.}~\bibnamefont {Rohart}},\ }\bibfield  {title} {\bibinfo
  {title} {Micromagnetics of antiskyrmions in ultrathin films},\ }\href
  {https://doi.org/10.1103/PhysRevB.97.134404} {\bibfield  {journal} {\bibinfo
  {journal} {Phys. Rev. B}\ }\textbf {\bibinfo {volume} {97}},\ \bibinfo
  {pages} {134404} (\bibinfo {year} {2018})}\BibitemShut {NoStop}%
\bibitem [{\citenamefont {Mori}\ \emph {et~al.}(2025)\citenamefont {Mori},
  \citenamefont {Ii}, \citenamefont {Moronaga}, \citenamefont {Hara},
  \citenamefont {Karube}, \citenamefont {Taguchi}, \citenamefont {Tokura},\
  and\ \citenamefont {Yu}}]{Mori2025}%
  \BibitemOpen
  \bibfield  {author} {\bibinfo {author} {\bibfnamefont {S.}~\bibnamefont
  {Mori}}, \bibinfo {author} {\bibfnamefont {S.}~\bibnamefont {Ii}}, \bibinfo
  {author} {\bibfnamefont {T.}~\bibnamefont {Moronaga}}, \bibinfo {author}
  {\bibfnamefont {T.}~\bibnamefont {Hara}}, \bibinfo {author} {\bibfnamefont
  {K.}~\bibnamefont {Karube}}, \bibinfo {author} {\bibfnamefont
  {Y.}~\bibnamefont {Taguchi}}, \bibinfo {author} {\bibfnamefont
  {Y.}~\bibnamefont {Tokura}},\ and\ \bibinfo {author} {\bibfnamefont
  {X.}~\bibnamefont {Yu}},\ }\bibfield  {title} {\bibinfo {title} {Direct
  {Observations} of {Mechanical} {Strain}-{Induced} {Wavevector} {Switching} in
  a ({Fe},{Ni},{Pd}){3P} {Magnet} with {Anisotropic}
  {Dzyaloshinskii}–{Moriya} {Interaction}},\ }\href
  {https://doi.org/10.1021/acsami.5c00625} {\bibfield  {journal} {\bibinfo
  {journal} {ACS Appl. Mater. Interfaces}\ }\textbf {\bibinfo {volume} {17}},\
  \bibinfo {pages} {22921} (\bibinfo {year} {2025})},\ \bibinfo {note}
  {publisher: American Chemical Society}\BibitemShut {NoStop}%
\bibitem [{\citenamefont {Hemmida}\ \emph {et~al.}(2024)\citenamefont
  {Hemmida}, \citenamefont {Masell}, \citenamefont {Karube}, \citenamefont
  {Ehlers}, \citenamefont {von Nidda}, \citenamefont {Tsurkan}, \citenamefont
  {Tokura}, \citenamefont {Taguchi},\ and\ \citenamefont
  {K\'ezsm\'arki}}]{Hemmida2024}%
  \BibitemOpen
  \bibfield  {author} {\bibinfo {author} {\bibfnamefont {M.}~\bibnamefont
  {Hemmida}}, \bibinfo {author} {\bibfnamefont {J.}~\bibnamefont {Masell}},
  \bibinfo {author} {\bibfnamefont {K.}~\bibnamefont {Karube}}, \bibinfo
  {author} {\bibfnamefont {D.}~\bibnamefont {Ehlers}}, \bibinfo {author}
  {\bibfnamefont {H.-A.~K.}\ \bibnamefont {von Nidda}}, \bibinfo {author}
  {\bibfnamefont {V.}~\bibnamefont {Tsurkan}}, \bibinfo {author} {\bibfnamefont
  {Y.}~\bibnamefont {Tokura}}, \bibinfo {author} {\bibfnamefont
  {Y.}~\bibnamefont {Taguchi}},\ and\ \bibinfo {author} {\bibfnamefont
  {I.}~\bibnamefont {K\'ezsm\'arki}},\ }\bibfield  {title} {\bibinfo {title}
  {Role of magnetic anisotropy in the antiskyrmion-host schreibersite
  magnets},\ }\href {https://doi.org/10.1103/PhysRevB.110.054416} {\bibfield
  {journal} {\bibinfo  {journal} {Phys. Rev. B}\ }\textbf {\bibinfo {volume}
  {110}},\ \bibinfo {pages} {054416} (\bibinfo {year} {2024})}\BibitemShut
  {NoStop}%
\bibitem [{\citenamefont {Karube}\ \emph {et~al.}(2022)\citenamefont {Karube},
  \citenamefont {Peng}, \citenamefont {Masell}, \citenamefont {Hemmida},
  \citenamefont {{Krug von Nidda}}, \citenamefont {K{\'e}zsm{\'a}rki},
  \citenamefont {Yu}, \citenamefont {Tokura},\ and\ \citenamefont
  {Taguchi}}]{karube2022doping}%
  \BibitemOpen
  \bibfield  {author} {\bibinfo {author} {\bibfnamefont {K.}~\bibnamefont
  {Karube}}, \bibinfo {author} {\bibfnamefont {L.}~\bibnamefont {Peng}},
  \bibinfo {author} {\bibfnamefont {J.}~\bibnamefont {Masell}}, \bibinfo
  {author} {\bibfnamefont {M.}~\bibnamefont {Hemmida}}, \bibinfo {author}
  {\bibfnamefont {H.-A.}\ \bibnamefont {{Krug von Nidda}}}, \bibinfo {author}
  {\bibfnamefont {I.}~\bibnamefont {K{\'e}zsm{\'a}rki}}, \bibinfo {author}
  {\bibfnamefont {X.}~\bibnamefont {Yu}}, \bibinfo {author} {\bibfnamefont
  {Y.}~\bibnamefont {Tokura}},\ and\ \bibinfo {author} {\bibfnamefont
  {Y.}~\bibnamefont {Taguchi}},\ }\bibfield  {title} {\bibinfo {title} {Doping
  {{Control}} of {{Magnetic Anisotropy}} for {{Stable Antiskyrmion Formation}}
  in {{Schreibersite}} ({{Fe}},{{Ni}}){{3P}} with {{S4}} symmetry},\ }\href
  {https://doi.org/10.1002/adma.202108770} {\bibfield  {journal} {\bibinfo
  {journal} {Adv. Mater.}\ }\textbf {\bibinfo {volume} {34}},\ \bibinfo {pages}
  {2108770} (\bibinfo {year} {2022})}\BibitemShut {NoStop}%
\bibitem [{\citenamefont {Yu}\ \emph {et~al.}(2022)\citenamefont {Yu},
  \citenamefont {Iakoubovskii}, \citenamefont {Yasin}, \citenamefont {Peng},
  \citenamefont {Nakajima}, \citenamefont {Schneider}, \citenamefont {Karube},
  \citenamefont {Arima}, \citenamefont {Taguchi},\ and\ \citenamefont
  {Tokura}}]{Yu2022}%
  \BibitemOpen
  \bibfield  {author} {\bibinfo {author} {\bibfnamefont {X.}~\bibnamefont
  {Yu}}, \bibinfo {author} {\bibfnamefont {K.~V.}\ \bibnamefont
  {Iakoubovskii}}, \bibinfo {author} {\bibfnamefont {F.~S.}\ \bibnamefont
  {Yasin}}, \bibinfo {author} {\bibfnamefont {L.}~\bibnamefont {Peng}},
  \bibinfo {author} {\bibfnamefont {K.}~\bibnamefont {Nakajima}}, \bibinfo
  {author} {\bibfnamefont {S.}~\bibnamefont {Schneider}}, \bibinfo {author}
  {\bibfnamefont {K.}~\bibnamefont {Karube}}, \bibinfo {author} {\bibfnamefont
  {T.}~\bibnamefont {Arima}}, \bibinfo {author} {\bibfnamefont
  {Y.}~\bibnamefont {Taguchi}},\ and\ \bibinfo {author} {\bibfnamefont
  {Y.}~\bibnamefont {Tokura}},\ }\bibfield  {title} {\bibinfo {title}
  {Real-{Space} {Observations} of {Three}-{Dimensional} {Antiskyrmions} and
  {Skyrmion} {Strings}},\ }\href {https://doi.org/10.1021/acs.nanolett.2c03142}
  {\bibfield  {journal} {\bibinfo  {journal} {Nano Lett.}\ }\textbf {\bibinfo
  {volume} {22}},\ \bibinfo {pages} {9358} (\bibinfo {year}
  {2022})}\BibitemShut {NoStop}%
\bibitem [{\citenamefont {Peng}\ \emph {et~al.}(2025)\citenamefont {Peng},
  \citenamefont {Yasin}, \citenamefont {Karube}, \citenamefont {Kanazawa},
  \citenamefont {Taguchi}, \citenamefont {Tokura},\ and\ \citenamefont
  {Yu}}]{Peng2025}%
  \BibitemOpen
  \bibfield  {author} {\bibinfo {author} {\bibfnamefont {L.}~\bibnamefont
  {Peng}}, \bibinfo {author} {\bibfnamefont {F.~S.}\ \bibnamefont {Yasin}},
  \bibinfo {author} {\bibfnamefont {K.}~\bibnamefont {Karube}}, \bibinfo
  {author} {\bibfnamefont {N.}~\bibnamefont {Kanazawa}}, \bibinfo {author}
  {\bibfnamefont {Y.}~\bibnamefont {Taguchi}}, \bibinfo {author} {\bibfnamefont
  {Y.}~\bibnamefont {Tokura}},\ and\ \bibinfo {author} {\bibfnamefont
  {X.}~\bibnamefont {Yu}},\ }\bibfield  {title} {\bibinfo {title} {In-situ
  l-tem observations of dynamics of nanometric skyrmions and antiskyrmions},\
  }\href {https://doi.org/10.1016/j.nantod.2025.102698} {\bibfield  {journal}
  {\bibinfo  {journal} {Nano Today}\ }\textbf {\bibinfo {volume} {62}},\
  \bibinfo {pages} {102698} (\bibinfo {year} {2025})}\BibitemShut {NoStop}%
\bibitem [{\citenamefont {Han}\ \emph {et~al.}(2019)\citenamefont {Han},
  \citenamefont {Garlow}, \citenamefont {Liu}, \citenamefont {Zhang},
  \citenamefont {Li}, \citenamefont {DiMarzio}, \citenamefont {Knight},
  \citenamefont {Petrovic}, \citenamefont {Jariwala},\ and\ \citenamefont
  {Zhu}}]{han2019topological}%
  \BibitemOpen
  \bibfield  {author} {\bibinfo {author} {\bibfnamefont {M.-G.}\ \bibnamefont
  {Han}}, \bibinfo {author} {\bibfnamefont {J.~A.}\ \bibnamefont {Garlow}},
  \bibinfo {author} {\bibfnamefont {Y.}~\bibnamefont {Liu}}, \bibinfo {author}
  {\bibfnamefont {H.}~\bibnamefont {Zhang}}, \bibinfo {author} {\bibfnamefont
  {J.}~\bibnamefont {Li}}, \bibinfo {author} {\bibfnamefont {D.}~\bibnamefont
  {DiMarzio}}, \bibinfo {author} {\bibfnamefont {M.~W.}\ \bibnamefont
  {Knight}}, \bibinfo {author} {\bibfnamefont {C.}~\bibnamefont {Petrovic}},
  \bibinfo {author} {\bibfnamefont {D.}~\bibnamefont {Jariwala}},\ and\
  \bibinfo {author} {\bibfnamefont {Y.}~\bibnamefont {Zhu}},\ }\bibfield
  {title} {\bibinfo {title} {Topological {{Magnetic-Spin Textures}} in
  {{Two-Dimensional}} van der {{Waals Cr2Ge2Te6}}},\ }\href
  {https://doi.org/10.1021/acs.nanolett.9b02849} {\bibfield  {journal}
  {\bibinfo  {journal} {Nano Lett.}\ }\textbf {\bibinfo {volume} {19}},\
  \bibinfo {pages} {7859} (\bibinfo {year} {2019})}\BibitemShut {NoStop}%
\bibitem [{\citenamefont {Liu}\ \emph {et~al.}(2023)\citenamefont {Liu},
  \citenamefont {Jiang}, \citenamefont {Zhang}, \citenamefont {Wang},
  \citenamefont {Zhang}, \citenamefont {Zheng}, \citenamefont {Li},
  \citenamefont {Ma}, \citenamefont {Algaidi}, \citenamefont {Gao} \emph
  {et~al.}}]{liu2023controllable}%
  \BibitemOpen
  \bibfield  {author} {\bibinfo {author} {\bibfnamefont {C.}~\bibnamefont
  {Liu}}, \bibinfo {author} {\bibfnamefont {J.}~\bibnamefont {Jiang}}, \bibinfo
  {author} {\bibfnamefont {C.}~\bibnamefont {Zhang}}, \bibinfo {author}
  {\bibfnamefont {Q.}~\bibnamefont {Wang}}, \bibinfo {author} {\bibfnamefont
  {H.}~\bibnamefont {Zhang}}, \bibinfo {author} {\bibfnamefont
  {D.}~\bibnamefont {Zheng}}, \bibinfo {author} {\bibfnamefont
  {Y.}~\bibnamefont {Li}}, \bibinfo {author} {\bibfnamefont {Y.}~\bibnamefont
  {Ma}}, \bibinfo {author} {\bibfnamefont {H.}~\bibnamefont {Algaidi}},
  \bibinfo {author} {\bibfnamefont {X.}~\bibnamefont {Gao}}, \emph {et~al.},\
  }\bibfield  {title} {\bibinfo {title} {Controllable skyrmionic phase
  transition between n{\'e}el skyrmions and bloch skyrmionic bubbles in van der
  waals ferromagnet fe3-$\delta$gete2},\ }\href@noop {} {\bibfield  {journal}
  {\bibinfo  {journal} {Advanced Science}\ }\textbf {\bibinfo {volume} {10}},\
  \bibinfo {pages} {2303443} (\bibinfo {year} {2023})}\BibitemShut {NoStop}%
\bibitem [{\citenamefont {Ding}\ \emph {et~al.}(2020)\citenamefont {Ding},
  \citenamefont {Li}, \citenamefont {Xu}, \citenamefont {Li}, \citenamefont
  {Hou}, \citenamefont {Liu}, \citenamefont {Xi}, \citenamefont {Xu},
  \citenamefont {Yao},\ and\ \citenamefont {Wang}}]{ding2020observation}%
  \BibitemOpen
  \bibfield  {author} {\bibinfo {author} {\bibfnamefont {B.}~\bibnamefont
  {Ding}}, \bibinfo {author} {\bibfnamefont {Z.}~\bibnamefont {Li}}, \bibinfo
  {author} {\bibfnamefont {G.}~\bibnamefont {Xu}}, \bibinfo {author}
  {\bibfnamefont {H.}~\bibnamefont {Li}}, \bibinfo {author} {\bibfnamefont
  {Z.}~\bibnamefont {Hou}}, \bibinfo {author} {\bibfnamefont {E.}~\bibnamefont
  {Liu}}, \bibinfo {author} {\bibfnamefont {X.}~\bibnamefont {Xi}}, \bibinfo
  {author} {\bibfnamefont {F.}~\bibnamefont {Xu}}, \bibinfo {author}
  {\bibfnamefont {Y.}~\bibnamefont {Yao}},\ and\ \bibinfo {author}
  {\bibfnamefont {W.}~\bibnamefont {Wang}},\ }\bibfield  {title} {\bibinfo
  {title} {Observation of {{Magnetic Skyrmion Bubbles}} in a van der {{Waals
  Ferromagnet Fe3GeTe2}}},\ }\href
  {https://doi.org/10.1021/acs.nanolett.9b03453} {\bibfield  {journal}
  {\bibinfo  {journal} {Nano Lett.}\ }\textbf {\bibinfo {volume} {20}},\
  \bibinfo {pages} {868} (\bibinfo {year} {2020})}\BibitemShut {NoStop}%
\bibitem [{\citenamefont {Park}\ \emph {et~al.}(2021)\citenamefont {Park},
  \citenamefont {Peng}, \citenamefont {Liang}, \citenamefont {Hallal},
  \citenamefont {Yasin}, \citenamefont {Zhang}, \citenamefont {Song},
  \citenamefont {Kim}, \citenamefont {Kim}, \citenamefont {Weigand},
  \citenamefont {Sch{\"u}tz}, \citenamefont {Finizio}, \citenamefont {Raabe},
  \citenamefont {Garcia}, \citenamefont {Xia}, \citenamefont {Zhou},
  \citenamefont {Ezawa}, \citenamefont {Liu}, \citenamefont {Chang},
  \citenamefont {Koo}, \citenamefont {Kim}, \citenamefont {Chshiev},
  \citenamefont {Fert}, \citenamefont {Yang}, \citenamefont {Yu},\ and\
  \citenamefont {Woo}}]{park2021neeltype}%
  \BibitemOpen
  \bibfield  {author} {\bibinfo {author} {\bibfnamefont {T.-E.}\ \bibnamefont
  {Park}}, \bibinfo {author} {\bibfnamefont {L.}~\bibnamefont {Peng}}, \bibinfo
  {author} {\bibfnamefont {J.}~\bibnamefont {Liang}}, \bibinfo {author}
  {\bibfnamefont {A.}~\bibnamefont {Hallal}}, \bibinfo {author} {\bibfnamefont
  {F.~S.}\ \bibnamefont {Yasin}}, \bibinfo {author} {\bibfnamefont
  {X.}~\bibnamefont {Zhang}}, \bibinfo {author} {\bibfnamefont {K.~M.}\
  \bibnamefont {Song}}, \bibinfo {author} {\bibfnamefont {S.~J.}\ \bibnamefont
  {Kim}}, \bibinfo {author} {\bibfnamefont {K.}~\bibnamefont {Kim}}, \bibinfo
  {author} {\bibfnamefont {M.}~\bibnamefont {Weigand}}, \bibinfo {author}
  {\bibfnamefont {G.}~\bibnamefont {Sch{\"u}tz}}, \bibinfo {author}
  {\bibfnamefont {S.}~\bibnamefont {Finizio}}, \bibinfo {author} {\bibfnamefont
  {J.}~\bibnamefont {Raabe}}, \bibinfo {author} {\bibfnamefont
  {K.}~\bibnamefont {Garcia}}, \bibinfo {author} {\bibfnamefont
  {J.}~\bibnamefont {Xia}}, \bibinfo {author} {\bibfnamefont {Y.}~\bibnamefont
  {Zhou}}, \bibinfo {author} {\bibfnamefont {M.}~\bibnamefont {Ezawa}},
  \bibinfo {author} {\bibfnamefont {X.}~\bibnamefont {Liu}}, \bibinfo {author}
  {\bibfnamefont {J.}~\bibnamefont {Chang}}, \bibinfo {author} {\bibfnamefont
  {H.~C.}\ \bibnamefont {Koo}}, \bibinfo {author} {\bibfnamefont {Y.~D.}\
  \bibnamefont {Kim}}, \bibinfo {author} {\bibfnamefont {M.}~\bibnamefont
  {Chshiev}}, \bibinfo {author} {\bibfnamefont {A.}~\bibnamefont {Fert}},
  \bibinfo {author} {\bibfnamefont {H.}~\bibnamefont {Yang}}, \bibinfo {author}
  {\bibfnamefont {X.}~\bibnamefont {Yu}},\ and\ \bibinfo {author}
  {\bibfnamefont {S.}~\bibnamefont {Woo}},\ }\bibfield  {title} {\bibinfo
  {title} {N{\textbackslash}'eel-type skyrmions and their current-induced
  motion in van der {{Waals}} ferromagnet-based heterostructures},\ }\href
  {https://doi.org/10.1103/PhysRevB.103.104410} {\bibfield  {journal} {\bibinfo
   {journal} {Phys. Rev. B}\ }\textbf {\bibinfo {volume} {103}},\ \bibinfo
  {pages} {104410} (\bibinfo {year} {2021})}\BibitemShut {NoStop}%
\bibitem [{\citenamefont {May}\ \emph {et~al.}(2019)\citenamefont {May},
  \citenamefont {Ovchinnikov}, \citenamefont {Zheng}, \citenamefont {Hermann},
  \citenamefont {Calder}, \citenamefont {Huang}, \citenamefont {Fei},
  \citenamefont {Liu}, \citenamefont {Xu},\ and\ \citenamefont
  {McGuire}}]{may2019ferromagnetism}%
  \BibitemOpen
  \bibfield  {author} {\bibinfo {author} {\bibfnamefont {A.~F.}\ \bibnamefont
  {May}}, \bibinfo {author} {\bibfnamefont {D.}~\bibnamefont {Ovchinnikov}},
  \bibinfo {author} {\bibfnamefont {Q.}~\bibnamefont {Zheng}}, \bibinfo
  {author} {\bibfnamefont {R.}~\bibnamefont {Hermann}}, \bibinfo {author}
  {\bibfnamefont {S.}~\bibnamefont {Calder}}, \bibinfo {author} {\bibfnamefont
  {B.}~\bibnamefont {Huang}}, \bibinfo {author} {\bibfnamefont
  {Z.}~\bibnamefont {Fei}}, \bibinfo {author} {\bibfnamefont {Y.}~\bibnamefont
  {Liu}}, \bibinfo {author} {\bibfnamefont {X.}~\bibnamefont {Xu}},\ and\
  \bibinfo {author} {\bibfnamefont {M.~A.}\ \bibnamefont {McGuire}},\
  }\bibfield  {title} {\bibinfo {title} {Ferromagnetism {{Near Room
  Temperature}} in the {{Cleavable}} van der {{Waals Crystal Fe5GeTe2}}},\
  }\href {https://doi.org/10.1021/acsnano.8b09660} {\bibfield  {journal}
  {\bibinfo  {journal} {ACS Nano}\ }\textbf {\bibinfo {volume} {13}},\ \bibinfo
  {pages} {4436} (\bibinfo {year} {2019})}\BibitemShut {NoStop}%
\bibitem [{\citenamefont {Gao}\ \emph {et~al.}(2020)\citenamefont {Gao},
  \citenamefont {Yin}, \citenamefont {Wang}, \citenamefont {Li}, \citenamefont
  {Cai}, \citenamefont {Zhao}, \citenamefont {Lei}, \citenamefont {Wang},
  \citenamefont {Zhang},\ and\ \citenamefont {Shen}}]{gao2020spontaneous}%
  \BibitemOpen
  \bibfield  {author} {\bibinfo {author} {\bibfnamefont {Y.}~\bibnamefont
  {Gao}}, \bibinfo {author} {\bibfnamefont {Q.}~\bibnamefont {Yin}}, \bibinfo
  {author} {\bibfnamefont {Q.}~\bibnamefont {Wang}}, \bibinfo {author}
  {\bibfnamefont {Z.}~\bibnamefont {Li}}, \bibinfo {author} {\bibfnamefont
  {J.}~\bibnamefont {Cai}}, \bibinfo {author} {\bibfnamefont {T.}~\bibnamefont
  {Zhao}}, \bibinfo {author} {\bibfnamefont {H.}~\bibnamefont {Lei}}, \bibinfo
  {author} {\bibfnamefont {S.}~\bibnamefont {Wang}}, \bibinfo {author}
  {\bibfnamefont {Y.}~\bibnamefont {Zhang}},\ and\ \bibinfo {author}
  {\bibfnamefont {B.}~\bibnamefont {Shen}},\ }\bibfield  {title} {\bibinfo
  {title} {Spontaneous ({{Anti}})meron {{Chains}} in the {{Domain Walls}} of
  van der {{Waals Ferromagnetic Fe5}}-{{xGeTe2}}},\ }\href
  {https://doi.org/10.1002/adma.202005228} {\bibfield  {journal} {\bibinfo
  {journal} {Adv. Mater.}\ }\textbf {\bibinfo {volume} {32}},\ \bibinfo {pages}
  {2005228} (\bibinfo {year} {2020})}\BibitemShut {NoStop}%
\bibitem [{\citenamefont {Zhang}\ \emph
  {et~al.}(2022{\natexlab{a}})\citenamefont {Zhang}, \citenamefont {Liu},
  \citenamefont {Zhang}, \citenamefont {Zhou}, \citenamefont {Guan},
  \citenamefont {Ma}, \citenamefont {Algaidi}, \citenamefont {Zheng},
  \citenamefont {Li}, \citenamefont {He}, \citenamefont {Zhang}, \citenamefont
  {Li}, \citenamefont {Hou}, \citenamefont {Yin}, \citenamefont {Liu},
  \citenamefont {Peng},\ and\ \citenamefont {Zhang}}]{zhang2022magnetic}%
  \BibitemOpen
  \bibfield  {author} {\bibinfo {author} {\bibfnamefont {C.}~\bibnamefont
  {Zhang}}, \bibinfo {author} {\bibfnamefont {C.}~\bibnamefont {Liu}}, \bibinfo
  {author} {\bibfnamefont {S.}~\bibnamefont {Zhang}}, \bibinfo {author}
  {\bibfnamefont {B.}~\bibnamefont {Zhou}}, \bibinfo {author} {\bibfnamefont
  {C.}~\bibnamefont {Guan}}, \bibinfo {author} {\bibfnamefont {Y.}~\bibnamefont
  {Ma}}, \bibinfo {author} {\bibfnamefont {H.}~\bibnamefont {Algaidi}},
  \bibinfo {author} {\bibfnamefont {D.}~\bibnamefont {Zheng}}, \bibinfo
  {author} {\bibfnamefont {Y.}~\bibnamefont {Li}}, \bibinfo {author}
  {\bibfnamefont {X.}~\bibnamefont {He}}, \bibinfo {author} {\bibfnamefont
  {J.}~\bibnamefont {Zhang}}, \bibinfo {author} {\bibfnamefont
  {P.}~\bibnamefont {Li}}, \bibinfo {author} {\bibfnamefont {Z.}~\bibnamefont
  {Hou}}, \bibinfo {author} {\bibfnamefont {G.}~\bibnamefont {Yin}}, \bibinfo
  {author} {\bibfnamefont {K.}~\bibnamefont {Liu}}, \bibinfo {author}
  {\bibfnamefont {Y.}~\bibnamefont {Peng}},\ and\ \bibinfo {author}
  {\bibfnamefont {X.-X.}\ \bibnamefont {Zhang}},\ }\bibfield  {title} {\bibinfo
  {title} {Magnetic {{Skyrmions}} with {{Unconventional Helicity Polarization}}
  in a {{Van Der Waals Ferromagnet}}},\ }\href
  {https://doi.org/10.1002/adma.202204163} {\bibfield  {journal} {\bibinfo
  {journal} {Adv. Mater.}\ }\textbf {\bibinfo {volume} {34}},\ \bibinfo {pages}
  {2204163} (\bibinfo {year} {2022}{\natexlab{a}})}\BibitemShut {NoStop}%
\bibitem [{\citenamefont {Zhang}\ \emph {et~al.}(2023)\citenamefont {Zhang},
  \citenamefont {Liu}, \citenamefont {Zhang}, \citenamefont {Yuan},
  \citenamefont {Wen}, \citenamefont {Li}, \citenamefont {Zheng}, \citenamefont
  {Zhang}, \citenamefont {Hou}, \citenamefont {Yin}, \citenamefont {Liu},
  \citenamefont {Peng},\ and\ \citenamefont
  {Zhang}}]{zhang2023roomtemperature}%
  \BibitemOpen
  \bibfield  {author} {\bibinfo {author} {\bibfnamefont {C.}~\bibnamefont
  {Zhang}}, \bibinfo {author} {\bibfnamefont {C.}~\bibnamefont {Liu}}, \bibinfo
  {author} {\bibfnamefont {J.}~\bibnamefont {Zhang}}, \bibinfo {author}
  {\bibfnamefont {Y.}~\bibnamefont {Yuan}}, \bibinfo {author} {\bibfnamefont
  {Y.}~\bibnamefont {Wen}}, \bibinfo {author} {\bibfnamefont {Y.}~\bibnamefont
  {Li}}, \bibinfo {author} {\bibfnamefont {D.}~\bibnamefont {Zheng}}, \bibinfo
  {author} {\bibfnamefont {Q.}~\bibnamefont {Zhang}}, \bibinfo {author}
  {\bibfnamefont {Z.}~\bibnamefont {Hou}}, \bibinfo {author} {\bibfnamefont
  {G.}~\bibnamefont {Yin}}, \bibinfo {author} {\bibfnamefont {K.}~\bibnamefont
  {Liu}}, \bibinfo {author} {\bibfnamefont {Y.}~\bibnamefont {Peng}},\ and\
  \bibinfo {author} {\bibfnamefont {X.-X.}\ \bibnamefont {Zhang}},\ }\bibfield
  {title} {\bibinfo {title} {Room-{{Temperature Magnetic Skyrmions}} and
  {{Large Topological Hall Effect}} in {{Chromium Telluride Engineered}} by
  {{Self-Intercalation}}},\ }\href {https://doi.org/10.1002/adma.202205967}
  {\bibfield  {journal} {\bibinfo  {journal} {Adv. Mater.}\ }\textbf {\bibinfo
  {volume} {35}},\ \bibinfo {pages} {2205967} (\bibinfo {year}
  {2023})}\BibitemShut {NoStop}%
\bibitem [{\citenamefont {Li}\ \emph {et~al.}(2024)\citenamefont {Li},
  \citenamefont {Zhang}, \citenamefont {Li}, \citenamefont {Guo}, \citenamefont
  {Wang}, \citenamefont {Deng}, \citenamefont {Hu}, \citenamefont {Hu},
  \citenamefont {Liu}, \citenamefont {Qin}, \citenamefont {Shen}, \citenamefont
  {Yu}, \citenamefont {Gao}, \citenamefont {Liao}, \citenamefont {Liu},
  \citenamefont {Hou}, \citenamefont {Zhu},\ and\ \citenamefont
  {Fu}}]{li2024roomtemperature}%
  \BibitemOpen
  \bibfield  {author} {\bibinfo {author} {\bibfnamefont {Z.}~\bibnamefont
  {Li}}, \bibinfo {author} {\bibfnamefont {H.}~\bibnamefont {Zhang}}, \bibinfo
  {author} {\bibfnamefont {G.}~\bibnamefont {Li}}, \bibinfo {author}
  {\bibfnamefont {J.}~\bibnamefont {Guo}}, \bibinfo {author} {\bibfnamefont
  {Q.}~\bibnamefont {Wang}}, \bibinfo {author} {\bibfnamefont {Y.}~\bibnamefont
  {Deng}}, \bibinfo {author} {\bibfnamefont {Y.}~\bibnamefont {Hu}}, \bibinfo
  {author} {\bibfnamefont {X.}~\bibnamefont {Hu}}, \bibinfo {author}
  {\bibfnamefont {C.}~\bibnamefont {Liu}}, \bibinfo {author} {\bibfnamefont
  {M.}~\bibnamefont {Qin}}, \bibinfo {author} {\bibfnamefont {X.}~\bibnamefont
  {Shen}}, \bibinfo {author} {\bibfnamefont {R.}~\bibnamefont {Yu}}, \bibinfo
  {author} {\bibfnamefont {X.}~\bibnamefont {Gao}}, \bibinfo {author}
  {\bibfnamefont {Z.}~\bibnamefont {Liao}}, \bibinfo {author} {\bibfnamefont
  {J.}~\bibnamefont {Liu}}, \bibinfo {author} {\bibfnamefont {Z.}~\bibnamefont
  {Hou}}, \bibinfo {author} {\bibfnamefont {Y.}~\bibnamefont {Zhu}},\ and\
  \bibinfo {author} {\bibfnamefont {X.}~\bibnamefont {Fu}},\ }\bibfield
  {title} {\bibinfo {title} {Room-temperature sub-100 nm {{N{\'e}el-type}}
  skyrmions in non-stoichiometric van der {{Waals}} ferromagnet {{Fe3-xGaTe2}}
  with ultrafast laser writability},\ }\href
  {https://doi.org/10.1038/s41467-024-45310-2} {\bibfield  {journal} {\bibinfo
  {journal} {Nat. Commun.}\ }\textbf {\bibinfo {volume} {15}},\ \bibinfo
  {pages} {1017} (\bibinfo {year} {2024})}\BibitemShut {NoStop}%
\bibitem [{\citenamefont {Zhang}\ \emph
  {et~al.}(2024{\natexlab{b}})\citenamefont {Zhang}, \citenamefont {Jiang},
  \citenamefont {Jiang}, \citenamefont {He}, \citenamefont {Zhang},
  \citenamefont {Hu}, \citenamefont {Zhao}, \citenamefont {Yang}, \citenamefont
  {Liu}, \citenamefont {Peng}, \citenamefont {Yang},\ and\ \citenamefont
  {Yang}}]{zhang2024aboveroomtemperature}%
  \BibitemOpen
  \bibfield  {author} {\bibinfo {author} {\bibfnamefont {C.}~\bibnamefont
  {Zhang}}, \bibinfo {author} {\bibfnamefont {Z.}~\bibnamefont {Jiang}},
  \bibinfo {author} {\bibfnamefont {J.}~\bibnamefont {Jiang}}, \bibinfo
  {author} {\bibfnamefont {W.}~\bibnamefont {He}}, \bibinfo {author}
  {\bibfnamefont {J.}~\bibnamefont {Zhang}}, \bibinfo {author} {\bibfnamefont
  {F.}~\bibnamefont {Hu}}, \bibinfo {author} {\bibfnamefont {S.}~\bibnamefont
  {Zhao}}, \bibinfo {author} {\bibfnamefont {D.}~\bibnamefont {Yang}}, \bibinfo
  {author} {\bibfnamefont {Y.}~\bibnamefont {Liu}}, \bibinfo {author}
  {\bibfnamefont {Y.}~\bibnamefont {Peng}}, \bibinfo {author} {\bibfnamefont
  {H.}~\bibnamefont {Yang}},\ and\ \bibinfo {author} {\bibfnamefont
  {H.}~\bibnamefont {Yang}},\ }\bibfield  {title} {\bibinfo {title}
  {Above-room-temperature chiral skyrmion lattice and
  {{Dzyaloshinskii}}--{{Moriya}} interaction in a van der {{Waals}} ferromagnet
  {{Fe3}}-{{xGaTe2}}},\ }\href {https://doi.org/10.1038/s41467-024-48799-9}
  {\bibfield  {journal} {\bibinfo  {journal} {Nat. Commun.}\ }\textbf {\bibinfo
  {volume} {15}},\ \bibinfo {pages} {4472} (\bibinfo {year}
  {2024}{\natexlab{b}})}\BibitemShut {NoStop}%
\bibitem [{\citenamefont {Saha}\ \emph {et~al.}(2022)\citenamefont {Saha},
  \citenamefont {Meyerheim}, \citenamefont {G{\"o}bel}, \citenamefont {Hazra},
  \citenamefont {Deniz}, \citenamefont {Mohseni}, \citenamefont {Antonov},
  \citenamefont {Ernst}, \citenamefont {Knyazev}, \citenamefont
  {{Bedoya-Pinto}}, \citenamefont {Mertig},\ and\ \citenamefont
  {Parkin}}]{saha2022observation}%
  \BibitemOpen
  \bibfield  {author} {\bibinfo {author} {\bibfnamefont {R.}~\bibnamefont
  {Saha}}, \bibinfo {author} {\bibfnamefont {H.~L.}\ \bibnamefont {Meyerheim}},
  \bibinfo {author} {\bibfnamefont {B.}~\bibnamefont {G{\"o}bel}}, \bibinfo
  {author} {\bibfnamefont {B.~K.}\ \bibnamefont {Hazra}}, \bibinfo {author}
  {\bibfnamefont {H.}~\bibnamefont {Deniz}}, \bibinfo {author} {\bibfnamefont
  {K.}~\bibnamefont {Mohseni}}, \bibinfo {author} {\bibfnamefont
  {V.}~\bibnamefont {Antonov}}, \bibinfo {author} {\bibfnamefont
  {A.}~\bibnamefont {Ernst}}, \bibinfo {author} {\bibfnamefont
  {D.}~\bibnamefont {Knyazev}}, \bibinfo {author} {\bibfnamefont
  {A.}~\bibnamefont {{Bedoya-Pinto}}}, \bibinfo {author} {\bibfnamefont
  {I.}~\bibnamefont {Mertig}},\ and\ \bibinfo {author} {\bibfnamefont
  {S.~S.~P.}\ \bibnamefont {Parkin}},\ }\bibfield  {title} {\bibinfo {title}
  {Observation of {{N{\'e}el-type}} skyrmions in acentric self-intercalated
  {{Cr1}}+{{$\delta$Te2}}},\ }\href
  {https://doi.org/10.1038/s41467-022-31319-y} {\bibfield  {journal} {\bibinfo
  {journal} {Nat. Commun.}\ }\textbf {\bibinfo {volume} {13}},\ \bibinfo
  {pages} {3965} (\bibinfo {year} {2022})}\BibitemShut {NoStop}%
\bibitem [{\citenamefont {Zhang}\ \emph
  {et~al.}(2022{\natexlab{b}})\citenamefont {Zhang}, \citenamefont {Raftrey},
  \citenamefont {Chan}, \citenamefont {Shao}, \citenamefont {Chen},
  \citenamefont {Chen}, \citenamefont {Huang}, \citenamefont {Reichanadter},
  \citenamefont {Dong}, \citenamefont {Susarla}, \citenamefont {Caretta},
  \citenamefont {Chen}, \citenamefont {Yao}, \citenamefont {Fischer},
  \citenamefont {Neaton}, \citenamefont {Wu}, \citenamefont {Muller},
  \citenamefont {Birgeneau},\ and\ \citenamefont
  {Ramesh}}]{zhang2022roomtemperature}%
  \BibitemOpen
  \bibfield  {author} {\bibinfo {author} {\bibfnamefont {H.}~\bibnamefont
  {Zhang}}, \bibinfo {author} {\bibfnamefont {D.}~\bibnamefont {Raftrey}},
  \bibinfo {author} {\bibfnamefont {Y.-T.}\ \bibnamefont {Chan}}, \bibinfo
  {author} {\bibfnamefont {Y.-T.}\ \bibnamefont {Shao}}, \bibinfo {author}
  {\bibfnamefont {R.}~\bibnamefont {Chen}}, \bibinfo {author} {\bibfnamefont
  {X.}~\bibnamefont {Chen}}, \bibinfo {author} {\bibfnamefont {X.}~\bibnamefont
  {Huang}}, \bibinfo {author} {\bibfnamefont {J.~T.}\ \bibnamefont
  {Reichanadter}}, \bibinfo {author} {\bibfnamefont {K.}~\bibnamefont {Dong}},
  \bibinfo {author} {\bibfnamefont {S.}~\bibnamefont {Susarla}}, \bibinfo
  {author} {\bibfnamefont {L.}~\bibnamefont {Caretta}}, \bibinfo {author}
  {\bibfnamefont {Z.}~\bibnamefont {Chen}}, \bibinfo {author} {\bibfnamefont
  {J.}~\bibnamefont {Yao}}, \bibinfo {author} {\bibfnamefont {P.}~\bibnamefont
  {Fischer}}, \bibinfo {author} {\bibfnamefont {J.~B.}\ \bibnamefont {Neaton}},
  \bibinfo {author} {\bibfnamefont {W.}~\bibnamefont {Wu}}, \bibinfo {author}
  {\bibfnamefont {D.~A.}\ \bibnamefont {Muller}}, \bibinfo {author}
  {\bibfnamefont {R.~J.}\ \bibnamefont {Birgeneau}},\ and\ \bibinfo {author}
  {\bibfnamefont {R.}~\bibnamefont {Ramesh}},\ }\bibfield  {title} {\bibinfo
  {title} {Room-temperature skyrmion lattice in a layered magnet
  ({{Fe0}}.{{5Co0}}.5){{5GeTe2}}},\ }\href
  {https://doi.org/10.1126/sciadv.abm7103} {\bibfield  {journal} {\bibinfo
  {journal} {Sci. Adv.}\ }\textbf {\bibinfo {volume} {8}},\ \bibinfo {pages}
  {eabm7103} (\bibinfo {year} {2022}{\natexlab{b}})}\BibitemShut {NoStop}%
\bibitem [{\citenamefont {Lv}\ \emph {et~al.}(2024)\citenamefont {Lv},
  \citenamefont {Lv}, \citenamefont {Huang}, \citenamefont {Zhang},
  \citenamefont {Qin}, \citenamefont {Dong}, \citenamefont {Liu}, \citenamefont
  {Pei}, \citenamefont {Cao}, \citenamefont {Zhang}, \citenamefont {Lai},\ and\
  \citenamefont {Che}}]{lv2024distinct}%
  \BibitemOpen
  \bibfield  {author} {\bibinfo {author} {\bibfnamefont {X.}~\bibnamefont
  {Lv}}, \bibinfo {author} {\bibfnamefont {H.}~\bibnamefont {Lv}}, \bibinfo
  {author} {\bibfnamefont {Y.}~\bibnamefont {Huang}}, \bibinfo {author}
  {\bibfnamefont {R.}~\bibnamefont {Zhang}}, \bibinfo {author} {\bibfnamefont
  {G.}~\bibnamefont {Qin}}, \bibinfo {author} {\bibfnamefont {Y.}~\bibnamefont
  {Dong}}, \bibinfo {author} {\bibfnamefont {M.}~\bibnamefont {Liu}}, \bibinfo
  {author} {\bibfnamefont {K.}~\bibnamefont {Pei}}, \bibinfo {author}
  {\bibfnamefont {G.}~\bibnamefont {Cao}}, \bibinfo {author} {\bibfnamefont
  {J.}~\bibnamefont {Zhang}}, \bibinfo {author} {\bibfnamefont
  {Y.}~\bibnamefont {Lai}},\ and\ \bibinfo {author} {\bibfnamefont
  {R.}~\bibnamefont {Che}},\ }\bibfield  {title} {\bibinfo {title} {Distinct
  skyrmion phases at room temperature in two-dimensional ferromagnet
  {{Fe3GaTe2}}},\ }\href {https://doi.org/10.1038/s41467-024-47579-9}
  {\bibfield  {journal} {\bibinfo  {journal} {Nat. Commun.}\ }\textbf {\bibinfo
  {volume} {15}},\ \bibinfo {pages} {3278} (\bibinfo {year}
  {2024})}\BibitemShut {NoStop}%
\bibitem [{\citenamefont {Tong}\ \emph {et~al.}(2018)\citenamefont {Tong},
  \citenamefont {Liu}, \citenamefont {Xiao},\ and\ \citenamefont
  {Yao}}]{tong2018skyrmions}%
  \BibitemOpen
  \bibfield  {author} {\bibinfo {author} {\bibfnamefont {Q.}~\bibnamefont
  {Tong}}, \bibinfo {author} {\bibfnamefont {F.}~\bibnamefont {Liu}}, \bibinfo
  {author} {\bibfnamefont {J.}~\bibnamefont {Xiao}},\ and\ \bibinfo {author}
  {\bibfnamefont {W.}~\bibnamefont {Yao}},\ }\bibfield  {title} {\bibinfo
  {title} {Skyrmions in the {{Moir{\'e}}} of van der {{Waals 2D Magnets}}},\
  }\href {https://doi.org/10.1021/acs.nanolett.8b03315} {\bibfield  {journal}
  {\bibinfo  {journal} {Nano Lett.}\ }\textbf {\bibinfo {volume} {18}},\
  \bibinfo {pages} {7194} (\bibinfo {year} {2018})}\BibitemShut {NoStop}%
\bibitem [{\citenamefont {Wang}\ \emph
  {et~al.}(2020{\natexlab{a}})\citenamefont {Wang}, \citenamefont {Liu},
  \citenamefont {Wu}, \citenamefont {Hou}, \citenamefont {Jiang}, \citenamefont
  {Li}, \citenamefont {Pandey}, \citenamefont {Chen}, \citenamefont {Yang},
  \citenamefont {Wang}, \citenamefont {Wei}, \citenamefont {Lei}, \citenamefont
  {Kang}, \citenamefont {Wen}, \citenamefont {Nie}, \citenamefont {Zhao},\ and\
  \citenamefont {Wang}}]{wang2020roomtemperature}%
  \BibitemOpen
  \bibfield  {author} {\bibinfo {author} {\bibfnamefont {H.}~\bibnamefont
  {Wang}}, \bibinfo {author} {\bibfnamefont {Y.}~\bibnamefont {Liu}}, \bibinfo
  {author} {\bibfnamefont {P.}~\bibnamefont {Wu}}, \bibinfo {author}
  {\bibfnamefont {W.}~\bibnamefont {Hou}}, \bibinfo {author} {\bibfnamefont
  {Y.}~\bibnamefont {Jiang}}, \bibinfo {author} {\bibfnamefont
  {X.}~\bibnamefont {Li}}, \bibinfo {author} {\bibfnamefont {C.}~\bibnamefont
  {Pandey}}, \bibinfo {author} {\bibfnamefont {D.}~\bibnamefont {Chen}},
  \bibinfo {author} {\bibfnamefont {Q.}~\bibnamefont {Yang}}, \bibinfo {author}
  {\bibfnamefont {H.}~\bibnamefont {Wang}}, \bibinfo {author} {\bibfnamefont
  {D.}~\bibnamefont {Wei}}, \bibinfo {author} {\bibfnamefont {N.}~\bibnamefont
  {Lei}}, \bibinfo {author} {\bibfnamefont {W.}~\bibnamefont {Kang}}, \bibinfo
  {author} {\bibfnamefont {L.}~\bibnamefont {Wen}}, \bibinfo {author}
  {\bibfnamefont {T.}~\bibnamefont {Nie}}, \bibinfo {author} {\bibfnamefont
  {W.}~\bibnamefont {Zhao}},\ and\ \bibinfo {author} {\bibfnamefont {K.~L.}\
  \bibnamefont {Wang}},\ }\bibfield  {title} {\bibinfo {title} {Above
  {{Room-Temperature Ferromagnetism}} in {{Wafer-Scale Two-Dimensional}} van
  der {{Waals Fe3GeTe2 Tailored}} by a {{Topological Insulator}}},\ }\href
  {https://doi.org/10.1021/acsnano.0c03152} {\bibfield  {journal} {\bibinfo
  {journal} {ACS Nano}\ }\textbf {\bibinfo {volume} {14}},\ \bibinfo {pages}
  {10045} (\bibinfo {year} {2020}{\natexlab{a}})}\BibitemShut {NoStop}%
\bibitem [{\citenamefont {Wu}\ \emph {et~al.}(2020)\citenamefont {Wu},
  \citenamefont {Zhang}, \citenamefont {Zhang}, \citenamefont {Wang},
  \citenamefont {Zhu}, \citenamefont {Hu}, \citenamefont {Yin}, \citenamefont
  {Wong}, \citenamefont {Fang}, \citenamefont {Wan}, \citenamefont {Han},
  \citenamefont {Shao}, \citenamefont {Taniguchi}, \citenamefont {Watanabe},
  \citenamefont {Zang}, \citenamefont {Mao}, \citenamefont {Zhang},\ and\
  \citenamefont {Wang}}]{wu2020neeltype}%
  \BibitemOpen
  \bibfield  {author} {\bibinfo {author} {\bibfnamefont {Y.}~\bibnamefont
  {Wu}}, \bibinfo {author} {\bibfnamefont {S.}~\bibnamefont {Zhang}}, \bibinfo
  {author} {\bibfnamefont {J.}~\bibnamefont {Zhang}}, \bibinfo {author}
  {\bibfnamefont {W.}~\bibnamefont {Wang}}, \bibinfo {author} {\bibfnamefont
  {Y.~L.}\ \bibnamefont {Zhu}}, \bibinfo {author} {\bibfnamefont
  {J.}~\bibnamefont {Hu}}, \bibinfo {author} {\bibfnamefont {G.}~\bibnamefont
  {Yin}}, \bibinfo {author} {\bibfnamefont {K.}~\bibnamefont {Wong}}, \bibinfo
  {author} {\bibfnamefont {C.}~\bibnamefont {Fang}}, \bibinfo {author}
  {\bibfnamefont {C.}~\bibnamefont {Wan}}, \bibinfo {author} {\bibfnamefont
  {X.}~\bibnamefont {Han}}, \bibinfo {author} {\bibfnamefont {Q.}~\bibnamefont
  {Shao}}, \bibinfo {author} {\bibfnamefont {T.}~\bibnamefont {Taniguchi}},
  \bibinfo {author} {\bibfnamefont {K.}~\bibnamefont {Watanabe}}, \bibinfo
  {author} {\bibfnamefont {J.}~\bibnamefont {Zang}}, \bibinfo {author}
  {\bibfnamefont {Z.}~\bibnamefont {Mao}}, \bibinfo {author} {\bibfnamefont
  {X.}~\bibnamefont {Zhang}},\ and\ \bibinfo {author} {\bibfnamefont {K.~L.}\
  \bibnamefont {Wang}},\ }\bibfield  {title} {\bibinfo {title} {N{\'e}el-type
  skyrmion in {{WTe2}}/{{Fe3GeTe2}} van der {{Waals}} heterostructure},\ }\href
  {https://doi.org/10.1038/s41467-020-17566-x} {\bibfield  {journal} {\bibinfo
  {journal} {Nat. Commun.}\ }\textbf {\bibinfo {volume} {11}},\ \bibinfo
  {pages} {3860} (\bibinfo {year} {2020})}\BibitemShut {NoStop}%
\bibitem [{\citenamefont {Liang}\ \emph {et~al.}(2020)\citenamefont {Liang},
  \citenamefont {Wang}, \citenamefont {Du}, \citenamefont {Hallal},
  \citenamefont {Garcia}, \citenamefont {Chshiev}, \citenamefont {Fert},\ and\
  \citenamefont {Yang}}]{liang2020very}%
  \BibitemOpen
  \bibfield  {author} {\bibinfo {author} {\bibfnamefont {J.}~\bibnamefont
  {Liang}}, \bibinfo {author} {\bibfnamefont {W.}~\bibnamefont {Wang}},
  \bibinfo {author} {\bibfnamefont {H.}~\bibnamefont {Du}}, \bibinfo {author}
  {\bibfnamefont {A.}~\bibnamefont {Hallal}}, \bibinfo {author} {\bibfnamefont
  {K.}~\bibnamefont {Garcia}}, \bibinfo {author} {\bibfnamefont
  {M.}~\bibnamefont {Chshiev}}, \bibinfo {author} {\bibfnamefont
  {A.}~\bibnamefont {Fert}},\ and\ \bibinfo {author} {\bibfnamefont
  {H.}~\bibnamefont {Yang}},\ }\bibfield  {title} {\bibinfo {title} {Very large
  {{Dzyaloshinskii-Moriya}} interaction in two-dimensional {{Janus}} manganese
  dichalcogenides and its application to realize skyrmion states},\ }\href
  {https://doi.org/10.1103/PhysRevB.101.184401} {\bibfield  {journal} {\bibinfo
   {journal} {Phys. Rev. B}\ }\textbf {\bibinfo {volume} {101}},\ \bibinfo
  {pages} {184401} (\bibinfo {year} {2020})}\BibitemShut {NoStop}%
\bibitem [{\citenamefont {Rybakov}\ \emph {et~al.}(2013)\citenamefont
  {Rybakov}, \citenamefont {Borisov},\ and\ \citenamefont
  {Bogdanov}}]{rybakov2013three}%
  \BibitemOpen
  \bibfield  {author} {\bibinfo {author} {\bibfnamefont {F.~N.}\ \bibnamefont
  {Rybakov}}, \bibinfo {author} {\bibfnamefont {A.~B.}\ \bibnamefont
  {Borisov}},\ and\ \bibinfo {author} {\bibfnamefont {A.~N.}\ \bibnamefont
  {Bogdanov}},\ }\bibfield  {title} {\bibinfo {title} {Three-dimensional
  skyrmion states in thin films of cubic helimagnets},\ }\href
  {https://doi.org/10.1103/physrevb.87.094424} {\bibfield  {journal} {\bibinfo
  {journal} {Phys. Rev. B}\ }\textbf {\bibinfo {volume} {87}},\ \bibinfo
  {pages} {094424} (\bibinfo {year} {2013})}\BibitemShut {NoStop}%
\bibitem [{\citenamefont {Zhang}\ \emph
  {et~al.}(2018{\natexlab{b}})\citenamefont {Zhang}, \citenamefont {van~der
  Laan}, \citenamefont {Wang}, \citenamefont {Haghighirad},\ and\ \citenamefont
  {Hesjedal}}]{zhang2018direct}%
  \BibitemOpen
  \bibfield  {author} {\bibinfo {author} {\bibfnamefont {S.~L.}\ \bibnamefont
  {Zhang}}, \bibinfo {author} {\bibfnamefont {G.}~\bibnamefont {van~der Laan}},
  \bibinfo {author} {\bibfnamefont {W.~W.}\ \bibnamefont {Wang}}, \bibinfo
  {author} {\bibfnamefont {A.~A.}\ \bibnamefont {Haghighirad}},\ and\ \bibinfo
  {author} {\bibfnamefont {T.}~\bibnamefont {Hesjedal}},\ }\bibfield  {title}
  {\bibinfo {title} {Direct observation of twisted surface skyrmions in bulk
  crystals},\ }\href {https://doi.org/10.1103/physrevlett.120.227202}
  {\bibfield  {journal} {\bibinfo  {journal} {Phys. Rev. Lett.}\ }\textbf
  {\bibinfo {volume} {120}},\ \bibinfo {pages} {227202} (\bibinfo {year}
  {2018}{\natexlab{b}})}\BibitemShut {NoStop}%
\bibitem [{\citenamefont {Zhang}\ \emph
  {et~al.}(2018{\natexlab{c}})\citenamefont {Zhang}, \citenamefont {van~der
  Laan}, \citenamefont {M{\"u}ller}, \citenamefont {Heinen}, \citenamefont
  {Garst}, \citenamefont {Bauer}, \citenamefont {Berger}, \citenamefont
  {Pfleiderer},\ and\ \citenamefont {Hesjedal}}]{zhang2018reciprocal}%
  \BibitemOpen
  \bibfield  {author} {\bibinfo {author} {\bibfnamefont {S.}~\bibnamefont
  {Zhang}}, \bibinfo {author} {\bibfnamefont {G.}~\bibnamefont {van~der Laan}},
  \bibinfo {author} {\bibfnamefont {J.}~\bibnamefont {M{\"u}ller}}, \bibinfo
  {author} {\bibfnamefont {L.}~\bibnamefont {Heinen}}, \bibinfo {author}
  {\bibfnamefont {M.}~\bibnamefont {Garst}}, \bibinfo {author} {\bibfnamefont
  {A.}~\bibnamefont {Bauer}}, \bibinfo {author} {\bibfnamefont
  {H.}~\bibnamefont {Berger}}, \bibinfo {author} {\bibfnamefont
  {C.}~\bibnamefont {Pfleiderer}},\ and\ \bibinfo {author} {\bibfnamefont
  {T.}~\bibnamefont {Hesjedal}},\ }\bibfield  {title} {\bibinfo {title}
  {{Reciprocal space tomography of 3D skyrmion lattice order in a chiral
  magnet}},\ }\href {https://doi.org/10.1073/pnas.1803367115} {\bibfield
  {journal} {\bibinfo  {journal} {Proc. Natl. Acad. Sci. U.S.A.}\ }\textbf
  {\bibinfo {volume} {115}},\ \bibinfo {pages} {6386} (\bibinfo {year}
  {2018}{\natexlab{c}})}\BibitemShut {NoStop}%
\bibitem [{\citenamefont {Zheng}\ \emph {et~al.}(2018)\citenamefont {Zheng},
  \citenamefont {Rybakov}, \citenamefont {Borisov}, \citenamefont {Song},
  \citenamefont {Wang}, \citenamefont {Li}, \citenamefont {Du}, \citenamefont
  {Kiselev}, \citenamefont {Caron}, \citenamefont {Kovacs}, \citenamefont
  {Tian}, \citenamefont {Zhang}, \citenamefont {Bl\"ugel},\ and\ \citenamefont
  {Dunin-Borkowski}}]{zheng2018experimental}%
  \BibitemOpen
  \bibfield  {author} {\bibinfo {author} {\bibfnamefont {F.}~\bibnamefont
  {Zheng}}, \bibinfo {author} {\bibfnamefont {F.~N.}\ \bibnamefont {Rybakov}},
  \bibinfo {author} {\bibfnamefont {A.~B.}\ \bibnamefont {Borisov}}, \bibinfo
  {author} {\bibfnamefont {D.}~\bibnamefont {Song}}, \bibinfo {author}
  {\bibfnamefont {S.}~\bibnamefont {Wang}}, \bibinfo {author} {\bibfnamefont
  {Z.-A.}\ \bibnamefont {Li}}, \bibinfo {author} {\bibfnamefont
  {H.}~\bibnamefont {Du}}, \bibinfo {author} {\bibfnamefont {N.~S.}\
  \bibnamefont {Kiselev}}, \bibinfo {author} {\bibfnamefont {J.}~\bibnamefont
  {Caron}}, \bibinfo {author} {\bibfnamefont {A.}~\bibnamefont {Kovacs}},
  \bibinfo {author} {\bibfnamefont {M.}~\bibnamefont {Tian}}, \bibinfo {author}
  {\bibfnamefont {Y.}~\bibnamefont {Zhang}}, \bibinfo {author} {\bibfnamefont
  {S.}~\bibnamefont {Bl\"ugel}},\ and\ \bibinfo {author} {\bibfnamefont
  {R.~E.}\ \bibnamefont {Dunin-Borkowski}},\ }\bibfield  {title} {\bibinfo
  {title} {{Experimental observation of chiral magnetic bobbers in B20-type
  FeGe}},\ }\href {https://doi.org/10.1038/s41565-018-0093-3} {\bibfield
  {journal} {\bibinfo  {journal} {Nature Nanotech.}\ }\textbf {\bibinfo
  {volume} {13}},\ \bibinfo {pages} {451} (\bibinfo {year} {2018})}\BibitemShut
  {NoStop}%
\bibitem [{\citenamefont {Birch}\ \emph {et~al.}(2020)\citenamefont {Birch},
  \citenamefont {Cort\'{e}s-Ortu\~{n}o}, \citenamefont {Turnbull},
  \citenamefont {Wilson}, \citenamefont {Gro{\ss}}, \citenamefont {Tr\"ager},
  \citenamefont {Laurenson}, \citenamefont {Bukin}, \citenamefont {Moody},
  \citenamefont {Weigand}, \citenamefont {Sch\"utz}, \citenamefont {Popescu},
  \citenamefont {Fan}, \citenamefont {Steadman}, \citenamefont {Verezhak},
  \citenamefont {Balakrishnan}, \citenamefont {Loudon}, \citenamefont
  {Twitchett-Harrison}, \citenamefont {Hovorka}, \citenamefont {Fangohr},
  \citenamefont {Ogrin}, \citenamefont {Gr\"afe},\ and\ \citenamefont
  {Hatton}}]{birch2020real}%
  \BibitemOpen
  \bibfield  {author} {\bibinfo {author} {\bibfnamefont {M.~T.}\ \bibnamefont
  {Birch}}, \bibinfo {author} {\bibfnamefont {D.}~\bibnamefont
  {Cort\'{e}s-Ortu\~{n}o}}, \bibinfo {author} {\bibfnamefont {L.~A.}\
  \bibnamefont {Turnbull}}, \bibinfo {author} {\bibfnamefont {M.~N.}\
  \bibnamefont {Wilson}}, \bibinfo {author} {\bibfnamefont {F.}~\bibnamefont
  {Gro{\ss}}}, \bibinfo {author} {\bibfnamefont {N.}~\bibnamefont {Tr\"ager}},
  \bibinfo {author} {\bibfnamefont {A.}~\bibnamefont {Laurenson}}, \bibinfo
  {author} {\bibfnamefont {N.}~\bibnamefont {Bukin}}, \bibinfo {author}
  {\bibfnamefont {S.~H.}\ \bibnamefont {Moody}}, \bibinfo {author}
  {\bibfnamefont {M.}~\bibnamefont {Weigand}}, \bibinfo {author} {\bibfnamefont
  {G.}~\bibnamefont {Sch\"utz}}, \bibinfo {author} {\bibfnamefont
  {H.}~\bibnamefont {Popescu}}, \bibinfo {author} {\bibfnamefont
  {R.}~\bibnamefont {Fan}}, \bibinfo {author} {\bibfnamefont {P.}~\bibnamefont
  {Steadman}}, \bibinfo {author} {\bibfnamefont {J.~A.~T.}\ \bibnamefont
  {Verezhak}}, \bibinfo {author} {\bibfnamefont {G.}~\bibnamefont
  {Balakrishnan}}, \bibinfo {author} {\bibfnamefont {J.~C.}\ \bibnamefont
  {Loudon}}, \bibinfo {author} {\bibfnamefont {A.~C.}\ \bibnamefont
  {Twitchett-Harrison}}, \bibinfo {author} {\bibfnamefont {O.}~\bibnamefont
  {Hovorka}}, \bibinfo {author} {\bibfnamefont {H.}~\bibnamefont {Fangohr}},
  \bibinfo {author} {\bibfnamefont {F.~Y.}\ \bibnamefont {Ogrin}}, \bibinfo
  {author} {\bibfnamefont {J.}~\bibnamefont {Gr\"afe}},\ and\ \bibinfo {author}
  {\bibfnamefont {P.~D.}\ \bibnamefont {Hatton}},\ }\bibfield  {title}
  {\bibinfo {title} {Real-space imaging of confined magnetic skyrmion tubes},\
  }\href {https://doi.org/10.1038/s41467-020-15474-8} {\bibfield  {journal}
  {\bibinfo  {journal} {Nat. Commun.}\ }\textbf {\bibinfo {volume} {11}},\
  \bibinfo {pages} {1726} (\bibinfo {year} {2020})}\BibitemShut {NoStop}%
\bibitem [{\citenamefont {Ran}\ \emph {et~al.}(2021)\citenamefont {Ran},
  \citenamefont {Liu}, \citenamefont {Guang}, \citenamefont {Burn},
  \citenamefont {van~der Laan}, \citenamefont {Hesjedal}, \citenamefont {Du},
  \citenamefont {Yu},\ and\ \citenamefont {Zhang}}]{ran2021creation}%
  \BibitemOpen
  \bibfield  {author} {\bibinfo {author} {\bibfnamefont {K.}~\bibnamefont
  {Ran}}, \bibinfo {author} {\bibfnamefont {Y.}~\bibnamefont {Liu}}, \bibinfo
  {author} {\bibfnamefont {Y.}~\bibnamefont {Guang}}, \bibinfo {author}
  {\bibfnamefont {D.~M.}\ \bibnamefont {Burn}}, \bibinfo {author}
  {\bibfnamefont {G.}~\bibnamefont {van~der Laan}}, \bibinfo {author}
  {\bibfnamefont {T.}~\bibnamefont {Hesjedal}}, \bibinfo {author}
  {\bibfnamefont {H.}~\bibnamefont {Du}}, \bibinfo {author} {\bibfnamefont
  {G.}~\bibnamefont {Yu}},\ and\ \bibinfo {author} {\bibfnamefont
  {S.}~\bibnamefont {Zhang}},\ }\bibfield  {title} {\bibinfo {title} {Creation
  of a chiral bobber lattice in helimagnet-multilayer heterostructures},\
  }\href {https://doi.org/10.1103/PhysRevLett.126.017204} {\bibfield  {journal}
  {\bibinfo  {journal} {Phys. Rev. Lett.}\ }\textbf {\bibinfo {volume} {126}},\
  \bibinfo {pages} {017204} (\bibinfo {year} {2021})}\BibitemShut {NoStop}%
\bibitem [{\citenamefont {Ran}\ \emph {et~al.}(2022)\citenamefont {Ran},
  \citenamefont {Liu}, \citenamefont {Jin}, \citenamefont {Shangguan},
  \citenamefont {Guang}, \citenamefont {Wen}, \citenamefont {Yu}, \citenamefont
  {van~der Laan}, \citenamefont {Hesjedal},\ and\ \citenamefont
  {Zhang}}]{ran2022axially}%
  \BibitemOpen
  \bibfield  {author} {\bibinfo {author} {\bibfnamefont {K.}~\bibnamefont
  {Ran}}, \bibinfo {author} {\bibfnamefont {Y.}~\bibnamefont {Liu}}, \bibinfo
  {author} {\bibfnamefont {H.}~\bibnamefont {Jin}}, \bibinfo {author}
  {\bibfnamefont {Y.}~\bibnamefont {Shangguan}}, \bibinfo {author}
  {\bibfnamefont {Y.}~\bibnamefont {Guang}}, \bibinfo {author} {\bibfnamefont
  {J.}~\bibnamefont {Wen}}, \bibinfo {author} {\bibfnamefont {G.}~\bibnamefont
  {Yu}}, \bibinfo {author} {\bibfnamefont {G.}~\bibnamefont {van~der Laan}},
  \bibinfo {author} {\bibfnamefont {T.}~\bibnamefont {Hesjedal}},\ and\
  \bibinfo {author} {\bibfnamefont {S.}~\bibnamefont {Zhang}},\ }\bibfield
  {title} {\bibinfo {title} {Axially bound magnetic skyrmions: Glueing
  topological strings across an interface},\ }\href
  {https://doi.org/10.1021/acs.nanolett.2c00689} {\bibfield  {journal}
  {\bibinfo  {journal} {Nano Lett.}\ }\textbf {\bibinfo {volume} {22}},\
  \bibinfo {pages} {3737} (\bibinfo {year} {2022})}\BibitemShut {NoStop}%
\bibitem [{\citenamefont {Birch}\ \emph {et~al.}(2022)\citenamefont {Birch},
  \citenamefont {{Cort\'es-Ortu\~{n}o}}, \citenamefont {Litzius}, \citenamefont
  {Wintz}, \citenamefont {Schulz}, \citenamefont {Weigand}, \citenamefont
  {Mayoh}, \citenamefont {Balakrishnan}, \citenamefont {Hatton},\ and\
  \citenamefont {Sch\"utz}}]{birch2022toggle}%
  \BibitemOpen
  \bibfield  {author} {\bibinfo {author} {\bibfnamefont {M.~T.}\ \bibnamefont
  {Birch}}, \bibinfo {author} {\bibfnamefont {D.}~\bibnamefont
  {{Cort\'es-Ortu\~{n}o}}}, \bibinfo {author} {\bibfnamefont {K.}~\bibnamefont
  {Litzius}}, \bibinfo {author} {\bibfnamefont {S.}~\bibnamefont {Wintz}},
  \bibinfo {author} {\bibfnamefont {F.}~\bibnamefont {Schulz}}, \bibinfo
  {author} {\bibfnamefont {M.}~\bibnamefont {Weigand}}, \bibinfo {author}
  {\bibfnamefont {D.~A.}\ \bibnamefont {Mayoh}}, \bibinfo {author}
  {\bibfnamefont {G.}~\bibnamefont {Balakrishnan}}, \bibinfo {author}
  {\bibfnamefont {P.~D.}\ \bibnamefont {Hatton}},\ and\ \bibinfo {author}
  {\bibfnamefont {G.}~\bibnamefont {Sch\"utz}},\ }\bibfield  {title} {\bibinfo
  {title} {{Toggle-like current-induced Bloch point dynamics of 3D skyrmion
  strings in a room temperature nanowire}},\ }\href
  {https://doi.org/10.1038/s41467-022-31335-y} {\bibfield  {journal} {\bibinfo
  {journal} {Nat. Commun.}\ }\textbf {\bibinfo {volume} {13}},\ \bibinfo
  {pages} {3630} (\bibinfo {year} {2022})}\BibitemShut {NoStop}%
\bibitem [{\citenamefont {Jin}\ \emph {et~al.}(2023)\citenamefont {Jin},
  \citenamefont {Tan}, \citenamefont {Liu}, \citenamefont {Ran}, \citenamefont
  {Fan}, \citenamefont {Shangguan}, \citenamefont {Guang}, \citenamefont
  {van~der Laan}, \citenamefont {Hesjedal}, \citenamefont {Wen}, \citenamefont
  {Yu},\ and\ \citenamefont {Zhang}}]{jin2023evolution}%
  \BibitemOpen
  \bibfield  {author} {\bibinfo {author} {\bibfnamefont {H.}~\bibnamefont
  {Jin}}, \bibinfo {author} {\bibfnamefont {W.}~\bibnamefont {Tan}}, \bibinfo
  {author} {\bibfnamefont {Y.}~\bibnamefont {Liu}}, \bibinfo {author}
  {\bibfnamefont {K.}~\bibnamefont {Ran}}, \bibinfo {author} {\bibfnamefont
  {R.}~\bibnamefont {Fan}}, \bibinfo {author} {\bibfnamefont {Y.}~\bibnamefont
  {Shangguan}}, \bibinfo {author} {\bibfnamefont {Y.}~\bibnamefont {Guang}},
  \bibinfo {author} {\bibfnamefont {G.}~\bibnamefont {van~der Laan}}, \bibinfo
  {author} {\bibfnamefont {T.}~\bibnamefont {Hesjedal}}, \bibinfo {author}
  {\bibfnamefont {J.}~\bibnamefont {Wen}}, \bibinfo {author} {\bibfnamefont
  {G.}~\bibnamefont {Yu}},\ and\ \bibinfo {author} {\bibfnamefont
  {S.}~\bibnamefont {Zhang}},\ }\bibfield  {title} {\bibinfo {title} {Evolution
  of emergent monopoles into magnetic skyrmion strings},\ }\href
  {https://doi.org/10.1021/acs.nanolett.3c01117} {\bibfield  {journal}
  {\bibinfo  {journal} {Nano Lett.}\ }\textbf {\bibinfo {volume} {23}},\
  \bibinfo {pages} {5164} (\bibinfo {year} {2023})}\BibitemShut {NoStop}%
\bibitem [{\citenamefont {Xie}\ \emph {et~al.}(2023)\citenamefont {Xie},
  \citenamefont {Ran}, \citenamefont {Liu}, \citenamefont {Fan}, \citenamefont
  {Tan}, \citenamefont {Jin}, \citenamefont {Valvidares}, \citenamefont
  {Jaouen}, \citenamefont {Du}, \citenamefont {van~der Laan}, \citenamefont
  {Hesjedal},\ and\ \citenamefont {Zhang}}]{xie2023observation}%
  \BibitemOpen
  \bibfield  {author} {\bibinfo {author} {\bibfnamefont {X.}~\bibnamefont
  {Xie}}, \bibinfo {author} {\bibfnamefont {K.}~\bibnamefont {Ran}}, \bibinfo
  {author} {\bibfnamefont {Y.}~\bibnamefont {Liu}}, \bibinfo {author}
  {\bibfnamefont {R.}~\bibnamefont {Fan}}, \bibinfo {author} {\bibfnamefont
  {W.}~\bibnamefont {Tan}}, \bibinfo {author} {\bibfnamefont {H.}~\bibnamefont
  {Jin}}, \bibinfo {author} {\bibfnamefont {M.}~\bibnamefont {Valvidares}},
  \bibinfo {author} {\bibfnamefont {N.}~\bibnamefont {Jaouen}}, \bibinfo
  {author} {\bibfnamefont {H.}~\bibnamefont {Du}}, \bibinfo {author}
  {\bibfnamefont {G.}~\bibnamefont {van~der Laan}}, \bibinfo {author}
  {\bibfnamefont {T.}~\bibnamefont {Hesjedal}},\ and\ \bibinfo {author}
  {\bibfnamefont {S.}~\bibnamefont {Zhang}},\ }\bibfield  {title} {\bibinfo
  {title} {Observation of the skyrmion side-face state in a chiral magnet},\
  }\href {https://doi.org/10.1103/PhysRevB.107.L060405} {\bibfield  {journal}
  {\bibinfo  {journal} {Phys. Rev. B}\ }\textbf {\bibinfo {volume} {107}},\
  \bibinfo {pages} {L060405} (\bibinfo {year} {2023})}\BibitemShut {NoStop}%
\bibitem [{\citenamefont {Yu}\ \emph {et~al.}(2024)\citenamefont {Yu},
  \citenamefont {Nakanishi}, \citenamefont {Chiew}, \citenamefont {Liu},
  \citenamefont {Nakajima}, \citenamefont {Kanazawa}, \citenamefont {Karube},
  \citenamefont {Taguchi}, \citenamefont {Nagaosa},\ and\ \citenamefont
  {Tokura}}]{yu2024skyrmion}%
  \BibitemOpen
  \bibfield  {author} {\bibinfo {author} {\bibfnamefont {X.}~\bibnamefont
  {Yu}}, \bibinfo {author} {\bibfnamefont {N.}~\bibnamefont {Nakanishi}},
  \bibinfo {author} {\bibfnamefont {Y.-L.}\ \bibnamefont {Chiew}}, \bibinfo
  {author} {\bibfnamefont {Y.}~\bibnamefont {Liu}}, \bibinfo {author}
  {\bibfnamefont {K.}~\bibnamefont {Nakajima}}, \bibinfo {author}
  {\bibfnamefont {N.}~\bibnamefont {Kanazawa}}, \bibinfo {author}
  {\bibfnamefont {K.}~\bibnamefont {Karube}}, \bibinfo {author} {\bibfnamefont
  {Y.}~\bibnamefont {Taguchi}}, \bibinfo {author} {\bibfnamefont
  {N.}~\bibnamefont {Nagaosa}},\ and\ \bibinfo {author} {\bibfnamefont
  {Y.}~\bibnamefont {Tokura}},\ }\bibfield  {title} {\bibinfo {title} {{3D
  skyrmion strings and their melting dynamics revealed via scalar-field
  electron tomography}},\ }\href {https://doi.org/10.1038/s43246-024-00512-5}
  {\bibfield  {journal} {\bibinfo  {journal} {Commun. Mater.}\ }\textbf
  {\bibinfo {volume} {5}},\ \bibinfo {pages} {80} (\bibinfo {year}
  {2024})}\BibitemShut {NoStop}%
\bibitem [{\citenamefont {Zhang}\ \emph
  {et~al.}(2020{\natexlab{a}})\citenamefont {Zhang}, \citenamefont {Burn},
  \citenamefont {Jaouen}, \citenamefont {Chauleau}, \citenamefont
  {Haghighirad}, \citenamefont {Liu}, \citenamefont {Wang}, \citenamefont
  {van~der Laan},\ and\ \citenamefont {Hesjedal}}]{zhang2020robust}%
  \BibitemOpen
  \bibfield  {author} {\bibinfo {author} {\bibfnamefont {S.}~\bibnamefont
  {Zhang}}, \bibinfo {author} {\bibfnamefont {D.~M.}\ \bibnamefont {Burn}},
  \bibinfo {author} {\bibfnamefont {N.}~\bibnamefont {Jaouen}}, \bibinfo
  {author} {\bibfnamefont {J.-Y.}\ \bibnamefont {Chauleau}}, \bibinfo {author}
  {\bibfnamefont {A.~A.}\ \bibnamefont {Haghighirad}}, \bibinfo {author}
  {\bibfnamefont {Y.}~\bibnamefont {Liu}}, \bibinfo {author} {\bibfnamefont
  {W.}~\bibnamefont {Wang}}, \bibinfo {author} {\bibfnamefont {G.}~\bibnamefont
  {van~der Laan}},\ and\ \bibinfo {author} {\bibfnamefont {T.}~\bibnamefont
  {Hesjedal}},\ }\bibfield  {title} {\bibinfo {title} {Robust perpendicular
  skyrmions and their surface confinement},\ }\href
  {https://doi.org/10.1021/acs.nanolett.9b05141} {\bibfield  {journal}
  {\bibinfo  {journal} {Nano Lett.}\ }\textbf {\bibinfo {volume} {20}},\
  \bibinfo {pages} {1428} (\bibinfo {year} {2020}{\natexlab{a}})}\BibitemShut
  {NoStop}%
\bibitem [{\citenamefont {Rybakov}\ \emph {et~al.}(2015)\citenamefont
  {Rybakov}, \citenamefont {Borisov}, \citenamefont {Bl\"ugel},\ and\
  \citenamefont {Kiselev}}]{rybakov2015new}%
  \BibitemOpen
  \bibfield  {author} {\bibinfo {author} {\bibfnamefont {F.~N.}\ \bibnamefont
  {Rybakov}}, \bibinfo {author} {\bibfnamefont {A.~B.}\ \bibnamefont
  {Borisov}}, \bibinfo {author} {\bibfnamefont {S.}~\bibnamefont {Bl\"ugel}},\
  and\ \bibinfo {author} {\bibfnamefont {N.~S.}\ \bibnamefont {Kiselev}},\
  }\bibfield  {title} {\bibinfo {title} {New type of stable particlelike states
  in chiral magnets},\ }\href {https://doi.org/10.1103/physrevlett.115.117201}
  {\bibfield  {journal} {\bibinfo  {journal} {Phys. Rev. Lett.}\ }\textbf
  {\bibinfo {volume} {115}},\ \bibinfo {pages} {117201} (\bibinfo {year}
  {2015})}\BibitemShut {NoStop}%
\bibitem [{\citenamefont {Donnelly}\ \emph {et~al.}(2020)\citenamefont
  {Donnelly}, \citenamefont {Finizio}, \citenamefont {Gliga}, \citenamefont
  {Holler}, \citenamefont {Hrabec}, \citenamefont {Odstr\u{c}il}, \citenamefont
  {Mayr}, \citenamefont {Scagnoli}, \citenamefont {Heyderman}, \citenamefont
  {Guizar-Sicairos},\ and\ \citenamefont {Raabe}}]{donnelly2020time}%
  \BibitemOpen
  \bibfield  {author} {\bibinfo {author} {\bibfnamefont {C.}~\bibnamefont
  {Donnelly}}, \bibinfo {author} {\bibfnamefont {S.}~\bibnamefont {Finizio}},
  \bibinfo {author} {\bibfnamefont {S.}~\bibnamefont {Gliga}}, \bibinfo
  {author} {\bibfnamefont {M.}~\bibnamefont {Holler}}, \bibinfo {author}
  {\bibfnamefont {A.}~\bibnamefont {Hrabec}}, \bibinfo {author} {\bibfnamefont
  {M.}~\bibnamefont {Odstr\u{c}il}}, \bibinfo {author} {\bibfnamefont
  {S.}~\bibnamefont {Mayr}}, \bibinfo {author} {\bibfnamefont {V.}~\bibnamefont
  {Scagnoli}}, \bibinfo {author} {\bibfnamefont {L.~J.}\ \bibnamefont
  {Heyderman}}, \bibinfo {author} {\bibfnamefont {M.}~\bibnamefont
  {Guizar-Sicairos}},\ and\ \bibinfo {author} {\bibfnamefont {J.}~\bibnamefont
  {Raabe}},\ }\bibfield  {title} {\bibinfo {title} {Time-resolved imaging of
  three-dimensional nanoscale magnetization dynamics},\ }\href@noop {}
  {\bibfield  {journal} {\bibinfo  {journal} {Nat. Nanotech.}\ }\textbf
  {\bibinfo {volume} {15}},\ \bibinfo {pages} {356} (\bibinfo {year}
  {2020})}\BibitemShut {NoStop}%
\bibitem [{\citenamefont {Nagaosa}\ and\ \citenamefont
  {Tokura}(2012)}]{nagaosa2012emergent}%
  \BibitemOpen
  \bibfield  {author} {\bibinfo {author} {\bibfnamefont {N.}~\bibnamefont
  {Nagaosa}}\ and\ \bibinfo {author} {\bibfnamefont {Y.}~\bibnamefont
  {Tokura}},\ }\bibfield  {title} {\bibinfo {title} {Emergent electromagnetism
  in solids},\ }\href@noop {} {\bibfield  {journal} {\bibinfo  {journal} {Phys.
  Scr.}\ }\textbf {\bibinfo {volume} {2012}},\ \bibinfo {pages} {014020}
  (\bibinfo {year} {2012})}\BibitemShut {NoStop}%
\bibitem [{\citenamefont {Volovik}(2013)}]{volovik2013spinmotive}%
  \BibitemOpen
  \bibfield  {author} {\bibinfo {author} {\bibfnamefont {G.~E.}\ \bibnamefont
  {Volovik}},\ }\bibfield  {title} {\bibinfo {title} {Spin-motive force and
  orbital-motive force: From magnon {{Bose-Einstein}} condensation to chiral
  {{Weyl}} superfluids},\ }\href {https://doi.org/10.1134/S0021364013210145}
  {\bibfield  {journal} {\bibinfo  {journal} {Jetp Lett.}\ }\textbf {\bibinfo
  {volume} {98}},\ \bibinfo {pages} {480} (\bibinfo {year} {2013})}\BibitemShut
  {NoStop}%
\bibitem [{\citenamefont {Yang}\ \emph {et~al.}(2009)\citenamefont {Yang},
  \citenamefont {Beach}, \citenamefont {Knutson}, \citenamefont {Xiao},
  \citenamefont {Niu}, \citenamefont {Tsoi},\ and\ \citenamefont
  {Erskine}}]{yang2009universal}%
  \BibitemOpen
  \bibfield  {author} {\bibinfo {author} {\bibfnamefont {S.~A.}\ \bibnamefont
  {Yang}}, \bibinfo {author} {\bibfnamefont {G.~S.~D.}\ \bibnamefont {Beach}},
  \bibinfo {author} {\bibfnamefont {C.}~\bibnamefont {Knutson}}, \bibinfo
  {author} {\bibfnamefont {D.}~\bibnamefont {Xiao}}, \bibinfo {author}
  {\bibfnamefont {Q.}~\bibnamefont {Niu}}, \bibinfo {author} {\bibfnamefont
  {M.}~\bibnamefont {Tsoi}},\ and\ \bibinfo {author} {\bibfnamefont {J.~L.}\
  \bibnamefont {Erskine}},\ }\bibfield  {title} {\bibinfo {title} {Universal
  {{Electromotive Force Induced}} by {{Domain Wall Motion}}},\ }\href
  {https://doi.org/10.1103/PhysRevLett.102.067201} {\bibfield  {journal}
  {\bibinfo  {journal} {Phys. Rev. Lett.}\ }\textbf {\bibinfo {volume} {102}},\
  \bibinfo {pages} {067201} (\bibinfo {year} {2009})}\BibitemShut {NoStop}%
\bibitem [{\citenamefont {Schulz}\ \emph {et~al.}(2012)\citenamefont {Schulz},
  \citenamefont {Ritz}, \citenamefont {Bauer}, \citenamefont {Halder},
  \citenamefont {Wagner}, \citenamefont {Franz}, \citenamefont {Pfleiderer},
  \citenamefont {Everschor}, \citenamefont {Garst},\ and\ \citenamefont
  {Rosch}}]{schulz2012emergent}%
  \BibitemOpen
  \bibfield  {author} {\bibinfo {author} {\bibfnamefont {T.}~\bibnamefont
  {Schulz}}, \bibinfo {author} {\bibfnamefont {R.}~\bibnamefont {Ritz}},
  \bibinfo {author} {\bibfnamefont {A.}~\bibnamefont {Bauer}}, \bibinfo
  {author} {\bibfnamefont {M.}~\bibnamefont {Halder}}, \bibinfo {author}
  {\bibfnamefont {M.}~\bibnamefont {Wagner}}, \bibinfo {author} {\bibfnamefont
  {C.}~\bibnamefont {Franz}}, \bibinfo {author} {\bibfnamefont
  {C.}~\bibnamefont {Pfleiderer}}, \bibinfo {author} {\bibfnamefont
  {K.}~\bibnamefont {Everschor}}, \bibinfo {author} {\bibfnamefont
  {M.}~\bibnamefont {Garst}},\ and\ \bibinfo {author} {\bibfnamefont
  {A.}~\bibnamefont {Rosch}},\ }\bibfield  {title} {\bibinfo {title} {Emergent
  electrodynamics of skyrmions in a chiral magnet},\ }\href
  {https://doi.org/10.1038/nphys2231} {\bibfield  {journal} {\bibinfo
  {journal} {Nat. Phys.}\ }\textbf {\bibinfo {volume} {8}},\ \bibinfo {pages}
  {301} (\bibinfo {year} {2012})}\BibitemShut {NoStop}%
\bibitem [{\citenamefont {Birch}\ \emph {et~al.}(2024)\citenamefont {Birch},
  \citenamefont {Belopolski}, \citenamefont {Fujishiro}, \citenamefont
  {Kawamura}, \citenamefont {Kikkawa}, \citenamefont {Taguchi}, \citenamefont
  {Hirschberger}, \citenamefont {Nagaosa},\ and\ \citenamefont
  {Tokura}}]{birch2024dynamic}%
  \BibitemOpen
  \bibfield  {author} {\bibinfo {author} {\bibfnamefont {M.~T.}\ \bibnamefont
  {Birch}}, \bibinfo {author} {\bibfnamefont {I.}~\bibnamefont {Belopolski}},
  \bibinfo {author} {\bibfnamefont {Y.}~\bibnamefont {Fujishiro}}, \bibinfo
  {author} {\bibfnamefont {M.}~\bibnamefont {Kawamura}}, \bibinfo {author}
  {\bibfnamefont {A.}~\bibnamefont {Kikkawa}}, \bibinfo {author} {\bibfnamefont
  {Y.}~\bibnamefont {Taguchi}}, \bibinfo {author} {\bibfnamefont
  {M.}~\bibnamefont {Hirschberger}}, \bibinfo {author} {\bibfnamefont
  {N.}~\bibnamefont {Nagaosa}},\ and\ \bibinfo {author} {\bibfnamefont
  {Y.}~\bibnamefont {Tokura}},\ }\bibfield  {title} {\bibinfo {title} {Dynamic
  transition and {{Galilean}} relativity of current-driven skyrmions},\ }\href
  {https://doi.org/10.1038/s41586-024-07859-2} {\bibfield  {journal} {\bibinfo
  {journal} {Nature}\ }\textbf {\bibinfo {volume} {633}},\ \bibinfo {pages}
  {554} (\bibinfo {year} {2024})}\BibitemShut {NoStop}%
\bibitem [{\citenamefont {Littlehales}\ \emph {et~al.}(2025)\citenamefont
  {Littlehales}, \citenamefont {Birch}, \citenamefont {Kikkawa}, \citenamefont
  {Taguchi}, \citenamefont {Venero}, \citenamefont {Hatton}, \citenamefont
  {Nagaosa}, \citenamefont {Tokura},\ and\ \citenamefont
  {Yokouchi}}]{littlehales2025emergent}%
  \BibitemOpen
  \bibfield  {author} {\bibinfo {author} {\bibfnamefont {M.~T.}\ \bibnamefont
  {Littlehales}}, \bibinfo {author} {\bibfnamefont {M.~T.}\ \bibnamefont
  {Birch}}, \bibinfo {author} {\bibfnamefont {A.}~\bibnamefont {Kikkawa}},
  \bibinfo {author} {\bibfnamefont {Y.}~\bibnamefont {Taguchi}}, \bibinfo
  {author} {\bibfnamefont {D.~A.}\ \bibnamefont {Venero}}, \bibinfo {author}
  {\bibfnamefont {P.~D.}\ \bibnamefont {Hatton}}, \bibinfo {author}
  {\bibfnamefont {N.}~\bibnamefont {Nagaosa}}, \bibinfo {author} {\bibfnamefont
  {Y.}~\bibnamefont {Tokura}},\ and\ \bibinfo {author} {\bibfnamefont
  {T.}~\bibnamefont {Yokouchi}},\ }\bibfield  {title} {\bibinfo {title}
  {Emergent reactance induced by the deformation of a current-driven skyrmion
  lattice},\ }\href@noop {} {\bibfield  {journal} {\bibinfo  {journal} {arXiv
  preprint arXiv:2505.18029}\ } (\bibinfo {year} {2025})}\BibitemShut {NoStop}%
\bibitem [{\citenamefont {Yokouchi}\ \emph
  {et~al.}(2020{\natexlab{a}})\citenamefont {Yokouchi}, \citenamefont {Kagawa},
  \citenamefont {Hirschberger}, \citenamefont {Otani}, \citenamefont
  {Nagaosa},\ and\ \citenamefont {Tokura}}]{yokouchi2020emergent}%
  \BibitemOpen
  \bibfield  {author} {\bibinfo {author} {\bibfnamefont {T.}~\bibnamefont
  {Yokouchi}}, \bibinfo {author} {\bibfnamefont {F.}~\bibnamefont {Kagawa}},
  \bibinfo {author} {\bibfnamefont {M.}~\bibnamefont {Hirschberger}}, \bibinfo
  {author} {\bibfnamefont {Y.}~\bibnamefont {Otani}}, \bibinfo {author}
  {\bibfnamefont {N.}~\bibnamefont {Nagaosa}},\ and\ \bibinfo {author}
  {\bibfnamefont {Y.}~\bibnamefont {Tokura}},\ }\bibfield  {title} {\bibinfo
  {title} {Emergent electromagnetic induction in a helical-spin magnet},\
  }\href {https://doi.org/10.1038/s41586-020-2775-x} {\bibfield  {journal}
  {\bibinfo  {journal} {Nature}\ }\textbf {\bibinfo {volume} {586}},\ \bibinfo
  {pages} {232} (\bibinfo {year} {2020}{\natexlab{a}})}\BibitemShut {NoStop}%
\bibitem [{\citenamefont {Hirschberger}\ \emph {et~al.}(2020)\citenamefont
  {Hirschberger}, \citenamefont {Spitz}, \citenamefont {Nomoto}, \citenamefont
  {Kurumaji}, \citenamefont {Gao}, \citenamefont {Masell}, \citenamefont
  {Nakajima}, \citenamefont {Kikkawa}, \citenamefont {Yamasaki}, \citenamefont
  {Sagayama}, \citenamefont {Nakao}, \citenamefont {Taguchi}, \citenamefont
  {Arita}, \citenamefont {Arima},\ and\ \citenamefont
  {Tokura}}]{hirschberger2020topological}%
  \BibitemOpen
  \bibfield  {author} {\bibinfo {author} {\bibfnamefont {M.}~\bibnamefont
  {Hirschberger}}, \bibinfo {author} {\bibfnamefont {L.}~\bibnamefont {Spitz}},
  \bibinfo {author} {\bibfnamefont {T.}~\bibnamefont {Nomoto}}, \bibinfo
  {author} {\bibfnamefont {T.}~\bibnamefont {Kurumaji}}, \bibinfo {author}
  {\bibfnamefont {S.}~\bibnamefont {Gao}}, \bibinfo {author} {\bibfnamefont
  {J.}~\bibnamefont {Masell}}, \bibinfo {author} {\bibfnamefont
  {T.}~\bibnamefont {Nakajima}}, \bibinfo {author} {\bibfnamefont
  {A.}~\bibnamefont {Kikkawa}}, \bibinfo {author} {\bibfnamefont
  {Y.}~\bibnamefont {Yamasaki}}, \bibinfo {author} {\bibfnamefont
  {H.}~\bibnamefont {Sagayama}}, \bibinfo {author} {\bibfnamefont
  {H.}~\bibnamefont {Nakao}}, \bibinfo {author} {\bibfnamefont
  {Y.}~\bibnamefont {Taguchi}}, \bibinfo {author} {\bibfnamefont
  {R.}~\bibnamefont {Arita}}, \bibinfo {author} {\bibfnamefont {T.-h.}\
  \bibnamefont {Arima}},\ and\ \bibinfo {author} {\bibfnamefont
  {Y.}~\bibnamefont {Tokura}},\ }\bibfield  {title} {\bibinfo {title}
  {Topological {{Nernst Effect}} of the {{Two-Dimensional Skyrmion Lattice}}},\
  }\href {https://doi.org/10.1103/PhysRevLett.125.076602} {\bibfield  {journal}
  {\bibinfo  {journal} {Phys. Rev. Lett.}\ }\textbf {\bibinfo {volume} {125}},\
  \bibinfo {pages} {076602} (\bibinfo {year} {2020})}\BibitemShut {NoStop}%
\bibitem [{\citenamefont {Kato}\ \emph {et~al.}(2023)\citenamefont {Kato},
  \citenamefont {Okamura}, \citenamefont {Hirschberger}, \citenamefont
  {Tokura},\ and\ \citenamefont {Takahashi}}]{kato2023topological}%
  \BibitemOpen
  \bibfield  {author} {\bibinfo {author} {\bibfnamefont {Y.~D.}\ \bibnamefont
  {Kato}}, \bibinfo {author} {\bibfnamefont {Y.}~\bibnamefont {Okamura}},
  \bibinfo {author} {\bibfnamefont {M.}~\bibnamefont {Hirschberger}}, \bibinfo
  {author} {\bibfnamefont {Y.}~\bibnamefont {Tokura}},\ and\ \bibinfo {author}
  {\bibfnamefont {Y.}~\bibnamefont {Takahashi}},\ }\bibfield  {title} {\bibinfo
  {title} {Topological magneto-optical effect from skyrmion lattice},\ }\href
  {https://doi.org/10.1038/s41467-023-41203-y} {\bibfield  {journal} {\bibinfo
  {journal} {Nat. Commun.}\ }\textbf {\bibinfo {volume} {14}},\ \bibinfo
  {pages} {5416} (\bibinfo {year} {2023})}\BibitemShut {NoStop}%
\bibitem [{\citenamefont {G{\"o}bel}\ \emph {et~al.}(2025)\citenamefont
  {G{\"o}bel}, \citenamefont {Schimpf},\ and\ \citenamefont
  {Mertig}}]{gobel2025topological}%
  \BibitemOpen
  \bibfield  {author} {\bibinfo {author} {\bibfnamefont {B.}~\bibnamefont
  {G{\"o}bel}}, \bibinfo {author} {\bibfnamefont {L.}~\bibnamefont {Schimpf}},\
  and\ \bibinfo {author} {\bibfnamefont {I.}~\bibnamefont {Mertig}},\
  }\bibfield  {title} {\bibinfo {title} {Topological orbital {{Hall}} effect
  caused by skyrmions and antiferromagnetic skyrmions},\ }\href
  {https://doi.org/10.1038/s42005-024-01925-x} {\bibfield  {journal} {\bibinfo
  {journal} {Commun Phys}\ }\textbf {\bibinfo {volume} {8}},\ \bibinfo {pages}
  {17} (\bibinfo {year} {2025})}\BibitemShut {NoStop}%
\bibitem [{\citenamefont {Matsui}\ \emph {et~al.}(2021)\citenamefont {Matsui},
  \citenamefont {Nomoto},\ and\ \citenamefont
  {Arita}}]{matsui2021skyrmionsize}%
  \BibitemOpen
  \bibfield  {author} {\bibinfo {author} {\bibfnamefont {A.}~\bibnamefont
  {Matsui}}, \bibinfo {author} {\bibfnamefont {T.}~\bibnamefont {Nomoto}},\
  and\ \bibinfo {author} {\bibfnamefont {R.}~\bibnamefont {Arita}},\ }\bibfield
   {title} {\bibinfo {title} {Skyrmion-size dependence of the topological
  {{Hall}} effect: {{A}} real-space calculation},\ }\href
  {https://doi.org/10.1103/PhysRevB.104.174432} {\bibfield  {journal} {\bibinfo
   {journal} {Phys. Rev. B}\ }\textbf {\bibinfo {volume} {104}},\ \bibinfo
  {pages} {174432} (\bibinfo {year} {2021})}\BibitemShut {NoStop}%
\bibitem [{\citenamefont {Garst}\ \emph {et~al.}(2017)\citenamefont {Garst},
  \citenamefont {Waizner},\ and\ \citenamefont
  {Grundler}}]{garst2017collective}%
  \BibitemOpen
  \bibfield  {author} {\bibinfo {author} {\bibfnamefont {M.}~\bibnamefont
  {Garst}}, \bibinfo {author} {\bibfnamefont {J.}~\bibnamefont {Waizner}},\
  and\ \bibinfo {author} {\bibfnamefont {D.}~\bibnamefont {Grundler}},\
  }\bibfield  {title} {\bibinfo {title} {Collective spin excitations of helices
  and magnetic skyrmions: review and perspectives of magnonics in
  non-centrosymmetric magnets},\ }\href@noop {} {\bibfield  {journal} {\bibinfo
   {journal} {J. Phys. D: Appl. Phys.}\ }\textbf {\bibinfo {volume} {50}},\
  \bibinfo {pages} {293002} (\bibinfo {year} {2017})}\BibitemShut {NoStop}%
\bibitem [{\citenamefont {Weber}\ \emph {et~al.}(2022)\citenamefont {Weber},
  \citenamefont {Fobes}, \citenamefont {Waizner}, \citenamefont {Steffens},
  \citenamefont {Tucker}, \citenamefont {B{\"o}hm}, \citenamefont {Beddrich},
  \citenamefont {Franz}, \citenamefont {Gabold}, \citenamefont {Bewley} \emph
  {et~al.}}]{weber2022topological}%
  \BibitemOpen
  \bibfield  {author} {\bibinfo {author} {\bibfnamefont {T.}~\bibnamefont
  {Weber}}, \bibinfo {author} {\bibfnamefont {D.~M.}\ \bibnamefont {Fobes}},
  \bibinfo {author} {\bibfnamefont {J.}~\bibnamefont {Waizner}}, \bibinfo
  {author} {\bibfnamefont {P.}~\bibnamefont {Steffens}}, \bibinfo {author}
  {\bibfnamefont {G.~S.}\ \bibnamefont {Tucker}}, \bibinfo {author}
  {\bibfnamefont {M.}~\bibnamefont {B{\"o}hm}}, \bibinfo {author}
  {\bibfnamefont {L.}~\bibnamefont {Beddrich}}, \bibinfo {author}
  {\bibfnamefont {C.}~\bibnamefont {Franz}}, \bibinfo {author} {\bibfnamefont
  {H.}~\bibnamefont {Gabold}}, \bibinfo {author} {\bibfnamefont
  {R.}~\bibnamefont {Bewley}}, \emph {et~al.},\ }\bibfield  {title} {\bibinfo
  {title} {Topological magnon band structure of emergent landau levels in a
  skyrmion lattice},\ }\href@noop {} {\bibfield  {journal} {\bibinfo  {journal}
  {Science}\ }\textbf {\bibinfo {volume} {375}},\ \bibinfo {pages} {1025}
  (\bibinfo {year} {2022})}\BibitemShut {NoStop}%
\bibitem [{\citenamefont {D{\'\i}az}\ \emph {et~al.}(2019)\citenamefont
  {D{\'\i}az}, \citenamefont {Klinovaja},\ and\ \citenamefont
  {Loss}}]{diaz2019topological}%
  \BibitemOpen
  \bibfield  {author} {\bibinfo {author} {\bibfnamefont {S.~A.}\ \bibnamefont
  {D{\'\i}az}}, \bibinfo {author} {\bibfnamefont {J.}~\bibnamefont
  {Klinovaja}},\ and\ \bibinfo {author} {\bibfnamefont {D.}~\bibnamefont
  {Loss}},\ }\bibfield  {title} {\bibinfo {title} {Topological magnons and edge
  states in antiferromagnetic skyrmion crystals},\ }\href@noop {} {\bibfield
  {journal} {\bibinfo  {journal} {Phys. Rev. Lett.}\ }\textbf {\bibinfo
  {volume} {122}},\ \bibinfo {pages} {187203} (\bibinfo {year}
  {2019})}\BibitemShut {NoStop}%
\bibitem [{\citenamefont {Schwarze}\ \emph {et~al.}(2015)\citenamefont
  {Schwarze}, \citenamefont {Waizner}, \citenamefont {Garst}, \citenamefont
  {Bauer}, \citenamefont {Stasinopoulos}, \citenamefont {Berger}, \citenamefont
  {Pfleiderer},\ and\ \citenamefont {Grundler}}]{schwarze2015universal}%
  \BibitemOpen
  \bibfield  {author} {\bibinfo {author} {\bibfnamefont {T.}~\bibnamefont
  {Schwarze}}, \bibinfo {author} {\bibfnamefont {J.}~\bibnamefont {Waizner}},
  \bibinfo {author} {\bibfnamefont {M.}~\bibnamefont {Garst}}, \bibinfo
  {author} {\bibfnamefont {A.}~\bibnamefont {Bauer}}, \bibinfo {author}
  {\bibfnamefont {I.}~\bibnamefont {Stasinopoulos}}, \bibinfo {author}
  {\bibfnamefont {H.}~\bibnamefont {Berger}}, \bibinfo {author} {\bibfnamefont
  {C.}~\bibnamefont {Pfleiderer}},\ and\ \bibinfo {author} {\bibfnamefont
  {D.}~\bibnamefont {Grundler}},\ }\bibfield  {title} {\bibinfo {title}
  {Universal helimagnon and skyrmion excitations in metallic, semiconducting
  and insulating chiral magnets},\ }\href@noop {} {\bibfield  {journal}
  {\bibinfo  {journal} {Nat. Mater.}\ }\textbf {\bibinfo {volume} {14}},\
  \bibinfo {pages} {478} (\bibinfo {year} {2015})}\BibitemShut {NoStop}%
\bibitem [{\citenamefont {Ogawa}\ \emph {et~al.}(2015)\citenamefont {Ogawa},
  \citenamefont {Seki},\ and\ \citenamefont {Tokura}}]{ogawa2015ultrafast}%
  \BibitemOpen
  \bibfield  {author} {\bibinfo {author} {\bibfnamefont {N.}~\bibnamefont
  {Ogawa}}, \bibinfo {author} {\bibfnamefont {S.}~\bibnamefont {Seki}},\ and\
  \bibinfo {author} {\bibfnamefont {Y.}~\bibnamefont {Tokura}},\ }\bibfield
  {title} {\bibinfo {title} {Ultrafast optical excitation of magnetic
  skyrmions},\ }\href {https://doi.org/10.1038/srep09552} {\bibfield  {journal}
  {\bibinfo  {journal} {Sci Rep}\ }\textbf {\bibinfo {volume} {5}},\ \bibinfo
  {pages} {9552} (\bibinfo {year} {2015})}\BibitemShut {NoStop}%
\bibitem [{\citenamefont {Che}\ \emph {et~al.}(2025)\citenamefont {Che},
  \citenamefont {Ciola}, \citenamefont {Garst}, \citenamefont {Kravchuk},
  \citenamefont {Baral}, \citenamefont {Magrez}, \citenamefont {Berger},
  \citenamefont {Sch{\"o}nenberger}, \citenamefont {R{\o}nnow},\ and\
  \citenamefont {Grundler}}]{che2025short}%
  \BibitemOpen
  \bibfield  {author} {\bibinfo {author} {\bibfnamefont {P.}~\bibnamefont
  {Che}}, \bibinfo {author} {\bibfnamefont {R.}~\bibnamefont {Ciola}}, \bibinfo
  {author} {\bibfnamefont {M.}~\bibnamefont {Garst}}, \bibinfo {author}
  {\bibfnamefont {V.}~\bibnamefont {Kravchuk}}, \bibinfo {author}
  {\bibfnamefont {P.~R.}\ \bibnamefont {Baral}}, \bibinfo {author}
  {\bibfnamefont {A.}~\bibnamefont {Magrez}}, \bibinfo {author} {\bibfnamefont
  {H.}~\bibnamefont {Berger}}, \bibinfo {author} {\bibfnamefont
  {T.}~\bibnamefont {Sch{\"o}nenberger}}, \bibinfo {author} {\bibfnamefont
  {H.~M.}\ \bibnamefont {R{\o}nnow}},\ and\ \bibinfo {author} {\bibfnamefont
  {D.}~\bibnamefont {Grundler}},\ }\bibfield  {title} {\bibinfo {title}
  {Short-wave magnons with multipole spin precession detected in the
  topological bands of a skyrmion lattice},\ }\href@noop {} {\bibfield
  {journal} {\bibinfo  {journal} {Communications Materials}\ }\textbf {\bibinfo
  {volume} {6}},\ \bibinfo {pages} {139} (\bibinfo {year} {2025})}\BibitemShut
  {NoStop}%
\bibitem [{\citenamefont {Takagi}\ \emph {et~al.}(2021)\citenamefont {Takagi},
  \citenamefont {Garst}, \citenamefont {Sahliger}, \citenamefont {Back},
  \citenamefont {Tokura},\ and\ \citenamefont {Seki}}]{takagi2021hybridized}%
  \BibitemOpen
  \bibfield  {author} {\bibinfo {author} {\bibfnamefont {R.}~\bibnamefont
  {Takagi}}, \bibinfo {author} {\bibfnamefont {M.}~\bibnamefont {Garst}},
  \bibinfo {author} {\bibfnamefont {J.}~\bibnamefont {Sahliger}}, \bibinfo
  {author} {\bibfnamefont {C.~H.}\ \bibnamefont {Back}}, \bibinfo {author}
  {\bibfnamefont {Y.}~\bibnamefont {Tokura}},\ and\ \bibinfo {author}
  {\bibfnamefont {S.}~\bibnamefont {Seki}},\ }\bibfield  {title} {\bibinfo
  {title} {Hybridized magnon modes in the quenched skyrmion crystal},\
  }\href@noop {} {\bibfield  {journal} {\bibinfo  {journal} {Phys. Rev. B}\
  }\textbf {\bibinfo {volume} {104}},\ \bibinfo {pages} {144410} (\bibinfo
  {year} {2021})}\BibitemShut {NoStop}%
\bibitem [{\citenamefont {Okamura}\ \emph {et~al.}(2013)\citenamefont
  {Okamura}, \citenamefont {Kagawa}, \citenamefont {Mochizuki}, \citenamefont
  {Kubota}, \citenamefont {Seki}, \citenamefont {Ishiwata}, \citenamefont
  {Kawasaki}, \citenamefont {Onose},\ and\ \citenamefont
  {Tokura}}]{okamura2013microwave}%
  \BibitemOpen
  \bibfield  {author} {\bibinfo {author} {\bibfnamefont {Y.}~\bibnamefont
  {Okamura}}, \bibinfo {author} {\bibfnamefont {F.}~\bibnamefont {Kagawa}},
  \bibinfo {author} {\bibfnamefont {M.}~\bibnamefont {Mochizuki}}, \bibinfo
  {author} {\bibfnamefont {M.}~\bibnamefont {Kubota}}, \bibinfo {author}
  {\bibfnamefont {S.}~\bibnamefont {Seki}}, \bibinfo {author} {\bibfnamefont
  {S.}~\bibnamefont {Ishiwata}}, \bibinfo {author} {\bibfnamefont
  {M.}~\bibnamefont {Kawasaki}}, \bibinfo {author} {\bibfnamefont
  {Y.}~\bibnamefont {Onose}},\ and\ \bibinfo {author} {\bibfnamefont
  {Y.}~\bibnamefont {Tokura}},\ }\bibfield  {title} {\bibinfo {title}
  {Microwave magnetoelectric effect via skyrmion resonance modes in a
  helimagnetic multiferroic},\ }\href@noop {} {\bibfield  {journal} {\bibinfo
  {journal} {Nat. Commun.}\ }\textbf {\bibinfo {volume} {4}},\ \bibinfo {pages}
  {2391} (\bibinfo {year} {2013})}\BibitemShut {NoStop}%
\bibitem [{\citenamefont {Habel}\ \emph {et~al.}(2024)\citenamefont {Habel},
  \citenamefont {Mook}, \citenamefont {Willsher},\ and\ \citenamefont
  {Knolle}}]{habel2024breakdown}%
  \BibitemOpen
  \bibfield  {author} {\bibinfo {author} {\bibfnamefont {J.}~\bibnamefont
  {Habel}}, \bibinfo {author} {\bibfnamefont {A.}~\bibnamefont {Mook}},
  \bibinfo {author} {\bibfnamefont {J.}~\bibnamefont {Willsher}},\ and\
  \bibinfo {author} {\bibfnamefont {J.}~\bibnamefont {Knolle}},\ }\bibfield
  {title} {\bibinfo {title} {Breakdown of chiral edge modes in topological
  magnon insulators},\ }\href@noop {} {\bibfield  {journal} {\bibinfo
  {journal} {Phys. Rev. B}\ }\textbf {\bibinfo {volume} {109}},\ \bibinfo
  {pages} {024441} (\bibinfo {year} {2024})}\BibitemShut {NoStop}%
\bibitem [{\citenamefont {Stasinopoulos}\ \emph {et~al.}(2017)\citenamefont
  {Stasinopoulos}, \citenamefont {Weichselbaumer}, \citenamefont {Bauer},
  \citenamefont {Waizner}, \citenamefont {Berger}, \citenamefont {Maendl},
  \citenamefont {Garst}, \citenamefont {Pfleiderer},\ and\ \citenamefont
  {Grundler}}]{stasinopoulos2017low}%
  \BibitemOpen
  \bibfield  {author} {\bibinfo {author} {\bibfnamefont {I.}~\bibnamefont
  {Stasinopoulos}}, \bibinfo {author} {\bibfnamefont {S.}~\bibnamefont
  {Weichselbaumer}}, \bibinfo {author} {\bibfnamefont {A.}~\bibnamefont
  {Bauer}}, \bibinfo {author} {\bibfnamefont {J.}~\bibnamefont {Waizner}},
  \bibinfo {author} {\bibfnamefont {H.}~\bibnamefont {Berger}}, \bibinfo
  {author} {\bibfnamefont {S.}~\bibnamefont {Maendl}}, \bibinfo {author}
  {\bibfnamefont {M.}~\bibnamefont {Garst}}, \bibinfo {author} {\bibfnamefont
  {C.}~\bibnamefont {Pfleiderer}},\ and\ \bibinfo {author} {\bibfnamefont
  {D.}~\bibnamefont {Grundler}},\ }\bibfield  {title} {\bibinfo {title} {Low
  spin wave damping in the insulating chiral magnet cu2oseo3},\ }\href@noop {}
  {\bibfield  {journal} {\bibinfo  {journal} {Appl. Phys. Lett.}\ }\textbf
  {\bibinfo {volume} {111}} (\bibinfo {year} {2017})}\BibitemShut {NoStop}%
\bibitem [{\citenamefont {del Ser}\ \emph {et~al.}(2021)\citenamefont {del
  Ser}, \citenamefont {Heinen},\ and\ \citenamefont
  {Rosch}}]{del2021archimedean}%
  \BibitemOpen
  \bibfield  {author} {\bibinfo {author} {\bibfnamefont {N.}~\bibnamefont {del
  Ser}}, \bibinfo {author} {\bibfnamefont {L.}~\bibnamefont {Heinen}},\ and\
  \bibinfo {author} {\bibfnamefont {A.}~\bibnamefont {Rosch}},\ }\bibfield
  {title} {\bibinfo {title} {Archimedean screw in driven chiral magnets},\
  }\href@noop {} {\bibfield  {journal} {\bibinfo  {journal} {SciPost Phys.}\
  }\textbf {\bibinfo {volume} {11}},\ \bibinfo {pages} {009} (\bibinfo {year}
  {2021})}\BibitemShut {NoStop}%
\bibitem [{\citenamefont {Hirosawa}\ \emph {et~al.}(2022)\citenamefont
  {Hirosawa}, \citenamefont {Mook}, \citenamefont {Klinovaja},\ and\
  \citenamefont {Loss}}]{hirosawa2022magnetoelectric}%
  \BibitemOpen
  \bibfield  {author} {\bibinfo {author} {\bibfnamefont {T.}~\bibnamefont
  {Hirosawa}}, \bibinfo {author} {\bibfnamefont {A.}~\bibnamefont {Mook}},
  \bibinfo {author} {\bibfnamefont {J.}~\bibnamefont {Klinovaja}},\ and\
  \bibinfo {author} {\bibfnamefont {D.}~\bibnamefont {Loss}},\ }\bibfield
  {title} {\bibinfo {title} {Magnetoelectric cavity magnonics in skyrmion
  crystals},\ }\href@noop {} {\bibfield  {journal} {\bibinfo  {journal} {PRX
  Quantum}\ }\textbf {\bibinfo {volume} {3}},\ \bibinfo {pages} {040321}
  (\bibinfo {year} {2022})}\BibitemShut {NoStop}%
\bibitem [{\citenamefont {Tabuchi}\ \emph {et~al.}(2015)\citenamefont
  {Tabuchi}, \citenamefont {Ishino}, \citenamefont {Noguchi}, \citenamefont
  {Ishikawa}, \citenamefont {Yamazaki}, \citenamefont {Usami},\ and\
  \citenamefont {Nakamura}}]{tabuchi2015coherent}%
  \BibitemOpen
  \bibfield  {author} {\bibinfo {author} {\bibfnamefont {Y.}~\bibnamefont
  {Tabuchi}}, \bibinfo {author} {\bibfnamefont {S.}~\bibnamefont {Ishino}},
  \bibinfo {author} {\bibfnamefont {A.}~\bibnamefont {Noguchi}}, \bibinfo
  {author} {\bibfnamefont {T.}~\bibnamefont {Ishikawa}}, \bibinfo {author}
  {\bibfnamefont {R.}~\bibnamefont {Yamazaki}}, \bibinfo {author}
  {\bibfnamefont {K.}~\bibnamefont {Usami}},\ and\ \bibinfo {author}
  {\bibfnamefont {Y.}~\bibnamefont {Nakamura}},\ }\bibfield  {title} {\bibinfo
  {title} {Coherent coupling between a ferromagnetic magnon and a
  superconducting qubit},\ }\href@noop {} {\bibfield  {journal} {\bibinfo
  {journal} {Science}\ }\textbf {\bibinfo {volume} {349}},\ \bibinfo {pages}
  {405} (\bibinfo {year} {2015})}\BibitemShut {NoStop}%
\bibitem [{\citenamefont {Tengdin}\ \emph {et~al.}(2022)\citenamefont
  {Tengdin}, \citenamefont {Truc}, \citenamefont {Sapozhnik}, \citenamefont
  {Kong}, \citenamefont {del Ser}, \citenamefont {Gargiulo}, \citenamefont
  {Madan}, \citenamefont {Sch{\"o}nenberger}, \citenamefont {Baral},
  \citenamefont {Che} \emph {et~al.}}]{tengdin2022imaging}%
  \BibitemOpen
  \bibfield  {author} {\bibinfo {author} {\bibfnamefont {P.}~\bibnamefont
  {Tengdin}}, \bibinfo {author} {\bibfnamefont {B.}~\bibnamefont {Truc}},
  \bibinfo {author} {\bibfnamefont {A.}~\bibnamefont {Sapozhnik}}, \bibinfo
  {author} {\bibfnamefont {L.}~\bibnamefont {Kong}}, \bibinfo {author}
  {\bibfnamefont {N.}~\bibnamefont {del Ser}}, \bibinfo {author} {\bibfnamefont
  {S.}~\bibnamefont {Gargiulo}}, \bibinfo {author} {\bibfnamefont
  {I.}~\bibnamefont {Madan}}, \bibinfo {author} {\bibfnamefont
  {T.}~\bibnamefont {Sch{\"o}nenberger}}, \bibinfo {author} {\bibfnamefont
  {P.~R.}\ \bibnamefont {Baral}}, \bibinfo {author} {\bibfnamefont
  {P.}~\bibnamefont {Che}}, \emph {et~al.},\ }\bibfield  {title} {\bibinfo
  {title} {Imaging the ultrafast coherent control of a skyrmion crystal},\
  }\href@noop {} {\bibfield  {journal} {\bibinfo  {journal} {Phys. Rev. X}\
  }\textbf {\bibinfo {volume} {12}},\ \bibinfo {pages} {041030} (\bibinfo
  {year} {2022})}\BibitemShut {NoStop}%
\bibitem [{\citenamefont {Chen}\ \emph
  {et~al.}(2025{\natexlab{b}})\citenamefont {Chen}, \citenamefont {Xu},
  \citenamefont {Wang}, \citenamefont {Wagner}, \citenamefont {Sheng},
  \citenamefont {Jia}, \citenamefont {Wei}, \citenamefont {Zhang},
  \citenamefont {Zhang}, \citenamefont {Wang} \emph
  {et~al.}}]{chen2025deterministic}%
  \BibitemOpen
  \bibfield  {author} {\bibinfo {author} {\bibfnamefont {J.}~\bibnamefont
  {Chen}}, \bibinfo {author} {\bibfnamefont {M.}~\bibnamefont {Xu}}, \bibinfo
  {author} {\bibfnamefont {J.}~\bibnamefont {Wang}}, \bibinfo {author}
  {\bibfnamefont {K.}~\bibnamefont {Wagner}}, \bibinfo {author} {\bibfnamefont
  {L.}~\bibnamefont {Sheng}}, \bibinfo {author} {\bibfnamefont
  {H.}~\bibnamefont {Jia}}, \bibinfo {author} {\bibfnamefont {W.}~\bibnamefont
  {Wei}}, \bibinfo {author} {\bibfnamefont {H.}~\bibnamefont {Zhang}}, \bibinfo
  {author} {\bibfnamefont {Y.}~\bibnamefont {Zhang}}, \bibinfo {author}
  {\bibfnamefont {H.}~\bibnamefont {Wang}}, \emph {et~al.},\ }\bibfield
  {title} {\bibinfo {title} {Deterministic switching of antiferromagnetic spin
  textures by nonlinear magnons},\ }\href@noop {} {\bibfield  {journal}
  {\bibinfo  {journal} {Nat. Commun.}\ }\textbf {\bibinfo {volume} {16}},\
  \bibinfo {pages} {5794} (\bibinfo {year} {2025}{\natexlab{b}})}\BibitemShut
  {NoStop}%
\bibitem [{\citenamefont {Marrows}\ and\ \citenamefont
  {Zeissler}(2021)}]{Marrows2021}%
  \BibitemOpen
  \bibfield  {author} {\bibinfo {author} {\bibfnamefont {C.~H.}\ \bibnamefont
  {Marrows}}\ and\ \bibinfo {author} {\bibfnamefont {K.}~\bibnamefont
  {Zeissler}},\ }\bibfield  {title} {\bibinfo {title} {Perspective on skyrmion
  spintronics},\ }\href {https://doi.org/https://doi.org/10.1063/5.0072735}
  {\bibfield  {journal} {\bibinfo  {journal} {Appl. Phys. Lett.}\ }\textbf
  {\bibinfo {volume} {119}},\ \bibinfo {pages} {250502} (\bibinfo {year}
  {2021})}\BibitemShut {NoStop}%
\bibitem [{\citenamefont {G{\"o}bel}\ \emph {et~al.}(2021)\citenamefont
  {G{\"o}bel}, \citenamefont {Mertig},\ and\ \citenamefont
  {Tretiakov}}]{gobel2021skyrmions}%
  \BibitemOpen
  \bibfield  {author} {\bibinfo {author} {\bibfnamefont {B.}~\bibnamefont
  {G{\"o}bel}}, \bibinfo {author} {\bibfnamefont {I.}~\bibnamefont {Mertig}},\
  and\ \bibinfo {author} {\bibfnamefont {O.~A.}\ \bibnamefont {Tretiakov}},\
  }\bibfield  {title} {\bibinfo {title} {Beyond skyrmions: {{Review}} and
  perspectives of alternative magnetic quasiparticles},\ }\href
  {https://doi.org/10.1016/j.physrep.2020.10.001} {\bibfield  {journal}
  {\bibinfo  {journal} {Phys. Rep.}\ }\bibinfo {series} {Beyond Skyrmions:
  {{Review}} and Perspectives of Alternative Magnetic Quasiparticles},\ \textbf
  {\bibinfo {volume} {895}},\ \bibinfo {pages} {1} (\bibinfo {year}
  {2021})}\BibitemShut {NoStop}%
\bibitem [{\citenamefont {Caretta}\ \emph {et~al.}(2018)\citenamefont
  {Caretta}, \citenamefont {Mann}, \citenamefont {B\"{u}ttner}, \citenamefont
  {Ueda}, \citenamefont {Pfau}, \citenamefont {G\"{u}nther}, \citenamefont
  {Hessing}, \citenamefont {Churikova}, \citenamefont {Klose}, \citenamefont
  {Schneider}, \citenamefont {Engel}, \citenamefont {Marcus}, \citenamefont
  {Bono}, \citenamefont {Bagschik}, \citenamefont {Eisebitt},\ and\
  \citenamefont {Beach}}]{Caretta2018}%
  \BibitemOpen
  \bibfield  {author} {\bibinfo {author} {\bibfnamefont {L.}~\bibnamefont
  {Caretta}}, \bibinfo {author} {\bibfnamefont {M.}~\bibnamefont {Mann}},
  \bibinfo {author} {\bibfnamefont {F.}~\bibnamefont {B\"{u}ttner}}, \bibinfo
  {author} {\bibfnamefont {K.}~\bibnamefont {Ueda}}, \bibinfo {author}
  {\bibfnamefont {B.}~\bibnamefont {Pfau}}, \bibinfo {author} {\bibfnamefont
  {C.~M.}\ \bibnamefont {G\"{u}nther}}, \bibinfo {author} {\bibfnamefont
  {P.}~\bibnamefont {Hessing}}, \bibinfo {author} {\bibfnamefont
  {A.}~\bibnamefont {Churikova}}, \bibinfo {author} {\bibfnamefont
  {C.}~\bibnamefont {Klose}}, \bibinfo {author} {\bibfnamefont
  {M.}~\bibnamefont {Schneider}}, \bibinfo {author} {\bibfnamefont
  {D.}~\bibnamefont {Engel}}, \bibinfo {author} {\bibfnamefont
  {C.}~\bibnamefont {Marcus}}, \bibinfo {author} {\bibfnamefont
  {D.}~\bibnamefont {Bono}}, \bibinfo {author} {\bibfnamefont {K.}~\bibnamefont
  {Bagschik}}, \bibinfo {author} {\bibfnamefont {S.}~\bibnamefont {Eisebitt}},\
  and\ \bibinfo {author} {\bibfnamefont {G.~S.~D.}\ \bibnamefont {Beach}},\
  }\bibfield  {title} {\bibinfo {title} {Fast current-driven domain walls and
  small skyrmions in a compensated ferrimagnet},\ }\href
  {https://doi.org/https://doi.org/10.1038/s41565-018-0255-3} {\bibfield
  {journal} {\bibinfo  {journal} {Nat. Nanotechnol.}\ }\textbf {\bibinfo
  {volume} {13}},\ \bibinfo {pages} {1154–1160} (\bibinfo {year}
  {2018})}\BibitemShut {NoStop}%
\bibitem [{\citenamefont {Romming}\ \emph {et~al.}(2015)\citenamefont
  {Romming}, \citenamefont {Kubetzka}, \citenamefont {Hanneken}, \citenamefont
  {von Bergmann},\ and\ \citenamefont {Wiesendanger}}]{Romming2015}%
  \BibitemOpen
  \bibfield  {author} {\bibinfo {author} {\bibfnamefont {N.}~\bibnamefont
  {Romming}}, \bibinfo {author} {\bibfnamefont {A.}~\bibnamefont {Kubetzka}},
  \bibinfo {author} {\bibfnamefont {C.}~\bibnamefont {Hanneken}}, \bibinfo
  {author} {\bibfnamefont {K.}~\bibnamefont {von Bergmann}},\ and\ \bibinfo
  {author} {\bibfnamefont {R.}~\bibnamefont {Wiesendanger}},\ }\bibfield
  {title} {\bibinfo {title} {Field-dependent size and shape of single magnetic
  skyrmions},\ }\href
  {https://doi.org/https://doi.org/10.1103/PhysRevLett.114.177203} {\bibfield
  {journal} {\bibinfo  {journal} {Phys. Rev. Lett.}\ }\textbf {\bibinfo
  {volume} {114}},\ \bibinfo {pages} {177203} (\bibinfo {year}
  {2015})}\BibitemShut {NoStop}%
\bibitem [{\citenamefont {Baruffaldi}\ \emph {et~al.}(2025)\citenamefont
  {Baruffaldi}, \citenamefont {Bergamaschi}, \citenamefont {Boscardin},
  \citenamefont {Br{\"u}ckner}, \citenamefont {Butcher}, \citenamefont
  {Carulla}, \citenamefont {Centis~Vignali}, \citenamefont {Dinapoli},
  \citenamefont {Finizio}, \citenamefont {Fr{\"o}jdh} \emph
  {et~al.}}]{Baruffaldi2025}%
  \BibitemOpen
  \bibfield  {author} {\bibinfo {author} {\bibfnamefont {F.}~\bibnamefont
  {Baruffaldi}}, \bibinfo {author} {\bibfnamefont {A.}~\bibnamefont
  {Bergamaschi}}, \bibinfo {author} {\bibfnamefont {M.}~\bibnamefont
  {Boscardin}}, \bibinfo {author} {\bibfnamefont {M.}~\bibnamefont
  {Br{\"u}ckner}}, \bibinfo {author} {\bibfnamefont {T.~A.}\ \bibnamefont
  {Butcher}}, \bibinfo {author} {\bibfnamefont {M.}~\bibnamefont {Carulla}},
  \bibinfo {author} {\bibfnamefont {M.}~\bibnamefont {Centis~Vignali}},
  \bibinfo {author} {\bibfnamefont {R.}~\bibnamefont {Dinapoli}}, \bibinfo
  {author} {\bibfnamefont {S.}~\bibnamefont {Finizio}}, \bibinfo {author}
  {\bibfnamefont {E.}~\bibnamefont {Fr{\"o}jdh}}, \emph {et~al.},\ }\bibfield
  {title} {\bibinfo {title} {Single-photon counting pixel detector for soft
  x-rays},\ }\href@noop {} {\bibfield  {journal} {\bibinfo  {journal} {Commun.
  Phys.}\ }\textbf {\bibinfo {volume} {8}},\ \bibinfo {pages} {321} (\bibinfo
  {year} {2025})}\BibitemShut {NoStop}%
\bibitem [{\citenamefont {Neethirajan}\ \emph {et~al.}(2024)\citenamefont
  {Neethirajan}, \citenamefont {Daurer}, \citenamefont {Mart\'{\i}nez},
  \citenamefont {Hrabec}, \citenamefont {Turnbull}, \citenamefont {Yamamoto},
  \citenamefont {Ferreira}, \citenamefont {\ifmmode \check{S}\else
  \v{S}\fi{}tefan\ifmmode \check{c}\else \v{c}\fi{}i\ifmmode~\check{c}\else
  \v{c}\fi{}}, \citenamefont {Mayoh}, \citenamefont {Balakrishnan},
  \citenamefont {Pei}, \citenamefont {Xue}, \citenamefont {Chang},
  \citenamefont {Ringe}, \citenamefont {Harrison}, \citenamefont {Valencia},
  \citenamefont {Kazemian}, \citenamefont {Kaulich},\ and\ \citenamefont
  {Donnelly}}]{Neethirajan2024}%
  \BibitemOpen
  \bibfield  {author} {\bibinfo {author} {\bibfnamefont {J.}~\bibnamefont
  {Neethirajan}}, \bibinfo {author} {\bibfnamefont {B.~J.}\ \bibnamefont
  {Daurer}}, \bibinfo {author} {\bibfnamefont {M.~D.~P.}\ \bibnamefont
  {Mart\'{\i}nez}}, \bibinfo {author} {\bibfnamefont {A.~c.~v.}\ \bibnamefont
  {Hrabec}}, \bibinfo {author} {\bibfnamefont {L.}~\bibnamefont {Turnbull}},
  \bibinfo {author} {\bibfnamefont {R.}~\bibnamefont {Yamamoto}}, \bibinfo
  {author} {\bibfnamefont {M.~R.}\ \bibnamefont {Ferreira}}, \bibinfo {author}
  {\bibfnamefont {A.~c.~v.}\ \bibnamefont {\ifmmode \check{S}\else
  \v{S}\fi{}tefan\ifmmode \check{c}\else \v{c}\fi{}i\ifmmode~\check{c}\else
  \v{c}\fi{}}}, \bibinfo {author} {\bibfnamefont {D.~A.}\ \bibnamefont
  {Mayoh}}, \bibinfo {author} {\bibfnamefont {G.}~\bibnamefont {Balakrishnan}},
  \bibinfo {author} {\bibfnamefont {Z.}~\bibnamefont {Pei}}, \bibinfo {author}
  {\bibfnamefont {P.}~\bibnamefont {Xue}}, \bibinfo {author} {\bibfnamefont
  {L.}~\bibnamefont {Chang}}, \bibinfo {author} {\bibfnamefont
  {E.}~\bibnamefont {Ringe}}, \bibinfo {author} {\bibfnamefont
  {R.}~\bibnamefont {Harrison}}, \bibinfo {author} {\bibfnamefont
  {S.}~\bibnamefont {Valencia}}, \bibinfo {author} {\bibfnamefont
  {M.}~\bibnamefont {Kazemian}}, \bibinfo {author} {\bibfnamefont
  {B.}~\bibnamefont {Kaulich}},\ and\ \bibinfo {author} {\bibfnamefont
  {C.}~\bibnamefont {Donnelly}},\ }\bibfield  {title} {\bibinfo {title} {Soft
  x-ray phase nanomicroscopy of micrometer-thick magnets},\ }\href
  {https://doi.org/10.1103/PhysRevX.14.031028} {\bibfield  {journal} {\bibinfo
  {journal} {Phys. Rev. X}\ }\textbf {\bibinfo {volume} {14}},\ \bibinfo
  {pages} {031028} (\bibinfo {year} {2024})}\BibitemShut {NoStop}%
\bibitem [{\citenamefont {Panigrahy}\ \emph {et~al.}(2022)\citenamefont
  {Panigrahy}, \citenamefont {Mallick}, \citenamefont {Sampaio},\ and\
  \citenamefont {Rohart}}]{Panigrahy2022}%
  \BibitemOpen
  \bibfield  {author} {\bibinfo {author} {\bibfnamefont {S.}~\bibnamefont
  {Panigrahy}}, \bibinfo {author} {\bibfnamefont {S.}~\bibnamefont {Mallick}},
  \bibinfo {author} {\bibfnamefont {J.}~\bibnamefont {Sampaio}},\ and\ \bibinfo
  {author} {\bibfnamefont {S.}~\bibnamefont {Rohart}},\ }\bibfield  {title}
  {\bibinfo {title} {Skyrmion inertia in synthetic antiferromagnets},\ }\href
  {https://doi.org/https://doi.org/10.1103/PhysRevB.106.144405} {\bibfield
  {journal} {\bibinfo  {journal} {Phys. Rev. B}\ }\textbf {\bibinfo {volume}
  {106}},\ \bibinfo {pages} {144405} (\bibinfo {year} {2022})}\BibitemShut
  {NoStop}%
\bibitem [{\citenamefont {Llopart}\ \emph {et~al.}(2022)\citenamefont
  {Llopart}, \citenamefont {Alozy}, \citenamefont {Ballabriga}, \citenamefont
  {Campbell}, \citenamefont {Casanova}, \citenamefont {Gromov}, \citenamefont
  {Heijne}, \citenamefont {Poikela}, \citenamefont {Santin}, \citenamefont
  {Sriskaran}, \citenamefont {Tlustos},\ and\ \citenamefont
  {Vitkovskiy}}]{Llopart2022}%
  \BibitemOpen
  \bibfield  {author} {\bibinfo {author} {\bibfnamefont {X.}~\bibnamefont
  {Llopart}}, \bibinfo {author} {\bibfnamefont {J.}~\bibnamefont {Alozy}},
  \bibinfo {author} {\bibfnamefont {R.}~\bibnamefont {Ballabriga}}, \bibinfo
  {author} {\bibfnamefont {M.}~\bibnamefont {Campbell}}, \bibinfo {author}
  {\bibfnamefont {R.}~\bibnamefont {Casanova}}, \bibinfo {author}
  {\bibfnamefont {V.}~\bibnamefont {Gromov}}, \bibinfo {author} {\bibfnamefont
  {E.}~\bibnamefont {Heijne}}, \bibinfo {author} {\bibfnamefont
  {T.}~\bibnamefont {Poikela}}, \bibinfo {author} {\bibfnamefont
  {E.}~\bibnamefont {Santin}}, \bibinfo {author} {\bibfnamefont
  {V.}~\bibnamefont {Sriskaran}}, \bibinfo {author} {\bibfnamefont
  {L.}~\bibnamefont {Tlustos}},\ and\ \bibinfo {author} {\bibfnamefont
  {A.}~\bibnamefont {Vitkovskiy}},\ }\bibfield  {title} {\bibinfo {title}
  {Timepix4, a large area pixel detector readout chip which can be tiled on 4
  sides providing sub-200 ps timestamp binning},\ }\href
  {https://doi.org/10.1088/1748-0221/17/01/C01044} {\bibfield  {journal}
  {\bibinfo  {journal} {J. Instrum.}\ }\textbf {\bibinfo {volume} {17}}\bibinfo
   {number} { (01)},\ \bibinfo {pages} {C01044}}\BibitemShut {NoStop}%
\bibitem [{\citenamefont {Pancaldi}\ \emph {et~al.}(2024)\citenamefont
  {Pancaldi}, \citenamefont {Guzzi}, \citenamefont {Bevis}, \citenamefont
  {Manfredda}, \citenamefont {Barolak}, \citenamefont {Bonetti}, \citenamefont
  {Bykova}, \citenamefont {Angelis}, \citenamefont {Ninno}, \citenamefont
  {Fanciulli}, \citenamefont {Novinec}, \citenamefont {Pedersoli},
  \citenamefont {Ravindran}, \citenamefont {R\"{o}sner}, \citenamefont {David},
  \citenamefont {Ruchon}, \citenamefont {Simoncig}, \citenamefont {Zangrando},
  \citenamefont {Adams}, \citenamefont {Vavassori}, \citenamefont {Sacchi},
  \citenamefont {Kourousias}, \citenamefont {Mancini},\ and\ \citenamefont
  {Capotondi}}]{Pancaldi2024}%
  \BibitemOpen
\bibfield  {number} {  }\bibfield  {author} {\bibinfo {author} {\bibfnamefont
  {M.}~\bibnamefont {Pancaldi}}, \bibinfo {author} {\bibfnamefont
  {F.}~\bibnamefont {Guzzi}}, \bibinfo {author} {\bibfnamefont {C.~S.}\
  \bibnamefont {Bevis}}, \bibinfo {author} {\bibfnamefont {M.}~\bibnamefont
  {Manfredda}}, \bibinfo {author} {\bibfnamefont {J.}~\bibnamefont {Barolak}},
  \bibinfo {author} {\bibfnamefont {S.}~\bibnamefont {Bonetti}}, \bibinfo
  {author} {\bibfnamefont {I.}~\bibnamefont {Bykova}}, \bibinfo {author}
  {\bibfnamefont {D.~D.}\ \bibnamefont {Angelis}}, \bibinfo {author}
  {\bibfnamefont {G.~D.}\ \bibnamefont {Ninno}}, \bibinfo {author}
  {\bibfnamefont {M.}~\bibnamefont {Fanciulli}}, \bibinfo {author}
  {\bibfnamefont {L.}~\bibnamefont {Novinec}}, \bibinfo {author} {\bibfnamefont
  {E.}~\bibnamefont {Pedersoli}}, \bibinfo {author} {\bibfnamefont
  {A.}~\bibnamefont {Ravindran}}, \bibinfo {author} {\bibfnamefont
  {B.}~\bibnamefont {R\"{o}sner}}, \bibinfo {author} {\bibfnamefont
  {C.}~\bibnamefont {David}}, \bibinfo {author} {\bibfnamefont
  {T.}~\bibnamefont {Ruchon}}, \bibinfo {author} {\bibfnamefont
  {A.}~\bibnamefont {Simoncig}}, \bibinfo {author} {\bibfnamefont
  {M.}~\bibnamefont {Zangrando}}, \bibinfo {author} {\bibfnamefont {D.~E.}\
  \bibnamefont {Adams}}, \bibinfo {author} {\bibfnamefont {P.}~\bibnamefont
  {Vavassori}}, \bibinfo {author} {\bibfnamefont {M.}~\bibnamefont {Sacchi}},
  \bibinfo {author} {\bibfnamefont {G.}~\bibnamefont {Kourousias}}, \bibinfo
  {author} {\bibfnamefont {G.~F.}\ \bibnamefont {Mancini}},\ and\ \bibinfo
  {author} {\bibfnamefont {F.}~\bibnamefont {Capotondi}},\ }\bibfield  {title}
  {\bibinfo {title} {High-resolution ptychographic imaging at a seeded
  free-electron laser source using oam beams},\ }\href
  {https://doi.org/10.1364/OPTICA.509745} {\bibfield  {journal} {\bibinfo
  {journal} {Optica}\ }\textbf {\bibinfo {volume} {11}},\ \bibinfo {pages}
  {403} (\bibinfo {year} {2024})}\BibitemShut {NoStop}%
\bibitem [{\citenamefont {Witte}\ \emph {et~al.}(2020)\citenamefont {Witte},
  \citenamefont {Späth}, \citenamefont {Finizio}, \citenamefont {Donnelly},
  \citenamefont {Watts}, \citenamefont {Sarafimov}, \citenamefont {Odstrcil},
  \citenamefont {Guizar-Sicairos}, \citenamefont {Holler}, \citenamefont
  {Fink},\ and\ \citenamefont {Raabe}}]{Witte2020}%
  \BibitemOpen
  \bibfield  {author} {\bibinfo {author} {\bibfnamefont {K.}~\bibnamefont
  {Witte}}, \bibinfo {author} {\bibfnamefont {A.}~\bibnamefont {Späth}},
  \bibinfo {author} {\bibfnamefont {S.}~\bibnamefont {Finizio}}, \bibinfo
  {author} {\bibfnamefont {C.}~\bibnamefont {Donnelly}}, \bibinfo {author}
  {\bibfnamefont {B.}~\bibnamefont {Watts}}, \bibinfo {author} {\bibfnamefont
  {B.}~\bibnamefont {Sarafimov}}, \bibinfo {author} {\bibfnamefont
  {M.}~\bibnamefont {Odstrcil}}, \bibinfo {author} {\bibfnamefont
  {M.}~\bibnamefont {Guizar-Sicairos}}, \bibinfo {author} {\bibfnamefont
  {M.}~\bibnamefont {Holler}}, \bibinfo {author} {\bibfnamefont {R.~H.}\
  \bibnamefont {Fink}},\ and\ \bibinfo {author} {\bibfnamefont
  {J.}~\bibnamefont {Raabe}},\ }\bibfield  {title} {\bibinfo {title} {From 2d
  stxm to 3d imaging: Soft x-ray laminography of thin specimens},\ }\href
  {https://doi.org/10.1021/acs.nanolett.9b04782} {\bibfield  {journal}
  {\bibinfo  {journal} {Nano Lett.}\ }\textbf {\bibinfo {volume} {20}},\
  \bibinfo {pages} {1305} (\bibinfo {year} {2020})}\BibitemShut {NoStop}%
\bibitem [{\citenamefont {Iwasaki}\ \emph {et~al.}(2013)\citenamefont
  {Iwasaki}, \citenamefont {Mochizuki},\ and\ \citenamefont
  {Nagaosa}}]{iwasaki2013universal}%
  \BibitemOpen
  \bibfield  {author} {\bibinfo {author} {\bibfnamefont {J.}~\bibnamefont
  {Iwasaki}}, \bibinfo {author} {\bibfnamefont {M.}~\bibnamefont {Mochizuki}},\
  and\ \bibinfo {author} {\bibfnamefont {N.}~\bibnamefont {Nagaosa}},\
  }\bibfield  {title} {\bibinfo {title} {Universal current-velocity relation of
  skyrmion motion in chiral magnets},\ }\href
  {https://doi.org/10.1038/ncomms2442} {\bibfield  {journal} {\bibinfo
  {journal} {Nat. Commun.}\ }\textbf {\bibinfo {volume} {4}},\ \bibinfo {pages}
  {1463} (\bibinfo {year} {2013})}\BibitemShut {NoStop}%
\bibitem [{\citenamefont {Peng}\ \emph {et~al.}(2021)\citenamefont {Peng},
  \citenamefont {Karube}, \citenamefont {Taguchi}, \citenamefont {Nagaosa},
  \citenamefont {Tokura},\ and\ \citenamefont {Yu}}]{peng2021dynamic}%
  \BibitemOpen
  \bibfield  {author} {\bibinfo {author} {\bibfnamefont {L.}~\bibnamefont
  {Peng}}, \bibinfo {author} {\bibfnamefont {K.}~\bibnamefont {Karube}},
  \bibinfo {author} {\bibfnamefont {Y.}~\bibnamefont {Taguchi}}, \bibinfo
  {author} {\bibfnamefont {N.}~\bibnamefont {Nagaosa}}, \bibinfo {author}
  {\bibfnamefont {Y.}~\bibnamefont {Tokura}},\ and\ \bibinfo {author}
  {\bibfnamefont {X.}~\bibnamefont {Yu}},\ }\bibfield  {title} {\bibinfo
  {title} {Dynamic transition of current-driven single-skyrmion motion in a
  room-temperature chiral-lattice magnet},\ }\href
  {https://doi.org/10.1038/s41467-021-27073-2} {\bibfield  {journal} {\bibinfo
  {journal} {Nat. Commun.}\ }\textbf {\bibinfo {volume} {12}},\ \bibinfo
  {pages} {6797} (\bibinfo {year} {2021})}\BibitemShut {NoStop}%
\bibitem [{\citenamefont {Peng}\ \emph {et~al.}(2020)\citenamefont {Peng},
  \citenamefont {Takagi}, \citenamefont {Koshibae}, \citenamefont {Shibata},
  \citenamefont {Nakajima}, \citenamefont {Arima}, \citenamefont {Nagaosa},
  \citenamefont {Seki}, \citenamefont {Yu},\ and\ \citenamefont
  {Tokura}}]{peng2020controlled}%
  \BibitemOpen
  \bibfield  {author} {\bibinfo {author} {\bibfnamefont {L.}~\bibnamefont
  {Peng}}, \bibinfo {author} {\bibfnamefont {R.}~\bibnamefont {Takagi}},
  \bibinfo {author} {\bibfnamefont {W.}~\bibnamefont {Koshibae}}, \bibinfo
  {author} {\bibfnamefont {K.}~\bibnamefont {Shibata}}, \bibinfo {author}
  {\bibfnamefont {K.}~\bibnamefont {Nakajima}}, \bibinfo {author}
  {\bibfnamefont {T.-h.}\ \bibnamefont {Arima}}, \bibinfo {author}
  {\bibfnamefont {N.}~\bibnamefont {Nagaosa}}, \bibinfo {author} {\bibfnamefont
  {S.}~\bibnamefont {Seki}}, \bibinfo {author} {\bibfnamefont {X.}~\bibnamefont
  {Yu}},\ and\ \bibinfo {author} {\bibfnamefont {Y.}~\bibnamefont {Tokura}},\
  }\bibfield  {title} {\bibinfo {title} {Controlled transformation of skyrmions
  and antiskyrmions in a non-centrosymmetric magnet},\ }\href
  {https://doi.org/10.1038/s41565-019-0616-6} {\bibfield  {journal} {\bibinfo
  {journal} {Nat. Nanotechnol.}\ }\textbf {\bibinfo {volume} {15}},\ \bibinfo
  {pages} {181} (\bibinfo {year} {2020})}\BibitemShut {NoStop}%
\bibitem [{\citenamefont {G{\"o}bel}\ and\ \citenamefont
  {Mertig}(2021)}]{gobel2021skyrmion}%
  \BibitemOpen
  \bibfield  {author} {\bibinfo {author} {\bibfnamefont {B.}~\bibnamefont
  {G{\"o}bel}}\ and\ \bibinfo {author} {\bibfnamefont {I.}~\bibnamefont
  {Mertig}},\ }\bibfield  {title} {\bibinfo {title} {Skyrmion ratchet
  propagation: utilizing the skyrmion {H}all effect in ac racetrack storage
  devices},\ }\href@noop {} {\bibfield  {journal} {\bibinfo  {journal} {Sci.
  Rep.}\ }\textbf {\bibinfo {volume} {11}},\ \bibinfo {pages} {3020} (\bibinfo
  {year} {2021})}\BibitemShut {NoStop}%
\bibitem [{\citenamefont {Shimojima}\ \emph {et~al.}(2021)\citenamefont
  {Shimojima}, \citenamefont {Nakamura}, \citenamefont {Yu}, \citenamefont
  {Karube}, \citenamefont {Taguchi}, \citenamefont {Tokura},\ and\
  \citenamefont {Ishizaka}}]{shimojima2021nano}%
  \BibitemOpen
  \bibfield  {author} {\bibinfo {author} {\bibfnamefont {T.}~\bibnamefont
  {Shimojima}}, \bibinfo {author} {\bibfnamefont {A.}~\bibnamefont {Nakamura}},
  \bibinfo {author} {\bibfnamefont {X.}~\bibnamefont {Yu}}, \bibinfo {author}
  {\bibfnamefont {K.}~\bibnamefont {Karube}}, \bibinfo {author} {\bibfnamefont
  {Y.}~\bibnamefont {Taguchi}}, \bibinfo {author} {\bibfnamefont
  {Y.}~\bibnamefont {Tokura}},\ and\ \bibinfo {author} {\bibfnamefont
  {K.}~\bibnamefont {Ishizaka}},\ }\bibfield  {title} {\bibinfo {title}
  {Nano-to-micro spatiotemporal imaging of magnetic skyrmion’s life cycle},\
  }\href@noop {} {\bibfield  {journal} {\bibinfo  {journal} {Sci. Adv.}\
  }\textbf {\bibinfo {volume} {7}},\ \bibinfo {pages} {eabg1322} (\bibinfo
  {year} {2021})}\BibitemShut {NoStop}%
\bibitem [{\citenamefont {Bo}\ \emph {et~al.}(2024)\citenamefont {Bo},
  \citenamefont {Zhang}, \citenamefont {Mochizuki},\ and\ \citenamefont
  {Zhang}}]{bo2024suppression}%
  \BibitemOpen
  \bibfield  {author} {\bibinfo {author} {\bibfnamefont {L.}~\bibnamefont
  {Bo}}, \bibinfo {author} {\bibfnamefont {X.}~\bibnamefont {Zhang}}, \bibinfo
  {author} {\bibfnamefont {M.}~\bibnamefont {Mochizuki}},\ and\ \bibinfo
  {author} {\bibfnamefont {X.}~\bibnamefont {Zhang}},\ }\bibfield  {title}
  {\bibinfo {title} {Suppression of the skyrmion {H}all effect in synthetic
  ferrimagnets with gradient magnetization},\ }\href@noop {} {\bibfield
  {journal} {\bibinfo  {journal} {Phys. Rev. Res.}\ }\textbf {\bibinfo {volume}
  {6}},\ \bibinfo {pages} {023199} (\bibinfo {year} {2024})}\BibitemShut
  {NoStop}%
\bibitem [{\citenamefont {Jin}\ \emph {et~al.}(2024)\citenamefont {Jin},
  \citenamefont {Zeng}, \citenamefont {Cao},\ and\ \citenamefont
  {Yan}}]{Jin2024}%
  \BibitemOpen
  \bibfield  {author} {\bibinfo {author} {\bibfnamefont {Z.}~\bibnamefont
  {Jin}}, \bibinfo {author} {\bibfnamefont {Z.}~\bibnamefont {Zeng}}, \bibinfo
  {author} {\bibfnamefont {Y.}~\bibnamefont {Cao}},\ and\ \bibinfo {author}
  {\bibfnamefont {P.}~\bibnamefont {Yan}},\ }\bibfield  {title} {\bibinfo
  {title} {{Skyrmion Hall effect in altermagnets}},\ }\href
  {https://link.aps.org/doi/10.1103/PhysRevLett.133.196701} {\bibfield
  {journal} {\bibinfo  {journal} {Phys. Rev. Lett.}\ }\textbf {\bibinfo
  {volume} {133}},\ \bibinfo {pages} {196701} (\bibinfo {year}
  {2024})}\BibitemShut {NoStop}%
\bibitem [{\citenamefont {Yang}\ \emph {et~al.}(2024)\citenamefont {Yang},
  \citenamefont {Zhao}, \citenamefont {Zhang}, \citenamefont {Xing},
  \citenamefont {Du}, \citenamefont {Li}, \citenamefont {Mochizuki},
  \citenamefont {Xu}, \citenamefont {{\AA}kerman},\ and\ \citenamefont
  {Zhou}}]{Yang2024_Fundamentals_applications}%
  \BibitemOpen
  \bibfield  {author} {\bibinfo {author} {\bibfnamefont {S.}~\bibnamefont
  {Yang}}, \bibinfo {author} {\bibfnamefont {Y.}~\bibnamefont {Zhao}}, \bibinfo
  {author} {\bibfnamefont {X.}~\bibnamefont {Zhang}}, \bibinfo {author}
  {\bibfnamefont {X.}~\bibnamefont {Xing}}, \bibinfo {author} {\bibfnamefont
  {H.}~\bibnamefont {Du}}, \bibinfo {author} {\bibfnamefont {X.}~\bibnamefont
  {Li}}, \bibinfo {author} {\bibfnamefont {M.}~\bibnamefont {Mochizuki}},
  \bibinfo {author} {\bibfnamefont {X.}~\bibnamefont {Xu}}, \bibinfo {author}
  {\bibfnamefont {J.}~\bibnamefont {{\AA}kerman}},\ and\ \bibinfo {author}
  {\bibfnamefont {Y.}~\bibnamefont {Zhou}},\ }\bibfield  {title} {\bibinfo
  {title} {{Fundamentals and applications of the skyrmion Hall effect}},\
  }\href
  {https://pubs.aip.org/apr/article/11/4/041335/3325797/Fundamentals-and-applications-of-the-skyrmion-Hall}
  {\bibfield  {journal} {\bibinfo  {journal} {Applied Physics Reviews}\
  }\textbf {\bibinfo {volume} {11}} (\bibinfo {year} {2024})}\BibitemShut
  {NoStop}%
\bibitem [{\citenamefont {Litzius}\ \emph {et~al.}(2020)\citenamefont
  {Litzius}, \citenamefont {Leliaert}, \citenamefont {Bassirian}, \citenamefont
  {Rodrigues}, \citenamefont {Kromin}, \citenamefont {Lemesh}, \citenamefont
  {Zazvorka}, \citenamefont {Lee}, \citenamefont {Mulkers}, \citenamefont
  {Kerber} \emph {et~al.}}]{litzius2020role}%
  \BibitemOpen
  \bibfield  {author} {\bibinfo {author} {\bibfnamefont {K.}~\bibnamefont
  {Litzius}}, \bibinfo {author} {\bibfnamefont {J.}~\bibnamefont {Leliaert}},
  \bibinfo {author} {\bibfnamefont {P.}~\bibnamefont {Bassirian}}, \bibinfo
  {author} {\bibfnamefont {D.}~\bibnamefont {Rodrigues}}, \bibinfo {author}
  {\bibfnamefont {S.}~\bibnamefont {Kromin}}, \bibinfo {author} {\bibfnamefont
  {I.}~\bibnamefont {Lemesh}}, \bibinfo {author} {\bibfnamefont
  {J.}~\bibnamefont {Zazvorka}}, \bibinfo {author} {\bibfnamefont {K.-J.}\
  \bibnamefont {Lee}}, \bibinfo {author} {\bibfnamefont {J.}~\bibnamefont
  {Mulkers}}, \bibinfo {author} {\bibfnamefont {N.}~\bibnamefont {Kerber}},
  \emph {et~al.},\ }\bibfield  {title} {\bibinfo {title} {The role of
  temperature and drive current in skyrmion dynamics},\ }\href@noop {}
  {\bibfield  {journal} {\bibinfo  {journal} {Nat. Electron.}\ }\textbf
  {\bibinfo {volume} {3}},\ \bibinfo {pages} {30} (\bibinfo {year}
  {2020})}\BibitemShut {NoStop}%
\bibitem [{\citenamefont {Juge}\ \emph {et~al.}(2021)\citenamefont {Juge},
  \citenamefont {Bairagi}, \citenamefont {Rana}, \citenamefont {Vogel},
  \citenamefont {Sall}, \citenamefont {Mailly}, \citenamefont {Pham},
  \citenamefont {Zhang}, \citenamefont {Sisodia}, \citenamefont {Foerster}
  \emph {et~al.}}]{juge2021helium}%
  \BibitemOpen
  \bibfield  {author} {\bibinfo {author} {\bibfnamefont {R.}~\bibnamefont
  {Juge}}, \bibinfo {author} {\bibfnamefont {K.}~\bibnamefont {Bairagi}},
  \bibinfo {author} {\bibfnamefont {K.~G.}\ \bibnamefont {Rana}}, \bibinfo
  {author} {\bibfnamefont {J.}~\bibnamefont {Vogel}}, \bibinfo {author}
  {\bibfnamefont {M.}~\bibnamefont {Sall}}, \bibinfo {author} {\bibfnamefont
  {D.}~\bibnamefont {Mailly}}, \bibinfo {author} {\bibfnamefont {V.~T.}\
  \bibnamefont {Pham}}, \bibinfo {author} {\bibfnamefont {Q.}~\bibnamefont
  {Zhang}}, \bibinfo {author} {\bibfnamefont {N.}~\bibnamefont {Sisodia}},
  \bibinfo {author} {\bibfnamefont {M.}~\bibnamefont {Foerster}}, \emph
  {et~al.},\ }\bibfield  {title} {\bibinfo {title} {Helium ions put magnetic
  skyrmions on the track},\ }\href@noop {} {\bibfield  {journal} {\bibinfo
  {journal} {Nano Lett.}\ }\textbf {\bibinfo {volume} {21}},\ \bibinfo {pages}
  {2989} (\bibinfo {year} {2021})}\BibitemShut {NoStop}%
\bibitem [{\citenamefont {Ahrens}\ \emph {et~al.}(2023)\citenamefont {Ahrens},
  \citenamefont {Kiesselbach}, \citenamefont {Gnoli}, \citenamefont {Giuliano},
  \citenamefont {Mendisch}, \citenamefont {Kiechle}, \citenamefont {Riente},\
  and\ \citenamefont {Becherer}}]{ahrens2023skyrmions}%
  \BibitemOpen
  \bibfield  {author} {\bibinfo {author} {\bibfnamefont {V.}~\bibnamefont
  {Ahrens}}, \bibinfo {author} {\bibfnamefont {C.}~\bibnamefont {Kiesselbach}},
  \bibinfo {author} {\bibfnamefont {L.}~\bibnamefont {Gnoli}}, \bibinfo
  {author} {\bibfnamefont {D.}~\bibnamefont {Giuliano}}, \bibinfo {author}
  {\bibfnamefont {S.}~\bibnamefont {Mendisch}}, \bibinfo {author}
  {\bibfnamefont {M.}~\bibnamefont {Kiechle}}, \bibinfo {author} {\bibfnamefont
  {F.}~\bibnamefont {Riente}},\ and\ \bibinfo {author} {\bibfnamefont
  {M.}~\bibnamefont {Becherer}},\ }\bibfield  {title} {\bibinfo {title}
  {Skyrmions under control—{FIB} irradiation as a versatile tool for skyrmion
  circuits},\ }\href@noop {} {\bibfield  {journal} {\bibinfo  {journal} {Adv.
  Mater.}\ }\textbf {\bibinfo {volume} {35}},\ \bibinfo {pages} {2207321}
  (\bibinfo {year} {2023})}\BibitemShut {NoStop}%
\bibitem [{\citenamefont {He}\ \emph {et~al.}(2024{\natexlab{a}})\citenamefont
  {He}, \citenamefont {Yu}, \citenamefont {Zhu}, \citenamefont {Qiu},\ and\
  \citenamefont {Zhou}}]{he2024guided}%
  \BibitemOpen
  \bibfield  {author} {\bibinfo {author} {\bibfnamefont {X.}~\bibnamefont
  {He}}, \bibinfo {author} {\bibfnamefont {G.}~\bibnamefont {Yu}}, \bibinfo
  {author} {\bibfnamefont {M.}~\bibnamefont {Zhu}}, \bibinfo {author}
  {\bibfnamefont {Y.}~\bibnamefont {Qiu}},\ and\ \bibinfo {author}
  {\bibfnamefont {H.}~\bibnamefont {Zhou}},\ }\bibfield  {title} {\bibinfo
  {title} {Guided motion of a magnetic skyrmion by a voltage-controlled
  strained channel, and logic applications},\ }\href@noop {} {\bibfield
  {journal} {\bibinfo  {journal} {Phys. Rev. Appl.}\ }\textbf {\bibinfo
  {volume} {22}},\ \bibinfo {pages} {054047} (\bibinfo {year}
  {2024}{\natexlab{a}})}\BibitemShut {NoStop}%
\bibitem [{\citenamefont {Ma}\ \emph {et~al.}(2025{\natexlab{b}})\citenamefont
  {Ma}, \citenamefont {Yang}, \citenamefont {Wang}, \citenamefont {Tian},
  \citenamefont {Zhang}, \citenamefont {Luo},\ and\ \citenamefont
  {Piao}}]{ma2025deterministic}%
  \BibitemOpen
  \bibfield  {author} {\bibinfo {author} {\bibfnamefont {X.-P.}\ \bibnamefont
  {Ma}}, \bibinfo {author} {\bibfnamefont {X.-X.}\ \bibnamefont {Yang}},
  \bibinfo {author} {\bibfnamefont {Q.-S.}\ \bibnamefont {Wang}}, \bibinfo
  {author} {\bibfnamefont {K.}~\bibnamefont {Tian}}, \bibinfo {author}
  {\bibfnamefont {H.}~\bibnamefont {Zhang}}, \bibinfo {author} {\bibfnamefont
  {Z.}~\bibnamefont {Luo}},\ and\ \bibinfo {author} {\bibfnamefont {H.-G.}\
  \bibnamefont {Piao}},\ }\bibfield  {title} {\bibinfo {title} {Deterministic
  control of magnetic skyrmion motions via engineered potential wells},\
  }\href@noop {} {\bibfield  {journal} {\bibinfo  {journal} {Appl. Phys.
  Lett.}\ }\textbf {\bibinfo {volume} {126}} (\bibinfo {year}
  {2025}{\natexlab{b}})}\BibitemShut {NoStop}%
\bibitem [{\citenamefont {Chang}\ \emph {et~al.}(2025)\citenamefont {Chang},
  \citenamefont {Hao}, \citenamefont {Wu}, \citenamefont {Zuo}, \citenamefont
  {Wu}, \citenamefont {Hou}, \citenamefont {Yu}, \citenamefont {Liu},
  \citenamefont {Xi}, \citenamefont {Zhang} \emph
  {et~al.}}]{chang2025suppressed}%
  \BibitemOpen
  \bibfield  {author} {\bibinfo {author} {\bibfnamefont {Y.}~\bibnamefont
  {Chang}}, \bibinfo {author} {\bibfnamefont {H.}~\bibnamefont {Hao}}, \bibinfo
  {author} {\bibfnamefont {H.}~\bibnamefont {Wu}}, \bibinfo {author}
  {\bibfnamefont {Y.}~\bibnamefont {Zuo}}, \bibinfo {author} {\bibfnamefont
  {H.}~\bibnamefont {Wu}}, \bibinfo {author} {\bibfnamefont {Z.}~\bibnamefont
  {Hou}}, \bibinfo {author} {\bibfnamefont {G.}~\bibnamefont {Yu}}, \bibinfo
  {author} {\bibfnamefont {X.}~\bibnamefont {Liu}}, \bibinfo {author}
  {\bibfnamefont {L.}~\bibnamefont {Xi}}, \bibinfo {author} {\bibfnamefont
  {S.}~\bibnamefont {Zhang}}, \emph {et~al.},\ }\bibfield  {title} {\bibinfo
  {title} {Suppressed skyrmion hall effect of hybrid magnetic skyrmions by
  helicity engineering},\ }\href@noop {} {\bibfield  {journal} {\bibinfo
  {journal} {Adv. Funct. Mater.}\ ,\ \bibinfo {pages} {2421771}} (\bibinfo
  {year} {2025})}\BibitemShut {NoStop}%
\bibitem [{\citenamefont {Yu}\ \emph {et~al.}(2020)\citenamefont {Yu},
  \citenamefont {Morikawa}, \citenamefont {Nakajima}, \citenamefont {Shibata},
  \citenamefont {Kanazawa}, \citenamefont {Arima}, \citenamefont {Nagaosa},\
  and\ \citenamefont {Tokura}}]{yu2020motion}%
  \BibitemOpen
  \bibfield  {author} {\bibinfo {author} {\bibfnamefont {X.}~\bibnamefont
  {Yu}}, \bibinfo {author} {\bibfnamefont {D.}~\bibnamefont {Morikawa}},
  \bibinfo {author} {\bibfnamefont {K.}~\bibnamefont {Nakajima}}, \bibinfo
  {author} {\bibfnamefont {K.}~\bibnamefont {Shibata}}, \bibinfo {author}
  {\bibfnamefont {N.}~\bibnamefont {Kanazawa}}, \bibinfo {author}
  {\bibfnamefont {T.-h.}\ \bibnamefont {Arima}}, \bibinfo {author}
  {\bibfnamefont {N.}~\bibnamefont {Nagaosa}},\ and\ \bibinfo {author}
  {\bibfnamefont {Y.}~\bibnamefont {Tokura}},\ }\bibfield  {title} {\bibinfo
  {title} {Motion tracking of 80-nm-size skyrmions upon directional current
  injections},\ }\href@noop {} {\bibfield  {journal} {\bibinfo  {journal} {Sci.
  Adv.}\ }\textbf {\bibinfo {volume} {6}},\ \bibinfo {pages} {eaaz9744}
  (\bibinfo {year} {2020})}\BibitemShut {NoStop}%
\bibitem [{\citenamefont {Song}\ \emph {et~al.}(2024)\citenamefont {Song},
  \citenamefont {Wang}, \citenamefont {Zhang}, \citenamefont {Liu},
  \citenamefont {Wang}, \citenamefont {Zheng}, \citenamefont {Tian},
  \citenamefont {{Dunin-Borkowski}}, \citenamefont {Zang},\ and\ \citenamefont
  {Du}}]{song2024steady}%
  \BibitemOpen
  \bibfield  {author} {\bibinfo {author} {\bibfnamefont {D.}~\bibnamefont
  {Song}}, \bibinfo {author} {\bibfnamefont {W.}~\bibnamefont {Wang}}, \bibinfo
  {author} {\bibfnamefont {S.}~\bibnamefont {Zhang}}, \bibinfo {author}
  {\bibfnamefont {Y.}~\bibnamefont {Liu}}, \bibinfo {author} {\bibfnamefont
  {N.}~\bibnamefont {Wang}}, \bibinfo {author} {\bibfnamefont {F.}~\bibnamefont
  {Zheng}}, \bibinfo {author} {\bibfnamefont {M.}~\bibnamefont {Tian}},
  \bibinfo {author} {\bibfnamefont {R.~E.}\ \bibnamefont {{Dunin-Borkowski}}},
  \bibinfo {author} {\bibfnamefont {J.}~\bibnamefont {Zang}},\ and\ \bibinfo
  {author} {\bibfnamefont {H.}~\bibnamefont {Du}},\ }\bibfield  {title}
  {\bibinfo {title} {Steady motion of 80-nm-size skyrmions in a 100-nm-wide
  track},\ }\href {https://doi.org/10.1038/s41467-024-49976-6} {\bibfield
  {journal} {\bibinfo  {journal} {Nat. Commun.}\ }\textbf {\bibinfo {volume}
  {15}},\ \bibinfo {pages} {5614} (\bibinfo {year} {2024})}\BibitemShut
  {NoStop}%
\bibitem [{\citenamefont {Mallick}\ \emph {et~al.}(2024)\citenamefont
  {Mallick}, \citenamefont {Sassi}, \citenamefont {Prestes}, \citenamefont
  {Krishnia}, \citenamefont {Gallego}, \citenamefont {M.~Vicente~Arche},
  \citenamefont {Denneulin}, \citenamefont {Collin}, \citenamefont
  {Bouzehouane}, \citenamefont {Thiaville}, \citenamefont {{Dunin-Borkowski}},
  \citenamefont {Jeudy}, \citenamefont {Fert}, \citenamefont {Reyren},\ and\
  \citenamefont {Cros}}]{mallick2024driving}%
  \BibitemOpen
  \bibfield  {author} {\bibinfo {author} {\bibfnamefont {S.}~\bibnamefont
  {Mallick}}, \bibinfo {author} {\bibfnamefont {Y.}~\bibnamefont {Sassi}},
  \bibinfo {author} {\bibfnamefont {N.~F.}\ \bibnamefont {Prestes}}, \bibinfo
  {author} {\bibfnamefont {S.}~\bibnamefont {Krishnia}}, \bibinfo {author}
  {\bibfnamefont {F.}~\bibnamefont {Gallego}}, \bibinfo {author} {\bibfnamefont
  {L.}~\bibnamefont {M.~Vicente~Arche}}, \bibinfo {author} {\bibfnamefont
  {T.}~\bibnamefont {Denneulin}}, \bibinfo {author} {\bibfnamefont
  {S.}~\bibnamefont {Collin}}, \bibinfo {author} {\bibfnamefont
  {K.}~\bibnamefont {Bouzehouane}}, \bibinfo {author} {\bibfnamefont
  {A.}~\bibnamefont {Thiaville}}, \bibinfo {author} {\bibfnamefont {R.~E.}\
  \bibnamefont {{Dunin-Borkowski}}}, \bibinfo {author} {\bibfnamefont
  {V.}~\bibnamefont {Jeudy}}, \bibinfo {author} {\bibfnamefont
  {A.}~\bibnamefont {Fert}}, \bibinfo {author} {\bibfnamefont {N.}~\bibnamefont
  {Reyren}},\ and\ \bibinfo {author} {\bibfnamefont {V.}~\bibnamefont {Cros}},\
  }\bibfield  {title} {\bibinfo {title} {Driving skyrmions in flow regime in
  synthetic ferrimagnets},\ }\href {https://doi.org/10.1038/s41467-024-52210-y}
  {\bibfield  {journal} {\bibinfo  {journal} {Nat. Commun.}\ }\textbf {\bibinfo
  {volume} {15}},\ \bibinfo {pages} {8472} (\bibinfo {year}
  {2024})}\BibitemShut {NoStop}%
\bibitem [{\citenamefont {Song}\ \emph {et~al.}(2022)\citenamefont {Song},
  \citenamefont {Wang}, \citenamefont {Yu}, \citenamefont {Zhang},
  \citenamefont {Pershoguba}, \citenamefont {Yin}, \citenamefont {Wei},
  \citenamefont {Jiang}, \citenamefont {Ge}, \citenamefont {Fan} \emph
  {et~al.}}]{song2022experimental}%
  \BibitemOpen
  \bibfield  {author} {\bibinfo {author} {\bibfnamefont {D.}~\bibnamefont
  {Song}}, \bibinfo {author} {\bibfnamefont {W.}~\bibnamefont {Wang}}, \bibinfo
  {author} {\bibfnamefont {J.-X.}\ \bibnamefont {Yu}}, \bibinfo {author}
  {\bibfnamefont {P.}~\bibnamefont {Zhang}}, \bibinfo {author} {\bibfnamefont
  {S.~S.}\ \bibnamefont {Pershoguba}}, \bibinfo {author} {\bibfnamefont
  {G.}~\bibnamefont {Yin}}, \bibinfo {author} {\bibfnamefont {W.}~\bibnamefont
  {Wei}}, \bibinfo {author} {\bibfnamefont {J.}~\bibnamefont {Jiang}}, \bibinfo
  {author} {\bibfnamefont {B.}~\bibnamefont {Ge}}, \bibinfo {author}
  {\bibfnamefont {X.}~\bibnamefont {Fan}}, \emph {et~al.},\ }\bibfield  {title}
  {\bibinfo {title} {Experimental observation of one-dimensional motion of
  interstitial skyrmion in fege},\ }\href@noop {} {\bibfield  {journal}
  {\bibinfo  {journal} {arXiv preprint arXiv:2212.08991}\ } (\bibinfo {year}
  {2022})}\BibitemShut {NoStop}%
\bibitem [{\citenamefont {He}\ \emph {et~al.}(2024{\natexlab{b}})\citenamefont
  {He}, \citenamefont {Li}, \citenamefont {Chen}, \citenamefont {Wang},
  \citenamefont {Shen}, \citenamefont {Wang}, \citenamefont {Song},
  \citenamefont {Zhao}, \citenamefont {Cai}, \citenamefont {Lin}, \citenamefont
  {Zhang},\ and\ \citenamefont {Shen}}]{he2024experimental}%
  \BibitemOpen
  \bibfield  {author} {\bibinfo {author} {\bibfnamefont {Z.}~\bibnamefont
  {He}}, \bibinfo {author} {\bibfnamefont {Z.}~\bibnamefont {Li}}, \bibinfo
  {author} {\bibfnamefont {Z.}~\bibnamefont {Chen}}, \bibinfo {author}
  {\bibfnamefont {Z.}~\bibnamefont {Wang}}, \bibinfo {author} {\bibfnamefont
  {J.}~\bibnamefont {Shen}}, \bibinfo {author} {\bibfnamefont {S.}~\bibnamefont
  {Wang}}, \bibinfo {author} {\bibfnamefont {C.}~\bibnamefont {Song}}, \bibinfo
  {author} {\bibfnamefont {T.}~\bibnamefont {Zhao}}, \bibinfo {author}
  {\bibfnamefont {J.}~\bibnamefont {Cai}}, \bibinfo {author} {\bibfnamefont
  {S.-Z.}\ \bibnamefont {Lin}}, \bibinfo {author} {\bibfnamefont
  {Y.}~\bibnamefont {Zhang}},\ and\ \bibinfo {author} {\bibfnamefont
  {B.}~\bibnamefont {Shen}},\ }\bibfield  {title} {\bibinfo {title}
  {Experimental observation of current-driven antiskyrmion sliding in stripe
  domains},\ }\href {https://doi.org/10.1038/s41563-024-01870-8} {\bibfield
  {journal} {\bibinfo  {journal} {Nat. Mater.}\ }\textbf {\bibinfo {volume}
  {23}},\ \bibinfo {pages} {1048} (\bibinfo {year}
  {2024}{\natexlab{b}})}\BibitemShut {NoStop}%
\bibitem [{\citenamefont {Wang}\ \emph {et~al.}(2022)\citenamefont {Wang},
  \citenamefont {Song}, \citenamefont {Wei}, \citenamefont {Nan}, \citenamefont
  {Zhang}, \citenamefont {Ge}, \citenamefont {Tian}, \citenamefont {Zang},\
  and\ \citenamefont {Du}}]{wang2022electrical}%
  \BibitemOpen
  \bibfield  {author} {\bibinfo {author} {\bibfnamefont {W.}~\bibnamefont
  {Wang}}, \bibinfo {author} {\bibfnamefont {D.}~\bibnamefont {Song}}, \bibinfo
  {author} {\bibfnamefont {W.}~\bibnamefont {Wei}}, \bibinfo {author}
  {\bibfnamefont {P.}~\bibnamefont {Nan}}, \bibinfo {author} {\bibfnamefont
  {S.}~\bibnamefont {Zhang}}, \bibinfo {author} {\bibfnamefont
  {B.}~\bibnamefont {Ge}}, \bibinfo {author} {\bibfnamefont {M.}~\bibnamefont
  {Tian}}, \bibinfo {author} {\bibfnamefont {J.}~\bibnamefont {Zang}},\ and\
  \bibinfo {author} {\bibfnamefont {H.}~\bibnamefont {Du}},\ }\bibfield
  {title} {\bibinfo {title} {Electrical manipulation of skyrmions in a chiral
  magnet},\ }\href {https://doi.org/10.1038/s41467-022-29217-4} {\bibfield
  {journal} {\bibinfo  {journal} {Nat. Commun.}\ }\textbf {\bibinfo {volume}
  {13}},\ \bibinfo {pages} {1593} (\bibinfo {year} {2022})}\BibitemShut
  {NoStop}%
\bibitem [{\citenamefont {Lonsky}\ and\ \citenamefont
  {Hoffmann}(2020)}]{lonsky2020dynamic}%
  \BibitemOpen
  \bibfield  {author} {\bibinfo {author} {\bibfnamefont {M.}~\bibnamefont
  {Lonsky}}\ and\ \bibinfo {author} {\bibfnamefont {A.}~\bibnamefont
  {Hoffmann}},\ }\bibfield  {title} {\bibinfo {title} {Dynamic excitations of
  chiral magnetic textures},\ }\href {https://doi.org/10.1063/5.0027042}
  {\bibfield  {journal} {\bibinfo  {journal} {APL Mater.}\ }\textbf {\bibinfo
  {volume} {8}},\ \bibinfo {pages} {100903} (\bibinfo {year}
  {2020})}\BibitemShut {NoStop}%
\bibitem [{\citenamefont {Martin}\ and\ \citenamefont
  {Axel}(2020)}]{Lonsky2020coupled}%
  \BibitemOpen
  \bibfield  {author} {\bibinfo {author} {\bibfnamefont {L.}~\bibnamefont
  {Martin}}\ and\ \bibinfo {author} {\bibfnamefont {H.}~\bibnamefont {Axel}},\
  }\bibfield  {title} {\bibinfo {title} {Coupled skyrmion breathing modes in
  synthetic ferri- and antiferromagnets},\ }\href
  {https://doi.org/10.1103/physrevb.102.104403} {\bibfield  {journal} {\bibinfo
   {journal} {Phys. Rev. B}\ }\textbf {\bibinfo {volume} {102}},\ \bibinfo
  {pages} {104403} (\bibinfo {year} {2020})}\BibitemShut {NoStop}%
\bibitem [{\citenamefont {Yu}\ \emph {et~al.}(2021)\citenamefont {Yu},
  \citenamefont {Xiao},\ and\ \citenamefont {Schultheiss}}]{Yu_2021}%
  \BibitemOpen
  \bibfield  {author} {\bibinfo {author} {\bibfnamefont {H.}~\bibnamefont
  {Yu}}, \bibinfo {author} {\bibfnamefont {J.}~\bibnamefont {Xiao}},\ and\
  \bibinfo {author} {\bibfnamefont {H.}~\bibnamefont {Schultheiss}},\
  }\bibfield  {title} {\bibinfo {title} {Magnetic texture based magnonics},\
  }\href {https://doi.org/https://doi.org/10.1016/j.physrep.2020.12.004}
  {\bibfield  {journal} {\bibinfo  {journal} {Phys. Rep.}\ }\textbf {\bibinfo
  {volume} {905}},\ \bibinfo {pages} {1} (\bibinfo {year} {2021})}\BibitemShut
  {NoStop}%
\bibitem [{\citenamefont {Pan}\ \emph {et~al.}(2024)\citenamefont {Pan},
  \citenamefont {Li}, \citenamefont {Hei}, \citenamefont {Zhang}, \citenamefont
  {Mochizuki}, \citenamefont {Li},\ and\ \citenamefont {Nori}}]{Pan_2024}%
  \BibitemOpen
  \bibfield  {author} {\bibinfo {author} {\bibfnamefont {X.-F.}\ \bibnamefont
  {Pan}}, \bibinfo {author} {\bibfnamefont {P.-B.}\ \bibnamefont {Li}},
  \bibinfo {author} {\bibfnamefont {X.-L.}\ \bibnamefont {Hei}}, \bibinfo
  {author} {\bibfnamefont {X.}~\bibnamefont {Zhang}}, \bibinfo {author}
  {\bibfnamefont {M.}~\bibnamefont {Mochizuki}}, \bibinfo {author}
  {\bibfnamefont {F.-L.}\ \bibnamefont {Li}},\ and\ \bibinfo {author}
  {\bibfnamefont {F.}~\bibnamefont {Nori}},\ }\bibfield  {title} {\bibinfo
  {title} {Magnon-skyrmion hybrid quantum systems: Tailoring interactions via
  magnons},\ }\href {https://doi.org/10.1103/PhysRevLett.132.193601} {\bibfield
   {journal} {\bibinfo  {journal} {Phys. Rev. Lett.}\ }\textbf {\bibinfo
  {volume} {132}},\ \bibinfo {pages} {193601} (\bibinfo {year}
  {2024})}\BibitemShut {NoStop}%
\bibitem [{\citenamefont {Seki}\ \emph {et~al.}(2020)\citenamefont {Seki},
  \citenamefont {Garst}, \citenamefont {Waizner}, \citenamefont {Takagi},
  \citenamefont {Khanh}, \citenamefont {Okamura}, \citenamefont {Kondou},
  \citenamefont {Kagawa}, \citenamefont {Otani},\ and\ \citenamefont
  {Tokura}}]{Seki_2020}%
  \BibitemOpen
  \bibfield  {author} {\bibinfo {author} {\bibfnamefont {S.}~\bibnamefont
  {Seki}}, \bibinfo {author} {\bibfnamefont {M.}~\bibnamefont {Garst}},
  \bibinfo {author} {\bibfnamefont {J.}~\bibnamefont {Waizner}}, \bibinfo
  {author} {\bibfnamefont {R.}~\bibnamefont {Takagi}}, \bibinfo {author}
  {\bibfnamefont {N.~D.}\ \bibnamefont {Khanh}}, \bibinfo {author}
  {\bibfnamefont {Y.}~\bibnamefont {Okamura}}, \bibinfo {author} {\bibfnamefont
  {K.}~\bibnamefont {Kondou}}, \bibinfo {author} {\bibfnamefont
  {F.}~\bibnamefont {Kagawa}}, \bibinfo {author} {\bibfnamefont
  {Y.}~\bibnamefont {Otani}},\ and\ \bibinfo {author} {\bibfnamefont
  {Y.}~\bibnamefont {Tokura}},\ }\bibfield  {title} {\bibinfo {title}
  {Propagation dynamics of spin excitations along skyrmion strings},\ }\href
  {https://doi.org/10.1038/s41467-019-14095-0} {\bibfield  {journal} {\bibinfo
  {journal} {Nat. Commun.}\ }\textbf {\bibinfo {volume} {11}},\ \bibinfo
  {pages} {256} (\bibinfo {year} {2020})}\BibitemShut {NoStop}%
\bibitem [{\citenamefont {Satywali}\ \emph {et~al.}(2021)\citenamefont
  {Satywali}, \citenamefont {Kravchuk}, \citenamefont {Pan}, \citenamefont
  {Raju}, \citenamefont {He}, \citenamefont {Ma}, \citenamefont {Petrovi{\'c}},
  \citenamefont {Garst},\ and\ \citenamefont
  {Panagopoulos}}]{satywali2021microwave}%
  \BibitemOpen
  \bibfield  {author} {\bibinfo {author} {\bibfnamefont {B.}~\bibnamefont
  {Satywali}}, \bibinfo {author} {\bibfnamefont {V.~P.}\ \bibnamefont
  {Kravchuk}}, \bibinfo {author} {\bibfnamefont {L.}~\bibnamefont {Pan}},
  \bibinfo {author} {\bibfnamefont {M.}~\bibnamefont {Raju}}, \bibinfo {author}
  {\bibfnamefont {S.}~\bibnamefont {He}}, \bibinfo {author} {\bibfnamefont
  {F.}~\bibnamefont {Ma}}, \bibinfo {author} {\bibfnamefont {A.~P.}\
  \bibnamefont {Petrovi{\'c}}}, \bibinfo {author} {\bibfnamefont
  {M.}~\bibnamefont {Garst}},\ and\ \bibinfo {author} {\bibfnamefont
  {C.}~\bibnamefont {Panagopoulos}},\ }\bibfield  {title} {\bibinfo {title}
  {Microwave resonances of magnetic skyrmions in thin film multilayers},\
  }\href {https://doi.org/10.1038/s41467-021-22220-1} {\bibfield  {journal}
  {\bibinfo  {journal} {Nat. Commun.}\ }\textbf {\bibinfo {volume} {12}},\
  \bibinfo {pages} {1909} (\bibinfo {year} {2021})}\BibitemShut {NoStop}%
\bibitem [{\citenamefont {Srivastava}\ \emph {et~al.}(2023)\citenamefont
  {Srivastava}, \citenamefont {Sassi}, \citenamefont {Ajejas}, \citenamefont
  {Vecchiola}, \citenamefont {Ngouagnia~Yemeli}, \citenamefont {Hurdequint},
  \citenamefont {Bouzehouane}, \citenamefont {Reyren}, \citenamefont {Cros},
  \citenamefont {Devolder}, \citenamefont {Kim},\ and\ \citenamefont
  {de~Loubens}}]{Srivastava_2023}%
  \BibitemOpen
  \bibfield  {author} {\bibinfo {author} {\bibfnamefont {T.}~\bibnamefont
  {Srivastava}}, \bibinfo {author} {\bibfnamefont {Y.}~\bibnamefont {Sassi}},
  \bibinfo {author} {\bibfnamefont {F.}~\bibnamefont {Ajejas}}, \bibinfo
  {author} {\bibfnamefont {A.}~\bibnamefont {Vecchiola}}, \bibinfo {author}
  {\bibfnamefont {I.}~\bibnamefont {Ngouagnia~Yemeli}}, \bibinfo {author}
  {\bibfnamefont {H.}~\bibnamefont {Hurdequint}}, \bibinfo {author}
  {\bibfnamefont {K.}~\bibnamefont {Bouzehouane}}, \bibinfo {author}
  {\bibfnamefont {N.}~\bibnamefont {Reyren}}, \bibinfo {author} {\bibfnamefont
  {V.}~\bibnamefont {Cros}}, \bibinfo {author} {\bibfnamefont {T.}~\bibnamefont
  {Devolder}}, \bibinfo {author} {\bibfnamefont {J.-V.}\ \bibnamefont {Kim}},\
  and\ \bibinfo {author} {\bibfnamefont {G.}~\bibnamefont {de~Loubens}},\
  }\bibfield  {title} {\bibinfo {title} {Resonant dynamics of three-dimensional
  skyrmionic textures in thin film multilayers},\ }\href
  {https://doi.org/10.1063/5.0150265} {\bibfield  {journal} {\bibinfo
  {journal} {APL Mater.}\ }\textbf {\bibinfo {volume} {11}},\ \bibinfo {pages}
  {061110} (\bibinfo {year} {2023})}\BibitemShut {NoStop}%
\bibitem [{\citenamefont {Zhang}\ \emph {et~al.}(2017)\citenamefont {Zhang},
  \citenamefont {Müller}, \citenamefont {Xia}, \citenamefont {Garst},
  \citenamefont {Liu},\ and\ \citenamefont {Zhou}}]{Zhang_2017}%
  \BibitemOpen
  \bibfield  {author} {\bibinfo {author} {\bibfnamefont {X.}~\bibnamefont
  {Zhang}}, \bibinfo {author} {\bibfnamefont {J.}~\bibnamefont {Müller}},
  \bibinfo {author} {\bibfnamefont {J.}~\bibnamefont {Xia}}, \bibinfo {author}
  {\bibfnamefont {M.}~\bibnamefont {Garst}}, \bibinfo {author} {\bibfnamefont
  {X.}~\bibnamefont {Liu}},\ and\ \bibinfo {author} {\bibfnamefont
  {Y.}~\bibnamefont {Zhou}},\ }\bibfield  {title} {\bibinfo {title} {Motion of
  skyrmions in nanowires driven by magnonic momentum-transfer forces},\ }\href
  {https://doi.org/10.1088/1367-2630/aa6b70} {\bibfield  {journal} {\bibinfo
  {journal} {New J. Phys.}\ }\textbf {\bibinfo {volume} {19}},\ \bibinfo
  {pages} {065001} (\bibinfo {year} {2017})}\BibitemShut {NoStop}%
\bibitem [{\citenamefont {Chen}\ and\ \citenamefont {Ma}(2021)}]{Chen_2021}%
  \BibitemOpen
  \bibfield  {author} {\bibinfo {author} {\bibfnamefont {Z.}~\bibnamefont
  {Chen}}\ and\ \bibinfo {author} {\bibfnamefont {F.}~\bibnamefont {Ma}},\
  }\bibfield  {title} {\bibinfo {title} {Skyrmion based magnonic crystals},\
  }\href {https://doi.org/10.1063/5.0061832} {\bibfield  {journal} {\bibinfo
  {journal} {J. Appl. Phys.}\ }\textbf {\bibinfo {volume} {130}},\ \bibinfo
  {pages} {090901} (\bibinfo {year} {2021})}\BibitemShut {NoStop}%
\bibitem [{\citenamefont {D{\'{\i}}az}\ \emph {et~al.}(2020)\citenamefont
  {D{\'{\i}}az}, \citenamefont {Hirosawa}, \citenamefont {Klinovaja},\ and\
  \citenamefont {Loss}}]{Diaz_2020}%
  \BibitemOpen
  \bibfield  {author} {\bibinfo {author} {\bibfnamefont {S.~A.}\ \bibnamefont
  {D{\'{\i}}az}}, \bibinfo {author} {\bibfnamefont {T.}~\bibnamefont
  {Hirosawa}}, \bibinfo {author} {\bibfnamefont {J.}~\bibnamefont
  {Klinovaja}},\ and\ \bibinfo {author} {\bibfnamefont {D.}~\bibnamefont
  {Loss}},\ }\bibfield  {title} {\bibinfo {title} {Chiral magnonic edge states
  in ferromagnetic skyrmion crystals controlled by magnetic fields},\
  }\bibfield  {journal} {\bibinfo  {journal} {Phys. Rev. Res.}\ }\textbf
  {\bibinfo {volume} {2}},\ \href
  {https://doi.org/10.1103/PhysRevResearch.2.013231}
  {10.1103/PhysRevResearch.2.013231} (\bibinfo {year} {2020})\BibitemShut
  {NoStop}%
\bibitem [{\citenamefont {Sebastian}\ \emph {et~al.}(2015)\citenamefont
  {Sebastian}, \citenamefont {Schultheiss}, \citenamefont {Obry}, \citenamefont
  {Hillebrands},\ and\ \citenamefont {Schultheiss}}]{Sebastian_2015}%
  \BibitemOpen
  \bibfield  {author} {\bibinfo {author} {\bibfnamefont {T.}~\bibnamefont
  {Sebastian}}, \bibinfo {author} {\bibfnamefont {K.}~\bibnamefont
  {Schultheiss}}, \bibinfo {author} {\bibfnamefont {B.}~\bibnamefont {Obry}},
  \bibinfo {author} {\bibfnamefont {B.}~\bibnamefont {Hillebrands}},\ and\
  \bibinfo {author} {\bibfnamefont {H.}~\bibnamefont {Schultheiss}},\
  }\bibfield  {title} {\bibinfo {title} {Micro-focused brillouin light
  scattering: imaging spin waves at the nanoscale},\ }\href
  {https://www.frontiersin.org/journals/physics/articles/10.3389/fphy.2015.00035}
  {\bibfield  {journal} {\bibinfo  {journal} {Front. Phys.}\ }\textbf {\bibinfo
  {volume} {3}},\ \bibinfo {pages} {35} (\bibinfo {year} {2015})}\BibitemShut
  {NoStop}%
\bibitem [{\citenamefont {Casola}\ \emph {et~al.}(2018)\citenamefont {Casola},
  \citenamefont {Van Der~Sar},\ and\ \citenamefont {Yacoby}}]{Casola_2018}%
  \BibitemOpen
  \bibfield  {author} {\bibinfo {author} {\bibfnamefont {F.}~\bibnamefont
  {Casola}}, \bibinfo {author} {\bibfnamefont {T.}~\bibnamefont {Van
  Der~Sar}},\ and\ \bibinfo {author} {\bibfnamefont {A.}~\bibnamefont
  {Yacoby}},\ }\bibfield  {title} {\bibinfo {title} {Probing condensed matter
  physics with magnetometry based on nitrogen-vacancy centres in diamond},\
  }\href {https://doi.org/10.1038/natrevmats.2017.88} {\bibfield  {journal}
  {\bibinfo  {journal} {Nat. Rev. Mater.}\ }\textbf {\bibinfo {volume} {3}},\
  \bibinfo {pages} {1} (\bibinfo {year} {2018})}\BibitemShut {NoStop}%
\bibitem [{\citenamefont {Szulc}\ \emph {et~al.}(2025)\citenamefont {Szulc},
  \citenamefont {Zelent},\ and\ \citenamefont
  {Krawczyk}}]{szulc2025multifunctional}%
  \BibitemOpen
  \bibfield  {author} {\bibinfo {author} {\bibfnamefont {K.}~\bibnamefont
  {Szulc}}, \bibinfo {author} {\bibfnamefont {M.}~\bibnamefont {Zelent}},\ and\
  \bibinfo {author} {\bibfnamefont {M.}~\bibnamefont {Krawczyk}},\ }\bibfield
  {title} {\bibinfo {title} {Multifunctional magnonic platform based on the
  interplay between spin-wave waveguide and nanodots with pma and dmi},\
  }\href@noop {} {\bibfield  {journal} {\bibinfo  {journal} {APL Mater.}\
  }\textbf {\bibinfo {volume} {13}} (\bibinfo {year} {2025})}\BibitemShut
  {NoStop}%
\bibitem [{\citenamefont {Wang}\ \emph
  {et~al.}(2021{\natexlab{b}})\citenamefont {Wang}, \citenamefont {Yuan},
  \citenamefont {Cao}, \citenamefont {Li}, \citenamefont {Duine},\ and\
  \citenamefont {Yan}}]{Wang_2021}%
  \BibitemOpen
  \bibfield  {author} {\bibinfo {author} {\bibfnamefont {Z.}~\bibnamefont
  {Wang}}, \bibinfo {author} {\bibfnamefont {H.~Y.}\ \bibnamefont {Yuan}},
  \bibinfo {author} {\bibfnamefont {Y.}~\bibnamefont {Cao}}, \bibinfo {author}
  {\bibfnamefont {Z.-X.}\ \bibnamefont {Li}}, \bibinfo {author} {\bibfnamefont
  {R.~A.}\ \bibnamefont {Duine}},\ and\ \bibinfo {author} {\bibfnamefont
  {P.}~\bibnamefont {Yan}},\ }\bibfield  {title} {\bibinfo {title} {Magnonic
  frequency comb through nonlinear magnon-skyrmion scattering},\ }\href
  {https://doi.org/10.1103/PhysRevLett.127.037202} {\bibfield  {journal}
  {\bibinfo  {journal} {Phys. Rev. Lett.}\ }\textbf {\bibinfo {volume} {127}},\
  \bibinfo {pages} {037202} (\bibinfo {year} {2021}{\natexlab{b}})}\BibitemShut
  {NoStop}%
\bibitem [{\citenamefont {Lee}\ and\ \citenamefont
  {Mochizuki}(2022)}]{Lee_2022}%
  \BibitemOpen
  \bibfield  {author} {\bibinfo {author} {\bibfnamefont {M.-K.}\ \bibnamefont
  {Lee}}\ and\ \bibinfo {author} {\bibfnamefont {M.}~\bibnamefont
  {Mochizuki}},\ }\bibfield  {title} {\bibinfo {title} {Reservoir computing
  with spin waves in a skyrmion crystal},\ }\href
  {https://doi.org/10.1103/PhysRevApplied.18.014074} {\bibfield  {journal}
  {\bibinfo  {journal} {Phys. Rev. Appl.}\ }\textbf {\bibinfo {volume} {18}},\
  \bibinfo {pages} {014074} (\bibinfo {year} {2022})}\BibitemShut {NoStop}%
\bibitem [{\citenamefont {Li}\ \emph {et~al.}(2022{\natexlab{a}})\citenamefont
  {Li}, \citenamefont {Ma}, \citenamefont {Chen}, \citenamefont {Xie},\ and\
  \citenamefont {Ma}}]{Li_2022}%
  \BibitemOpen
  \bibfield  {author} {\bibinfo {author} {\bibfnamefont {Z.}~\bibnamefont
  {Li}}, \bibinfo {author} {\bibfnamefont {M.}~\bibnamefont {Ma}}, \bibinfo
  {author} {\bibfnamefont {Z.}~\bibnamefont {Chen}}, \bibinfo {author}
  {\bibfnamefont {K.}~\bibnamefont {Xie}},\ and\ \bibinfo {author}
  {\bibfnamefont {F.}~\bibnamefont {Ma}},\ }\bibfield  {title} {\bibinfo
  {title} {Interaction between magnon and skyrmion: Toward quantum magnonics},\
  }\href {https://doi.org/10.1063/5.0121314} {\bibfield  {journal} {\bibinfo
  {journal} {J. Appl. Phys.}\ }\textbf {\bibinfo {volume} {132}},\ \bibinfo
  {pages} {210702} (\bibinfo {year} {2022}{\natexlab{a}})}\BibitemShut
  {NoStop}%
\bibitem [{\citenamefont {Je}\ \emph {et~al.}(2018)\citenamefont {Je},
  \citenamefont {Vallobra}, \citenamefont {Srivastava}, \citenamefont
  {{Rojas-S{\'a}nchez}}, \citenamefont {Pham}, \citenamefont {Hehn},
  \citenamefont {Malinowski}, \citenamefont {Baraduc}, \citenamefont {Auffret},
  \citenamefont {Gaudin}, \citenamefont {Mangin}, \citenamefont {B{\'e}a},\
  and\ \citenamefont {Boulle}}]{je2018creation}%
  \BibitemOpen
  \bibfield  {author} {\bibinfo {author} {\bibfnamefont {S.-G.}\ \bibnamefont
  {Je}}, \bibinfo {author} {\bibfnamefont {P.}~\bibnamefont {Vallobra}},
  \bibinfo {author} {\bibfnamefont {T.}~\bibnamefont {Srivastava}}, \bibinfo
  {author} {\bibfnamefont {J.-C.}\ \bibnamefont {{Rojas-S{\'a}nchez}}},
  \bibinfo {author} {\bibfnamefont {T.~H.}\ \bibnamefont {Pham}}, \bibinfo
  {author} {\bibfnamefont {M.}~\bibnamefont {Hehn}}, \bibinfo {author}
  {\bibfnamefont {G.}~\bibnamefont {Malinowski}}, \bibinfo {author}
  {\bibfnamefont {C.}~\bibnamefont {Baraduc}}, \bibinfo {author} {\bibfnamefont
  {S.}~\bibnamefont {Auffret}}, \bibinfo {author} {\bibfnamefont
  {G.}~\bibnamefont {Gaudin}}, \bibinfo {author} {\bibfnamefont
  {S.}~\bibnamefont {Mangin}}, \bibinfo {author} {\bibfnamefont
  {H.}~\bibnamefont {B{\'e}a}},\ and\ \bibinfo {author} {\bibfnamefont
  {O.}~\bibnamefont {Boulle}},\ }\bibfield  {title} {\bibinfo {title} {Creation
  of {{Magnetic Skyrmion Bubble Lattices}} by {{Ultrafast Laser}} in
  {{Ultrathin Films}}},\ }\href {https://doi.org/10.1021/acs.nanolett.8b03653}
  {\bibfield  {journal} {\bibinfo  {journal} {Nano Lett.}\ }\textbf {\bibinfo
  {volume} {18}},\ \bibinfo {pages} {7362} (\bibinfo {year}
  {2018})}\BibitemShut {NoStop}%
\bibitem [{\citenamefont {B{\"u}ttner}\ \emph {et~al.}(2021)\citenamefont
  {B{\"u}ttner}, \citenamefont {Pfau}, \citenamefont {B{\"o}ttcher},
  \citenamefont {Schneider}, \citenamefont {Mercurio}, \citenamefont
  {G{\"u}nther}, \citenamefont {Hessing}, \citenamefont {Klose}, \citenamefont
  {Wittmann}, \citenamefont {Gerlinger}, \citenamefont {Kern}, \citenamefont
  {Str{\"u}ber}, \citenamefont {{von Korff Schmising}}, \citenamefont {Fuchs},
  \citenamefont {Engel}, \citenamefont {Churikova}, \citenamefont {Huang},
  \citenamefont {Suzuki}, \citenamefont {Lemesh}, \citenamefont {Huang},
  \citenamefont {Caretta}, \citenamefont {Weder}, \citenamefont {Gaida},
  \citenamefont {M{\"o}ller}, \citenamefont {Harvey}, \citenamefont {Zayko},
  \citenamefont {Bagschik}, \citenamefont {Carley}, \citenamefont {Mercadier},
  \citenamefont {Schlappa}, \citenamefont {Yaroslavtsev}, \citenamefont
  {Le~Guyarder}, \citenamefont {Gerasimova}, \citenamefont {Scherz},
  \citenamefont {Deiter}, \citenamefont {Gort}, \citenamefont {Hickin},
  \citenamefont {Zhu}, \citenamefont {Turcato}, \citenamefont {Lomidze},
  \citenamefont {Erdinger}, \citenamefont {Castoldi}, \citenamefont
  {Maffessanti}, \citenamefont {Porro}, \citenamefont {Samartsev},
  \citenamefont {Sinova}, \citenamefont {Ropers}, \citenamefont {Mentink},
  \citenamefont {Dup{\'e}}, \citenamefont {Beach},\ and\ \citenamefont
  {Eisebitt}}]{buttner2021observation}%
  \BibitemOpen
  \bibfield  {author} {\bibinfo {author} {\bibfnamefont {F.}~\bibnamefont
  {B{\"u}ttner}}, \bibinfo {author} {\bibfnamefont {B.}~\bibnamefont {Pfau}},
  \bibinfo {author} {\bibfnamefont {M.}~\bibnamefont {B{\"o}ttcher}}, \bibinfo
  {author} {\bibfnamefont {M.}~\bibnamefont {Schneider}}, \bibinfo {author}
  {\bibfnamefont {G.}~\bibnamefont {Mercurio}}, \bibinfo {author}
  {\bibfnamefont {C.~M.}\ \bibnamefont {G{\"u}nther}}, \bibinfo {author}
  {\bibfnamefont {P.}~\bibnamefont {Hessing}}, \bibinfo {author} {\bibfnamefont
  {C.}~\bibnamefont {Klose}}, \bibinfo {author} {\bibfnamefont
  {A.}~\bibnamefont {Wittmann}}, \bibinfo {author} {\bibfnamefont
  {K.}~\bibnamefont {Gerlinger}}, \bibinfo {author} {\bibfnamefont {L.-M.}\
  \bibnamefont {Kern}}, \bibinfo {author} {\bibfnamefont {C.}~\bibnamefont
  {Str{\"u}ber}}, \bibinfo {author} {\bibfnamefont {C.}~\bibnamefont {{von
  Korff Schmising}}}, \bibinfo {author} {\bibfnamefont {J.}~\bibnamefont
  {Fuchs}}, \bibinfo {author} {\bibfnamefont {D.}~\bibnamefont {Engel}},
  \bibinfo {author} {\bibfnamefont {A.}~\bibnamefont {Churikova}}, \bibinfo
  {author} {\bibfnamefont {S.}~\bibnamefont {Huang}}, \bibinfo {author}
  {\bibfnamefont {D.}~\bibnamefont {Suzuki}}, \bibinfo {author} {\bibfnamefont
  {I.}~\bibnamefont {Lemesh}}, \bibinfo {author} {\bibfnamefont
  {M.}~\bibnamefont {Huang}}, \bibinfo {author} {\bibfnamefont
  {L.}~\bibnamefont {Caretta}}, \bibinfo {author} {\bibfnamefont
  {D.}~\bibnamefont {Weder}}, \bibinfo {author} {\bibfnamefont {J.~H.}\
  \bibnamefont {Gaida}}, \bibinfo {author} {\bibfnamefont {M.}~\bibnamefont
  {M{\"o}ller}}, \bibinfo {author} {\bibfnamefont {T.~R.}\ \bibnamefont
  {Harvey}}, \bibinfo {author} {\bibfnamefont {S.}~\bibnamefont {Zayko}},
  \bibinfo {author} {\bibfnamefont {K.}~\bibnamefont {Bagschik}}, \bibinfo
  {author} {\bibfnamefont {R.}~\bibnamefont {Carley}}, \bibinfo {author}
  {\bibfnamefont {L.}~\bibnamefont {Mercadier}}, \bibinfo {author}
  {\bibfnamefont {J.}~\bibnamefont {Schlappa}}, \bibinfo {author}
  {\bibfnamefont {A.}~\bibnamefont {Yaroslavtsev}}, \bibinfo {author}
  {\bibfnamefont {L.}~\bibnamefont {Le~Guyarder}}, \bibinfo {author}
  {\bibfnamefont {N.}~\bibnamefont {Gerasimova}}, \bibinfo {author}
  {\bibfnamefont {A.}~\bibnamefont {Scherz}}, \bibinfo {author} {\bibfnamefont
  {C.}~\bibnamefont {Deiter}}, \bibinfo {author} {\bibfnamefont
  {R.}~\bibnamefont {Gort}}, \bibinfo {author} {\bibfnamefont {D.}~\bibnamefont
  {Hickin}}, \bibinfo {author} {\bibfnamefont {J.}~\bibnamefont {Zhu}},
  \bibinfo {author} {\bibfnamefont {M.}~\bibnamefont {Turcato}}, \bibinfo
  {author} {\bibfnamefont {D.}~\bibnamefont {Lomidze}}, \bibinfo {author}
  {\bibfnamefont {F.}~\bibnamefont {Erdinger}}, \bibinfo {author}
  {\bibfnamefont {A.}~\bibnamefont {Castoldi}}, \bibinfo {author}
  {\bibfnamefont {S.}~\bibnamefont {Maffessanti}}, \bibinfo {author}
  {\bibfnamefont {M.}~\bibnamefont {Porro}}, \bibinfo {author} {\bibfnamefont
  {A.}~\bibnamefont {Samartsev}}, \bibinfo {author} {\bibfnamefont
  {J.}~\bibnamefont {Sinova}}, \bibinfo {author} {\bibfnamefont
  {C.}~\bibnamefont {Ropers}}, \bibinfo {author} {\bibfnamefont {J.~H.}\
  \bibnamefont {Mentink}}, \bibinfo {author} {\bibfnamefont {B.}~\bibnamefont
  {Dup{\'e}}}, \bibinfo {author} {\bibfnamefont {G.~S.~D.}\ \bibnamefont
  {Beach}},\ and\ \bibinfo {author} {\bibfnamefont {S.}~\bibnamefont
  {Eisebitt}},\ }\bibfield  {title} {\bibinfo {title} {Observation of
  fluctuation-mediated picosecond nucleation of a topological phase},\ }\href
  {https://doi.org/10.1038/s41563-020-00807-1} {\bibfield  {journal} {\bibinfo
  {journal} {Nat. Mater.}\ }\textbf {\bibinfo {volume} {20}},\ \bibinfo {pages}
  {30} (\bibinfo {year} {2021})}\BibitemShut {NoStop}%
\bibitem [{\citenamefont {Gerlinger}\ \emph {et~al.}(2021)\citenamefont
  {Gerlinger}, \citenamefont {Pfau}, \citenamefont {B{\"u}ttner}, \citenamefont
  {Schneider}, \citenamefont {Kern}, \citenamefont {Fuchs}, \citenamefont
  {Engel}, \citenamefont {G{\"u}nther}, \citenamefont {Huang}, \citenamefont
  {Lemesh}, \citenamefont {Caretta}, \citenamefont {Churikova}, \citenamefont
  {Hessing}, \citenamefont {Klose}, \citenamefont {Str{\"u}ber}, \citenamefont
  {Schmising}, \citenamefont {Huang}, \citenamefont {Wittmann}, \citenamefont
  {Litzius}, \citenamefont {Metternich}, \citenamefont {Battistelli},
  \citenamefont {Bagschik}, \citenamefont {Sadovnikov}, \citenamefont {Beach},\
  and\ \citenamefont {Eisebitt}}]{gerlinger2021application}%
  \BibitemOpen
  \bibfield  {author} {\bibinfo {author} {\bibfnamefont {K.}~\bibnamefont
  {Gerlinger}}, \bibinfo {author} {\bibfnamefont {B.}~\bibnamefont {Pfau}},
  \bibinfo {author} {\bibfnamefont {F.}~\bibnamefont {B{\"u}ttner}}, \bibinfo
  {author} {\bibfnamefont {M.}~\bibnamefont {Schneider}}, \bibinfo {author}
  {\bibfnamefont {L.-M.}\ \bibnamefont {Kern}}, \bibinfo {author}
  {\bibfnamefont {J.}~\bibnamefont {Fuchs}}, \bibinfo {author} {\bibfnamefont
  {D.}~\bibnamefont {Engel}}, \bibinfo {author} {\bibfnamefont {C.~M.}\
  \bibnamefont {G{\"u}nther}}, \bibinfo {author} {\bibfnamefont
  {M.}~\bibnamefont {Huang}}, \bibinfo {author} {\bibfnamefont
  {I.}~\bibnamefont {Lemesh}}, \bibinfo {author} {\bibfnamefont
  {L.}~\bibnamefont {Caretta}}, \bibinfo {author} {\bibfnamefont
  {A.}~\bibnamefont {Churikova}}, \bibinfo {author} {\bibfnamefont
  {P.}~\bibnamefont {Hessing}}, \bibinfo {author} {\bibfnamefont
  {C.}~\bibnamefont {Klose}}, \bibinfo {author} {\bibfnamefont
  {C.}~\bibnamefont {Str{\"u}ber}}, \bibinfo {author} {\bibfnamefont
  {C.~v.~K.}\ \bibnamefont {Schmising}}, \bibinfo {author} {\bibfnamefont
  {S.}~\bibnamefont {Huang}}, \bibinfo {author} {\bibfnamefont
  {A.}~\bibnamefont {Wittmann}}, \bibinfo {author} {\bibfnamefont
  {K.}~\bibnamefont {Litzius}}, \bibinfo {author} {\bibfnamefont
  {D.}~\bibnamefont {Metternich}}, \bibinfo {author} {\bibfnamefont
  {R.}~\bibnamefont {Battistelli}}, \bibinfo {author} {\bibfnamefont
  {K.}~\bibnamefont {Bagschik}}, \bibinfo {author} {\bibfnamefont
  {A.}~\bibnamefont {Sadovnikov}}, \bibinfo {author} {\bibfnamefont {G.~S.~D.}\
  \bibnamefont {Beach}},\ and\ \bibinfo {author} {\bibfnamefont
  {S.}~\bibnamefont {Eisebitt}},\ }\bibfield  {title} {\bibinfo {title}
  {Application concepts for ultrafast laser-induced skyrmion creation and
  annihilation},\ }\href {https://doi.org/10.1063/5.0046033} {\bibfield
  {journal} {\bibinfo  {journal} {Appl. Phys. Lett.}\ }\textbf {\bibinfo
  {volume} {118}},\ \bibinfo {pages} {192403} (\bibinfo {year}
  {2021})}\BibitemShut {NoStop}%
\bibitem [{\citenamefont {Zhu}\ \emph {et~al.}(2024)\citenamefont {Zhu},
  \citenamefont {Bi}, \citenamefont {Zhang}, \citenamefont {Zheng},
  \citenamefont {Yang}, \citenamefont {Li}, \citenamefont {Tian}, \citenamefont
  {Cai}, \citenamefont {Yang}, \citenamefont {Zhang},\ and\ \citenamefont
  {Li}}]{zhu2024ultrafast}%
  \BibitemOpen
  \bibfield  {author} {\bibinfo {author} {\bibfnamefont {K.}~\bibnamefont
  {Zhu}}, \bibinfo {author} {\bibfnamefont {L.}~\bibnamefont {Bi}}, \bibinfo
  {author} {\bibfnamefont {Y.}~\bibnamefont {Zhang}}, \bibinfo {author}
  {\bibfnamefont {D.}~\bibnamefont {Zheng}}, \bibinfo {author} {\bibfnamefont
  {D.}~\bibnamefont {Yang}}, \bibinfo {author} {\bibfnamefont {J.}~\bibnamefont
  {Li}}, \bibinfo {author} {\bibfnamefont {H.}~\bibnamefont {Tian}}, \bibinfo
  {author} {\bibfnamefont {J.}~\bibnamefont {Cai}}, \bibinfo {author}
  {\bibfnamefont {H.}~\bibnamefont {Yang}}, \bibinfo {author} {\bibfnamefont
  {Y.}~\bibnamefont {Zhang}},\ and\ \bibinfo {author} {\bibfnamefont
  {J.}~\bibnamefont {Li}},\ }\bibfield  {title} {\bibinfo {title} {Ultrafast
  switching to zero field topological spin textures in ferrimagnetic {{TbFeCo}}
  films},\ }\href {https://doi.org/10.1039/D3NR04529C} {\bibfield  {journal}
  {\bibinfo  {journal} {Nanoscale}\ }\textbf {\bibinfo {volume} {16}},\
  \bibinfo {pages} {3133} (\bibinfo {year} {2024})}\BibitemShut {NoStop}%
\bibitem [{\citenamefont {Titze}\ \emph
  {et~al.}(2024{\natexlab{a}})\citenamefont {Titze}, \citenamefont {Koraltan},
  \citenamefont {Schmidt}, \citenamefont {M{\"o}ller}, \citenamefont
  {Bruckner}, \citenamefont {Abert}, \citenamefont {Suess}, \citenamefont
  {Ropers}, \citenamefont {Steil}, \citenamefont {Albrecht},\ and\
  \citenamefont {Mathias}}]{titze2024laserinduced}%
  \BibitemOpen
  \bibfield  {author} {\bibinfo {author} {\bibfnamefont {T.}~\bibnamefont
  {Titze}}, \bibinfo {author} {\bibfnamefont {S.}~\bibnamefont {Koraltan}},
  \bibinfo {author} {\bibfnamefont {T.}~\bibnamefont {Schmidt}}, \bibinfo
  {author} {\bibfnamefont {M.}~\bibnamefont {M{\"o}ller}}, \bibinfo {author}
  {\bibfnamefont {F.}~\bibnamefont {Bruckner}}, \bibinfo {author}
  {\bibfnamefont {C.}~\bibnamefont {Abert}}, \bibinfo {author} {\bibfnamefont
  {D.}~\bibnamefont {Suess}}, \bibinfo {author} {\bibfnamefont
  {C.}~\bibnamefont {Ropers}}, \bibinfo {author} {\bibfnamefont
  {D.}~\bibnamefont {Steil}}, \bibinfo {author} {\bibfnamefont
  {M.}~\bibnamefont {Albrecht}},\ and\ \bibinfo {author} {\bibfnamefont
  {S.}~\bibnamefont {Mathias}},\ }\bibfield  {title} {\bibinfo {title}
  {Laser-{{Induced Real-Space Topology Control}} of {{Spin Wave Resonances}}},\
  }\href {https://doi.org/10.1002/adfm.202313619} {\bibfield  {journal}
  {\bibinfo  {journal} {Adv. Funct. Mater.}\ }\textbf {\bibinfo {volume}
  {34}},\ \bibinfo {pages} {2313619} (\bibinfo {year}
  {2024}{\natexlab{a}})}\BibitemShut {NoStop}%
\bibitem [{\citenamefont {Titze}\ \emph {et~al.}(2025)\citenamefont {Titze},
  \citenamefont {Koraltan}, \citenamefont {Matthies}, \citenamefont {Schmidt},
  \citenamefont {Suess}, \citenamefont {Albrecht}, \citenamefont {Mathias},\
  and\ \citenamefont {Steil}}]{titze2025pathways}%
  \BibitemOpen
  \bibfield  {author} {\bibinfo {author} {\bibfnamefont {T.}~\bibnamefont
  {Titze}}, \bibinfo {author} {\bibfnamefont {S.}~\bibnamefont {Koraltan}},
  \bibinfo {author} {\bibfnamefont {M.}~\bibnamefont {Matthies}}, \bibinfo
  {author} {\bibfnamefont {T.}~\bibnamefont {Schmidt}}, \bibinfo {author}
  {\bibfnamefont {D.}~\bibnamefont {Suess}}, \bibinfo {author} {\bibfnamefont
  {M.}~\bibnamefont {Albrecht}}, \bibinfo {author} {\bibfnamefont
  {S.}~\bibnamefont {Mathias}},\ and\ \bibinfo {author} {\bibfnamefont
  {D.}~\bibnamefont {Steil}},\ }\bibfield  {title} {\bibinfo {title} {Pathways
  to bubble and skyrmion lattice formation in fe/gd multilayers},\ }\href@noop
  {} {\bibfield  {journal} {\bibinfo  {journal} {Phys. Rev. B}\ }\textbf
  {\bibinfo {volume} {112}},\ \bibinfo {pages} {064413} (\bibinfo {year}
  {2025})}\BibitemShut {NoStop}%
\bibitem [{\citenamefont {Titze}\ \emph
  {et~al.}(2024{\natexlab{b}})\citenamefont {Titze}, \citenamefont {Koraltan},
  \citenamefont {Schmidt}, \citenamefont {Suess}, \citenamefont {Albrecht},
  \citenamefont {Mathias},\ and\ \citenamefont {Steil}}]{titze2024alloptical}%
  \BibitemOpen
  \bibfield  {author} {\bibinfo {author} {\bibfnamefont {T.}~\bibnamefont
  {Titze}}, \bibinfo {author} {\bibfnamefont {S.}~\bibnamefont {Koraltan}},
  \bibinfo {author} {\bibfnamefont {T.}~\bibnamefont {Schmidt}}, \bibinfo
  {author} {\bibfnamefont {D.}~\bibnamefont {Suess}}, \bibinfo {author}
  {\bibfnamefont {M.}~\bibnamefont {Albrecht}}, \bibinfo {author}
  {\bibfnamefont {S.}~\bibnamefont {Mathias}},\ and\ \bibinfo {author}
  {\bibfnamefont {D.}~\bibnamefont {Steil}},\ }\bibfield  {title} {\bibinfo
  {title} {All-{{Optical Control}} of {{Bubble}} and {{Skyrmion Breathing}}},\
  }\href {https://doi.org/10.1103/PhysRevLett.133.156701} {\bibfield  {journal}
  {\bibinfo  {journal} {Phys. Rev. Lett.}\ }\textbf {\bibinfo {volume} {133}},\
  \bibinfo {pages} {156701} (\bibinfo {year} {2024}{\natexlab{b}})}\BibitemShut
  {NoStop}%
\bibitem [{\citenamefont {Padmanabhan}\ \emph {et~al.}(2019)\citenamefont
  {Padmanabhan}, \citenamefont {Sekiguchi}, \citenamefont {Versteeg},
  \citenamefont {Slivina}, \citenamefont {Tsurkan}, \citenamefont
  {Bord{\'a}cs}, \citenamefont {K{\'e}zsm{\'a}rki},\ and\ \citenamefont {{van
  Loosdrecht}}}]{padmanabhan2019optically}%
  \BibitemOpen
  \bibfield  {author} {\bibinfo {author} {\bibfnamefont {P.}~\bibnamefont
  {Padmanabhan}}, \bibinfo {author} {\bibfnamefont {F.}~\bibnamefont
  {Sekiguchi}}, \bibinfo {author} {\bibfnamefont {R.~B.}\ \bibnamefont
  {Versteeg}}, \bibinfo {author} {\bibfnamefont {E.}~\bibnamefont {Slivina}},
  \bibinfo {author} {\bibfnamefont {V.}~\bibnamefont {Tsurkan}}, \bibinfo
  {author} {\bibfnamefont {S.}~\bibnamefont {Bord{\'a}cs}}, \bibinfo {author}
  {\bibfnamefont {I.}~\bibnamefont {K{\'e}zsm{\'a}rki}},\ and\ \bibinfo
  {author} {\bibfnamefont {P.~H.~M.}\ \bibnamefont {{van Loosdrecht}}},\
  }\bibfield  {title} {\bibinfo {title} {Optically {{Driven Collective Spin
  Excitations}} and {{Magnetization Dynamics}} in the
  {{N}}{\textbackslash}'eel-type {{Skyrmion Host}}
  \$\{{\textbackslash}mathrm\{\vphantom{\}\}}{{GaV}}\vphantom\{\}\vphantom\{\}\_\{4\}\{{\textbackslash}mathrm\{\vphantom{\}\}}{{S}}\vphantom\{\}\vphantom\{\}\_\{8\}\$},\
  }\href {https://doi.org/10.1103/PhysRevLett.122.107203} {\bibfield  {journal}
  {\bibinfo  {journal} {Phys. Rev. Lett.}\ }\textbf {\bibinfo {volume} {122}},\
  \bibinfo {pages} {107203} (\bibinfo {year} {2019})}\BibitemShut {NoStop}%
\bibitem [{\citenamefont {Kalin}\ \emph {et~al.}(2022)\citenamefont {Kalin},
  \citenamefont {Sievers}, \citenamefont {F{\"u}ser}, \citenamefont
  {Schumacher}, \citenamefont {Bieler}, \citenamefont
  {{Garc{\'i}a-S{\'a}nchez}}, \citenamefont {Bauer},\ and\ \citenamefont
  {Pfleiderer}}]{kalin2022optically}%
  \BibitemOpen
  \bibfield  {author} {\bibinfo {author} {\bibfnamefont {J.}~\bibnamefont
  {Kalin}}, \bibinfo {author} {\bibfnamefont {S.}~\bibnamefont {Sievers}},
  \bibinfo {author} {\bibfnamefont {H.}~\bibnamefont {F{\"u}ser}}, \bibinfo
  {author} {\bibfnamefont {H.~W.}\ \bibnamefont {Schumacher}}, \bibinfo
  {author} {\bibfnamefont {M.}~\bibnamefont {Bieler}}, \bibinfo {author}
  {\bibfnamefont {F.}~\bibnamefont {{Garc{\'i}a-S{\'a}nchez}}}, \bibinfo
  {author} {\bibfnamefont {A.}~\bibnamefont {Bauer}},\ and\ \bibinfo {author}
  {\bibfnamefont {C.}~\bibnamefont {Pfleiderer}},\ }\bibfield  {title}
  {\bibinfo {title} {Optically excited spin dynamics of thermally metastable
  skyrmions in
  \$\{{\textbackslash}mathrm\{\vphantom{\}\}}{{Fe}}\vphantom\{\}\vphantom\{\}\_\{0.75\}\{{\textbackslash}mathrm\{\vphantom{\}\}}{{Co}}\vphantom\{\}\vphantom\{\}\_\{0.25\}{\textbackslash}mathrm\{\vphantom\}{{Si}}\vphantom\{\}\$},\
  }\href {https://doi.org/10.1103/PhysRevB.106.054430} {\bibfield  {journal}
  {\bibinfo  {journal} {Phys. Rev. B}\ }\textbf {\bibinfo {volume} {106}},\
  \bibinfo {pages} {054430} (\bibinfo {year} {2022})}\BibitemShut {NoStop}%
\bibitem [{\citenamefont {Chumak}\ \emph {et~al.}(2017)\citenamefont {Chumak},
  \citenamefont {Serga},\ and\ \citenamefont
  {Hillebrands}}]{chumak2017magnonic}%
  \BibitemOpen
  \bibfield  {author} {\bibinfo {author} {\bibfnamefont {A.~V.}\ \bibnamefont
  {Chumak}}, \bibinfo {author} {\bibfnamefont {A.~A.}\ \bibnamefont {Serga}},\
  and\ \bibinfo {author} {\bibfnamefont {B.}~\bibnamefont {Hillebrands}},\
  }\bibfield  {title} {\bibinfo {title} {Magnonic crystals for data
  processing},\ }\href {https://doi.org/10.1088/1361-6463/aa6a65} {\bibfield
  {journal} {\bibinfo  {journal} {J. Phys. D: Appl. Phys.}\ }\textbf {\bibinfo
  {volume} {50}},\ \bibinfo {pages} {244001} (\bibinfo {year}
  {2017})}\BibitemShut {NoStop}%
\bibitem [{\citenamefont {Tokura}\ and\ \citenamefont
  {Kanazawa}(2021)}]{tokura2021magnetic}%
  \BibitemOpen
  \bibfield  {author} {\bibinfo {author} {\bibfnamefont {Y.}~\bibnamefont
  {Tokura}}\ and\ \bibinfo {author} {\bibfnamefont {N.}~\bibnamefont
  {Kanazawa}},\ }\bibfield  {title} {\bibinfo {title} {Magnetic {{Skyrmion
  Materials}}},\ }\href {https://doi.org/10.1021/acs.chemrev.0c00297}
  {\bibfield  {journal} {\bibinfo  {journal} {Chem. Rev.}\ }\textbf {\bibinfo
  {volume} {121}},\ \bibinfo {pages} {2857} (\bibinfo {year}
  {2021})}\BibitemShut {NoStop}%
\bibitem [{\citenamefont {Kern}\ \emph
  {et~al.}(2022{\natexlab{b}})\citenamefont {Kern}, \citenamefont {Pfau},
  \citenamefont {Schneider}, \citenamefont {Gerlinger}, \citenamefont
  {Deinhart}, \citenamefont {Wittrock}, \citenamefont {Sidiropoulos},
  \citenamefont {Engel}, \citenamefont {Will}, \citenamefont {G{\"u}nther},
  \citenamefont {Litzius}, \citenamefont {Wintz}, \citenamefont {Weigand},
  \citenamefont {B{\"u}ttner},\ and\ \citenamefont
  {Eisebitt}}]{kern2022tailoring}%
  \BibitemOpen
  \bibfield  {author} {\bibinfo {author} {\bibfnamefont {L.-M.}\ \bibnamefont
  {Kern}}, \bibinfo {author} {\bibfnamefont {B.}~\bibnamefont {Pfau}}, \bibinfo
  {author} {\bibfnamefont {M.}~\bibnamefont {Schneider}}, \bibinfo {author}
  {\bibfnamefont {K.}~\bibnamefont {Gerlinger}}, \bibinfo {author}
  {\bibfnamefont {V.}~\bibnamefont {Deinhart}}, \bibinfo {author}
  {\bibfnamefont {S.}~\bibnamefont {Wittrock}}, \bibinfo {author}
  {\bibfnamefont {T.}~\bibnamefont {Sidiropoulos}}, \bibinfo {author}
  {\bibfnamefont {D.}~\bibnamefont {Engel}}, \bibinfo {author} {\bibfnamefont
  {I.}~\bibnamefont {Will}}, \bibinfo {author} {\bibfnamefont {C.~M.}\
  \bibnamefont {G{\"u}nther}}, \bibinfo {author} {\bibfnamefont
  {K.}~\bibnamefont {Litzius}}, \bibinfo {author} {\bibfnamefont
  {S.}~\bibnamefont {Wintz}}, \bibinfo {author} {\bibfnamefont
  {M.}~\bibnamefont {Weigand}}, \bibinfo {author} {\bibfnamefont
  {F.}~\bibnamefont {B{\"u}ttner}},\ and\ \bibinfo {author} {\bibfnamefont
  {S.}~\bibnamefont {Eisebitt}},\ }\bibfield  {title} {\bibinfo {title}
  {Tailoring optical excitation to control magnetic skyrmion nucleation},\
  }\href {https://doi.org/10.1103/PhysRevB.106.054435} {\bibfield  {journal}
  {\bibinfo  {journal} {Phys. Rev. B}\ }\textbf {\bibinfo {volume} {106}},\
  \bibinfo {pages} {054435} (\bibinfo {year} {2022}{\natexlab{b}})}\BibitemShut
  {NoStop}%
\bibitem [{\citenamefont {Woo}\ \emph {et~al.}(2018{\natexlab{b}})\citenamefont
  {Woo}, \citenamefont {Song}, \citenamefont {Zhang}, \citenamefont {Ezawa},
  \citenamefont {Zhou}, \citenamefont {Liu}, \citenamefont {Weigand},
  \citenamefont {Finizio}, \citenamefont {Raabe}, \citenamefont {Park} \emph
  {et~al.}}]{woo2018deterministic}%
  \BibitemOpen
  \bibfield  {author} {\bibinfo {author} {\bibfnamefont {S.}~\bibnamefont
  {Woo}}, \bibinfo {author} {\bibfnamefont {K.~M.}\ \bibnamefont {Song}},
  \bibinfo {author} {\bibfnamefont {X.}~\bibnamefont {Zhang}}, \bibinfo
  {author} {\bibfnamefont {M.}~\bibnamefont {Ezawa}}, \bibinfo {author}
  {\bibfnamefont {Y.}~\bibnamefont {Zhou}}, \bibinfo {author} {\bibfnamefont
  {X.}~\bibnamefont {Liu}}, \bibinfo {author} {\bibfnamefont {M.}~\bibnamefont
  {Weigand}}, \bibinfo {author} {\bibfnamefont {S.}~\bibnamefont {Finizio}},
  \bibinfo {author} {\bibfnamefont {J.}~\bibnamefont {Raabe}}, \bibinfo
  {author} {\bibfnamefont {M.-C.}\ \bibnamefont {Park}}, \emph {et~al.},\
  }\bibfield  {title} {\bibinfo {title} {Deterministic creation and deletion of
  a single magnetic skyrmion observed by direct time-resolved x-ray
  microscopy},\ }\href@noop {} {\bibfield  {journal} {\bibinfo  {journal} {Nat.
  Electron.}\ }\textbf {\bibinfo {volume} {1}},\ \bibinfo {pages} {288}
  (\bibinfo {year} {2018}{\natexlab{b}})}\BibitemShut {NoStop}%
\bibitem [{\citenamefont {Powalla}\ \emph {et~al.}(2022)\citenamefont
  {Powalla}, \citenamefont {Birch}, \citenamefont {Litzius}, \citenamefont
  {Wintz}, \citenamefont {Schulz}, \citenamefont {Weigand}, \citenamefont
  {Scholz}, \citenamefont {Lotsch}, \citenamefont {Kern}, \citenamefont
  {Sch{\"u}tz},\ and\ \citenamefont {Burghard}}]{powalla2022single}%
  \BibitemOpen
  \bibfield  {author} {\bibinfo {author} {\bibfnamefont {L.}~\bibnamefont
  {Powalla}}, \bibinfo {author} {\bibfnamefont {M.~T.}\ \bibnamefont {Birch}},
  \bibinfo {author} {\bibfnamefont {K.}~\bibnamefont {Litzius}}, \bibinfo
  {author} {\bibfnamefont {S.}~\bibnamefont {Wintz}}, \bibinfo {author}
  {\bibfnamefont {F.}~\bibnamefont {Schulz}}, \bibinfo {author} {\bibfnamefont
  {M.}~\bibnamefont {Weigand}}, \bibinfo {author} {\bibfnamefont
  {T.}~\bibnamefont {Scholz}}, \bibinfo {author} {\bibfnamefont {B.~V.}\
  \bibnamefont {Lotsch}}, \bibinfo {author} {\bibfnamefont {K.}~\bibnamefont
  {Kern}}, \bibinfo {author} {\bibfnamefont {G.}~\bibnamefont {Sch{\"u}tz}},\
  and\ \bibinfo {author} {\bibfnamefont {M.}~\bibnamefont {Burghard}},\
  }\bibfield  {title} {\bibinfo {title} {Single {{Skyrmion Generation}} via a
  {{Vertical Nanocontact}} in a {{2D Magnet-Based Heterostructure}}},\ }\href
  {https://doi.org/10.1021/acs.nanolett.2c01944} {\bibfield  {journal}
  {\bibinfo  {journal} {Nano Lett.}\ }\textbf {\bibinfo {volume} {22}},\
  \bibinfo {pages} {9236} (\bibinfo {year} {2022})}\BibitemShut {NoStop}%
\bibitem [{\citenamefont {Yokouchi}\ \emph
  {et~al.}(2020{\natexlab{b}})\citenamefont {Yokouchi}, \citenamefont
  {Sugimoto}, \citenamefont {Rana}, \citenamefont {Seki}, \citenamefont
  {Ogawa}, \citenamefont {Kasai},\ and\ \citenamefont
  {Otani}}]{yokouchi2020creation}%
  \BibitemOpen
  \bibfield  {author} {\bibinfo {author} {\bibfnamefont {T.}~\bibnamefont
  {Yokouchi}}, \bibinfo {author} {\bibfnamefont {S.}~\bibnamefont {Sugimoto}},
  \bibinfo {author} {\bibfnamefont {B.}~\bibnamefont {Rana}}, \bibinfo {author}
  {\bibfnamefont {S.}~\bibnamefont {Seki}}, \bibinfo {author} {\bibfnamefont
  {N.}~\bibnamefont {Ogawa}}, \bibinfo {author} {\bibfnamefont
  {S.}~\bibnamefont {Kasai}},\ and\ \bibinfo {author} {\bibfnamefont
  {Y.}~\bibnamefont {Otani}},\ }\bibfield  {title} {\bibinfo {title} {Creation
  of magnetic skyrmions by surface acoustic waves},\ }\href
  {https://doi.org/10.1038/s41565-020-0661-1} {\bibfield  {journal} {\bibinfo
  {journal} {Nat. Nanotechnol.}\ }\textbf {\bibinfo {volume} {15}},\ \bibinfo
  {pages} {361} (\bibinfo {year} {2020}{\natexlab{b}})}\BibitemShut {NoStop}%
\bibitem [{\citenamefont {Berruto}\ \emph {et~al.}(2018)\citenamefont
  {Berruto}, \citenamefont {Madan}, \citenamefont {Murooka}, \citenamefont
  {Vanacore}, \citenamefont {Pomarico}, \citenamefont {Rajeswari},
  \citenamefont {Lamb}, \citenamefont {Huang}, \citenamefont {Kruchkov},
  \citenamefont {Togawa}, \citenamefont {LaGrange}, \citenamefont {McGrouther},
  \citenamefont {R{\o}nnow},\ and\ \citenamefont
  {Carbone}}]{berruto2018laserinduced}%
  \BibitemOpen
  \bibfield  {author} {\bibinfo {author} {\bibfnamefont {G.}~\bibnamefont
  {Berruto}}, \bibinfo {author} {\bibfnamefont {I.}~\bibnamefont {Madan}},
  \bibinfo {author} {\bibfnamefont {Y.}~\bibnamefont {Murooka}}, \bibinfo
  {author} {\bibfnamefont {G.~M.}\ \bibnamefont {Vanacore}}, \bibinfo {author}
  {\bibfnamefont {E.}~\bibnamefont {Pomarico}}, \bibinfo {author}
  {\bibfnamefont {J.}~\bibnamefont {Rajeswari}}, \bibinfo {author}
  {\bibfnamefont {R.}~\bibnamefont {Lamb}}, \bibinfo {author} {\bibfnamefont
  {P.}~\bibnamefont {Huang}}, \bibinfo {author} {\bibfnamefont {A.~J.}\
  \bibnamefont {Kruchkov}}, \bibinfo {author} {\bibfnamefont {Y.}~\bibnamefont
  {Togawa}}, \bibinfo {author} {\bibfnamefont {T.}~\bibnamefont {LaGrange}},
  \bibinfo {author} {\bibfnamefont {D.}~\bibnamefont {McGrouther}}, \bibinfo
  {author} {\bibfnamefont {H.~M.}\ \bibnamefont {R{\o}nnow}},\ and\ \bibinfo
  {author} {\bibfnamefont {F.}~\bibnamefont {Carbone}},\ }\bibfield  {title}
  {\bibinfo {title} {Laser-{{Induced Skyrmion Writing}} and {{Erasing}} in an
  {{Ultrafast Cryo-Lorentz Transmission Electron Microscope}}},\ }\href
  {https://doi.org/10.1103/PhysRevLett.120.117201} {\bibfield  {journal}
  {\bibinfo  {journal} {Phys. Rev. Lett.}\ }\textbf {\bibinfo {volume} {120}},\
  \bibinfo {pages} {117201} (\bibinfo {year} {2018})}\BibitemShut {NoStop}%
\bibitem [{\citenamefont {Khela}\ \emph {et~al.}(2023)\citenamefont {Khela},
  \citenamefont {D{\c a}browski}, \citenamefont {Khan}, \citenamefont
  {Keatley}, \citenamefont {Verzhbitskiy}, \citenamefont {Eda}, \citenamefont
  {Hicken}, \citenamefont {Kurebayashi},\ and\ \citenamefont
  {Santos}}]{khela2023laserinduced}%
  \BibitemOpen
  \bibfield  {author} {\bibinfo {author} {\bibfnamefont {M.}~\bibnamefont
  {Khela}}, \bibinfo {author} {\bibfnamefont {M.}~\bibnamefont {D{\c
  a}browski}}, \bibinfo {author} {\bibfnamefont {S.}~\bibnamefont {Khan}},
  \bibinfo {author} {\bibfnamefont {P.~S.}\ \bibnamefont {Keatley}}, \bibinfo
  {author} {\bibfnamefont {I.}~\bibnamefont {Verzhbitskiy}}, \bibinfo {author}
  {\bibfnamefont {G.}~\bibnamefont {Eda}}, \bibinfo {author} {\bibfnamefont
  {R.~J.}\ \bibnamefont {Hicken}}, \bibinfo {author} {\bibfnamefont
  {H.}~\bibnamefont {Kurebayashi}},\ and\ \bibinfo {author} {\bibfnamefont
  {E.~J.~G.}\ \bibnamefont {Santos}},\ }\bibfield  {title} {\bibinfo {title}
  {Laser-induced topological spin switching in a {{2D}} van der {{Waals}}
  magnet},\ }\href {https://doi.org/10.1038/s41467-023-37082-y} {\bibfield
  {journal} {\bibinfo  {journal} {Nat. Commun.}\ }\textbf {\bibinfo {volume}
  {14}},\ \bibinfo {pages} {1378} (\bibinfo {year} {2023})}\BibitemShut
  {NoStop}%
\bibitem [{\citenamefont {Hsu}\ \emph {et~al.}(2017)\citenamefont {Hsu},
  \citenamefont {Kubetzka}, \citenamefont {Finco}, \citenamefont {Romming},
  \citenamefont {{von Bergmann}},\ and\ \citenamefont
  {Wiesendanger}}]{hsu2017electricfielddriven}%
  \BibitemOpen
  \bibfield  {author} {\bibinfo {author} {\bibfnamefont {P.-J.}\ \bibnamefont
  {Hsu}}, \bibinfo {author} {\bibfnamefont {A.}~\bibnamefont {Kubetzka}},
  \bibinfo {author} {\bibfnamefont {A.}~\bibnamefont {Finco}}, \bibinfo
  {author} {\bibfnamefont {N.}~\bibnamefont {Romming}}, \bibinfo {author}
  {\bibfnamefont {K.}~\bibnamefont {{von Bergmann}}},\ and\ \bibinfo {author}
  {\bibfnamefont {R.}~\bibnamefont {Wiesendanger}},\ }\bibfield  {title}
  {\bibinfo {title} {Electric-field-driven switching of individual magnetic
  skyrmions},\ }\href {https://doi.org/10.1038/nnano.2016.234} {\bibfield
  {journal} {\bibinfo  {journal} {Nature Nanotech}\ }\textbf {\bibinfo {volume}
  {12}},\ \bibinfo {pages} {123} (\bibinfo {year} {2017})}\BibitemShut
  {NoStop}%
\bibitem [{\citenamefont {Bhattacharya}\ \emph {et~al.}(2020)\citenamefont
  {Bhattacharya}, \citenamefont {Razavi}, \citenamefont {Wu}, \citenamefont
  {Dai}, \citenamefont {Wang},\ and\ \citenamefont
  {Atulasimha}}]{bhattacharya2020creation}%
  \BibitemOpen
  \bibfield  {author} {\bibinfo {author} {\bibfnamefont {D.}~\bibnamefont
  {Bhattacharya}}, \bibinfo {author} {\bibfnamefont {S.~A.}\ \bibnamefont
  {Razavi}}, \bibinfo {author} {\bibfnamefont {H.}~\bibnamefont {Wu}}, \bibinfo
  {author} {\bibfnamefont {B.}~\bibnamefont {Dai}}, \bibinfo {author}
  {\bibfnamefont {K.~L.}\ \bibnamefont {Wang}},\ and\ \bibinfo {author}
  {\bibfnamefont {J.}~\bibnamefont {Atulasimha}},\ }\bibfield  {title}
  {\bibinfo {title} {Creation and annihilation of non-volatile fixed magnetic
  skyrmions using voltage control of magnetic anisotropy},\ }\href
  {https://doi.org/10.1038/s41928-020-0432-x} {\bibfield  {journal} {\bibinfo
  {journal} {Nat. Electron.}\ }\textbf {\bibinfo {volume} {3}},\ \bibinfo
  {pages} {539} (\bibinfo {year} {2020})}\BibitemShut {NoStop}%
\bibitem [{\citenamefont {Metternich}\ \emph {et~al.}(2025)\citenamefont
  {Metternich}, \citenamefont {Litzius}, \citenamefont {Wintz}, \citenamefont
  {Gerlinger}, \citenamefont {Petz}, \citenamefont {Engel}, \citenamefont
  {Sidiropoulos}, \citenamefont {Battistelli}, \citenamefont {Steinbach},
  \citenamefont {Weigand}, \citenamefont {Wittrock}, \citenamefont {{von Korff
  Schmising}},\ and\ \citenamefont {B{\"u}ttner}}]{metternich2025defects}%
  \BibitemOpen
  \bibfield  {author} {\bibinfo {author} {\bibfnamefont {D.}~\bibnamefont
  {Metternich}}, \bibinfo {author} {\bibfnamefont {K.}~\bibnamefont {Litzius}},
  \bibinfo {author} {\bibfnamefont {S.}~\bibnamefont {Wintz}}, \bibinfo
  {author} {\bibfnamefont {K.}~\bibnamefont {Gerlinger}}, \bibinfo {author}
  {\bibfnamefont {S.}~\bibnamefont {Petz}}, \bibinfo {author} {\bibfnamefont
  {D.}~\bibnamefont {Engel}}, \bibinfo {author} {\bibfnamefont
  {T.}~\bibnamefont {Sidiropoulos}}, \bibinfo {author} {\bibfnamefont
  {R.}~\bibnamefont {Battistelli}}, \bibinfo {author} {\bibfnamefont
  {F.}~\bibnamefont {Steinbach}}, \bibinfo {author} {\bibfnamefont
  {M.}~\bibnamefont {Weigand}}, \bibinfo {author} {\bibfnamefont
  {S.}~\bibnamefont {Wittrock}}, \bibinfo {author} {\bibfnamefont
  {C.}~\bibnamefont {{von Korff Schmising}}},\ and\ \bibinfo {author}
  {\bibfnamefont {F.}~\bibnamefont {B{\"u}ttner}},\ }\bibfield  {title}
  {\bibinfo {title} {Defects in magnetic domain walls after single-shot
  all-optical switching},\ }\href {https://doi.org/10.1063/4.0000287}
  {\bibfield  {journal} {\bibinfo  {journal} {Struct. Dyn.}\ }\textbf {\bibinfo
  {volume} {12}},\ \bibinfo {pages} {024504} (\bibinfo {year}
  {2025})}\BibitemShut {NoStop}%
\bibitem [{\citenamefont {Fert}\ \emph {et~al.}(2017)\citenamefont {Fert},
  \citenamefont {Reyren},\ and\ \citenamefont {Cros}}]{fert2017magnetic}%
  \BibitemOpen
  \bibfield  {author} {\bibinfo {author} {\bibfnamefont {A.}~\bibnamefont
  {Fert}}, \bibinfo {author} {\bibfnamefont {N.}~\bibnamefont {Reyren}},\ and\
  \bibinfo {author} {\bibfnamefont {V.}~\bibnamefont {Cros}},\ }\bibfield
  {title} {\bibinfo {title} {Magnetic skyrmions: Advances in physics and
  potential applications},\ }\href {https://doi.org/10.1038/natrevmats.2017.31}
  {\bibfield  {journal} {\bibinfo  {journal} {Nat. Rev. Mater.}\ }\textbf
  {\bibinfo {volume} {2}},\ \bibinfo {pages} {17031} (\bibinfo {year}
  {2017})}\BibitemShut {NoStop}%
\bibitem [{\citenamefont {Zhang}\ \emph
  {et~al.}(2020{\natexlab{b}})\citenamefont {Zhang}, \citenamefont {Zhou},
  \citenamefont {Song}, \citenamefont {Park}, \citenamefont {Xia},
  \citenamefont {Ezawa}, \citenamefont {Liu}, \citenamefont {Zhao},
  \citenamefont {Zhao},\ and\ \citenamefont {Woo}}]{zhang2020skyrmion}%
  \BibitemOpen
  \bibfield  {author} {\bibinfo {author} {\bibfnamefont {X.}~\bibnamefont
  {Zhang}}, \bibinfo {author} {\bibfnamefont {Y.}~\bibnamefont {Zhou}},
  \bibinfo {author} {\bibfnamefont {K.~M.}\ \bibnamefont {Song}}, \bibinfo
  {author} {\bibfnamefont {T.-E.}\ \bibnamefont {Park}}, \bibinfo {author}
  {\bibfnamefont {J.}~\bibnamefont {Xia}}, \bibinfo {author} {\bibfnamefont
  {M.}~\bibnamefont {Ezawa}}, \bibinfo {author} {\bibfnamefont
  {X.}~\bibnamefont {Liu}}, \bibinfo {author} {\bibfnamefont {W.}~\bibnamefont
  {Zhao}}, \bibinfo {author} {\bibfnamefont {G.}~\bibnamefont {Zhao}},\ and\
  \bibinfo {author} {\bibfnamefont {S.}~\bibnamefont {Woo}},\ }\bibfield
  {title} {\bibinfo {title} {Skyrmion-electronics: writing, deleting, reading
  and processing magnetic skyrmions toward spintronic applications},\
  }\href@noop {} {\bibfield  {journal} {\bibinfo  {journal} {J. Phys.: Condens.
  Matter}\ }\textbf {\bibinfo {volume} {32}},\ \bibinfo {pages} {143001}
  (\bibinfo {year} {2020}{\natexlab{b}})}\BibitemShut {NoStop}%
\bibitem [{\citenamefont {Vakili}\ \emph {et~al.}(2021)\citenamefont {Vakili},
  \citenamefont {Xu}, \citenamefont {Zhou}, \citenamefont {Sakib},
  \citenamefont {Morshed}, \citenamefont {Hartnett}, \citenamefont {Quessab},
  \citenamefont {Litzius}, \citenamefont {Ma}, \citenamefont {Ganguly} \emph
  {et~al.}}]{vakili2021skyrmionics}%
  \BibitemOpen
  \bibfield  {author} {\bibinfo {author} {\bibfnamefont {H.}~\bibnamefont
  {Vakili}}, \bibinfo {author} {\bibfnamefont {J.-W.}\ \bibnamefont {Xu}},
  \bibinfo {author} {\bibfnamefont {W.}~\bibnamefont {Zhou}}, \bibinfo {author}
  {\bibfnamefont {M.~N.}\ \bibnamefont {Sakib}}, \bibinfo {author}
  {\bibfnamefont {M.~G.}\ \bibnamefont {Morshed}}, \bibinfo {author}
  {\bibfnamefont {T.}~\bibnamefont {Hartnett}}, \bibinfo {author}
  {\bibfnamefont {Y.}~\bibnamefont {Quessab}}, \bibinfo {author} {\bibfnamefont
  {K.}~\bibnamefont {Litzius}}, \bibinfo {author} {\bibfnamefont {C.~T.}\
  \bibnamefont {Ma}}, \bibinfo {author} {\bibfnamefont {S.}~\bibnamefont
  {Ganguly}}, \emph {et~al.},\ }\bibfield  {title} {\bibinfo {title}
  {Skyrmionics—computing and memory technologies based on topological
  excitations in magnets},\ }\href@noop {} {\bibfield  {journal} {\bibinfo
  {journal} {J. Appl. Phys.}\ }\textbf {\bibinfo {volume} {130}} (\bibinfo
  {year} {2021})}\BibitemShut {NoStop}%
\bibitem [{\citenamefont {Fillion}\ \emph {et~al.}(2022)\citenamefont
  {Fillion}, \citenamefont {Fischer}, \citenamefont {Kumar}, \citenamefont
  {Fassatoui}, \citenamefont {Pizzini}, \citenamefont {Ranno}, \citenamefont
  {Ourdani}, \citenamefont {Belmeguenai}, \citenamefont {Roussign{\'e}},
  \citenamefont {Ch{\'e}rif} \emph {et~al.}}]{fillion2022gate}%
  \BibitemOpen
  \bibfield  {author} {\bibinfo {author} {\bibfnamefont {C.-E.}\ \bibnamefont
  {Fillion}}, \bibinfo {author} {\bibfnamefont {J.}~\bibnamefont {Fischer}},
  \bibinfo {author} {\bibfnamefont {R.}~\bibnamefont {Kumar}}, \bibinfo
  {author} {\bibfnamefont {A.}~\bibnamefont {Fassatoui}}, \bibinfo {author}
  {\bibfnamefont {S.}~\bibnamefont {Pizzini}}, \bibinfo {author} {\bibfnamefont
  {L.}~\bibnamefont {Ranno}}, \bibinfo {author} {\bibfnamefont
  {D.}~\bibnamefont {Ourdani}}, \bibinfo {author} {\bibfnamefont
  {M.}~\bibnamefont {Belmeguenai}}, \bibinfo {author} {\bibfnamefont
  {Y.}~\bibnamefont {Roussign{\'e}}}, \bibinfo {author} {\bibfnamefont {S.-M.}\
  \bibnamefont {Ch{\'e}rif}}, \emph {et~al.},\ }\bibfield  {title} {\bibinfo
  {title} {Gate-controlled skyrmion and domain wall chirality},\ }\href@noop {}
  {\bibfield  {journal} {\bibinfo  {journal} {Nat. Commun.}\ }\textbf {\bibinfo
  {volume} {13}},\ \bibinfo {pages} {5257} (\bibinfo {year}
  {2022})}\BibitemShut {NoStop}%
\bibitem [{\citenamefont {Juge}\ \emph {et~al.}(2022)\citenamefont {Juge},
  \citenamefont {Sisodia}, \citenamefont {Larra{\~n}aga}, \citenamefont
  {Zhang}, \citenamefont {Pham}, \citenamefont {Rana}, \citenamefont {Sarpi},
  \citenamefont {Mille}, \citenamefont {Stanescu}, \citenamefont {Belkhou},
  \citenamefont {Mawass}, \citenamefont {{Novakovic-Marinkovic}}, \citenamefont
  {Kronast}, \citenamefont {Weigand}, \citenamefont {Gr{\"a}fe}, \citenamefont
  {Wintz}, \citenamefont {Finizio}, \citenamefont {Raabe}, \citenamefont
  {Aballe}, \citenamefont {Foerster}, \citenamefont {Belmeguenai},
  \citenamefont {{Buda-Prejbeanu}}, \citenamefont {{Pelloux-Prayer}},
  \citenamefont {Shaw}, \citenamefont {Nembach}, \citenamefont {Ranno},
  \citenamefont {Gaudin},\ and\ \citenamefont {Boulle}}]{juge2022skyrmions}%
  \BibitemOpen
  \bibfield  {author} {\bibinfo {author} {\bibfnamefont {R.}~\bibnamefont
  {Juge}}, \bibinfo {author} {\bibfnamefont {N.}~\bibnamefont {Sisodia}},
  \bibinfo {author} {\bibfnamefont {J.~U.}\ \bibnamefont {Larra{\~n}aga}},
  \bibinfo {author} {\bibfnamefont {Q.}~\bibnamefont {Zhang}}, \bibinfo
  {author} {\bibfnamefont {V.~T.}\ \bibnamefont {Pham}}, \bibinfo {author}
  {\bibfnamefont {K.~G.}\ \bibnamefont {Rana}}, \bibinfo {author}
  {\bibfnamefont {B.}~\bibnamefont {Sarpi}}, \bibinfo {author} {\bibfnamefont
  {N.}~\bibnamefont {Mille}}, \bibinfo {author} {\bibfnamefont
  {S.}~\bibnamefont {Stanescu}}, \bibinfo {author} {\bibfnamefont
  {R.}~\bibnamefont {Belkhou}}, \bibinfo {author} {\bibfnamefont {M.-A.}\
  \bibnamefont {Mawass}}, \bibinfo {author} {\bibfnamefont {N.}~\bibnamefont
  {{Novakovic-Marinkovic}}}, \bibinfo {author} {\bibfnamefont {F.}~\bibnamefont
  {Kronast}}, \bibinfo {author} {\bibfnamefont {M.}~\bibnamefont {Weigand}},
  \bibinfo {author} {\bibfnamefont {J.}~\bibnamefont {Gr{\"a}fe}}, \bibinfo
  {author} {\bibfnamefont {S.}~\bibnamefont {Wintz}}, \bibinfo {author}
  {\bibfnamefont {S.}~\bibnamefont {Finizio}}, \bibinfo {author} {\bibfnamefont
  {J.}~\bibnamefont {Raabe}}, \bibinfo {author} {\bibfnamefont
  {L.}~\bibnamefont {Aballe}}, \bibinfo {author} {\bibfnamefont
  {M.}~\bibnamefont {Foerster}}, \bibinfo {author} {\bibfnamefont
  {M.}~\bibnamefont {Belmeguenai}}, \bibinfo {author} {\bibfnamefont {L.~D.}\
  \bibnamefont {{Buda-Prejbeanu}}}, \bibinfo {author} {\bibfnamefont
  {J.}~\bibnamefont {{Pelloux-Prayer}}}, \bibinfo {author} {\bibfnamefont
  {J.~M.}\ \bibnamefont {Shaw}}, \bibinfo {author} {\bibfnamefont {H.~T.}\
  \bibnamefont {Nembach}}, \bibinfo {author} {\bibfnamefont {L.}~\bibnamefont
  {Ranno}}, \bibinfo {author} {\bibfnamefont {G.}~\bibnamefont {Gaudin}},\ and\
  \bibinfo {author} {\bibfnamefont {O.}~\bibnamefont {Boulle}},\ }\bibfield
  {title} {\bibinfo {title} {Skyrmions in synthetic antiferromagnets and their
  nucleation via electrical current and ultra-fast laser illumination},\ }\href
  {https://doi.org/10.1038/s41467-022-32525-4} {\bibfield  {journal} {\bibinfo
  {journal} {Nat. Commun.}\ }\textbf {\bibinfo {volume} {13}},\ \bibinfo
  {pages} {4807} (\bibinfo {year} {2022})}\BibitemShut {NoStop}%
\bibitem [{\citenamefont {Urrestarazu~Larra{\~n}aga}\ \emph
  {et~al.}(2024)\citenamefont {Urrestarazu~Larra{\~n}aga}, \citenamefont
  {Sisodia}, \citenamefont {Guedas}, \citenamefont {Pham}, \citenamefont
  {Di~Manici}, \citenamefont {Masseboeuf}, \citenamefont {Garello},
  \citenamefont {Disdier}, \citenamefont {Fernandez}, \citenamefont {Wintz},
  \citenamefont {Weigand}, \citenamefont {Belmeguenai}, \citenamefont
  {Pizzini}, \citenamefont {Sousa}, \citenamefont {{Buda-Prejbeanu}},
  \citenamefont {Gaudin},\ and\ \citenamefont
  {Boulle}}]{urrestarazu2024electrical}%
  \BibitemOpen
  \bibfield  {author} {\bibinfo {author} {\bibfnamefont {J.}~\bibnamefont
  {Urrestarazu~Larra{\~n}aga}}, \bibinfo {author} {\bibfnamefont
  {N.}~\bibnamefont {Sisodia}}, \bibinfo {author} {\bibfnamefont
  {R.}~\bibnamefont {Guedas}}, \bibinfo {author} {\bibfnamefont {V.~T.}\
  \bibnamefont {Pham}}, \bibinfo {author} {\bibfnamefont {I.}~\bibnamefont
  {Di~Manici}}, \bibinfo {author} {\bibfnamefont {A.}~\bibnamefont
  {Masseboeuf}}, \bibinfo {author} {\bibfnamefont {K.}~\bibnamefont {Garello}},
  \bibinfo {author} {\bibfnamefont {F.}~\bibnamefont {Disdier}}, \bibinfo
  {author} {\bibfnamefont {B.}~\bibnamefont {Fernandez}}, \bibinfo {author}
  {\bibfnamefont {S.}~\bibnamefont {Wintz}}, \bibinfo {author} {\bibfnamefont
  {M.}~\bibnamefont {Weigand}}, \bibinfo {author} {\bibfnamefont
  {M.}~\bibnamefont {Belmeguenai}}, \bibinfo {author} {\bibfnamefont
  {S.}~\bibnamefont {Pizzini}}, \bibinfo {author} {\bibfnamefont {R.~C.}\
  \bibnamefont {Sousa}}, \bibinfo {author} {\bibfnamefont {L.~D.}\ \bibnamefont
  {{Buda-Prejbeanu}}}, \bibinfo {author} {\bibfnamefont {G.}~\bibnamefont
  {Gaudin}},\ and\ \bibinfo {author} {\bibfnamefont {O.}~\bibnamefont
  {Boulle}},\ }\bibfield  {title} {\bibinfo {title} {Electrical {{Detection}}
  and {{Nucleation}} of a {{Magnetic Skyrmion}} in a {{Magnetic Tunnel Junction
  Observed}} via {{Operando Magnetic Microscopy}}},\ }\href
  {https://doi.org/10.1021/acs.nanolett.4c00316} {\bibfield  {journal}
  {\bibinfo  {journal} {Nano Lett.}\ }\textbf {\bibinfo {volume} {24}},\
  \bibinfo {pages} {3557} (\bibinfo {year} {2024})}\BibitemShut {NoStop}%
\bibitem [{\citenamefont {Srivastava}\ \emph {et~al.}(2018)\citenamefont
  {Srivastava}, \citenamefont {Schott}, \citenamefont {Juge}, \citenamefont
  {Krizakova}, \citenamefont {Belmeguenai}, \citenamefont {Roussign{\'e}},
  \citenamefont {Bernand-Mantel}, \citenamefont {Ranno}, \citenamefont
  {Pizzini}, \citenamefont {Ch{\'e}rif} \emph {et~al.}}]{srivastava2018large}%
  \BibitemOpen
  \bibfield  {author} {\bibinfo {author} {\bibfnamefont {T.}~\bibnamefont
  {Srivastava}}, \bibinfo {author} {\bibfnamefont {M.}~\bibnamefont {Schott}},
  \bibinfo {author} {\bibfnamefont {R.}~\bibnamefont {Juge}}, \bibinfo {author}
  {\bibfnamefont {V.}~\bibnamefont {Krizakova}}, \bibinfo {author}
  {\bibfnamefont {M.}~\bibnamefont {Belmeguenai}}, \bibinfo {author}
  {\bibfnamefont {Y.}~\bibnamefont {Roussign{\'e}}}, \bibinfo {author}
  {\bibfnamefont {A.}~\bibnamefont {Bernand-Mantel}}, \bibinfo {author}
  {\bibfnamefont {L.}~\bibnamefont {Ranno}}, \bibinfo {author} {\bibfnamefont
  {S.}~\bibnamefont {Pizzini}}, \bibinfo {author} {\bibfnamefont {S.-M.}\
  \bibnamefont {Ch{\'e}rif}}, \emph {et~al.},\ }\bibfield  {title} {\bibinfo
  {title} {Large-voltage tuning of dzyaloshinskii--moriya interactions: A route
  toward dynamic control of skyrmion chirality},\ }\href@noop {} {\bibfield
  {journal} {\bibinfo  {journal} {Nano Lett.}\ }\textbf {\bibinfo {volume}
  {18}},\ \bibinfo {pages} {4871} (\bibinfo {year} {2018})}\BibitemShut
  {NoStop}%
\bibitem [{\citenamefont {Ameziane}\ \emph {et~al.}(2023)\citenamefont
  {Ameziane}, \citenamefont {Huhtasalo}, \citenamefont {Flajsman},
  \citenamefont {Mansell},\ and\ \citenamefont {van
  Dijken}}]{ameziane2023solid}%
  \BibitemOpen
  \bibfield  {author} {\bibinfo {author} {\bibfnamefont {M.}~\bibnamefont
  {Ameziane}}, \bibinfo {author} {\bibfnamefont {J.}~\bibnamefont {Huhtasalo}},
  \bibinfo {author} {\bibfnamefont {L.}~\bibnamefont {Flajsman}}, \bibinfo
  {author} {\bibfnamefont {R.}~\bibnamefont {Mansell}},\ and\ \bibinfo {author}
  {\bibfnamefont {S.}~\bibnamefont {van Dijken}},\ }\bibfield  {title}
  {\bibinfo {title} {Solid-state lithium ion supercapacitor for voltage control
  of skyrmions},\ }\href@noop {} {\bibfield  {journal} {\bibinfo  {journal}
  {Nano Lett.}\ }\textbf {\bibinfo {volume} {23}},\ \bibinfo {pages} {3167}
  (\bibinfo {year} {2023})}\BibitemShut {NoStop}%
\bibitem [{\citenamefont {Dai}\ \emph {et~al.}(2023)\citenamefont {Dai},
  \citenamefont {Wu}, \citenamefont {Razavi}, \citenamefont {Xu}, \citenamefont
  {He}, \citenamefont {Shu}, \citenamefont {Jackson}, \citenamefont {Mahfouzi},
  \citenamefont {Huang}, \citenamefont {Pan} \emph {et~al.}}]{dai2023electric}%
  \BibitemOpen
  \bibfield  {author} {\bibinfo {author} {\bibfnamefont {B.}~\bibnamefont
  {Dai}}, \bibinfo {author} {\bibfnamefont {D.}~\bibnamefont {Wu}}, \bibinfo
  {author} {\bibfnamefont {S.~A.}\ \bibnamefont {Razavi}}, \bibinfo {author}
  {\bibfnamefont {S.}~\bibnamefont {Xu}}, \bibinfo {author} {\bibfnamefont
  {H.}~\bibnamefont {He}}, \bibinfo {author} {\bibfnamefont {Q.}~\bibnamefont
  {Shu}}, \bibinfo {author} {\bibfnamefont {M.}~\bibnamefont {Jackson}},
  \bibinfo {author} {\bibfnamefont {F.}~\bibnamefont {Mahfouzi}}, \bibinfo
  {author} {\bibfnamefont {H.}~\bibnamefont {Huang}}, \bibinfo {author}
  {\bibfnamefont {Q.}~\bibnamefont {Pan}}, \emph {et~al.},\ }\bibfield  {title}
  {\bibinfo {title} {Electric field manipulation of spin chirality and skyrmion
  dynamic},\ }\href@noop {} {\bibfield  {journal} {\bibinfo  {journal} {Sci.
  Adv.}\ }\textbf {\bibinfo {volume} {9}},\ \bibinfo {pages} {eade6836}
  (\bibinfo {year} {2023})}\BibitemShut {NoStop}%
\bibitem [{\citenamefont {Yang}\ \emph {et~al.}(2023)\citenamefont {Yang},
  \citenamefont {Son}, \citenamefont {Ju}, \citenamefont {Tran}, \citenamefont
  {Han}, \citenamefont {Park}, \citenamefont {Park}, \citenamefont {Moon},\
  and\ \citenamefont {Hwang}}]{yang2023magnetic}%
  \BibitemOpen
  \bibfield  {author} {\bibinfo {author} {\bibfnamefont {S.}~\bibnamefont
  {Yang}}, \bibinfo {author} {\bibfnamefont {J.~W.}\ \bibnamefont {Son}},
  \bibinfo {author} {\bibfnamefont {T.-S.}\ \bibnamefont {Ju}}, \bibinfo
  {author} {\bibfnamefont {D.~M.}\ \bibnamefont {Tran}}, \bibinfo {author}
  {\bibfnamefont {H.-S.}\ \bibnamefont {Han}}, \bibinfo {author} {\bibfnamefont
  {S.}~\bibnamefont {Park}}, \bibinfo {author} {\bibfnamefont {B.~H.}\
  \bibnamefont {Park}}, \bibinfo {author} {\bibfnamefont {K.-W.}\ \bibnamefont
  {Moon}},\ and\ \bibinfo {author} {\bibfnamefont {C.}~\bibnamefont {Hwang}},\
  }\bibfield  {title} {\bibinfo {title} {Magnetic skyrmion transistor gated
  with voltage-controlled magnetic anisotropy},\ }\href@noop {} {\bibfield
  {journal} {\bibinfo  {journal} {Adv. Mater.}\ }\textbf {\bibinfo {volume}
  {35}},\ \bibinfo {pages} {2208881} (\bibinfo {year} {2023})}\BibitemShut
  {NoStop}%
\bibitem [{\citenamefont {Quessab}\ \emph {et~al.}(2022)\citenamefont
  {Quessab}, \citenamefont {Xu}, \citenamefont {Cogulu}, \citenamefont
  {Finizio}, \citenamefont {Raabe},\ and\ \citenamefont
  {Kent}}]{quessab2022zero}%
  \BibitemOpen
  \bibfield  {author} {\bibinfo {author} {\bibfnamefont {Y.}~\bibnamefont
  {Quessab}}, \bibinfo {author} {\bibfnamefont {J.-W.}\ \bibnamefont {Xu}},
  \bibinfo {author} {\bibfnamefont {E.}~\bibnamefont {Cogulu}}, \bibinfo
  {author} {\bibfnamefont {S.}~\bibnamefont {Finizio}}, \bibinfo {author}
  {\bibfnamefont {J.}~\bibnamefont {Raabe}},\ and\ \bibinfo {author}
  {\bibfnamefont {A.~D.}\ \bibnamefont {Kent}},\ }\bibfield  {title} {\bibinfo
  {title} {Zero-field nucleation and fast motion of skyrmions induced by
  nanosecond current pulses in a ferrimagnetic thin film},\ }\href@noop {}
  {\bibfield  {journal} {\bibinfo  {journal} {Nano Lett.}\ }\textbf {\bibinfo
  {volume} {22}},\ \bibinfo {pages} {6091} (\bibinfo {year}
  {2022})}\BibitemShut {NoStop}%
\bibitem [{\citenamefont {Tremsina}\ and\ \citenamefont
  {Beach}(2022)}]{tremsina2022atomistic}%
  \BibitemOpen
  \bibfield  {author} {\bibinfo {author} {\bibfnamefont {E.~A.}\ \bibnamefont
  {Tremsina}}\ and\ \bibinfo {author} {\bibfnamefont {G.~S.}\ \bibnamefont
  {Beach}},\ }\bibfield  {title} {\bibinfo {title} {Atomistic simulations of
  distortion-limited high-speed dynamics of antiferromagnetic skyrmions},\
  }\href@noop {} {\bibfield  {journal} {\bibinfo  {journal} {Phys. Rev. B}\
  }\textbf {\bibinfo {volume} {106}},\ \bibinfo {pages} {L220402} (\bibinfo
  {year} {2022})}\BibitemShut {NoStop}%
\bibitem [{\citenamefont {Higo}\ and\ \citenamefont
  {Nakatsuji}(2022)}]{higo2022thin}%
  \BibitemOpen
  \bibfield  {author} {\bibinfo {author} {\bibfnamefont {T.}~\bibnamefont
  {Higo}}\ and\ \bibinfo {author} {\bibfnamefont {S.}~\bibnamefont
  {Nakatsuji}},\ }\bibfield  {title} {\bibinfo {title} {Thin film properties of
  the non-collinear weyl antiferromagnet mn3sn},\ }\href@noop {} {\bibfield
  {journal} {\bibinfo  {journal} {J. Magn. Magn. Mater.}\ }\textbf {\bibinfo
  {volume} {564}},\ \bibinfo {pages} {170176} (\bibinfo {year}
  {2022})}\BibitemShut {NoStop}%
\bibitem [{\citenamefont {Song}\ \emph {et~al.}(2025)\citenamefont {Song},
  \citenamefont {Bai}, \citenamefont {Zhou}, \citenamefont {Han}, \citenamefont
  {Reichlova}, \citenamefont {Dil}, \citenamefont {Liu}, \citenamefont {Chen},\
  and\ \citenamefont {Pan}}]{song2025altermagnets}%
  \BibitemOpen
  \bibfield  {author} {\bibinfo {author} {\bibfnamefont {C.}~\bibnamefont
  {Song}}, \bibinfo {author} {\bibfnamefont {H.}~\bibnamefont {Bai}}, \bibinfo
  {author} {\bibfnamefont {Z.}~\bibnamefont {Zhou}}, \bibinfo {author}
  {\bibfnamefont {L.}~\bibnamefont {Han}}, \bibinfo {author} {\bibfnamefont
  {H.}~\bibnamefont {Reichlova}}, \bibinfo {author} {\bibfnamefont {J.~H.}\
  \bibnamefont {Dil}}, \bibinfo {author} {\bibfnamefont {J.}~\bibnamefont
  {Liu}}, \bibinfo {author} {\bibfnamefont {X.}~\bibnamefont {Chen}},\ and\
  \bibinfo {author} {\bibfnamefont {F.}~\bibnamefont {Pan}},\ }\bibfield
  {title} {\bibinfo {title} {Altermagnets as a new class of functional
  materials},\ }\href@noop {} {\bibfield  {journal} {\bibinfo  {journal}
  {Nature Reviews Materials}\ ,\ \bibinfo {pages} {1}} (\bibinfo {year}
  {2025})}\BibitemShut {NoStop}%
\bibitem [{\citenamefont {Chen}\ \emph {et~al.}(2023)\citenamefont {Chen},
  \citenamefont {Higo}, \citenamefont {Tanaka}, \citenamefont {Nomoto},
  \citenamefont {Tsai}, \citenamefont {Idzuchi}, \citenamefont {Shiga},
  \citenamefont {Sakamoto}, \citenamefont {Ando}, \citenamefont {Kosaki} \emph
  {et~al.}}]{chen2023octupole}%
  \BibitemOpen
  \bibfield  {author} {\bibinfo {author} {\bibfnamefont {X.}~\bibnamefont
  {Chen}}, \bibinfo {author} {\bibfnamefont {T.}~\bibnamefont {Higo}}, \bibinfo
  {author} {\bibfnamefont {K.}~\bibnamefont {Tanaka}}, \bibinfo {author}
  {\bibfnamefont {T.}~\bibnamefont {Nomoto}}, \bibinfo {author} {\bibfnamefont
  {H.}~\bibnamefont {Tsai}}, \bibinfo {author} {\bibfnamefont {H.}~\bibnamefont
  {Idzuchi}}, \bibinfo {author} {\bibfnamefont {M.}~\bibnamefont {Shiga}},
  \bibinfo {author} {\bibfnamefont {S.}~\bibnamefont {Sakamoto}}, \bibinfo
  {author} {\bibfnamefont {R.}~\bibnamefont {Ando}}, \bibinfo {author}
  {\bibfnamefont {H.}~\bibnamefont {Kosaki}}, \emph {et~al.},\ }\bibfield
  {title} {\bibinfo {title} {Octupole-driven magnetoresistance in an
  antiferromagnetic tunnel junction},\ }\href@noop {} {\bibfield  {journal}
  {\bibinfo  {journal} {Nature}\ }\textbf {\bibinfo {volume} {613}},\ \bibinfo
  {pages} {490} (\bibinfo {year} {2023})}\BibitemShut {NoStop}%
\bibitem [{\citenamefont {Qin}\ \emph {et~al.}(2023)\citenamefont {Qin},
  \citenamefont {Yan}, \citenamefont {Wang}, \citenamefont {Chen},
  \citenamefont {Meng}, \citenamefont {Dong}, \citenamefont {Zhu},
  \citenamefont {Cai}, \citenamefont {Feng}, \citenamefont {Zhou} \emph
  {et~al.}}]{qin2023room}%
  \BibitemOpen
  \bibfield  {author} {\bibinfo {author} {\bibfnamefont {P.}~\bibnamefont
  {Qin}}, \bibinfo {author} {\bibfnamefont {H.}~\bibnamefont {Yan}}, \bibinfo
  {author} {\bibfnamefont {X.}~\bibnamefont {Wang}}, \bibinfo {author}
  {\bibfnamefont {H.}~\bibnamefont {Chen}}, \bibinfo {author} {\bibfnamefont
  {Z.}~\bibnamefont {Meng}}, \bibinfo {author} {\bibfnamefont {J.}~\bibnamefont
  {Dong}}, \bibinfo {author} {\bibfnamefont {M.}~\bibnamefont {Zhu}}, \bibinfo
  {author} {\bibfnamefont {J.}~\bibnamefont {Cai}}, \bibinfo {author}
  {\bibfnamefont {Z.}~\bibnamefont {Feng}}, \bibinfo {author} {\bibfnamefont
  {X.}~\bibnamefont {Zhou}}, \emph {et~al.},\ }\bibfield  {title} {\bibinfo
  {title} {Room-temperature magnetoresistance in an all-antiferromagnetic
  tunnel junction},\ }\href@noop {} {\bibfield  {journal} {\bibinfo  {journal}
  {Nature}\ }\textbf {\bibinfo {volume} {613}},\ \bibinfo {pages} {485}
  (\bibinfo {year} {2023})}\BibitemShut {NoStop}%
\bibitem [{\citenamefont {Li}\ \emph {et~al.}(2022{\natexlab{b}})\citenamefont
  {Li}, \citenamefont {Du}, \citenamefont {Wang}, \citenamefont {Wang},
  \citenamefont {Zhang}, \citenamefont {Cheng}, \citenamefont {Cai},
  \citenamefont {Lu}, \citenamefont {Cao}, \citenamefont {Pan} \emph
  {et~al.}}]{li2022experimental}%
  \BibitemOpen
  \bibfield  {author} {\bibinfo {author} {\bibfnamefont {S.}~\bibnamefont
  {Li}}, \bibinfo {author} {\bibfnamefont {A.}~\bibnamefont {Du}}, \bibinfo
  {author} {\bibfnamefont {Y.}~\bibnamefont {Wang}}, \bibinfo {author}
  {\bibfnamefont {X.}~\bibnamefont {Wang}}, \bibinfo {author} {\bibfnamefont
  {X.}~\bibnamefont {Zhang}}, \bibinfo {author} {\bibfnamefont
  {H.}~\bibnamefont {Cheng}}, \bibinfo {author} {\bibfnamefont
  {W.}~\bibnamefont {Cai}}, \bibinfo {author} {\bibfnamefont {S.}~\bibnamefont
  {Lu}}, \bibinfo {author} {\bibfnamefont {K.}~\bibnamefont {Cao}}, \bibinfo
  {author} {\bibfnamefont {B.}~\bibnamefont {Pan}}, \emph {et~al.},\ }\bibfield
   {title} {\bibinfo {title} {Experimental demonstration of skyrmionic magnetic
  tunnel junction at room temperature},\ }\href@noop {} {\bibfield  {journal}
  {\bibinfo  {journal} {Sci. Bull.}\ }\textbf {\bibinfo {volume} {67}},\
  \bibinfo {pages} {691} (\bibinfo {year} {2022}{\natexlab{b}})}\BibitemShut
  {NoStop}%
\bibitem [{\citenamefont {Sisodia}\ \emph
  {et~al.}(2022{\natexlab{a}})\citenamefont {Sisodia}, \citenamefont
  {Pelloux-Prayer}, \citenamefont {Buda-Prejbeanu}, \citenamefont {Anghel},
  \citenamefont {Gaudin},\ and\ \citenamefont {Boulle}}]{sisodia2022robust}%
  \BibitemOpen
  \bibfield  {author} {\bibinfo {author} {\bibfnamefont {N.}~\bibnamefont
  {Sisodia}}, \bibinfo {author} {\bibfnamefont {J.}~\bibnamefont
  {Pelloux-Prayer}}, \bibinfo {author} {\bibfnamefont {L.~D.}\ \bibnamefont
  {Buda-Prejbeanu}}, \bibinfo {author} {\bibfnamefont {L.}~\bibnamefont
  {Anghel}}, \bibinfo {author} {\bibfnamefont {G.}~\bibnamefont {Gaudin}},\
  and\ \bibinfo {author} {\bibfnamefont {O.}~\bibnamefont {Boulle}},\
  }\bibfield  {title} {\bibinfo {title} {Robust and programmable
  logic-in-memory devices exploiting skyrmion confinement and channeling using
  local energy barriers},\ }\href@noop {} {\bibfield  {journal} {\bibinfo
  {journal} {Phys. Rev. Appl.}\ }\textbf {\bibinfo {volume} {18}},\ \bibinfo
  {pages} {014025} (\bibinfo {year} {2022}{\natexlab{a}})}\BibitemShut
  {NoStop}%
\bibitem [{\citenamefont {Sisodia}\ \emph
  {et~al.}(2022{\natexlab{b}})\citenamefont {Sisodia}, \citenamefont
  {Pelloux-Prayer}, \citenamefont {Buda-Prejbeanu}, \citenamefont {Anghel},
  \citenamefont {Gaudin},\ and\ \citenamefont
  {Boulle}}]{sisodia2022programmable}%
  \BibitemOpen
  \bibfield  {author} {\bibinfo {author} {\bibfnamefont {N.}~\bibnamefont
  {Sisodia}}, \bibinfo {author} {\bibfnamefont {J.}~\bibnamefont
  {Pelloux-Prayer}}, \bibinfo {author} {\bibfnamefont {L.~D.}\ \bibnamefont
  {Buda-Prejbeanu}}, \bibinfo {author} {\bibfnamefont {L.}~\bibnamefont
  {Anghel}}, \bibinfo {author} {\bibfnamefont {G.}~\bibnamefont {Gaudin}},\
  and\ \bibinfo {author} {\bibfnamefont {O.}~\bibnamefont {Boulle}},\
  }\bibfield  {title} {\bibinfo {title} {Programmable skyrmion logic gates
  based on skyrmion tunneling},\ }\href@noop {} {\bibfield  {journal} {\bibinfo
   {journal} {Phys. Rev. Appl.}\ }\textbf {\bibinfo {volume} {17}},\ \bibinfo
  {pages} {064035} (\bibinfo {year} {2022}{\natexlab{b}})}\BibitemShut
  {NoStop}%
\bibitem [{\citenamefont {Albisetti}\ \emph {et~al.}(2016)\citenamefont
  {Albisetti}, \citenamefont {Petti}, \citenamefont {Pancaldi}, \citenamefont
  {Madami}, \citenamefont {Tacchi}, \citenamefont {Curtis}, \citenamefont
  {King}, \citenamefont {Papp}, \citenamefont {Csaba}, \citenamefont {Porod}
  \emph {et~al.}}]{albisetti2016nanopatterning}%
  \BibitemOpen
  \bibfield  {author} {\bibinfo {author} {\bibfnamefont {E.}~\bibnamefont
  {Albisetti}}, \bibinfo {author} {\bibfnamefont {D.}~\bibnamefont {Petti}},
  \bibinfo {author} {\bibfnamefont {M.}~\bibnamefont {Pancaldi}}, \bibinfo
  {author} {\bibfnamefont {M.}~\bibnamefont {Madami}}, \bibinfo {author}
  {\bibfnamefont {S.}~\bibnamefont {Tacchi}}, \bibinfo {author} {\bibfnamefont
  {J.}~\bibnamefont {Curtis}}, \bibinfo {author} {\bibfnamefont
  {W.}~\bibnamefont {King}}, \bibinfo {author} {\bibfnamefont {A.}~\bibnamefont
  {Papp}}, \bibinfo {author} {\bibfnamefont {G.}~\bibnamefont {Csaba}},
  \bibinfo {author} {\bibfnamefont {W.}~\bibnamefont {Porod}}, \emph {et~al.},\
  }\bibfield  {title} {\bibinfo {title} {Nanopatterning reconfigurable magnetic
  landscapes via thermally assisted scanning probe lithography},\ }\href@noop
  {} {\bibfield  {journal} {\bibinfo  {journal} {Nat. Nanotechnol.}\ }\textbf
  {\bibinfo {volume} {11}},\ \bibinfo {pages} {545} (\bibinfo {year}
  {2016})}\BibitemShut {NoStop}%
\bibitem [{\citenamefont {Akhtar}\ \emph {et~al.}(2019)\citenamefont {Akhtar},
  \citenamefont {Hrabec}, \citenamefont {Chouaieb}, \citenamefont {Haykal},
  \citenamefont {Gross}, \citenamefont {Belmeguenai}, \citenamefont {Gabor},
  \citenamefont {Shields}, \citenamefont {Maletinsky}, \citenamefont
  {Thiaville}, \citenamefont {Rohart},\ and\ \citenamefont
  {Jacques}}]{akhtar2019currentinduced}%
  \BibitemOpen
  \bibfield  {author} {\bibinfo {author} {\bibfnamefont {W.}~\bibnamefont
  {Akhtar}}, \bibinfo {author} {\bibfnamefont {A.}~\bibnamefont {Hrabec}},
  \bibinfo {author} {\bibfnamefont {S.}~\bibnamefont {Chouaieb}}, \bibinfo
  {author} {\bibfnamefont {A.}~\bibnamefont {Haykal}}, \bibinfo {author}
  {\bibfnamefont {I.}~\bibnamefont {Gross}}, \bibinfo {author} {\bibfnamefont
  {M.}~\bibnamefont {Belmeguenai}}, \bibinfo {author} {\bibfnamefont
  {M.}~\bibnamefont {Gabor}}, \bibinfo {author} {\bibfnamefont
  {B.}~\bibnamefont {Shields}}, \bibinfo {author} {\bibfnamefont
  {P.}~\bibnamefont {Maletinsky}}, \bibinfo {author} {\bibfnamefont
  {A.}~\bibnamefont {Thiaville}}, \bibinfo {author} {\bibfnamefont
  {S.}~\bibnamefont {Rohart}},\ and\ \bibinfo {author} {\bibfnamefont
  {V.}~\bibnamefont {Jacques}},\ }\bibfield  {title} {\bibinfo {title}
  {Current-{{Induced Nucleation}} and {{Dynamics}} of {{Skyrmions}} in a
  \${\textbackslash}mathrm\{\vphantom\}{{Co}}\vphantom\{\}\$-based {{Heusler
  Alloy}}},\ }\href {https://doi.org/10.1103/PhysRevApplied.11.034066}
  {\bibfield  {journal} {\bibinfo  {journal} {Phys. Rev. Appl.}\ }\textbf
  {\bibinfo {volume} {11}},\ \bibinfo {pages} {034066} (\bibinfo {year}
  {2019})}\BibitemShut {NoStop}%
\bibitem [{\citenamefont {Finazzi}\ \emph {et~al.}(2013)\citenamefont
  {Finazzi}, \citenamefont {Savoini}, \citenamefont {Khorsand}, \citenamefont
  {Tsukamoto}, \citenamefont {Itoh}, \citenamefont {Du{\`o}}, \citenamefont
  {Kirilyuk}, \citenamefont {Rasing},\ and\ \citenamefont
  {Ezawa}}]{finazzi2013laserinduced}%
  \BibitemOpen
  \bibfield  {author} {\bibinfo {author} {\bibfnamefont {M.}~\bibnamefont
  {Finazzi}}, \bibinfo {author} {\bibfnamefont {M.}~\bibnamefont {Savoini}},
  \bibinfo {author} {\bibfnamefont {A.~R.}\ \bibnamefont {Khorsand}}, \bibinfo
  {author} {\bibfnamefont {A.}~\bibnamefont {Tsukamoto}}, \bibinfo {author}
  {\bibfnamefont {A.}~\bibnamefont {Itoh}}, \bibinfo {author} {\bibfnamefont
  {L.}~\bibnamefont {Du{\`o}}}, \bibinfo {author} {\bibfnamefont
  {A.}~\bibnamefont {Kirilyuk}}, \bibinfo {author} {\bibfnamefont {{\relax
  Th}.}~\bibnamefont {Rasing}},\ and\ \bibinfo {author} {\bibfnamefont
  {M.}~\bibnamefont {Ezawa}},\ }\bibfield  {title} {\bibinfo {title}
  {Laser-{{Induced Magnetic Nanostructures}} with {{Tunable Topological
  Properties}}},\ }\href {https://doi.org/10.1103/PhysRevLett.110.177205}
  {\bibfield  {journal} {\bibinfo  {journal} {Phys. Rev. Lett.}\ }\textbf
  {\bibinfo {volume} {110}},\ \bibinfo {pages} {177205} (\bibinfo {year}
  {2013})}\BibitemShut {NoStop}%
\bibitem [{\citenamefont {Z{\'a}zvorka}\ \emph {et~al.}(2019)\citenamefont
  {Z{\'a}zvorka}, \citenamefont {Jakobs}, \citenamefont {Heinze}, \citenamefont
  {Keil}, \citenamefont {Kromin}, \citenamefont {Jaiswal}, \citenamefont
  {Litzius}, \citenamefont {Jakob}, \citenamefont {Virnau}, \citenamefont
  {Pinna}, \citenamefont {{Everschor-Sitte}}, \citenamefont {R{\'o}zsa},
  \citenamefont {Donges}, \citenamefont {Nowak},\ and\ \citenamefont
  {Kl{\"a}ui}}]{zazvorka2019thermal}%
  \BibitemOpen
  \bibfield  {author} {\bibinfo {author} {\bibfnamefont {J.}~\bibnamefont
  {Z{\'a}zvorka}}, \bibinfo {author} {\bibfnamefont {F.}~\bibnamefont
  {Jakobs}}, \bibinfo {author} {\bibfnamefont {D.}~\bibnamefont {Heinze}},
  \bibinfo {author} {\bibfnamefont {N.}~\bibnamefont {Keil}}, \bibinfo {author}
  {\bibfnamefont {S.}~\bibnamefont {Kromin}}, \bibinfo {author} {\bibfnamefont
  {S.}~\bibnamefont {Jaiswal}}, \bibinfo {author} {\bibfnamefont
  {K.}~\bibnamefont {Litzius}}, \bibinfo {author} {\bibfnamefont
  {G.}~\bibnamefont {Jakob}}, \bibinfo {author} {\bibfnamefont
  {P.}~\bibnamefont {Virnau}}, \bibinfo {author} {\bibfnamefont
  {D.}~\bibnamefont {Pinna}}, \bibinfo {author} {\bibfnamefont
  {K.}~\bibnamefont {{Everschor-Sitte}}}, \bibinfo {author} {\bibfnamefont
  {L.}~\bibnamefont {R{\'o}zsa}}, \bibinfo {author} {\bibfnamefont
  {A.}~\bibnamefont {Donges}}, \bibinfo {author} {\bibfnamefont
  {U.}~\bibnamefont {Nowak}},\ and\ \bibinfo {author} {\bibfnamefont
  {M.}~\bibnamefont {Kl{\"a}ui}},\ }\bibfield  {title} {\bibinfo {title}
  {Thermal skyrmion diffusion used in a reshuffler device},\ }\href
  {https://doi.org/10.1038/s41565-019-0436-8} {\bibfield  {journal} {\bibinfo
  {journal} {Nat. Nanotechnol.}\ }\textbf {\bibinfo {volume} {14}},\ \bibinfo
  {pages} {658} (\bibinfo {year} {2019})}\BibitemShut {NoStop}%
\bibitem [{\citenamefont {Nakajima}\ and\ \citenamefont
  {Fischer}(2021)}]{nakajima2021reservoir}%
  \BibitemOpen
  \bibfield  {author} {\bibinfo {author} {\bibfnamefont {K.}~\bibnamefont
  {Nakajima}}\ and\ \bibinfo {author} {\bibfnamefont {I.}~\bibnamefont
  {Fischer}},\ }\href@noop {} {\emph {\bibinfo {title} {Reservoir computing}}}\
  (\bibinfo  {publisher} {Springer},\ \bibinfo {year} {2021})\BibitemShut
  {NoStop}%
\bibitem [{\citenamefont {Yokouchi}\ \emph {et~al.}(2022)\citenamefont
  {Yokouchi}, \citenamefont {Sugimoto}, \citenamefont {Rana}, \citenamefont
  {Seki}, \citenamefont {Ogawa}, \citenamefont {Shiomi}, \citenamefont
  {Kasai},\ and\ \citenamefont {Otani}}]{yokouchi2022pattern}%
  \BibitemOpen
  \bibfield  {author} {\bibinfo {author} {\bibfnamefont {T.}~\bibnamefont
  {Yokouchi}}, \bibinfo {author} {\bibfnamefont {S.}~\bibnamefont {Sugimoto}},
  \bibinfo {author} {\bibfnamefont {B.}~\bibnamefont {Rana}}, \bibinfo {author}
  {\bibfnamefont {S.}~\bibnamefont {Seki}}, \bibinfo {author} {\bibfnamefont
  {N.}~\bibnamefont {Ogawa}}, \bibinfo {author} {\bibfnamefont
  {Y.}~\bibnamefont {Shiomi}}, \bibinfo {author} {\bibfnamefont
  {S.}~\bibnamefont {Kasai}},\ and\ \bibinfo {author} {\bibfnamefont
  {Y.}~\bibnamefont {Otani}},\ }\bibfield  {title} {\bibinfo {title} {Pattern
  recognition with neuromorphic computing using magnetic field--induced
  dynamics of skyrmions},\ }\href {https://doi.org/10.1126/sciadv.abq5652}
  {\bibfield  {journal} {\bibinfo  {journal} {Sci. Adv.}\ }\textbf {\bibinfo
  {volume} {8}},\ \bibinfo {pages} {eabq5652} (\bibinfo {year}
  {2022})}\BibitemShut {NoStop}%
\bibitem [{\citenamefont {Sun}\ \emph {et~al.}(2023)\citenamefont {Sun},
  \citenamefont {Lin}, \citenamefont {Lei}, \citenamefont {Chen}, \citenamefont
  {Kang}, \citenamefont {Zhao}, \citenamefont {Wei}, \citenamefont {Chen},
  \citenamefont {Pang}, \citenamefont {Hu}, \citenamefont {Yang}, \citenamefont
  {Dong}, \citenamefont {Zhao}, \citenamefont {Liu}, \citenamefont {Yuan},
  \citenamefont {Ullrich}, \citenamefont {Back}, \citenamefont {Zhang},
  \citenamefont {Pan}, \citenamefont {Zhao}, \citenamefont {Feng},
  \citenamefont {Fert},\ and\ \citenamefont {Zhao}}]{sun2023experimental}%
  \BibitemOpen
  \bibfield  {author} {\bibinfo {author} {\bibfnamefont {Y.}~\bibnamefont
  {Sun}}, \bibinfo {author} {\bibfnamefont {T.}~\bibnamefont {Lin}}, \bibinfo
  {author} {\bibfnamefont {N.}~\bibnamefont {Lei}}, \bibinfo {author}
  {\bibfnamefont {X.}~\bibnamefont {Chen}}, \bibinfo {author} {\bibfnamefont
  {W.}~\bibnamefont {Kang}}, \bibinfo {author} {\bibfnamefont {Z.}~\bibnamefont
  {Zhao}}, \bibinfo {author} {\bibfnamefont {D.}~\bibnamefont {Wei}}, \bibinfo
  {author} {\bibfnamefont {C.}~\bibnamefont {Chen}}, \bibinfo {author}
  {\bibfnamefont {S.}~\bibnamefont {Pang}}, \bibinfo {author} {\bibfnamefont
  {L.}~\bibnamefont {Hu}}, \bibinfo {author} {\bibfnamefont {L.}~\bibnamefont
  {Yang}}, \bibinfo {author} {\bibfnamefont {E.}~\bibnamefont {Dong}}, \bibinfo
  {author} {\bibfnamefont {L.}~\bibnamefont {Zhao}}, \bibinfo {author}
  {\bibfnamefont {L.}~\bibnamefont {Liu}}, \bibinfo {author} {\bibfnamefont
  {Z.}~\bibnamefont {Yuan}}, \bibinfo {author} {\bibfnamefont {A.}~\bibnamefont
  {Ullrich}}, \bibinfo {author} {\bibfnamefont {C.~H.}\ \bibnamefont {Back}},
  \bibinfo {author} {\bibfnamefont {J.}~\bibnamefont {Zhang}}, \bibinfo
  {author} {\bibfnamefont {D.}~\bibnamefont {Pan}}, \bibinfo {author}
  {\bibfnamefont {J.}~\bibnamefont {Zhao}}, \bibinfo {author} {\bibfnamefont
  {M.}~\bibnamefont {Feng}}, \bibinfo {author} {\bibfnamefont {A.}~\bibnamefont
  {Fert}},\ and\ \bibinfo {author} {\bibfnamefont {W.}~\bibnamefont {Zhao}},\
  }\bibfield  {title} {\bibinfo {title} {Experimental demonstration of a
  skyrmion-enhanced strain-mediated physical reservoir computing system},\
  }\href {https://doi.org/10.1038/s41467-023-39207-9} {\bibfield  {journal}
  {\bibinfo  {journal} {Nat. Commun.}\ }\textbf {\bibinfo {volume} {14}},\
  \bibinfo {pages} {3434} (\bibinfo {year} {2023})}\BibitemShut {NoStop}%
\bibitem [{\citenamefont {Beneke}\ \emph {et~al.}(2024)\citenamefont {Beneke},
  \citenamefont {Winkler}, \citenamefont {Raab}, \citenamefont {Brems},
  \citenamefont {Kammerbauer}, \citenamefont {Gerhards}, \citenamefont
  {Knobloch}, \citenamefont {Krishnia}, \citenamefont {Mentink},\ and\
  \citenamefont {Kl{\"a}ui}}]{beneke2024gesture}%
  \BibitemOpen
  \bibfield  {author} {\bibinfo {author} {\bibfnamefont {G.}~\bibnamefont
  {Beneke}}, \bibinfo {author} {\bibfnamefont {T.~B.}\ \bibnamefont {Winkler}},
  \bibinfo {author} {\bibfnamefont {K.}~\bibnamefont {Raab}}, \bibinfo {author}
  {\bibfnamefont {M.~A.}\ \bibnamefont {Brems}}, \bibinfo {author}
  {\bibfnamefont {F.}~\bibnamefont {Kammerbauer}}, \bibinfo {author}
  {\bibfnamefont {P.}~\bibnamefont {Gerhards}}, \bibinfo {author}
  {\bibfnamefont {K.}~\bibnamefont {Knobloch}}, \bibinfo {author}
  {\bibfnamefont {S.}~\bibnamefont {Krishnia}}, \bibinfo {author}
  {\bibfnamefont {J.~H.}\ \bibnamefont {Mentink}},\ and\ \bibinfo {author}
  {\bibfnamefont {M.}~\bibnamefont {Kl{\"a}ui}},\ }\bibfield  {title} {\bibinfo
  {title} {Gesture recognition with {{Brownian}} reservoir computing using
  geometrically confined skyrmion dynamics},\ }\href
  {https://doi.org/10.1038/s41467-024-52345-y} {\bibfield  {journal} {\bibinfo
  {journal} {Nat. Commun.}\ }\textbf {\bibinfo {volume} {15}},\ \bibinfo
  {pages} {8103} (\bibinfo {year} {2024})}\BibitemShut {NoStop}%
\bibitem [{\citenamefont {Zheng}\ \emph {et~al.}(2017)\citenamefont {Zheng},
  \citenamefont {Li}, \citenamefont {Wang}, \citenamefont {Song}, \citenamefont
  {Jin}, \citenamefont {Wei}, \citenamefont {Kov{\'a}cs}, \citenamefont {Zang},
  \citenamefont {Tian}, \citenamefont {Zhang}, \citenamefont {Du},\ and\
  \citenamefont {{Dunin-Borkowski}}}]{zheng2017direct}%
  \BibitemOpen
  \bibfield  {author} {\bibinfo {author} {\bibfnamefont {F.}~\bibnamefont
  {Zheng}}, \bibinfo {author} {\bibfnamefont {H.}~\bibnamefont {Li}}, \bibinfo
  {author} {\bibfnamefont {S.}~\bibnamefont {Wang}}, \bibinfo {author}
  {\bibfnamefont {D.}~\bibnamefont {Song}}, \bibinfo {author} {\bibfnamefont
  {C.}~\bibnamefont {Jin}}, \bibinfo {author} {\bibfnamefont {W.}~\bibnamefont
  {Wei}}, \bibinfo {author} {\bibfnamefont {A.}~\bibnamefont {Kov{\'a}cs}},
  \bibinfo {author} {\bibfnamefont {J.}~\bibnamefont {Zang}}, \bibinfo {author}
  {\bibfnamefont {M.}~\bibnamefont {Tian}}, \bibinfo {author} {\bibfnamefont
  {Y.}~\bibnamefont {Zhang}}, \bibinfo {author} {\bibfnamefont
  {H.}~\bibnamefont {Du}},\ and\ \bibinfo {author} {\bibfnamefont {R.~E.}\
  \bibnamefont {{Dunin-Borkowski}}},\ }\bibfield  {title} {\bibinfo {title}
  {Direct {{Imaging}} of a {{Zero-Field Target Skyrmion}} and {{Its Polarity
  Switch}} in a {{Chiral Magnetic Nanodisk}}},\ }\href
  {https://doi.org/10.1103/PhysRevLett.119.197205} {\bibfield  {journal}
  {\bibinfo  {journal} {Phys. Rev. Lett.}\ }\textbf {\bibinfo {volume} {119}},\
  \bibinfo {pages} {197205} (\bibinfo {year} {2017})}\BibitemShut {NoStop}%
\bibitem [{\citenamefont {Borisov}\ \emph {et~al.}(2002)\citenamefont
  {Borisov}, \citenamefont {Zykov}, \citenamefont {Mikushina},\ and\
  \citenamefont {Moskvin}}]{borisov2002vortices}%
  \BibitemOpen
  \bibfield  {author} {\bibinfo {author} {\bibfnamefont {A.~B.}\ \bibnamefont
  {Borisov}}, \bibinfo {author} {\bibfnamefont {S.~A.}\ \bibnamefont {Zykov}},
  \bibinfo {author} {\bibfnamefont {N.~A.}\ \bibnamefont {Mikushina}},\ and\
  \bibinfo {author} {\bibfnamefont {A.~S.}\ \bibnamefont {Moskvin}},\
  }\bibfield  {title} {\bibinfo {title} {Vortices and magnetic structures of
  the target type in a two-dimensional ferromagnet with anisotropic exchange},\
  }\href {https://doi.org/10.1134/1.1451023} {\bibfield  {journal} {\bibinfo
  {journal} {Phys. Solid State}\ }\textbf {\bibinfo {volume} {44}},\ \bibinfo
  {pages} {324} (\bibinfo {year} {2002})}\BibitemShut {NoStop}%
\bibitem [{\citenamefont {Seki}\ \emph {et~al.}(2022)\citenamefont {Seki},
  \citenamefont {Suzuki}, \citenamefont {Ishibashi}, \citenamefont {Takagi},
  \citenamefont {Khanh}, \citenamefont {Shiota}, \citenamefont {Shibata},
  \citenamefont {Koshibae}, \citenamefont {Tokura},\ and\ \citenamefont
  {Ono}}]{seki2022direct}%
  \BibitemOpen
  \bibfield  {author} {\bibinfo {author} {\bibfnamefont {S.}~\bibnamefont
  {Seki}}, \bibinfo {author} {\bibfnamefont {M.}~\bibnamefont {Suzuki}},
  \bibinfo {author} {\bibfnamefont {M.}~\bibnamefont {Ishibashi}}, \bibinfo
  {author} {\bibfnamefont {R.}~\bibnamefont {Takagi}}, \bibinfo {author}
  {\bibfnamefont {N.~D.}\ \bibnamefont {Khanh}}, \bibinfo {author}
  {\bibfnamefont {Y.}~\bibnamefont {Shiota}}, \bibinfo {author} {\bibfnamefont
  {K.}~\bibnamefont {Shibata}}, \bibinfo {author} {\bibfnamefont
  {W.}~\bibnamefont {Koshibae}}, \bibinfo {author} {\bibfnamefont
  {Y.}~\bibnamefont {Tokura}},\ and\ \bibinfo {author} {\bibfnamefont
  {T.}~\bibnamefont {Ono}},\ }\bibfield  {title} {\bibinfo {title} {Direct
  visualization of the three-dimensional shape of skyrmion strings in a
  noncentrosymmetric magnet},\ }\href
  {https://doi.org/10.1038/s41563-021-01141-w} {\bibfield  {journal} {\bibinfo
  {journal} {Nat. Mater.}\ }\textbf {\bibinfo {volume} {21}},\ \bibinfo {pages}
  {181} (\bibinfo {year} {2022})}\BibitemShut {NoStop}%
\bibitem [{\citenamefont {Leonov}(2021)}]{leonov2021surface}%
  \BibitemOpen
  \bibfield  {author} {\bibinfo {author} {\bibfnamefont {A.~O.}\ \bibnamefont
  {Leonov}},\ }\bibfield  {title} {\bibinfo {title} {Surface anchoring as a
  control parameter for shaping skyrmion or toron properties in thin layers of
  chiral nematic liquid crystals and noncentrosymmetric magnets},\ }\href
  {https://doi.org/10.1103/PhysRevE.104.044701} {\bibfield  {journal} {\bibinfo
   {journal} {Phys. Rev. E}\ }\textbf {\bibinfo {volume} {104}},\ \bibinfo
  {pages} {044701} (\bibinfo {year} {2021})}\BibitemShut {NoStop}%
\bibitem [{\citenamefont {Chen}\ \emph {et~al.}(2020)\citenamefont {Chen},
  \citenamefont {Li}, \citenamefont {Pavlidis},\ and\ \citenamefont
  {Moutafis}}]{chen2020skyrmionic}%
  \BibitemOpen
  \bibfield  {author} {\bibinfo {author} {\bibfnamefont {R.}~\bibnamefont
  {Chen}}, \bibinfo {author} {\bibfnamefont {Y.}~\bibnamefont {Li}}, \bibinfo
  {author} {\bibfnamefont {V.~F.}\ \bibnamefont {Pavlidis}},\ and\ \bibinfo
  {author} {\bibfnamefont {C.}~\bibnamefont {Moutafis}},\ }\bibfield  {title}
  {\bibinfo {title} {Skyrmionic interconnect device},\ }\href
  {https://doi.org/10.1103/PhysRevResearch.2.043312} {\bibfield  {journal}
  {\bibinfo  {journal} {Phys. Rev. Res.}\ }\textbf {\bibinfo {volume} {2}},\
  \bibinfo {pages} {043312} (\bibinfo {year} {2020})}\BibitemShut {NoStop}%
\bibitem [{\citenamefont {Zeng}\ \emph {et~al.}(2013)\citenamefont {Zeng},
  \citenamefont {Finocchio},\ and\ \citenamefont {Jiang}}]{zeng2013spin}%
  \BibitemOpen
  \bibfield  {author} {\bibinfo {author} {\bibfnamefont {Z.}~\bibnamefont
  {Zeng}}, \bibinfo {author} {\bibfnamefont {G.}~\bibnamefont {Finocchio}},\
  and\ \bibinfo {author} {\bibfnamefont {H.}~\bibnamefont {Jiang}},\ }\bibfield
   {title} {\bibinfo {title} {Spin transfer nano-oscillators},\ }\href
  {https://doi.org/10.1039/C2NR33407K} {\bibfield  {journal} {\bibinfo
  {journal} {Nanoscale}\ }\textbf {\bibinfo {volume} {5}},\ \bibinfo {pages}
  {2219} (\bibinfo {year} {2013})}\BibitemShut {NoStop}%
\bibitem [{\citenamefont {Finocchio}\ \emph {et~al.}(2021)\citenamefont
  {Finocchio}, \citenamefont {Tomasello}, \citenamefont {Fang}, \citenamefont
  {Giordano}, \citenamefont {Puliafito}, \citenamefont {Carpentieri},\ and\
  \citenamefont {Zeng}}]{finocchio2021perspectives}%
  \BibitemOpen
  \bibfield  {author} {\bibinfo {author} {\bibfnamefont {G.}~\bibnamefont
  {Finocchio}}, \bibinfo {author} {\bibfnamefont {R.}~\bibnamefont
  {Tomasello}}, \bibinfo {author} {\bibfnamefont {B.}~\bibnamefont {Fang}},
  \bibinfo {author} {\bibfnamefont {A.}~\bibnamefont {Giordano}}, \bibinfo
  {author} {\bibfnamefont {V.}~\bibnamefont {Puliafito}}, \bibinfo {author}
  {\bibfnamefont {M.}~\bibnamefont {Carpentieri}},\ and\ \bibinfo {author}
  {\bibfnamefont {Z.}~\bibnamefont {Zeng}},\ }\bibfield  {title} {\bibinfo
  {title} {Perspectives on spintronic diodes},\ }\href
  {https://doi.org/10.1063/5.0048947} {\bibfield  {journal} {\bibinfo
  {journal} {Appl. Phys. Lett.}\ }\textbf {\bibinfo {volume} {118}},\ \bibinfo
  {pages} {160502} (\bibinfo {year} {2021})}\BibitemShut {NoStop}%
\bibitem [{\citenamefont {Zhou}\ \emph {et~al.}(2015)\citenamefont {Zhou},
  \citenamefont {Iacocca}, \citenamefont {Awad}, \citenamefont {Dumas},
  \citenamefont {Zhang}, \citenamefont {Braun},\ and\ \citenamefont
  {{\AA}kerman}}]{zhou2015dynamically}%
  \BibitemOpen
  \bibfield  {author} {\bibinfo {author} {\bibfnamefont {Y.}~\bibnamefont
  {Zhou}}, \bibinfo {author} {\bibfnamefont {E.}~\bibnamefont {Iacocca}},
  \bibinfo {author} {\bibfnamefont {A.~A.}\ \bibnamefont {Awad}}, \bibinfo
  {author} {\bibfnamefont {R.~K.}\ \bibnamefont {Dumas}}, \bibinfo {author}
  {\bibfnamefont {F.~C.}\ \bibnamefont {Zhang}}, \bibinfo {author}
  {\bibfnamefont {H.~B.}\ \bibnamefont {Braun}},\ and\ \bibinfo {author}
  {\bibfnamefont {J.}~\bibnamefont {{\AA}kerman}},\ }\bibfield  {title}
  {\bibinfo {title} {Dynamically stabilized magnetic skyrmions},\ }\href
  {https://doi.org/10.1038/ncomms9193} {\bibfield  {journal} {\bibinfo
  {journal} {Nat. Commun.}\ }\textbf {\bibinfo {volume} {6}},\ \bibinfo {pages}
  {8193} (\bibinfo {year} {2015})}\BibitemShut {NoStop}%
\bibitem [{\citenamefont {Carpentieri}\ \emph {et~al.}(2015)\citenamefont
  {Carpentieri}, \citenamefont {Tomasello}, \citenamefont {Zivieri},\ and\
  \citenamefont {Finocchio}}]{carpentieri2015topological}%
  \BibitemOpen
  \bibfield  {author} {\bibinfo {author} {\bibfnamefont {M.}~\bibnamefont
  {Carpentieri}}, \bibinfo {author} {\bibfnamefont {R.}~\bibnamefont
  {Tomasello}}, \bibinfo {author} {\bibfnamefont {R.}~\bibnamefont {Zivieri}},\
  and\ \bibinfo {author} {\bibfnamefont {G.}~\bibnamefont {Finocchio}},\
  }\bibfield  {title} {\bibinfo {title} {Topological, non-topological and
  instanton droplets driven by spin-transfer torque in materials with
  perpendicular magnetic anisotropy and {{Dzyaloshinskii}}--{{Moriya
  Interaction}}},\ }\href {https://doi.org/10.1038/srep16184} {\bibfield
  {journal} {\bibinfo  {journal} {Sci Rep}\ }\textbf {\bibinfo {volume} {5}},\
  \bibinfo {pages} {16184} (\bibinfo {year} {2015})}\BibitemShut {NoStop}%
\bibitem [{\citenamefont {Zhang}\ \emph {et~al.}(2015)\citenamefont {Zhang},
  \citenamefont {Wang}, \citenamefont {Zheng}, \citenamefont {Zhu},
  \citenamefont {Liu}, \citenamefont {Chen}, \citenamefont {Jin}, \citenamefont
  {Liu}, \citenamefont {Jia},\ and\ \citenamefont {Xue}}]{zhang2015current}%
  \BibitemOpen
  \bibfield  {author} {\bibinfo {author} {\bibfnamefont {S.}~\bibnamefont
  {Zhang}}, \bibinfo {author} {\bibfnamefont {J.}~\bibnamefont {Wang}},
  \bibinfo {author} {\bibfnamefont {Q.}~\bibnamefont {Zheng}}, \bibinfo
  {author} {\bibfnamefont {Q.}~\bibnamefont {Zhu}}, \bibinfo {author}
  {\bibfnamefont {X.}~\bibnamefont {Liu}}, \bibinfo {author} {\bibfnamefont
  {S.}~\bibnamefont {Chen}}, \bibinfo {author} {\bibfnamefont {C.}~\bibnamefont
  {Jin}}, \bibinfo {author} {\bibfnamefont {Q.}~\bibnamefont {Liu}}, \bibinfo
  {author} {\bibfnamefont {C.}~\bibnamefont {Jia}},\ and\ \bibinfo {author}
  {\bibfnamefont {D.}~\bibnamefont {Xue}},\ }\bibfield  {title} {\bibinfo
  {title} {Current-induced magnetic skyrmions oscillator},\ }\href@noop {}
  {\bibfield  {journal} {\bibinfo  {journal} {New J. Phys.}\ }\textbf {\bibinfo
  {volume} {17}},\ \bibinfo {pages} {023061} (\bibinfo {year}
  {2015})}\BibitemShut {NoStop}%
\bibitem [{\citenamefont {Giordano}\ \emph {et~al.}(2016)\citenamefont
  {Giordano}, \citenamefont {Verba}, \citenamefont {Zivieri}, \citenamefont
  {Laudani}, \citenamefont {Puliafito}, \citenamefont {Gubbiotti},
  \citenamefont {Tomasello}, \citenamefont {Siracusano}, \citenamefont
  {Azzerboni}, \citenamefont {Carpentieri}, \citenamefont {Slavin},\ and\
  \citenamefont {Finocchio}}]{giordano2016spinhall}%
  \BibitemOpen
  \bibfield  {author} {\bibinfo {author} {\bibfnamefont {A.}~\bibnamefont
  {Giordano}}, \bibinfo {author} {\bibfnamefont {R.}~\bibnamefont {Verba}},
  \bibinfo {author} {\bibfnamefont {R.}~\bibnamefont {Zivieri}}, \bibinfo
  {author} {\bibfnamefont {A.}~\bibnamefont {Laudani}}, \bibinfo {author}
  {\bibfnamefont {V.}~\bibnamefont {Puliafito}}, \bibinfo {author}
  {\bibfnamefont {G.}~\bibnamefont {Gubbiotti}}, \bibinfo {author}
  {\bibfnamefont {R.}~\bibnamefont {Tomasello}}, \bibinfo {author}
  {\bibfnamefont {G.}~\bibnamefont {Siracusano}}, \bibinfo {author}
  {\bibfnamefont {B.}~\bibnamefont {Azzerboni}}, \bibinfo {author}
  {\bibfnamefont {M.}~\bibnamefont {Carpentieri}}, \bibinfo {author}
  {\bibfnamefont {A.}~\bibnamefont {Slavin}},\ and\ \bibinfo {author}
  {\bibfnamefont {G.}~\bibnamefont {Finocchio}},\ }\bibfield  {title} {\bibinfo
  {title} {Spin-{{Hall}} nano-oscillator with oblique magnetization and
  {{Dzyaloshinskii-Moriya}} interaction as generator of skyrmions and
  nonreciprocal spin-waves},\ }\href {https://doi.org/10.1038/srep36020}
  {\bibfield  {journal} {\bibinfo  {journal} {Sci Rep}\ }\textbf {\bibinfo
  {volume} {6}},\ \bibinfo {pages} {36020} (\bibinfo {year}
  {2016})}\BibitemShut {NoStop}%
\bibitem [{\citenamefont {Garcia-Sanchez}\ \emph {et~al.}(2016)\citenamefont
  {Garcia-Sanchez}, \citenamefont {Sampaio}, \citenamefont {Reyren},
  \citenamefont {Cros},\ and\ \citenamefont {Kim}}]{garcia2016skyrmion}%
  \BibitemOpen
  \bibfield  {author} {\bibinfo {author} {\bibfnamefont {F.}~\bibnamefont
  {Garcia-Sanchez}}, \bibinfo {author} {\bibfnamefont {J.}~\bibnamefont
  {Sampaio}}, \bibinfo {author} {\bibfnamefont {N.}~\bibnamefont {Reyren}},
  \bibinfo {author} {\bibfnamefont {V.}~\bibnamefont {Cros}},\ and\ \bibinfo
  {author} {\bibfnamefont {J.}~\bibnamefont {Kim}},\ }\bibfield  {title}
  {\bibinfo {title} {A skyrmion-based spin-torque nano-oscillator},\
  }\href@noop {} {\bibfield  {journal} {\bibinfo  {journal} {New J. Phys.}\
  }\textbf {\bibinfo {volume} {18}},\ \bibinfo {pages} {075011} (\bibinfo
  {year} {2016})}\BibitemShut {NoStop}%
\bibitem [{\citenamefont {Zhao}\ \emph
  {et~al.}(2024{\natexlab{b}})\citenamefont {Zhao}, \citenamefont {Chen},
  \citenamefont {Huang}, \citenamefont {Liu}, \citenamefont {Shen},
  \citenamefont {Liu}, \citenamefont {Zhao}, \citenamefont {Fang},
  \citenamefont {Yue}, \citenamefont {Zheng}, \citenamefont {Wang},
  \citenamefont {Bai}, \citenamefont {Shen}, \citenamefont {Zhou},
  \citenamefont {Wang}, \citenamefont {Liu}, \citenamefont {He}, \citenamefont
  {Wang}, \citenamefont {Zhang},\ and\ \citenamefont
  {Jiang}}]{zhao2024electrical}%
  \BibitemOpen
  \bibfield  {author} {\bibinfo {author} {\bibfnamefont {M.}~\bibnamefont
  {Zhao}}, \bibinfo {author} {\bibfnamefont {A.}~\bibnamefont {Chen}}, \bibinfo
  {author} {\bibfnamefont {P.-Y.}\ \bibnamefont {Huang}}, \bibinfo {author}
  {\bibfnamefont {C.}~\bibnamefont {Liu}}, \bibinfo {author} {\bibfnamefont
  {L.}~\bibnamefont {Shen}}, \bibinfo {author} {\bibfnamefont {J.}~\bibnamefont
  {Liu}}, \bibinfo {author} {\bibfnamefont {L.}~\bibnamefont {Zhao}}, \bibinfo
  {author} {\bibfnamefont {B.}~\bibnamefont {Fang}}, \bibinfo {author}
  {\bibfnamefont {W.-C.}\ \bibnamefont {Yue}}, \bibinfo {author} {\bibfnamefont
  {D.}~\bibnamefont {Zheng}}, \bibinfo {author} {\bibfnamefont
  {L.}~\bibnamefont {Wang}}, \bibinfo {author} {\bibfnamefont {H.}~\bibnamefont
  {Bai}}, \bibinfo {author} {\bibfnamefont {K.}~\bibnamefont {Shen}}, \bibinfo
  {author} {\bibfnamefont {Y.}~\bibnamefont {Zhou}}, \bibinfo {author}
  {\bibfnamefont {S.}~\bibnamefont {Wang}}, \bibinfo {author} {\bibfnamefont
  {E.}~\bibnamefont {Liu}}, \bibinfo {author} {\bibfnamefont {S.}~\bibnamefont
  {He}}, \bibinfo {author} {\bibfnamefont {Y.-L.}\ \bibnamefont {Wang}},
  \bibinfo {author} {\bibfnamefont {X.}~\bibnamefont {Zhang}},\ and\ \bibinfo
  {author} {\bibfnamefont {W.}~\bibnamefont {Jiang}},\ }\bibfield  {title}
  {\bibinfo {title} {Electrical detection of mobile skyrmions with 100\%
  tunneling magnetoresistance in a racetrack-like device},\ }\href
  {https://doi.org/10.1038/s41535-024-00655-1} {\bibfield  {journal} {\bibinfo
  {journal} {npj Quantum Mater.}\ }\textbf {\bibinfo {volume} {9}},\ \bibinfo
  {pages} {50} (\bibinfo {year} {2024}{\natexlab{b}})}\BibitemShut {NoStop}%
\bibitem [{\citenamefont {Wang}\ \emph
  {et~al.}(2020{\natexlab{b}})\citenamefont {Wang}, \citenamefont {Guo},
  \citenamefont {Zhou}, \citenamefont {Zhao}, \citenamefont {Xu}, \citenamefont
  {Tomasello}, \citenamefont {Bai}, \citenamefont {Dong}, \citenamefont {Je},
  \citenamefont {Chao}, \citenamefont {Han}, \citenamefont {Lee}, \citenamefont
  {Lee}, \citenamefont {Yao}, \citenamefont {Han}, \citenamefont {Song},
  \citenamefont {Wu}, \citenamefont {Carpentieri}, \citenamefont {Finocchio},
  \citenamefont {Im}, \citenamefont {Lin},\ and\ \citenamefont
  {Jiang}}]{wang2020thermal}%
  \BibitemOpen
  \bibfield  {author} {\bibinfo {author} {\bibfnamefont {Z.}~\bibnamefont
  {Wang}}, \bibinfo {author} {\bibfnamefont {M.}~\bibnamefont {Guo}}, \bibinfo
  {author} {\bibfnamefont {H.-A.}\ \bibnamefont {Zhou}}, \bibinfo {author}
  {\bibfnamefont {L.}~\bibnamefont {Zhao}}, \bibinfo {author} {\bibfnamefont
  {T.}~\bibnamefont {Xu}}, \bibinfo {author} {\bibfnamefont {R.}~\bibnamefont
  {Tomasello}}, \bibinfo {author} {\bibfnamefont {H.}~\bibnamefont {Bai}},
  \bibinfo {author} {\bibfnamefont {Y.}~\bibnamefont {Dong}}, \bibinfo {author}
  {\bibfnamefont {S.-G.}\ \bibnamefont {Je}}, \bibinfo {author} {\bibfnamefont
  {W.}~\bibnamefont {Chao}}, \bibinfo {author} {\bibfnamefont {H.-S.}\
  \bibnamefont {Han}}, \bibinfo {author} {\bibfnamefont {S.}~\bibnamefont
  {Lee}}, \bibinfo {author} {\bibfnamefont {K.-S.}\ \bibnamefont {Lee}},
  \bibinfo {author} {\bibfnamefont {Y.}~\bibnamefont {Yao}}, \bibinfo {author}
  {\bibfnamefont {W.}~\bibnamefont {Han}}, \bibinfo {author} {\bibfnamefont
  {C.}~\bibnamefont {Song}}, \bibinfo {author} {\bibfnamefont {H.}~\bibnamefont
  {Wu}}, \bibinfo {author} {\bibfnamefont {M.}~\bibnamefont {Carpentieri}},
  \bibinfo {author} {\bibfnamefont {G.}~\bibnamefont {Finocchio}}, \bibinfo
  {author} {\bibfnamefont {M.-Y.}\ \bibnamefont {Im}}, \bibinfo {author}
  {\bibfnamefont {S.-Z.}\ \bibnamefont {Lin}},\ and\ \bibinfo {author}
  {\bibfnamefont {W.}~\bibnamefont {Jiang}},\ }\bibfield  {title} {\bibinfo
  {title} {Thermal generation, manipulation and thermoelectric detection of
  skyrmions},\ }\href {https://doi.org/10.1038/s41928-020-00489-2} {\bibfield
  {journal} {\bibinfo  {journal} {Nat. Electron.}\ }\textbf {\bibinfo {volume}
  {3}},\ \bibinfo {pages} {672} (\bibinfo {year}
  {2020}{\natexlab{b}})}\BibitemShut {NoStop}%
\bibitem [{\citenamefont {Fern{\'a}ndez~Scarioni}\ \emph
  {et~al.}(2021)\citenamefont {Fern{\'a}ndez~Scarioni}, \citenamefont {Barton},
  \citenamefont {{Corte-Le{\'o}n}}, \citenamefont {Sievers}, \citenamefont
  {Hu}, \citenamefont {Ajejas}, \citenamefont {Legrand}, \citenamefont
  {Reyren}, \citenamefont {Cros}, \citenamefont {Kazakova},\ and\ \citenamefont
  {Schumacher}}]{fernandezscarioni2021thermoelectric}%
  \BibitemOpen
  \bibfield  {author} {\bibinfo {author} {\bibfnamefont {A.}~\bibnamefont
  {Fern{\'a}ndez~Scarioni}}, \bibinfo {author} {\bibfnamefont {C.}~\bibnamefont
  {Barton}}, \bibinfo {author} {\bibfnamefont {H.}~\bibnamefont
  {{Corte-Le{\'o}n}}}, \bibinfo {author} {\bibfnamefont {S.}~\bibnamefont
  {Sievers}}, \bibinfo {author} {\bibfnamefont {X.}~\bibnamefont {Hu}},
  \bibinfo {author} {\bibfnamefont {F.}~\bibnamefont {Ajejas}}, \bibinfo
  {author} {\bibfnamefont {W.}~\bibnamefont {Legrand}}, \bibinfo {author}
  {\bibfnamefont {N.}~\bibnamefont {Reyren}}, \bibinfo {author} {\bibfnamefont
  {V.}~\bibnamefont {Cros}}, \bibinfo {author} {\bibfnamefont {O.}~\bibnamefont
  {Kazakova}},\ and\ \bibinfo {author} {\bibfnamefont {H.~W.}\ \bibnamefont
  {Schumacher}},\ }\bibfield  {title} {\bibinfo {title} {Thermoelectric
  {{Signature}} of {{Individual Skyrmions}}},\ }\href
  {https://doi.org/10.1103/PhysRevLett.126.077202} {\bibfield  {journal}
  {\bibinfo  {journal} {Phys. Rev. Lett.}\ }\textbf {\bibinfo {volume} {126}},\
  \bibinfo {pages} {077202} (\bibinfo {year} {2021})}\BibitemShut {NoStop}%
\bibitem [{\citenamefont {Yu}\ \emph {et~al.}(2017)\citenamefont {Yu},
  \citenamefont {Upadhyaya}, \citenamefont {Shao}, \citenamefont {Wu},
  \citenamefont {Yin}, \citenamefont {Li}, \citenamefont {He}, \citenamefont
  {Jiang}, \citenamefont {Han}, \citenamefont {Amiri},\ and\ \citenamefont
  {Wang}}]{yu2017roomtemperature}%
  \BibitemOpen
  \bibfield  {author} {\bibinfo {author} {\bibfnamefont {G.}~\bibnamefont
  {Yu}}, \bibinfo {author} {\bibfnamefont {P.}~\bibnamefont {Upadhyaya}},
  \bibinfo {author} {\bibfnamefont {Q.}~\bibnamefont {Shao}}, \bibinfo {author}
  {\bibfnamefont {H.}~\bibnamefont {Wu}}, \bibinfo {author} {\bibfnamefont
  {G.}~\bibnamefont {Yin}}, \bibinfo {author} {\bibfnamefont {X.}~\bibnamefont
  {Li}}, \bibinfo {author} {\bibfnamefont {C.}~\bibnamefont {He}}, \bibinfo
  {author} {\bibfnamefont {W.}~\bibnamefont {Jiang}}, \bibinfo {author}
  {\bibfnamefont {X.}~\bibnamefont {Han}}, \bibinfo {author} {\bibfnamefont
  {P.~K.}\ \bibnamefont {Amiri}},\ and\ \bibinfo {author} {\bibfnamefont
  {K.~L.}\ \bibnamefont {Wang}},\ }\bibfield  {title} {\bibinfo {title}
  {Room-{{Temperature Skyrmion Shift Device}} for {{Memory Application}}},\
  }\href {https://doi.org/10.1021/acs.nanolett.6b04010} {\bibfield  {journal}
  {\bibinfo  {journal} {Nano Lett.}\ }\textbf {\bibinfo {volume} {17}},\
  \bibinfo {pages} {261} (\bibinfo {year} {2017})}\BibitemShut {NoStop}%
\bibitem [{\citenamefont {Yan}\ \emph {et~al.}(2021)\citenamefont {Yan},
  \citenamefont {Liu}, \citenamefont {Guang}, \citenamefont {Yue},
  \citenamefont {Feng}, \citenamefont {Lake}, \citenamefont {Yu},\ and\
  \citenamefont {Han}}]{yan2021skyrmionbased}%
  \BibitemOpen
  \bibfield  {author} {\bibinfo {author} {\bibfnamefont {Z.}~\bibnamefont
  {Yan}}, \bibinfo {author} {\bibfnamefont {Y.}~\bibnamefont {Liu}}, \bibinfo
  {author} {\bibfnamefont {Y.}~\bibnamefont {Guang}}, \bibinfo {author}
  {\bibfnamefont {K.}~\bibnamefont {Yue}}, \bibinfo {author} {\bibfnamefont
  {J.}~\bibnamefont {Feng}}, \bibinfo {author} {\bibfnamefont {R.}~\bibnamefont
  {Lake}}, \bibinfo {author} {\bibfnamefont {G.}~\bibnamefont {Yu}},\ and\
  \bibinfo {author} {\bibfnamefont {X.}~\bibnamefont {Han}},\ }\bibfield
  {title} {\bibinfo {title} {Skyrmion-{{Based Programmable Logic Device}} with
  {{Complete Boolean Logic Functions}}},\ }\href
  {https://doi.org/10.1103/PhysRevApplied.15.064004} {\bibfield  {journal}
  {\bibinfo  {journal} {Phys. Rev. Appl.}\ }\textbf {\bibinfo {volume} {15}},\
  \bibinfo {pages} {064004} (\bibinfo {year} {2021})}\BibitemShut {NoStop}%
\bibitem [{\citenamefont {He}\ \emph {et~al.}(2023)\citenamefont {He},
  \citenamefont {Tomasello}, \citenamefont {Luo}, \citenamefont {Zhang},
  \citenamefont {Nie}, \citenamefont {Carpentieri}, \citenamefont {Han},
  \citenamefont {Finocchio},\ and\ \citenamefont {Yu}}]{he2023allelectrical}%
  \BibitemOpen
  \bibfield  {author} {\bibinfo {author} {\bibfnamefont {B.}~\bibnamefont
  {He}}, \bibinfo {author} {\bibfnamefont {R.}~\bibnamefont {Tomasello}},
  \bibinfo {author} {\bibfnamefont {X.}~\bibnamefont {Luo}}, \bibinfo {author}
  {\bibfnamefont {R.}~\bibnamefont {Zhang}}, \bibinfo {author} {\bibfnamefont
  {Z.}~\bibnamefont {Nie}}, \bibinfo {author} {\bibfnamefont {M.}~\bibnamefont
  {Carpentieri}}, \bibinfo {author} {\bibfnamefont {X.}~\bibnamefont {Han}},
  \bibinfo {author} {\bibfnamefont {G.}~\bibnamefont {Finocchio}},\ and\
  \bibinfo {author} {\bibfnamefont {G.}~\bibnamefont {Yu}},\ }\bibfield
  {title} {\bibinfo {title} {All-{{Electrical}} 9-{{Bit Skyrmion-Based
  Racetrack Memory Designed}} with {{Laser Irradiation}}},\ }\href
  {https://doi.org/10.1021/acs.nanolett.3c02978} {\bibfield  {journal}
  {\bibinfo  {journal} {Nano Lett.}\ }\textbf {\bibinfo {volume} {23}},\
  \bibinfo {pages} {9482} (\bibinfo {year} {2023})}\BibitemShut {NoStop}%
\end{thebibliography}
\end{document}